\documentclass[11pt,twoside,a4paper]{cernyrep}

\usepackage{graphicx}%
\usepackage{multirow}%
\usepackage{amsmath,amssymb,amsfonts}%
\usepackage{amsthm}%
\usepackage{mathrsfs}%
\usepackage[title]{appendix}%
\usepackage[table,xcdraw]{xcolor}
\usepackage{textcomp}%
\usepackage{manyfoot}%
\usepackage{booktabs}%
\usepackage{algorithm}%
\usepackage{algorithmicx}%
\usepackage{algpseudocode}%
\usepackage{listings}%

\usepackage{subcaption}  
\usepackage{wasysym}

\usepackage{subfiles}%
\usepackage{textgreek}

\usepackage{wrapfig}

\usepackage{comment}
\usepackage{siunitx}
\usepackage{tikz} 
\usepackage{physics}

\usepackage{tabularx}
\usepackage{makecell}

\usepackage{todonotes}

\usepackage{qrcode}

\usepackage{hyperref}
\hypersetup{%
        colorlinks,
        breaklinks=true,
        plainpages=false,%
        citecolor=blue,
        linkcolor=blue,
        urlcolor=blue,
        bookmarksopen=true,%
        bookmarksnumbered=false,%
        bookmarksdepth=5%
}

\counterwithin{section}{chapter}

\begin{document}

\title{Future Circular Collider}

\pagenumbering{roman}
\setcounter{page}{1}

\newcommand{\main}{.}
\def\biblio{}

\begin{titlepage}
        \begin{center}
		\textbf{\Large Future Circular Collider}\\
        \vspace{0.5cm}
        \textbf{\LARGE Feasibility Study Report}\\
        \vspace{8 cm}
        \textbf{\Huge Volume 3}\\
        \vspace{1 cm}
        \textbf{\Huge Civil Engineering, Implementation}\\
        ~\par
        \textbf{\Huge and Sustainability}\\
        \vspace{3 cm}
        \textbf{March 31, 2025}\\
        \vspace{\fill}
Submitted to the European Physics Journal ST, a joint publication of EDP Sciences,\\Springer Science+Business Media, and the Società Italiana di Fisica.
        
        \end{center}

\end{titlepage}
\section*{Note from the Editors}

\noindent One of the recommendations of the 2020 update of the European Strategy for Particle Physics was that “Europe, together with its international partners, should investigate the technical and financial feasibility of a future hadron collider at CERN with a centre-of-mass energy of at least 100 TeV and with an electron-positron Higgs and electroweak factory as a possible first stage.”

\noindent  In June 2021, the CERN Council launched the FCC Feasibility Study to be completed by 2025, in time for the next update of the European Strategy for Particle Physics. The study results are made publicly available through this FCC Feasibility Study Report, as input to the European Particle Physics Strategy update process, initiated by the CERN Council in March 2024.The studies presented in this FCC Feasibility Study Report do not imply any commitment by the CERN Member or Associate Member States to build the Future Circular Collider.

\noindent  This report and the assumptions contained in it do not prejudge further territorial feasibility analysis by the Host States, France and Switzerland, as well as the outcome of their respective public debate and concertation processes, and future decisions of their relevant authorities.

\newpage

\section*{Acknowledgements}
We would like to thank the \textbf{International Steering Committee members:}\\

\begin{table}[!ht]
	\centering
	\begin{tabular}{l} 
		\toprule
		F.~Gianotti (Chair), CERN \\ 
		R.~Bello, CERN \\ 
		P.~Chomaz, CEA, France  \\ 
                 M.~Cobal, INFN and University of Udine, Italy \\
		B.~Heinemann, DESY, Germany \\ 
		T.~Koseki, KEK, Japan \\ 
		M.~Lamont, CERN \\ 
		L.~Merminga, FNAL, United States \\ 
		J.~Mnich, CERN \\ 
		M.~Seidel, PSI and EPFL, Switzerland \\
		C.~Warakaulle, CERN \\
		\bottomrule
	\end{tabular}
\end{table}

\noindent and the \textbf{Scientific Advisory Committee members:}\\
\vspace*{-2mm}

\begin{table}[!ht]
	\centering
	\vspace*{-3mm}
	\begin{tabular}{l}
		\toprule
		A.~Parker (Chair), Cambridge University, UK \\ 
		R.~Bartolini, DESY, Germany \\  
		A.~Chabert, SFTRF, France \\ 
		H.~Ehrbar, Heinz Ehrbar Partners LLC, Switzerland \\ 
		B.~Gavela Legazpi, UAM Madrid, Spain \\
		G.~Hiller, TU Dortmund, Germany \\
        S.~Krishnagopal, FNAL, U.S. \\		
        P.~Križan, University of Ljubljana, Slovenia\\
        P.~Lebrun, ESI, France\\
        P.~McIntosh, STFC, ASTeC,  UKRI, UK\\ 
        M.~Minty, BNL, U.S. \\  
        R.~Tenchini, INFN Sezione di Pisa, Italy  \\  
\bottomrule 
	\end{tabular}
\end{table}

\noindent for their continued guidance and careful reviewing that helped to complete this report successfully.\\

\newpage

\begin{wrapfigure}[4]{l}[0cm]{2cm}
  \begin{center}
  \vspace{-0.5cm}
    \includegraphics[width=2cm]{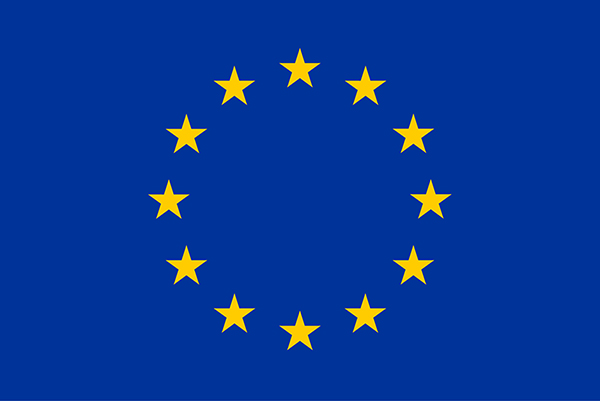}
  \end{center}
\end{wrapfigure}
\noindent The research carried out by the international FCC collaboration hosted by CERN, which led to this publication, has received funding from the European Union's Horizon 2020 research and innovation programme under the grant numbers 951754 (FCCIS), 654305 (EuroCirCol), 764879 (EASITrain), 730871 (ARIES), 777563 (RI-Paths) and from FP7 under grant number 312453 (EuCARD-2). 
\vspace{1em}

\begin{wrapfigure}[4]{l}[0cm]{2cm}
  \begin{center}
  \vspace{-0.85cm}
    \includegraphics[width=2cm]{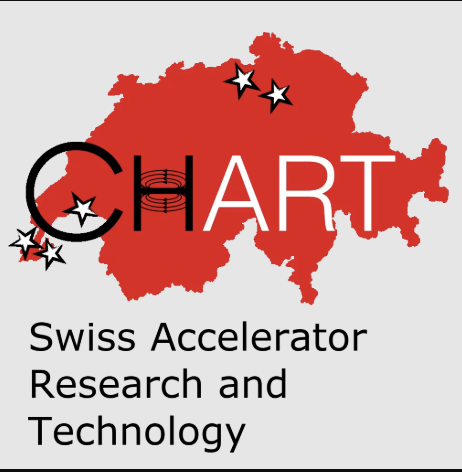}
  \end{center}
\end{wrapfigure}

\noindent This work has also benefited from the support of CHART (Swiss Accelerator Research and Technology, founded in 2016 as an umbrella collaboration for accelerator research and technology activities. Present partners in CHART are CERN, PSI, EPFL, ETH-Zurich and the University of Geneva.

\vspace*{\fill}
\noindent\textbf{Trademark notice:} All trademarks appearing in this report are acknowledged as such.\\

\newpage
~~
\\
This report was edited with the Overleaf.com collaborative writing and publishing system. Typesetting and final print preparation were performed using pdf\TeX 3.14159265-2.6-1.40.17\\
\\
\noindent Copyright CERN for the benefit of the FCC collaboration 2024\\
Creative Commons Attribution 4.0\\
\\
Knowledge transfer is an integral part of CERN's mission.\\
CERN publishes this volume Open Access under the Creative Commons Attribution 4.0 licence\\ (\mbox{\url{http://creativecommons.org/licenses/by/4.0/}}) in order to permit its wide dissemination and use.
The submission of a contribution to the CERN document server shall be deemed to constitute the contributor's agreement to this copyright and licence statement. Contributors are requested to obtain any clearances that may be necessary for this purpose.

\bigskip
\noindent This volume is indexed in: CERN Document Server (CDS):

\smallskip
\noindent CERN-FCC-ACC-2025-0003\\
\noindent DOI 10.17181/CERN.I26X.V4VF\\
\noindent \url{http://cds.cern.ch/record/2928194}\\

\bigskip
\noindent This report edition should be cited as:

\smallskip
\noindent Future Circular Collider Feasibility Study Report Volume 3: Civil Engineering, Implementation and Sustainability, preprint edition edited by 
\mbox{M.~Benedikt et al.}, CERN accelerator reports,\\
\mbox{CERN-FCC-ACC-2025-0003},\mbox{DOI 10.17181/CERN.I26X.V4VF}, Geneva, 2025.\\
\noindent Available online: \url{https://cds.cern.ch/record/2928194}
\clearpage


\begingroup
\raggedright %

\section*{List of Editors at 31 March 2025}
M.~Benedikt$^{{1}}$~(Study Leader),
F.~Zimmermann$^{{1}}$~(Deputy Study Leader),
B.~Auchmann$^{{1, 2}}$,
W.~Bartmann$^{{1}}$,
J.P.~Burnet$^{{1}}$,
C.~Carli$^{{1}}$,
A.~Chanc{\'e}$^{{3}}$,
P.~Craievich$^{{2}}$,
M.~Giovannozzi$^{{1}}$,
C.~Grojean$^{{4, 5}}$,
J.~Gutleber$^{{1}}$,
K.~Hanke$^{{1}}$,
A.~Henriques$^{{1}}$,
P.~Janot$^{{1}}$,
C.~Louren{\c c}o$^{{1}}$,
M.~Mangano$^{{1}}$,
T.~Otto$^{{1}}$,
J.~Poole$^{{1}}$,
S.~Rajagopalan$^{{6}}$,
T.~Raubenheimer$^{{7}}$,
E.~Todesco$^{{1}}$,
L.~Ulrici$^{{1}}$,
T.~Watson$^{{1}}$,
G.~Wilkinson$^{{1,8}}$.
\section*{List of Contributors at 31 March 2025}
A.~Abada$^{{9, 10, 11}}$,
M.~Abbrescia$^{{12, 13}}$,
H.~Abdolmaleki$^{{14, 15}}$,
S.H.~Abidi$^{{6}}$,
A.~Abramov$^{{1}}$,
C.~Adam$^{{9, 16, 17}}$,
M.~Ady$^{{1}}$,
P.R.~Ad{\u z}i{\'c}$^{{18}}$,
I.~Agapov$^{{4}}$,
D.~Aguglia$^{{1}}$
% ...
\begin{itemize}
\item[${}^\dag$] deceased 
\item[$^{1}$] Switzerland - CERN, European Organization for Nuclear Research
\item[$^{2}$] Switzerland - PSI, Paul Scherrer Institute
\item[$^{3}$] France - CEA/Irfu, Commissariat {\`a} l'Energie Atomique et aux Energies Alternatives, Institut de recherche sur les lois fondamentales de l'Univers
\item[$^{4}$] Germany - DESY, Deutsches Elektronen-Synchrotron 
\item[$^{5}$] Germany - Humboldt-Universit\"at zu Berlin
\item[$^{6}$] United States - BNL, Brookhaven National Laboratory
\item[$^{7}$] United States - SLAC National Accelerator Laboratory
\item[$^{8}$] United Kingdom - University of Oxford
\item[$^{9}$] France - CNRS/IN2P3, Centre National de la Recherche Scientifique, Institut National de Physique Nucl{\'e}aire et de Physique des Particules
\item[$^{10}$] France - IJCLab, Laboratoire de Physique des 2 Infinis Ir{\`e}ne Joliot Curie
\item[$^{11}$] France - Universit{\'e} Paris-Saclay et Universit{\'e} Paris-Cit{\'e}
\item[$^{12}$] Italy - INFN, Istituto Nazionale di Fisica Nucleare, Sezione di Bari
\item[$^{13}$] Italy - Universit{\`a} di Bari
\item[$^{14}$] Iran - IPM, Institute for Research in Fundamental Science
\item[$^{15}$] Iran - Malayer University
\item[$^{16}$] France - LAPP, Laboratoire d'Annecy de Physique des Particules
\item[$^{17}$] France - Universit{\'e} Savoie Mont Blanc
\item[$^{18}$] Serbia - University of Belgrade
\item[$^{19}$] Spain - ICMAB/CISC, Institut de Ci{\`e}ncia de Materials de Barcelona, Consejo Superior de Investigaciones Cient{\'\i}ificas
\item[$^{20}$] Italy - INFN, Istituto Nazionale di Fisica Nucleare, Sezione di Roma Tor Vergata
\item[$^{21}$] Italy - Universit{\`a} Roma Tor Vergata
\item[$^{22}$] T{\"u}rkiye - Yozgat Bozok {\"U}niversitesi
\item[$^{23}$] United States - Texas Tech University
\item[$^{24}$] T{\"u}rkiye - TOBB ETU, TOBB Ekonomi ve Teknoloji {\"U}niversitesi
\item[$^{25}$] Italy - INFN, Istituto Nazionale di Fisica Nucleare, Sezione di Bologna
\item[$^{26}$] Italy - Universit{\`a} di Bologna
\item[$^{27}$] Spain - CIEMAT, Centro de Investigaciones Energ{\'e}ticas, Medioambientales y Tecnol{\'o}gicas
\item[$^{28}$] Saudi Arabia - KACST, King Abdulaziz City for Science and Technology
\item[$^{29}$] Italy - Universit{\`a} Roma Tre
\item[$^{30}$] Italy - INFN, Istituto Nazionale di Fisica Nucleare, Sezione di Milano-Bicocca
\item[$^{31}$] Italy - Universit{\`a} di Milano-Bicocca
\item[$^{32}$] United Kingdom - University of Cambridge
\item[$^{33}$] T{\"u}rkiye - {\.{I}}stanbul {\"U}niversitesi
\item[$^{34}$] T{\"u}rkiye - Eski{\c s}ehir Teknik {\"U}niversitesi
\item[$^{35}$] Italy - INFN, Istituto Nazionale di Fisica Nucleare, Sezione di Napoli
\item[$^{36}$] Italy - Universit{\`a} di Napoli Federico II
\item[$^{37}$] United States - FNAL, Fermi National Accelerator Laboratory
\item[$^{38}$] Iran - University of Tehran
\item[$^{39}$] Iran- FUM, Ferdowsi University of Mashhad
\item[$^{40}$] Italy - INFN, Istituto Nazionale di Fisica Nucleare, Sezione di Pisa
\item[$^{41}$] United States - University of Texas Austin
\item[$^{42}$] France - IPHC, Institut Pluridisciplinaire Hubert Curien
\item[$^{43}$] France - Universit{\'e} de Strasbourg
\item[$^{44}$] Italy - INFN, Istituto Nazionale di Fisica Nucleare, Sezione di Milano
\item[$^{45}$] Italy - Universit{\`a} di Milano
\item[$^{46}$] Switzerland - PIBG, P{\^o}le Invert{\'e}br{\'e}s du Basin Genevois
\item[$^{47}$] Brazil - UFRN, Universidade Federal do Rio Grande do Norte
\item[$^{48}$] Italy - INFN, Istituto Nazionale di Fisica Nucleare, Laboratori Nazionali di Frascati
\item[$^{49}$] Switzerland - UNIBAS, University of Basel
\item[$^{50}$] Italy - Politecnico di Bari
\item[$^{51}$] Portugal - LIP, Laborat{\'o}rio de Instrumenta{\c c}{\~a}o e F{\'\i}sica Experimental de Part{\'\i}culas
\item[$^{52}$] United Kingdom - University of Manchester
\item[$^{53}$] United Kingdom - CI, Cockcroft Institute
\item[$^{54}$] United States - Brandeis University
\item[$^{55}$] Armenia - A. Alikhanyan National Laboratory
\item[$^{56}$] France - IP2I, Institut de Physique des 2 Infinis de Lyon
\item[$^{57}$] France - Universit{\'e} Claude Bernard Lyon 1
\item[$^{58}$] T{\"u}rkiye - Ankara {\"U}niversitesi
\item[$^{59}$] Sweden - Lund University
\item[$^{60}$] Italy - INFN, Istituto Nazionale di Fisica Nucleare, Sezione di Torino
\item[$^{61}$] United Arab Emirates - New York University Abu Dhabi
\item[$^{62}$] United States - Stony Brook University
\item[$^{63}$] Italy - INFN, Istituto Nazionale di Fisica Nucleare, Sezione di Perugia
\item[$^{64}$] Italy - Universit{\`a} di Perugia
\item[$^{65}$] Germany - Technische Universit{\"a}t M{\"u}nchen
\item[$^{66}$] T{\"u}rkiye - Hatay Mustafa Kemal {\"U}niversitesi
\item[$^{67}$] T{\"u}rkiye - Do{\u g}u{\c s} {\"U}niversitesi
\item[$^{68}$] Italy - INFN, Istituto Nazionale di Fisica Nucleare, Sezione di Padova
\item[$^{69}$] People's Republic of China - Harbin Institute of Technology
\item[$^{70}$] United States - University of Wisconsin-Madison
\item[$^{71}$] United Kingdom - RAL, Rutherford Appleton Laboratory, Science and Technology Facilities Council
\item[$^{72}$] India - IMSc, Institute of Mathematical Sciences, Chennai
\item[$^{73}$] Switzerland - Universit{\"a}t Z{\"u}rich 
\item[$^{74}$] United States - University of New Mexico
\item[$^{75}$] France - CPPM, Centre de Physique des Particules de Marseille
\item[$^{76}$] France - Aix-Marseille Universit{\'e}
\item[$^{77}$] Italy - Universit{\`a} di Pisa
\item[$^{78}$] Hungary - HUN-REN Wigner Research Centre for Physics
\item[$^{79}$] United States - Catholic University of America
\item[$^{80}$] United States - Duke University
\item[$^{81}$] France - LLR, Laboratoire Leprince-Ringuet
\item[$^{82}$] France - {\'E}cole Polytechnique, Institut Polytechnique de Paris
\item[$^{83}$] Canada - TRIUMF, Canada's National Laboratory for Particle and Nuclear Physics
\item[$^{84}$] Canada - Simon Fraser University
\item[$^{85}$] Italy - FEEM, Fondazione Ente Nazionale Idrocarburi (ENI) Enrico Mattei
\item[$^{86}$] France - BRGM, Bureau de Recherches G{\'e}ologiques et Mini{\`e}res
\item[$^{87}$] T{\"u}rkiye - I{\c s}{\i}k {\"U}niversitesi
\item[$^{88}$] T{\"u}rkiye - {\.{I}}stanbul Teknik {\"U}niversitesi
\item[$^{89}$] Switzerland - UNIBE, University of Bern
\item[$^{90}$] Italy - Universit{\`a} di Roma la Sapienza
\item[$^{91}$] Italy - CNR-SPIN, Consiglio Nazionale delle Ricerche
\item[$^{92}$] France - Expert naturaliste et entomologiste
\item[$^{93}$] United States - ORNL, Oak Ridge National Laboratory
\item[$^{94}$] Austria - HEPHY, Institut f{\"u}r Hochenergiephysik
\item[$^{95}$] Switzerland - Geos, Bureau d'ing{\'e}nieurs conseils en g{\'e}otechnique, g{\'e}nie civil, hydraulique et environnement
\item[$^{96}$] France - APC, Laboratoire AstroParticule et Cosmologie
\item[$^{97}$] France - Universit{\'e} Paris Cit{\'e}
\item[$^{98}$] Austria - TUWIEN, Technische Universit{\"a}t Wien
\item[$^{99}$] Switzerland - HEPIA, Haute {\'E}cole du Paysage, d'Ing{\'e}nierie et d'Architecture de Gen{\`e}ve
\item[$^{100}$] Switzerland - HES-SO University of Applied Sciences and Arts Western Switzerland
\item[$^{101}$] Switzerland - UNIGE, Universit{\'e} de Gen{\`e}ve
\item[$^{102}$] France - SETEC ALS, Soci{\'e}t{\'e} d'ing{\'e}nierie conseil en infrastructures de transport, g{\'e}nie civil et environnement
\item[$^{103}$] United States - East Texas A\&M University
\item[$^{104}$] Switzerland - Amberg Engineering Ltd
\item[$^{105}$] Switzerland - ECOTEC Environnement SA, Bureau d'{\'e}tudes et de conseil en environnement
\item[$^{106}$] United States - Southern Methodist University
\item[$^{107}$] Poland - IFJ PAN, Institute of Nuclear Physics, Polish Academy of Sciences
\item[$^{108}$] France - Cerema, {\'e}tablissement public pour l'{\'e}laboration, le d{\'e}ploiement et l'{\'e}valuation de politiques publiques d'am{\'e}nagement et de transport
\item[$^{109}$] United Kingdom - Rendel Ltd, Engineering design consultancy firm
\item[$^{110}$] Italy - INFN, Istituto Nazionale di Fisica Nucleare, Sezione di Roma Tre
\item[$^{111}$] T{\"u}rkiye - {\.{I}}stanbul Beykent {\"U}niversitesi
\item[$^{112}$] United States - University of Iowa
\item[$^{113}$] India - Indian Institute of Technology Kanpur
\item[$^{114}$] Switzerland - EPFL, {\'E}cole Polytechnique F{\'e}d{\'e}rale de Lausanne
\item[$^{115}$] Germany - Universit{\"a}t Hamburg, Fakult{\"a}t f{\"u}r Mathematik, Informatik und Naturwissenschaften
\item[$^{116}$] Belgium - VUB, Vrije Universiteit Brussel 
\item[$^{117}$] France - LPNHE, Laboratoire de Physique Nucl{\'e}aire et de Hautes {\'E}nergies
\item[$^{118}$] Italy - Scuola Superiore Meridionale
\item[$^{119}$] Affiliated with an institute formerly covered by a cooperation agreement with CERN
\item[$^{120}$] Canada - University of Saskatchewan and the Canadian Light Source
\item[$^{121}$] United Kingdom - University of Bristol
\item[$^{122}$] United States - Northwestern University
\item[$^{123}$] United States - University of Florida
\item[$^{124}$] Canada - York University
\item[$^{125}$] Italy - INFN, Istituto Nazionale di Fisica Nucleare, Sezione di Cagliari
\item[$^{126}$] Italy - Universit{\`a} di Cagliari
\item[$^{127}$] Italy - INFN, Istituto Nazionale di Fisica Nucleare, Sezione di Pavia
\item[$^{128}$] Canada - Queen's University
\item[$^{129}$] Portugal - CFTP-IST, Centro de F{\'\i}sica T{\'e}orica de Part{\'\i}culas, Instituto Superior Tecnico, Universidade de Lisboa
\item[$^{130}$] Sweden - Uppsala University
\item[$^{131}$] Germany - MPP, Max-Planck-Institut f{\"u}r Physik Garching
\item[$^{132}$] Germany - KIT, Karlsruher Institut f{\"u}r Technologie
\item[$^{133}$] Ukraine - NSC KIPT, National Science Center Kharkiv Institute of Physics and Technology
\item[$^{134}$] Italy - Universit{\`a} degli Studi dell'Insubria
\item[$^{135}$] Germany - Albert-Ludwigs-Universit{\"a}t Freiburg
\item[$^{136}$] United Kingdom - JAI, John Adams Institute for Accelerator Science, University of Oxford
\item[$^{137}$] France - CNRS/INP, Centre National de la Recherche Scientifique, Institut de Physique
\item[$^{138}$] France - LPTHE, Laboratoire de Physique Th{\'e}orique et Hautes Energies
\item[$^{139}$] France - Sorbonne Universit{\'e}
\item[$^{140}$] Italy - INFN, Istituto Nazionale di Fisica Nucleare, Sezione di Genova
\item[$^{141}$] United States - LBNL, Lawrence Berkeley National Laboratory
\item[$^{142}$] Italy - IIT, Instituto Italiano di Tecnologia
\item[$^{143}$] T{\"u}rkiye - G{\"u}m{\"u}{\c s}hane {\"U}niversitesi
\item[$^{144}$] Switzerland - WSP Ing{\'e}nieurs Conseils SA
\item[$^{145}$] Mexico - UADY, Autonomous University of Yucatan
\item[$^{146}$] Italy - INFN, Istituto Nazionale di Fisica Nucleare, Sezione di Ferrara
\item[$^{147}$] Italy - Universit{\`a} di Ferrara
\item[$^{148}$] France - MARCELEON, Cabinet d'ing{\'e}nierie juridique et fonci{\`e}re
\item[$^{149}$] Italy - CSIL (Economic Research Institute)
\item[$^{150}$] Switzerland - ETHZ, Swiss Federal Institute of Technology Zurich
\item[$^{151}$] Spain - UAH, Universidad de Alcal\'a Madrid
\item[$^{152}$] France - CIA, Conseil Ing{\'e}nierie Acoustique
\item[$^{153}$] France - Evinerude, Bureau d'{\'e}tudes environnementales
\item[$^{154}$] France - LPCA, Laboratoire de Physique de Clermont Auvergne
\item[$^{155}$] France - Universit{\'e} Clermont Auvergne
\item[$^{156}$] Australia - ANSTO, Australian Synchrotron
\item[$^{157}$] India - Brahmananda Keshab Chandra College
\item[$^{158}$] United States - ANL, Argonne National Laboratory
\item[$^{159}$] Italy - INFN, Istituto Nazionale di Fisica Nucleare, Sezione di Lecce
\item[$^{160}$] Italy - Universit{\`a} di Palermo
\item[$^{161}$] Poland - Wroc{\l}aw University of Science and Technology
\item[$^{162}$] United States - Rutgers University
\item[$^{163}$] United States - Princeton University
\item[$^{164}$] Italy - Universit{\`a} di Padova
\item[$^{165}$] T{\"u}rkiye - IUE, {\.{I}}zmir Ekonomi {\"U}niversitesi
\item[$^{166}$] T{\"u}rkiye - Ege {\"U}niversitesi
\item[$^{167}$] Italy - INFN, Istituto Nazionale di Fisica Nucleare, Gruppo Collegato di Udine
\item[$^{168}$] Italy - Universit{\`a} di Udine
\item[$^{169}$] United States - University of Massachusetts Amherst
\item[$^{170}$] Italy - INFN, Istituto Nazionale di Fisica Nucleare, Sezione di Roma
\item[$^{171}$] Spain - CELLS/ALBA, Consortium for the Construction, Equipment and Exploitation of the Synchrotron Light Laboratory
\item[$^{172}$] France - LPSC, Laboratoire de Physique Subatomique et de Cosmologie
\item[$^{173}$] France - Universit{\'e} Grenoble Alpes
\item[$^{174}$] Italy - Universit{\`a} degli Studi di Napoli Parthenope
\item[$^{175}$] United States - National High Magnetic Field Laboratory
\item[$^{176}$] United States - Florida State University
\item[$^{177}$] South Africa - University of Johannesburg
\item[$^{178}$] Spain - IGFAE, Instituto Galego de Fisica de Altas Enerx{\'\i}as, Universidade de Santiago de Compostela
\item[$^{179}$] United Kingdom - LSE, London School of Economics
\item[$^{180}$] Spain - Universidade de Santiago de Compostela
\item[$^{181}$] United Kingdom - Imperial College London
\item[$^{182}$] United States - MIT, Massachusetts Institute of Technology
\item[$^{183}$] France - CETU, Centre d'Etude des Tunnels
\item[$^{184}$] United Kingdom - University of Liverpool
\item[$^{185}$] Switzerland - Linde Kryotechnik AG
\item[$^{186}$] Switzerland - ILF Consulting Engineers
\item[$^{187}$] Denmark - NBI, Niels Bohr Institute
\item[$^{188}$] Japan - Hokkaido University
\item[$^{189}$] Czech Republic - CUNI, Charles University
\item[$^{190}$] Spain - Universidad de Granada
\item[$^{191}$] Italy - INFN, Istituto Nazionale di Fisica Nucleare, Sezione di Firenze
\item[$^{192}$] Malta - University of Malta
\item[$^{193}$] United States - BU, Boston University
\item[$^{194}$] T{\"u}rkiye - IBU, Bolu Abant {\.{I}}zzet Baysal {\"U}niversitesi
\item[$^{195}$] Germany - Julius-Maximilians-Universit{\"a}t W{\"u}rzburg
\item[$^{196}$] Australia - University of Adelaide
\item[$^{197}$] Finland - HIP, Helsinki Institute of Physics, University of Helsinki
\item[$^{198}$] Belgium - ULB, Universit{\'e} Libre de Bruxelles
\item[$^{199}$] Germany - IMA, Institut f{\"u}r Maschinenelemente, Universit{\"a}t Stuttgart
\item[$^{200}$] Belgium - CP3, Centre de Cosmologie, de Physique des Particules et de Ph{\'e}nom{\'e}nologie, Universit{\'e} Catholique de Louvain
\item[$^{201}$] Netherlands - NIKHEF, Nationaal instituut voor subatomaire fysica
\item[$^{202}$] People's Republic of China - IHEP, Chinese Academy of Sciences
\item[$^{203}$] Mexico - UAS, Universidad Aut{\'o}noma de Sinaloa
\item[$^{204}$] Austria - BOKU, Universit{\"a}t f{\"u}r Bodenkultur Wien
\item[$^{205}$] United States - Purdue University
\item[$^{206}$] India - University of Delhi
\item[$^{207}$] France - Ginger BURGEAP, bureau d'{\'e}tudes en environnement
\item[$^{208}$] United Kingdom - King's College London
\item[$^{209}$] United States - University of Maryland
\item[$^{210}$] Japan - KEK, High Energy Accelerator Research Organization
\item[$^{211}$] Switzerland - Geoenergy, Reservoir Geology and Basin Analysis Group
\item[$^{212}$] France - SETEC International, Soci{\'e}t{\'e} d'ing{\'e}nierie en charge des transports et des infrastructures
\item[$^{213}$] T{\"u}rkiye - KKU, K{\i}r{\i}kkale {\"U}niversitesi
\item[$^{214}$] France - SETEC LERM, Soci{\'e}t{\'e} d'ing{\'e}nierie conseil en mat{\'e}riaux de construction
\item[$^{215}$] Mexico - BUAP, Benem{\'e}rita Universidad Aut{\'o}noma de Puebla
\item[$^{216}$] United States - Stanford University
\item[$^{217}$] United States - University of Pittsburgh
\item[$^{218}$] Austria - MUL, Montanuniversit{\"a}t Leoben, Lehrstuhl f{\"u}r Subsurface Engineering, Geotechnik und unterirdisches Bauen
\item[$^{219}$] Austria - MUL-ZaB, Underground Research Center, Zentrum am Berg
\item[$^{220}$] Spain - IFAE, Institut de F{\'\i}sica d'Altes Energies
\item[$^{221}$] Switzerland - FHNW, University of Applied Sciences Northwestern Switzerland
\item[$^{222}$] India - UPES, University of Petroleum and Energy Studies
\item[$^{223}$] France - GANIL, Grand Acc{\'e}l{\'e}rateur National d'Ions Lourds
\item[$^{224}$] France - Universit{\'e} Caen Normandie
\item[$^{225}$] Brazil - UFRGS, Universidade Federal do Rio Grande do Sul
\item[$^{226}$] Germany - Technische Universit{\"a}t Darmstadt
\item[$^{227}$] Poland - University of Silesia in Katowice
\item[$^{228}$] Portugal - Universidade de Coimbra
\item[$^{229}$] Brazil - UFPel, Universidade Federal de Pelotas
\item[$^{230}$] United States - University of California Santa Cruz
\item[$^{231}$] Italy - Universit{\`a} del Salento
\item[$^{232}$] United States - Brown University
\item[$^{233}$] United States - University of California Berkeley
\item[$^{234}$] Italy - Universit{\`a} di Torino
\item[$^{235}$] United States - Johns Hopkins University
\item[$^{236}$] People's Republic of China - Fudan University
\item[$^{237}$] France - LAPTh, Laboratoire d'Annecy-le-Vieux de Physique Th{\'e}orique
\item[$^{238}$] People's Republic of China - Dongguan University of Technology
\item[$^{239}$] T{\"u}rkiye - Kahramanmara{\c s} S{\"u}t{\c c}{\"u} {\.{I}}mam {\"U}niversitesi
\item[$^{240}$] Mexico - UNACH, Universidad Aut{\'o}noma de Chiapas
\item[$^{241}$] Mexico - MCTP, Mesoamerican Centre for Theoretical Physics
\item[$^{242}$] Mexico - UAZ, Universidad Aut\'onoma de Zacatecas
\item[$^{243}$] Finland - University of Jyv{\"a}skyl{\"a}
\item[$^{244}$] Germany - Technische Universit{\"a}t Dortmund
\item[$^{245}$] United Kingdom - Overleaf
\item[$^{246}$] United States - University of Virginia
\item[$^{247}$] United States - Cornell University
\item[$^{248}$] United States - FIT, Florida Institute of Technology
\item[$^{249}$] Switzerland - Shirokuma GmbH
\item[$^{250}$] Pakistan - National Centre for Physics
\item[$^{251}$] Germany - Johannes Gutenberg Universit{\"a}t Mainz
\item[$^{252}$] Pakistan - PAEC, Pakistan Atomic Energy Commission
\item[$^{253}$] India - Mathabhanga College
\item[$^{254}$] India - Harish-Chandra Research Institute
\item[$^{255}$] Germany - MPIK, Max-Planck-Institut f{\"u}r Kernphysik Heidelberg
\item[$^{256}$] Sweden - European Spallation Source ERIC
\item[$^{257}$] France - Centre de calcul de l'IN2P3
\item[$^{258}$] Germany - GSI, Helmholtzzentrum f{\"u}r Schwerionenforschung GmbH
\item[$^{259}$] Republic of Korea - IBS, Institute for Basic Science, Center for Theoretical Physics of the Universe
\item[$^{260}$] Poland - University of Warsaw
\item[$^{261}$] Slovenia - University of Ljubljana
\item[$^{262}$] Slovenia - Jozef Stefan Institute
\item[$^{263}$] France - Microhumus, Bureau d'{\'e}tude et d'ing{\'e}nierie sp{\'e}cialis{\'e} dans la gestion des sols d{\'e}grad{\'e}s
\item[$^{264}$] Germany - Fakult{\"a}t f{\"u}r Physik und Astronomie, Universit{\"a}t Heidelberg
\item[$^{265}$] T{\"u}rkiye - Ni{\u g}de {\"O}mer Halisdemir {\"U}niversitesi
\item[$^{266}$] T{\"u}rkiye - Giresun {\"U}niversitesi
\item[$^{267}$] Hungary - University of Miskolc
\item[$^{268}$] Greece - NTUA, National Technical University of Athens
\item[$^{269}$] Switzerland - Transmutex SA
\item[$^{270}$] Ireland - DIAS, Dublin Institute for Advanced Studies, School of Theoretical Physics
\item[$^{271}$] Poland - AGH, University of Science and Technology
\item[$^{272}$] Iran - University of Science and Technology of Mazandaran
\item[$^{273}$] United Kingdom - IPPP, Institute for Particle Physics Phenomenology, Durham University
\item[$^{274}$] T{\"u}rkiye - Bursa Uluda{\u g} {\"U}niversitesi
\item[$^{275}$] Republic of Korea - YU, Yonsei University
\item[$^{276}$] Affiliated with an international laboratory covered by a cooperation agreement with CERN
\item[$^{277}$] France - CPT, Centre de Physique Th\'eorique
\item[$^{278}$] France - Aix-Marseille Universit{\'e} et Universit{\'e} du Sud Toulon Var
\item[$^{279}$] Republic of Korea - KIAS, Korea Institute for Advanced Study
\item[$^{280}$] Canada - Carleton University
\item[$^{281}$] Greece - FEAC~Engineering~P.C.
\item[$^{282}$] Greece - UPATRAS, University of Patras
\item[$^{283}$] United States - University of Kansas
\item[$^{284}$] Greece - AUTH, Aristotle University of Thessaloniki
\item[$^{285}$] Germany - IML, Fraunhofer-Institut f{\"u}r Materialfluss und Logistik 
\item[$^{286}$] Germany - RWTH~Aachen, Rheinisch-Westf{\"a}lische Technische Hochschule Aachen
\item[$^{287}$] Switzerland - ZHAW, Zurich University of Applied Sciences
\item[$^{288}$] Gernany - Universit{\"a}t M{\"u}nster
\item[$^{289}$] South Africa - University of the Witwatersrand
\item[$^{290}$] Austria - Fachhochschule Technikum Wien
\item[$^{291}$] United States - University of California Irvine
\item[$^{292}$] United States - Northeastern University
\item[$^{293}$] France - ESI, European Scientific Institute
\item[$^{294}$] Republic of Korea - UOS, University of Seoul
\item[$^{295}$] Republic of Korea - KNU Kyungpook National University
\item[$^{296}$] Republic of Korea - KU, Korea University
\item[$^{297}$] Finland - Tampere University
\item[$^{298}$] United Kingdom - University of Edinburgh
\item[$^{299}$] Switzerland - BG Ing{\'e}nieurs Conseils
\item[$^{300}$] People's Republic of China - T.-D.~Lee Institute
\item[$^{301}$] People's Republic of China - Shanghai Jiao Tong University
\item[$^{302}$] Italy - CNR, Consiglio Nazionale delle Ricerche
\item[$^{303}$] United States - University of Michigan
\item[$^{304}$] United States - University of Pennsylvania
\item[$^{305}$] United States - University of Minnesota
\item[$^{306}$] France - ESRF, European Synchrotron Radiation Facility
\item[$^{307}$] United Kingdom - SUSSEX, University of Sussex
\item[$^{308}$] Italy - Universit{\`a} di Bari Aldo Moro
\item[$^{309}$] United Kingdom - University of Bath
\item[$^{310}$] Italy - Scuola Normale Superiore di Pisa
\item[$^{311}$] Brazil - Universidade de S{\~a}o Paulo
\item[$^{312}$] Austria - Universit{\"a}t Graz
\item[$^{313}$] Egypt - Center for High Energy Physics, Fayoum University
\item[$^{314}$] Egypt - Center of theoretical physics, British University in Egypt
\item[$^{315}$] Egypt - Cairo University
\item[$^{316}$] France - Sorbonne Universit{\'e} et Universit{\'e} Paris Cit\'e
\item[$^{317}$] Italy - Universit{\`a} di Genova
\item[$^{318}$] Spain - IFIC-CSIC/UV, Instituto de F\'{i}sica Corpuscular, Consejo Superior de Investigaciones Cient{\'\i}ficas/Universidad de Valencia
\item[$^{319}$] Switzerland - Service de g{\'e}ologie, sols et d{\'e}chets du canton de Gen{\`e}ve
\item[$^{320}$] Estonia - NICPB, National Institute for Chemical Physics and Biophysics
\item[$^{321}$] Estonia - UT, University of Tartu
\item[$^{322}$] Spain - Universidad de Salamanca
\item[$^{323}$] Mexico - UGTO, Universidad de Guanajuato
\item[$^{324}$] Switzerland - Edaphos engineering
\item[$^{325}$] Germany - Helmholtz-Zentrum Dresden-Rossendorf
\item[$^{326}$] India - Tata Institute of Fundamental Research Mumbai
\item[$^{327}$] Belgium - Universiteit Gent
\item[$^{328}$] Italy - CNR-IOM, Consiglio Nazionale delle Ricerche
\item[$^{329}$] Austria - JKU, Johannes Kepler Universit{\"a}t Linz
\item[$^{330}$] Norway - University of Stavanger
\item[$^{331}$] Colombia - Universidad Nacional de Colombia
\item[$^{332}$] Italy - Trento Institute for Fundamental Physics and Applications
\item[$^{333}$] United States - Caltech, California Institute of Technology
\item[$^{334}$] Italy - Universit{\`a} dalla Calabria
\item[$^{335}$] Spain - IFT, Instituto de F{\'\i}sica Te{\'o}rica, Universidad Aut{\'o}noma de Madrid
\item[$^{336}$] Spain - UPC, Universitat Polit\`{e}cnica de Catalunya
\item[$^{337}$] Portugal - Departamento de F{\'\i}sica, Universidade do Minho
\item[$^{338}$] Portugal - Centro de F{\'\i}sica das Universidades do Minho e do Porto
\item[$^{339}$] Portugal - LaPMET, Laboratory of Physics for Materials and Emergent Technologies
\item[$^{340}$] Norway - University of Bergen
\item[$^{341}$] Chile - SAPHIR, Instituto Milenio de F{\'\i}sica Subat{\'o}mica en la Frontera de Altas Energ{\'\i}as
\item[$^{342}$] Chile - Universidad Andres Bello
\item[$^{343}$] Brazil - IIP, International Institute of Physics
\item[$^{344}$] Japan - Tokyo International University
\item[$^{345}$] T{\"u}rkiye - {\.{I}}zmir Bak{\i }r{\c c}ay {\"U}niversitesi
\item[$^{346}$] Italy - INFN, Istituto Nazionale di Fisica Nucleare, Laboratori Nazionali del Gran Sasso
\item[$^{347}$] Italy - Universit{\'a} degli Studi di Cassino e del Lazio Meridionale
\item[$^{348}$] Serbia - Vin{\u{c}}a Institute of Nuclear Sciences
\item[$^{349}$] United States - Kennesaw State University
\item[$^{350}$] Italy - Universit{\`a} di Pavia
\item[$^{351}$] United States - Columbia University
\item[$^{352}$] United States - DOE, Department of Energy of the United States of America
\item[$^{353}$] France - IPSA, Institut Polytechnique des Sciences Avanc{\'e}es 
\item[$^{354}$] Italy - Universit{\`a} degli Studi del Sannio
\item[$^{355}$] Poland - UJ, Jagiellonian University
\item[$^{356}$] Austria - Universit{\"a}t Wien
\item[$^{357}$] Germany - Goethe-Universit{\"a}t Frankfurt, Institut f{\"u}r Angewandte Physik
\item[$^{358}$] Germany - HFFH, Helmholtz Forschungsakademie Hessen f{\"u}r FAIR
\item[$^{359}$] Romania - INCDTIM, National Institute for Research and Development of Isotopic and Molecular Technologies
\item[$^{360}$] United States - University of Arizona
\item[$^{361}$] Germany - Technische Universit{\"a}t Dresden
\item[$^{362}$] Latvia - RTU, Riga Technical University
\item[$^{363}$] United Kingdom - Lancaster University
\item[$^{364}$] Cyprus - University of Cyprus
\item[$^{365}$] Cyprus - Cosmos Open University
\item[$^{366}$] Brazil - CBPF, Centro Brasileiro de Pesquisas F{\'\i}sicas
\item[$^{367}$] Germany - Universit{\"a}t Bonn
\item[$^{368}$] Thailand - CMU, Chiang Mai University
\item[$^{369}$] United States - JLAB, Thomas Jefferson National Accelerator Facility
\item[$^{370}$] Ecuador - ESPOL, Escuela Superior Polit{\'e}cnica del Litoral
\item[$^{371}$] Croatia - IRB, Rudjer Boskovic Institute
\item[$^{372}$] France - Air Liquide Advanced Technologies
\item[$^{373}$] Netherlands - VU Amsterdam
\item[$^{374}$] France - ING{\'E}ROP ,Groupe d'ing{\'e}nierie et de conseil en mobilit{\'e} durable, transition {\'e}nerg{\'e}tique et cadre de vie
\item[$^{375}$] Poland - Warsaw University of Technology
\item[$^{376}$] Spain - IFCA, Instituto de F{\'\i}sica de Cantabria
\item[$^{377}$] Germany - Institut f{\"u}r Beschleunigerphysik und Technologie
\item[$^{378}$] T{\"u}rkiye - U{\c s}ak {\"U}niversitesi
\item[$^{379}$] Japan - ICEPP, International Center for Elementary Particle Physics, University of Tokyo
\item[$^{380}$] United Kingdom - Rudolf Peierls Centre for Theoretical Physics, University of Oxford
\item[$^{381}$] United Kingdom - All Souls College, University of Oxford
\item[$^{382}$] T{\"u}rkiye - Akdeniz {\"U}niversitesi
\item[$^{383}$] Switzerland - Latitude Durable SARL
\item[$^{384}$] Spain - USAL, Universidad de Salamanca
\item[$^{385}$] Germany - {PRISMA+} Cluster of Excellence
\item[$^{386}$] United States - Michigan State University
\item[$^{387}$] Spain - Universidad Complutense Madrid
\item[$^{388}$] Switzerland - scMetrology SARL
\item[$^{389}$] Portugal - IST, Instituto Superior Tecnico, Universidade de Lisboa
\item[$^{390}$] Portugal - CeFEMA, Center of Physics and Engineering of Advanced Materials
\item[$^{391}$] India - Indian Institute of Science Education and Research Mohali
\item[$^{392}$] United States - NIU, Northern Illinois University
\item[$^{393}$] India - Banaras Hindu University
\item[$^{394}$] Slovakia - Comenius University
\item[$^{395}$] Australia - Monash University
\item[$^{396}$] Slovakia - Slovak Academy of Sciences
\item[$^{397}$] Republic of Korea - KAIST, Korea Advanced Institute of Science and Technology
\item[$^{398}$] France - Amberg Engineering Chamb{\'e}ry
\item[$^{399}$] United Arab Emirates - Khalifa University of Science and Technology
\item[$^{400}$] United States - University of Tennessee
\item[$^{401}$] Austria - WIFO, {\"O}sterreichisches Institut f{\"u}r Wirtschaftsforschung
\item[$^{402}$] Brazil - Universidade do Estado do Rio de Janeiro
\item[$^{403}$] France - M{\'e}lica, NATURA SCOP, {\'E}tudes et expertises environnementales
\item[$^{404}$] Italy - Universit{\`a} LUM, Casamassima
\item[$^{405}$] Netherlands - University of Twente
\item[$^{406}$] Iran - Arak University
\item[$^{407}$] Italy - Universit{\`a} di Trieste
\item[$^{408}$] France - ForestAllia, Cabinet de gestion et d'expertise foresti{\`e}res
\item[$^{409}$] Germany - Universit{\"a}t Siegen
\item[$^{410}$] United States - University of Oregon
\item[$^{411}$] Germany - Universit{\"a}t Rostock
\item[$^{412}$] Switzerland - CEGELEC SA
\item[$^{413}$] Sweden - KTH, Royal Institute of Technology, Stockholm
\item[$^{414}$] Sweden - OKC, Oskar Klein Centre for Cosmoparticle Physics
\item[$^{415}$] United States - Brigham Young University
\item[$^{416}$] France - Expert foncier et agricole
\item[$^{417}$] Germany - ITSM, Institut f{\"u}r Thermische Str{\"o}mungsmaschinen und Maschinenlaboratorium, Universit{\"a}t Stuttgart
\item[$^{418}$] France - {\'E}cole Normale Sup{\'e}rieure de Lyon
\item[$^{419}$] Czech Republic - CTU, Czech Technical University
\item[$^{420}$] United States - University of Chicago
\item[$^{421}$] United States - Baylor University
\item[$^{422}$] United Kingdom - University of Birmingham
\item[$^{423}$] United Kingdom - University of Southampton
\item[$^{424}$] Switzerland - Swisstopo, Federal Office of Topography
\item[$^{425}$] United Kingdom - Daresbury Laboratory, Science and Technology Facilities Council
\item[$^{426}$] T{\"u}rkiye - IZTECH, {\.{I}}zmir Y{\"u}ksek Teknoloji Enstit{\"u}s{\"u}
\item[$^{427}$] Hong Kong - City University of Hong Kong
\end{itemize}

\endgroup

\clearpage

\section*{Abstract}

Volume 3 of the FCC Feasibility Report presents studies related to civil engineering, the development of a project implementation scenario, and environmental and sustainability aspects. The report details the iterative improvements made to the civil engineering concepts since 2018, taking into account subsurface conditions, accelerator and experiment requirements, and territorial considerations. It outlines a technically feasible and economically viable civil engineering configuration that serves as the baseline for detailed subsurface investigations, construction design, cost estimation, and project implementation planning. Additionally, the report highlights ongoing subsurface investigations in key areas to support the development of an improved 3D subsurface model of the region. 

The report describes development of the project scenario based on the `avoid-reduce-compensate' iterative optimisation approach. The reference scenario balances optimal physics performance with territorial compatibility, implementation risks, and costs. Environmental field investigations covering almost 600 hectares of terrain—including numerous urban, economic, social, and technical aspects—confirmed the project's technical feasibility and contributed to the preparation of essential input documents for the formal project authorisation phase. The summary also highlights the initiation of public dialogue as part of the authorisation process. The results of a comprehensive socio-economic impact assessment, which included significant environmental effects, are presented. Even under the most conservative and stringent conditions, a positive benefit-cost ratio for the FCC-ee is obtained. 
Finally, the report provides a concise summary of the studies conducted to document the current state of the environment.

\clearpage

\section*{Preface from CERN's Director-General}

In 2021, in response to the 2020 update of the European Strategy for Particle Physics, the CERN Council initiated the Future Circular Collider (FCC) Feasibility Study. 

This report summarises an immense amount of work carried out by the international FCC collaboration over several years. It covers, inter alia, physics objectives and potential, geology, civil engineering, technical infrastructure, territorial implementation, environmental aspects, R\&D needs for the accelerators and detectors, socio-economic benefits and cost. It constitutes important input for the ongoing update of the European Strategy for Particle Physics.

The Feasibility Study required engagement with a broad range of stakeholders. In particular, throughout the Study, CERN has been accompanied by its two Host States, France and Switzerland, and has been working with entities at local, regional and national level. I am very grateful to the Host State authorities and teams for their invaluable help. Furthermore, significant sections of the Study were supported by the European Union under the Horizon 2020 and Horizon Europe framework programmes. The Study also greatly benefited from contributions from accelerator laboratories and universities from across Europe, such as the Swiss Accelerator Research and Technology (CHART) initiative, and from the Americas, Asia, Africa and Australia. 

The proposed FCC integrated programme consists of two possible stages: an electron–positron collider serving as a Higgs-boson, electroweak and top-quark factory running at different centre-of-mass energies, followed at a later stage by a proton–proton collider operating at an unprecedented collision energy of around 100 TeV. The complementary physics programmes of each stage match the physics priorities expressed in the 2020 update of the European Strategy for Particle Physics. 

A major achievement of the Feasibility Study is the choice of placement of the collider ring and the entire infrastructure, including the surface sites and the access shafts, which was developed and optimised over several years following the principle `avoid, reduce, compensate'. Sustainability studies have assessed energy efficiency, land use, water and resource management, and socio-economic impact, ensuring that the FCC is designed in accordance with the latest environmental and societal standards. 

I would like to thank all contributors to this report for their hard work and commitment, which allowed the outstanding results presented here to be achieved.

\begin{flushright}
\textbf{Fabiola Gianotti}\\
CERN, Director-General
\end{flushright}

\clearpage

\section*{Preface from the FCC Collaboration Board Chair}

Building on the earlier Future Circular Collider (FCC) Conceptual Design Study conducted between 2014 and 2018, the FCC Feasibility Study (2021–2025) has been undertaken by a robust international collaboration, now comprising over 160 institutes worldwide.  The FCC `integrated programme', developed in the framework of the Feasibility Study, consists of an initial electron-positron collider, the FCC-ee, which could be followed by a proton-proton collider, the FCC-hh. This staging takes into account the physics priorities as formulated in the updates of the European Strategy for Particle Physics of 2012 and 2020, as well as the relative technology readiness and costs of the FCC-ee and FCC-hh.

Over the years, I have closely followed the steady progress of the study, representing the FCC collaboration at the international steering committee and participating in annual FCC Week meetings, which include sessions of the International Collaboration Board. The commitment and enthusiasm of the members of the collaboration has always been impressive. The collective effort is clearly visible. Participation by students and early-career researchers is increasing. There is a shared determination and momentum to move forward.

The strong international collaboration around the FCC and its global network provide a solid foundation for the future of this project. The FCC community continues to grow, with increasing engagement from new institutes and partners worldwide. This broad support will be essential as the project enters its next phase.

The FCC Feasibility Study demonstrates not only the technical viability of the project, but also the strength of the international community that supports it. As we move towards the next step in the decision-making phase, this collective effort is key to showing a possible path forward. The FCC promises far-reaching scientific opportunities and long-term benefits for innovation, training, and global collaboration in science and technology.

\begin{flushright}
\textbf{Philippe Chomaz}\\
CEA, Chair of the FCC International Collaboration Board
\end{flushright}

\clearpage

 \setcounter{tocdepth}{1}
 
\tableofcontents

\cleardoublepage

\pagenumbering{arabic}

\chapter{Civil engineering}\label{MTR-D2}

\section*{Introduction}\label{secintroduction}

Since the completion of the FCC conceptual design in 2018, several significant modifications have been made to civil engineering. They derive from the improved maturity of requirements for the systems that interface with civil engineering, such as the accelerators, the detectors, and the technical infrastructure to be housed within the underground and surface structures. Furthermore, more precise localisation of the surface sites has facilitated a greater understanding of the geographical and environmental constraints to be accounted for in the development of the civil engineering infrastructure.  
The main improvements that have occurred since the conceptual design of the underground civil engineering required for the FCC-ee collider was completed are as follows:
\begin{itemize}
\item A reduction in the overall circumference of the accelerator tunnel from 97.8\,km to 90.6\,km.

\item A reduction in the number of surface sites (and access points to the underground) from 12 to 8.

\item A reduction in the number of permanent shafts needed for operation from 18 to 12.

\item A reduction of the depth to the deepest shaft from 578\,m to 400\,m.

\item Additional underground civil engineering for the RF systems at technical sites PH and PL.
\item A simplification of the civil engineering needed for the beam absorber system. 
\item Simplification of the underground infrastructure required for the beam transfer lines, with the use of a single tunnel to house both clockwise and anti-clockwise transfer to the FCC.

\end{itemize}

In addition to the above, a more detailed—yet still evolving—understanding of the requirements for surface civil engineering has been built up. Initial spatial arrangements for all eight surface sites have been developed, with preliminary dimensioning of the necessary buildings, roads, and other surface infrastructure providing a framework for further refinement. For two of the eight sites, and under a collaboration with the U.S. Fermi National Accelerator Laboratory, preliminary design studies for buildings have been carried out. The results offer valuable insights into space requirements, user access needs and cost envelopes. They were mainly informed by the technical needs of the interfacing systems.
The integration of the ‘Avoid, Reduce, Compensate’ approach to develop the surface sites is an iterative process, requiring continued efforts to understand the technical needs, the territorial requirements and constraints and the project cost and risk implications. Understanding the technical system requirements is an essential guiding element. The analysis of the current state of the environment also carried out in the frame of the feasibility study, is another aspect that will inform the subsequent activities.

A staged approach to civil engineering has been studied. Infrastructure essential for the operation of the FCC-ee collider and its associated experiments is included within the initial stage, along with any civil engineering needed for FCC-hh for which it would be impossible, impractical, or inefficient to construct during a second stage. This staged approach primarily concerns the surface infrastructure since it is relatively straightforward to construct additional buildings in a second stage.
 
The civil engineering for the FCC-ee, as defined in this document, builds on the studies completed for the conceptual design phase and carried out for the feasibility phase. The current design demonstrates the technical feasibility of FCC civil engineering. It is to be noted, however, that there are still areas where further work needs to be done in order to reduce technical risks, in particular taking into account the remaining data that will be provided upon completion of phase 1 of the sub-surface site investigations that commenced in October 2024 and is expected to be completed before the end of 2025.

The work carried out since the completion of the conceptual design has largely maintained similar spatial elements for the underground works (sizes of caverns, shafts, tunnels, etc.), and therefore, it has not been necessary to undertake additional calculations for the overall structural stability of these elements. Furthermore, the construction methodologies remain largely as identified during the conceptual design phase, although the sequence for undertaking the works has now been revised, in particular, to account for the reduced number of access points to the underground civil engineering.
  
A key input to the feasibility study has come from the early interaction between the FCC civil engineering team and potential future stakeholders from the industry. CERN has held a number of meetings and workshops with several major contractors specialising in underground civil engineering. The feedback received has proved invaluable in developing not only robust technical solutions but also a realistic and efficient schedule for the execution of civil engineering. Feedback from the industry has also helped shape CERN's initial thinking on potential contractual routes for the delivery of the design and construction of civil engineering. The civil engineering schedule developed as part of the feasibility study has been integrated with the subsequent infrastructure and machine installation activities. The resulting integrated schedule results in a gradual handover of the civil engineering for each of the eight sectors, thus allowing parallel and more efficient execution of installation activities in the main tunnel. 

\section{Underground structures}\label{secundergroundstructures}

Several improvements have been made to the underground structures since the conceptual design report was completed. A thorough identification has been made of the underground structures necessary for the FCC-ee collider as well as those for the FCC-hh collider, which cannot be deferred. The latter will be constructed as part of the civil engineering necessary for the FCC-ee collider and associated experiments. 

\begin{figure}[!ht]
    \centering
    \includegraphics[scale=0.28]{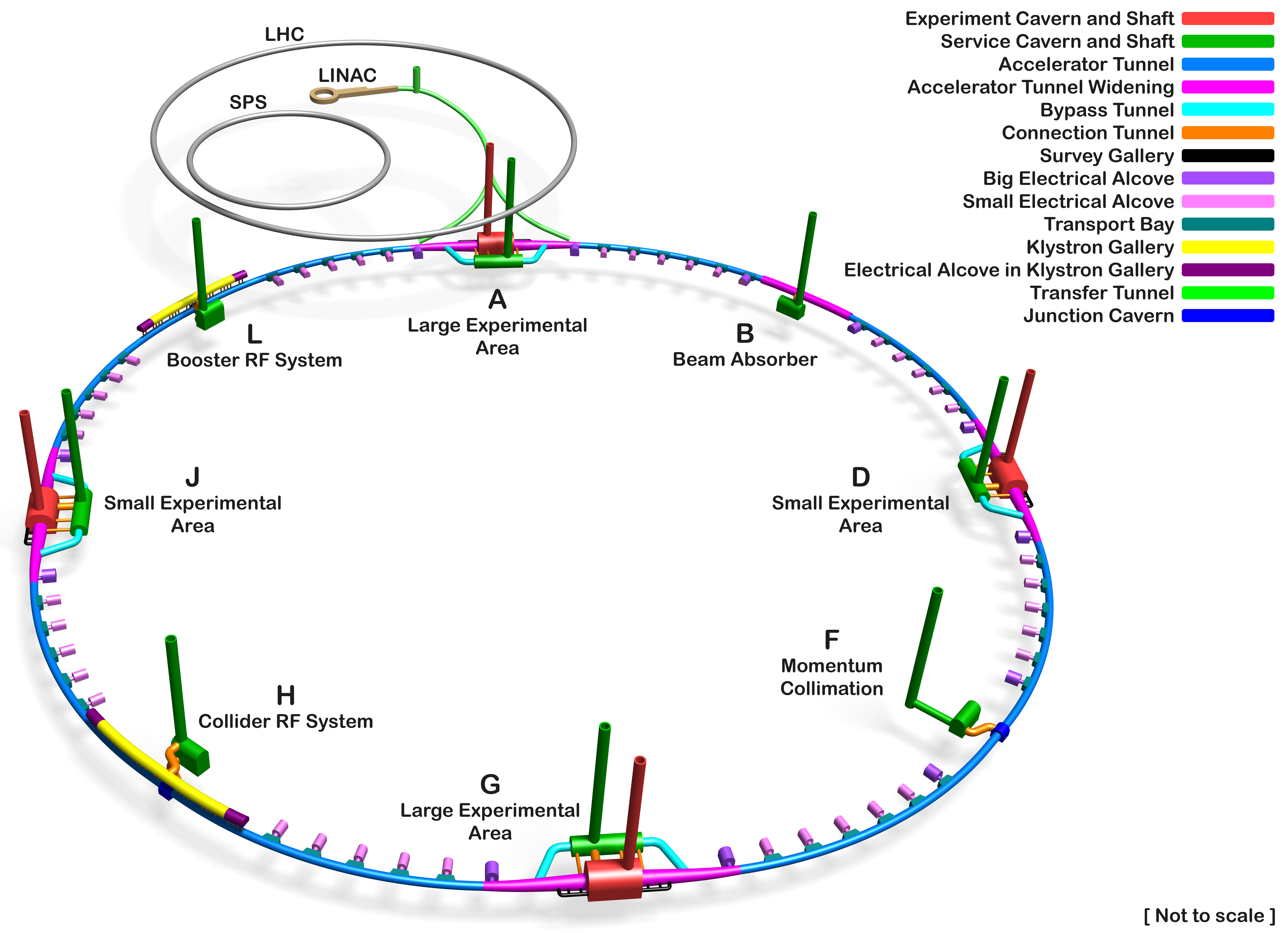}
    \caption{\label{undergroundschematic}Schematic layout of FCC-ee underground civil engineering.}
\end{figure}

A Product Breakdown Structure (PBS) has been produced down to level 4, containing 205 uniquely identified structures.  Table~\ref{PBS for PA Underground Civil Engineering} lists the PBS and associated structures for the sector PA to PB.  The PBS of other sectors follow a similar structure. In total, the underground civil works consist of twelve permanent shafts for operation, one temporary shaft required only during the civil engineering construction, twelve large caverns with spans exceeding 20\,m as well as numerous smaller caverns, alcoves, connection and bypass tunnels that collectively make up the underground civil engineering. Figure~\ref{undergroundschematic} shows a schematic arrangement of underground civil engineering.

\begin{table}[!ht]
    \centering
\caption{PBS for sector PA to PB underground structures.}
\label{PBS for PA Underground Civil Engineering}
   \begin{tabular}{lccccl}
    \toprule
        \multicolumn{5}{c}{\textbf{{PBS Level}}} & \multicolumn{1}{c}{\textbf{{PBS Description}}} \\ 
        \textbf{0} & \textbf{1} & \textbf{2} & \textbf{3} & \textbf{4} & ~ \\ \midrule
        ~ & \textbf{2} & ~ & ~ & ~ & \textbf{Civil engineering} \\ 
        ~ & \textbf{2} & \textbf{3} & ~ & ~ & \textbf{Underground structures} \\ 
        ~ & \textbf{2} & \textbf{3} & \textbf{1} & ~ & \textbf{Site PA} \\ 
        ~ & 2 & 3 & 1 & 1 & Experiment shaft \\ 
        ~ & 2 & 3 & 1 & 2 & Service shaft \\ 
        ~ & 2 & 3 & 1 & 3 & Experiment cavern \\ 
        ~ & 2 & 3 & 1 & 4 & Service cavern \\ 
        ~ & 2 & 3 & 1 & 5 & Alcoves \\ 
        ~ & 2 & 3 & 1 & 6 & Beam tunnel sector AB \\ 
        ~ & 2 & 3 & 1 & 7 & Bypass tunnel AL \\ 
        ~ & 2 & 3 & 1 & 8 & Bypass tunnel AB \\
        ~ & 2 & 3 & 1 & 9 & Connection tunnels and galleries \\ 
        ~ & 2 & 3 & 1 & 10 & Tunnel widenings \\ \bottomrule
    \end{tabular}
\end{table}

The overall circumference of the accelerator tunnel that will house the colliders is 90.6\,km, a reduction of about 7\% compared to the 97.8\,km tunnel developed at the conceptual design phase. The accelerator tunnel, which will house the collider, has an internal nominal diameter of 5.5\,m. On the basis of the pre-existing and recent site investigation data, the average elevation is 300\,m above datum and the inclination of the tunnel plane has been maintained at 0.1\% and 0.4\% in the two axes.  This results in a tunnel depth that varies between 30\,m where the tunnel passes under the Rh\^{o}ne river and 560\,m where the tunnel passes under the Borne plateau on the eastern side of the overall study site. The average depth of the tunnel is approximately 240\,m below the ground surface. 

Two sizes of experiment cavern complexes are envisaged; these are similar in layout and function. The first type includes a cavern to house the largest planned FCC-hh detector with a 35\,m span (similar to that of the existing ATLAS detector cavern) and the second type includes a cavern to house the smaller FCC-hh detectors with a span of 25\,m (similar to that of the existing CMS detector cavern). 

A single transfer tunnel connecting the injection system to the FCC is envisaged. This tunnel starts close to the surface of the existing CERN Pr\'{e}vessin site where it will connect to the cut and cover tunnel that will house the high-energy LINAC. The tunnel will descend over a distance of about 5\,km at which point it will bifurcate at a location close to PA to allow symmetrical clockwise and anticlockwise injection into the FCC. The injection will take place on either side of the experiment area located in PA.

Expectations regarding the geology to be encountered during the civil engineering of the underground structures are consistent with those established during the conceptual design phase of the FCC-ee study, providing a solid foundation for further refinement and risk mitigation in subsequent phases. The current expectation is that the majority of the tunnels and the cavern complexes will be located within the molasse rock, a low to medium-strength sedimentary rock made up of complex sequences of marls and sandstones. This material is well suited for tunnelling since it is typically watertight and can be supported through the implementation of a range of standard rock support measures such as rock bolts, shotcrete, reinforced concrete segments etc. Larger caverns can become more challenging and will require more complex methods for the excavation and support of the rock mass. This is further detailed in Section~\ref{subseccaverns}. 

The accelerator tunnel will pass through approximately 4.4\,km of limestone rock. Whilst it will certainly be possible to excavate the rock by either tunnel boring machine or traditional drill and blast methods, this rock mass may contain large interconnected voids (karsts) with water and silt, potentially under high pressure. Specific construction measures may be needed in this area, as presented in Section~\ref{subsecmainbeamtunnel}. Placement optimisation of the underground structures has ensured that the location of larger underground structures, such as caverns and shafts, avoids areas where limestone formations are likely to be encountered. This will require careful review at the conclusion of the ongoing site investigation campaign.
The shafts will need to be excavated through varying depths of so-called moraine strata before the molasse rock is reached. This stratum is a mix of clays, sands, gravels and boulders. It is known to contain aquifers.

CERN has over 40 years of experience in managing projects that involve the construction of tunnels, shafts, and caverns in the molasse rock and, therefore, has the experience and knowledge to undertake the FCC civil engineering. Some aspects, as listed below, will be challenging but nonetheless well within the capabilities of many Member States civil engineering contractors.

\begin{itemize}
\item The average depth at which the underground structures will be constructed is about three times greater than the average depth of the LHC tunnel. This will present higher ground stresses that will need to be considered in the design of the underground structures and when selecting the construction methodologies to be used.  Proven engineering solutions will be applied to effectively mitigate potential challenges, such as ground deformation, and ensure the smooth operation of the tunnel boring machine.
\item The tunnel will need to pass through a zone of so-called `molasse charri\'{e}' a geological formation with different characteristics compared to the molasse rock in which CERN's existing underground structures were constructed. These factors will be carefully addressed in the design and construction methodology to ensure a safe and efficient excavation process.
\item The tunnel will pass through several kilometres of limestone, which may contain water at high pressure. Similar conditions were encountered during the construction of the LEP tunnel, providing valuable insights into mitigation measures. Careful consideration will need to be given to the tunnel design, selection of tunnelling methods and the provision of ground treatment ahead of the tunnelling face in this zone of limestone rock. Modern engineering solutions, coupled with CERN’s experience, ensure that potential challenges are well understood and can be managed effectively.
\item The tunnel will pass under Lac L\'{e}man. As far as possible, the tunnel horizon will be kept within the molasse rock. If this is not possible then alternative tunnelling techniques for traversing water-bearing sands/gravel/silts will need to be used, such as earth pressure balance tunnel boring machines or so-called slurry tunnel boring machines).
\item It is likely that during the construction of some shafts, water-bearing moraine strata will need to be traversed before reaching the more favourable moraine rock. Again, this will require the use of specialised construction methodologies such as diaphragm walls or ground freezing.
\end{itemize}

Although some aspects of the civil engineering underground construction will be technically more challenging than previous CERN construction projects, it is considered that the underground civil engineering structures can be designed and constructed using existing, proven, conventional techniques, including CERN's extensive experience in underground construction. This consideration is supported by the technical discussions that CERN has held with tunnelling contractors. 

The constraints arising from the large-scale geological environment on underground civil engineering, as described in the conceptual design report, are still largely valid. It is to be noted, however, that the reduction in the circumference of the collider and accelerator tunnel has resulted in a higher probability that the tunnel can be predominantly located in the favourable molasse rock due to the general displacement of the tunnel away from the limestone regions associated with the Jura, Vuache and Mandallaz outcrops. Again, this should be confirmed with the completion of the ongoing site investigations.

\subsection{Accelerator tunnel}\label{subsecmainbeamtunnel}
\begin{figure}[ht]
    \centering
    \includegraphics[width=\linewidth]{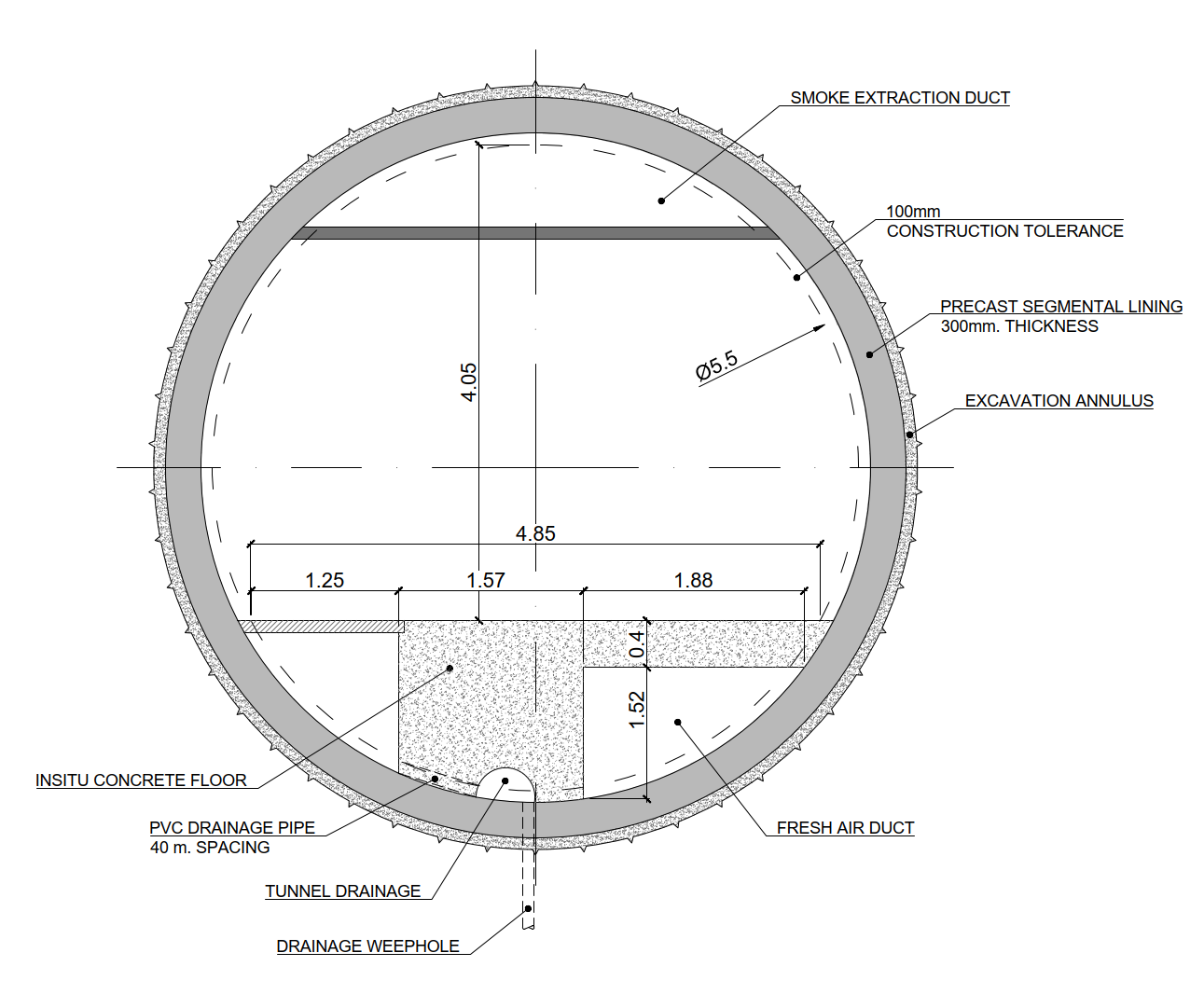}
    \caption{Accelerator tunnel cross-section.}
    \label{Accelerator Tunnel Cross-Section}
\end{figure}

The majority of the 90.6\,km circumference FCC tunnel alignment consists of a 5.5\,m internal diameter tunnel, as illustrated in Fig.~\ref{Accelerator Tunnel Cross-Section}. There are eight sectors, each approximately 11.3\,km in length. The majority of each sector consists of an arc of a radius of 14.5\,km. 

The accelerator tunnel houses the beam and service infrastructure, as well as a transport corridor. The current design of the tunnel assumes that where tunnel boring machines (TBMs) are used for excavation, a precast reinforced concrete segmental lining will be used to support the tunnel. In areas where TBMs will not be employed, primary support consisting of rock bolts and fibre-reinforced shotcrete will be used with a secondary cast in-situ concrete lining. The tunnel floor is to be cast in-situ concrete and installed over void formers for the tunnel drainage and the fresh air duct. Access chambers for the drainage network will be spaced every 100\,metres along the tunnel. To ensure that the internal diameter is at least 5.5\,m for the integration of technical infrastructure, the accelerator tunnel will be constructed with a tolerance envelope of 100\,mm.

The installation of the precast segmental lining that supports the ground is carried out automatically from within the TBM. CERN has discussed with a European TBM manufacturer the potential machine requirements for the construction of FCC. They stated that either a double shield or single shield TBM could be utilised.
\begin{figure}[ht]
    \centering
    \includegraphics[scale=0.2]{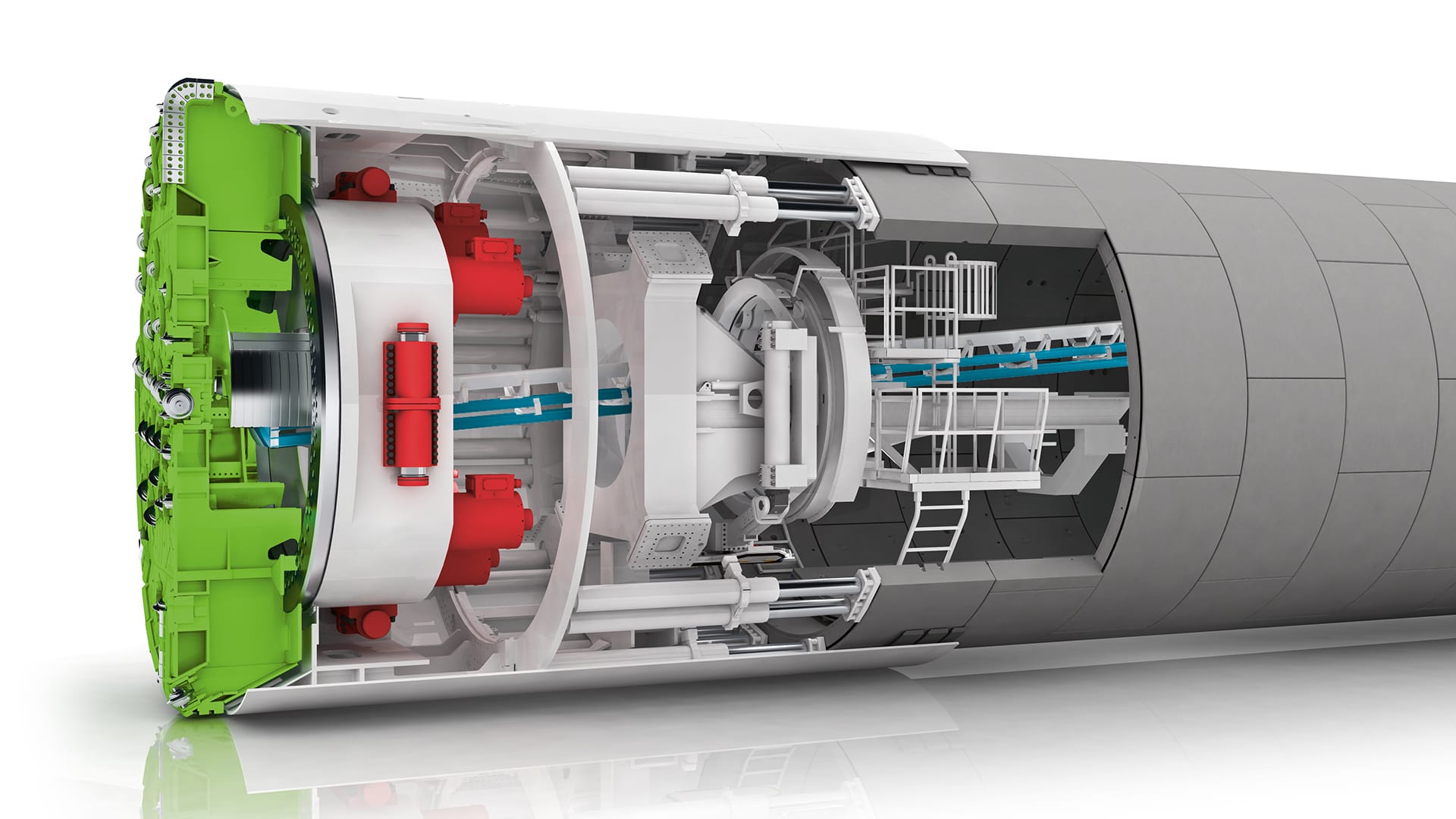}
    \caption{\label{Single Shield TBM}Single shield TBM. Source: Herrenknecht}
\end{figure}

A typical benefit of a double shield TBM over a single shield is the increased speed of construction, this is because the machine can simultaneously excavate the ground and install the segmental tunnel lining. Conversely, a single-shield machine requires these two phases to occur sequentially. A double-shield machine is also better suited and more adaptable to variable geology and unstable ground.

However, the increased complexity of the double shield machine TBM leads to greater capital cost, higher maintenance demand and increased risk of TBM breakdown, when compared to single shield TBM.

Feedback from civil engineering contractors specialised in tunnelling indicates that in the relatively stable molasse rock of the FCC a single shield TBM could be a more cost-effective solution. Furthermore, the increased speed of construction offered by a double shield machine may not actually be achieved, since the logistics of material delivery and spoil extraction at the shafts, and over the long tunnelling distances, would be more of a constraint to the TBM advance rate than the capability of the machine itself.

Whilst the majority of tunnel excavation will be within the molasse rock, which lends itself to TBM excavation, as detailed above, a specifically designed TBM may be used for the 4.4\,km section of Mandallaz limestone likely to be encountered in the sector PG to PH. This TBM, which would be designed for soft and hard rock conditions, would also have the capability to probe ahead of the tunnel face and provide a real-time assessment of the ground conditions in front of the TBM. The machine would have the capability to inject cement-based grout ahead of the excavation to reduce the risk of ground collapse and water inflow. To take account of the potential difficulties that may be encountered in the limestone, the TBM advance rate has been reduced from an average of 16\,m/day to 9\,m/ day in the construction schedule.

If not excavated by a TBM, this section of limestone would need to be excavated by drill and blast. This is a conventional tunnelling technique involving the use of explosives to excavate the rock face. Depending on the condition of the limestone in this area, additional ground treatment may be required ahead of the tunnel face to prevent water ingress during excavation. The ongoing site investigations will characterise the composition of the limestone along the FCC alignment and, therefore, provide greater certainty of the tunnelling conditions.

For the current feasibility study, it has been assumed that a single-pass precast lining will be the most efficient and effective ground support system, as this is the fastest and most cost-effective construction method. In the case that future site investigations reveal more challenging ground conditions than those that have been assumed, then a review of the tunnel support system will need to be carried out. Where necessary, a more appropriate ground support system, consisting of a drained, reinforced concrete in-situ lining, may be required. 

Table~\ref{5.5mTBMtable} shows the excavation and lining parameters assumed for the feasibility study.
\begin{table}[ht]
    \centering
 \caption{Proposed TBM excavation and lining parameters.}
    \begin{tabular}{p{9cm}p{2.5cm}}
    \toprule
        \textbf{Parameter} & \textbf{Properties}\\ \midrule
        Minimum internal diameter (m) & 5.5 \\ 
        
        Characteristic concrete compressive strength for pre-cast concrete, fck (MPa) & 50 \\ 
        Pre-cast concrete thickness (m) & 0.30 \\ 
        Reinforcement density for steel fibre reinforced pre-cast concrete (kg/m$^{3}$) & 35 \\
        Reinforcement density for steel bar reinforced pre-cast concrete (kg/m$^{3}$) & 80 \\
        Gasketed segments & yes \\ 
        Total radial construction tolerance (m) & 0.10 \\ 
        Excavation diameter & 6.6\\ 
        \bottomrule
    \end{tabular}
 \label{5.5mTBMtable}
\end{table}

\begin{figure}[ht]
    \centering
    \includegraphics[scale=0.25]{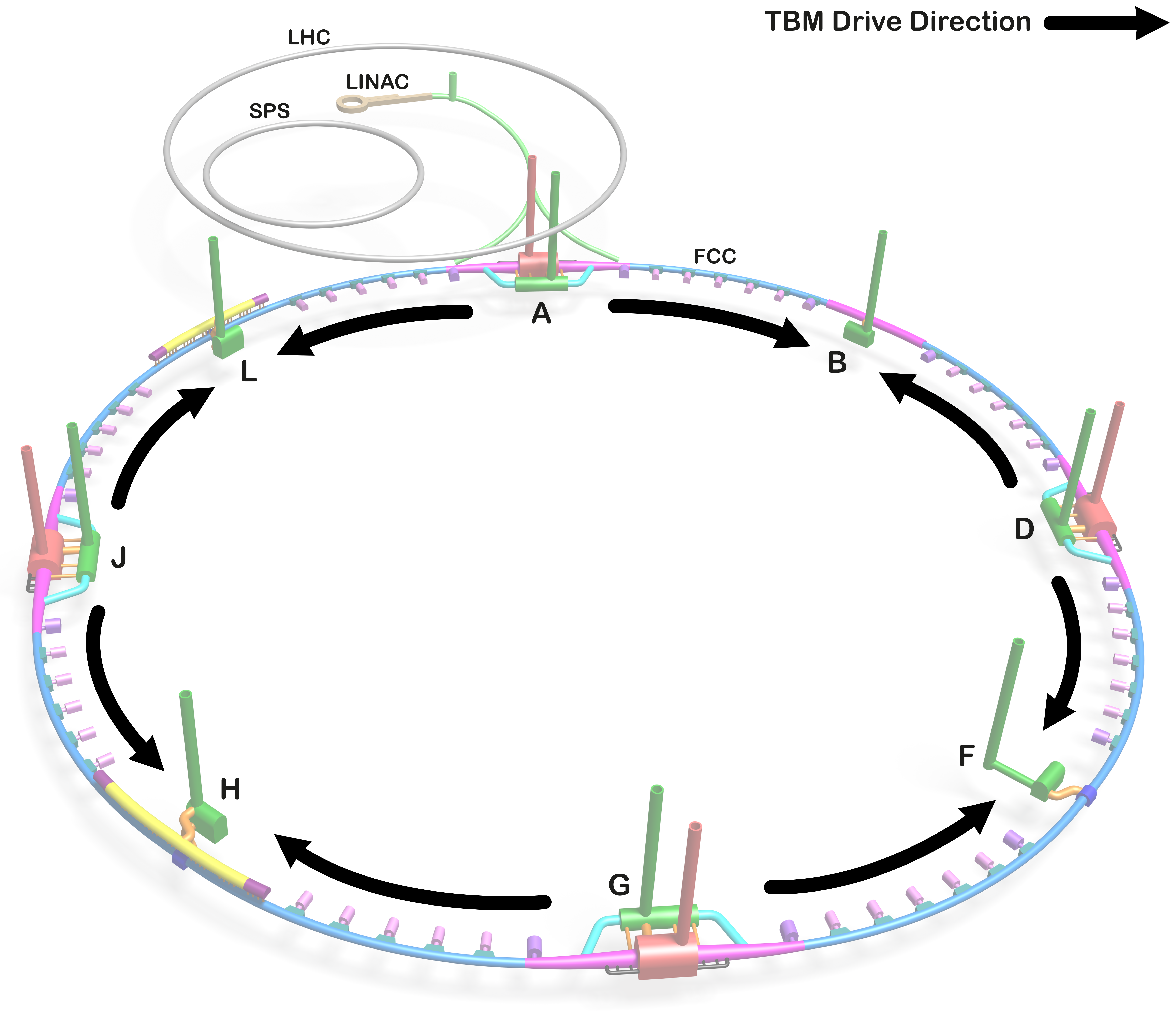}
    \caption{Proposed arrangement of TBM drives.}
    \label{TBM Arrangement}
\end{figure}

The current assumption for the start and finish sites for each of the TBMs is shown in Fig.~\ref{TBM Arrangement}. The main advantage of this arrangement is that it allows an earlier completion of the technical sites because these do not have TBM installation or deinstallation activities associated with them. This is a major benefit as it allows earlier installation of the infrastructure and accelerator components in the underground areas. Further benefits of this approach are:

\begin{itemize}
\item TBMs are large and heavy machines and as such cannot easily be manipulated in the confined spaces of an underground worksite. To install a TBM takes four to six months and requires a cavern large enough to allow the installation. At the four experiment sites, the two large caverns and the long, widened sections of the main beam tunnel can be used for the installation and commissioning of the TBMs.
\item The presence of a shaft located directly above the axis of the main beam tunnel gives a significant advantage for the installation of the TBM. It enables the TBM to be installed in relatively large pieces, thereby reducing installation time. 
\item Provision of two shafts, each capable of providing the necessary logistics and services to support a TBM (transport of people, spoil removal, material, power, water etc.), provides a level of redundancy since, if one shaft is out of action, work can continue via the second shaft.
\item Two shafts provide better provision for the evacuation of people in case of an accident or an emergency when access to one shaft may not be possible. 
\item Concentrating the underground civil engineering at the four experiment sites will allow the four technical sites to have a reduced impact on the local community and the local environment (less dust, noise, traffic etc.).
\end{itemize}
 
\begin{figure}[ht]
    \centering
    \includegraphics[scale=0.3]{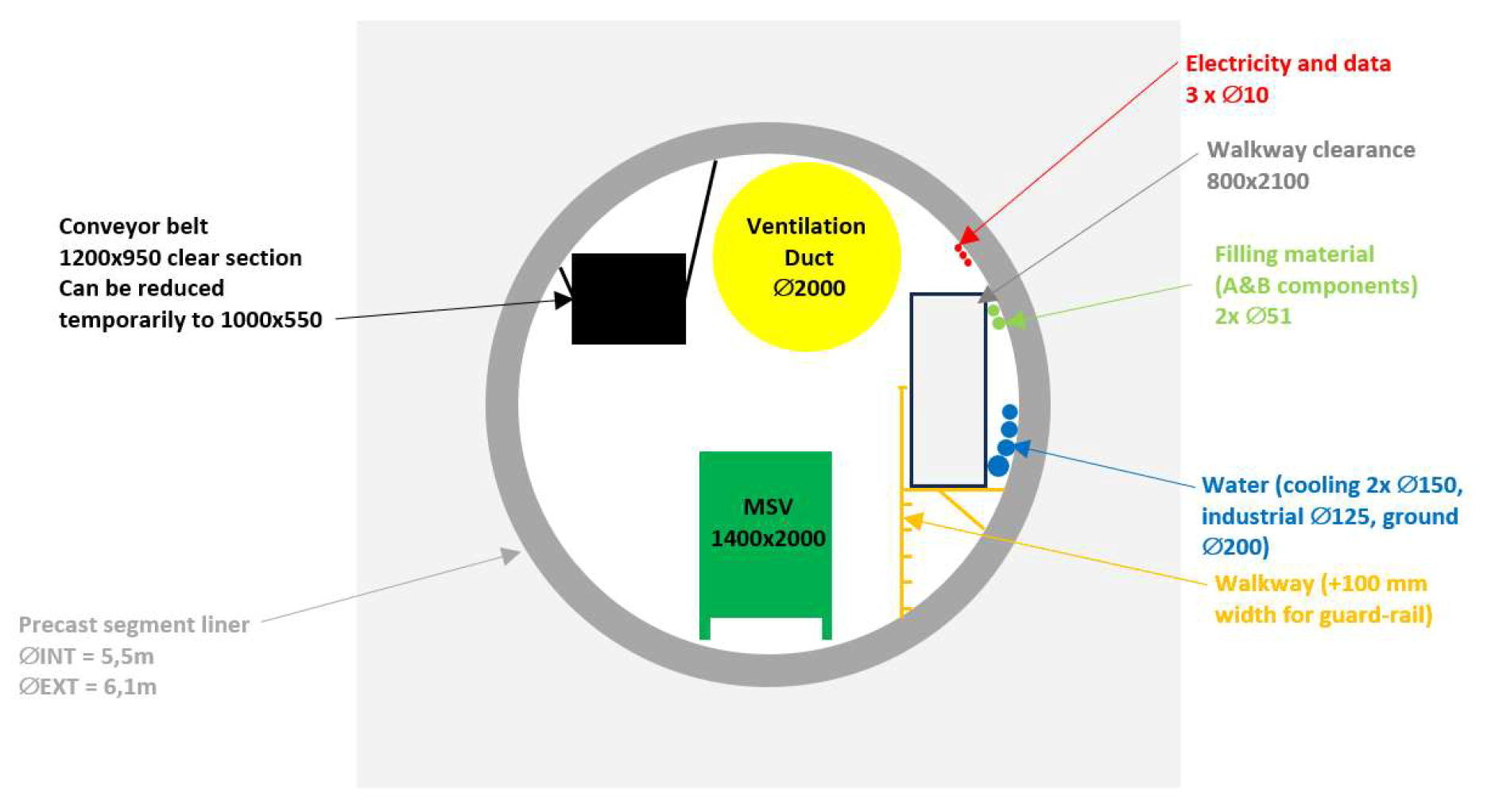}
    \caption{Cross-section of the accelerator tunnel during construction. Credit: Amberg}
    \label{Excavation Cross Section}
\end{figure}

\begin{figure}[ht]
    \centering
    \includegraphics[scale=0.3]{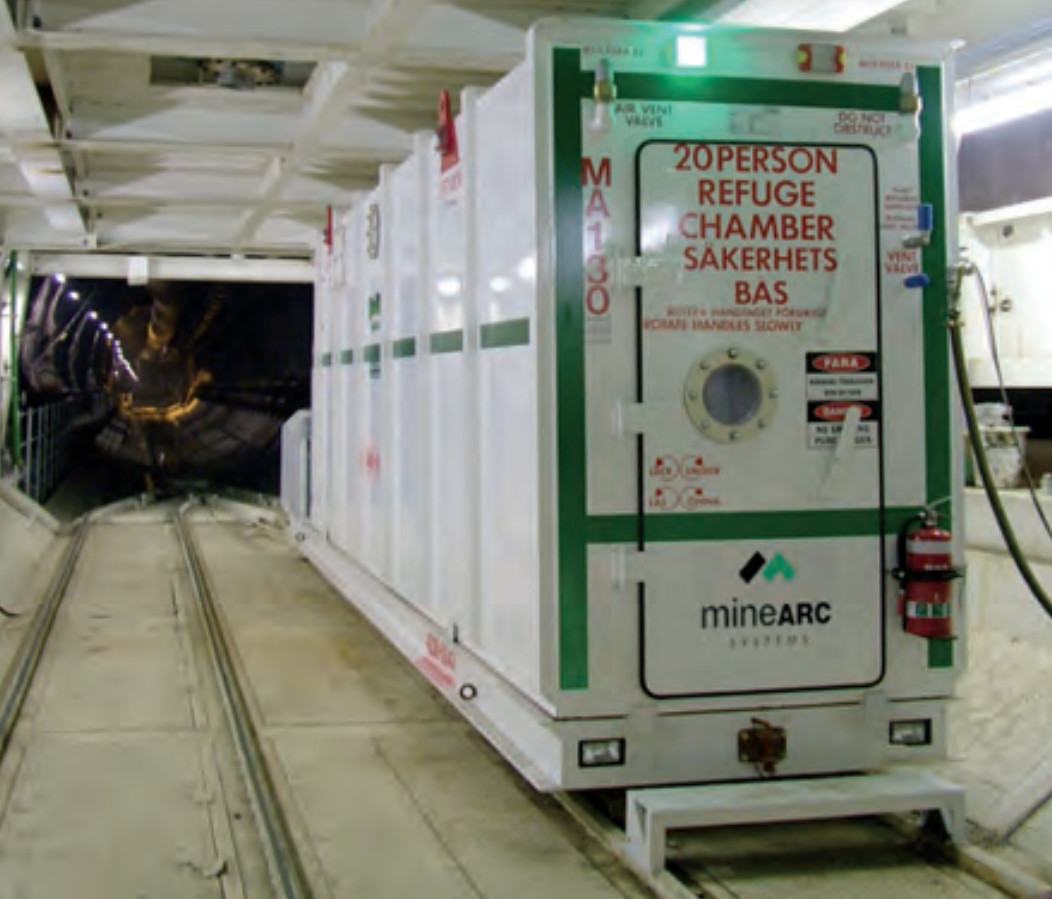}
    \caption{Example refuge chamber placed at the back of the TBM. Credit: mineARC}
    \label{Refuge}
\end{figure}

CERN commissioned a study into the safety, ventilation, and logistics aspects of the FCC construction to address concerns that the 5.5\,m internal diameter tunnel and 11\,km TBM drives would present significant safety and logistical risks, particularly with only a single means of access and egress. It was demonstrated that using currently available technology, the tunnel could be constructed safely, and the assumed TBM advance rate could be achieved. Ventilation requirements were shown to be met over the full length of the tunnel, with ducting and fan specifications calculated to supply adequate fresh air to the excavation front. Finally, safety measures were outlined to ensure that construction activities would meet the required standards both nationally and internationally. An example of the tunnel cross-section during construction is shown in Fig.~\ref{Excavation Cross Section}. An example of the safety refuge chambers specified in the study is shown in Fig.~\ref{Refuge}. 

\begin{figure}[ht]
    \centering
    \includegraphics[width=\linewidth]{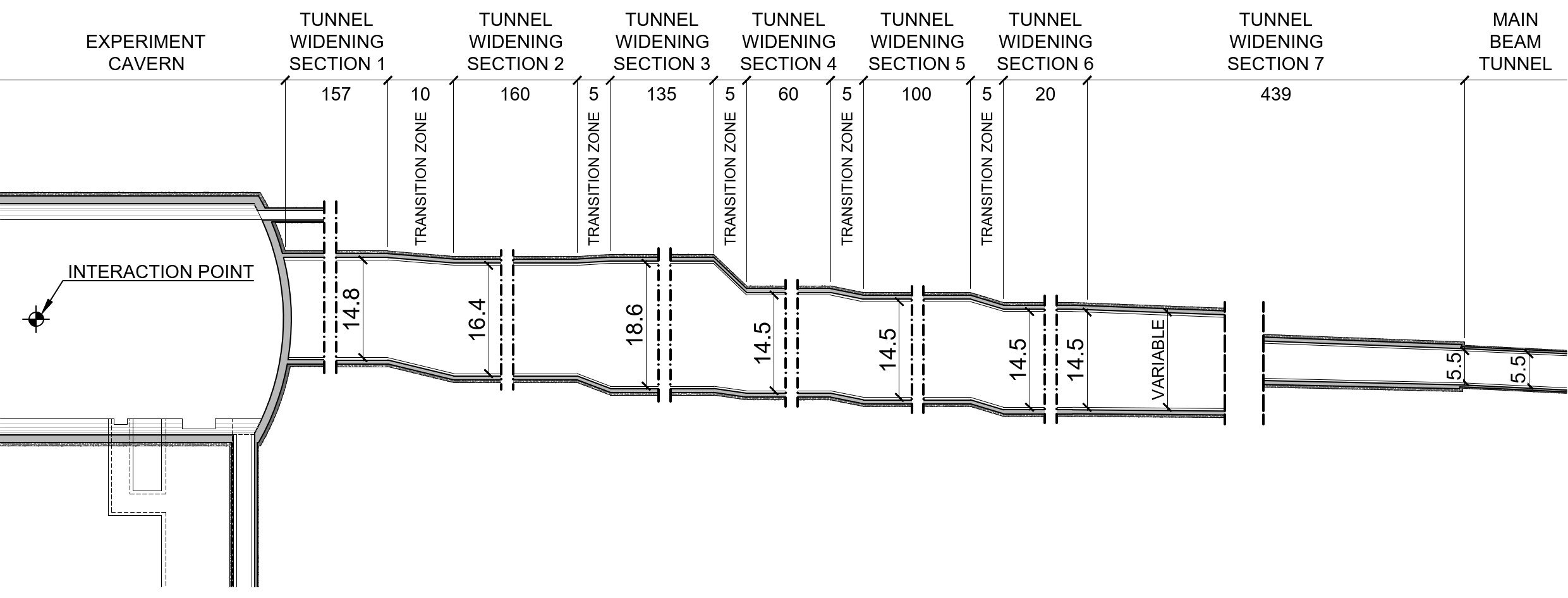}
    \caption{Plan view of the typical tunnel widening at an experiment cavern.}
    \label{TunnelWidening}
\end{figure}

To accommodate the separation of the e$^{+}$ and e$^{-}$ beams of the FCC-ee machine near the detector locations and the need to maintain space for the booster ring, the accelerator tunnel requires enlargement on each side of the experiment caverns at PA, PD, PG, and PJ. There are a total of 8 tunnel enlargement areas, which extend for 1.1\,km on either side of the experiment caverns. To minimise construction costs and optimise efficiency, the enlargements will be created in a stepped design, as shown in Fig.~\ref{TunnelWidening}. 

The widened tunnel sectors are split into 6 sections on either side of the experiment cavern, ranging from 18.6\,m to 14.5\,m in span and section lengths varying between 20 and 160\,m. A seventh sector of widening tapers from 14.5\,m span to the regular 5.5\,m diameter of the accelerator tunnel. This seventh sector commences from the end of the long straight section through into the arc section for a length of 438.5\,m. The beamstrahlung absorber will be located 500\,m from the interaction point (IP) within section 3 of the tunnel widening. A 20\,m section of increased tunnel height may be required to provide space for a crane around the beamstrahlung absorber. This is not currently incorporated in the feasibility study baseline for civil engineering. If confirmed as necessary, this will be incorporated in the baseline during the next design phase.

The construction of the tunnel-widening sections is proposed to be undertaken using conventional excavation methods such as roadheader and/or hydraulic rock-breaker machines. A shotcrete final lining is proposed with the aim of reducing construction time, thereby enabling an earlier installation of the TBMs, which will be carried out within the tunnel widening sections on either side of each experiment cavern. 

\subsection{Bypass tunnels}\label{subsecbypasstunnels}
Bypass tunnels are required at each of the four experiment areas to allow access for transport, personnel, and services directly from the service cavern to the accelerator tunnel, therefore bypassing the experiment cavern and detector areas. These tunnels will have an internal diameter of 5.5\,m, similar to the cross-section of the accelerator tunnel. The length of the bypass tunnels varies between 110 and 115\,m, depending on the experiment area. The 30\,m radius bends allow transport vehicle movements as well as provide radiation protection between the accelerator tunnel and the service cavern. The bypass tunnels connect the service cavern to the accelerator tunnel within the first section of the tunnel widening, approximately 74\,m from the experiment cavern. The junction between the accelerator tunnel and bypass tunnel is at an angle of 45 degrees, for the purpose of civil engineering constructibility. However, transport needs may require a shallower angle to be specified. In the next design phase, the details of the junctions between the two tunnels will be reassessed, and if required, a junction cavern or other more effective means of accommodating the connection will be implemented.

The bypass tunnels will be constructed using a roadheader machine and lined with in-situ concrete for the final lining. The construction of these tunnels will be completed in parallel with the works for the connection tunnels and tunnel-widening sections at each experiment point.

\subsection{Connection tunnels}\label{subsecconnectiontunnel}
\begin{figure}[ht]
    \centering
    \includegraphics[width=\linewidth]{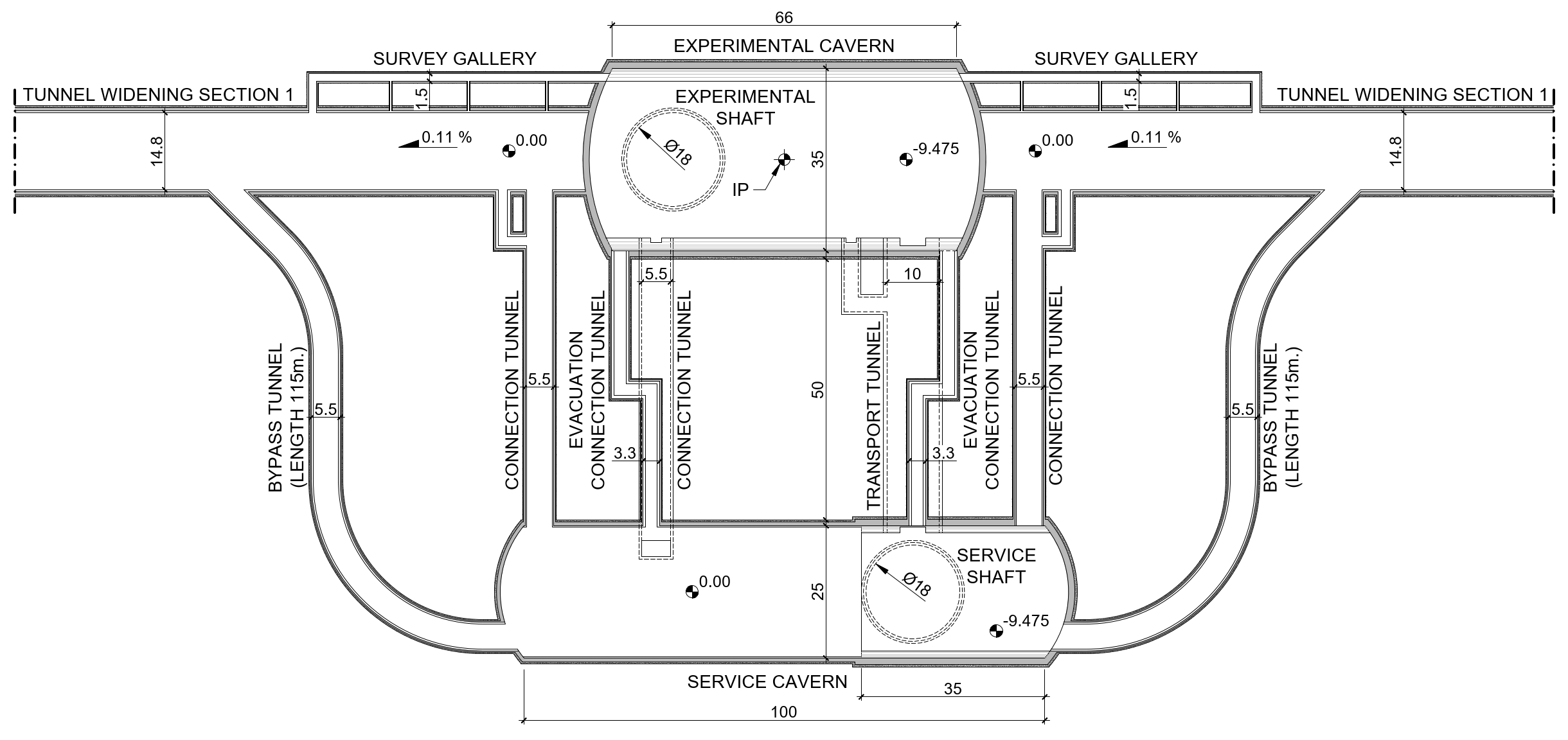}
    \caption{Plan view of PA showing the layout of connection tunnels between the service cavern and experiment cavern.}
    \label{PA}
\end{figure}

The connection tunnels between the service caverns and experiment caverns provide personnel access and materials/equipment transportation. These tunnels also house the service ducts, cables, and pipes linking the service caverns to the detectors and accelerator tunnels. Figure~\ref{PA} shows a typical layout of connection and survey galleries at an experiment point.
Two 5.5\,m diameter connection tunnels link either end of the service cavern directly to the accelerator tunnel. These tunnels have an additional forked section of tunnel, of 2.8\,m span, to allow personnel access whilst the radiation shielding doors remain closed across the full 5.5\,m tunnel section.
Two 3.3\,m span evacuation tunnels provide a safe means of escape for personnel from the experiment cavern directly to the service cavern. These tunnels are designed with a chicane to provide radiation protection between the two caverns.
At the lower level, a 5.5\,m diameter connection tunnel is required for service connections and personnel access to the floor of the experiment cavern. The transport tunnel is 10\,m in span to accommodate the movement of large equipment, and the connection tunnel has a span of 5.5\,m for personnel access and the conveyance of services between the two caverns. At each of the 4 experiment points, a 10\,m internal diameter connection tunnel between the bottom of the access shaft and the experiment cavern is required to transport large detector components to the areas inaccessible via the main shaft serving the experiment cavern.

\subsection{Shafts}\label{subsecshafts}
There are thirteen shafts proposed for accessing the underground structures:
\begin{itemize}
    \item Four 12\,m diameter shafts, one at each of the technical areas PB, PF, PH, and PL, for access and service requirements.
    \item Two 18\,m diameter shafts, one at each of the experiment areas, PA, and PG, for access and installation of the detector components in the experiment cavern.
    \item Two 15\,m diameter shafts, one at each of the experiment areas, PD, and PJ, for access and installation of the detector components in the experiment cavern. Note that these smaller experiment shafts are dimensioned to accommodate the smaller detectors planned for PD and PJ.
    \item Four 18\,m diameter shafts, one at each of the service caverns at points PA, PD, PG, and PJ, for access, service requirements and to facilitate the lowering of the largest accelerator components. 
    \item One 10\,m diameter shaft on the CERN Pr\'{e}vessin site, to enable the construction of the transfer tunnel from the Injection Complex to the FCC. This shaft will only be used for construction, in particular, to allow the assembly of the TBM to drive the 5\,km transfer tunnel length. 
\end{itemize}
\begin{figure}[ht]
    \centering
    \includegraphics[scale=0.25]{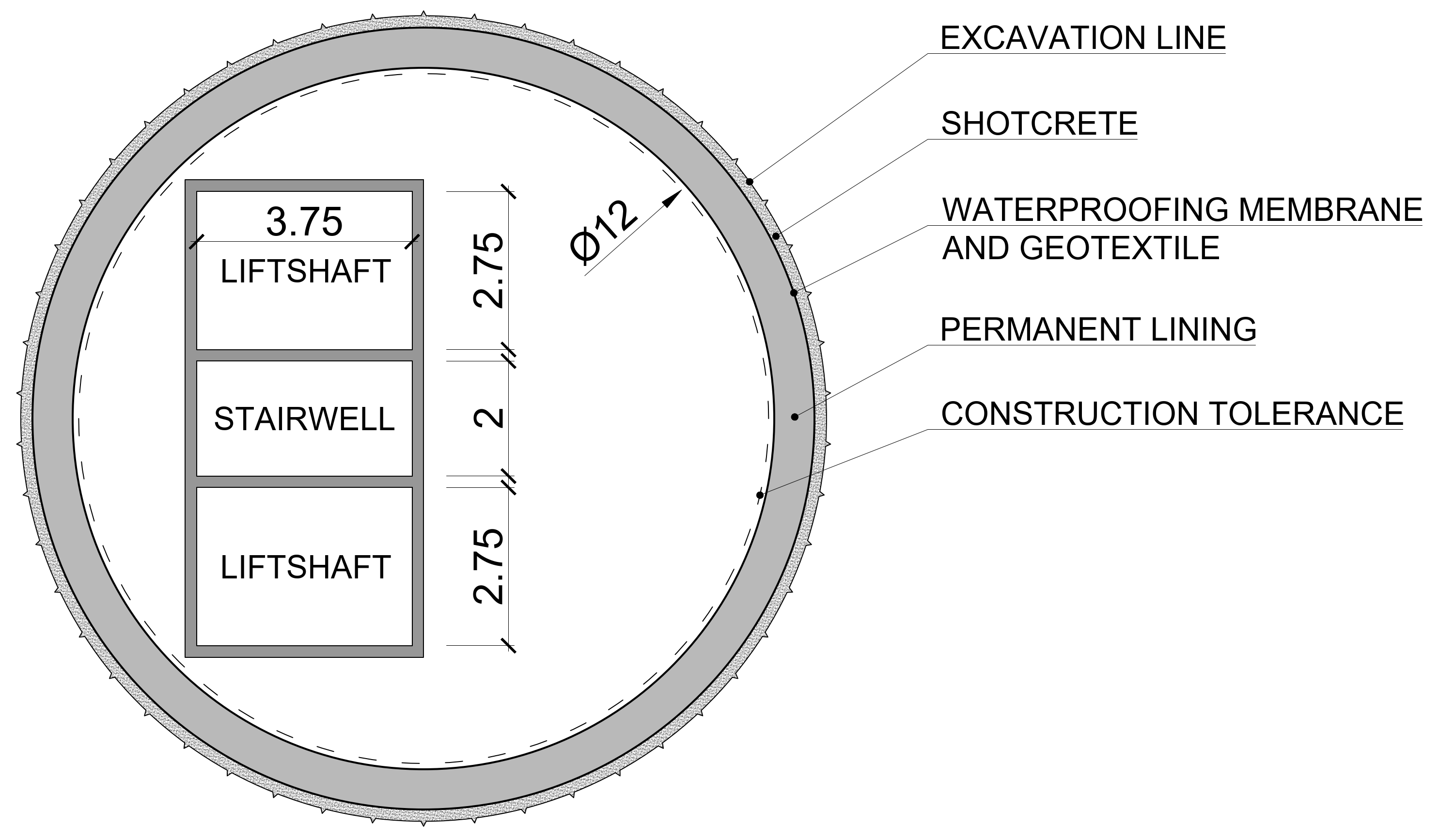}
    \caption{Cross-section through the 12\,m diameter service shaft, as proposed at the four technical areas.}
    \label{Service Shaft}
\end{figure}
The service shafts at each of the eight points provide access to the service caverns. During the construction phase, these will be used for the installation and commissioning of the infrastructure and accelerator components.  They will also be used during machine shutdowns for maintenance and upgrade of the accelerator and supporting infrastructure. They will be equipped with two lifts and a stairwell located within a pressurised inner shaft, which provides a safe escape route in case of fire. The shaft also contains a continuous vertical clear space for the crane to lower the equipment.

All four of the experiment caverns are served by either a 18\,m or 15\,m internal diameter shaft directly above the experiment cavern. These shafts primarily serve for the transportation of the detector components, which will be lowered down the shafts from the Surface hall(SX) for installation in the underground experiment caverns (UX). These shafts will also contain ventilation ducts and other services necessary to support the operation of the detector within the UX caverns. As PA and PG are the larger of the experiment areas, 18\,m diameter shafts are required. Being the smaller experiment areas, PD and PJ only require a 15\,m experiment shaft at each location. 

The construction shaft proposed for the TBM drives of the transfer tunnel may be used during the infrastructure installation phase to transfer equipment and/or components from the surface into the transfer tunnel.
\begin{table}[!ht]
    \centering
	\caption{Shaft depths at each site.}
    \begin{tabular}{lccccccccc}
    \toprule
       \textbf{Site} & \textbf{PA} & \textbf{PB} & \textbf{PD} & \textbf{PF} & \textbf{PG} & \textbf{PH} &\textbf{PJ} & \textbf{PL} & \textbf{Transfer Tunnel}\\ \midrule
        {Moraine Depth [m]} & 54 & 30 & 25 & <5& 24 & 20 & 31 & 40 & 12\\ 
        {Molasse Depth [m]} & 147 & 171 & 156 & 400 & 202 & 215 & 222 & 210 & 17\\ 
       {Total Depth [m]} & 201 & 201 & 181 & 400 & 226 & 235 & 253 & 250 & 29\\ 
        \bottomrule
    \end{tabular}
\label{ShaftTable}
\end{table}

\begin{figure}[!ht]
    \centering
    \includegraphics[width=0.8\linewidth]{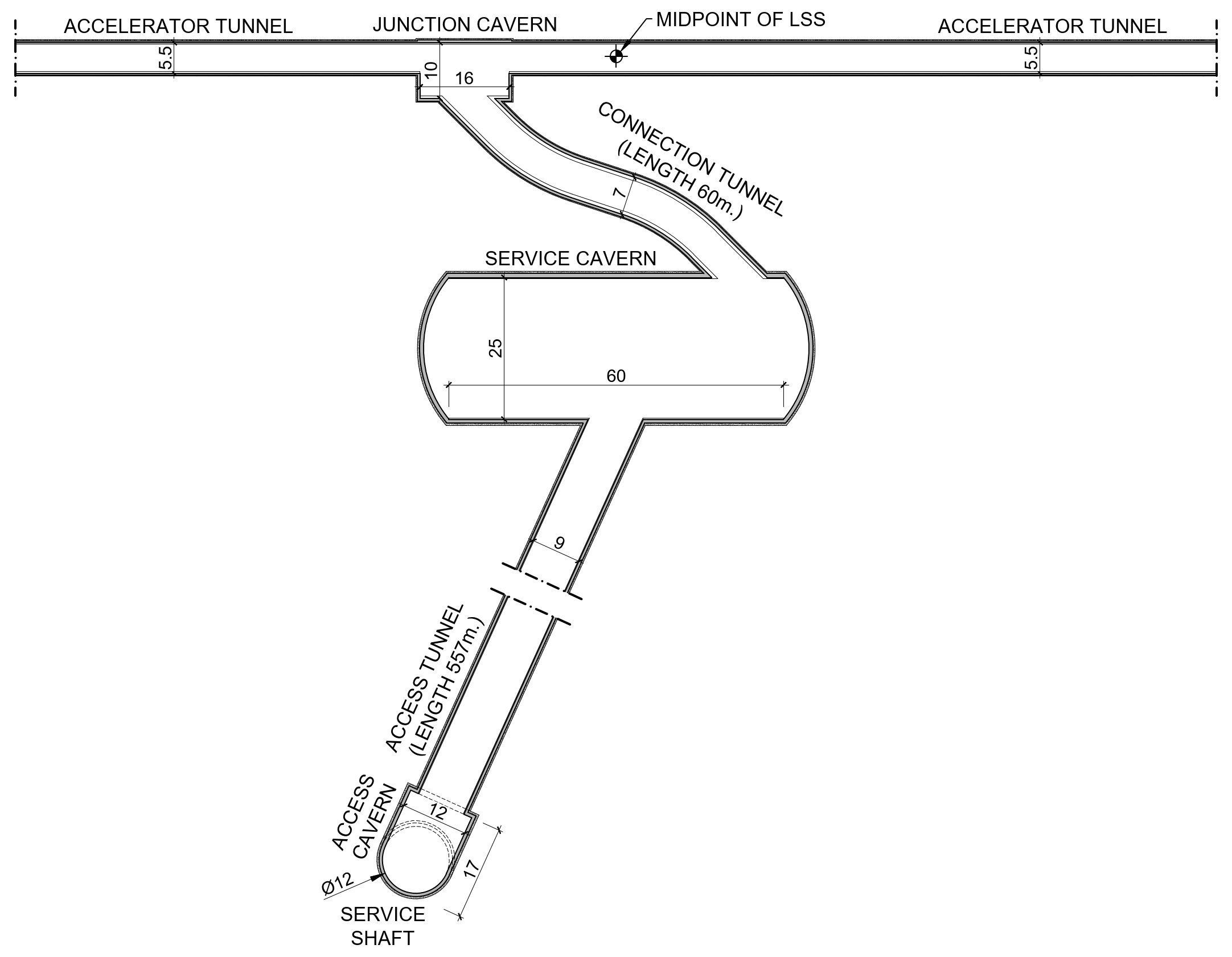}
    \caption{Plan view of the sub-surface arrangement at PF, showing the offset service shaft and access tunnel arrangement.}
    \label{PF Offset}
\end{figure}

The shafts have varying depths around the ring, ranging from 29\,m to 400\,m. Table~\ref{ShaftTable} summarises the depths of the shafts at each site, including the depths of the moraine and molasse geological layers. Generally, the shafts are located directly above the corresponding cavern. However, at PF and PB, surface constraints require the shaft to be offset from the service cavern. At PF, this is achieved by constructing a 9\,m internal diameter access tunnel 585\,m in length, as shown in Fig.~\ref{PF Offset}.  

All the shafts will be excavated through the moraine strata. Historically, supporting methods such as diaphragm walls, secant piles, and ground freezing have all been used on CERN projects and, therefore, will be appropriate for the construction of FCC shafts. 

The shaft depths below the moraine strata will be located in molasse rock. Shafts in molasse have historically been built most cost-effectively using conventional construction techniques (roadheader, hydraulic hammer) combined with shotcrete and rock bolts as primary support followed by a permanent cast in-situ reinforced concrete lining. 
\begin{figure}[ht]
    \centering
    \includegraphics[width = 0.7\textwidth]{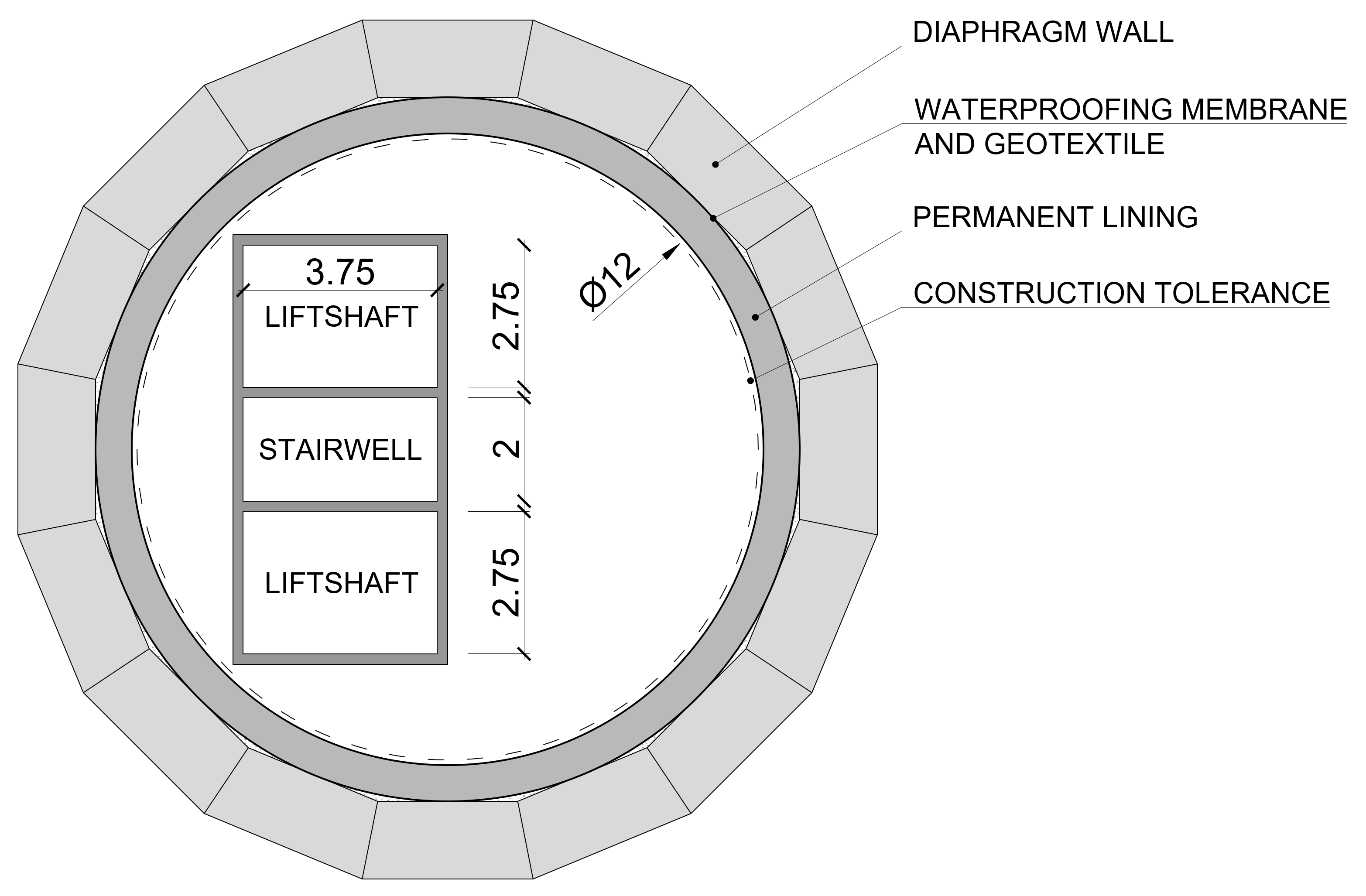}
    \caption{Cross-section of the 12\,m diameter service shaft with diaphragm wall construction.}
    \label{12mDWall}
\end{figure}
Diaphragm wall construction consists of excavating a number of straight sections of wall to form a quasi-circular shape. The sides of these excavations are supported with a temporary bentonite slurry mix, which is replaced by reinforcement cages and concrete to form the primary lining, inside which the shaft excavation can proceed. Figure~\ref{12mDWall} shows an example of the cross-section of a typical 12\,m diameter service shaft in the moraine using the diaphragm wall technique.

Once stable rock (i.e., the molasse) is reached, the excavation is supported temporarily by rock reinforcement such as rock bolts and shotcrete, before installing a waterproof membrane and the permanent cast in-situ reinforced concrete lining. The slip-forming technique can be used for the permanent lining, where the concrete is poured into a continuously climbing formwork. 

The staircases and lift shafts required within the service shafts for personnel and material access are constructed from prefabricated elements that are stacked on top of each other from the top of the shaft. These are placed as one of the final construction activities once all major civil engineering works are completed below ground and before the civil engineering structures are handed over to CERN.

As the shafts for FCC are deeper than any previously constructed at CERN, a specialist shaft-sinking consultant provided an assessment of possible excavation techniques suitable for these greater depths. The study focused on the logistical considerations of the construction and the servicing of shafts to up to 400\, m in depth (the deepest shaft at site PF). The assessment concluded that utilising current technologies and best working practices, the shafts as proposed are feasible to construct. They advised that with all shafts being constructed in parallel, procurement of sufficient specialised labour and equipment ahead of time would be key to ensuring that sufficient resources are available, thus minimising the risk of a delay to the construction schedule. The study also highlighted that there would be potential for equipment used during shaft sinking, also to be utilised during later cavern and tunnel excavation works. This could offer benefits to both the cost and the construction schedule. Furthermore, CERN was advised that advances in mechanisation and future shaft-sinking technologies could reduce construction durations as well as project risks, in particular through the recent development of a vertical shaft-sinking machine that does not require human intervention within the shaft itself.

\subsection{Caverns}\label{subseccaverns}
Large-span caverns are required at both experiments, PA and PG, to accommodate the FCC detectors and associated infrastructure. The proposed cavern dimensions are 66\,m\,$\times$\,35\,m\,$\times$\,35\,m (L$\times$W$\times$H) and the caverns will be constructed at a depth of up to 226\,m in the molasse rock. Although these will be the largest caverns ever constructed at CERN, they will not be significantly larger than the current ATLAS cavern of the LHC.
\begin{figure}[ht]
    \centering
    \includegraphics[width=0.7\linewidth]{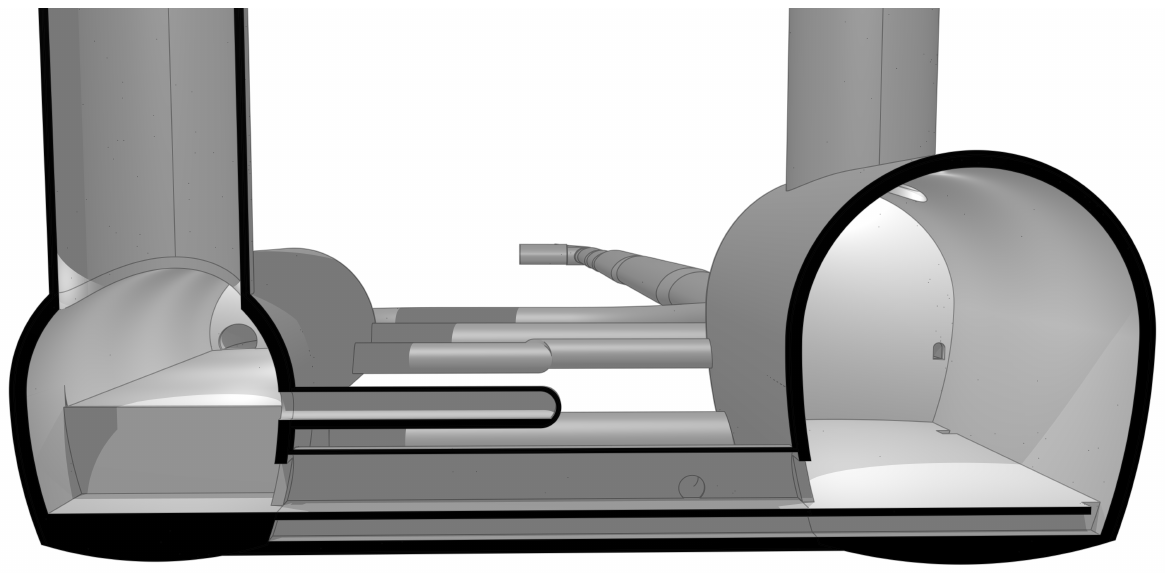}
    \caption{Cross-section through the 3D model at PA, showing the service cavern (left) and experiment cavern (right).}
    \label{PointACaverns}
\end{figure}

The construction sequence will consist of benched excavations using rock breaker and roadheader machines, with the primary support being provided by rock bolts, cable bolts and layers of steel-reinforced shotcrete. During the widening of the crown area of the experiment cavern, additional lattice girders and layers of steel-reinforced shotcrete will be installed. The lattice girders for the various excavation steps can be bolted together to ensure continuous rock support along the excavated area. The secondary lining will be constructed from cast in-situ concrete to provide additional strength and protection to the cavern walls. A waterproofing and drainage membrane will be installed between the primary and secondary linings to ensure that the cavern remains dry and that the structure is not subject to excessive water pressure arising from any groundwater that may be present.

Two smaller experiment caverns, 66\,m\,$\times$\,25\,m\,$\times$\,25\,m (L$\times$W$\times$H), are required at PD and PJ. These caverns will be constructed at up to 253\,m depth using the same techniques for excavation and support as the experiment caverns situated at PA and PG.
 
A service cavern at the same elevation as the accelerator tunnel with dimensions of 100\,m\,$\times$\,25\,m (L$\times$W) is required adjacent to each of the four experiment caverns. Below the service shaft, the height of the service caverns is 22.4\,m thereby providing direct access to the experiment cavern floor for large equipment and detector components. The remainder of the 100\,m long service cavern has a height of 15 m. Shorter service caverns of 60\,m length are necessary at the remaining four technical points. The service caverns will house infrastructure equipment such as electrical, cooling, ventilation and cryogenics. Furthermore, the caverns will provide a safe refuge in the event of an emergency, with a dedicated pressurised area at the bottom of the shaft. These caverns will be constructed in the same manner as the experiment caverns. At the experiment sites, the spacing between the two caverns is approximately 50\,m, to mitigate electromagnetic effects from the detector on the nearby electrical components. This also improves the overall structural efficiency by providing a sufficiently large rock pillar between the experiment and service caverns, thus minimising the structural support needed and reducing the risk and complexity of construction. 

The service caverns will contain three floor levels, with steel structures providing the frame for each level. This greatly increases the usable space for technical infrastructure and services. Steel structures will also be used to create the gallery levels around the detectors in the experiment caverns. These will be similar to the galleries currently in place within the LHC experiment caverns, such as ATLAS and CMS.

Where tunnels of similar cross-section dimensions connect, a junction cavern is required. PF, PH and PL each require a junction cavern where the 7\,m diameter connection tunnel from the service cavern intersects with the 5.5\,m accelerator tunnel. Each cavern is 16\,m long and at a span of 10\,m.  

An additional cavern for the FCC-ee machine beam absorber will be located at PB, and this will accommodate two beam absorbers, one for each of the two beam lines. Further details of this structure are provided in Section~\ref{subsecbeamdump}. 

Survey galleries are required at each of the four experiment caverns to survey the beam alignment on either side of the detectors. These consist of 1.5\,m span tunnels of 60\,m length, running parallel to the accelerator tunnel, with perpendicular connections made every 15\,m into the tunnel widening section 1. The possibility of incorporating these into the widened tunnel sections to reduce complexity and cost of construction will be studied in the next design phase.

\subsection{Underground structures for Radio Frequency Infrastructure}\label{subsecRF}
\begin{figure}[ht]
    \centering
    \includegraphics[width=0.9\linewidth]{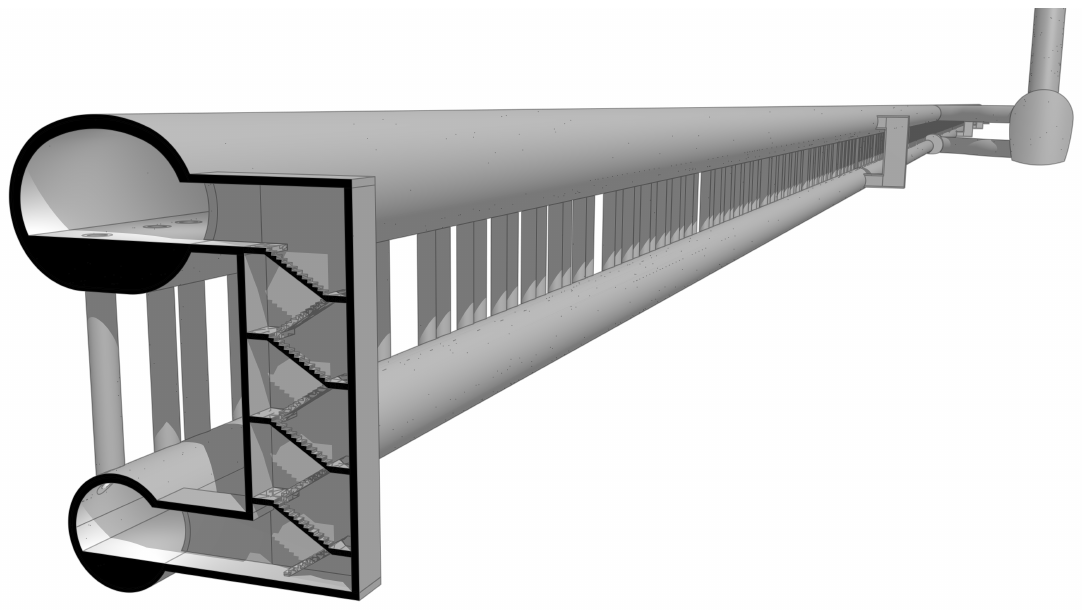}
    \caption{\label{PointH}3D model view of the klystron gallery arrangement at PH.}
\end{figure}
The klystron galleries are an essential part of the FCC-ee civil engineering, as they will house the klystrons and other components of the Radio Frequency (RF) system. To accommodate the necessary infrastructure, two separate galleries will be constructed, one at PH and the other at PL. The klystron gallery at PH (Fig.~\ref{PointH}) will be 2012\,m long for the collider RF system, while the gallery at PL will be 1446\,m long for the booster RF system. To allow access to the klystron galleries during the operation of the FCC-ee machine, sufficient radiation shielding needs to be provided. To achieve this, 10\,m of rock will be maintained between the RF galleries and the accelerator tunnel. The internal dimensions of the galleries are 10\,m in span and 5.5\,m high to accommodate the transport and placement of the equipment within the galleries, as well as allowing access for maintenance and repair during the operation phase. 

The galleries are sized with a 50\,m extension in length at either end to incorporate the space for service equipment. This is proposed instead of constructing the large alcoves along the accelerator tunnel, thereby enabling a more efficient use of tunnel excavation and space.

\begin{figure}[ht]
    \centering
    \includegraphics[scale=0.8]{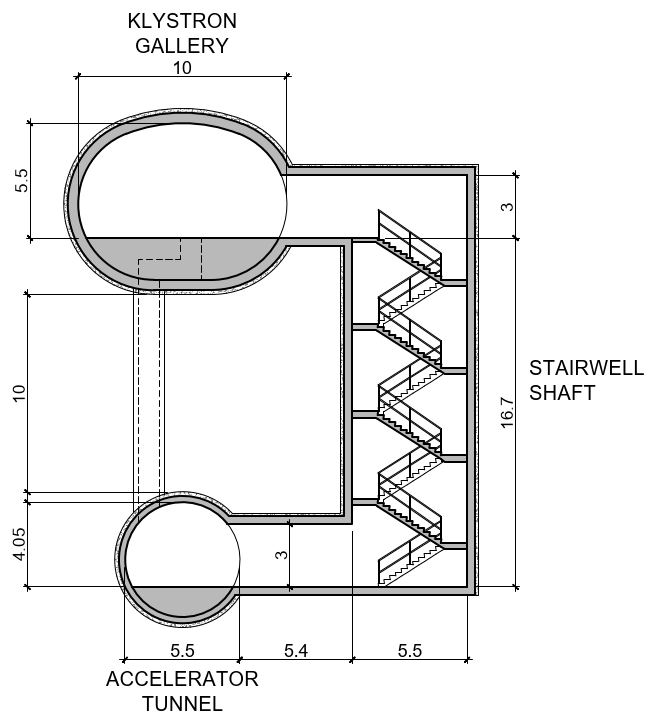}
    \caption{\label{KlystronCrossSection}Cross-section through the klystron gallery and accelerator tunnel, including wave-guide duct between.}
\end{figure}

The galleries are connected directly to the accelerator tunnel via 1\,m internal diameter wave-guide ducts, spaced every few metres along the gallery (see Fig.~\ref{KlystronCrossSection}. There will be 428 individual wave-guide ducts at PH and 312 at PL. 

The safety concept for FCC requires that for emergency egress, stairwell shafts are spaced at intervals of 341\,m along the gallery to connect the galleries to the accelerator tunnel. 

For access to the klystron galleries, a 7\,m internal diameter connection tunnel links the upper level of the service cavern and the klystron gallery. This connection tunnel has a length of 62\,m.  The size of the connection tunnel is designed to accommodate personnel access and the transport of the RF equipment into the gallery. 

The klystron galleries will be excavated using the same road header type machines that will be used for the excavation of the caverns and connection tunnels at points PH and PL. Roadheader excavation is efficient and accurate, allowing precise excavation of the galleries. 
The klystron gallery lining will consist of a shotcrete primary lining and a cast in-situ concrete secondary lining. It may be feasible to use shotcrete for the final lining of the gallery, as this would potentially offer faster construction and lower cost. This option will be studied during the next phase. The floor of the gallery will also be made of cast in-situ concrete and will include a drainage channel and the entry of the waveguide ducts up through the floor slab. The entry of the wave-guide ducts into the gallery will require the use of a chicane to provide radiation protection. This will likely be constructed with a precast concrete unit placed after the final installation of the cryogenic line (QRL) infrastructure through the duct.

The wave-guide ducts, which will contain the wave-guides linking the klystrons to the RF cavities in the accelerator tunnel, will be excavated by raise boring. This process involves first drilling a pilot hole and then attaching a reamer head at the bottom, which is pulled up, excavating the duct from below in a vertical direction. This method of excavation is efficient and minimises disruption to the surrounding rock. A steel or concrete lining will be grouted into the rock to provide a suitable finished surface and final lining.

\subsection{Alcoves}\label{subsecalcoves}
\begin{figure}[ht]
    \centering
    \includegraphics[scale = 0.55]{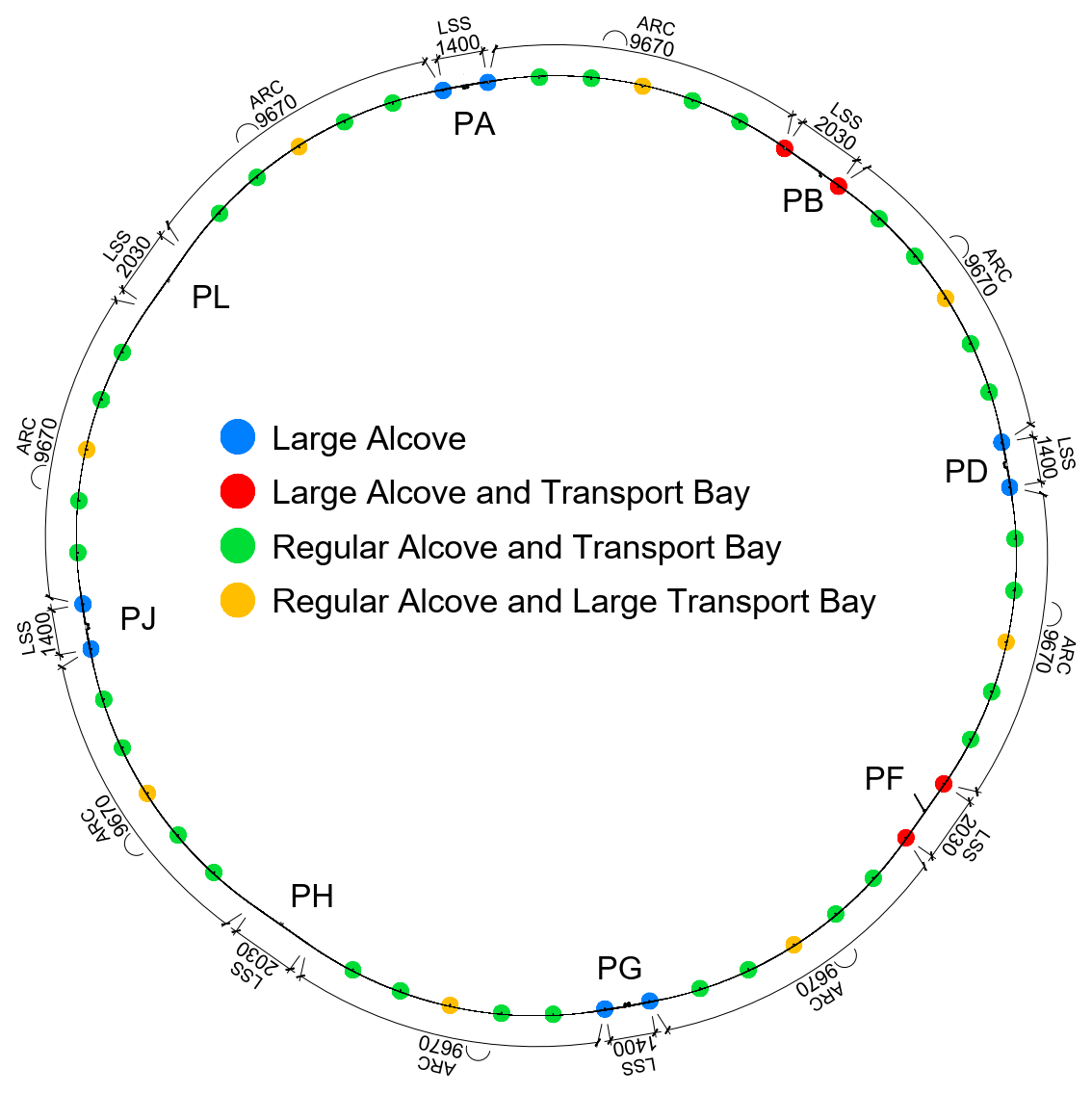}
    \caption{\label{regularalcove}Layout of alcoves around the FCC ring.}
\end{figure}

At 1.6\,km centres around the circumference of the machine, equipment alcoves, as illustrated in Fig.~\ref{regularalcove}, are required to accommodate electrical equipment, services and transport needs. The majority of these alcoves are considered \lq{regular}\rq, measuring 40\,m in length, 10.6\,m in width, and 4.6\,m in height. Positioned on the inside of the ring, the alcoves are arranged perpendicular to the accelerator tunnel. 

\begin{figure}[ht]
    \centering
    \includegraphics[width=\linewidth]{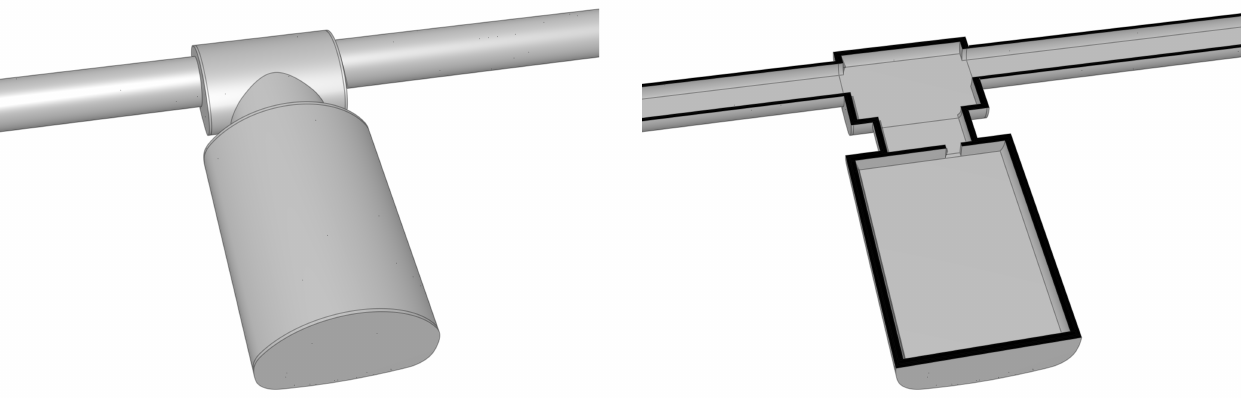}
    \caption{\label{LargeAlcove} Model view of a large alcove and transport passing bay.}
\end{figure}

In addition to the regular alcoves, twelve \lq{large}\rq alcoves are needed on either side of each of the FCC access points to provide extra space for electrical equipment. These larger alcoves are 29\,m in length, 18\,m in width and 8.3\,m in height. These large alcoves will be located at the end of the long straight sections on either side of the caverns. Two large alcoves are required at PA, PB, PD, PF, PG, and PJ, but not at PH or PL, because the additional size of the klystron galleries provides the necessary volume for the equipment. An example of a large alcove is shown in   Fig.~\ref{LargeAlcove}. In total, the project requires 40 regular alcoves and 12 large alcoves.

An access area to each alcove is provided to accommodate the radiation protection chicane walls. This section is 6\,m long and has a 10.6\,m span at the entrance to both large and regular alcove types.

\begin{figure}[ht]
    \centering
    \includegraphics[scale=0.6]{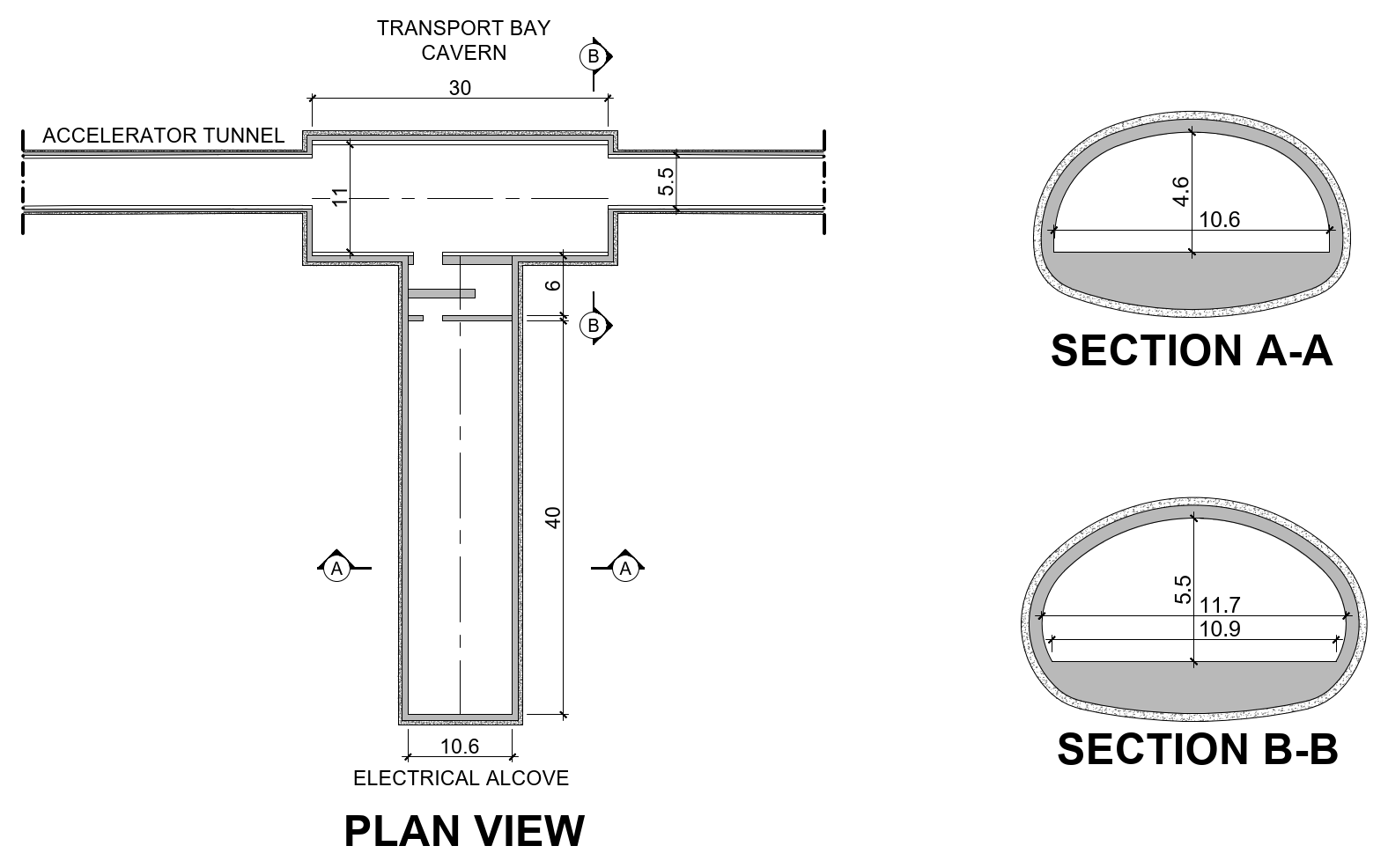}
    \caption{\label{TransportPassingBay}Example of a regular alcove and large passing bay located at the centre of each arc sector.}
\end{figure}

A transport-passing bay is required at the entrance to each alcove. This will facilitate the passing of transport vehicles travelling in opposite directions through the tunnel. The passing bay also provides space for the parking of transport vehicles during installation and shutdowns, with additional space for servicing/repairing the vehicles if necessary. The majority of the passing bays are accommodated in a cavern along the accelerator tunnel alignment and measure 16\,m in length and 11\,m in width. Additional larger passing bays are required at the mid-section of each accelerator tunnel arc. There will, therefore, be a total of 8 larger passing bay caverns, measuring 30\,m by 11\,m. These facilitate the passing of the magnet delivery vehicles during the installation phase of the FCC-ee machine. An example of the large passing bay is shown in  Fig.~\ref{TransportPassingBay}.

Unlike the caverns and tunnels in proximity to the FCC points, which can be at least partially excavated in parallel with the TBM drives, excavation of the alcoves and passing bays will need to be completed after the TBM drive is complete. This is because the full tunnel section is required to support the TBM (ventilation, conveyor for spoil removal, transport corridor for concrete segmented linings, corridor for personnel etc.). It will, therefore, be necessary to break out the concrete tunnel lining and excavate the passing bay cavern and alcoves using roadheader excavation. The inner lining works for the alcoves will then be carried out ahead of the relining of the accelerator tunnels. This will have to be coordinated with the installation of the tunnel floor. The overall construction sequence and structural design will be similar to the caverns, as detailed in  Section~\ref{subseccaverns}.

The complexity of the alcove construction and the associated logistical challenges have a major impact on the construction schedule, since each alcove is potentially on the critical path for civil engineering. CERN has commissioned an additional study to investigate the optimal method of constructing the alcoves in order to minimise the impact on the overall construction schedule. As a result, the current assumption in the schedule remains that alcoves can be constructed in parallel to one another, maintaining sufficient access, safety and logistical standards. By adopting an overlapping sequence of construction activities, all alcoves of each sector can be constructed within a 12-month period, thereby achieving the targeted handover date of each sector for the subsequent infrastructure and accelerator installation. 

\subsection{Beam absorber cavern}\label{subsecbeamdump}
The absorbers for the FCC-ee beams will be located at PB. The beam absorber cavern layout includes a 708\,m long cavern with a span of 13.6 m, to house both the e$^{+}$ and e$^{-}$ absorber infrastructure. The size of the cavern has been dictated by the beam extraction length and angle, which requires a separation from the accelerator line of at least 5.5\,m and a septum/kicker angle of 10\,mrad. The two beam extraction lines are arranged to cross at the centre of the cavern, adjacent to the centre of the PB long straight section. This ensures that the beam absorber cavern volume is optimised. 

To accommodate the separation of the two FCC-ee beams on the approach to PB, a series of tunnel widening sections is required either side of the cavern. Three sections of tunnel widening, of lengths from 83 to 330\,m will be constructed either side of the cavern to increase the span of the accelerator tunnel from 5.5 to 7.8\,m.

The absorber cavern will be excavated in the same manner as other caverns, using roadheader and/or rock breaker machines to excavate from the roof level down using a series of `benches'. Due to the constraints at the surface, the service shaft and cavern will need to be offset from the centre point of the long straight section (LSS). The offset service shaft will, therefore, be positioned 350\,m from the centre of the LSS and connect to the beam absorber cavern via a connection tunnel of 97\,m length. 

\begin{figure}[ht]
    \centering
    \includegraphics[width=\linewidth]{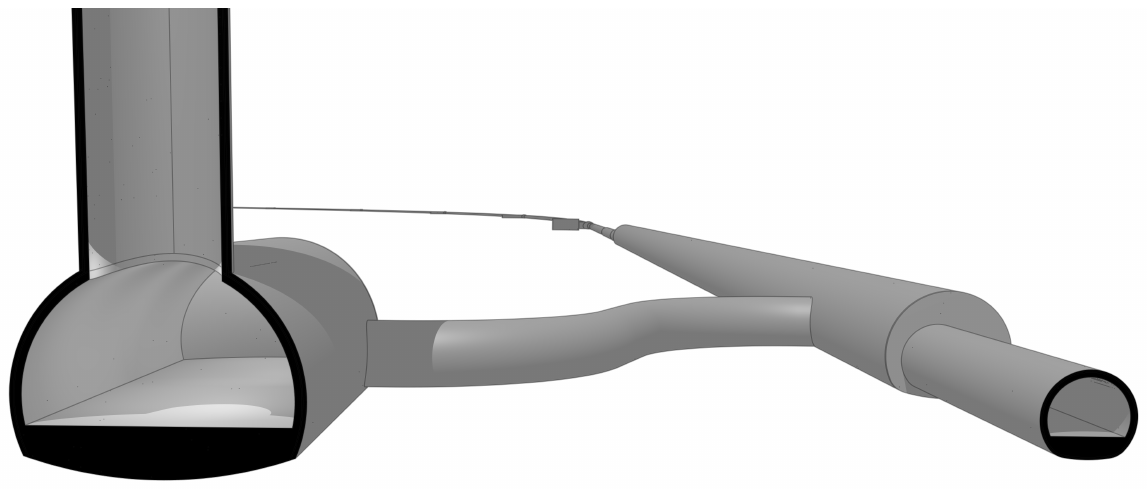}
    \caption{\label{beamabsorbermodel}Sub-surface structural layout at PB, service cavern (left) and beam absorber cavern (right).}
\end{figure}

The absorber cavern will be constructed as mentioned in Section~\ref{subseccaverns}. There will be a shotcrete and rock bolt primary lining with an in-situ concrete secondary lining. However, the option to create the final lining from shotcrete will be considered in future design development to potentially reduce costs and reduce the construction time. It is yet to be confirmed whether a shielding wall/structure is required within the cavern to shield the accelerator from the beam absorber apparatus; however, this would be constructed with precast concrete blocks well after the civil engineering has been completed. 
A schematic view of PB including the beam absorber cavern is shown in Fig.~\ref{beamabsorbermodel}.

\subsection{Excavated material}

To construct the subsurface structures, approximately 6.3 million\,m$^3$(in-situ volume) of rock will be excavated. This material will be extracted via the eight FCC access points. The predominant rock type will be molasse, accounting for 96\% of the total material. Moraine rock accounts for 1.5\% of the excavated material, and limestone makes up the remaining 2.5\%.

Figure~\ref{TBM Arrangement} in Section~\ref{subsecmainbeamtunnel} shows the baseline arrangement for the TBM drives. As mentioned, the accelerator tunnel sectors will be excavated using TBM. Shafts will be excavated by conventional mined excavation or vertical shaft-sinking machines. The use of diaphragm walls will be necessary to support the excavation of shafts through the initial moraine layer. All other excavations for subsurface structures will be constructed by hydraulic hammer (rock breaker) and roadheader machines. The physical characteristics of the excavated material will differ according to the excavation method utilised.

Table~\ref{ExcavationTable} provides a breakdown of the excavated material quantities extracted from each of the eight sites and the transfer line between Pr\'{e}vessin and the FCC tunnel. The volumes are expressed as the in-situ volume i.e., the volume prior to excavation, the bulked volume i.e., the volume after excavation and the compacted volume i.e., the volume after the material has been re-compacted for example in a spoil deposit zone. 

\begin{table}[!ht]
    \centering
    \caption{Excavation volumes for each FCC sector.}
    \label{ExcavationTable}
    \begin{tabular}{lcccr}
    \toprule
        ~ & In-situ Vol,  & Bulk Vol., & Compacted Vol.,  & \% of Total \\   
        ~ &  10\,$^{3}$\,m$^3$ & 10$^3$\,m$^3$ & 10$^3$\,m$^3$ & ~ \\ \midrule
        PA &  1378&  2205&  1791& 22\%\\ 
        PB &  148&  237&  192& 2\%\\ 
        PD &  1274&  2038&  1656& 20\%\\ 
        PF &  165&  264&  215& 3\%\\ 
        PG &  1365&  2184&  1775& 22\% \\ 
        PH &  312&  499&  405& 5\% \\ 
        PJ &  1289&  2062&  1675& 20\%\\ 
        PL &  241&  386&  313& 4\%\\ 
        Transfer Tunnel&  122&  195&  159& 2\%\\
        \bottomrule
    \end{tabular}
\end{table}

Two TBM drives will commence at PA, one driven towards PB and the other driven towards PL. The resulting total material extracted from PA will be almost 1.4 million m$^3$, of which, 62\,721\,m$^{3}$ is expected to be moraine.

No TBM drives are planned from PB, the volume of excavated material, therefore, arises from the subsurface structures at PB, including the beam absorber cavern. Of the total quantity of, 147\,852\,m$^3$, 10\,473 \,m$^3$ is expected to be moraine and the remainder is molasse.

Two TBM drives will commence at PD, one driven towards PB and the other towards PF. The resulting quantity of material to be extracted from PD is 1.3 million\,m$^3$ of which, 24\,925\,m$^3$ is expected to be moraine. 

No TBM drives are planned from PF. As with PB, the excavated material at PF only arises from the subsurface structures directly located at PF and not the accelerator tunnels either side. Almost all the excavated material at PF will be molasse as there are little or no quaternary deposits.

Two TBMs will be driven from PG, one driving towards PF and the other towards PH. The total quantity of material excavated from PG is almost 1.4 million\,m$^3$. Moraine makes up 30\,829\,m$^3$ of the total excavated material at PG. However, due to the 4.4\,km long sector of limestone along the accelerator tunnel between PG and PH, 141\,175\,m$^3$ of the total material will be limestone rock. The remaining 1.2\,million\,m$^3$ of excavated material will be molasse.

No TBM drives are planned from PH, therefore all the excavated material comes directly from subsurface structures at PH, including the klystron gallery and associated structures. 7482\,m$^3$ of the total excavated material at PH is expected to be moraine, and the remainder is molasse.

Two TBMs will be driven from PJ, one driven towards PH and the other towards PL. Therefore, the quantity of material extracted from PD is around 1.3\,million\,m$^3$ of which, 29\,910\,m$^3$ is expected to be moraine. 

No TBM drives are planned from PL, the volume of excavated material is therefore attributed to the subsurface structures at PL, including the klystron gallery and associated structures. Moraine rock is expected to make up 13\,468\,m$^3$ of the total excavated material at PL.

The Pr\'{e}vessin to SPS and SPS to FCC injection tunnels account for 122\,329\,m$^3$ of excavated material, the majority of this material will be extracted from the Pr\'{e}vessin site, by means of the proposed construction shaft. A small proportion of the excavated material will be moraine, from the initial excavation of the shaft, but the majority of the shaft and transfer tunnel excavation will be in molasse rock.

\subsection{Tunnelling in the molasse rock}

\begin{figure}[ht]
    \centering
    \includegraphics[scale=0.8]{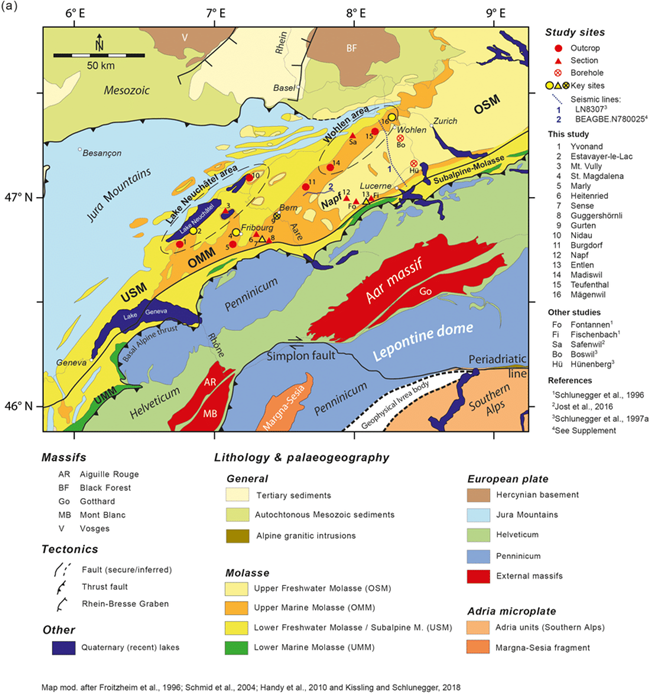}
    \caption{\label{Molasse Map}Geographical map of Switzerland. Credit: Philippos Garefalakis, F. S. (2019). Tectonic processes, variations in sediment flux, and eustatic sea level recorded by the 20\,Myr old Burdigalian transgression in the Swiss Molasse basin.}
\end{figure}

The large-scale geological environment within which the FCC underground infrastructure will be excavated is shown in Fig.~\ref{Molasse Map}, with most of the tunnel located within the Lower Freshwater Molasse (USM) and Lower Marine Molasse (UMM). The FCC will be excavated at depths of up to 560\,m, with a total combined length of all excavations exceeding 100\,km. Whilst the experience gained by CERN over the previous 50 years for the construction of the LEP, LHC and Hi-Luminosity LHC has given valuable insight into the likely characteristics and behaviour of the molasse rock, other projects excavated within the molasse rock have also been the subject of a desktop study in order to improve the understanding of the environment in which the FCC civil engineering will be constructed.

 Most of the projects reviewed were transport tunnels (road and rail) from a few hundred metres to a few kilometres in length at shallow depths. Tunnels were excavated by means of conventional excavation, cut and cover, and TBM. One project in particular, the Moutier tunnel, was excavated by a single shield TBM in the Alsace tertiary molasse (a mix of marl, sandstone and limestone). Due to an unforeseen geological environment, the TBM became stuck 190\,metres into the tunnel drive. As a result, the remainder of the tunnel was excavated using conventional methods. This project highlights the need for sufficient geotechnical investigations and appropriate analysis to be completed before construction begins. At the greater depth of FCC, the molasse is likely to be more consolidated and unlikely to have been altered due to weathering. However, appropriate geotechnical investigations and numerical analysis will be required to better predict the likely behaviour of rock and its influence on the performance of the tunnel boring machine.

Generally, the review confirms that excavations in molasse rock have been extensively conducted in Switzerland using both TBM and conventional methods. However, the FCC will require a comprehensive geotechnical investigation to assess the characteristics of molasse as well as the in-situ stress regime. The investigation will address potential risks, including those arising from changes in pore water pressure, rock squeezing, and geological faults.

A tunnelling project has been identified in the UK with characteristics similar to those of the FCC. It is currently under construction as part of a new Halite mine development. A single TBM excavates a 37\,km long tunnel with cross-sectional dimensions that are very similar to those of the FCC. Furthermore, the tunnel is being excavated at comparable depths and within sedimentary geology that is not dissimilar to that expected for the FCC. An average tunnelling excavation rate of 20\,m/day have been achieved. CERN will assess this project for lessons learned in more detail during the next phase of the FCC.

In conclusion, it is considered that underground civil engineering for the FCC is feasible and within the current experience and capabilities of many of the major designers and contractors currently active within the CERN member states. Additional site investigations will be required to finalise the precise depth and inclination of the FCC tunnel and to determine the geotechnical properties of the rock mass, which are necessary for CERN to move forward to the detailed design and construction phases.  

\section{Surface structures}\label{secsurfacestructures}

\begin{figure}[ht]
    \centering
    \includegraphics[scale=0.25]{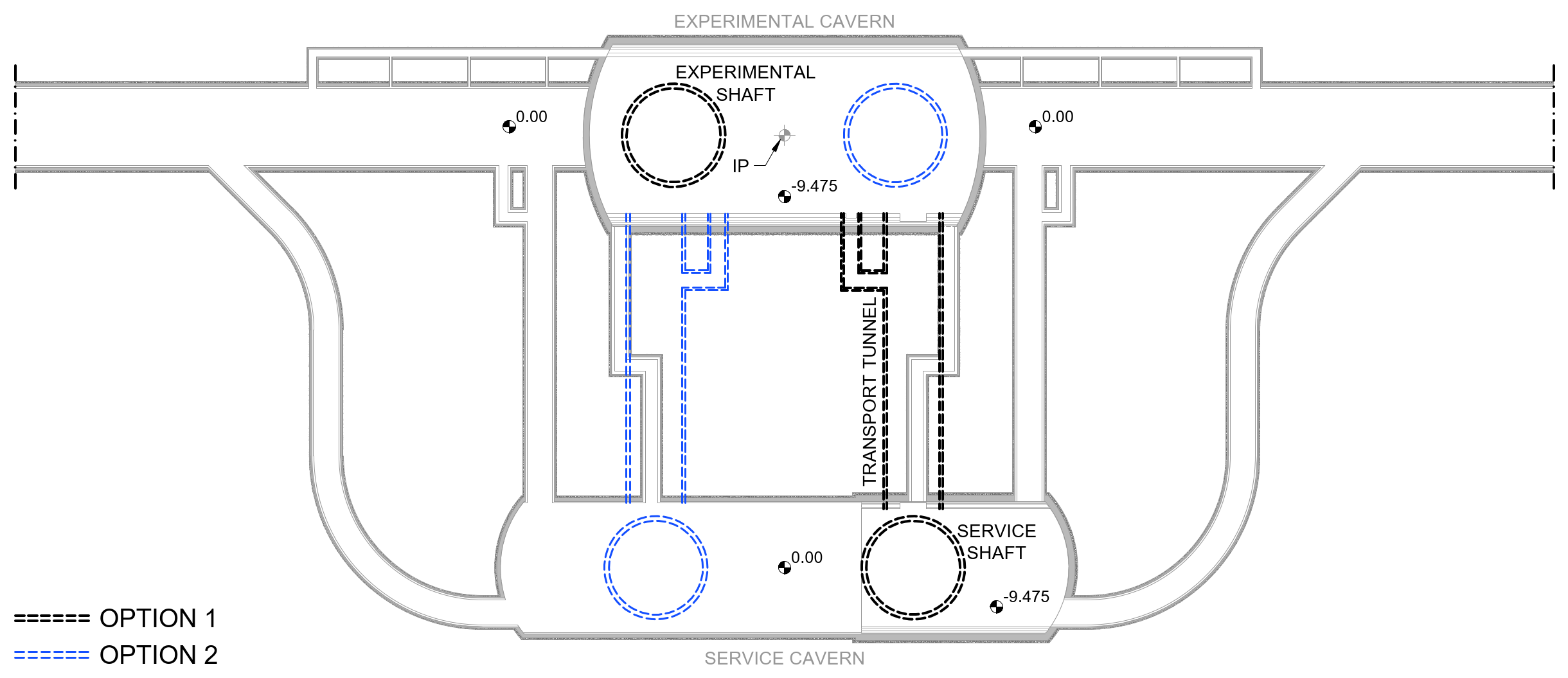}
    \caption{\label{relative_shaft_locations} Options for experiment and service shaft locations relative to the interaction point.}
\end{figure}

Since the completion of the conceptual design, there has been a rationalisation of the surface sites associated with the FCC-ee civil engineering, with the number of surface sites reduced from twelve to eight. The eight surface sites will comprise four areas suitable for siting experiments and four assigned as technical areas. The eight sites are spread evenly around the circumference of FCC-ee collider ring. The four experiment areas are symmetrically distributed such that each site is diametrically opposite another experiment site, as illustrated in Fig.~\ref{undergroundschematic}. The experiment sites are at sites PA, PD, PG and PJ with the technical sites at the remaining sites PB, PF, PH and PL.
The location of the four experiment surface sites are interdependent since the interaction points for each of the four sites define the location of the experiment and service shafts at each site as illustrated in Fig.~\ref{relative_shaft_locations}

The experiment access shafts can only be in one of two locations with respect to the interaction point, and the machine access shafts should ideally be placed directly above the service cavern, which itself needs to be located approximately 50\,m from the experiment cavern for structural stability of the two caverns and for electromagnetic shielding. Since each of the four interaction points associated with the experiment areas is located precisely with respect to each other, the locations of the shafts at each experiment point are also fixed with respect to the shafts at the other experiment points. This interdependency across the four experiment sites influences the layouts of the surface sites, since the two shafts are associated with specific building configurations. 

The selection of surface site locations for the experiment areas requires careful coordination, as each site is interconnected with the location of the other three. For the technical sites, while proximity to the centre of the associated long straight section (LSS) is a key factor, some flexibility exists. The single machine access shaft can be linked to the accelerator tunnel via an access gallery, which, if required, can extend several hundred metres in length.  

For a comprehensive discussion on environmental considerations, including landscape integration, readers are referred to the relevant sections in Chapters 2 and 3 in this Volume.

The engineering designs of the surface site constructions that are eventually part of the project authorisation files need respond to the requirements of the equipment that will eventually be housed at each site. Noisy equipment will be adequately noise insulated and placed in constructions that permit containing the residual noise within the applicable regulatory frameworks in place at the time. Examples of structures that may require such specific measures include:
\begin{itemize}
\item Cryogenic plants comprising compressors.
\item Cooling and ventilation equipment with pumps, motors and fans.
\item Electrical equipment such as transformers.
\end{itemize}

The choice of construction materials and techniques for surface site buildings is guided by multiple factors, including durability, safety, cost and environmental considerations such as landscape integration and regulatory constraints such as urbanism prescriptions. Newly emerging materials and technologies, as well as continuous monitoring of the environmental and regulatory prescriptions, including applicable energy efficiency standards, will be considered during the subsequent design phase. In line with CERN's governance framework, the territorial principle applies, i.e., the national laws and guidelines apply for infrastructure on the surface sites in the Host States.

At this stage, no architectural strategies have been adopted. This will be done in the next design phase. The development of the architectural solutions will include a robust consultative process with the Host States, regional stakeholders and local communities. A more detailed discussion can be found in section~\ref{sec:HostStateStudies}

For technical feasibility and costing purposes. the U.S. Fermi National Accelerator Laboratory has analysed the technical needs of two of the eight surface sites in the framework of the international FCC collaboration. The study used an experiment site (PA) and a technical site (PB) as generic examples. The work resulted in the production of detailed drawings of the buildings at the sites. These designs were used as the basis for an initial cost estimate for the surface works. It should be stressed that these designs represent preliminary input for technical and financial feasibility and construction planning purposes only and are not approved designs. 

The subsections below present simplified versions of these concepts that permit further development of site configurations and cost estimates.

The chapters ~\ref{sec:HostStateStudies} and \emph{Territorial implementation} provide a further detailed examination of the contextual considerations, planning strategies, and logistical challenges associated with the surface sites, offering further insight into how these elements are being harmonized within the broader project vision.

\subsection{Surface site - PA}\label{surfacePA}
The surface site PA will be the location for one of the four experiment areas.
The site is located close to the existing CERN LHC surface site P8 which allows the re-purposing of some of CERN's existing LHC facilities and infrastructure to support the FCC-hh activities at site PA.  

The buildings and other necessary civil engineering surface infrastructure for all surface sites have been identified and included within the civil engineering product breakdown (PBS) structure, which is shown for PA only in Table~\ref{PBSforPA}.

\begin{table}[!ht]

    \centering
\caption{PBS for PA surface structures.}
\label{PBSforPA}
    \begin{tabular}{cccccl}
    \toprule
    \multicolumn{5}{c}{\textbf{PBS Level}} & \multirow{2}{*}{{\textbf{PBS Description}}} \\
            \textbf{0} & \textbf{1} & \textbf{2} & \textbf{3} & \textbf{4} & ~ \\
            \midrule
        ~ & \textbf{2} & ~ & ~ & ~ & \textbf{Civil Engineering} \\ 
        ~ & \textbf{2} & \textbf{2} & ~ & ~ & \textbf{Surface Structures} \\
        ~ & \textbf{2} & \textbf{2} & \textbf{1} & ~ & \textbf{Site PA} \\ 
        ~ & 2 & 2 & 1 & 1 & Roads, parking spaces, footpaths, fences, gates, landscaping, drainage \\ 
        ~ & 2 & 2 & 1 & 2 & Technical galleries \\ 
        ~ & 2 & 2 & 1 & 3 & Fire fighting equipment and medical station \\ 
        ~ & 2 & 2 & 1 & 4 & Control building \\ 
        ~ & 2 & 2 & 1 & 5 & Assembly hall \\ 
        ~ & 2 & 2 & 1 & 6 & Magnet storage \\ 
        ~ & 2 & 2 & 1 & 7 & Shaft head building \\ 
        ~ & 2 & 2 & 1 & 8 & Tunnel ventilation \\ 
        ~ & 2 & 2 & 1 & 9 & Chilled water production facility \\ 
        ~ & 2 & 2 & 1 & 10 & Experiment cavern ventilation \\ 
        ~ & 2 & 2 & 1 & 11 & Service cavern ventilation \\ 
        ~ & 2 & 2 & 1 & 12 & Cooling plant and waste heat recovery \\ 
        ~ & 2 & 2 & 1 & 13 & SF annex \\ 
        ~ & 2 & 2 & 1 & 14 & Warm compressor \\ 
        ~ & 2 & 2 & 1 & 15 & Helium gas tanks foundations \\ 
        ~ & 2 & 2 & 1 & 16 & Liquid Nitrogen storage tank foundations \\ 
        ~ & 2 & 2 & 1 & 17 & Power converters building \\ 
        ~ & 2 & 2 & 1 & 18 & Electrical equipment building \\ 
        ~ & 2 & 2 & 1 & 19 & Electrical substation \\ 
        ~ & 2 & 2 & 1 & 20 & Waste material storage building \\
        \bottomrule
    \end{tabular}
\end{table}

A significant feature of this and all experiment area surface sites is the large assembly hall that will be used for the sub-assembly and preparation prior to the transfer of the detector components to the underground experiment cavern. It is to be noted that any future assembly hall for an FCC-hh detector may need to be larger than that for an FCC-ee detector, and therefore, a footprint is reserved on the site for such a future expansion/reconfiguration of the assembly hall. The final details of a conceptual design for this building at the FCC-hh phase will be determined in parallel with a future fabrication and assembly strategy for the associated detector.  
\begin{figure}[!ht]
    \centering
    \includegraphics[width=\linewidth]{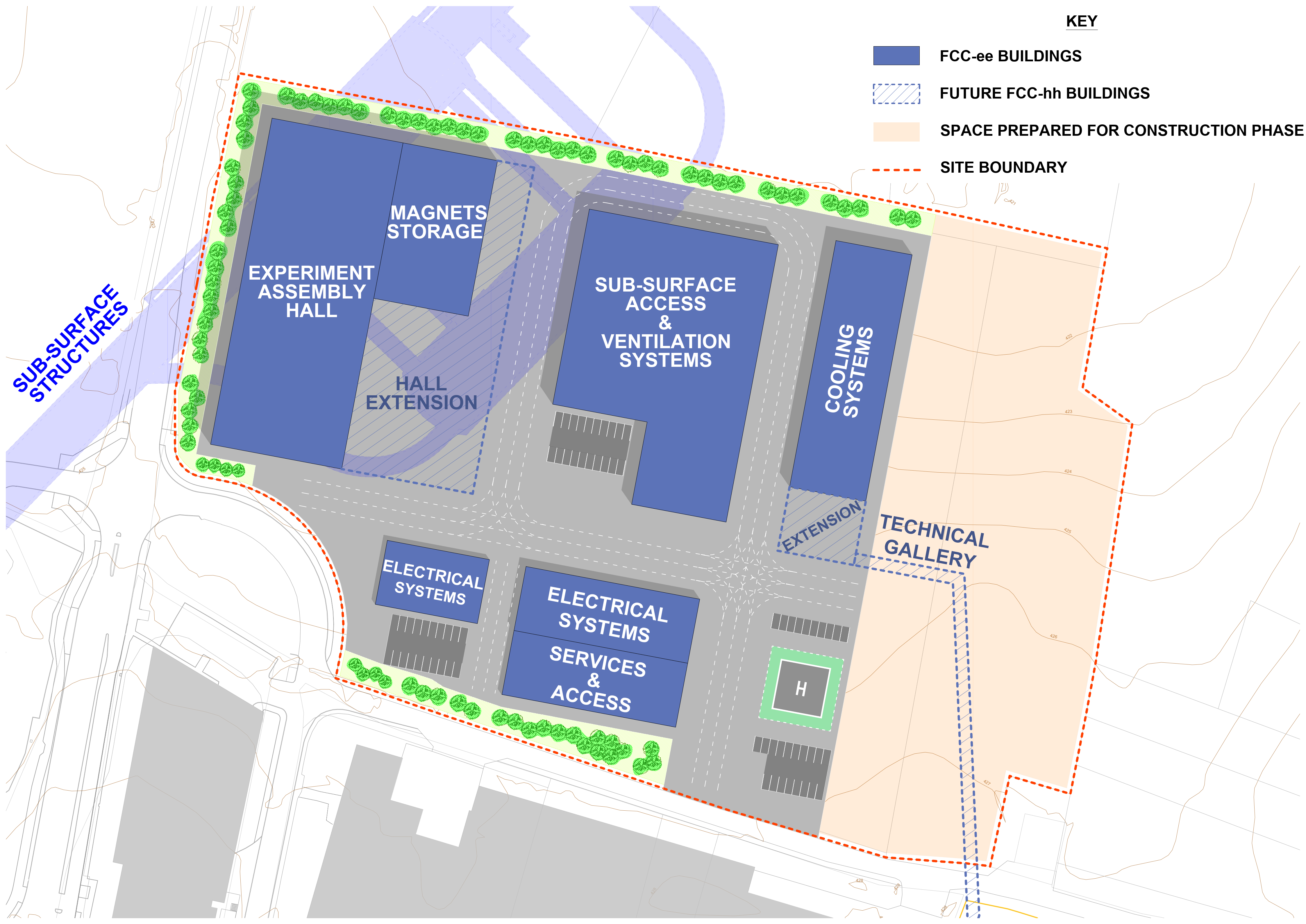}
     \caption{\label{pasurfacesimplified}Preliminary simplified surface requirements for PA site.}
\end{figure}

\begin{figure}[!ht]
    \centering
    \includegraphics[width=\linewidth]{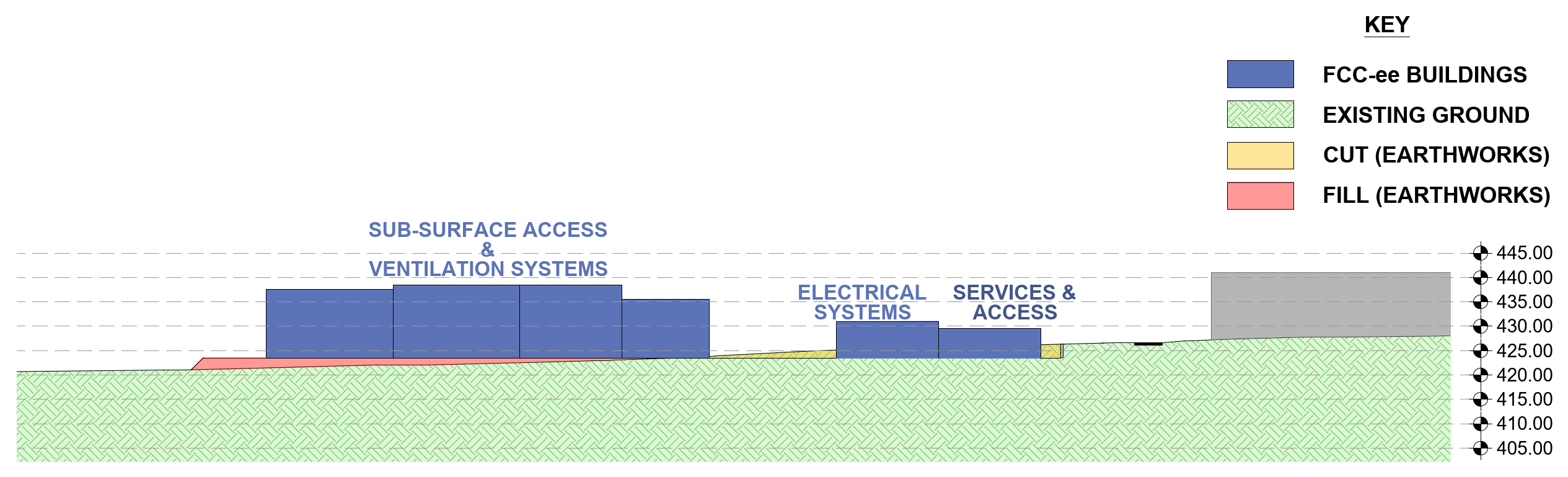}
     \caption{\label{PAsurfacesectionsimplified} Preliminary cross-sectional diagram of PA site.}
\end{figure}

The site location has some specific constraints, including:

\begin{itemize}
\item The presence of a gas pipeline running adjacent to but outside the site. The layout of the buildings within the site takes into account the presence of this pipeline and, in the future development of the design, how to cross the pipeline to allow services to connect to the existing LHC site P8 prior to the FCC-hh construction. 
\item Likely urban development in the close vicinity of the proposed site, with direct views across Geneva to Mont Blanc. These views have been taken into account and building heights have been kept as low as possible, and the overall elevation of the platform has been minimised. To this end, the requirements brief included a target roof elevation no greater than that of the buildings located adjacent to the proposed site.
\item The presence of a protected environmental compensation area immediately to the north of the site, an existing development to the south of the site and a major arterial road to the west of the site restricts the available area and layout of buildings within the surface site. 
\end{itemize}
The functional requirements of the surface buildings at PA are similar to those built for previous CERN experiment areas. Some key requirements are:
\begin{itemize}
\item A large assembly hall for the pre-assembly of the detector. This building will house at least one crane capable of lowering the detector subcomponents from the surface down to the underground cavern located about 200\,m below. The precise crane dimensions and weight will need to be considered during the detailed design phase.

\item A shaft head building to accommodate the lifts, cranes, and staircases needed for personnel access to the underground areas and for the movement of material, equipment, and services between the surface and underground areas. This building also houses an unloading bay for trucks transporting material and equipment for installation in the underground areas. 
\item Buildings to house the equipment necessary to ventilate the underground areas. The equipment for ventilating the experiment cavern is housed separately from that for ventilating the rest of the underground areas, since these two systems enter and exit the underground areas via different shafts. This equipment requires effective noise insulation, therefore these buildings will either be made of reinforced concrete or have specific noise reduction measures incorporated or attached to the building structure.
\item Buildings to accommodate the compressors and other equipment necessary for the cryogenic systems that will be used by the detector. Similar noise abatement measures to those used for the ventilation buildings will be necessary. 
\item Cooling plant buildings will include the cooling towers and associated pump houses, basins etc. Although traditionally constructed in reinforced concrete, the use of prefabricated fibreglass reinforced polyester (FRP) may be considered, in which case the civil engineering component would be reduced to the provision of the cooling basin and appropriate supporting foundations, with the towers themselves delivered as part of cooling, mechanical, and electrical infrastructure. In this case, additional noise abatement measures may be needed around these pre-fabricated cooling towers.
\item Control building to house offices and control room for the operation of the detector. This building would also house areas for receiving visitors, noting that as PA is the closest FCC experiment area to the CERN main campus, it is likely to be an important focal point for CERN visitors.
\item Buildings and support structures for transformers, switches, pylons etc. will be required for the electrical infrastructure. 
\item Other smaller buildings for various purposes as listed in the PBS given in Table~\ref{PBSforPA}.

\end{itemize}

Based on initial conceptual layouts developed by the Fermi National Accelerator Laboratory and evolving requirements for the PA surface site, the conceptual designs have been further refined to develop site layout scenarios, elevations, and earthwork estimates suitable for cost and schedule planning. It should be noted that these representations merely serve early-stage feasibility inputs for planning purposes and have not been reviewed, validated and approved. Fig.\ref{pasurfacesimplified} presents a simplified view of the proposed civil engineering buildings and associated works, while Fig.\ref{PAsurfacesectionsimplified} illustrates a streamlined cross-section of the site.

One aspect in which PA differs from the other three experiment sites, PD, PG and PJ, is the presence of an existing LHC surface site (P8) in the immediate vicinity. This presents an opportunity to reduce the land required for FCC-hh by re-purposing the existing LHC P8 site after the LHC finishes its physics  programme. A technical gallery from the existing site to PA would need to be constructed prior to FCC-hh to connect the services of the two sites together. 

\subsection{Surface site PB}\label{subsec_surfacePB}
Site PB is a technical site. At this site, there will be no experiment area, and therefore, civil engineering is limited to the provision of buildings necessary to support the FCC-ee machine and the associated beam absorber, which will be sited in the underground areas of PB. The site also reserves the necessary space for the civil engineering required to accommodate a future hh machine, although these buildings will not be constructed during the FCC-ee phase. The space reserved will be used temporarily during the construction phase to house the civil engineering contractors plant, equipment, material etc. This surface site is the only one planned to be located on Swiss territory. The collaboration with the Fermi National Accelerator Laboratory also studied the need for a technical site, taking PB as an example. 
Building on the initial work, CERN has developed further layout scenarios for cost and planning purposes.

The product breakdown structure (PBS) for the buildings and associated civil engineering works to be constructed at site PB is given in Table \ref{pointBpbstable}.

\begin{table}[h!]
    \centering
	\caption{\label{pointBpbstable}PBS structure for surface site PB.}

    \begin{tabular}{ccccccl}
	\toprule
	\multicolumn{6}{c}{\textbf{PBS Level}} & \multirow{2}{*}{{\textbf{PBS Title}}} \\
	\textbf{0} & \textbf{1} & \textbf{2} & \textbf{3} & \textbf{4} & \textbf{5} \\ 
	\cmidrule{1-6} 
        ~ & 5 & 3 & ~ & ~ & ~ & PB Construction, testing, commissioning \\ 
        ~ & 5 & 3 & 1 & ~ & ~ & Surface works  \\
        ~ & 5 & 3 & 1 & 1 & ~ & Roads, footpaths, parking, fences, gates \\
        ~ & 5 & 3 & 1 & 2 & ~ & Technical galleries \\
        ~ & 5 & 3 & 1 & 3 & ~ & Access control building\\
        ~ & 5 & 3 & 1 & 4 & ~ & Fire fighting equipment and medical station \\
        ~ & 5 & 3 & 1 & 5 & ~ & Shaft head building \\
        ~ & 5 & 3 & 1 & 6 & ~ & Ventilation building \\
        ~ & 5 & 3 & 1 & 7 & ~ & Chilled water production building \\
        ~ & 5 & 3 & 1 & 8 & ~ & Cold box and control Building \\
        ~ & 5 & 3 & 1 & 9 & ~ & Cooling plant Building \\ 
        ~ & 5 & 3 & 1 & 10 & ~ & SF Annex (demineralised water) \\ 
        ~ & 5 & 3 & 1 & 11 & ~ & Power converters building \\ 
        ~ & 5 & 3 & 1 & 12 & ~ & Electrical building \\ 
        ~ & 5 & 3 & 1 & 13 & ~ & Electrical substation (foundations) \\ 
        ~ & 5 & 3 & 1 & 14 & ~ & Waste material storage \\ 
        \bottomrule
     \end{tabular}
\end{table}

Site PB is one of the smaller technical sites. Nonetheless, its location close to a protected water-course and within sight of nearby residential and agricultural properties will require rigorous integration studies to be carried out. The smooth integration in the existing landscape will be further studied during a subsequent project preparatory phase. 

\begin{figure}[ht!]
    \centering
    \includegraphics[width=0.9\linewidth]{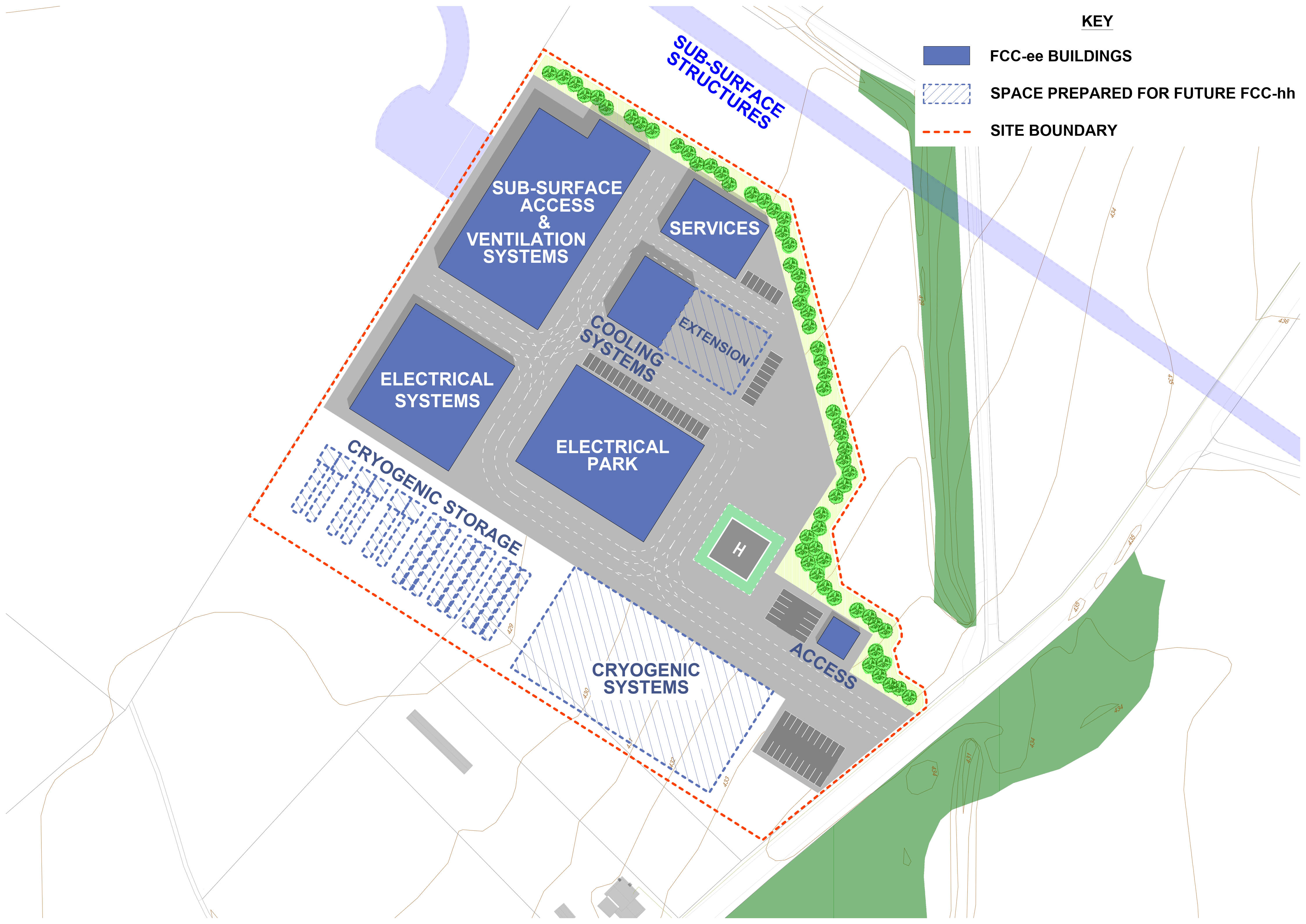}
    \caption{\label{PB_Plan}Preliminary simplified area requirements for PB surface site.}
\end{figure}

\begin{figure}[ht!]
    \centering
    \includegraphics[width=\linewidth]{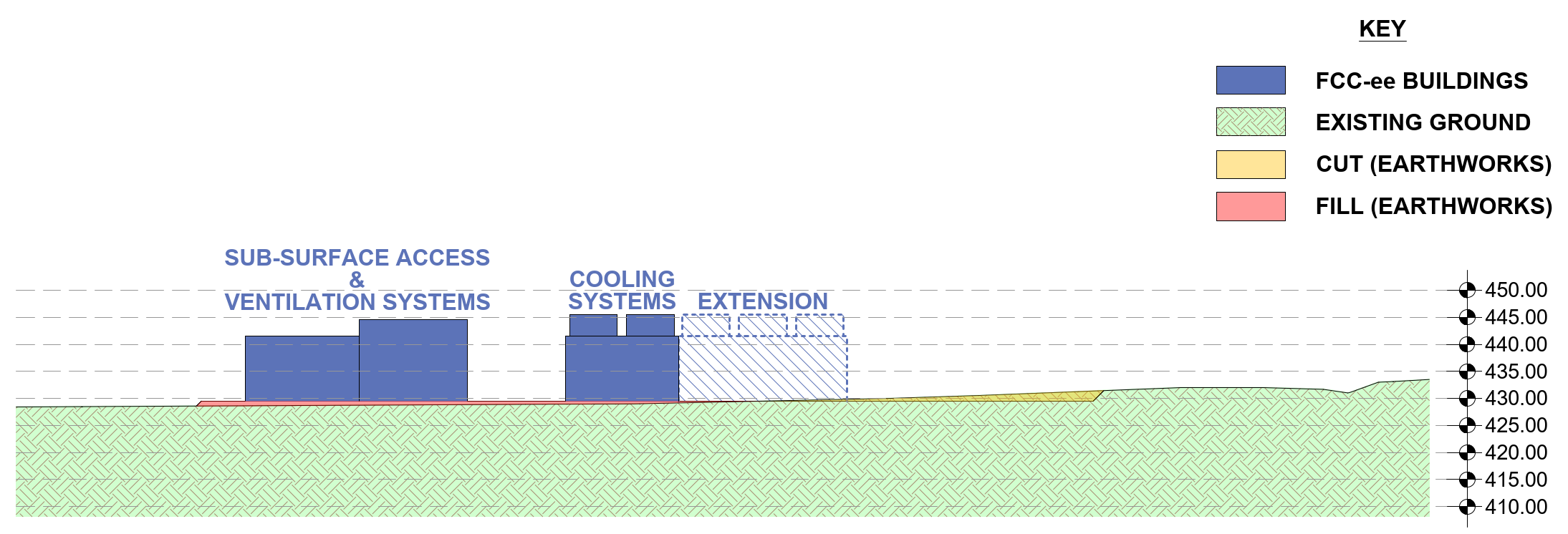}
    \caption{\label{PB_section}Preliminary simplified cross-section of PB surface site.}
\end{figure}

At FCC-ee phase, the principal surface infrastructure at site PB consists of:
\begin{itemize}
\item A shaft head building to accommodate the lifts, cranes, and staircases needed for personnel access to the underground areas and for the movement of material, equipment, and services between the surface and underground areas. This building also houses an unloading bay for trucks bringing material and equipment for transportation via the shaft to the underground areas. This building must be located directly over the associated shaft. The position of this building and the associated shaft could be adjusted within the site boundary to accommodate any changes or additions to the surface requirements and constraints. This building would typically be constructed as a steel-framed building on reinforced concrete foundations and a reinforced concrete slab to sustain the load imposed by the vehicles entering the building for unloading. The building would include a secure access area to ensure appropriate access control to the underground areas.
\item A building will be constructed to house the equipment necessary for ventilating the underground areas, including the shaft, access gallery, accelerator tunnel, and beam absorber cavern. As this equipment includes motorised components that may generate noise, the building design will incorporate appropriate noise-reduction measures to ensure that noise levels at the site boundary remain within the required limits. The choice of construction materials and noise mitigation solutions will be determined based on technical and environmental considerations.
\item Cooling plant buildings, including the cooling towers and associated pump houses, basins and the demineralised water plant building.
\item Various buildings and structures to house electrical equipment, including buildings for power converters and an emergency power system.
\item An access control building to house personnel undertaking security and safety operations at the site.
\end{itemize}

The location of site PB in flat, open countryside requires careful consideration of its integration into the surrounding landscape. To ensure minimal environmental impact, particular attention will be given to architectural and noise mitigation measures, both during the construction phase of FCC-ee and its subsequent operation.

As with site PA, the FCC-ee layout for site PB has taken into account the needs of a future FCC-hh collider, and space has been reserved for the necessary civil engineering. The site boundary fence at the FCC-ee phase will already encompass these future needs.

Figure~\ref{PB_Plan} shows a simplified view of the PB surface site. The specific buildings required for FCC-ee are illustrated, along with the necessary space reservations for future FCC-hh civil engineering. A simplified cross-section is shown in Fig.~\ref{PB_section}.

\subsection{Surface site PD}
Site PD has an identical function to site PA, namely, to house an experiment and provide support to the accelerator. As such, the requirements for civil engineering are almost identical, although the specificities of the local terrain and land plots require a different approach to the layout of the site. 

The layout of the PD site has a number of constraints that need to be addressed. The main constraint is the planned future extension of the public road network in these areas, which reduces the potential land space for FCC at PD and requires a site access scenario that is compatible with the future layout. After several iterations, a site layout that meets CERN requirements and is compatible with the road development project has been developed. 
The proposed PD site is located on sloping agricultural land. The elevation change from one end of the site to another is about 20\,m. This will require earthworks to be carried out to create a single flat platform on which to site the surface structures. 

\begin{figure}[!ht]
    \centering
    \includegraphics[width=\linewidth]{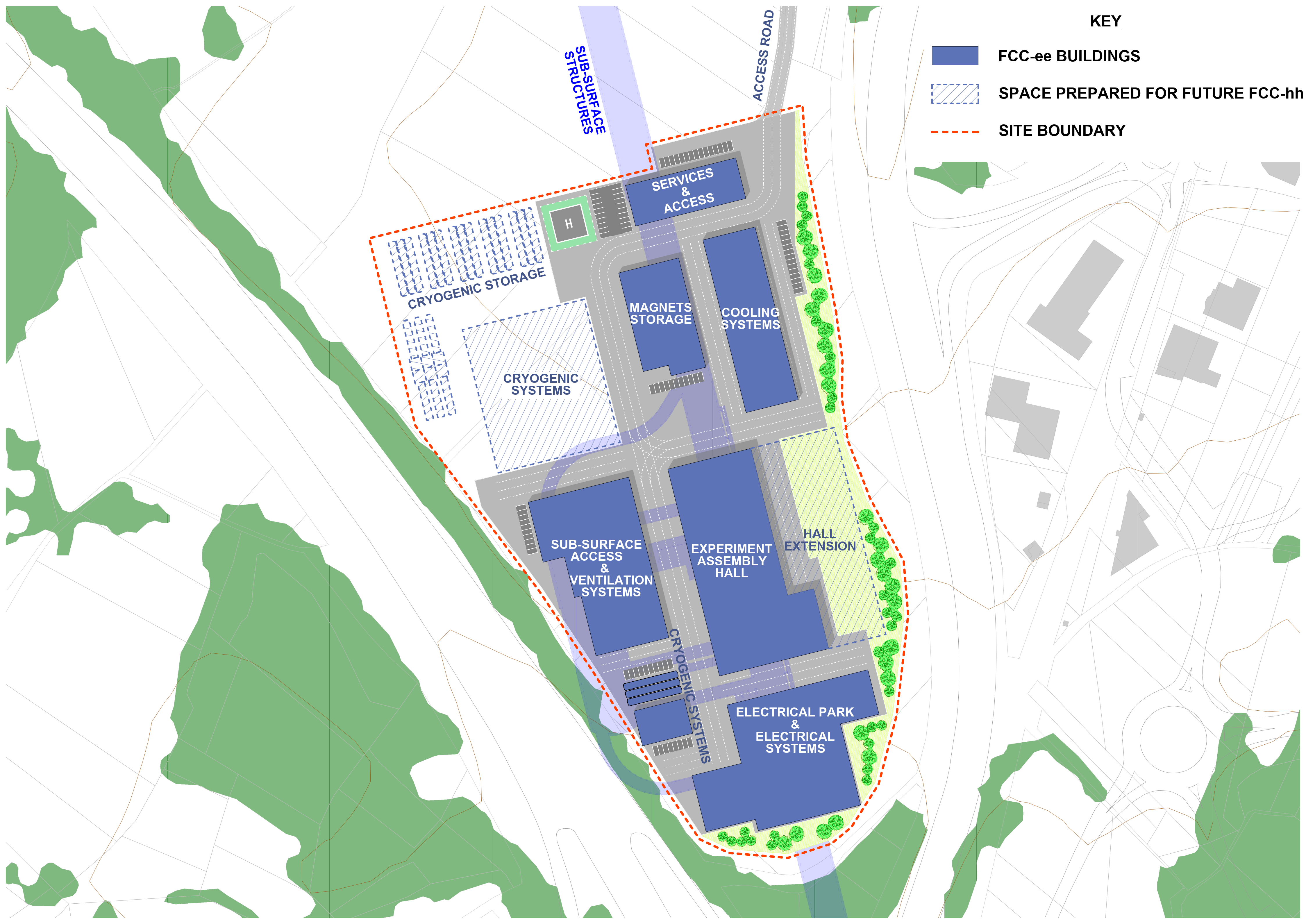}
    \caption{\label{PD_surface_simplified}Preliminary simplified area requirements for PD surface site.}
\end{figure}

\begin{figure}[!ht]
    \centering
    \includegraphics[width=\linewidth]{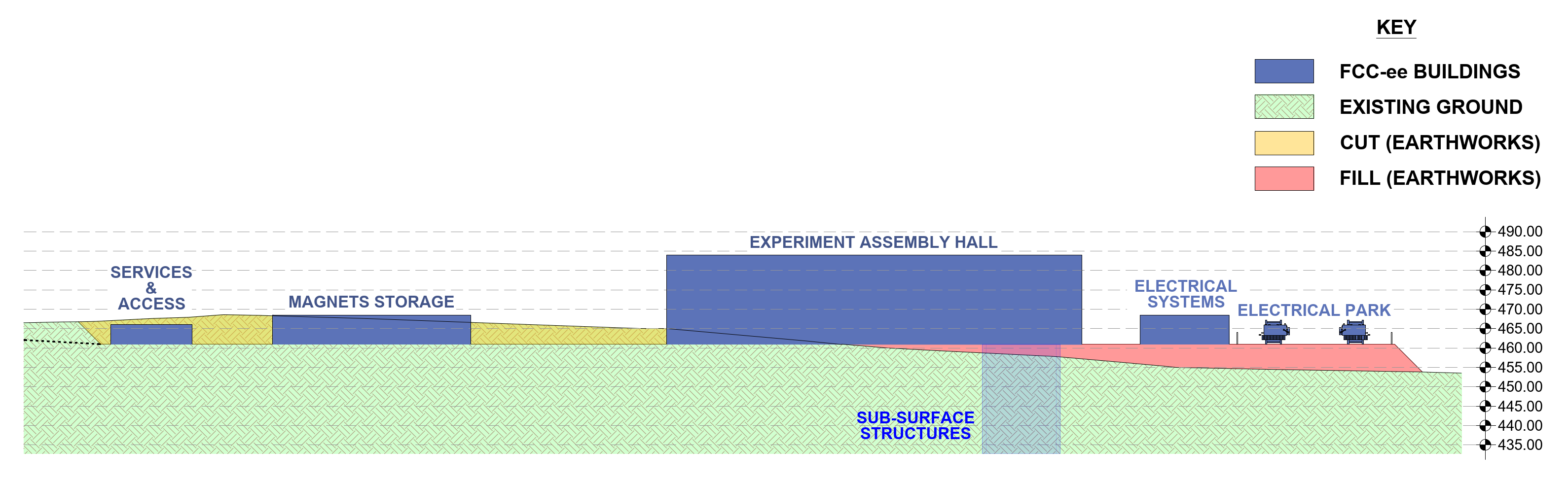}
    \caption{\label{PD_section_simplified}Preliminary simplified cross-section of PD surface site.}
\end{figure}

Other specific features of the site which will need to be addressed after the feasibility study include:

\begin{itemize}
\item The scheduling compatibility of the FCC-ee civil engineering works with the planned road extension project
\item The proximity of a medical facility to the site will be taken into account in the site’s design. Measures such as tree planting and landscaped features may be incorporated to ensure a well-integrated and visually harmonious environment.
\item The need to pass under bridges local to the site may result in a limitation on the height of oversized components that can be transported to the site. 
\end{itemize}

 A simplified view and section of the civil engineering structures at PD are given in Fig.~\ref{PD_surface_simplified} and Fig.~\ref{PD_section_simplified}. 

\clearpage
\subsection{ Surface site PF}

PF is a technical access point with surface civil engineering structures similar in function to PB. In order to avoid siting surface buildings within an existing village, the PF surface site is offset by about 600\,m from the axis of the accelerator tunnel. 

\begin{figure}[!ht]
    \centering
    \includegraphics[width=\linewidth]{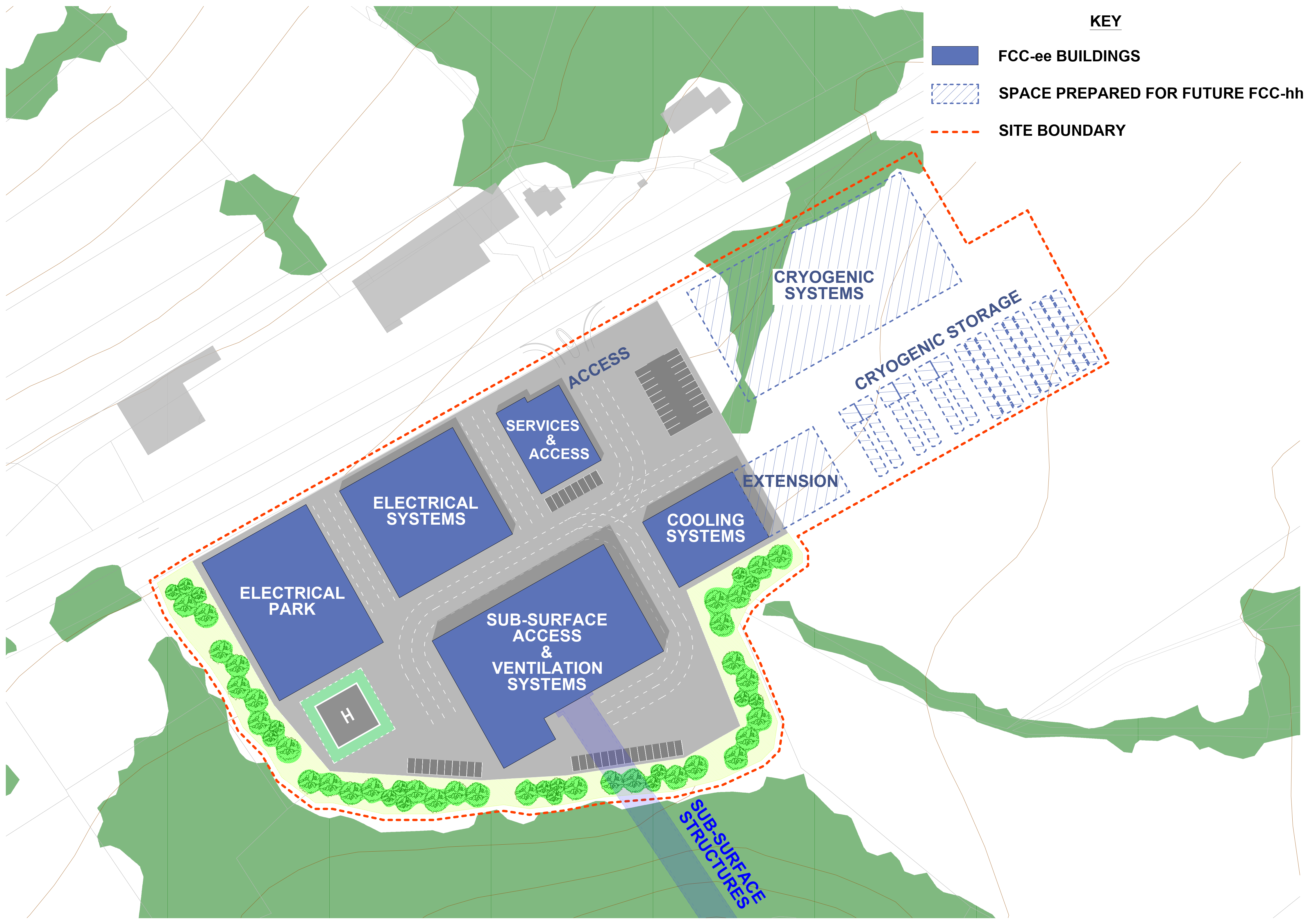}
    \caption{\label{PF_surface_simplified}Preliminary simplified area requirements for PF surface site.}
\end{figure}

\begin{figure}[!ht]
    \centering
    \includegraphics[width=\linewidth]{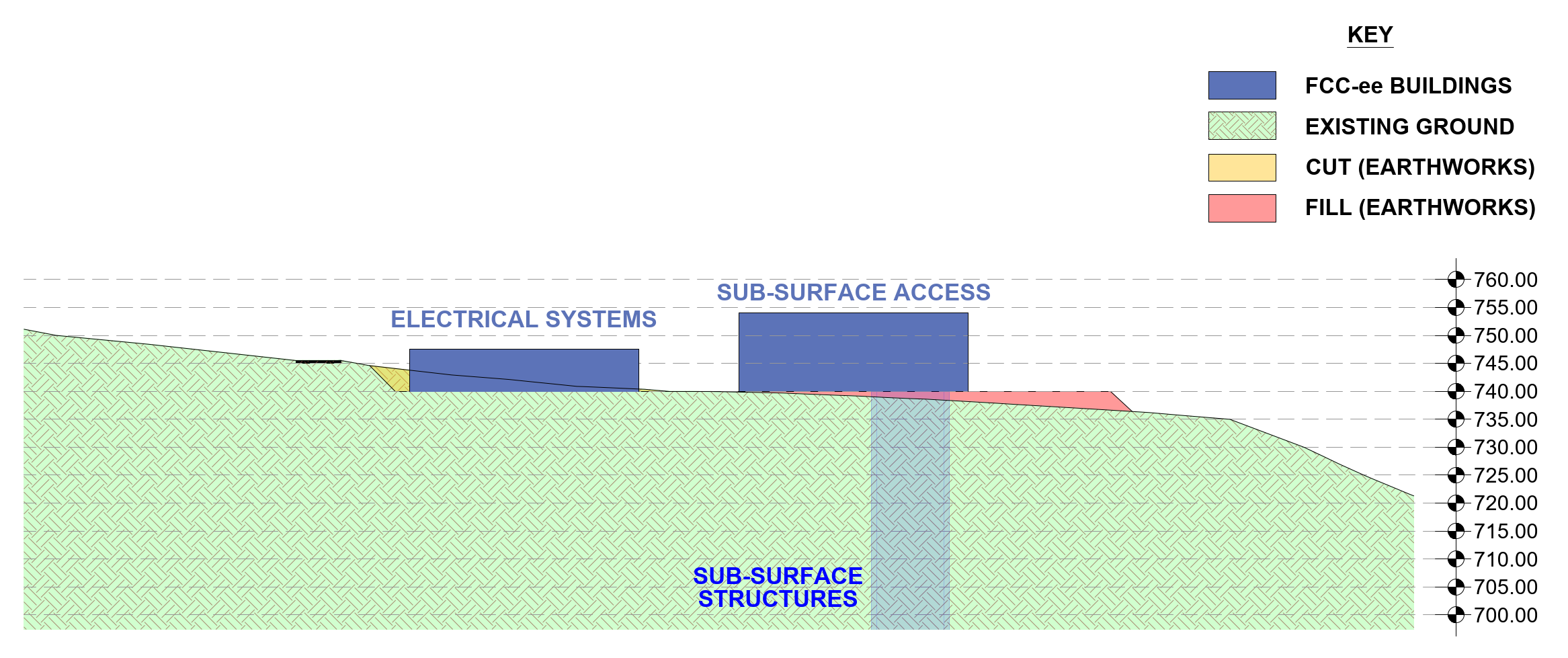}
    \caption{\label{PF_section_simplified}Preliminary simplified cross-section of PF surface site.}
\end{figure}

The site is adjacent to a major road, minimising the need for new external roads to access the site. The necessary platform needed for the surface buildings will be of sufficient size to accommodate the future FCC-hh buildings.  A simplified view and cross-section of the PF site are given in 
 Fig.~\ref{PF_surface_simplified} and Fig.~\ref{PF_section_simplified}. 

The main constraints at PF relate to the presence of a so-called \lq{humid zone}\rq in the vicinity of the proposed site. This will require specific attention during the detailed design of the civil engineering structures to ensure that water run-off during the construction and operational phases does not impact this zone. 

In summary, the construction of the surface civil engineering required for FCC-ee at PF is considered technically feasible. Particular attention will be given to implementing measures that ensure minimal environmental impact.

\subsection{Surface site PG}

PG is an experiment point with civil engineering requirements similar to those of PA. The site is situated on a plateau that is currently partially used for pasture and partially forested. To minimise the impact on the forested area, the surface buildings associated with the cooling plant and some of the electrical systems will be positioned in a secondary location approximately 300\,m from the main site. This approach follows the example of the existing CERN LHC Point 4, where cooling towers and pump stations are located in an area remote from the main site, in this case, to reduce their visual impact.

\begin{figure}[!ht]
    \centering
    \includegraphics[width=\linewidth]{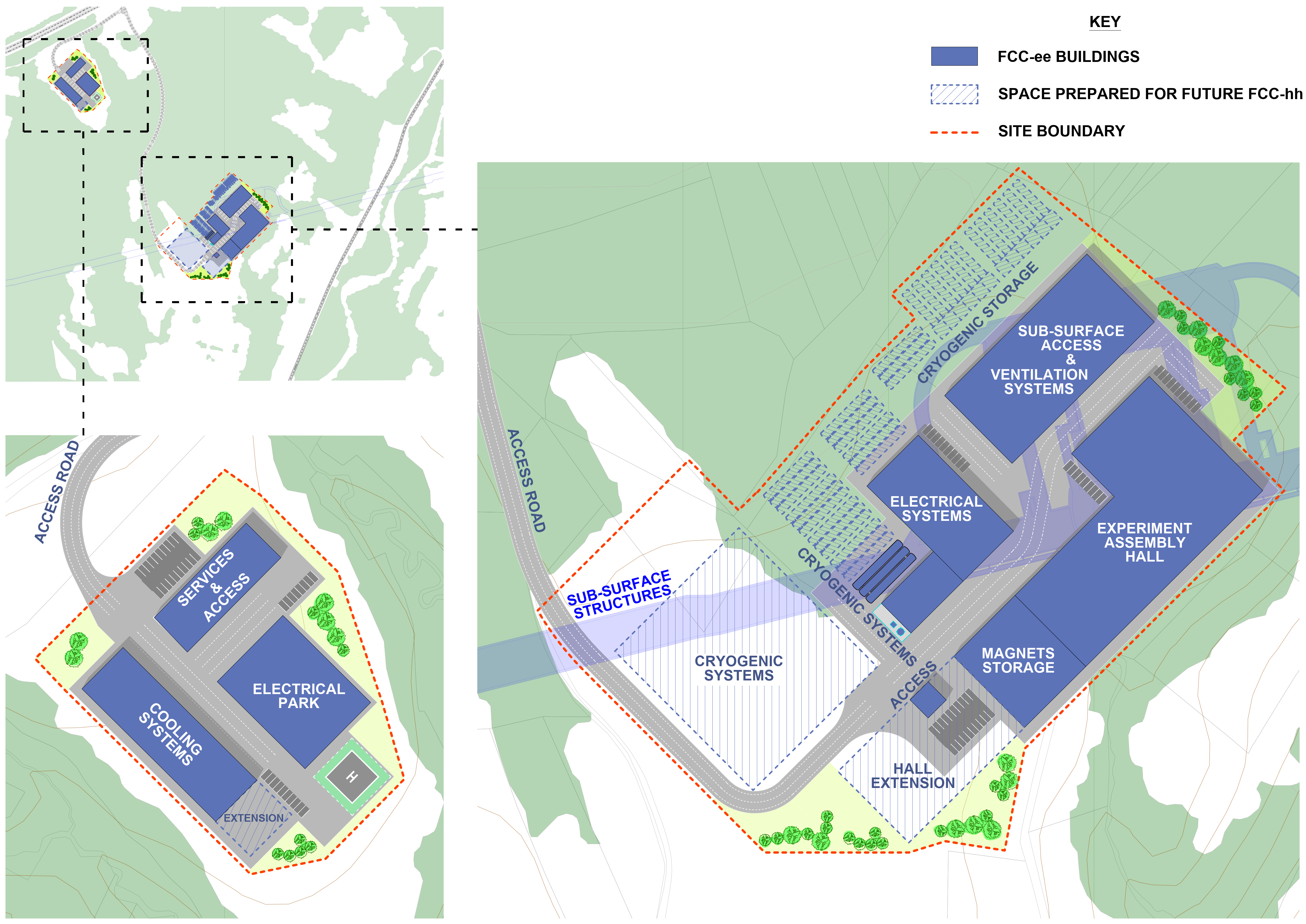}
    \caption{\label{PG_surface_simplified} Preliminary simplified area requirements for PG surface site.}
\end{figure}

\begin{figure}[!ht]
    \centering
    \includegraphics[width=\linewidth]{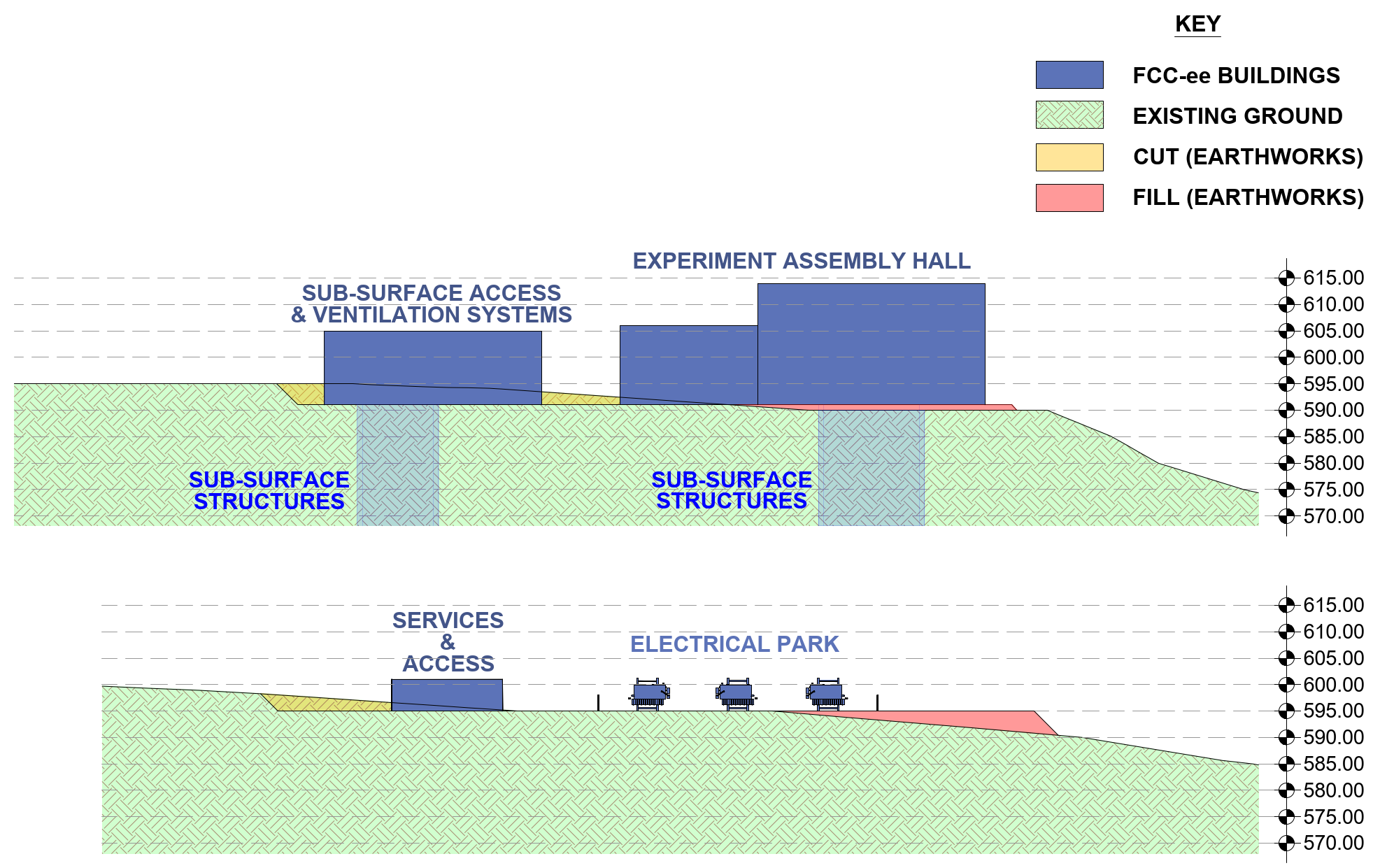}
    \caption{\label{PG_crossection_simplified} Preliminary simplified cross-section of PG surface site.}
\end{figure}

In addition to mitigating visual impact, the layout of PG will be designed to ensure efficient use of space while maintaining accessibility for operations and maintenance. Noise reduction measures will be implemented where necessary to minimise any potential disturbances to the surrounding environment. Furthermore, efforts will be made to integrate the site harmoniously into the landscape.

\subsection{Surface site PH}
PH will be a technical area that will support not only the access and ventilation systems necessary for the accelerator tunnel, but also the infrastructure associated with the RF systems installed at this site. 

\begin{figure}[!ht]
    \centering
    \includegraphics[width=\linewidth]{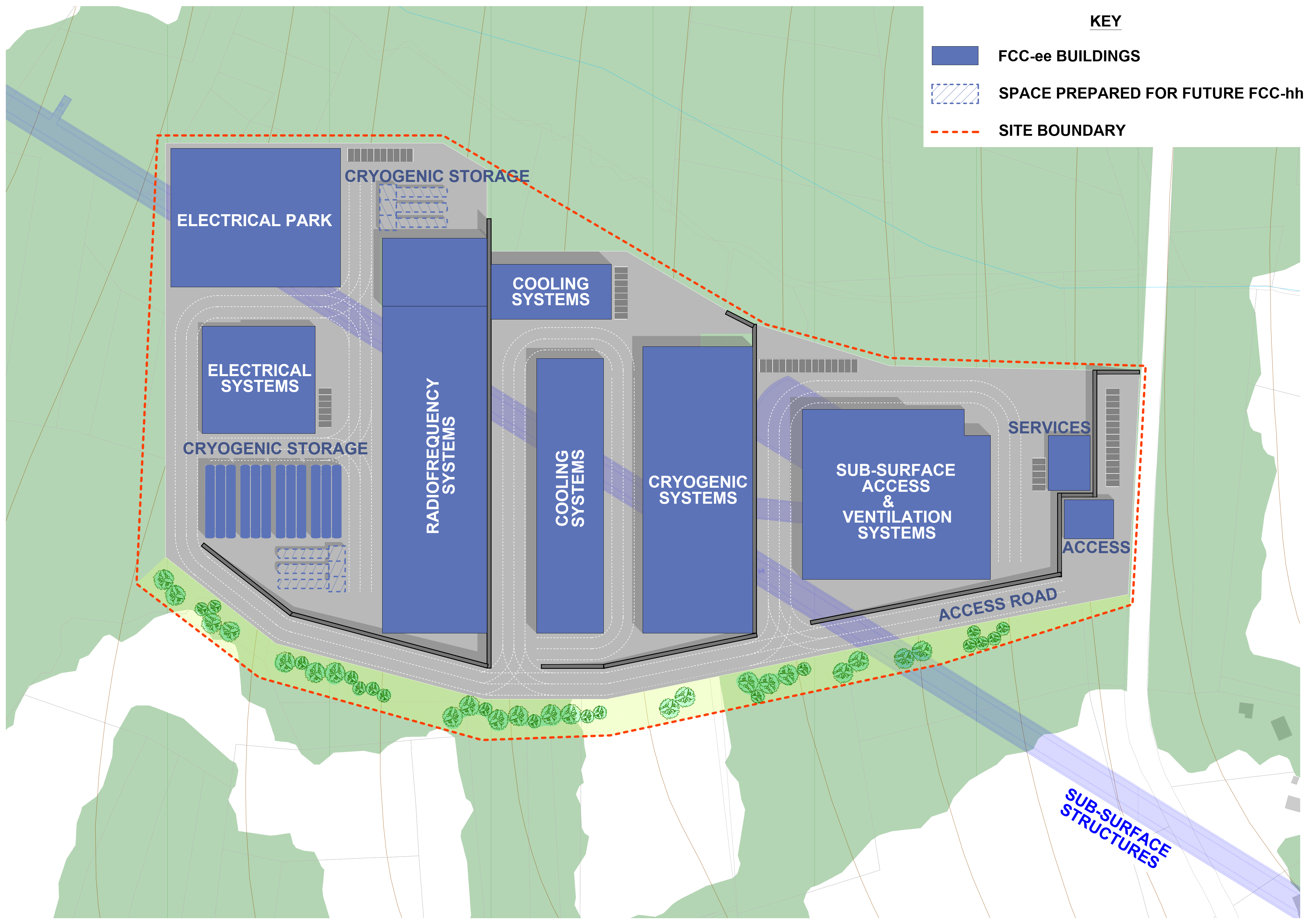}
    \caption{\label{PH_surface_simplified} Preliminary simplified area requirements for PH surface site.}
\end{figure}

The surface site will have a larger footprint than PB and PF, as the power and cryogenics associated with the RF systems require additional buildings and dedicated foundations for infrastructure such as tanks and transformers. 

\begin{figure}[!ht]
    \centering
    \includegraphics[width=\linewidth]{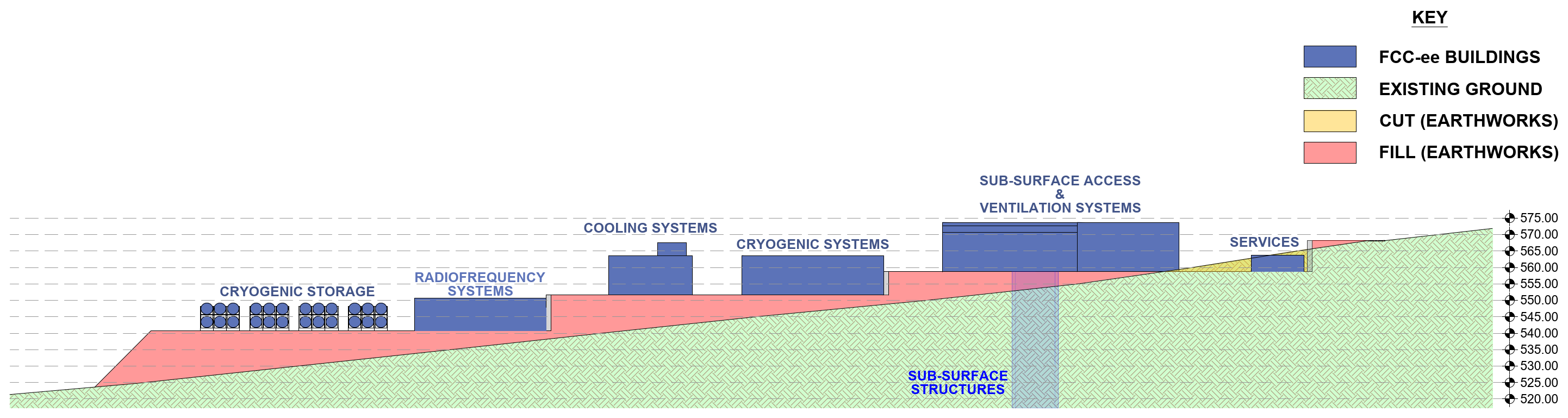}
    \caption{\label{PH_crossection_simplified} Preliminary simplified cross-section of PH surface site.}
\end{figure}

The layout will be designed to optimise space efficiency while ensuring accessibility for maintenance and operations. Additionally, consideration will be given to minimising environmental impact and integrating the site with its surroundings through appropriate architectural and landscaping measures.

A preliminary simplified layout of PH is given in Fig.~\ref{PH_surface_simplified} and a simplified cross-section is given in Fig.~\ref{PH_crossection_simplified}

Site PH is currently located on sloping ground within an agricultural and forested area, with some houses in the near vicinity. This site will require specific studies into the visual impact of the site in order to identify measures that could be taken to reduce its visual impact on the surrounding area.

\clearpage
\subsection{Surface site PJ}
\begin{figure}[!ht]
    \centering
    \includegraphics[width=\linewidth]{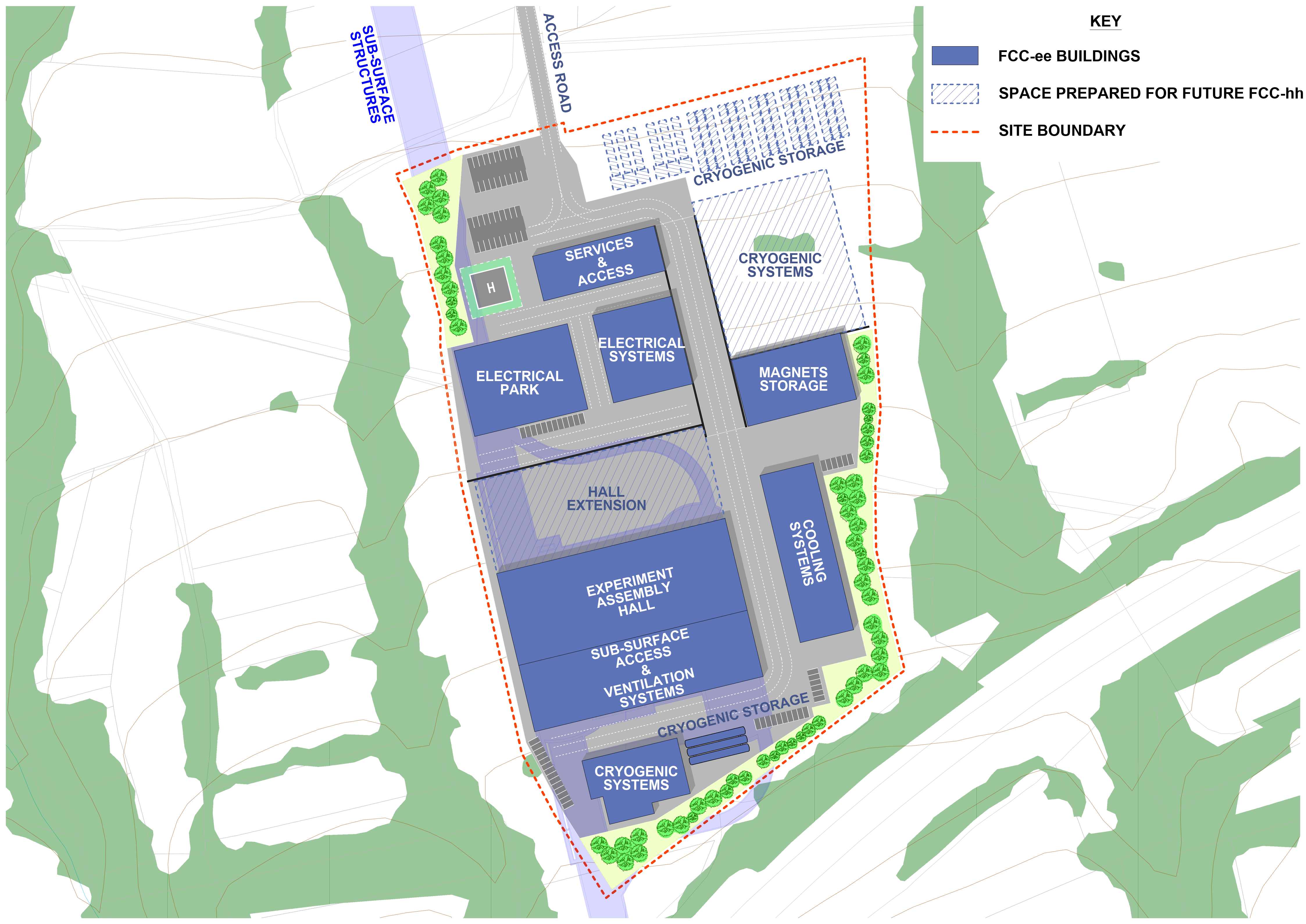}
    \caption{\label{PJ_surface_simplified} Preliminary simplified area requirements for surface site PJ.}
\end{figure}

\begin{figure}[!ht]
    \centering
    \includegraphics[width=\linewidth]{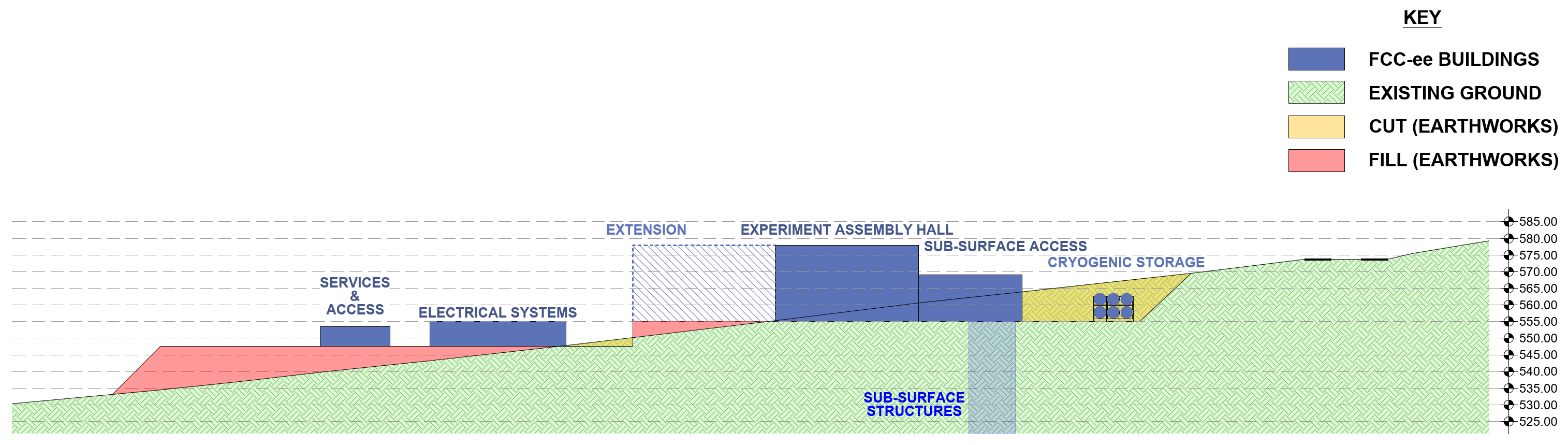}
    \caption{\label{PJ_surface_cross_Section_simplified} Preliminary simplified cross-section of site PJ.}
\end{figure}

PJ will be an experiment area with similar civil engineering structures to those described for site PD.  The site is located within agricultural land close to an autoroute. Beyond ensuring that the site is visually integrated into its surroundings, no significant challenges have been identified at this stage. A simplified layout for PJ is given in 
 Fig.~\ref{PJ_surface_simplified} and a simplified cross-section is given in Fig.~\ref{PJ_surface_cross_Section_simplified}

\subsection{Surface site PL}
\begin{figure}[!ht]
    \centering
    \includegraphics[width=\linewidth]{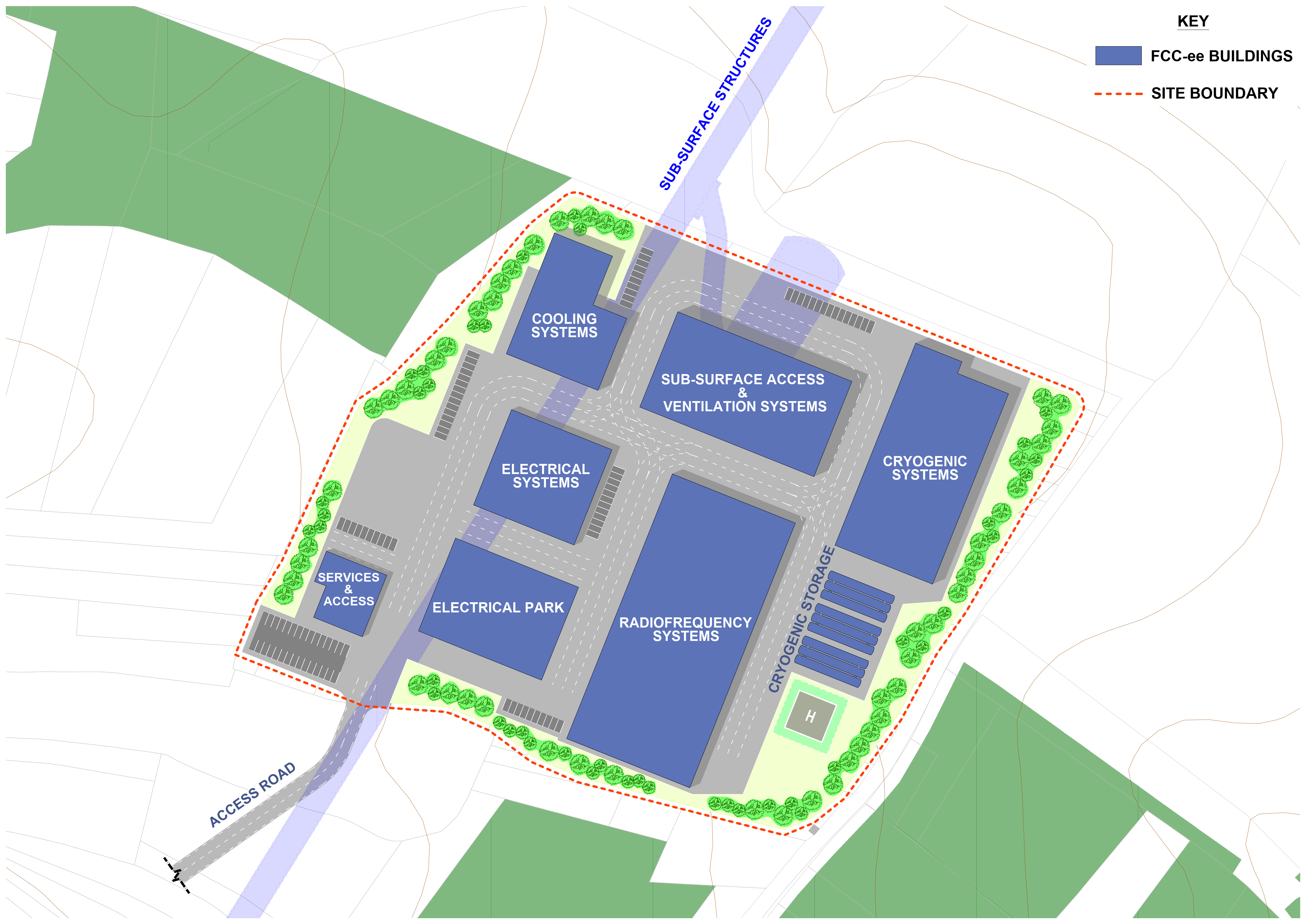}
    \caption{\label{PL_surface_simplified}Preliminary simplified area requirements for surface site PL.}
\end{figure}

\begin{figure}[!ht]
    \centering
    \includegraphics[width=\linewidth]{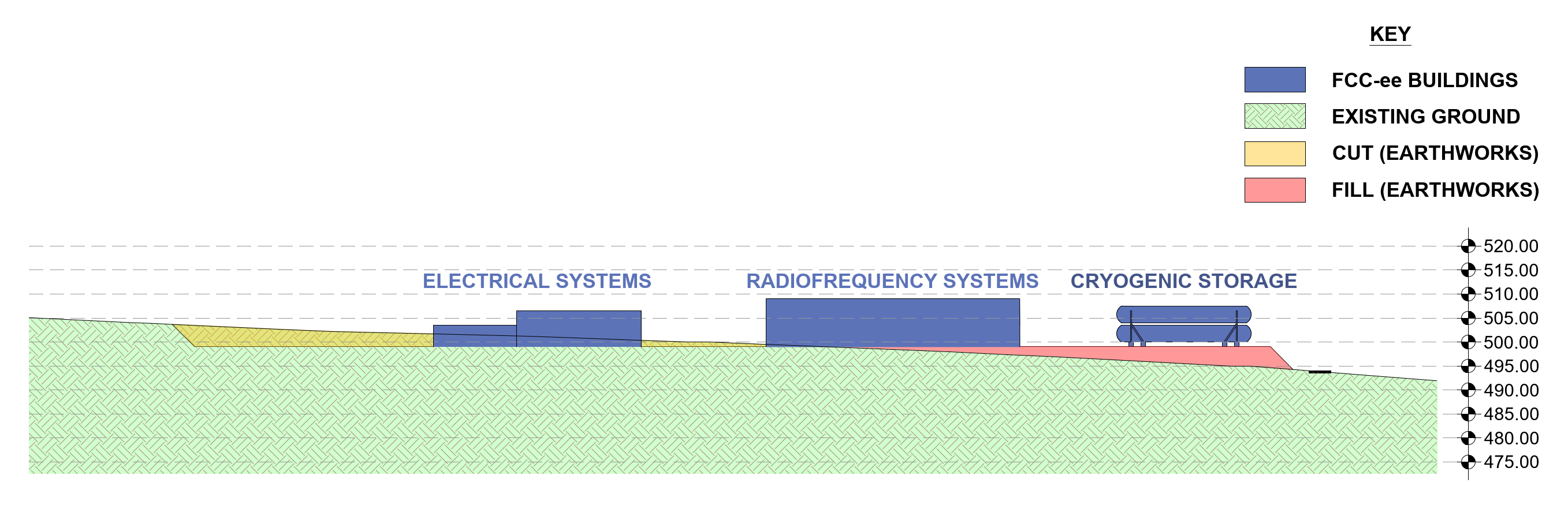}
    \caption{\label{PL_surface_cross_Section_simplified}Preliminary simplified cross-section through site PL.}
\end{figure}

PL will be a technical area with similar requirements to PH due to the location of the Booster RF systems at this point. The potential presence of an aquifer linked to a potable water supply will need future investigation and if necessary, construction techniques and processes will need to be selected to ensure the aquifer is protected. . A simplified layout for PL is given in 
 Fig.~\ref{PL_surface_simplified} and a simplified cross-section is given in Fig.~\ref{PL_surface_cross_Section_simplified}

In summary, preliminary requirements for all eight surface sites have been gathered and \textbf{preliminary} plan views and sections have been developed to a level sufficient for costing and planning purposes. The construction of these sites is considered technically feasible. However, a project preparatory phase will need to focus considerable effort to ensure that these sites are carefully integrated into their specific local environments. This process will need to be carried out in close collaboration with appropriate architectural specialists and must include a consultative process with the local inhabitants and other stakeholders. It may be necessary during this process to \textbf{ adjust and optimise the current preliminary layouts and locations} to achieve an acceptable outcome for all stakeholders that conforms with technical requirements. 

\section{Staged approach}\label{secstageapproach}
An assessment has been made of both the underground and surface civil engineering to identify structures that will only be needed for a subsequent FCC-hh machine and its associated detectors and to classify the identified structures into those for which construction can be deferred until after the completion of FCC-ee physics and those which need to be constructed simultaneously with the civil engineering for the \mbox{FCC-ee} machine. Consideration was given to both the cost and practicality of undertaking civil engineering for FCC-hh in a second stage after the operation of the FCC-ee machine.

\subsection{Underground structures}\label{subsecunderground}
Several underground civil engineering structures have been considered for staging as follows:

\subsubsection{Experiment detector caverns}
The experiment detector caverns required for the FCC-hh are almost 100\% larger by volume than those required for the FCC-ee detectors. The possibility of constructing the smaller experiment caverns in the initial FCC-ee phase and then enlarging them prior to FCC-hh was investigated. However, this scenario was eventually rejected as a feasible approach since very few of the first-phase structural elements of the cavern would be retained for the second phase, and it would be necessary to demolish the majority of the first-phase works. This would not only be prohibitively expensive and create significant quantities of waste material, but it would also represent a significant risk from a safety perspective since the reinforced rock around the periphery of the smaller FCC-ee cavern would need to be removed and re-supported for the larger FCC-hh caverns.

\subsubsection{Service caverns}
The possibility of undertaking the construction of the service caverns that are adjacent to the experiment caverns at points PA, PD, PG and PJ was also considered. The driver for this was the possibility of co-locating the systems that would traditionally be housed in the service cavern, such as air treatment and demineralised water systems, in the experiment cavern. Alternatively, it was considered only to construct a part of the service cavern for the FCC-ee phase and construct the remaining part necessary for FCC-hh at a second stage. Constructing the second phase would require a complete re-mobilisation of the construction site at the surface and removing all systems from the service cavern and partially, at least, from the machine access shaft. It was eventually concluded that neither a full nor partial staging of the service cavern would bring any major benefits since the cost of relocating the services from one cavern to another and the cost of remobilising the construction site would increase costs considerably compared to completing the service cavern in one stage only.

\subsubsection{Civil engineering for beam absorber}
The beam absorber system for FCC-ee requires only a localised enlargement of the beam tunnel over a length of 708\,m. This will be constructed at PB. 

The FCC-hh beam absorber system, although not yet fully designed, will be more complex and located at PF. It will require the construction of over two kilometres of additional tunnels, two beam dump caverns and several junction caverns. Since none of the FCC-hh beam absorber structures can be used at FCC-ee phase and given the criticality of PF for the overall construction schedule, it was deemed preferable to defer the construction of the FCC-hh beam absorber civil engineering to a second stage. The downside to this approach will be the need to re-establish a construction site and remove all services from the beam tunnel over several hundred metres on either side of PB. Furthermore, it will not be possible to transport personnel or equipment through the main tunnel in this area for a period of about two years while the FCC-hh beam absorber civil engineering is being carried out.

\subsubsection{FCC-hh transfer tunnels}
The FCC-ee civil engineering includes a single transfer tunnel from the Pr\'{e}vessin site down to the FCC machine. The tunnel bifurcates to provide clock-wise and anti-clockwise injection into the FCC-ee.
For a future FCC-hh machine, the pre-injection complex is likely to utilise either the current SPS or LHC tunnels with two new transfer tunnels for the clock-wise and anti-clockwise injection into FCC-hh. In the case that the SPS tunnel is re-configured to house the FCC-hh injector system, then a single additional transfer line with an approximate length of 3000\,m and shaft for constructing the tunnel will be required. In the case that the LHC tunnel is re-purposed then two transfer tunnels each of about 2500\,m length will be required, each with a single temporary shaft for construction purposes. A tunnel diameter of about 4\,m would be required, and connection caverns would be created to connect the transfer tunnels to the FCC main tunnel.

\subsubsection{Additional by-pass tunnels}
The FCC-ee civil engineering only includes by-pass tunnels at the four experiment sites PA, PD, PG, PJ. At the technical sites, by-pass tunnels are not required since a continuous transport corridor can be maintained through the accelerator tunnel.

For FCC-hh there is a possibility that at PB and PH new by-pass galleries may be required in order to avoid the need for personnel to frequently pass through areas of higher radiation (due to the presence of collimators). If a definitive need for these additional by-pass tunnels can be confirmed prior to FCC-ee civil engineering design works being completed, then it may be more efficient to include them at FCC-ee phase. In the case that the need is not fully confirmed,  they can be constructed after the FCC-ee machine operation is complete. These would be constructed similarly to the additional by-pass tunnels that were constructed for the LEP 200 upgrade in 1995.

\subsection{Surface structures}\label{subsecsurface}
Staging of surface civil engineering for a future FCC-hh presents less technical difficulty than underground civil engineering since access to the construction zones is readily achieved. 
Although detailed assessments of civil engineering requirements for all eight surface sites have not yet been fully completed, initial studies at points PA and PB suggest that the following buildings could be constructed after FCC-ee physics is completed and before FCC-hh commissioning and operation.

\subsubsection{Assembly halls for FCC-hh detectors elements}

Assessments of the assembly space needed for the larger FCC-hh detector elements, such as the superconducting magnet coil, indicate that for at least the two larger detectors, some components will have physical dimensions and weights that will make it impossible to transport them on the existing public roads without modifications to the existing public infrastructure. As such, it is likely that larger spaces and buildings will be required for the associated assembly activities. 

This additional space would be required for manufacturing activities that would not normally be carried out at the site, such as coil winding.
Since the precise needs for a potential FCC-hh detector are not known at this stage and since there will be at least a 20-year gap between the facilities needed for FCC-ee and FCC-hh detector assembly, it is not considered efficient or necessary to construct the larger facilities already at FCC-ee phase. However, to avoid unnecessary demolition and re-construction work in the future, the necessary space reservation for larger assembly halls will already be made at each of the four experiment sites. The space may be used for easily transferable facilities such as car parks or storage areas, but no complex facilities will be installed within these reserved areas, and, as far as possible, underground utilities will be avoided, including technical galleries.

\subsubsection{Cryoplants}
A future FCC-hh machine will require greater cryoplant capacity than during the \mbox{FCC-ee} phase. It is currently expected that the four experiment sites and sites housing RF systems will require cryoplant facilities for the FCC-ee phase. However, a future FCC-hh would require upgraded larger plants at sites PA, PD, PG and PJ as well as new plants at Points PB and PF where no cryoplants are planned for FCC-ee.
The same approach as the one adopted for the assembly hall will be taken, whereby space will be kept available and free of FCC-ee infrastructure for the currently expected FCC-hh cryogenics infrastructure. Furthermore, the landscaping and tree planting carried out for FCC-ee will, as far as possible, be executed in a manner that also serves the future FCC-hh needs.

\subsubsection{Cooling and ventilation buildings}
Although cooling and ventilation systems will be required at all surface points for the FCC-ee phase, it is currently anticipated that cooling needs for the machine and experiments at FCC-hh will require plant upgrades including a possible increase in the cooling capacity and therefore number of cooling towers. This will be accommodated in the same manner as the cryoplant, with due account taken of potential space needs and visual/acoustic screening for the FCC-ee phase.


\section{Subsurface site investigations}\label{secgeotechnicalinvestigations}

A dependable and comprehensive 3D geological model is essential for assessing the feasibility of underground civil engineering projects. Reliable geological and geotechnical data integrated into the model are required to design tunnels, caverns, and shafts. As the model evolves, the understanding of subsurface geology improves, enabling the identification of potential risks and constraints that could affect these structures. These factors may influence design, costs, and scheduling, making it vital to gather extensive subsurface information during the early stages of planning.

Since the autumn of 2024, CERN has been making a series of subsurface site investigations (SSI) using a combination of data acquisition methods. Once complete, the information obtained from these SSI will be used to enhance confidence in the 3D geological model for the FCC study area. The phase one SSI is currently about 40\% complete, and the preliminary results are positive when compared to the model's predictions.

\subsection{Geology in the region}\label{secgeology}

\begin{figure}[!ht]
    \centering
    \includegraphics[scale=0.8]{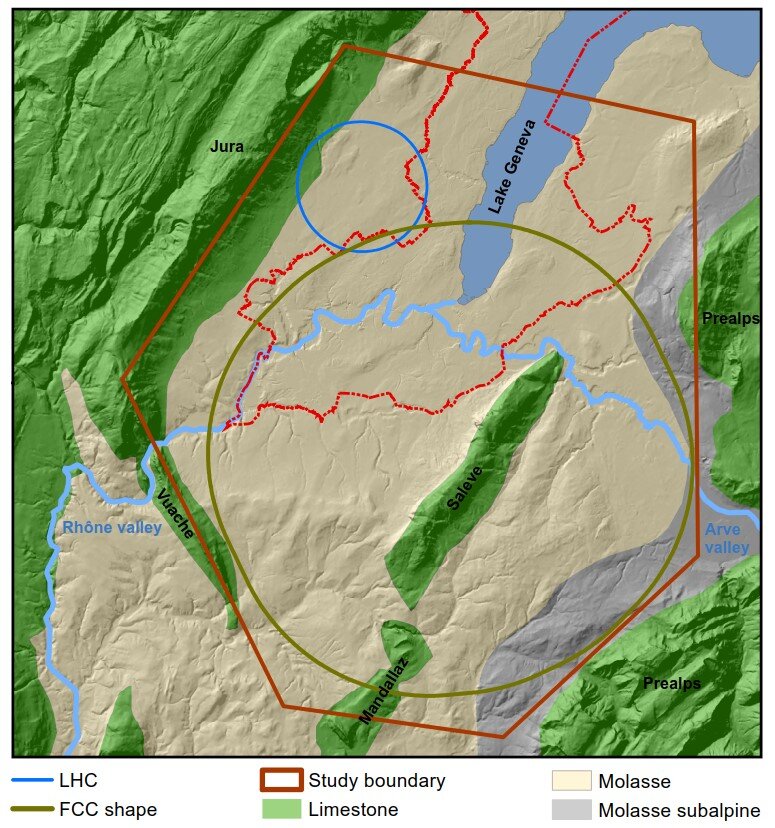}
    \caption{ Geology of the Geneva area.}
    \label{Geology of the Geneva area}
\end{figure}

For the purposes of the feasibility study and the targeted depths of the underground infrastructure, CERN has considered that the Geneva region features three primary geological strata: moraines, molasse, and limestone. A plan view of the Geneva basin geology is shown in Fig.~\ref{Geology of the Geneva area}.

The glacial moraines, characterised by their low strength, overlay the sedimentary molasse, which consists of horizontally bedded layers of marl and sandstone that vary significantly in strength. The thickness of the moraines ranges from just a few metres to over 100 metres. Bordering and intersecting the molasse are limestone formations, including the Alpine foothills and the Jura, Vuache and Sal\`{e}ve chains. Limestone in the Jura and Vuache foothills can potentially contain karsts caused by chemical weathering. These karsts, often filled with water and sediment, can lead to water inflow and structural instability if encountered during excavation. 

Beneath Lake Geneva, prior investigations have revealed very soft deposits, including lacustrine clayey silts and glacial-lacustrine silts and clays, with compressive strengths ranging between 2\,MPa and 10\,MPa. These deposits extend from the lakebed to approximately 260\,metres in depth. Limited data exists for the Arve and Rh\^{o}ne Valleys, but soft deposits, including alluvial and alluvial-glacial moraines, are expected to reach depths of up to 100\,metres. To mitigate construction risks and reduce water inflow challenges, the tunnel alignment has been situated at a depth that is expected to remain entirely below the moraines.

\begin{figure}[!ht]
    \centering
    \includegraphics[width=\linewidth]{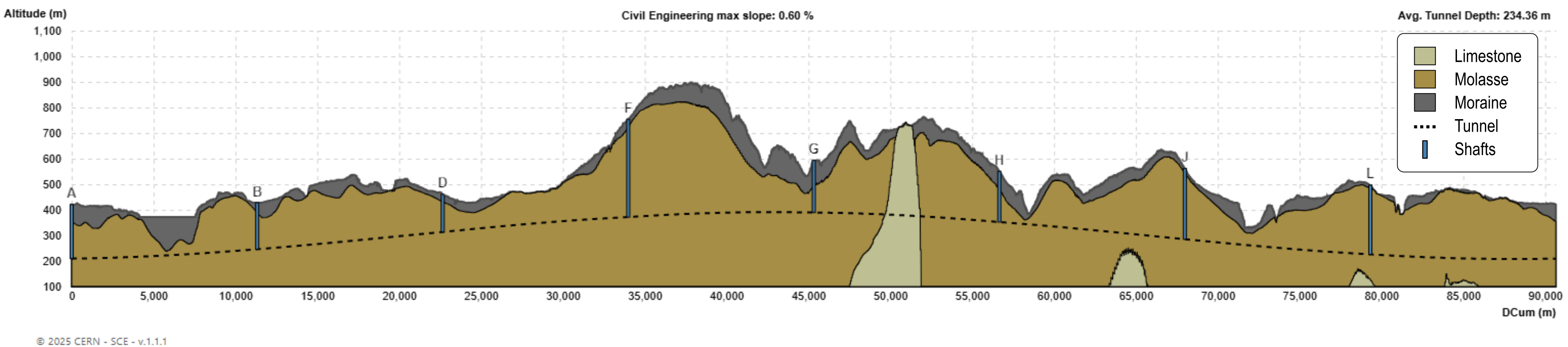}
    \caption{\label{FCC Long Geological Profile} FCC long geological profile.}
\end{figure}

The molasse is composed of horizontally bedded layers of marl and sandstones. These layers can vary in strength but are considered to be mostly stable and dry. Molasse is generally considered a favourable geological stratum for tunnel boring machine excavation. For large-span caverns, constructing in stronger sandstone layers is preferable. The objective throughout the feasibility study has been to locate underground infrastructure within the molasse, as can be seen in Fig.~\ref{FCC Long Geological Profile}, wherever possible.

\subsection{Development of geological 3D model}\label{sec3dmodel}

Within the feasibility study, one of the primary objectives was to identify and locate the interfaces between the moraines and the molasse, and between the molasse and the limestone more precisely. Although data on the individual sub-strata of each layer exists, its quality is inconsistent and does not currently allow a clear separation of the moraines, molasse, and limestone into distinct sub-layers.

In collaboration with the University of Geneva (UNIGE) and specialist consultants from CERN's Member States, CERN has been developing a 3D geological model. This model is based on data from previous borehole investigations, geophysical surveys in the study region and the ongoing SSI campaign. Urban areas such as Geneva and its suburbs benefit from numerous logged boreholes, which provide a high level of confidence for the upper 50\,metres of the subsurface. In contrast, rural areas have fewer logged boreholes, and certain locations, such as the foot of the Bornes and Mandallaz limestone outcrops, extend several kilometres without relevant geological data at the depths where the FCC underground infrastructure is proposed.

To improve understanding of the geological conditions along the proposed FCC tunnel, a targeted subsurface investigation campaign has been designed and is currently underway in areas of greatest geological uncertainty. This campaign, which commenced in October 2024 and will continue until December 2025, combines geophysical surveys with deep borehole drilling. Geophysical surveys are first conducted using seismic refraction to examine shallow strata and seismic reflection to explore deeper layers. The collected data is then processed and interpreted to guide targeted drilling at depths ranging from 70 to 500\,metres. By integrating these complementary methods, the reliability of the geological model in previously uncertain areas is significantly enhanced. This, in turn, increases confidence in the estimated construction costs and schedules, provides valuable information for the preliminary civil engineering design stage, and establishes key geological and geotechnical parameters prior to more targeted investigation campaigns.

Once the updated geological model is finalised, it, along with all individual borehole logs and geophysical interpretations, will be made publicly available. This data will contribute to a broader understanding of the geology in the region of the FCC, serve as a foundation for academic research, and potentially support future infrastructure projects beyond the FCC.

\subsection{Phase 1 subsurface site investigations}\label{secP1SSI}

To identify the site investigation to undertake in the first phase, three main criteria were considered:
\begin{enumerate}
    \item Areas where there is a risk of the tunnel crossing the molasse interface into moraines or limestones.
    \item Areas where crossing limestone is unavoidable.
    \item Areas where there is a complete lack of relevant geological information.
\end{enumerate}

Using these three criteria, the phase one SSI campaign has been divided into nine separate sections namely; Jura 1, Jura 2, Lake, Arve, Bornes, Mandallaz, Usses, Vuache and the Rh\^{o}ne.

\begin{figure}[!ht]
    \centering
    \includegraphics[width=\linewidth]{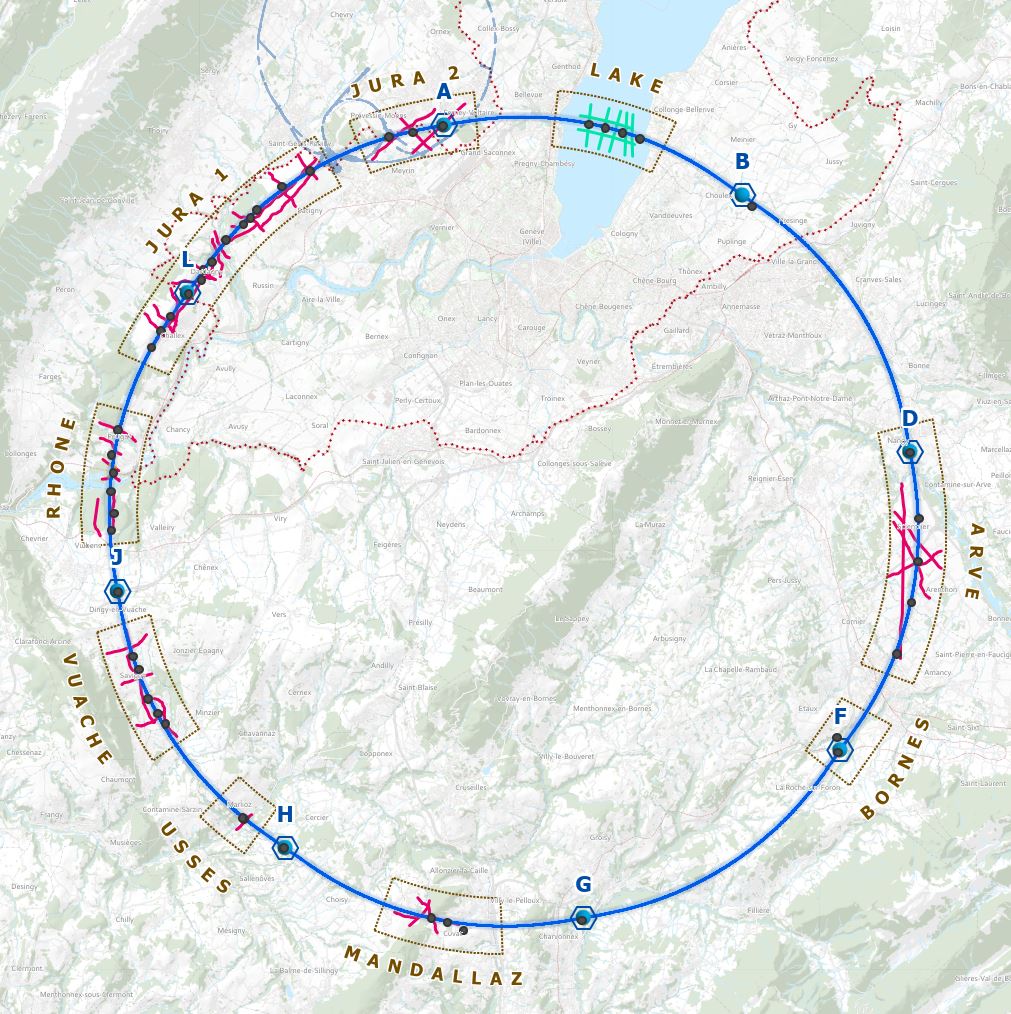}
    \caption{ Areas of geological uncertainty.}
    \label{Areas of geological uncertainty}
\end{figure}

The individual sections depicted in Fig.~\ref{Areas of geological uncertainty} each present unique uncertainties and challenges. The following sections will describe the current uncertainties in each area, explain the objectives of the subsurface investigations, and present the preliminary results from the investigations already completed.

\subsubsection{Jura 1}

Section Jura 1 is situated at the base of the predominantly limestone Jura mountain range, extending for approximately eight kilometres from Challex in France in the south to Satigny near CERN in Switzerland in the north. The proposed alignment is located at depths ranging from about 140\,metres beneath the Allondon Valley at its shallowest point to 250\,metres at its deepest on either side of the valley. 

The understanding of the subsurface conditions in this section has been enhanced through recent investigations by other parties. In 2021, an extensive campaign of 2D and 3D geophysical investigations was conducted by the Canton de Geneve and the Services Industrial de Geneve as part of the \textit{GEothermies} programme. Additionally, borehole data obtained near Satigny clearly identified the molasse-limestone interface, and encountered artesian water flows near this interface. The data from this campaign was incorporated into a 3D geological model developed in collaboration with UNIGE, which allowed a reduction in the planned scope of further investigations and provided an updated prediction that, in the northern part of the section, the limestone is located at a greater depth than previously expected.

Despite these advances, residual uncertainties remain—most notably in accurately defining the interface between the limestone and the molasse. As confidence in the updated data diminishes toward the southern portion of the section, further subsurface investigations remain necessary. Such investigations are required to confirm the current hypotheses and to ensure that the tunnel and associated underground infrastructure avoid limestone wherever feasible, thus mitigating the risks associated with unforeseen high water pressures and challenges associated with tunnelling in limestone.

\begin{figure}[!ht]
    \centering
    \includegraphics[width=\linewidth]{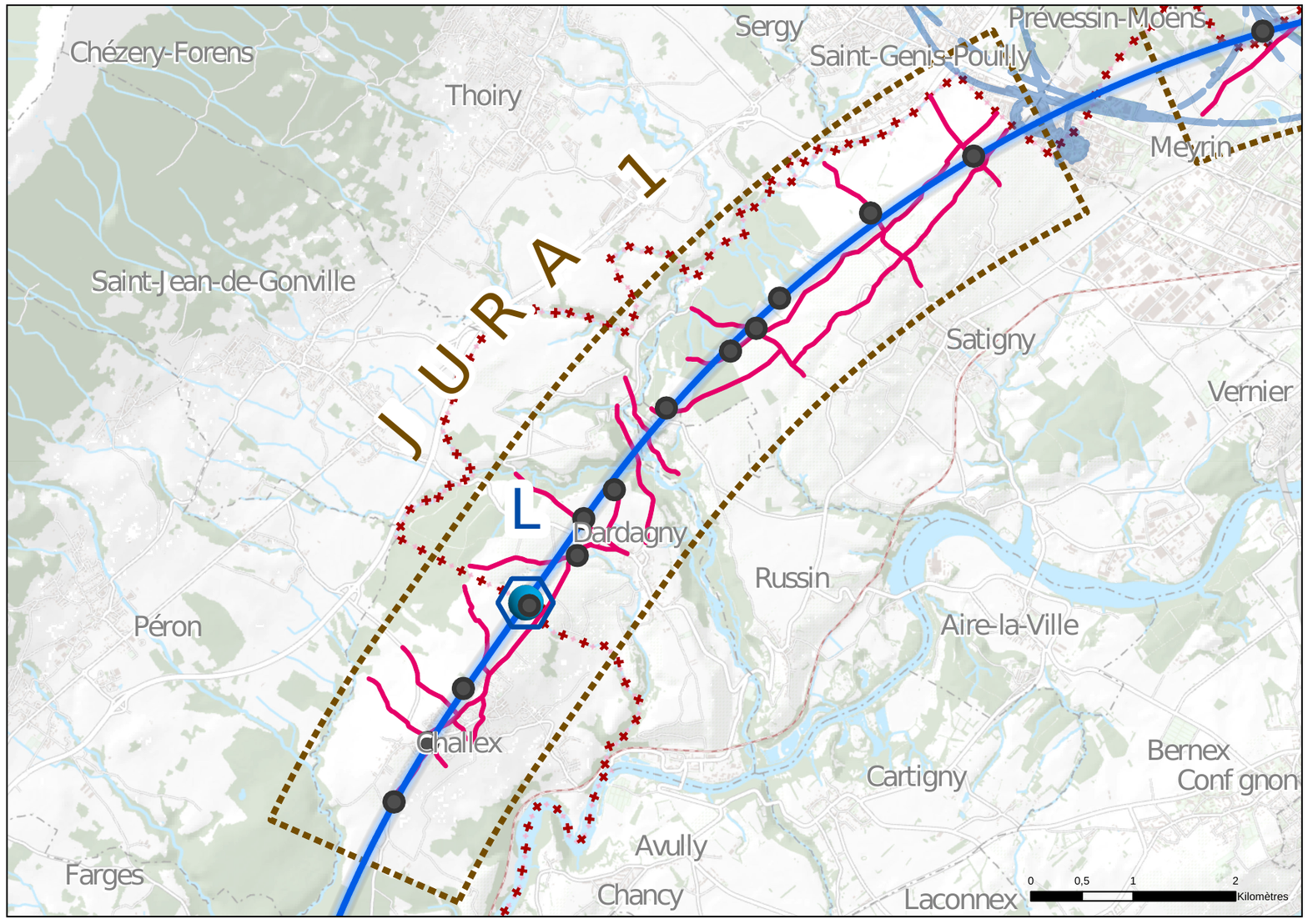}
    \caption{Jura 1 section showing the locations of the investigations.}
    \label{Sector Jura 1}
\end{figure}

The SSI in Jura 1 are scheduled to begin in mid-2025 and will continue until the end of 2025. The following is planned:  
\begin{itemize}
    \item thirteen boreholes ranging from 230-270\,m in depth:
        \begin{itemize}   
        \item[$\circ$]one fully cored
        \item[$\circ$]twelve destructive and cored 
        \item[$\circ$]three boreholes are optional, depending on the results achieved from geophysics and adjacent boreholes
    \end{itemize}
    \item 13 high-resolution seismic reflection profiles with a  total length of 23\,510\,m
\end{itemize}
The locations of the SSI in the section Jura 1 are shown in Fig.~\ref{Sector Jura 1}.

\subsubsection{Jura 2}

Section Jura 2 is located between Lake Geneva and the base of the Jura mountains, extending approximately three km from Meyrin in Switzerland to Ferney-Voltaire in France. The civil engineering infrastructure in this area is planned to be situated at about 200\,m below the surface. At this depth, 3D geological models indicate the potential for limestone to be present near the tunnel and primary infrastructure, similar to the risks identified in Jura 1.

Recent improvements in understanding have been achieved by incorporating seismic data from the nearby \textit{GEothermies} programme into the geological models. This updated information suggests that the top of the limestone is deeper than predicted. 

The targeted SSI will enable further improvements to the understanding of the molasse-limestone interface.

\begin{figure}[!ht]
    \centering
    \includegraphics[width=\linewidth]{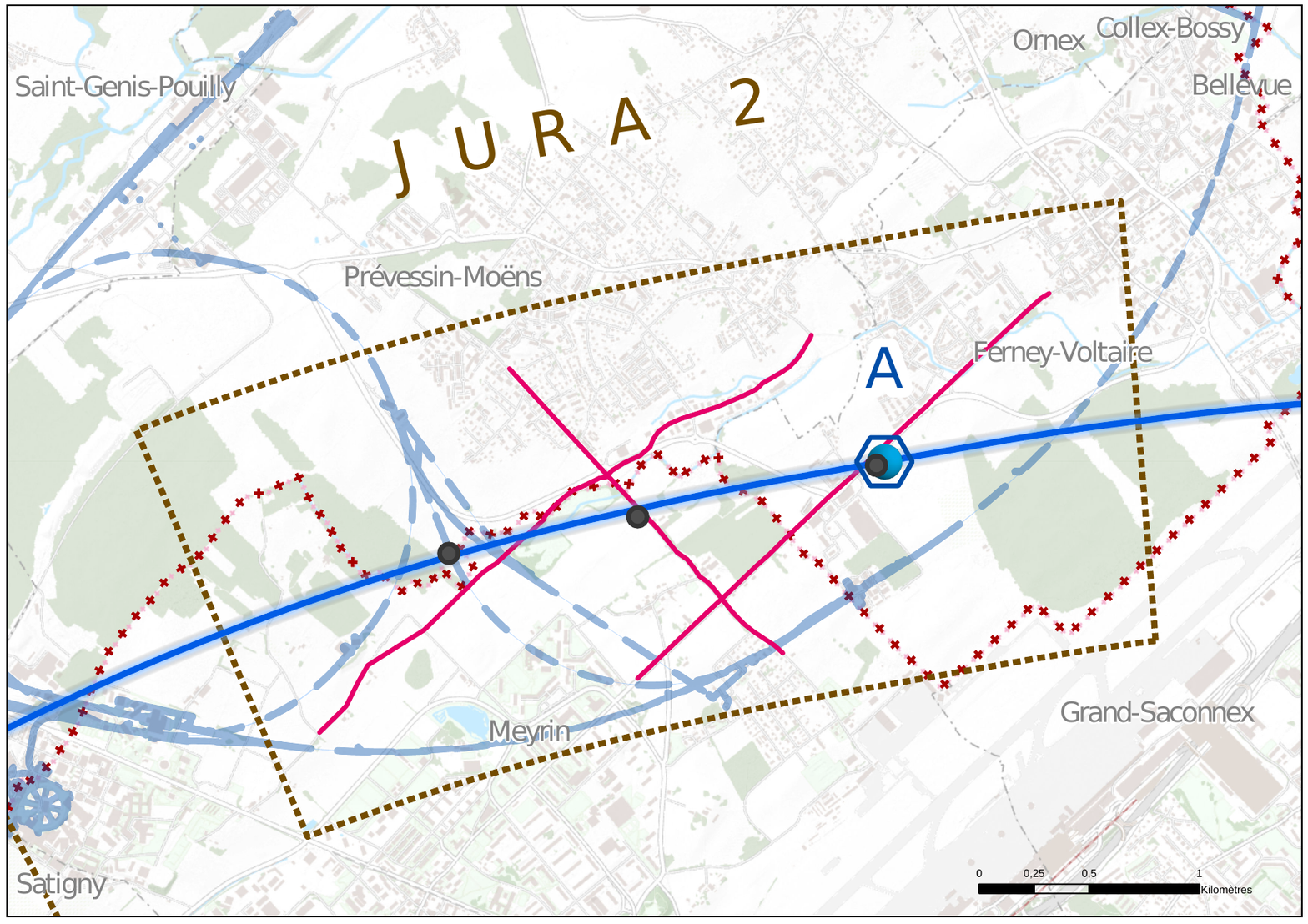}
    \caption{Jura 2 section showing the locations of the investigations.}
    \label{Sector Jura 2}
\end{figure}

The following SSI is planned to start in mid-2025 and will last until late 2025:
\begin{itemize}
    \item three boreholes ranging from 240-250\,m in depth.
    \begin{itemize} 
        \item[$\circ$]  one destructive and partially cored.
        \item[$\circ$]  one fully cored.
        \item[$\circ$] one optional destructive and partially cored depending on the results of the geophysics and adjacent boreholes.
    \end{itemize}
      
    \item 7300\,m of high-resolution reflection seismic geophysics.
\end{itemize}
The locations of the proposed SSI are shown in Fig.~\ref{Sector Jura 2}.

\subsubsection{Lake}

Under Lake Geneva, the proposed FCC tunnel is at a depth of approximately 100\,m below the bottom of the lake. The subsurface geology is characterised by moraines of mixed characteristics; however, the absence of pre-existing borehole data in this area has led to uncertainty in the geological model. 

Improvements in understanding have been achieved by incorporating data from the GEothermies programme (acquired south of the proposed alignment) into the 3D geological model. Although this data does not clearly define the interface between the moraines and the underlying molasse to the north where the proposed FCC tunnel is foreseen, it does indicate that the interface is located at a higher depth than previously considered.

Avoidance of the moraines is considered a key design priority, as tunnelling in water-bearing moraines can be challenging. This will be confirmed after the SSI campaign.

\begin{figure}[!ht]
    \centering
    \includegraphics[width=\linewidth]{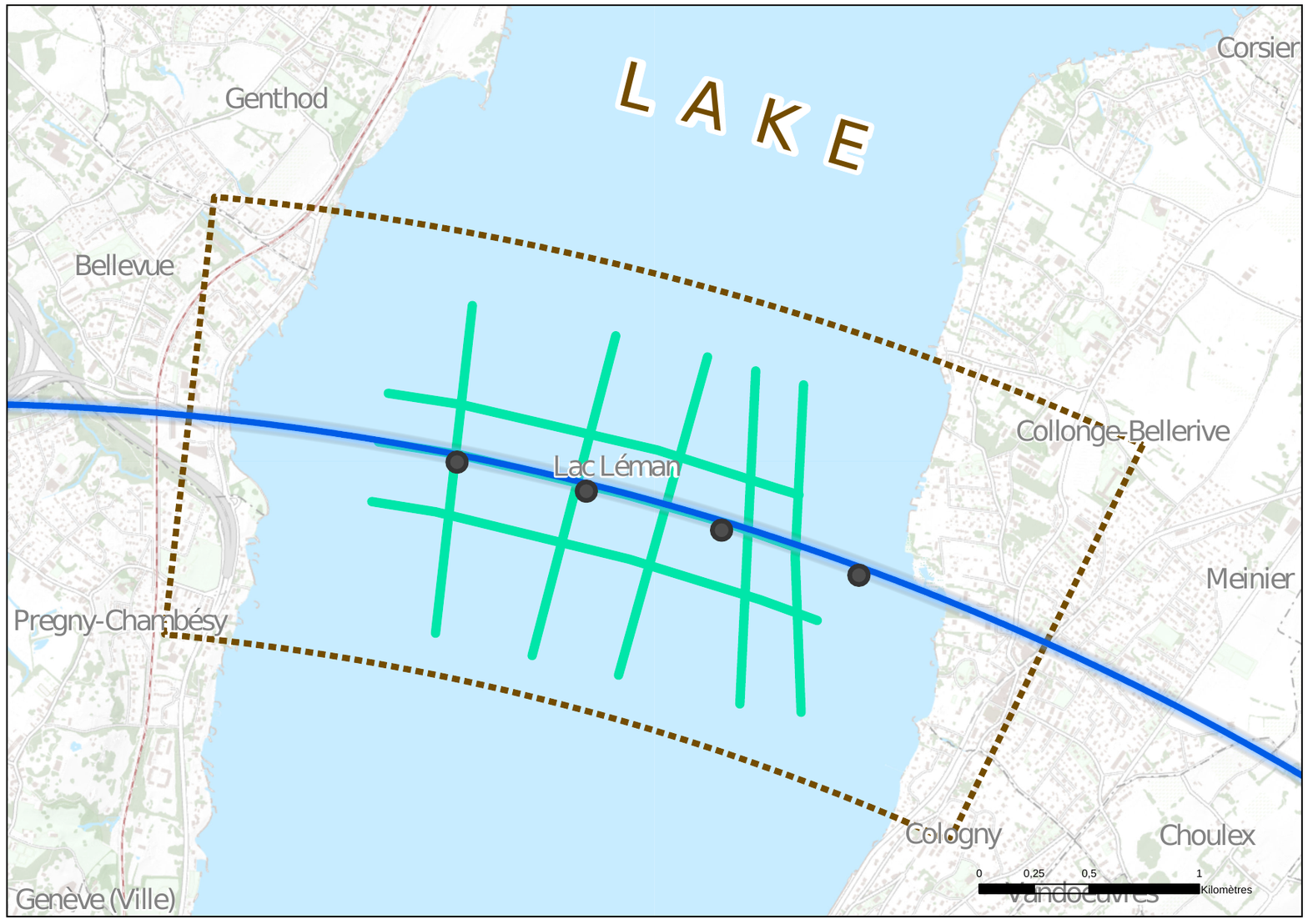}
    \caption{Lake section showing the locations of the investigations.}
    \label{Sector Lake}
\end{figure}

The following SSI is planned to start in mid-2025 and will last until late 2025:
\begin{itemize}
\item four fully cored boreholes ranging from 130-180\,m in depth.
\item 16\,340\,m of offshore seismic reflection.
\end{itemize}
The locations of the proposed SSI are shown in Fig.~\ref{Sector Lake}.

\subsubsection{Arve} 
The Arve section, located at the eastern extremity of the proposed FCC tunnel, extends approximately eight km in length, with the proposed FCC tunnel depth varying between 140 and 170\,m. Situated within the Arve Valley and adjacent to the Arve River, the area is known to be made up of a mixture of molasse and glacial moraines near the surface, although geological data at the proposed FCC tunnel depths is limited.

Further investigations will confirm the depth of the molasse-moraine interface and ensure that the tunnel remains within the molasse, thereby avoiding the potentially water-bearing moraines.

The following SSI is planned to start in early 2025 and will last until mid-2025:
\begin{itemize}
\item five destructive and cored boreholes totalling 990\,m in depth.
    \begin{itemize}
    \item[$\circ$]  Up to three of these boreholes may not be undertaken following results from geophysics.
    \end{itemize}
\item three high-resolution seismic reflection lines totalling, 13\,150\,m. 
 \end{itemize}
 
\begin{figure}[!ht]
    \centering
    \includegraphics[width=\linewidth]{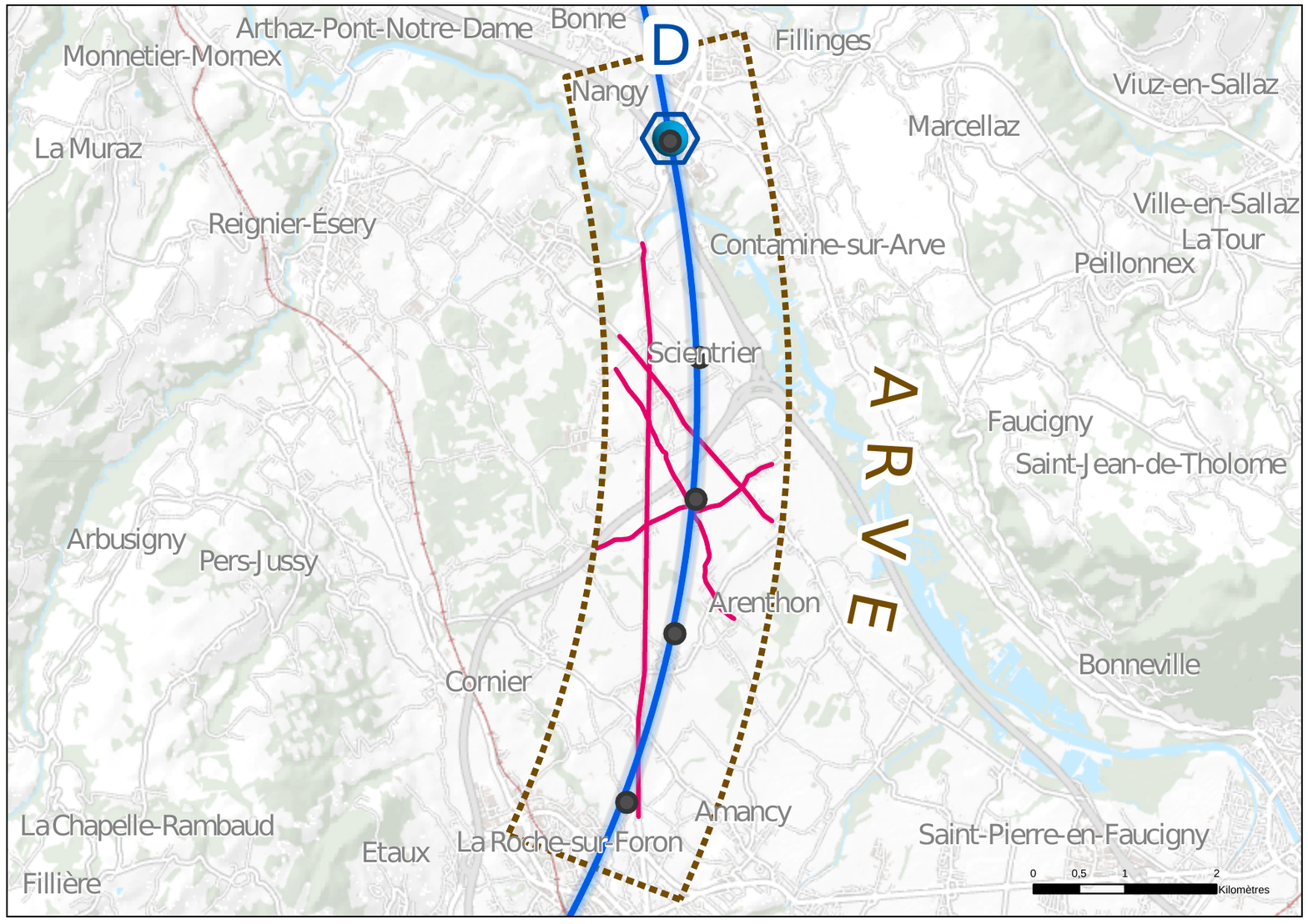}
    \caption{\label{Sector Arve}  Arve section showing the location of the investigations. }
\end{figure}

The locations of the proposed SSI are shown in Fig.~\ref{Sector Arve}.

\subsubsection{Bornes} 

The area adjacent to the Bornes Plateau is the deepest section of the project, with depths ranging from 500 to 560\,m over five\,km. The deepest access shaft at PF, approximately 400\,m deep, is also located within this section.

The proposed FCC tunnel will remain within the molasse, as the limestone is situated much deeper—nearly 1000\,m below the surface. However, little is known about the characteristics and quality of the molasse. Additionally, thrusts, faults, and potential tectonic materials have been identified in the region.

Given these factors and the high overburden, further investigations are required to obtain detailed information on the molasse and reduce geological uncertainties in this sector.

\begin{figure}[!ht]
    \centering
    \includegraphics[width=\linewidth]{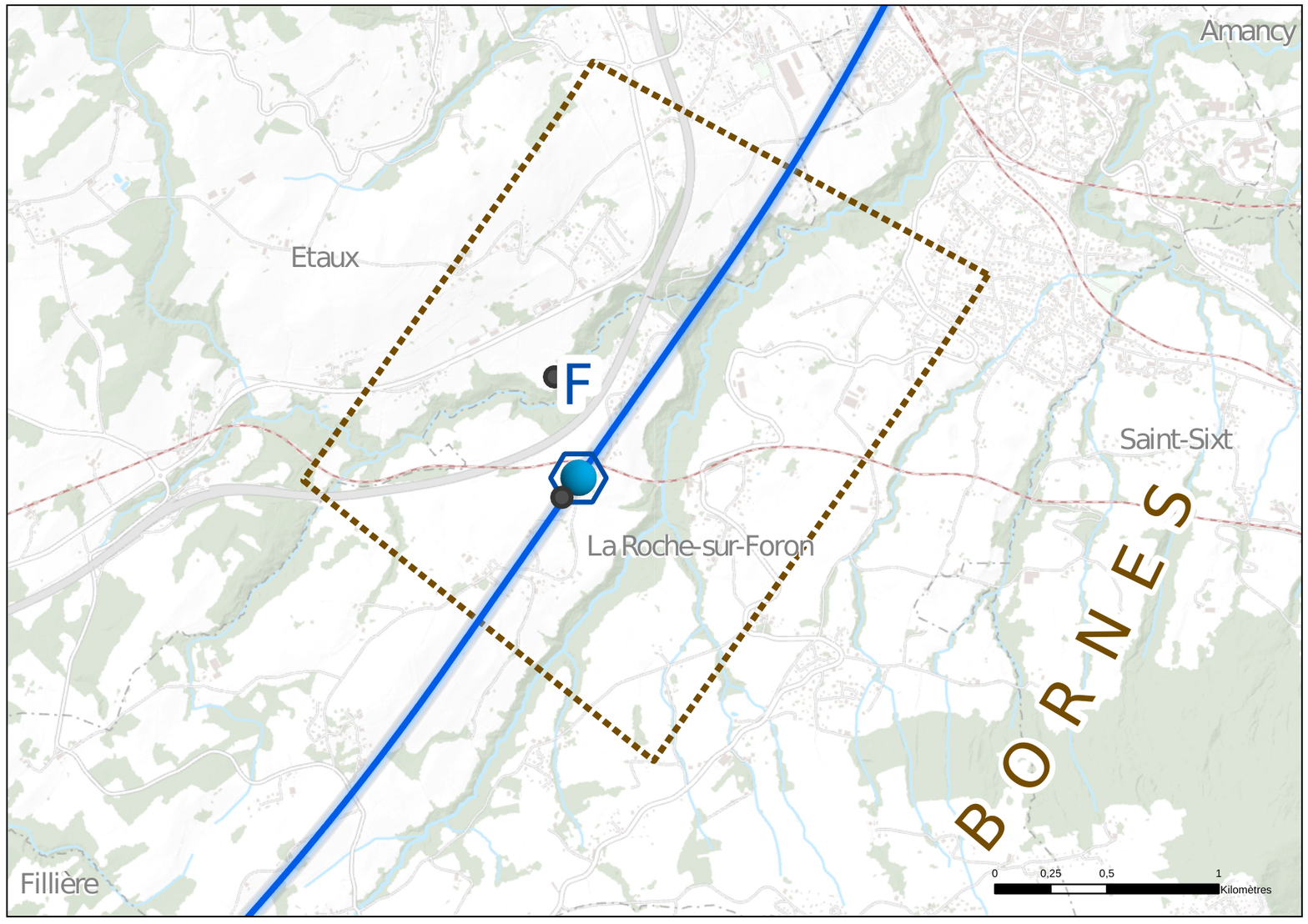}
    \caption{Bornes section showing the locations of the investigations. }
    \label{Sector Bornes}
\end{figure}

The following SSI is planned to start in early 2025 and will last until mid-2025:
\begin{itemize}
\item one fully cored borehole totalling 425\,m in depth.
 \end{itemize}
The locations for the SSI are shown in Fig.~\ref{Sector Bornes}.
 
\subsubsection{Mandallaz} 

The Mandallaz is an anticline limestone range characterised by faulting and thrusting over the underlying molasse on its western side. Little detailed geotechnical information was available prior to the start of the SSI campaign. This presented uncertainties regarding the extent of the limestone and the nature of its structure.

Improvements in understanding have been achieved through geophysical investigations as part of the SSI. Data acquisition has been completed, and provisional interpretations indicate that the limestone at the depth of the proposed FCC tunnel is narrower than modelled before the campaign. This revised interpretation suggests that the section requiring tunnelling through limestone is shorter than initially foreseen. Additionally, preliminary borehole data have revealed the presence of small karsts and traces of hydrocarbons, and provisional geophysical interpretations have identified an east-west strike-slip fault across the Mandallaz mountains, although this has yet to be confirmed by the boreholes.

\begin{figure}[!ht]
    \centering
    \includegraphics[width=\linewidth]{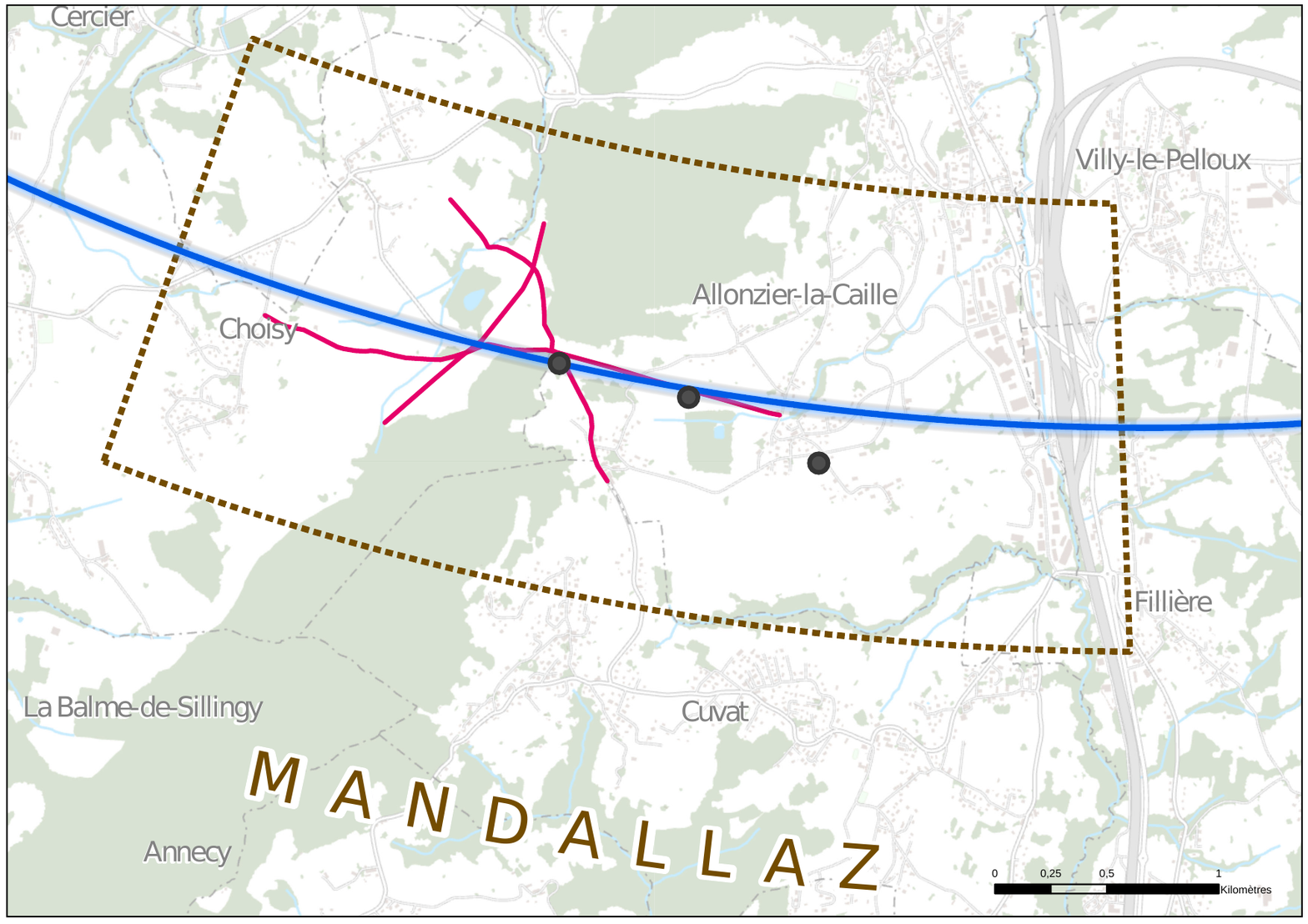}
    \caption{Mandallaz section showing the locations of the investigations.}
    \label{Sector Mandallaz}
\end{figure}

The following SSI has been carried out or is taking place:
\begin{itemize}
\item three fully cored boreholes totalling 1240\,m in depth
    \begin{itemize}
    \item[$\circ$] one of these boreholes is inclined at 30\textdegree from vertical
    \item[$\circ$] one of these boreholes may not be required to be undertaken following the results of the first two boreholes.
    \end{itemize}
\item three high-resolution seismic reflection lines totalling, 5100\,m. 
\end{itemize}
The locations for this SSI are shown in Fig.~\ref{Sector Mandallaz}.

\subsubsection{Usses} 

The Usses, a small section extending roughly one km, is where the proposed FCC tunnel is closest to the surface, with about 50\,m of cover above the tunnel at the lowest point in the Usses Valley.

 On-site investigations have improved the understanding of the subsurface conditions in this sector. All data acquisition has been completed, and the final interpretation of the geophysical data is underway. Borehole results indicate that the moraine-molasse interface is encountered at 15\,m, shallower than modelled, suggesting that the proposed tunnel does not intrude into water-bearing moraines beneath the Usses River. This will be confirmed once the final geophysical interpretations are complete and compared against the borehole log.

\begin{figure}[!ht]
    \centering
    \includegraphics[width=\linewidth]{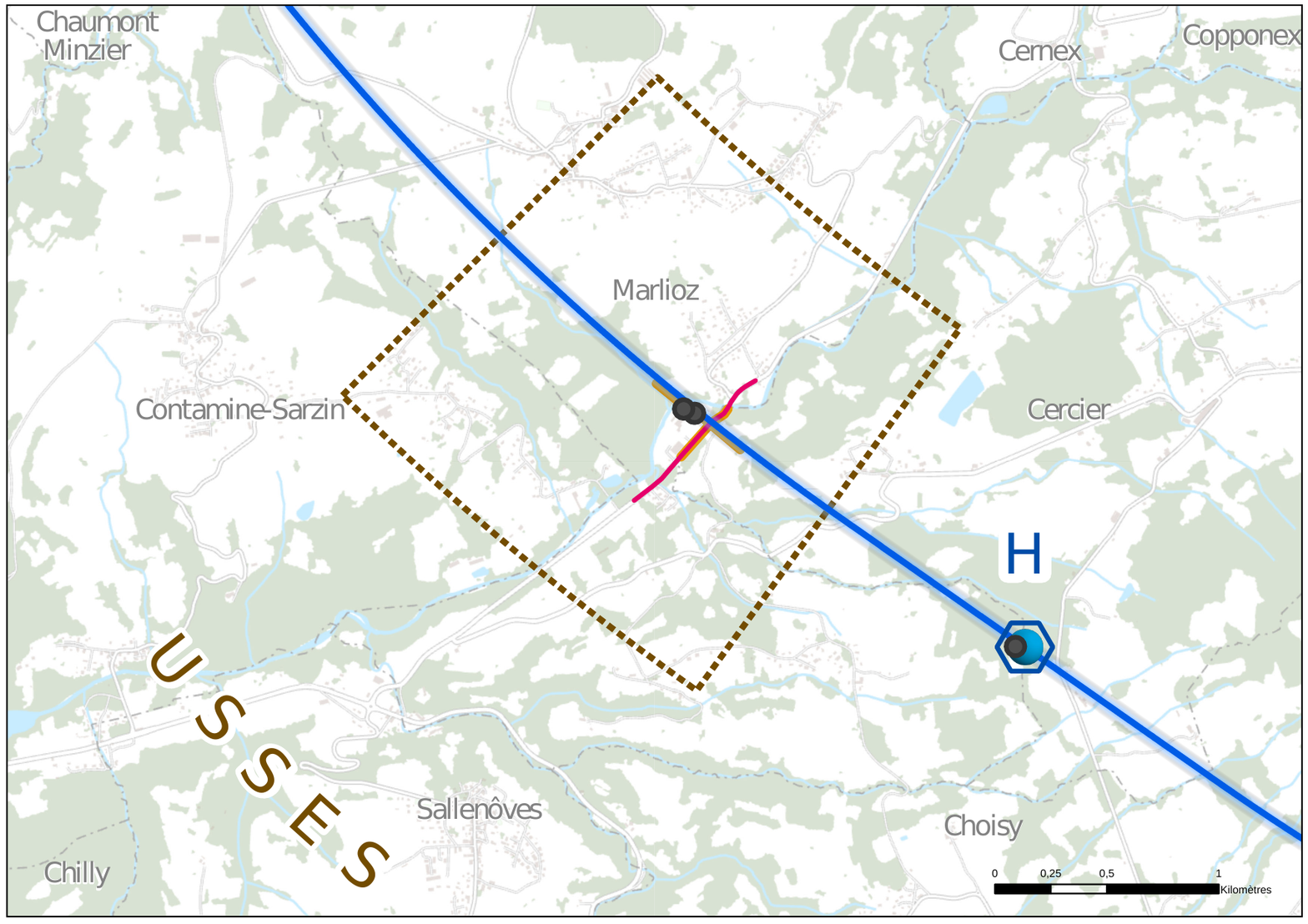}
    \caption{ Usses section showing the locations of the investigations.}
    \label{Sector Usses}
\end{figure}

The following SSI has been undertaken:
\begin{itemize}
\item one fully cored borehole of 70\,m.
\item two seismic refraction lines totalling 780\,m, the first performed using an explosive source and the second utilising a weight drop source.
\item one very high-resolution seismic reflection of 770\,m in length. 
\end{itemize}
The locations for the SSI are shown in Fig.~\ref{Sector Usses}.

\subsubsection{Vuache} 

The Vuache section at the southwestern extent of the proposed FCC tunnel extends over approximately four km, with proposed tunnel depths ranging from 190 to 300\,m and an average depth of around 250\,m. The Vuache mountain range, primarily an anticline structure bounded by thrust faults, underlies this section, with the prominent Vuache Fault located to its south-west. The geology is dominated by Jurassic-Cretaceous limestones and marls, overlain by the typical molasse found elsewhere in the Geneva basin.

Geo-physical investigations carried out as part of the SSI have improved the understanding of the subsurface conditions. One completed borehole has confirmed that molasse is present to about  30\,m below the tunnel. The ongoing analysis of the geophysical data, along with additional borehole results, will identify whether this is also the case for the remainder of the section.

\begin{figure}[!ht]
    \centering
    \includegraphics[width=\linewidth]{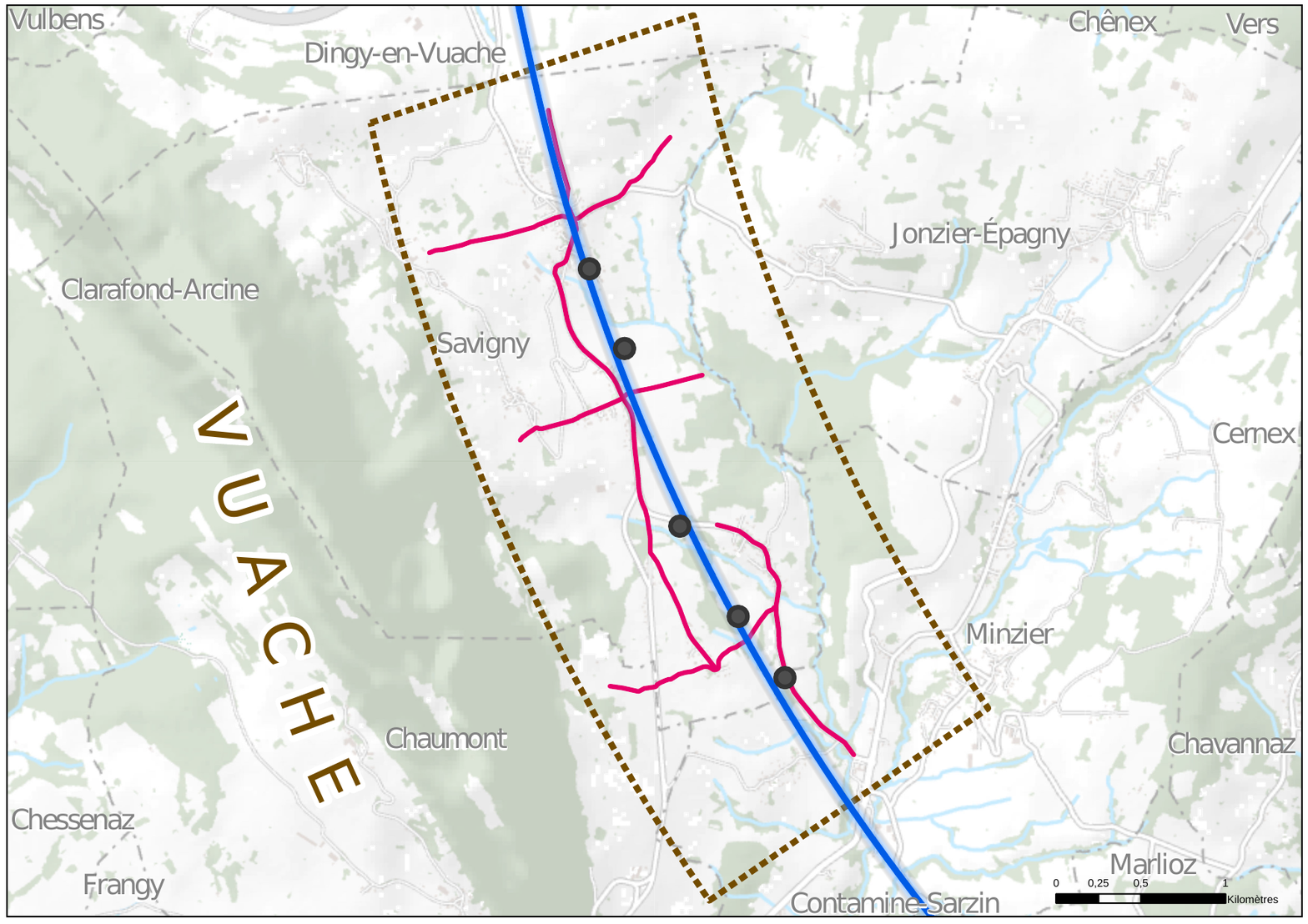}
    \caption{Vuache section showing the locations of the investigations.}
    \label{Sector Vuache}
\end{figure}

The following SSI has been carried out or is taking place:

\begin{itemize}
\item five destructive and partially cored boreholes totalling 1270\,m in depth.
    \begin{itemize}
    \item[$\circ$] two of these will be optional, depending on the conclusions of the geophysics.
    \end{itemize}
\item five high-resolution seismic reflection lines totalling, 9370\,m.
\end{itemize}
The locations of the SSI are shown in Fig.~\ref{Sector Vuache}.

\subsubsection{Rh\^{o}ne}

In the Rh\^{o}ne Valley, the proposed FCC tunnel is located approximately 50\,m below the surface under the Rh\^{o}ne River and the environmentally sensitive Marais de l’Etournel. The overlying soils are predominantly composed of moraine or alluvial sands, gravels and boulders, and the area was formerly quarried as a source of gravel. These layers are saturated due to the influence of the Rh\^{o}ne River.

The current geological model predicts that the tunnel is situated entirely within the molasse, although the interface between the molasse and the overlying moraine is predicted to be close. While limited available data has allowed this initial interpretation, the scarcity of detailed geotechnical information limits its precision.

As a result, further investigations are required to accurately define the moraine–molasse interface and confirm that the proposed FCC tunnel remains fully within the molasse, thereby mitigating potential risks associated with the water-bearing overlying layers.

\begin{figure}[!ht]
    \centering
    \includegraphics[width=\linewidth]{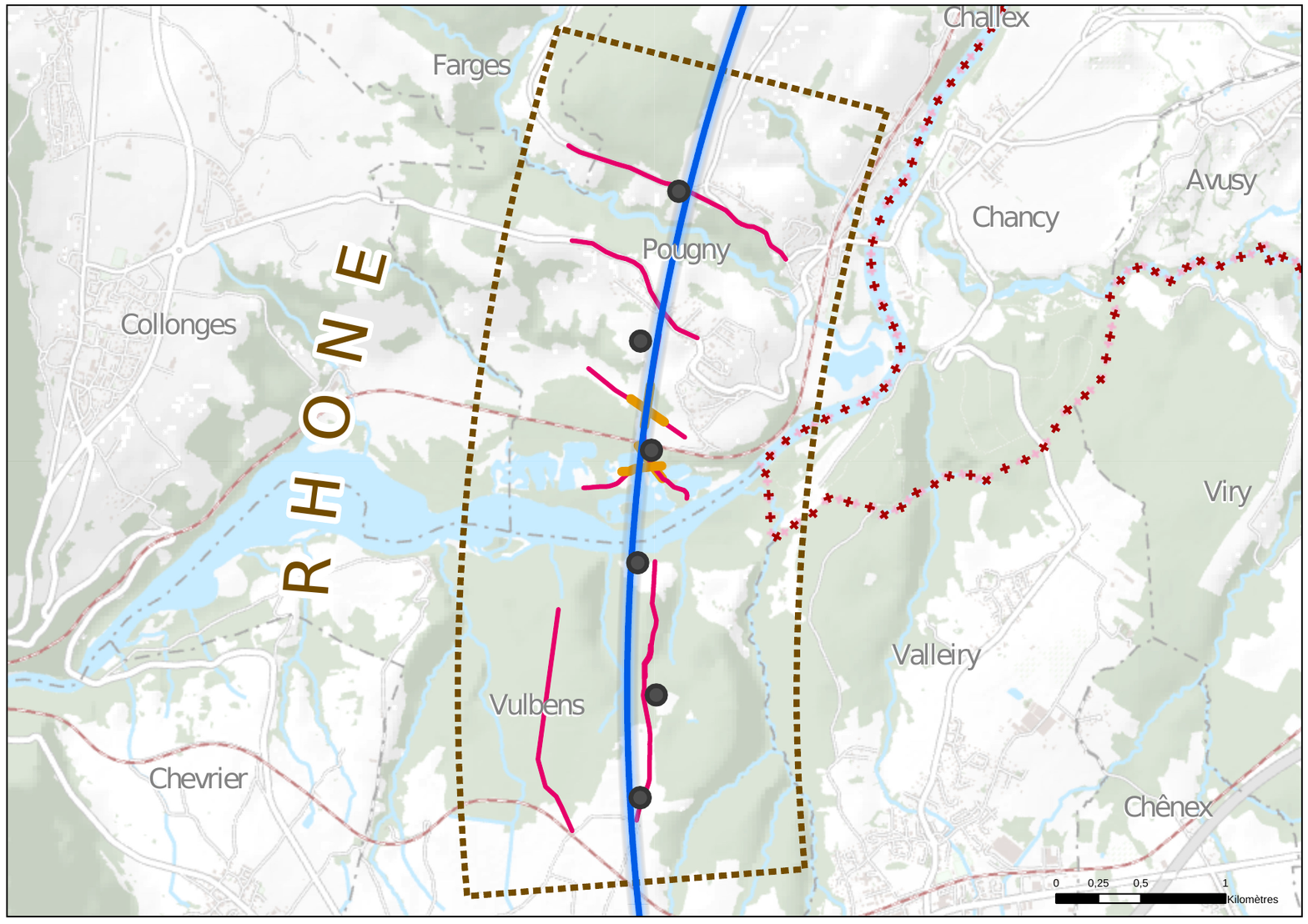}
    \caption{ Rh\^{o}ne section showing the location of the investigations.  }
    \label{Sector Rhone}
\end{figure}

The following SSI could start in mid-2025 and finish by late-2025:

\begin{itemize}
\item six destructive and partially cored boreholes totalling 765\,m in depth.
    \begin{itemize}
    \item[$\circ$] Four of these boreholes may not be required following results from geophysics.
    \end{itemize}
\item eleven geophysics seismic lines.
    \begin{itemize}
    \item[$\circ$] four profiles of high-resolution seismic reflection totalling 4920\,m towards the upper part of the valley.
    \item[$\circ$] four profiles of very high-resolution seismic reflection totalling 2990\,m towards the bottom of the valley.
    \item[$\circ$] four profiles of seismic refraction at the bottom of the valley using a weight drop source, totalling 240\,m.
    \end{itemize}
\item five high-resolution seismic reflection lines totalling 9370\,m.
\end{itemize}
The locations of the SSI are shown in Fig.~\ref{Sector Rhone}.

\subsection{Phase 2 sub-surface site investigations}

After the completion of the first phase of the SSI at the end of 2025, a geotechnical and civil engineering specialist may be contracted in early 2026—subject to the decision-making process—to analyze the campaign results and the corresponding geological model. Working in collaboration with CERN, this specialist would define the scope and technical specifications for the second phase of the sub-surface site investigation and oversee the on-site operations.

This second phase would focus primarily on obtaining the necessary geotechnical data for the detailed design of all underground and surface works to be undertaken. In particular, the second phase would focus on the sites where experiment and service caverns are required since these require very detailed knowledge of the lithology and rock engineering properties in order to undertake the complex numerical analysis required to define the rock support necessary to create stable underground structures. 

\subsubsection{Detailed Investigations Around Key Infrastructure}

\begin{figure}[!ht]
    \centering
    \includegraphics[width=\linewidth]{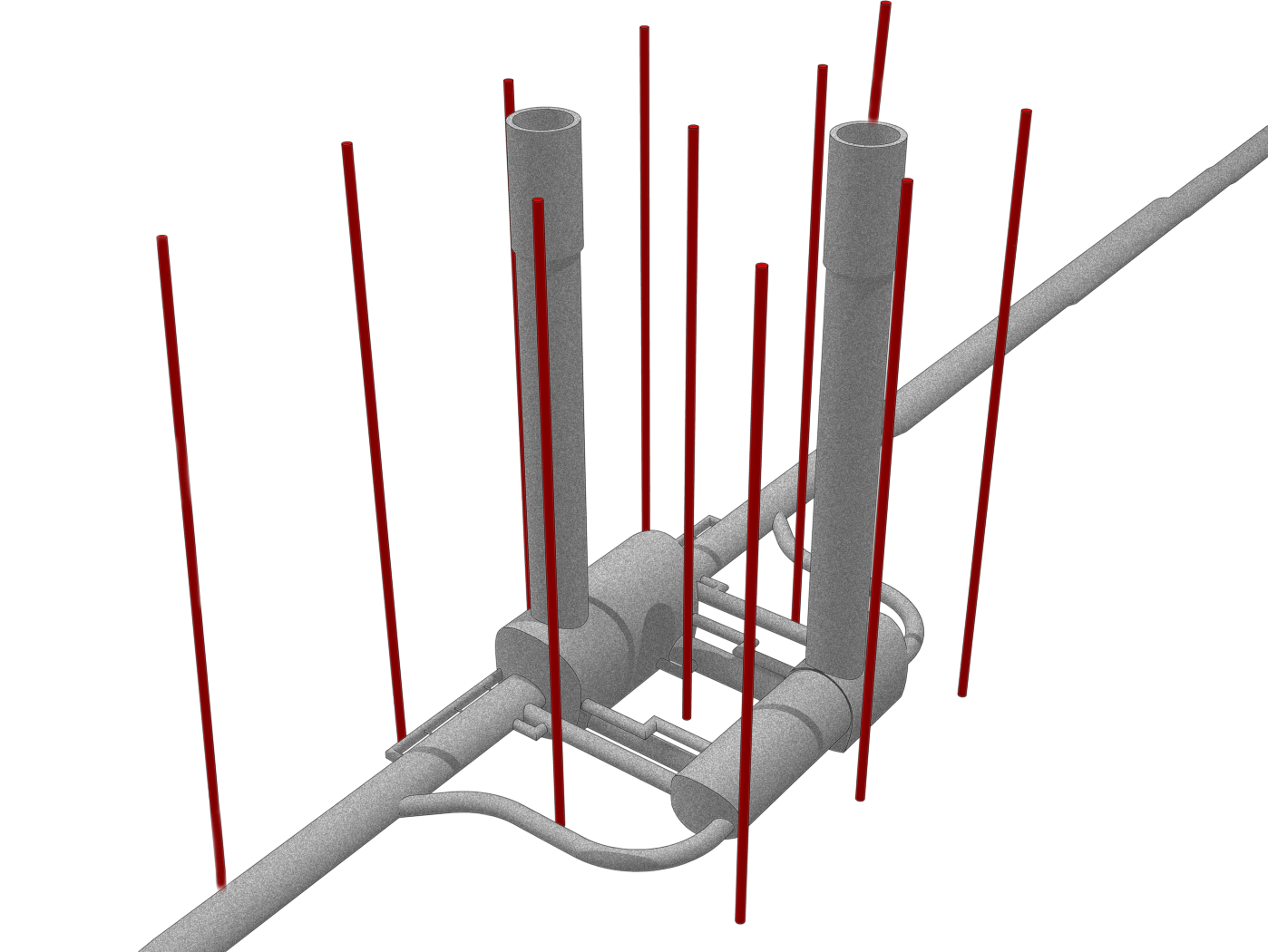}
    \caption{Example of targeted borehole investigations for experiment cavern complexes.}
    \label{boreholes for cavern complex}
\end{figure}

Precise geological, geotechnical, and hydrogeological data are essential for the effective design of both experiment and service caverns, which are typically located at depths of several hundred metres. This information ensures structural integrity and supports the selection of appropriate construction methodologies. An indicative example of a typical drilling configuration can be observed in Fig.~\ref{boreholes for cavern complex}.

These boreholes will accurately characterise the molasse at the cavern depth, identifying structural features such as fractures and weak zones that could affect cavern stability. Core samples and in-situ testing will determine the molasse's strength, deformability, and stress state, enabling precise structural analyses and the design of appropriate excavation and support systems. In addition, data on groundwater levels, permeability, and pore pressures will inform dewatering plans and help manage water ingress - factors critical to ensuring safety and stability during and after construction. Consequently, the borehole data from this campaign will be essential for optimising excavation methods, reducing risks, and estimating construction costs accurately.

\subsection{Fault Mapping and Seismic Modelling}

The FCC study region is known to contain active or potentially active faults, including the Vuache Fault. Historical seismic events have provided valuable insights into the area's seismic behaviour.

\begin{figure}[!ht]
    \centering
    \includegraphics[width=\linewidth]{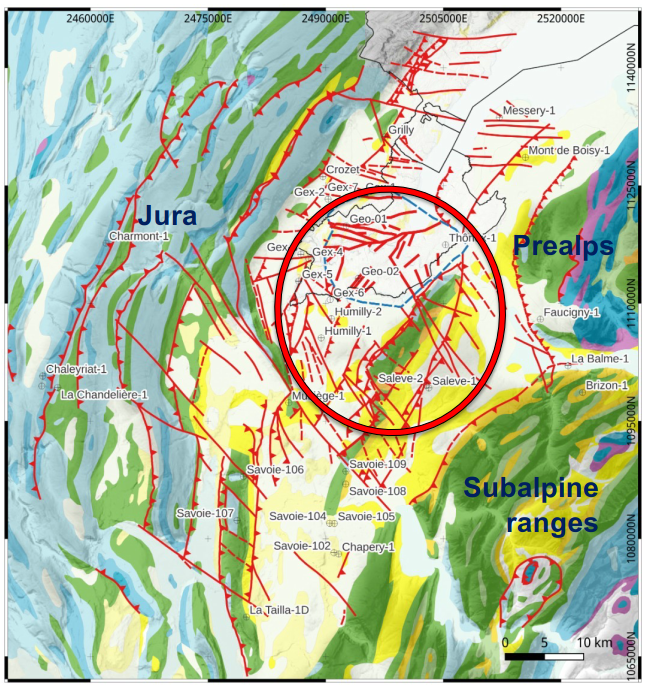}
    \caption{ Fault map of FCC study area.  }
    \label{Faulting Map}
\end{figure}

While detailed study of seismic activity and structural faulting was not the primary focus of the first phase of the SSI, CERN, in collaboration with the University of Geneva, has utilised historical data to identify the main fault zones in the region, as shown in  Fig.~\ref{Faulting Map}.  

\begin{figure}[!ht]
    \centering
    \includegraphics[width=\linewidth]{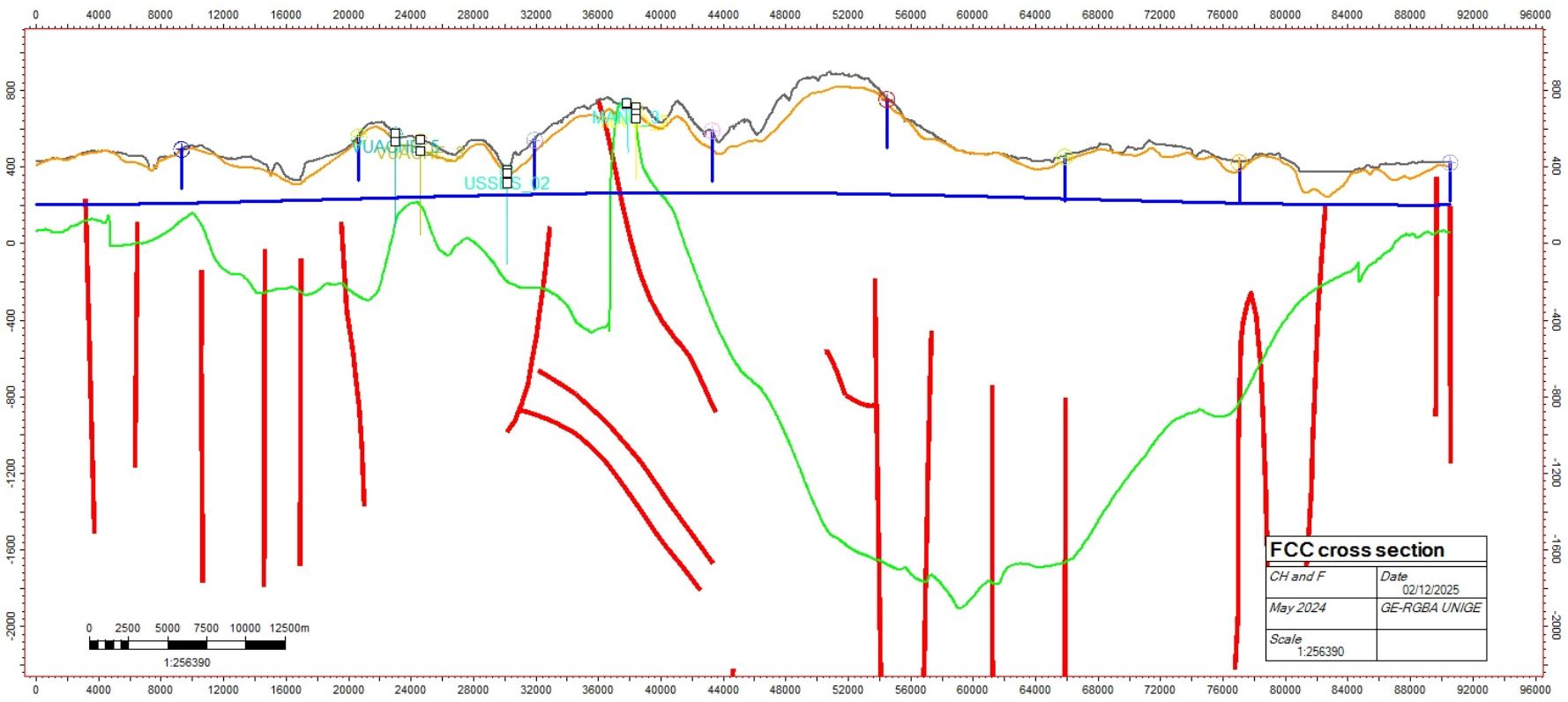}
    \caption{ FCC long profile with potential faults.  }
    \label{FCC Long Profile Faulting}
\end{figure}

The proposed FCC tunnel is expected to pass through a zone with faulting, as illustrated in Fig.~\ref{FCC Long Profile Faulting}. In areas such as the Mandallaz and Jura sections, where faults are known to be present, further refinement of fault characterisation will be possible once the fully processed results from the phase one SSI campaign become available. This will continue to be developed in collaboration with external partners over the coming years.

\subsection{Summary of the Subsurface Site Investigations}

The subsurface site investigations that have already been carried out as part of Phase One have improved the reliability of the 3D geological model for the proposed FCC tunnel. The preliminary results from investigations in the key sections of Mandallaz, Usses, Vuache and Arve have been integrated into the model, and early results show promising signs that the FCC tunnel would be located within the molasse except in the Mandallaz section, where crossing the limestone is unavoidable.

The results from the remainder of the campaign will need to be fully acquired and processed before any definitive conclusions on the precise elevation of the tunnel can be made. However, current indications are that the tunnel might be shallower than currently planned. This would allow an overall reduction in shaft depth, reducing the impact and, therefore, have positive cost and schedule implications


\section{Management of excavated materials}

\subsection{Introduction} \label{matexintro}
Building the underground infrastructures of the Future Circular Collider (FCC) in the Franco-Geneva Basin would produce approximately 6.3 million cubic metres of excavated materials, mainly constituted by molasse, a soft heterogeneous rock (96\%).

The management of excavated material stands as a key element in the realisation of the FCC within the local region. Furthemore, it exemplifies a visionary investment designed to reduce environmental impact and unlock benefits that extend well beyond the local context. Accordingly, from the earliest stages, an approach to the strategy for the excavated material management was incorporated into the feasibility study as part of the FCCIS project, co-funded by the EU under the H2020 programme (grant agreement no.\,951754).

The French tunnel design centre, (CETU), the Centre for studies and expertise on risks, environment, mobility and urban planning, (CEREMA), the Montanuniversit\"{a}t Leoben (Austria) collaborated with the FCC study on this subject, with the technical support of the Swiss cantonal `Service de g\'{e}ologie, sols et d\'{e}chets' (GESDEC). 

Current regulations still consider excavated materials as waste as soon as they exit the project boundaries, but the legislation is evolving in many European Countries. Excavated materials can be reused on-site (prevention of waste production) or recovered off-site. Final disposal as waste must be the last option when no reasonable recovery (even after suitable preparation or treatment) is possible. The successful management and use of excavated materials strongly depend on the early implementation of a management strategy shared and agreed upon with the Host States by the project owner.

The presentation of a management plan of the excavated materials is considered to be the responsibility of the future project owner, irrespective of who implements the actual tasks. To satisfy this requirement, a report outlining the approach to the strategy for the management of the excavated material (in French `Strat\'{e}gie de gestion et d'usage des matériaux excav\'{e}s' \cite{ulrici_2025_14923266} also available in English \cite{ulrici_2024_13785651}) was developed. The following aspects are included: 
\begin{itemize}
\item Quantities of excavated materials and current knowledge of the geological characteristics.
\item General principles for the management of the materials.
\item Identification of the potential risks.
\item Legal framework in the two Host States.
\item Potential reuse cases.
\end{itemize}

\subsection{The excavated material quantities} \label{matexquant}
The FCC excavation work is estimated to generate approximately 6.3 million cubic metres of excavated material, equal to approximately 8.2 million cubic metres of expanded material over nearly a decade. The quantities of excavated materials  (see Table \ref{ExcavationTable} ) were calculated on the planned design of the underground structures (shafts, tunnels, caverns, alcoves, etc.). They correspond to the materials in place and the expanded materials (applying an expansion factor of 1.3). 

\subsection{The excavated material characteristics} \label{matexcharact}

The material extracted will mostly comprise different types of heterogeneous soft rocks, called molasse (96\%), along with limestone (2.5\%) and quaternary deposits (1.5\%). 

The molasse is formed by a series of horizontal layers of cemented and silty sandstone interspersed with layers of marl and argillaceous rocks. Geogenic anomalies such as the presence of natural hydrocarbons or enhanced concentration of metals including nickel and chromium may exist naturally in varying quantities in parts of the various molasse layers.

\begin{figure}[!ht]
    \centering
    \includegraphics[width=0.95\linewidth]{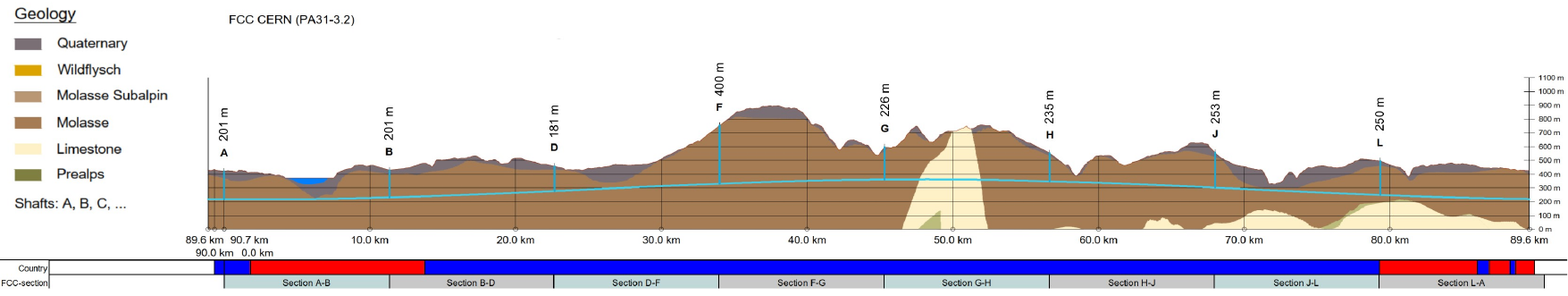}
    \caption{Geological profile along the FCC path. The part of the tunnel under French territory is shown in blue; the part under Swiss territory is shown in red.}
    \label{fig:FCC profile MATEX}
\end{figure}

The Lemanic basin features three main geological groups: quaternary formations (glaciolacustrine, fluvio-glacial and moraines), molasse and limestone bedrock. The reference placement proposed for the FCC has been optimised to place the tunnel infrastructure in the Franco-Geneva molasse basin (Fig.~\ref{fig:FCC profile MATEX}). The limestone elements of the Jura and Vuache mountains, along with the Mandallaz and Sal\`{e}ve ranges, skirt and intersect the molasse layers. It is hard to excavate limestone in the region due to its karstic characteristics caused by chemical alteration of the rock. These deep karsts are likely filled with water and unconsolidated materials, which could penetrate the excavation because of strong water pressure. CERN experienced these effects when conducting excavations in the limestone of the Geneva region during previous projects (e.g., LEP in the ’80s). Because of these difficulties, the design deliberately avoids the Jura and the limestone of the Vuache. However, despite every effort to minimise crossing limestone strata, one stretch of the proposed tunnel still passes through the Mandallaz limestone.

The molasse in the region is generally made up as follows:
\begin{itemize}
\item 30 to 50\% clay,
\item 10 to 15\% silt,
\item 10 to 15\% sand, with a grain size of between 63 μm and 4 mm,
\item 15 to 20\% sandstone particles larger than 4 mm.
\end{itemize}

The sandstones (detrital sedimentary rocks formed from grains of sand cemented by silica, calcite and iron oxide) in the Lake Geneva region are mainly made up of:
\begin{itemize}
\item 40 to 70\% quartz,
\item 20 to 45\% calcite,
\item 5 to 10\% feldspar,
\item 5 to 20\% phyllosilicates (micas, chlorites, serpentinite, etc.)
\end{itemize}

The marls (sedimentary rocks containing clay and limestone) present in the region exhibit a wide variety of compositions. Marls are ductile, micro-cracked and subject to swelling following contact with air or significant changes in soil moisture.

The nature of the excavated materials depends on the longitudinal profile of the geological structures encountered. Different formations and different facies may be encountered. Homogeneity in terms of the nature and properties of the excavated materials is a key consideration, as it will determine their subsequent use.

A strategy will have to be devised during the construction phase to select the materials to be reused, based on the following:
\begin{itemize}
\item Initial visual selection at the face.
\item Tests on the geological formations and updates in case of changes in petrographic properties or indications of lithological changes.
\item Rapid and effective material quality control as soon as possible after the extraction of the material, or at the latest when the material reaches the surface.  
\end{itemize}

Specific investigations would be required prior to the start of the excavation to determine the geological characteristics of the subsurface through which the FCC will pass and to assess the potential geological risks. CERN is conducting an initial investigation campaign (2024 - 2025) as part of the FCC feasibility studies. These first subsurface investigations could be later complemented by extensive subsurface investigations, devoted mainly to confirming the geological model and, therefore, to anticipating the possible risks and preparing adequate mitigation. Laboratory tests are being conducted on core samples extracted during the subsurface investigations to identify and characterise the rocks in view of their potential for reuse and to analyse the presence of geogenic anomalies.  

\subsection{Approach to the management of excavated materials and associated risks} \label{matexstratrisk}

Defining what could constitute a basis for a future strategy for managing and utilising excavated materials is a key feasibility criterion for the FCC study, reflecting both the alignment with the regulations of the Host States and a broader commitment to sustainability and environmental stewardship. The basic concept has been documented in a stand-alone deliverable of the FCCIS study \cite{ulrici_2025_14923266}. The core principles, summarised here, present the high-level aspects, related constraints and the opportunities identified for treating the material as a resource instead of disposing of it as waste. 

The approach to excavated materials management marks the initial step toward a comprehensive approach, conducted in accordance with the regulatory frameworks of both Host States. Its implementation, in the form of a preliminary operational management plan, can only begin once the first set of subsurface investigations is completed (by the end of 2025). Subsequently, this preliminary plan should be revised whenever significant new information is obtained, following an iterative process that would culminate in a final operational management plan to be shared and agreed with the Host States, and should be adopted before the start of excavation.

\subsubsection{Approach to the excavated material strategy} \label{matexstrat}

The studies carried out to develop an approach to the excavated material strategy were: 
\begin{itemize}
\item Iterative development of a 3D subsurface geological model (see Section~\ref{sec3dmodel}).
\item Inventory of the regional opportunities for reuse in France and Switzerland.
\item Investigation of the possible connection to the regional railway network.
\item Investigation of the possible connection of the railway sidings to the extraction sites by conveyor belts, to avoid local nuisances due to truck transport.
\end{itemize}

The FCC schedule allows the possibility of optimising potential reuse scenarios. Accordingly, rather than centring on fixed solutions, the excavated materials management focuses on flexible approaches and guiding principles that can be adapted to accommodate:

\begin{itemize}
\item Evolutions in technology that could result in improvements in the techniques applied to the excavation and separation and treatment of the excavated materials. 
\item New recovery pathways which are better suited to the characteristics of the extracted materials that may become available in the future.
\item Evolution of regulations governing the management of excavated materials for specific applications (e.g., use as soil improvers in agriculture).
\end{itemize}

The main parameters to be taken into consideration when building a strategy for the management of the excavated materials are: 
\begin{itemize}
\item TBM kinematics (Fig.~\ref{TBM Arrangement}) and the logistic aspects that determine the output and availability of excavated materials within the area as well as the nearby storage and treatment areas required.
\item Identified uses and streams: specifications, demand (quantities and variation over time), location, accessibility and service, criteria for acceptance of the material.
\item Modes of transport and distances, as well as whether existing infrastructures can be used or whether new infrastructures will need to be built.
\end{itemize}

\begin{figure}[!ht]
    \centering
    \includegraphics[width=\linewidth]{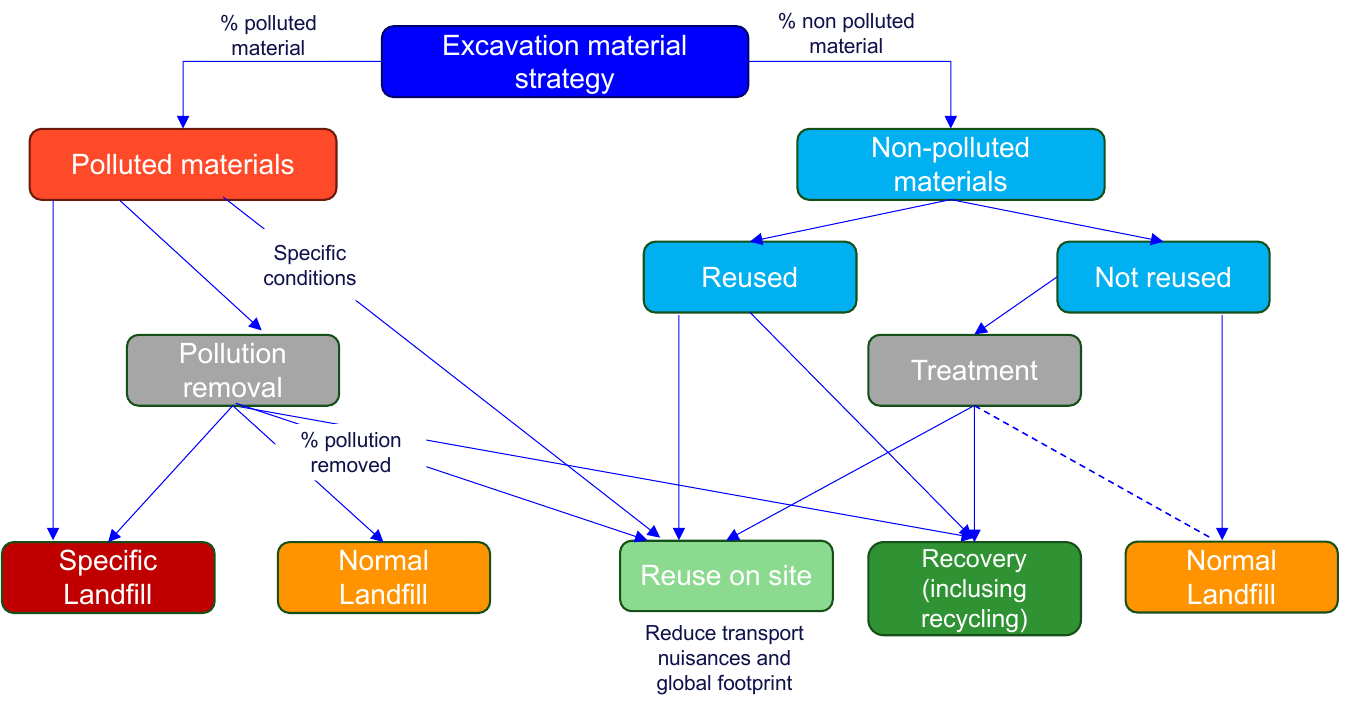}
    \caption{Schematic diagram of the scenarios defined for the excavated material strategy. (not including transport).} 
    \label{fig:Matex Strategy}
\end{figure}

A schematic overview of the approach to the excavated materials management appears in Fig.~\ref{fig:Matex Strategy}. Experience from previous CERN projects shows that the proportion of materials affected by geogenic anomalies (such as hydrocarbons or elevated chromium and nickel levels) may vary, ranging from approximately 15\% to 45\%, in part due to different regulatory thresholds of the Host States. For the record, exemptions have been granted for specific materials, for example, in 2019, during the HL-LHC project, to facilitate higher recovery rates in various recovery streams.

\subsubsection{Evaluation of risk in the management of excavated materials} \label{matexrisk}
The fraction of re-usable excavated materials depends on the geochemical, mineralogical and geotechnical properties of the materials excavated from the tunnel. When preparing a management plan for these materials, particular attention must be paid to the management of both technical and non-technical risks, which have a direct effect on limiting costs, optimising transport and avoiding compromising the schedule. 
Typical technical risks relating to the external context for excavated materials are associated with:
\begin{itemize}
\item The nature of the materials, which determines their classification as a specific type of material and their potential reuse;
\item The proportion of facies, in the case of a heterogeneous formation or a mixed facies with several types of material;
\item The position of the contacts between geological formations, which has an impact on the distribution of the volume of material per formation.
\end{itemize}

The technical risks of the internal context, regarded as points to keep in mind by the project owner, are associated with:
\begin{itemize}
\item The impact of the excavation methods, chosen according to the characteristics of the excavated materials and, in particular, the additives necessary for the excavation, which can potentially pollute the materials and hinder certain reuse cases.
\item The equipment used in the transport and in the treatment of the excavated materials, for the potential of unforeseen mixing and potential pollution of the excavated materials.
\item The availability and extent of the temporary and final storage areas for excavated materials.
\end{itemize}

Along with the technical risks, non-technical challenges, such as political or administrative factors, should also be taken into account in the development plan. For example, there may be modifications to environmental regulations. These non-technical challenges must be considered from the study phase.

Excavation methods can have an impact on the potential use of molasse and other excavated materials since they can alter their properties and quality. At this stage of the FCC study, the use of double-shield Tunnel Boring Machines (TBM) is likely to be possible for most of the main tunnel. Further studies should provide more detail for the information provided in Table~\ref{tab:matexExcavMeth}, specifying the impact of the chosen excavation techniques on the quality of the excavated materials (type of TBM, explosives or road header).

\begin{table}[!ht]
\centering
\caption{Quantity of excavated material by excavation method.}
\label{tab:matexExcavMeth}
\begin{tabular}{lcccr}
	\toprule
	&  Using TBMs& \makecell[c]{ Using \\conventional\\ method}&  Shafts 
	& \multicolumn{1}{c}{Total} \\ \midrule
	Collider infrastructure (m$^3$, in situ)&  2\,689\,500&  2\,828\,500&  
	652\,600& 6\,170\,600\\ \midrule
	\makecell[l]{Connecting infrastructure\\ to existing tunnels\\ (injection) (m$^3$, in situ)}&  97\,900&  9800&  14\,600& 122\,300\\ \midrule
	Total (m$^3$, in situ)&  2\,787\,400&  2\,838\,300&  667\,200& 6\,292\,900\\ \midrule
	Percentage of Total &  44\% &  45\% &  11\% & 100\% \\ \bottomrule
\end{tabular}
\end{table}

\subsection{Regulatory frameworks}
\label{matexREG}

In France, AFTES recommendation GT35R1F2 on the management and use of excavated materials \cite{AFTES_2019_GT35R1F2} and the CETU information document on naturally occurring geological materials excavated in underground structures \cite{CETU_2016_info-doc} are the main reference texts for the management of excavated materials.

In Switzerland, the ordinance on the avoidance and the disposal of waste \cite{ADWO_814-600}, complemented by the ordinance on the movement of waste \cite{VeVA_814-610}, the `Aide à l’execution relative à l’OLED' (help for the execution of the ADWO, not available in English) and the `Guide pour la r\'{e}utilisation des mat\'{e}riaux d’excavation non pollu\'{e}s' (guide to the reuse of unpolluted excavated materials, OCEV), provide the framework for the subsequent steps.

The main regulations in force in the two Host States regarding the application of international agreements on the export of excavated materials are: 

\begin{itemize}
\item In Switzerland, Article 15, paragraph 1 of VeVA 814.610 requires that anyone exporting waste (including unpolluted excavated materials) must obtain authorisation from the Federal Office for the Environment (VeVA 814.610, 2005).
\item In France, since 12 July 2007, the cross-border movement of waste has been subject to the provisions of Regulation (EC) No 1013/2006 of 14 June 2006, which incorporates  the provisions of the Basel Convention \cite{basel_convention}: 
 \begin{itemize}
     \item Article 43 permits the importing of waste for the purposes of recovery from countries which are Parties to the Basel Convention to a European country.
     \item Articles 40 and 41 prohibit the importing of waste for the purposes of disposal from countries which are Parties to the Basel Convention except where one of the countries (including Switzerland) presents a prior duly reasoned request (paragraph\,4). 
 \end{itemize}    
\end{itemize}

Article 4.2 of the Basel Convention \cite{basel_convention} states that each country must take measures to reduce the generation of waste and ensure the availability of adequate disposal facilities located, where possible, within the country, with a view to the environmentally sound management of dangerous waste and other waste, whatever the place of its disposal.

In the north-west sector of the FCC, its path crosses the Franco-Swiss border at several points. If the excavated materials were to be managed by each state, this would lead to the excavation of additional shafts for the specific purpose of material evacuation, which is not realistic and does not respect the `avoid, reduce and compensate' approach because it would increase the quantity of materials to be managed and the number of nuisance zones on the surface due to the extraction points. For this reason, it would be desirable to propose the adoption of the principle of each state managing the equivalent mass of spoil excavated on its territory. This adjustment, which remains to be agreed with the two Host States and in association with CERN, would make it possible to deal with the small enclaves that cannot be managed by the TBMs without changing the principle of division laid down in the Basel Convention.

A further point where an agreement between the project owner and the Host States seems appropriate concerns the country in which the materials are extracted. In principle, according to the Basel Convention, each country is responsible for extracting and managing its share of the mass on its territory and for dealing with the associated adverse effects (dust, noise, traffic, etc.). The distribution of the TBMs, however, does not necessarily correspond to this mass distribution (e.g., no TBM launched from site PB) for technical reasons. For such a scenario, compensatory measures will have to be agreed with and by the Host States.

Various tools will have to be developed to ensure that the management of the materials respects an established and agreed management plan that is part of the project authorisation files. These will have to be devised sufficiently early, in consultation with the Host States and relevant authorities, to ensure they are properly implemented when the time comes:

\begin{itemize}
\item Administrative tools for the authorisation procedures for spoil transport and storage to enable effective recovery of the material.
\item Regulatory tools to enable the validation of the innovative processes, particularly in terms of characterisation, and to ensure they can be standardised. 
\item Cross-border agreements: if relevant optimisations are identified, these will need to be made explicit, substantiated and submitted for joint approval by the two Host States.
\item Agreements with offtakers of materials and owners of land plots that would receive materials before the start of the excavation process.
\item Logistics tools: an adequate traceability system is the responsibility of the owner and it provides a useful way of improving the recovery of the materials and acts as a guarantee that the spoil is of sufficiently high quality.
\end{itemize}

The first three points could be governed by a bilateral international agreement.

Materials excavated during underground works are considered as ‘waste’ in the European Union (EU) countries and Switzerland unless they can be reused on the site of the project. The definition of the status of excavated soil and its classification as `waste' or `non-waste' are fundamental aspects, as they determine the legal framework to be followed for managing the excavated materials. The status of the excavated soil must be specified according to the various on- and off-site management scenarios with the French and Swiss authorities.

To develop an integrated plan for the management of the excavated materials that is part of the authorisation process, it would be advisable that the FCC is treated as a single indivisible and transnational project in the meaning of 1.c) of Article 2 of the `waste' directive 2008/98/EC of 19 November 2008 \cite{EU_directive}. The integrated approach facilitates the development of the agreements between the project owners and the customers for the materials that must be in place before the construction works start. It also enables the possibility to reuse materials on either side of the border, independent of their extraction site of the unique, indivisible project.

\subsection{Scenarios envisaged for the management of the excavated materials} \label{matexSCEN}

As a long-term project offering sufficient lead time, the FCC may represent an opportunity to develop innovative approaches for the management of excavated materials. For this reason, the study invested already in launching the research and development of new technologies and solutions that have the potential to reduce the amount of waste that needs to be transported to deposit sites.

The identification of uses must comply with the waste hierarchy put forward by the European Union parliament (`waste' Directive 2008/98/EC) and implemented in the French and Swiss regulations. The national policy for the prevention of waste production and reduction of waste impacts, outlined in the Environment Code (Art. L-541-1), supports the transition to a more sustainable management of the excavated materials, by fostering the application of the waste hierarchy: avoid waste production (including reuse), preparation for reuse, recovery and disposal. It also includes achieving a total `material' recovery of the waste stemming from the building and public works sector of 70\%. This target is evaluated at the regional level.

The general principle of avoid/reduce/compensate will also be applied to the management of excavated materials of the FCC: the ‘avoid’ and ‘reduce’ principles are applied mainly at the time of the subsurface infrastructure planning (i.e avoiding producing unnecessary spoil and reducing the overall excavations by optimising location, design, and depth of underground infrastructures) and in the choice of the excavation methods (minimisation of pollutants during the excavation, thus reducing the quantities of excavated materials that will need specific land filling). 

The various types of reuses and recovery call for material storage and treatment areas close to the extraction sites. The surface areas necessary depend on the detailed technical concept of the construction work, the construction schedule and the dimensional design of the treatment unit on the sites. These elements will have to be developed at a later stage.

Among the identified pathways, the most likely to be realised are: 
\begin{itemize}
\item \textbf{Use for development requirements within the project} (e.g., FCC worksite tracks, landscaping purposes, etc.)

\item \textbf{Use in earthwork (backfilling of quarries and mines and rewilding) and development projects}. This reuse case concerns most of the materials, as it can be applied for both the inert material and the polluted excavated materials, after appropriate decontamination. As each quarry is subject to specific environmental impact restrictions, specific acceptance criteria for the backfilling materials must be respected. In total, it is estimated that all material that will not be addressed to specific land filling due to pollution can, in principle, be reused for quarry landscaping, after treatment. However, in order not to saturate the region, other reuse cases are investigated at the same time. 

\item \textbf{Use of the limestone fraction} in concrete production and stabilisation of structures. The limestone, marl and clay deposits of the Jura and the northern Alps have provided the raw material for lime and cement since ancient times. Produced by calcining limestone, lime is used to improve soils and whitewash walls; mixed with sand and water, it produces a mortar that is very easy to produce and therefore very common. All limestone that will be extracted during the FCC construction project will be destined to these reuse cases. Due to the costs of production of these resulting construction materials, it is assumed that the reuse of limestone is at zero gain and zero cost for the project.

\item \textbf{Use of the sand fraction in concrete production}. Direct reuse of sand and gravel could be envisaged, eventually after treatment of sieving and /or washing of the excavated materials. The use of cyclones for the washing of excavated materials is an established treatment method that allows separation of the clay and other materials and obtains good-quality materials. The study on the efficiency of this separation method remains to be performed. 

\item \textbf{Transformation of the molasse into fertile soil} for applications in the development of brownfield sites, urban recreational areas and forest areas, as well as improving the fertility of acidified land and/or as technical areas along the verges of roads and motorways (pollutant filtration). The use of excavated materials transformed into reconstructed soil requires the identification of areas for backfill in agricultural areas. The locations desired include hollows, slopes, areas of poor ground quality such as polluted or acidic soils, or of poor agronomic quality (e.g., low water retention potential). The application may be different depending on the type of land and topography. This type of measure could be particularly suitable for certain rural areas in the Auvergne-Rhône-Alpes region and in several cantons in Switzerland and would also allow deposits to be made near drilling sites. Due to the unknown topography of the chosen areas, it is currently difficult to identify the location and total areas. It is estimated that about 2\,million cubic metres (about 4\,million tons) of excavated materials could be reused via this pathway, provided the appropriate areas are identified. 

\item \textbf{Use of a part of the molasses as technical materials}: trench cover (e.g., roads), acoustic screens, farm tracks, forest paths etc. Both the French and Swiss neighbouring landscapes are constituted by woods of different nature or of rural areas. In particular, the zone in the proximity of Annecy has approximately 500\,km of rural paths and tracks. In total, there are approximately 705\,000 km of rural and forest paths in France. These paths and tracks must be maintained regularly; for this, inert materials are necessary to ensure their stability.  Raw molasse is also suitable for the implementation of vegetation-free strips near roads. With the installation of noise-reducing and privacy-screening gabions, this scenario can generate significant savings in the development and maintenance of public roads. Support from the relevant authorities is required to identify suitable roads and locations near extraction sites, establish certifications, and grant permits necessary to carry out this type of application. A further application in the layout of roads is the use of reconstructed soil based on molasse as landscaping of trenches of covered streets or highways. 

\item \textbf{Development of building components} by compression (bricks or compressed earth using `sandwich' technology) for use within the scope of the project where possible or outside the project (opportunities within the region to be investigated).

\item \textbf{Development of new building materials} containing a part of the molasse (e.g., shotcrete ingredients, supporting materials, insulating panels for surface buildings), to be used within the scope of the project where technically feasible or outside the project (opportunities within the region to be investigated). Literature shows that considering excavated materials for the reuse as recycled concrete aggregates (RCA) in the structural concrete of the tunnel or of the basement of the tunnel could turn out to be not feasible because of the decreased strength and therefore further thorough studies are required. A prudent approach, until more complete studies are performed, could be to use RCA in shotcrete with fibre-reinforcement for non-structural parts only. 
\end{itemize}

Some of these pathways are listed in excavated material management guides used in France e.g., from Cerema/UMTM  (publication pending) and from the Minist\`ere de la transition \'ecologique et solidaire \cite{BRGM_2020} and in Switzerland from the OCEV \cite{Ecomat_2018_guide} and also in the AFTES recommendation GT35R1F2 \cite{AFTES_2019_GT35R1F2} (currently in revision). The other pathways must be identified and a framework for cross-border use could be proposed to the respective authorities in the Host States.

The first four in the list correspond to traditional reuse cases already in use in other tunnel construction projects. These are taken into consideration and will be applied in priority whenever the geo-mechanical, mineralogical and chemical characteristics of the material are suitable for these applications. 

In addition to these traditional reuse cases, innovative reuse pathways were identified during the `Mining the Future\textregistered' international competition carried out in 2021 and 2022. The competition aimed to identify innovative pathways for excavated material from tunnel construction which could be profitable not only to the FCC but to any further projects in the subalpine region. The competition was launched under the Horizon2020 project ‘Future Circular Collider-Innovation Study’ (Grant Agreement n. 951753) and was jointly organised by CERN and Montanuniversit\"{a}t Leoben. The submissions were scrutinised by a jury panel of international experts. The four finalists’ concepts ranged from the manufacturing of substrates for agriculture and forestry to the production of raw construction materials like concrete and shotcrete, compressed earth bricks and other hydraulically bound building materials. They all need to analyse and separate the materials during the tunnelling process in real-time, with subsequent on-site pre-processing directly on the excavation surface sites. Geogenic contamination by hydrocarbons and heavy metals must be removed or at least be reduced to levels that are compatible with the proposed processes and end-use conditions. The consortium led by BG Engineering was selected as the most innovative and comprehensive concept and won the competition. 

Some of the solutions are now being integrated into a unique design and evaluated in the field, in a project planned to reach maturity by 2030. The objectives of the evaluations are twofold. Firstly, to establish how to conduct the online identification, sorting and pre-treatment of the materials during the excavation process. Secondly, to prepare different reuse pathways to sort and pre-treat materials, including transforming molasse into usable soil for forestry and rewilding applications, in line with the principles of a circular economy. The quality-assured creation of the reconstructed soil is a lengthy process spanning several years and has been chosen as the first large-scale experiment with field tests at OpenSkyLab. 

The OpenSkyLab, is a project based on a plot of about 10\,000\,m$^2$ located near LHC Point 5 (CMS, Cessy, France) that has been made available by CERN. Molasse extracted during the HL-LHC excavations will be transported to this field to be used in the tests. Initial laboratory analyses will be performed off-site to identify the most suitable mix of molasse and other natural additives (compost). These will be followed by field tests in the OpenSkyLab’s controlled environment (monitoring of the field, weather, and plant growth conditions), using scientific protocols developed by a collaboration of universities working in this domain.

In keeping with CERN’s long-standing tradition, this project relies on an open collaboration with academia and industry. Currently, the collaboration includes university and research experts in agronomy, pedogenesis and geology (HEPIA, BOKU, BRGM, Montan University Leoben) and industrial partners in soil engineering and phytoremediation (Microhumus, Edaphos), soil treatment techniques (WSP-BG) and monitoring and supervisory control systems (BECC).

In order to facilitate acceptance into the reuse pathways of the excavated materials, it will be necessary to set up material-separating units during the tunnel construction phase,  at the starting point of two TBMs as a minimum. These units could combine the screening, sorting, sieving and cleaning operations of the extracted materials. 

Assuming a two-shift 24-hour working day over 240 days per year, the consortium that won the Mining the Future competition proposed a facility that could treat approximately 750\,000 tonnes per year at an hourly rate of 200\,tonnes. Each facility would employ a team of five people. A pilot project for the testing of the efficiency of these sorting and treating facilities could be envisaged during the next phase. 

The working hypothesis assumes that the previously listed solutions will account for a global reuse of about 70\% of the overall excavated materials, among which the reconstructed soil reuse pathway could amount to up to 2 million cubic metres (corresponding to about 25\% of the total). The backfilling of quarries as well as more traditional reuse pathways (e.g., limestone and sand direct use) should overall account for about 45\%. These quantities remain to be confirmed during the next phase.

\subsubsection{Regional opportunities} \label{matexREGopport}
CERN commissioned a study to identify the regional opportunities available in France and Switzerland for the evacuation of the excavated materials. This included: 

\begin{itemize}
\item A global inventory of the final storages and of the opportunities for landscaping of quarries and mines at the end of their operational lifetime, but holding a prefectural permit valid beyond 2033.
\item A study on potential industrial railway sidings for the evacuation of the materials.
\end{itemize}

\paragraph{Global inventory of regional opportunities}
This study started with a preliminary inventory of potential regional opportunities for landscaping (quarries, mines) and storage, which was carried out at the start of the feasibility study (2021-2022) in France and Switzerland to provide initial indications of potential host sites and the capacities for excavated materials of the various pathways. These detailed inventories list the existing and planned facilities (horizon 2030) with sufficient capacity to receive and treat the excavated materials from the FCC. They will need to be repeated once the excavated material management plan has been defined to take account of the final material extraction fluxes and the possible availability of new reuse cases. Subject to updating and optimisation with respect to transport, these studies show that it is possible to find a pathway for all the materials excavated from the potential future FCC project and provide an initial estimate that can be used as a basis to begin the optimisation process according to the `avoid, reduce, compensate' principle. It should be underlined that the inventory of the potential recipients of the excavated materials was carried out with the aim of evaluating the financial envelope and identifying potential showstoppers for the FCC feasibility.  It is not possible at this stage to identify the specific recipients without engaging in contractual agreements. This step will be performed during the establishment of the final material management plan. 

\begin{figure}[!ht]
    \centering
    \includegraphics[width=\linewidth]{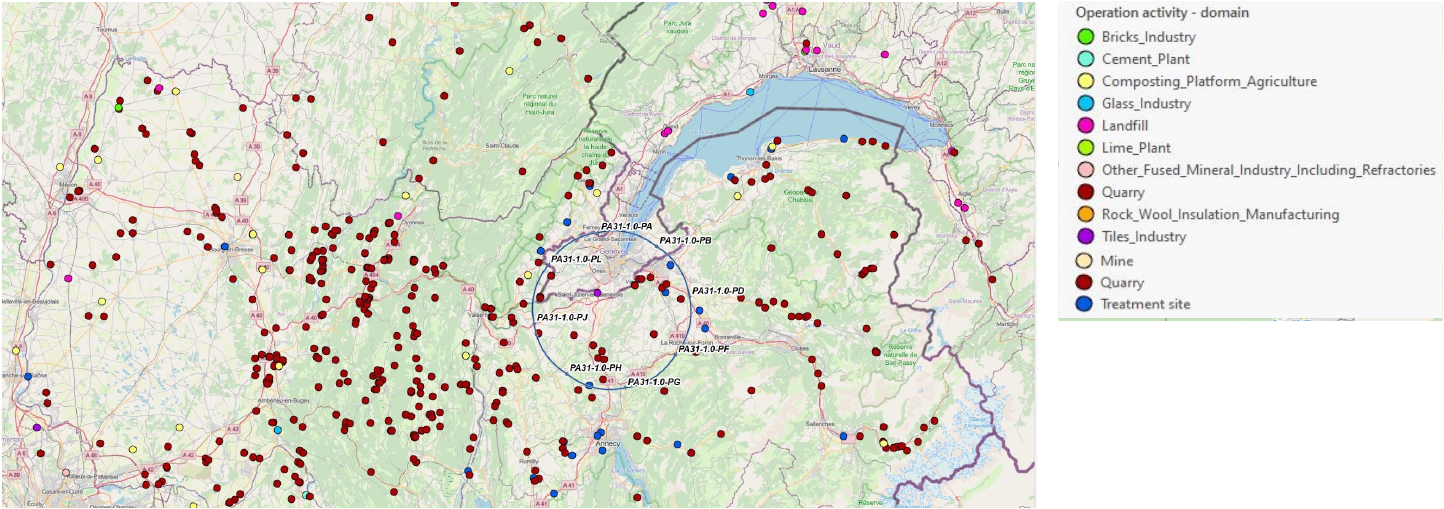}
    \caption{Schematic diagram of the inventory of the regional opportunities for material reuse (status 2021).} 
    \label{fig:Regional opportunities}
\end{figure}

\paragraph{Railway sidings}

The study investigated the option of constructing industrial rail sidings near the surface sites to take away the excavated materials. Potential locations near existing railway lines were identified, and the feasibility of creating sidings was examined. To supplement this study, the feasibility of a conveyor belt link has been assessed for the two sites deemed most appropriate for the installation of sidings: Vulbens (removal of materials from PJ) and Charvonnex (removal of materials from PG). This exploratory phase was completed by an initial assessment of greenhouse gas emissions from transporting excavated materials by road and/or by rail from the production site to the disposal site or to the location of the hypothetical reuse cases. 

The feasibility study of railway sidings only took technical factors into account. The weighing of certain factors linked to the acceptability of creating a railway siding (e.g., environmental aspects such as the nuisance factor in a densely populated area) must be carried out before a decision is reached.

The creation or refurbishment of railway access, already considered to lower the nuisances due to truck transport, would further facilitate the use of quarries and reuse cases not in the immediate vicinity. A technical study could be conducted to analyse the possibility of cross-border transport by conveyor belts from Challex (PL) to the railway network at La Plaine in Switzerland and from Ferney (PA) to the rail network at the Geneva Airport in order to use the existing railway infrastructure for the transport of the materials to their final destination. 

\begin{figure}[!ht]
    \centering
    \includegraphics[width=\linewidth]{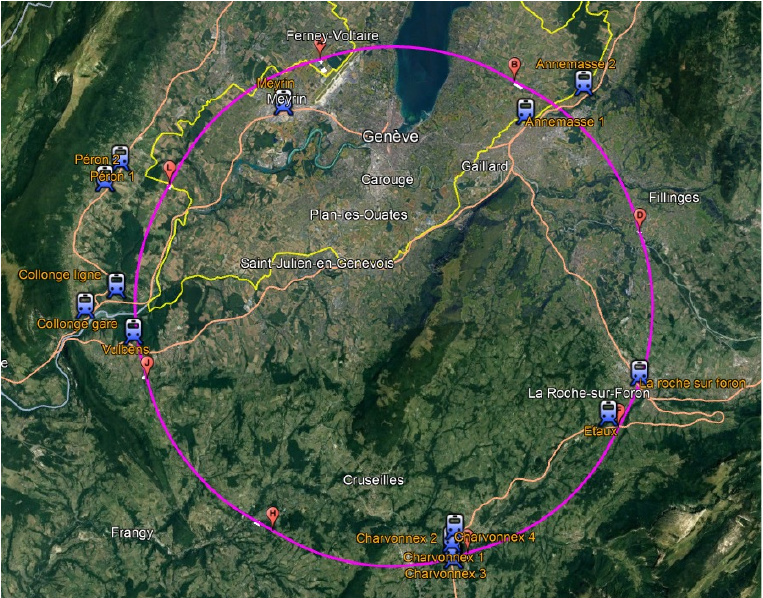}
    
    \caption{Study of the potential locations for railway sidings.} 
    \label{fig:Railway study}
\end{figure}

\subsection{Outline of the next studies} \label{matexNEXT}

The scenarios for the reuse of excavated materials are based on preliminary assumptions regarding the proportion of materials with geogenic or anthropogenic characteristics that may require specific handling. At this early stage, and in the absence of more detailed geochemical analyses, a working assumption of 30\% of materials exhibiting natural geogenic variations appears to be the most reasonable.

As part of the FCC feasibility study, additional insights may be gained by examining recent projects related to the enlargement of certain LHC accelerator caverns (HL-LHC project). Underground excavations for this project indicated the presence of hydrocarbons in approximately 15\% of the materials in France and nearly 50\% in Switzerland. This difference is influenced not only by actual variations in subsurface conditions but also by differences in environmental regulations between the two countries.

Based on these initial inventories, presented in Section~\ref{matexREGopport}, a preliminary approach to the management of the excavated material was drawn up and used to assess the transportation costs from the extraction sites to the storage facilities and backfilling sites. The average transport distances are estimated to be 20\,km in Switzerland and 40\,km in France for uncontaminated materials and approximately 100\,km in both countries for materials requiring specific treatment. This is because treatment facilities are generally located outside the Lake Geneva region, where the FCC will be installed. 

\begin{figure}[!ht]
    \centering
    \includegraphics[width=\linewidth]{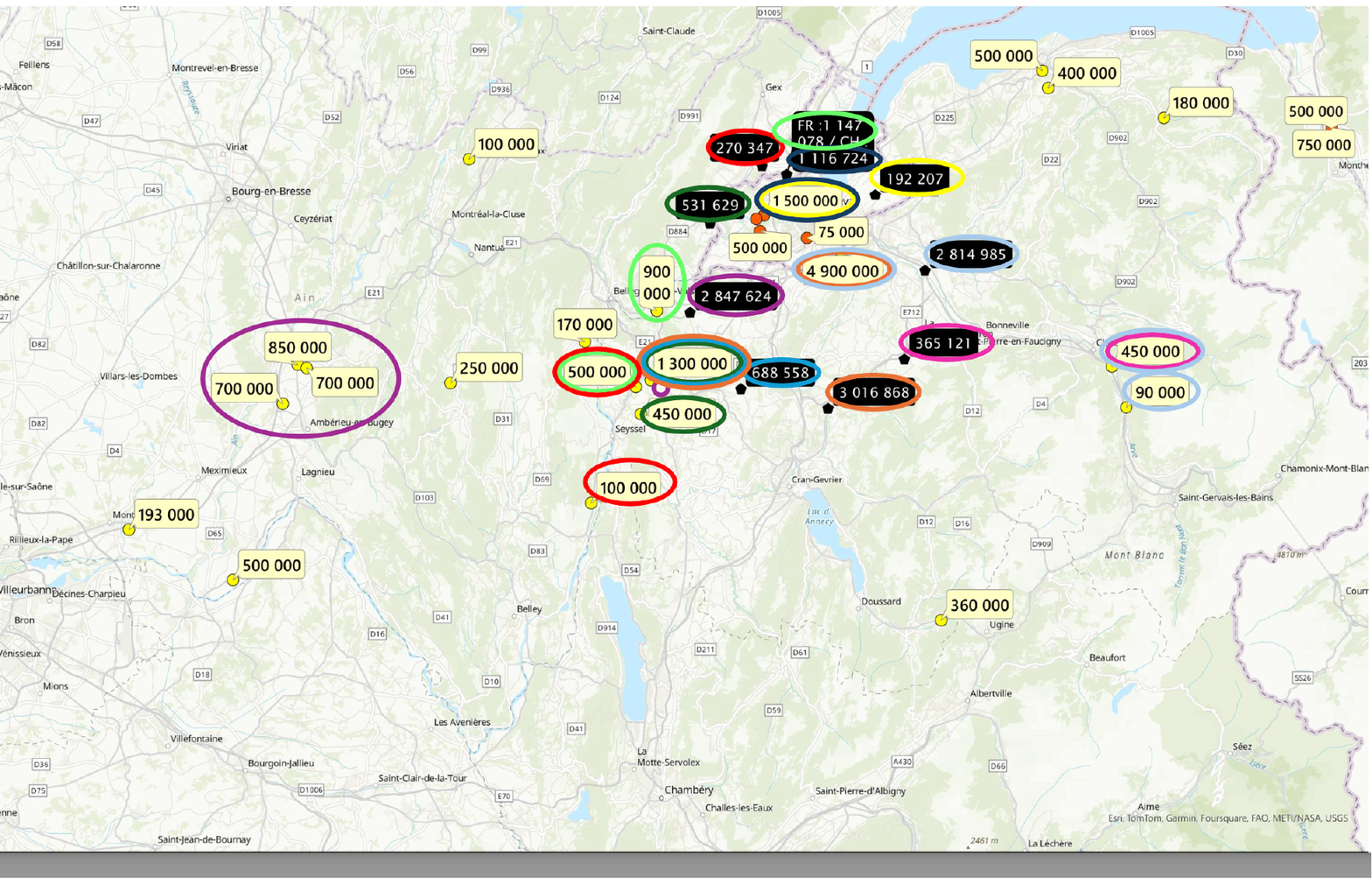}
    \caption{\small Example of possible distribution of unpolluted excavation materials to the quarries nearest to the material extraction sites (the figures in black are the value in tonnes). The colours of the circles link a given extraction site to one or more quarries or backfilling sites. The figures in yellow show the quantity of materials that can be accepted by the quarries during the excavation. Swiss pathways are represented by a red dot, and French pathways by a yellow dot.} 
    \label{fig:Quarries}
\end{figure}

Figure~\ref{fig:Quarries} presents the results obtained for uncontaminated materials. In this figure, the values displayed on black backgrounds indicate the quantities of material expected to be extracted per site, while the colours of the circles correspond to the quarries or backfilling sites designated for the disposal of outgoing materials. It should be noted that only quarries with a prefectoral permit valid beyond 2033 have been considered.

These scenarios will require updates, as the annual intake capacity of each site may be revised based on regional requirements. The extracted quantities are provided as an indication of each quarry’s backfill capacity. However, further studies will be necessary to assess the feasibility of this operation, and agreements with quarry operators should be established in due course.

Subject to updating and optimisation with respect to transportation, these studies show that it would be possible to find a pathway for all the materials excavated from the potential FCC project sites and provide an initial financial estimate which could be used as a basis to begin the optimisation process according to the `avoid, reduce, compensate' principle.

The distribution of excavated materials should integrate the identified reuse cases as soon as the technical information on their feasibility becomes available. Therefore, the current distribution should be viewed as a preliminary benchmark assessment, expected to evolve positively as studies progress. Close collaboration between the FCC and the Host State authorities is essential to refine and better adapt the principles of the strategy of the excavation material management in alignment with the regional framework. 

A subsequent project preparatory phase should include the development of a roadmap along the following points: 

\begin{itemize}
\item Confirmation of the characteristics of the excavated material and correlation to the potential reuses via the analysis of the ongoing and future subsurface investigations;
\item Study of the technical methods for the real-time analysis of the materials and their sorting according to their characteristics and potential reuse. 
\item Pilot projects on the potential reuse cases: on the example of the ongoing OpenSkyLab project for developing processes for using excavated materials for landscaping, other projects could be started for testing, for example, the possible reuse as construction or isolation materials. This study should include an evaluation of the environmental, economic and societal impact due to the potential injection on the market of the products based on excavated materials ;
\item Definition of the regulatory framework for the management of the excavated materials with the host States;
\item Update of the regional treatment and disposal opportunities (e.g., availability of quarries, final deposits, treatment facilities) and related regulation;
\item Study on the excavation material logistics (traceability, fluxes, conveyors etc.), including the evaluation of environmental and societal impacts and the potential limitations. 
\end{itemize}

Certain activities can be carried out in parallel. Others must follow a sequential process, leading to the development of a preliminary excavation material management plan. Based on this plan, more concrete discussions can be initiated with the administrations of the Host States and the owners of potential final deposit and reuse sites.

\chapter{Territorial implementation}
\label{sec:HostStateStudies}

\section{Introduction}
\label{imp:Introduction}

To be able to take an informed decision for a future particle-collider based research infrastructure and to undergo the necessary project authorisation process with national authorities, a specific implementation scenario needs to be conceived. Such a geo-localised scenario must be well-balanced, considering the three main dimensions:

\begin{enumerate}
\item Scientific excellence.
\item Territorial compatibility.
\item Risks related to the implementation that affect cost and schedule.
\end{enumerate}

While the exploratory phase of the FCC study between 2014 and 2018 \cite{FCC-eeCDR} focused on the `in principle' feasibility of the accelerator and the territorial boundary conditions, the studies between 2019 and 2024 included the development of a well-balanced project scenario considering the three above-cited dimensions.

This chapter summarises the methodology adopted to carry out this work as documented in \cite{gutleber_2025_14773243}, sheds light on the variants considered and the evolution towards a reference scenario. The resulting scenario presented serves as a reference to design the various elements of the future circular collider-based research infrastructure if the global science community decides to make such a facility their priority.

The reference scenario presented is the result of a total of 10 years study of a large variety of scientific, technical, cost, societal and environmental aspects. It represents an infrastructure with a circumference of approximately 91\,km, including eight surface sites. It is conceived to be able to host two distinct particle colliders, a high intensity lepton collider first and a high energy hadron collider in subsequent phases. The second machine and its experiments profit significantly from the assets put in place for the first phase due to the residual asset value of that infrastructure, contributing to the overall sustainability of this long-term science programme. The iterative process that was used to develop the reference implementation scenario is summarised in the following sections.

The presence of a sufficiently large community of users committed to carrying out scientific research with the particle colliders for several decades is a prerequisite for justifying the construction of a research infrastructure of this scale. Scenarios involving a much smaller collider with a circumference of less than 90\,km, would not allow the  provision of a performance and a research programme that could attract a critical mass of scientists for a sustained period of time. The doubling of the interaction regions from two to four reflects the aim of attracting as many scientists as possible to such an infrastructure. The size and the characteristics of the infrastructure also permit additional scientific activities with the injector and particle colliders, as is the case today at CERN with the LHC programme.

In terms of compatibility with the geological conditions that affect the construction risks and costs, only two types of configurations meet all requirements: an infrastructure with a circumference of between 89 and 91\,km and eight surface sites, and a layout with a circumference of between 97 and 98\,km and twelve surface sites. Concerning the availability of suitable surface site locations, however, layouts and placements suitable for scenarios involving a collider with a circumference well in excess of 91\,km and comprising more than eight surface sites turned out to entail unacceptable risks concerning their implementation. 

To date, only scenarios around 91\,km circumference and eight surface sites seem capable of satisfying the following three requirements:

\begin{enumerate}
\item Good scientific performance of the particle collider and four experiment sites.
\item Compatibility with territorial constraints at the surface and the subsurface.
\item Understanding of the implementation conditions from cost and risk perspectives.
\end{enumerate}

One of the scenarios named PA31 stands out from all of those iteratively developed following the `Avoid-reduce-compensate' methodology\cite{ERC:2021} and analysed with a multi-criteria approach recommended for industrial installations\cite{Cerema2020, UNIDO}. It has been further studied and optimised in depth involving bibliographic research, field work, and continuous discussions with the public administration services in both countries involved and with key stakeholders in the implementation area. In case of a decision to move forward with the implementation of a project, the conditions for this scenario will have to be further analysed, the scenario will have to be further improved, and detailed designs have to be developed before implementation can start. 

Territorial constraints and legal frameworks in host countries and within the European Union are constantly evolving. Between 2014 and 2023, several layout and placement scenarios had to be ruled out due to these changes. The information contained in this document and the working hypothesis described here are therefore provisional. For the project to come to fruition, the validation and definitive freeze of a scenario securing the required surface areas and subsurface volumes should be done as soon as possible.

\section{Methodology to develop a sustainable project}
\label{imp:Methodology}

\subsection{Approach}

\begin{figure}[!h]
    \centering
    \includegraphics[scale=0.9]{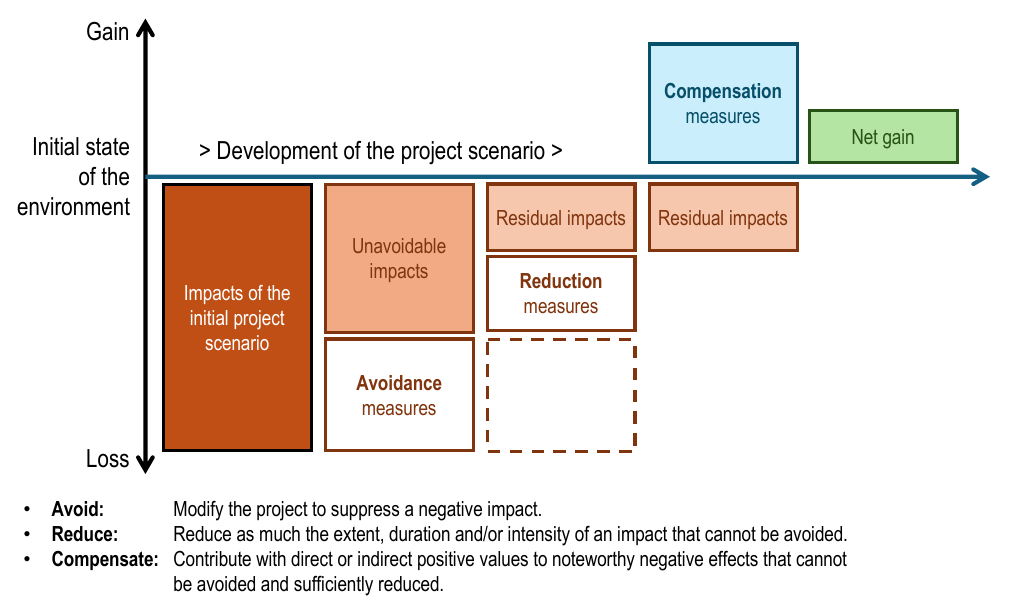}
    \caption{\label{fig:ERC}The "avoid-reduce-compensate" approach, known as "Éviter-réduire-compenser (ERC) in France and anchored in the French environmental law that determines the project authorisation process.}
\end{figure}

From the outset, the objective of the study was to draw up a scientific research infrastructure that reconciles, in an eco-design approach\cite{ISO14006:2020,ECecodesign}, i) scientific excellence, ii) territorial compatibility and iii) consideration of acceptable risks associated with project implementation. A systematic process therefore had to be set up to develop scenarios through an iterative approach, taking these three essential aspects into account at all times. The eco-design approach adopts the `Avoid-Reduce-Compensate' methodology (see Fig.~\ref{fig:ERC}), which takes into account both constraints and opportunities.

This process, which complies with the French Environmental Code, the Environmental Protection Act and the Swiss Federal Ordinance relating to the Environmental Impact Assessment Regulation, is perfectly in line with the desire and objective to arrive at a proposal based on a balanced scenario.

Ideally, all stakeholders are involved in the process from the outset. However, an iterative approach had to be taken given that in the beginning the placement of the infrastructure was not defined, the  knowledge of evolving territorial conditions was incomplete, the possibilities of effectively involving stakeholders at all levels of society were limited, the knowledge of the technological choices to be made for periods longer than twenty years is evolving, and the availability of resources for conceptual studies and personnel at this very early stage is limited.

Despite these constraints, the international science community is committed to transparency and public participation. The aim is to obtain a "social licence to operate"~\cite{Gunningham2004} by developing a scientific peaceful endeavour in the context of a systematic and structured public participation process. To this end, the feasibility study phase laid the foundation for the implementation of an approach in line with the regulatory frameworks and the best practices in France and in Switzerland. The approach goes beyond informing the public, engaging affected stakeholders in the discussion of the various segments of the project and, as far as possible, involving them in design. considerations that affect their territory.
A consultation process will therefore be carried out in an equitable manner on both sides of the border, respecting the approaches and procedures specific to each country.

The scenario development process first used the bibliographic information available. It gradually integrated further aspects as they became relevant and stakeholders as they were identified. This is the case, for example, for the participation of local elected representatives (municipalities, intercommunal structures, departments, regions). Depending on the layout, the number of surface sites and their location, between twelve and twenty communes were at any time directly concerned and had to be consulted. A further 30 or so municipalities could be indirectly affected, in terms of access needs or infrastructure connections, or simply because the tunnel passes under their territory. Other stakeholders will need to be added for the development of a detailed design scenario. These additional stakeholders include local infrastructure operators (e.g., water, canals or road networks), representatives of local authorities responsible for certain subjects (e.g., traffic and nature) and associations (hunting, fishing, tourism, environmental protection, heritage preservation, economic development, etc.). It is prudent to approach these representatives as soon as the likelihood of their community being affected in some way is sufficiently high, i.e., when a specific scenario and an intent for a project exist. In this way, their availability to be informed about the project vision can be taken into account, the relevant contacts can be designated and the limited availability of the study group members can be taken into account. This study has involved selected elected representatives of directly affected locations. A systematic involvement of a wider domain of stakeholders is potentially appropriate at a subsequent phase, when a project intent is formulated and when a reference scenario that permits a useful involvement exists.

\begin{figure}[!h]
    \centering
    \includegraphics[scale=1.0]{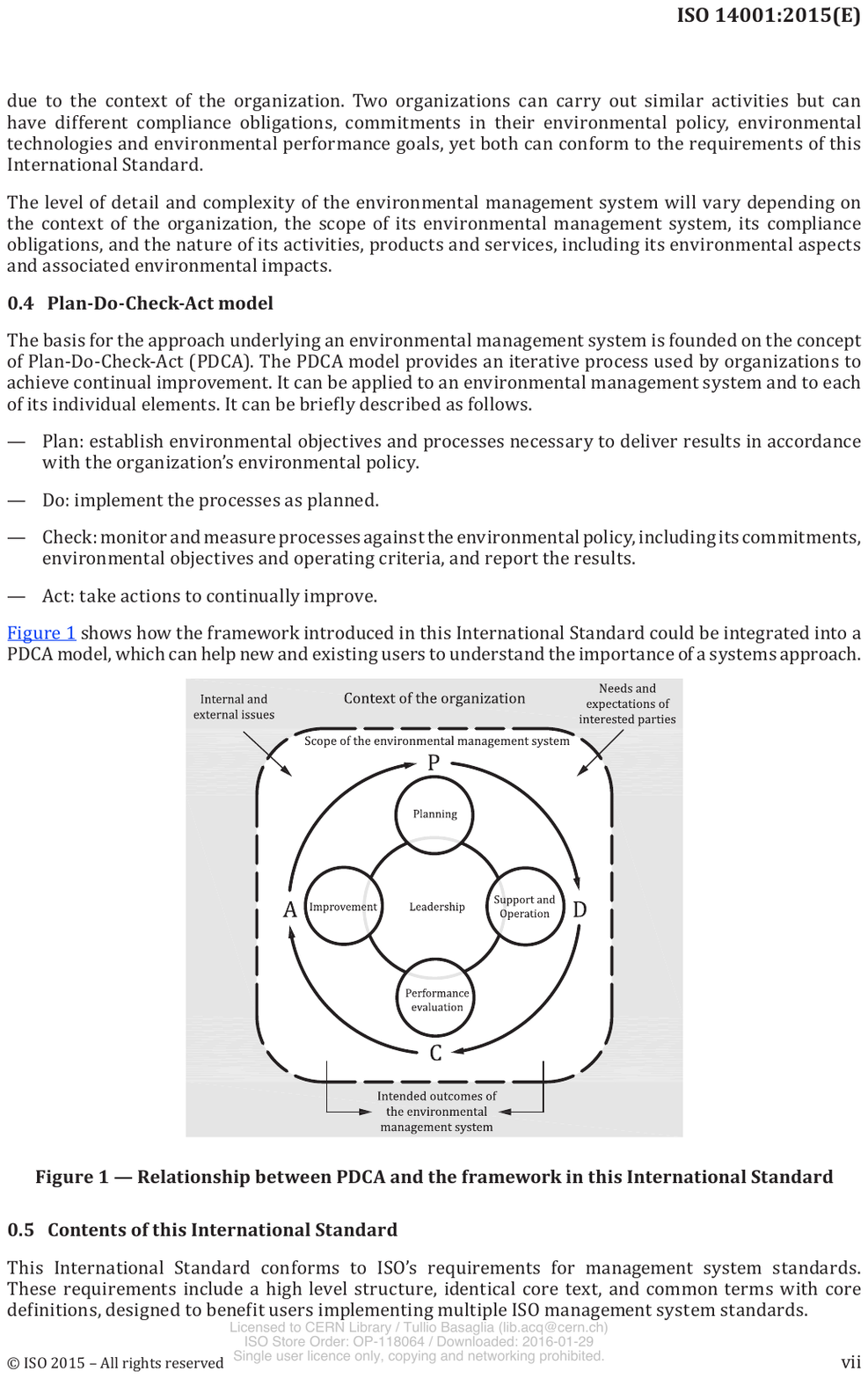}
    \caption{\label{fig:PDCA}The Plan-Do-Check-Act approach for environmental management, defined in standard NF EN ISO 14001, section 0.4, page vii, is the basis for `eco-design'.}
\end{figure}

The international standards applicable, NF\,EN\,ISO\,14001 (environmental management \cite{ISO-Enviro}) and (for eco-design) NF\,EN\,ISO\,14006 \cite{ISO14006:2020} (see Fig.~\ref{fig:PDCA}) require an iterative approach.

\begin{figure}[!h]
    \centering
    \includegraphics[scale=0.8]{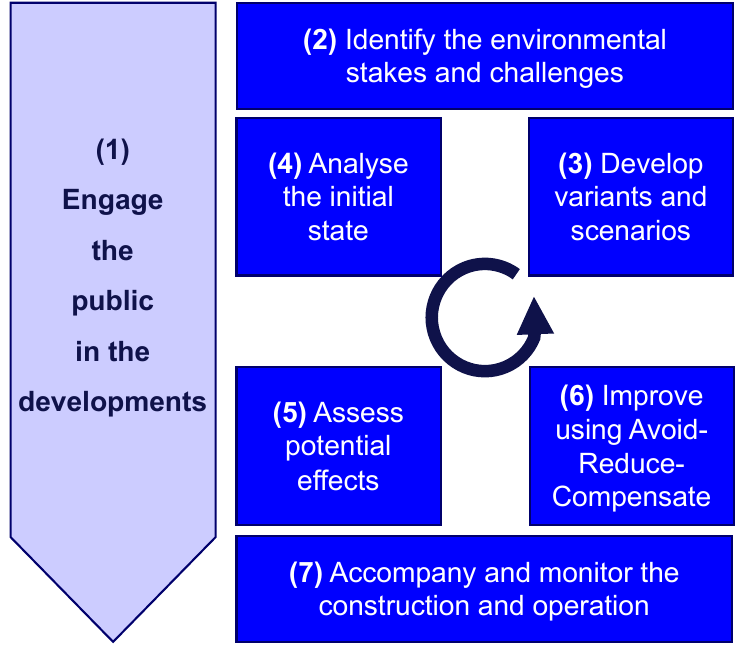}
    \caption{\label{fig:ERM}Diagram from iterative impact study, French Ministry of Territorial Development and the Environment, 2001, p. 27, see footnote 6). Although this guide is from 2001 and specific regulations have changed, the principles described in the diagram are still valid and recommended for use today.}
\end{figure}

This approach is also set out in greater detail in the good practices guidelines for environmental impact assessments, established by the French Ministry for the Ecological Transition, Biodiversity, Forests, Marine Affairs and Fisheries, and in the NF\,EN\,ISO\,31000 \cite{ISO31000:2018} standard on risk management (page 8 of the ISO standard) (see Fig.~\ref{fig:ERM}).

\subsection{Iterative process}

The scenario development process is based on the iterative Plan-Do-Check-Act (PDCA) approach, and incorporates the Avoid-Reduce-Compensate (ERC) approach.

The process starts at a macroscopic level to identify the major constraints and opportunities.
The constraints are recorded in a geographical information system (GIS) so that a map-based identification of areas to be avoided and areas in which a surface site or a subsurface passage would need to be limited can be rapidly identified. Such a system includes many data layers. Today, GIS-based Environmental Information System (EIS) for the FCC study comprises more than 120 layers that contain detailed information and are also used to build summary layers that are colour coded as shown in Table~\ref{tab:sensitivity-grid}. The individual territorial sensitivity grid levels differ for French and Swiss territories\cite{boillon_pierre_2023_8403173}. They are based on national legal and regulatory constraints and concern particular regional and local constraints that are the result of exchanges with expert companies in the areas of environmental impact studies and project development as well as with public administration services. Typically, the constraints considered for the development of the future collider layout and placement scenario are highly conservative and sometimes exceed required legal and regulatory constraints in several domains to ensure that good territorial compatibility can be achieved by respecting known local and cultural heritage.

Scenarios were explicitly checked against such different interest zones, and a weighting took place whether to continue considering locations with known constraints or to discard the scenario due to its particular local high-value nature. The same strict constraint was applied to enlarged drinking water perimeters that were excluded as surface site locations, despite the fact that from a regulatory point of view, surface site constructions would be allowed in such zones.

\begin{table}[!ht]
  \centering
   \caption{\label{tab:sensitivity-grid}Definition of the territorial sensitivity levels, representing constraints for determining the configuration and location. Each colour-coded level represents several data layers that can be shown on a map in a geographical information system (\url{http://cern.ch/fcc-sensitivity-grid}).}

  \begin{tabular}{p{0.05\linewidth}p{0.15\linewidth}p{0.7\linewidth}}
    \toprule
    \textbf{Level} & \textbf{Designation} & \textbf{Description} \\ \midrule
    4 & \cellcolor{red} Unacceptable & The level of constraint is such that the zone cannot be considered for a surface site. These zones are considered to be exclusion zones and should be avoided. \\ \midrule
    3 & \cellcolor{orange} High & The zone is not recommended for the location of a surface site, but may be considered if it is decisive for the feasibility of the project, with additional reduction, compensation, or mitigation measures. \\ \midrule
    2 & \cellcolor{yellow} Acceptable & The zone is acceptable for the location of a surface site with appropriate reduction, compensation or mitigation measures. \\ \midrule
    1 & \cellcolor{green} Low & The zone can be considered for a surface site without further significant measures. Compensation may still be necessary. \\ \bottomrule
  \end{tabular}
\end{table}

This step permits the definition of entire classes for particle collider layout and placement candidates. Picking a representative candidate of a class permits the choices between the various configurations and locations under consideration to be progressively refined, introducing additional information obtained from the study of promising variants and discarding variants as soon as obstacles are detected. If a scenario is discarded, the main exclusion reasons are analysed and the entire class of scenarios is analysed with respect to those conditions. In many cases, this permits the elimination of an entire class of scenarios and the establishment of major exclusion criteria and constraints at the macroscopic level. The approach helps to reduce the solution space and to avoid further consideration of unfeasible scenarios in the subsequent steps.

Scenarios that are considered potentially feasible are further analysed with more detailed information and further stakeholders are involved. The same elimination process as above is applied.

This approach leads to a gradual definition of exclusion criteria and zones, and to a selection of likely layouts and placements that can eventually be optimised.

With regard to the territorial analyses, the analysis process begins at a level that considers topography, bathymetry, geology, hydrography, protection zones and urban development. It then gradually expands to include additional aspects such as accessibility, transport, disturbances, potentially conflicting planned developments and the availability of technical and natural resources (e.g., electricity or water). It then incorporates additional elements, such as social factors, local preservation and development objectives, visibility, shared visibility, or disturbances for stakeholders affected directly (e.g., neighbours) and indirectly (e.g., communities affected by construction site traffic).

The scenario-building process requires a more detailed analysis phase, aimed at the direct participation of stakeholders and local players in order to draw up a scenario adapted to the specific plot of land, always taking into account the three main issues (science, territory, implementation) within the framework of the iteratively improving avoid-reduce-compensate project development approach.

With the choice of a reference scenario as a prerequisite for field studies, environmental impact assessment, and technical design work, also the dialogue with public stakeholders could be gradually defined and initiated. Considering the valuable accompaniment of the two CERN host states and their advice in engaging the public in territorial development projects, the study collaboration requested CERN to consult the French "Commission nationale du débat public (CNDP)" (english: National Commission for Public Debate) in 2024. A first analysis resulting in a list of recommendations were made public on 6 March 2025~\cite{CNDP2025}. Taking account of these recommendations and considering the trans-border context of the particle collider scenario, it was decided to extend the collaboration with the CNDP to prepare subsequent public engagement processes in case the international science collaboration expresses an intent to develop the study into a project and to enter a preparatory phase.

Informal and formal processes engaging the public are ideally carried out as early as possible, and the French government reminds the principle and need of anticipation on various occasions. At the same time, care must be taken to time engagement appropriately. This approach aims at assuring that the project scenario can be adapted based on the involvement of stakeholders and having sufficient technical information at hand to inform he engaged stakeholders about the needs and constraints that govern the project and the environment in which it would eventually be embedded.

\begin{figure}[!h]
    \centering
    \includegraphics[width=\textwidth]{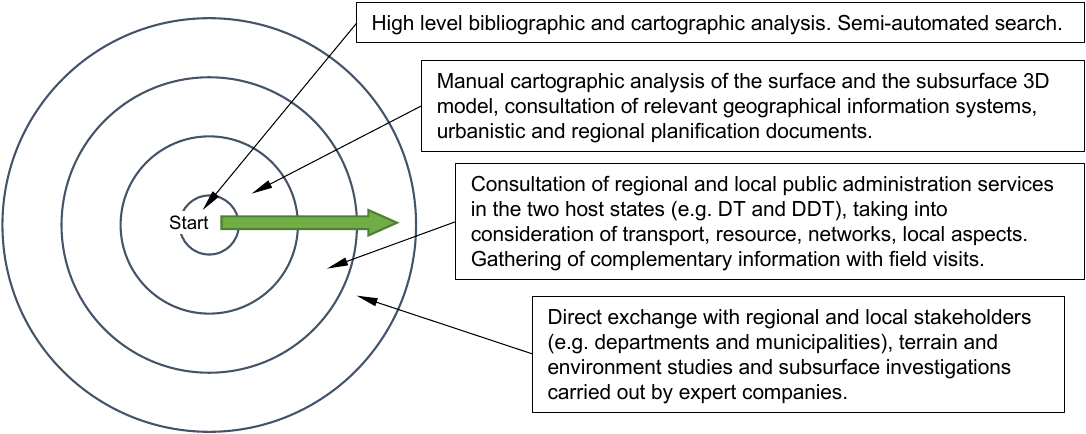}
    \caption{\label{fig:gradual}As the layout and placement studies progress, additional information and stakeholders are included in the process to create a balanced scenario that meets the needs of all stakeholders.}
\end{figure}

As new information is integrated into this iterative process, automated and cartographic research must be progressively completed, followed by manual research, interviews with people with good knowledge of the region, field visits and consultation with stakeholders (see Fig.~\ref{fig:gradual}).

\subsection{Scenario assessment}

To determine the value of a scenario, to assess if it should be further optimised or discarded and to be able to compare the merits of different scenarios among each other, a multi-criteria analysis scheme has been developed, based on an approach presented by the French organisation Cerema in its guidelines for the environmental analysis of linear transport infrastructures \cite{Cerema2020}. This approach was completed using the UNIDO International Guidelines for Industrial Parks, published in November 2019 \cite{UNIDO}.

As required by the regulatory environmental authorisation frameworks in France and Switzerland, a development project is to be considered in broad terms, extending the analysis to various non-technical relevant aspects such as indirectly affected stakeholders; legal, regulatory, social and economic factors; networks (roads, railways, water, electricity, canals, public service and safety infrastructures); heritage sites; visual aspects; disturbances (noise, dust, light, smells, pollution) and potential benefits. As recommended in the Cerema guide, this approach was used to assess the relative merits of each variant, and contribute to the development of reduction, compensation, and support measures in an open and transparent process.
In Switzerland, and more specifically in the canton of Geneva, a parallel can be drawn with the cantonal guide for the strategic environmental assessment tool (EES) \cite{Geneve2025}. When plans, programmes or projects (PPPs) are drawn up, the EES is sometimes used to ensure that environmental and human health issues are taken into account systematically and at an early stage, as defined in the Environmental Protection Act \cite{Switzerland1984}. As such, it can be used as a decision-making tool, assessing the merits of different scenarios.

The multi-criteria analysis results use standardised qualitative indicators to compare the suitability of different scenarios. Its application during the scenario-building process makes it possible to quickly identify the types of scenarios that present significant advantages or disadvantages, and to determine whether the differences between the scenarios are major or minor. This approach also offers the advantage of clarifying which elements have the greatest or least influence on the value of the scenario, and thus guides the development of new types of scenarios. This approach is then used in the subsequent optimisation stages, for example with regard to moving shaft and surface sites to more suitable locations and taking into account the availability of existing infrastructures (e.g., roads, railways, water supply and treatment, electricity), urban planning documents (PLU, PLUi, PADD, SCoT, PDcn, PDcom) and synergies and opportunities (e.g., supply of residual heat, sharing of technical infrastructures, reduced transport distances).

The criteria list is made up of nine topics, each comprising several detailed criteria. These criteria cover the various themes that are relevant for an implementation. 32 environmental factors, the scenario configuration determining the performance for scientific research, implementation costs and risks were taken into account:

\begin{enumerate}
\item Land status
    \begin{enumerate}
        \item Availability of plots
        \item Clearly defined ownership
        \item Plot price
        \item Acquisition time and expected difficulties in obtaining rights
        \item Plot development cost
    \end{enumerate}
\item Road connections
    \begin{enumerate}
        \item Distance from transport, industrial and other infrastructures
        \item Distance from populated areas
    \end{enumerate}
\item Raw materials and services
    \begin{enumerate}
        \item Availability of raw materials for construction and resources for operation 
        \item Proximity to service providers
    \end{enumerate}
\item Physical characteristics
        \begin{enumerate}
            \item Plot size and shape
            \item Topography
            \item Shaft depth
            \item Drainage and sanitation requirements for construction
            \item Surface soil condition
            \item Water resources
            \item Accessibility
            \item Subsurface conditions (physical)
            \item Subsurface conditions (regulatory)
            \end{enumerate}
\item Infrastructure
        \begin{enumerate}
            \item Accessibility of electrical power
            \item Communication network
            \item Water for industrial use
            \item Drinking water
            \item Waste water discharge, rainwater collection, disposal and treatment points
            \item Temporary storage and processing areas during construction
        \end{enumerate}
\item Environmental and social factors
        \begin{enumerate}
            \item Existing environmental constraints
            \item Fauna and flora
            \item Existence of construction constraints
            \item Adjacent constraints
            \item Disturbances
            \item Availability and accessibility of workforce
            \item Involvement of local authorities
            \item Support from civil society
        \end{enumerate}
\item Configuration
        \begin{enumerate}
            \item Geometry
            \item Size
            \item Transfer lines
        \end{enumerate}
\item Implementation cost
\item Risks related to the implementation
\end{enumerate}

For each scenario, six sets of criteria were evaluated individually for each of the surface site candidate locations for the surface sites in that scenario (land status, road connections, raw materials and services, physical characteristics, infrastructure, environmental and social factors) and three high-level criteria were considered for the overall scenario (configuration, costs, and risk).

Spreadsheets for each scenario were created to assign each criterion a qualitative score between -2 and +2, according to a pre-defined evaluation grid with standardised conditions. A score of `0' represents a neutral assessment of the indicator. For each high-level criterion, the scores of its sub-criteria were added together to provide an indicator for that criterion. The spreadsheet also shows the macro-criteria scores, and presents a final score for all criteria in the form of a percentage between 0 and 100. Lastly, the values of the criteria are synthesised to provide indicators for 1) scientific excellence, 2) territorial aspects and 3) appropriateness of project implementation. This approach not only makes it possible to quickly estimate the value of a scenario and compare it with others, but also to highlight criteria that are insufficiently known and require in-depth study.

Values were assigned for the qualitative indicators through a collaborative process by the multidisciplinary team, based on bibliographic and cartographic studies, database analysis, geographical information system queries, simulation and modelling. There were also field trips and field investigations; interviews to discuss the most suitable footprint with staff from the administrative departments of the two host states (e.g., DT, GESDEC, OCEV, OCAN, OCT in Switzerland; DDT 01 and DDT 74 in France); consultation with experts working in different technical fields (Cerema, Ecotec, HydroGéo, ILF, GADZ, particle accelerator designers working in many partner institutes, geologists employed in many partner universities); and feedback from local players during working meetings with municipalities, inter-communal entities, and elected representatives.

\begin{figure}[!h]
    \centering
    \includegraphics[width=0.95\textwidth]{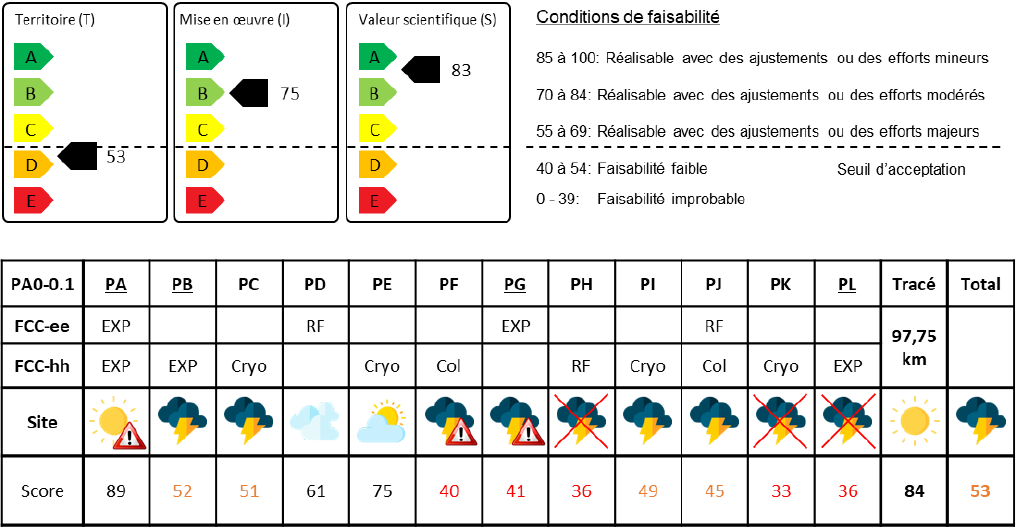}
    \caption{\label{fig:TIS-summary}Example of a summary view of the multi-criteria analysis of an scenario (PA0-0.1). The scenario seems feasible, even if the territorial feasibility would be low. However, a closer look at each site reveals many blocking points.}
\end{figure}

A disadvantage of this approach is that it simply adds up and averages all values. In some cases, there may be an obstacle related to territorial or scientific aspects or to project implementation. However, averaging means that a single low value will just lower the overall ranking but will not necessarily show the blocking point in the summary. The multi-criteria analysis was therefore complemented by an overall assessment of the scenario, which permitted the highlighting of the potential benefits of one scenario that make it particularly preferable to others or which indicate whether the scenario is difficult or not feasible. Thus, scenarios presenting obstacles continue to be analysed and recorded, but they are rejected even if the value of a thematic indicator (scientific performance, territorial compatibility or project implementation) is higher than the acceptable threshold value for that element. Examples of such situations include the strong probability of encountering geological features that would expose the project to an unacceptably high risk (karst, potentially conflicting water-bearing layers, crossing a major fault, presence of high-pressure aquifers), patrimony sites that represent incompatibilities with a technical installation, too dense residential zones, incompatibilities with local or regional development policies, a collider circumference that would not allow the scientific research programme to be conducted for technical reasons or would not allow its operation to be viable (for example, a collider circumference significantly smaller than 90\,km). The ranking is highlighted in summary diagrams such as the one outlined as an example in Fig.~\ref{fig:TIS-summary}. The illustration also shows the limitations of multi-criteria analysis: for example, two surface site locations in the PA0 scenario with twelve sites developed during the first exploratory phase are located in areas considered to be blocking points, but this is only visible in the individual sheet for each site. The three-column summary alone, however, does not show these exclusion criteria and must, therefore, be supplemented by a brief text description.

\begin{figure}[!h]
    \centering
    \includegraphics[width=0.95\textwidth]{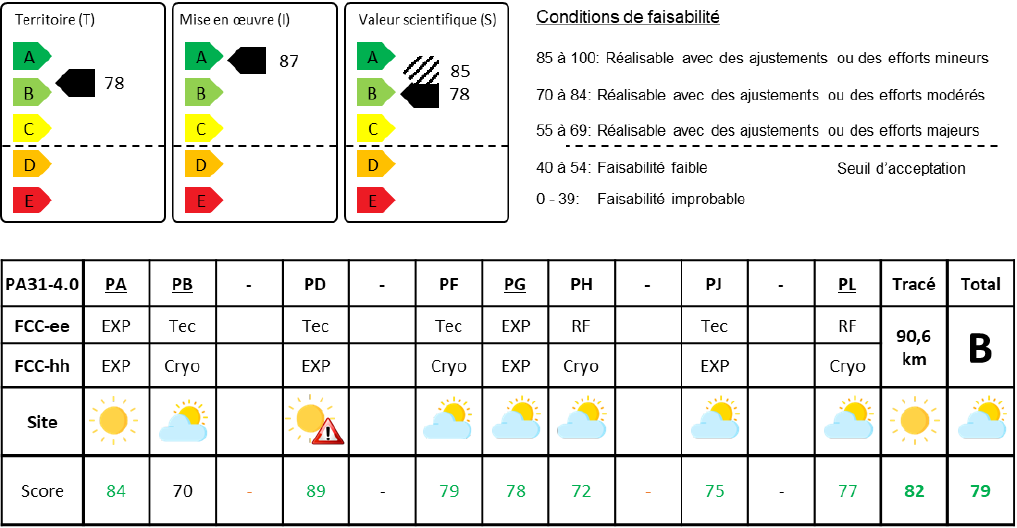}
    \caption{\label{fig:PA31TIS-summary}: Summary of the multi-criteria analysis of reference scenario PA31-4.0.}
\end{figure}

Territorial (T), implementation(I), and scientific (S) issues can, in fact, be easily visualised and compared thanks to this standardised qualitative approach. Figure~\ref{fig:PA31TIS-summary} highlights the merits of the current scenario working hypothesis PA31, which makes it a preferred scenario for the detailed territorial and technical analysis. Such overviews also helped to understand that scenarios involving twelve sites do not meet the required conditions to undergo an in-depth study. The approach permitted further optimisation by understanding how the improvement on one site could potentially lower the performance of another and thus retain scenarios for further optimisation that improved all sites individually and the entire scenario as a whole.

With additional information becoming available, the assessment changes and therefore further optimisation of the scenario will be required until a decision to proceed with a construction project can be taken. The advantage of the systematic approach is that it permits the demonstration that there is a continuous improvement of the project scenario based on the growing understanding of the boundary conditions and the consideration of stakeholder input during an environmental evaluation and project authorisation process.

\begin{figure}[!h]
    \centering
    \includegraphics[width=\textwidth]{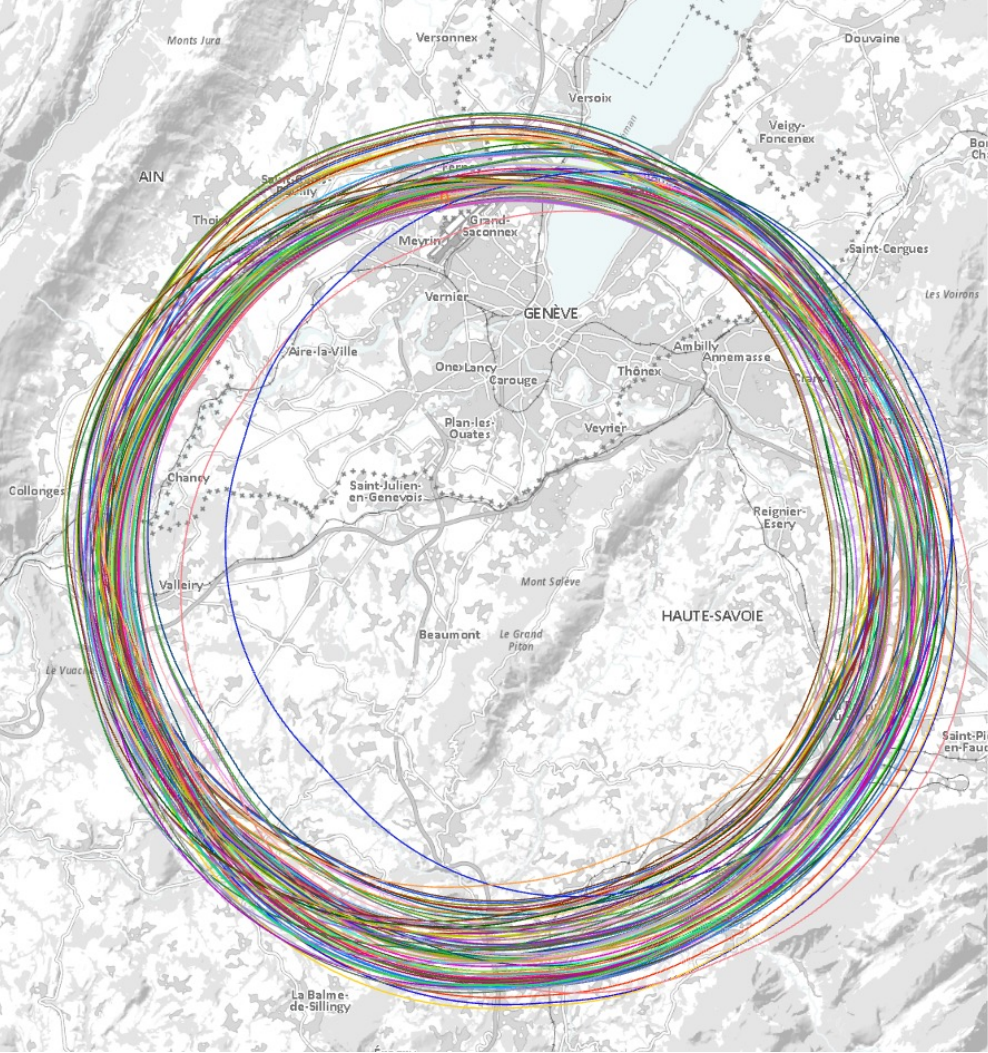}
    \caption{\label{fig:100-scenarios}: Around one hundred different layout scenarios were studied and individually analysed.}
\end{figure}

\begin{figure}[!h]
    \centering
    \includegraphics[width=0.98\textwidth]{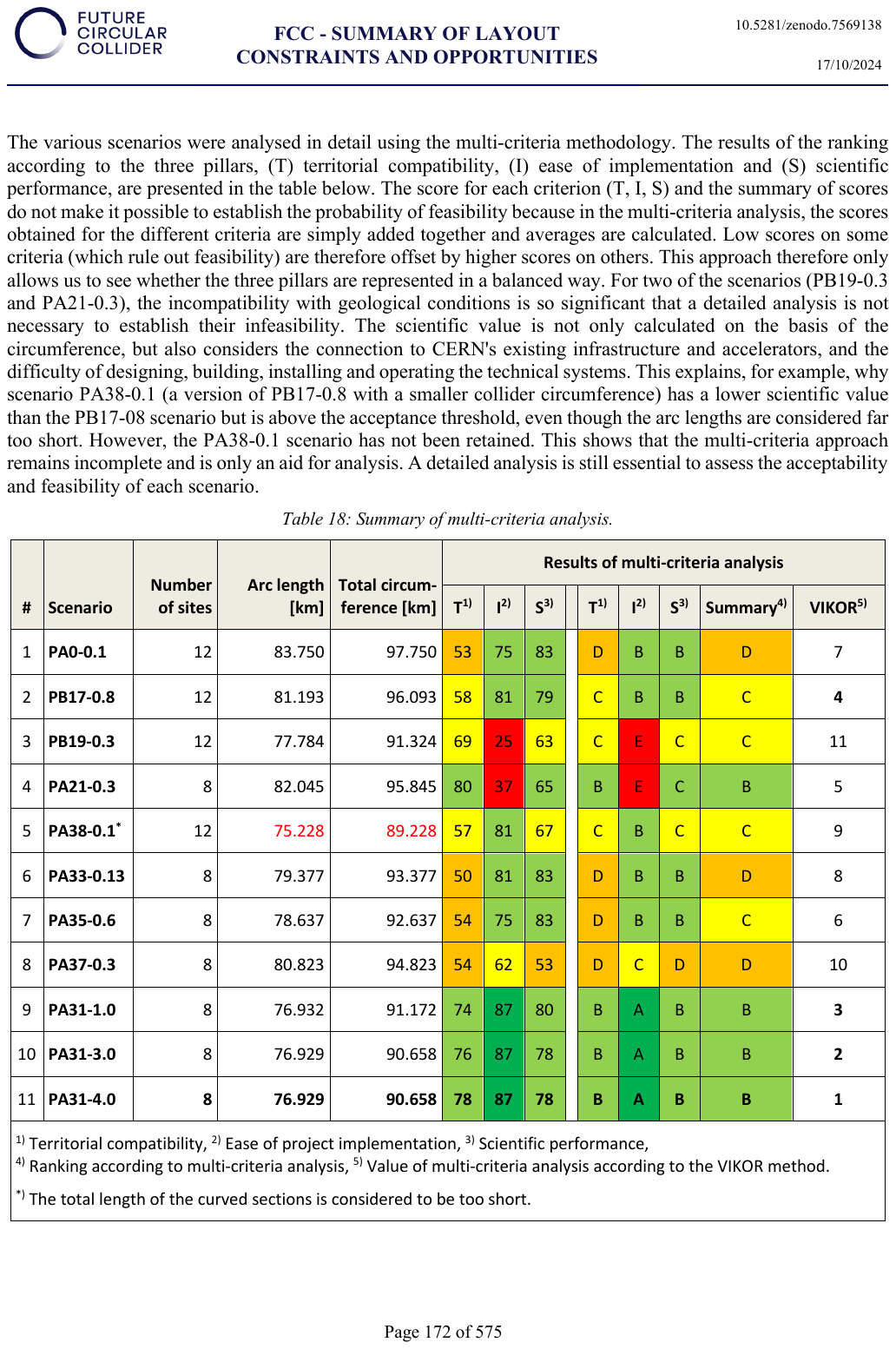}
    \caption{\label{fig:VIKOR-ranking} Multi-criteria analysis based ranking of the most promising implementation scenarios in 2022 with the performance of the gradual improvements of reference scenario class PA31.}
\end{figure}

Around a hundred different scenarios have been developed and individually analysed (see Fig.~\ref{fig:100-scenarios}). Eventually, the ten most promising scenarios (PA0-0.1, the scenario developed for the Conceptual Design Report, turned out to be not feasible and serves as a performance reference baseline only) have been retained for a review with experts from different project development domains. The scenarios were ranked using the VIKOR \cite{Opricovic2004} quantitative method that was applied to the established multi-criteria analysis (see Fig.~\ref{fig:VIKOR-ranking}). The algorithm permits the evaluation of different multi-criteria factors with different units and scales and with different, potentially even conflicting, minimisation and maximisation goals. The following VIKOR criteria were set for the establishment of the ranking:

\begin{enumerate}
\item \textbf{Number of sites:} An eight-site configuration offers four interaction points for the lepton collider, compared with just two interaction points for the twelve-site configuration. Therefore, the goal was to minimise the number of sites with a weight contribution of 5\%.
\item \textbf{Total length of curved sections:} A longer circumference length provides higher scientific performance or simpler design. Therefore, the goal was to maximise the circumference with a weight contribution of 5\%.
\item \textbf{Territorial compatibility:} Territorial compatibility, ease of implementation and scientific performance are given equal weights of 30\% each. Each of these three pillars must be maximised, and all three need to be balanced to ensure the feasibility of the scenario. The feasibility of each scenario can only be determined through detailed analysis.
\item \textbf{Compatibility of technical implementation and construction with geological constraints, configuration, and layout:} Maximise with a weight of 30\%.
\item \textbf{Scientific performance, taking into account the configuration and technical difficulties of the scenario:} Maximise with a weight of 30\%.

\end{enumerate}

The result was that the most promising scenario is an 8-site layout with a total circumference of the order of 91\,km. This scenario gave sufficient residual margins for further optimisations. Meanwhile, all the other scenarios developed are no longer considered feasible, either because of the loss of site locations during the study years, because of unacceptable incompatibilities with territorial or geological conditions, or because their scientific performance is not sufficient to attract researchers from all over the world over the long term. Therefore, PA31 was considered for further detailed studies, in particular involving subsurface investigations and detailed field studies concerning environmental aspects. Optimisations lead to the versions 3.0 and 4.0 for which the performance is also displayed.

\section{Requirements and invariants}
\label{imp:Requirements_and_invariants}

\subsection{Science and technology motivated requirements}

\subsubsection{Layout}
\begin{figure}[!h]
    \centering
    \includegraphics[scale=0.9]{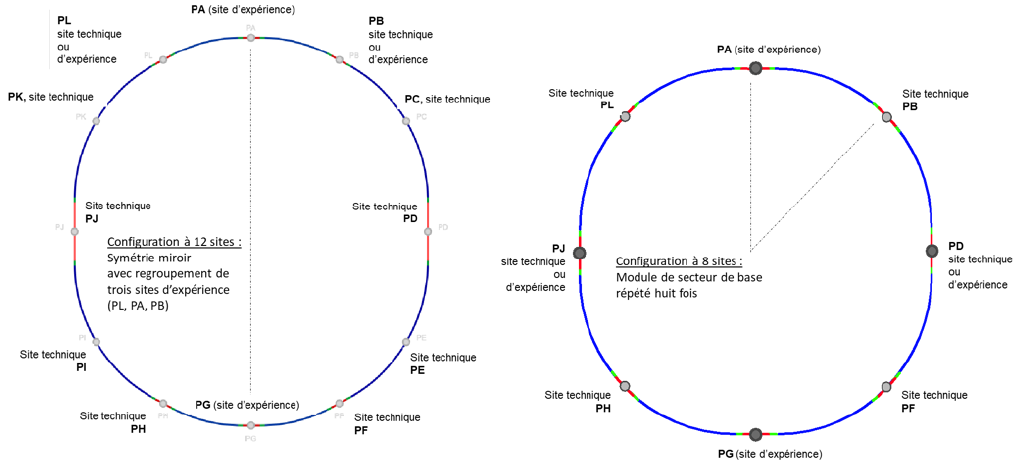}
    \caption{\label{fig:geometries}Two different collider layouts. The image on the left shows a configuration comprising twelve sites based on a geometry with mirror symmetry. The image on the right shows an eight-site configuration based on the repetition of a basic sector module. The arcs (blue) are separated by straight sections (green and red segments).}
\end{figure}

A set of parameters fundamental to particle accelerator design makes up the starting requirements for scenario development. The design parameters determining the overall size of the configuration cannot be chosen arbitrarily. These parameters include the geometry of the configuration (for example, a symmetry or periodicity of sectors involving a specific number of surface sites), the length of the basic curved cells, called `arcs' (which repeat like the parts of a chain, to build the curved sectors), the number of arc accelerator cells to be repeated and the lengths of the various types of straight sections between the arcs. Figure~\ref{fig:geometries} shows two basic configuration geometries used in configuration and layout studies. The first geometry (image on the left) serves as a starting point: it groups together three experiment sites in the upper part of the geometry. It offers greater freedom in terms of moving the various surface sites but requires twelve surface sites. The second geometry (image on the right) represents the current development: the four experiment sites are evenly distributed (top, right, bottom, left), and only eight surface sites are required. However, the freedom to move sites is less than with the first geometry.

The following requirements were initially set and had to be adapted during the development of a scenario that aimed at satisfying the territorial implementation constraints.

The initial simple mirror symmetry was discarded in favour of a 4 times repetition of a 90$^\circ$ sector, i.e., a four-fold symmetry. The initial configuration (12 sites) made it possible to group 3 experiment sites close to each other near the main CERN site, leading to shorter distances between surface sites and greater flexibility to lengthen and shorten the north and south sides. The current geometry (8 sites) enables a more regular design and thus facilitates performance optimisation processes. The lepton collider can also have four interaction points.

\subsubsection{Total circumference}
The initial objective was to design a particle collider infrastructure with a circumference of around 100\,km. A reduction of 10\%, to a circumference of 90\,km, is considered to be the lower acceptable limit, below which the performance obtained for scientific research would become too limited. The smaller the circumference, the smaller the radius of the curved sections and the more pronounced the curvature, resulting in greater energy losses, leading eventually to an unsustainable scenario from an energy efficiency perspective and the need for more powerful magnets for the hadron collider, which could then become technologically unattainable.

\subsubsection{Number of surface sites}
The initial 12 sites were reduced to 8 sites with the advent of the four-fold symmetry layout. It proved too difficult to find a suitable layout scenario with 12 sites featuring intermediate access points compatible with the territorial constraints and risks involved in implementing the project. A scenario with eight sites offers less flexibility for the layout study since the fixed interaction sites are geographically opposed. Still, the number of necessary suitable locations to be found is lower.

\subsubsection{Number of interaction points}
For the hadron collider, both geometries offer the possibility of four interaction points. However, only the four-fold symmetry layout provides the possibility of having four interaction points with the lepton collider. The 12-site scenario also proved difficult due to the need to combine experiments and injection at two points (PL and PB). 

\subsubsection{Arc cell length}
The hadron collider arc cell length drove the overall circumference of the layout. Initially, a cell length of 213.03\,m was considered. It was eventually extended to 275.79\,m. The total circumference depends on the multiples of that arc cell length and cannot, therefore, be arbitrarily chosen. In this case, the lepton collider adapts to the more stringent requirement of the hadron collider.

\subsubsection{Total number of arc cells}
The size of the hadron collider curve can only be increased or decreased by inserting or removing an arc cell in each curve. The overall configuration must remain symmetrical. The initial 12 site layout required 4 sectors of 19 short cells and 4 sectors of 73 long cells. The current 8 site layout requires 8 sectors with 26 cells, i.e., a total of 208 cells. This requirement determines the total circumference of the infrastructure. In this case, the lepton collider adapts to the more stringent requirement of the hadron collider.

\subsubsection{Total arc length}
The 12-site configuration featured four short arcs and four long arcs. The long arcs were too long to remain without intermediate access points for technical and personnel safety reasons. Therefore, an additional site had to be placed at the midway point, at a distance of around 8 to 10\,km from each end. The 8-site layout no longer has this requirement. The total arc length should, however, permit safe operation and the possibility to evacuate personnel in case of emergency efficiently. The current distance between access sites of about 11.5\,km is considered acceptable.

\subsubsection{Radio frequency system harmonics}
The stability and performance of the radio-frequency system used to accelerate the subatomic particles depend on the circumference of the collider. It also concerns the transfer of particle beams from the existing SPS and LHC tunnels. For a 91\,km long collider, 3 possibilities have been identified, which provide sufficient flexibility for the radiofrequency system design. A collider with a circumference of between 90.7 and 90.8\,km is the best option. A length of 90.6\,km is feasible but reduces the frequencies of the cavities in the radiofrequency system more than the others. In this case, the lepton collider adapts to the more stringent requirement of the hadron collider.

\subsubsection{Length of straight section at interaction points}
At least 700\,m are needed on either side of the interaction point to focus the collider beams to be able to achieve the high beam densities needed for high luminosities. The two colliders (lepton and hadron) have different space requirements for crossing the two beam lines, so the tunnel needs to be widened by around 10\,m to a distance of 1400\,m on either side of the interaction point.

\subsubsection{Length of straight section at radiofrequency locations}
A distance of 2800\,m between 2 straight sections was initially deemed adequate to accommodate the radiofrequency equipment that accelerates the beams. After various equipment integration studies, it was found that this value could be reduced. A distance of 2030\,m is now considered the acceptable lower limit required for the systems and the integration of the ancillary equipment.

\subsubsection{Cooling water requirements}
With the aim of conserving resources responsibly, ongoing studies have already helped to reduce requirements from an initial \SI{4.9}{million\, \cubic\metre} per year to a range between \SI{1.6}{million \cubic\metre} (Z mode) and \SI{3.0}{million\, \cubic\metre} per year ($\rm t \bar t$ mode). Further reductions can be expected due to the inclusion of the concept of heat reuse, taking into account technological developments and the development of water reuse synergies.

\subsubsection{Electricity requirements}
Initial powering requirements of the lepton collider ranged between 1.2 and 2.0 TWh per year, depending on the operation mode. Gradual advances in the concept development reduced these values to a range between 1.0 and 1.77 TWh per year. Since the current scenario has four interaction points rather than two, further reductions are possible by limiting the radiofrequency power if a reduction of the annual integrated luminosity is acceptable for the scientific research programme. Further optimisations are to be expected from an adaptive operation concept and energy consumption reduction of systems that are powered down or set to an energy-saving mode when not required for operation and maintenance.

For the hadron collider, an initial annual energy consumption estimate was based on current superconducting magnet technology at very low cryogenic temperatures. This estimate has been revised by developing a design that works at higher cryogenic temperatures to allow significantly lower energy consumption. Leveraging a high-temperature superconducting magnet technology would allow the cryogenic operation temperature to be further increased. Together with adapting the operation plan and introducing an annual energy consumption ceiling, this would bring the annual energy consumption into the range of the lepton collider. However, a focused R\&D programme is required to develop the technology. A window of opportunity in the range of more than 35 years is available to achieve this goal if a decision to proceed is taken.

\subsubsection{Space requirements of technical surface sites}
The space requirements of the technical surface sites are determined by the technical infrastructures that both particle colliders will need. Although the lepton collider has significantly fewer requirements than the subsequent hadron collider, the surface sites must be able to host the additional equipment that will only be put in place for the second phase.

The minimum equipment required on each site comprises an electrical substation, power converters to drive the particle accelerator equipment for the sectors in both directions of that site, power amplifiers where there is a radiofrequency system, tunnel and cavern ventilation, raw water-based accelerator cooling systems, raw water treatment systems, heat exchangers and refrigeration towers. Since the actual design of the particle accelerator is yet to come, it is challenging to estimate the space requirements for such systems. Therefore, the most advanced systems are used as a reference to estimate the space requirements.

In addition to the technical systems, space is required for temporary storage of the particle accelerator systems to be installed underground and for the handling of the technical systems at the surface. In the case of sites with a cryogenic refrigeration system, space is required to store the cryogens (liquid and gaseous helium as well as nitrogen). While the lepton collider will require only two sites with relevant cryogenic refrigeration systems for the radiofrequency systems, each site will host a cryogenic refrigeration plant for the hadron collider. 
Additional minor space requirements emerge from the roads on the site, such as the need to provide limited office space and storage of spare parts, cranes, and handling equipment.

The working hypothesis is that in the order of 100 technicians and engineers per day can be present during installation, major maintenance and upgrade interventions. During the operation phase, the presence of personnel is limited to the strict minimum, leveraging remote operation and intervention as much as possible. No permanent presence of personnel is planned during ordinary operations.

The current lepton collider net space requirements for construction elements on technical sites are:  2.2\,ha for PB, 2\,ha for PF, 4.4\,ha for PH and 2.6\,ha for PL. Additional space is required for the hadron collider cryogenic refrigeration plants and cryogen storage, as well as for additional technical equipment. Including the requirements for green buffers and landscape integration bring the requirements to about 3.5\,ha to 4\,ha for the technical sites PB, PF, PL. About 8\, ha are required for PH due to the initial large cryogenic refrigeration system and the topographical challenges that call for a terrace-based landscape integration. The indicated space requirements do not include the requirements emerging from the need to accommodate the different topographical constraints, the need for green buffers and landscape integration. Hence, the specific local requirements are higher.

Subsequent architectural designs and landscape integration will aim to reduce land consumption as much as possible.

\subsubsection{Space requirements of experiment surface sites}

In addition to the requirements of the technical sites, each experiment site also requires a pre-assembly hall for the experiment detector. To limit the surface space requirements, the working hypothesis is that the detector will be assembled underground. However, elements will need to be quality checked, pre-assembled and tested on the surface in a hall that measures about \SI{1250}{\metre\squared}. Sufficient space of about \SI{1000}{\metre\squared} for handling of parts and a temporary buffer for magnets to be installed is also required.

An experiment site typically also hosts a data centre, workspace for a group of about 50 scientists and engineers, meeting rooms and a few offices, small workshops, a data centre and visitor facilities.

For the hadron collider, the largest detector part, a superconducting magnet coil, cannot be transported directly to the sites. It needs to be manufactured on-site. For this process, about \SI{7500}{\metre\squared} are required in addition to each surface site. The space will remain untouched by construction during the first, lepton collider, phase, and it will be rewilded after the assembly and lowering of the magnet coil at the end of the hadron collider installation phase. 

The working hypothesis is that up to 300 scientists, technicians, and engineers can be present per day during installation and major maintenance and upgrade interventions. During operation, only a small team of less than 20 people would be present on site. Additional personnel would be present to operate the visitor facilities.

The current lepton collider net space requirements for construction elements on scientific sites are listed below.  3\,ha for site PA due to the possibility of leveraging the space of the existing LHC surface site Pt8. Including the temporary space required for the production of the superconducting magnet coil of the hadron collider experiment, the space requirement increases to about 4\,ha. The sites PD, PH and PJ require about 4\,ha for the lepton collider phase and 5\,ha for the hadron collider phase. The space requirements indicated do not include the requirements emerging from the need to accommodate the different topographical constraints, the need for green buffers and landscape integration. Hence, the specific local requirements are higher. 

Subsequent architectural designs and landscape integration will aim to reduce land consumption as much as possible.

\subsection{Initial invariants}

A set of invariants was established in the beginning to guide the development of an implementation scenario to ensure that a minimum set of criteria concerning scientific excellence, territorial compatibility and project risk control can be satisfied. They are a combination of avoidance and reduction constraints, goals and high-level requirements.

\begin{itemize}
\item Avoid karstic limestone formations, risks of high-pressure water penetration, known major faults and areas of seismic instability, as well as significant overburden.
\item Place the underground structures in the molasse layer as much as possible.
\item Provide a technically feasible connection to CERN's SPS underground infrastructure and, if possible, to the LHC infrastructure.
\item Dig the tunnel to a sufficient depth below the lake bed to ensure stability and long-term low maintenance. The alignment can be optimised by identifying the minimum depth based on subsurface geophysical and geotechnical investigations for a preferred scenario. The current assumption is that the tunnel alignment crossing the Geneva lake is 100\,m below the lake bed.
\item Dig the tunnel deep enough under the Arve, Rh\^{o}ne and Usses rivers to guarantee the micro-stability required for precise alignment of the accelerator and to avoid adverse effects with any, potentially existing, surface constructions.
\item Limit overburden under the Mandallaz, Fillière and Bornes zones (minimum depth to be defined by specific subsurface investigations).
\item Avoid elevations above 750\,m for surface sites to limit the depth of shafts and ensure good site accessibility.
\item Aim for shaft depths of less than 250\,m for scientific sites and 400\,m for technical sites to avoid excessive challenges for the installation of the accelerator equipment and the experiment detectors.
\item Avoid areas considered `exclusion zones' in an established territorial sensitivity grid drawn up with contracted organisation Cerema, Direction Départementale du Territoire in France, collecting information from the Direction Régionale de l'Environnement, de l'Aménagement et du Logement and with contracted company Ecotec in Switzerland with additional information collected from the cantonal DT services. This grid lists environmental, spatial planning, zoning and strategic constraints, as well as public health and safety constraints with different relevance levels that lead to the definition of zones to be avoided and zones at surface and subsurface constructions will need to be limited if they cannot be avoided.
\item Give preference to proximity to strategic project infrastructures (e.g., grid power lines, certain types of roads, autoroute stations, railway lines) and stay clear of certain other infrastructures (e.g., geothermal probes, gas pipelines, oil pipelines).
\item Remain outside of town and village centres and hamlets to limit potential nuisances associated with the construction and operation activities.
\item As far as possible, avoid areas subject to a set of less critical territorial constraints, and if avoidance is not possible, reduce surface areas in these zones.
\item Respect the fact that it is not possible to individually displace the locations planned for scientific sites where particle beams cross and where the experiment detectors are located.
\item Respect the limits for displacing technical sites along the straight sections of the ring within the lengths of those sections, according to the possibilities offered by the various technical systems required at the technical sites for the operation of the accelerators.
\item Consider the constraint for positioning the shafts inside the ring to facilitate access to the transport zone in the tunnel.
\item Limit the length to around 400\,m for the horizontal connection tunnels from the technical sites to the shafts inside the ring. Building a longer tunnel is not out of the question, but it would entail additional costs and difficulties.
\item If technically possible, limit the total occupied surface area of technical sites to 5\,ha and the surface area of experiment sites to 8\,ha. These areas include buffer zones, storage, and transport areas and take account of rewilding, reconstitution, replacement, or compensation measures. Wherever possible, seek synergies with existing infrastructures to reduce surface areas.
\item Avoid areas with difficult topography (steep and potentially unstable slopes, unstable soils, areas at risk of flooding, discontinuous terrain).
\item Ensure adequate road connections (7\,m wide, 2\,lanes, 20\,m curvature radius, gradient well below 10\%, minimum clearance height of 4.4\,m, compatibility with heavy goods vehicles of 44\,t to the main road network in accordance with article R312 of the French highway code).
\item Maintain a minimum distance of 100\,m between residential areas and sites. Depending on topographical conditions and the site's noise and visibility attenuation features, this distance can be as much as 250 to 300\,m.
\item Avoid protected agricultural areas unless they are essential to ensure technical feasibility.
\item Avoid natural and historic heritage sites, avoiding shared visibility between the sites.
\item Preserve the views and landscape protection zones as far as possible.
\item Avoid rivers and streams (to avoid having to modify or stabilise riverbanks).
\item Avoid wetlands and strictly protected areas.
\item Avoid protected forests because of the need to clear them and preserve existing biodiversity and because of their inaccessibility and topography, unless these areas are essential to ensure feasibility. In the latter case, the land occupation should be limited.
\item Give preference to areas in proximity to high-capacity power lines (225 kV and 400 kV).
\item Look for opportunities with a connection or access to the autoroute system.
\item Look for opportunities with a connection or access to the railway system.
\item Consider the use of known brownfield sites for the installation of technical infrastructures that may be remote from the sites (e.g., electrical substations, pumping stations, cooling towers) and for the development of compensation areas.
\item Look for scenarios that enable all surface sites to be reached within a reasonable time (30\,minutes by car) from strategic service and supply points (e.g., CERN and CNRS/LAPP).
\item Avoid creating new border crossings and do not consider sites that straddle the border.
\item Avoid vineyards and areas with fruit trees. 
\item Give preference to public land over private land, undeveloped land over developed land, unused land, overused land (special agricultural uses such as vineyards and protected fruit crops) and brownfields.
\end{itemize}

\subsection{Voluntary objectives}

The multi-year study also led to the development of voluntary objectives that would help make the project more territorially compatible. They have been taken into account in the evolutions of scenarios and in the development of a preferred layout scenario that serves as a reference, which is presented in later sections of this document.

\subsubsection{Sustainable operation and maintenance}
Ensure that the scenario can be operated and maintained sustainably. To achieve this, relevant technical facilities (high-tech workshops, offices) must be in proximity to the sites. Therefore, give preference to scenarios that take advantage of synergies with existing CERN sites and/or are close to the LAPP of CNRS/IN2P3.

\subsubsection{Protecting farmland}
Limit the consumption of farmlands in both host states. Avoid protected agricultural areas, unless they are essential for feasibility.

\subsubsection{Road access to surface sites}
To limit the need to build new roads and to avoid consuming land, surface sites must either be located directly on existing departmental roads or only require an improvement to existing access roads. If new roads are essential, then the requirements should be minimised.

\subsubsection{Disturbances}
Limit direct disturbances for local inhabitants. Avoid proximity to residential areas, visibility, and shared visibility. Avoid locations on roads passing through residential areas. Avoid proximity to natural and historical heritage sites. Provide means of evacuating spoil by conveyor belt to avoid truck traffic where possible.

\subsubsection{Landscape integration}
Aim to preserve views and landscape protection zones wherever possible. Specific studies by landscape architects are required for a preferred scenario during a subsequent design phase.

\subsubsection{Protection of water}
Avoid water protection zones, rivers, and streams. Do not take in water from water-bearing layers, and avoid the use of drinking water for systems that require raw water. Avoid having to stabilise or modify riverbanks. Specific studies are required for a preferred scenario during a subsequent design phase.

\subsubsection{Forests}
Respect protected forest zones and trees. Limit clearing forests or felling trees in both host countries to maintain biodiversity. Avoid high-quality forest zones whenever possible. Improve climate resilience by restoring impaired forests.

\subsubsection{Synergies}
\begin{figure}[!ht]
  \centering
  \includegraphics[width=\textwidth]{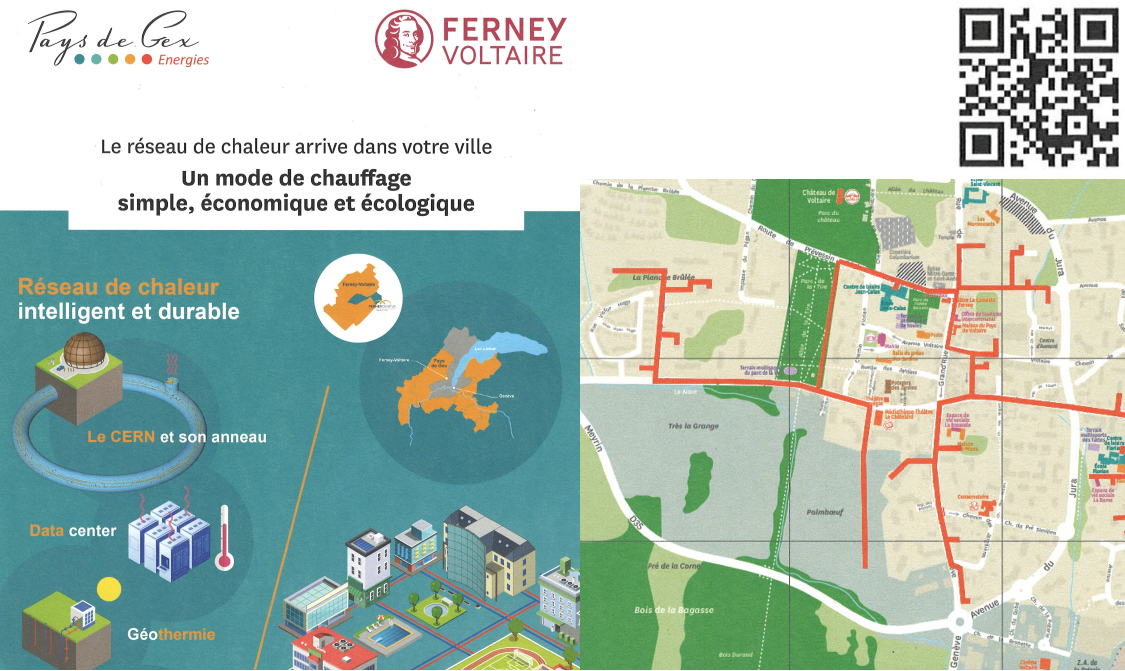}
  \caption{CERN powered heat network under construction in Ferney-Voltaire, France.}
  \label{fig:waste_heat_ferney}
\end{figure}

Give preference to locations that can create synergies with nearby private and public facilities and infrastructures. This includes, for instance, the availability of emergency rescue stations and fire brigades, hospitals, commercial and industrial development zones, food processing companies, infrastructures that support public works, electricity lines, major transport axes, autoroute service stations, train lines and school development projects.
Identify the possibility of supplying residual heat within a 5\,km radius, preferably to public facilities and major companies and residential housing. This is an approach that has already been put in place by CERN in Ferney-Voltaire and such an infrastructure serves as a demonstrator for the feasibility (see Fig.~\ref{fig:waste_heat_ferney}).
Examples include, but are not limited to, milk processing companies (cheese production), hospitals and health service infrastructures, schools, fire brigades, prisons, offices, airports, train stations, and leisure facilities (pools and spas). Locating surface sites in the proximity of public services can  lead to tangible added value for the region. For example, the significantly enlarged geographical extent triggered the need to develop an emergency intervention concept based on cooperation with regional fire departments and first-aid services. Other locations are particularly suited to foresee additional leisure activities around experiment surface sites that feature visitor centres. Additional services including exhibitions, restaurants, and meeting facilities have potential for indirect and induced economic value generation, but they need to be agreed with the local stakeholders to integrate with the local economic and tourism development strategies and plans. 

\subsubsection{Reduced need for electrical infrastructure}
To limit the effects and impacts of infrastructure development for electrical power, give preference to scenarios located near substations and high-capacity power lines. 

\subsubsection{Excavated materials}
Limit the annually excavated volume of materials excavated and reduce road transport wherever technically and economically feasible. Facilitate local reuse. The aim is to keep the volume of materials excavated annually to less than 10\% of the volumes generated in each region (Auvergne-Rhône-Alpes in France: 24.2 Mt in 2021~\cite{SRC_Auvergne_Rhone_Alpes_2021}, cantons of Geneva~\cite{ge2022constructionwaste} and Vaud~\cite{vd2023wastestats} in Switzerland: 3.071 Mt in 2022).

\subsubsection{Land use}

The goal for the scenario development is to develop a scenario that keeps land consumption as low as possible by still meeting the needs of the required surface site installations. The scenario development process avoids protected agricultural spaces, keeps interactions with multiple private landowners for land plot acquisition low, as a multitude of negotiations and complicated ownership situations can significantly impact project implementation preparation and lead to unsatisfactory results for them. Consequently, prioritise the use of public land over private property and steer clear of areas with planned development projects and prefer non-protected agricultural spaces over protected ones.

The surface site candidate location in Switzerland is mainly located on publicly owned land. The surface site candidate locations in France concern mainly privately owned land plots, subject to the typical urbanistic planning evolution. After exchanges with municipalities to determine the territorial conditions and constraints of possible locations for surface sites, CERN asked the Prefect of the Auvergne-Rhône-Alpes region in France to set aside the land plots to assure their availability for field studies. This process and the orders signed by the Prefect of the region mitigate the risk of unanticipated use of the land plots for other territorial developments for the study time period, permitting a comprehensive and exhaustive investigation of the environmental conditions. As a result, the land considered for the surface sites will not compete with other emerging projects. The solution was made possible thanks to the support of France. It contributes to securing the studies while ensuring compliance with national legislative frameworks.

\subsubsection{Biodiversity and nature}
Give preference to sites with low-quality nature characteristics. Through the ERC approach, preserve the elements of interest in terms of biodiversity and nature within the perimeter as much as possible. Study, together with design offices, the existing and projected conditions of the sites concerned, and carry out environmental, flora, and fauna surveys on the proposed site to identify valuable elements and the site's role in the ecological infrastructure. Propose appropriate restoration, replacement, and compensation measures based on the impacts identified by specialist consultants. Maintain or reinforce ecological infrastructure, particularly wildlife corridors.

The work on these voluntary goals was also considered during the development of the eco-design strategy and guidelines \cite{fcc-ecodesign-guideline-2024}, which all project participants are required to consider during the subsequent design phase and which is also published in the report on the current state of the environment \cite{EISA_2025}.

\section{Territorial constraints}
\label{imp:territorial_Constraints}

\subsection{Territorial sensitivity grid}
The development of a territorial sensitivity grid with its four levels (unacceptable `red', high `orange', acceptable `yellow', and low `green') served as a starting point to integrate the territorial constraints into the implementation scenario development from the onset.

\begin{figure}[!h]
    \centering
    \includegraphics[width=\textwidth]{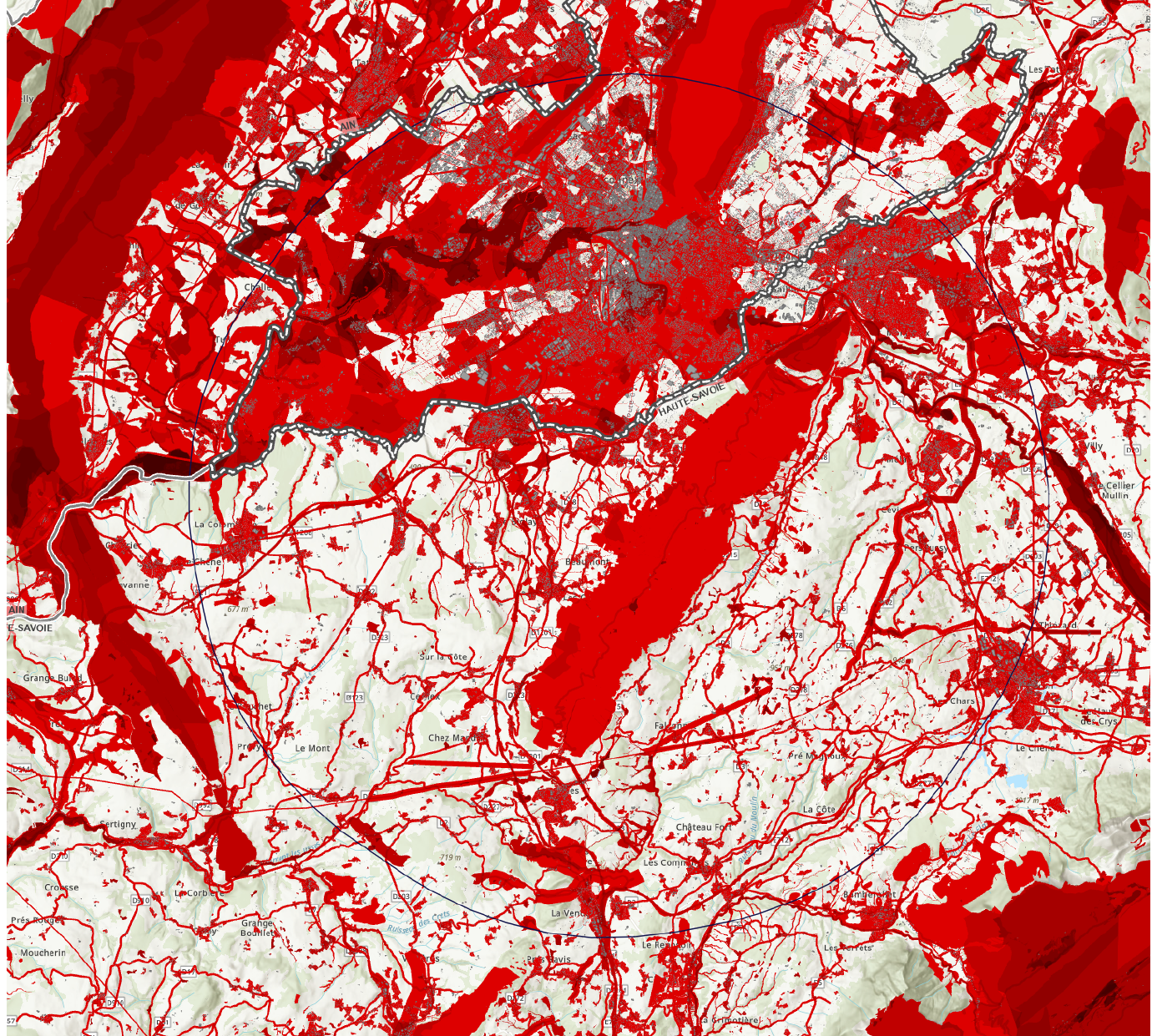}
    \caption{\label{fig:class-RED}: Zones in the scenario development perimeter classified as to be avoided (`red'). Note that this map does not consider numerous other exclusion criteria such as slopes and high altitudes (see also online at \url{http://cern.ch/fcc-sensitivity-grid}).}
\end{figure}

By applying the `Avoid' approach, the analysis eliminated all the unacceptable zones to be avoided, shown in red (see Fig.~\ref{fig:class-RED}), from the early development stages of the initial scenario at the macroscopic level. It should be emphasised that, in all cases, the scenario development process also aimed to avoid areas already developed if they were considered to be in active use (e.g., residences, farms, parking lots, industrial buildings and public infrastructures). Only brownfields or buildings that could be reliably described as unused (ruined houses, abandoned industrial warehouses, etc.) were examined on a case-by-case basis. In addition, for all the surface sites taken into consideration, choices have been made to avoid clearing protected forests, as well as avoiding the destruction of wetlands and hedgerows in agricultural areas.

\begin{figure}[!h]
    \centering
    \includegraphics[width=\textwidth]{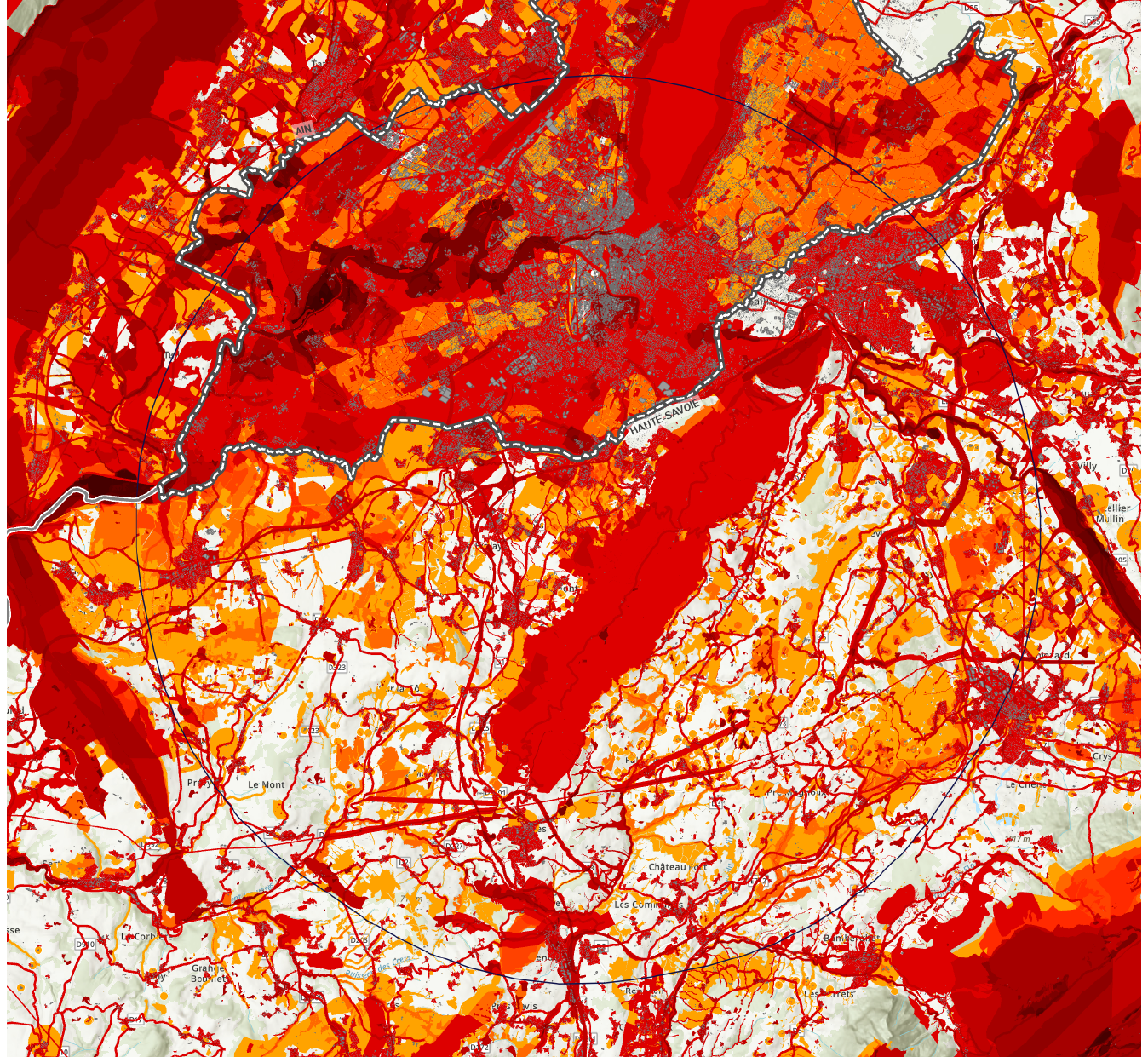}
    \caption{\label{fig:class-Orange}: Zones in the scenario development perimeter classified as to be avoided (`red') and with high constraints (`orange'). Note that this map does not consider numerous other exclusion criteria such as slopes and high altitudes (see also online at \url{http://cern.ch/fcc-sensitivity-grid}).}
\end{figure}

Zones classified with high constraints, shown as `orange' were generally avoided (see Fig.~\ref{fig:class-Orange}), but a more detailed manual analysis of the underlying rationale for this classification was always conducted to understand whether and under what conditions such a zone might be considered for certain parts of the research infrastructure. This step required a more detailed study of the different layers of constraint that make up the overall layers. It also required additional information that the regional and local planning authorities of the two host states were best placed to provide (for example, the various offices of the Département du Territoire and the Infrastructure Department of the canton of Geneva, the DREAL, the DDT 74 and the DDT 01 in France).

In France, the law of 22 August 2021, on combating climate change and strengthening resilience to its effects set a target of zero net artificialisation of soils (also known as ZAN). The ZAN initiative is a target set for 2050. It calls on local authorities, municipalities, departments, and regions to reduce the rate of artificialisation and consumption of natural, agricultural, and forest areas by 50\% by 2030, compared to the consumption rate measured between 2011 and 2020. A similar instrument for crop rotation areas in Switzerland, `surfaces d'assolement' known as SDA exists. 

The scenario development follows the regulatory frameworks in both host states regarding land use efficiency. This concerns in particular the consideration of the principle of "zéro artificialisation nette (ZAN)" (english: "zero net land take") in France and the principle of the "Surfaces d'Assolement (SDA)" (english: crop rotation surfaces) in Switzerland. By considering these constraints in the "avoid-reduce-compensate" based development cycle, they support the development of projects that exhibit a high degree of territorial responsibility. The SDA imposed constraint was addressed by the development of an authorisation process for the territorial development of CERN at the federal level.
Concerning ZAN it is noted that, on the recommendation of the prefect of the Auvergne-Rhône-Alpes region, the FCC segment on French territory is recognized among European and national large-scale projects (PENE)~\cite{ministry2024cartographie}.
Consequently, the land areas potentially consumed by surface sites would be outside the regional quota, in accordance with the decree of May 31, 2024~\cite{arrete2024mutualisation}, concerning the national pooling of the consumption of natural, agricultural, and forest spaces for projects of major general interest. However, this exclusion from the regional quota would not exempt the FCC from complying with the principles of land use efficiency and compensation. Through its transversal and eco-responsible approach, the study demonstrates its commitment in this area, thus affirming its intention to embed the project in a responsible and balanced dynamic.

Right from the start of the studies, it became clear that there were no areas with low constraints indicated with `green' colour in the two host countries. The minimum number of constraints in both countries corresponds to the `yellow' class, i.e., acceptable zones. In addition, Switzerland has special restrictions, imposed at the federal level, for a certain type of agricultural land, known as crop rotation areas (SDA), the best arable land in Switzerland. A federal sectoral plan sets a minimum surface area to be maintained per canton (distributed among the cantons by quota). The Confederation monitors the maintenance of the minimum SDA area. Under certain conditions, SDAs can be allocated to specific projects, provided that all stakeholders at the federal and cantonal levels agree and that the minimum is maintained. This constraint is high, but the SDA spaces have been classified as an `orange' zone to leave the possibility of downgrading open and to provide an option for the development of a layout and placement scenario. Specific coordination at the federal and cantonal levels is nevertheless required to be able to release such land for a development project. SDA spaces will have to be compensated on a 1:1 basis by removing the topsoil and re-constructing the agricultural space on lower-quality plots or wastelands. Agricultural areas in the canton of Geneva outside the SDAs were classified as acceptable `yellow' zones.

The initial classification of sensitivity layers is not fixed, but constantly changes as the territory evolves. Between 2014 and 2022, more than 1000\,ha became `red' exclusion zones (unacceptable constraints) in the Haute-Savoie department in France. In the canton of Geneva, around 100\,ha have been affected by new restrictions and are now part of the exclusion zones, mainly due to stricter surface and groundwater protection rules resulting from improved knowledge of the subsurface. This led to the rejection of several candidate sites proposing a configuration and location initially considered as possible. These changes have made it even more difficult to continue developing scenarios, while at the same time highlighting the iterative aspect of the process.

\begin{figure}[!h]
    \centering
    \includegraphics[width=\textwidth]{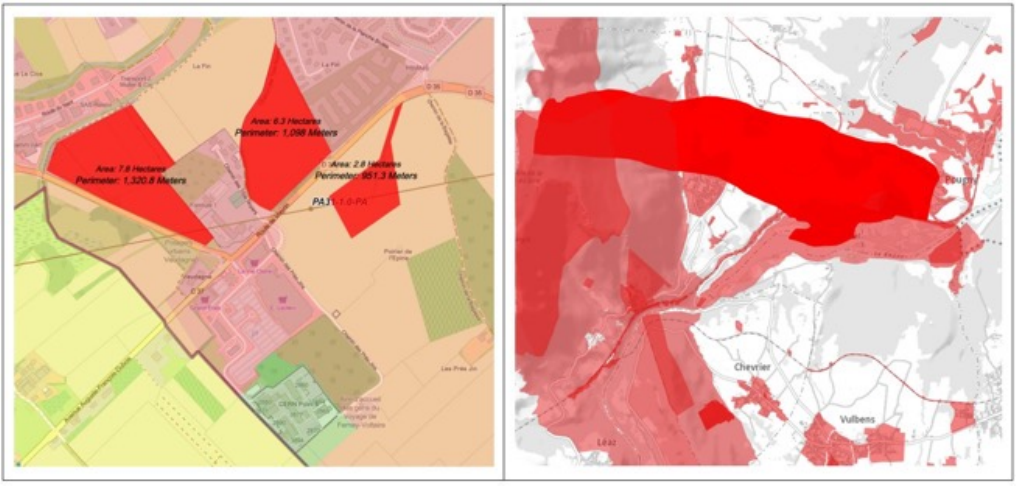}
    \caption{\label{fig:evolution-protection-zones}: Examples of regional changes. Left: 17\,ha of new exclusion zones in Ferney-Voltaire (Ain, France). Right: 790\,ha of peatland inventory on the banks of the Rhône in France: in light red, previously orange zones that have become red zones.}
\end{figure}

Figure~\ref{fig:evolution-protection-zones} shows two examples of these `lost spaces for layout scenarios'. The first concerns sectors of Ferney-Voltaire that are now intended for a development project (hospital project north of the D35, Route de Meyrin) and the classification of a sector as a compensation zone (south of the D35), which therefore cannot be compensated again. The second concerns a vast area north of the Rh\^{o}ne in France, which has been described as a `peatland inventory'. 

\subsection{Subsurface constraints}

Subsurface constraints, such as unfavourable geology, major faults, presence of strategic aquifers, drinking water catchment areas and buffer zones, pipelines, power lines and gas mains, as well as reserved areas of public interest such as geothermal exploration zones or no-drilling zones due to superimposed aquifers or other information, were considered from the onset. Geological exclusion zones have been defined by the geological consulting firms GADZ and ILF and the University of Geneva to take full account of geological conditions. These partners also examined the geological and hydrological situation based on the following elements:
\begin{enumerate}
\item Fault lines representing a high risk of seismic activity and that could cause tunnelling challenges.
\item The interface between limestone and molasse.
\item Potential high-pressure water penetrations or karstic formations that would expose the project to an unacceptable risk.
\end{enumerate}

\begin{figure}[!h]
    \centering
    \includegraphics[width=\textwidth]{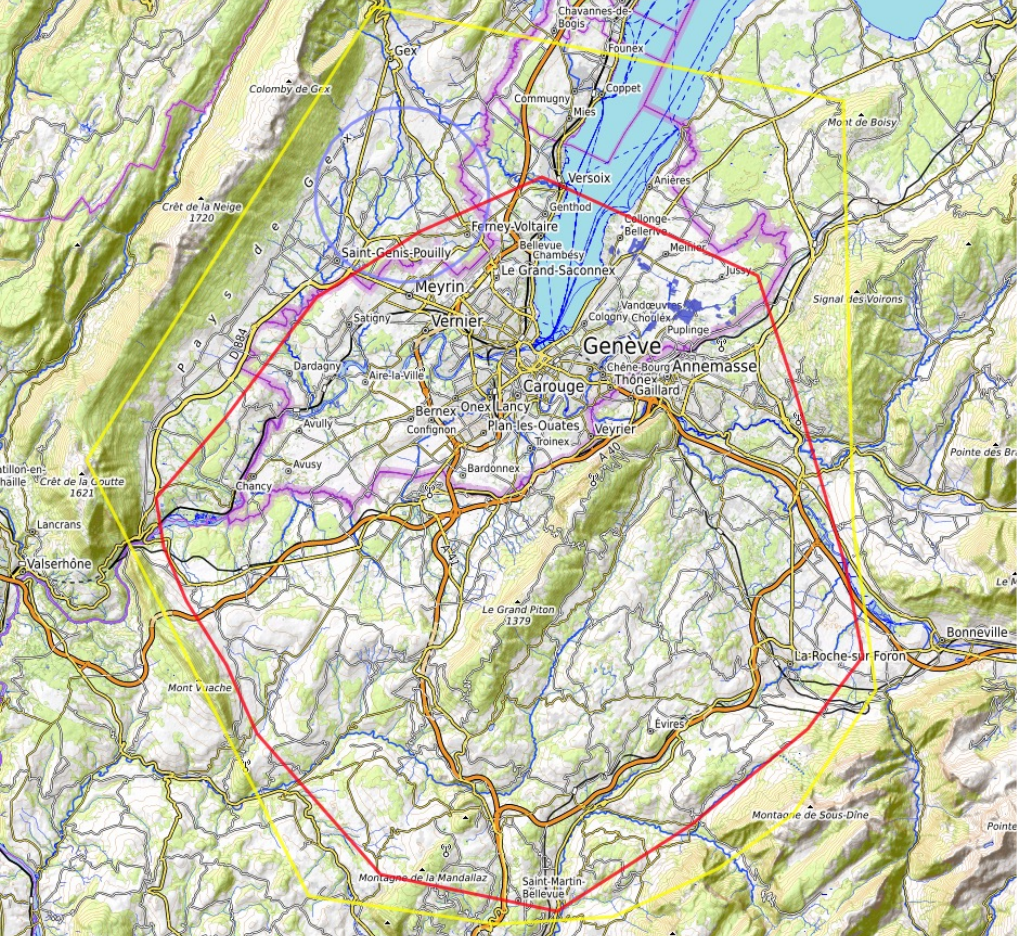}
    \caption{\label{fig:subsurface-limits}: After several years of analysing the subsurface conditions based on bibliographic data and modelling, the initial scenario development perimeter (yellow) had to be restricted (red) to assure compatibility with the known subsurface conditions.}
\end{figure}

\begin{figure}[!h]
    \centering
    \includegraphics[width=0.95\textwidth]{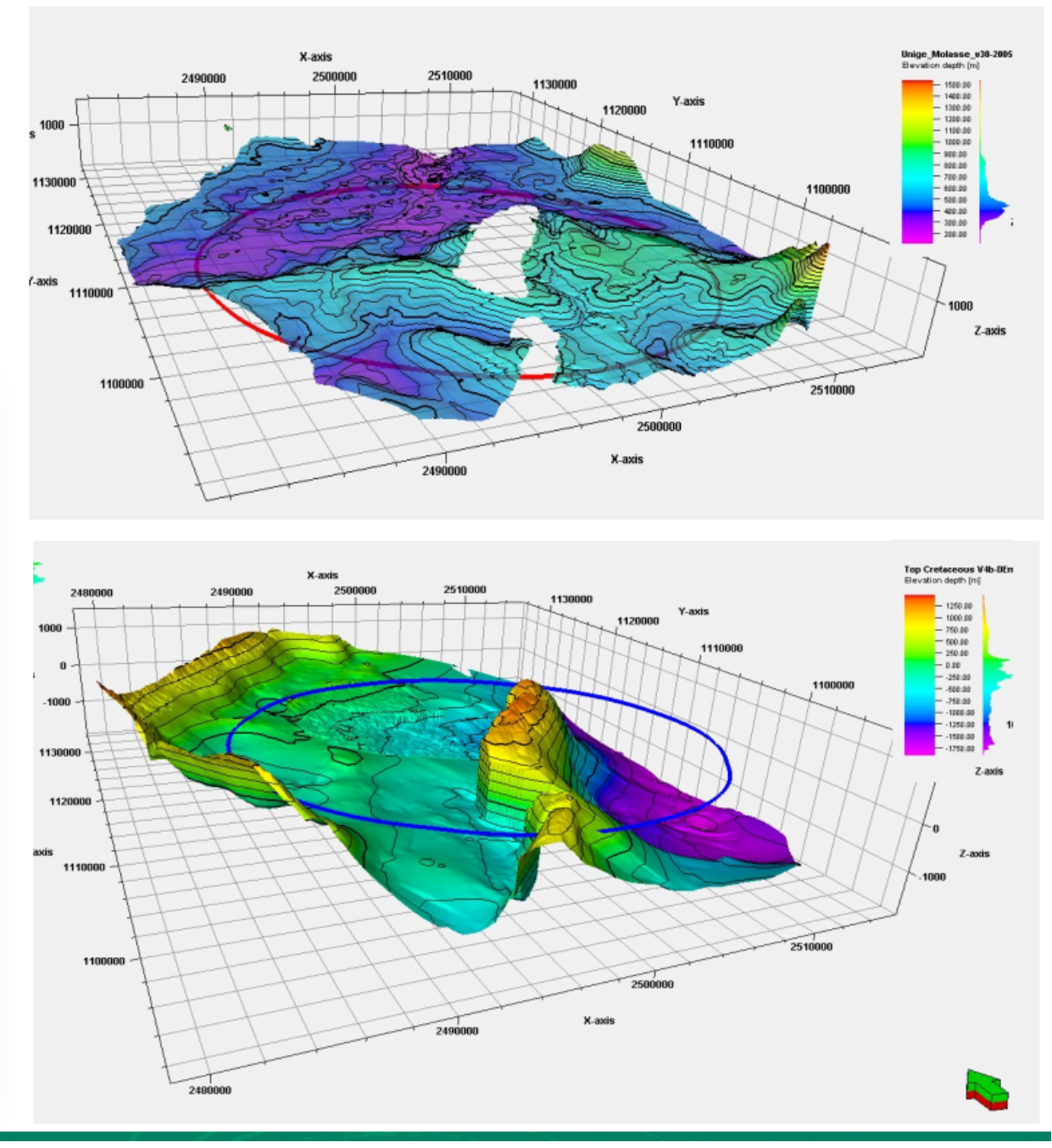}
    \caption{\label{fig:molasse-and-cretaceous}: Example images from the 3D model (here molasse and cretaceous formations) used to determine the geological constraints.}
\end{figure}

As a result, between 2014 and 2020, the project's initial study perimeter had to be significantly narrowed (see Fig.~\ref{fig:subsurface-limits}) as additional knowledge was acquired and integrated into a 3D model of the subsurface. Figure~\ref{fig:molasse-and-cretaceous} shows two example images from the 3D subsurface model that was established to support the analysis of the subsurface constraints, to identify zones that are definitely to be excluded, preferred locations for the subsurface structures, and to identify zones with particular challenges or insufficient data that require dedicated subsurface investigations.

Given the experience gained from the construction of the Large Electron Positron collider (LEP) underground structure and from the construction of underground transport structures in the region (e.g., the Vuache road tunnel), the risks of building a large tunnel and caverns, which could be unstable, move, collapse, or in which construction workers could be exposed to high-pressure water ingress that could lead to fatal accidents, are considered unacceptable and must be avoided. As far as possible, all underground structures should be located in the soft, stable molasse layer, which forms a reliable barrier against aquifers.

Although maps and modelling work have provided additional information, reliable data on the exact depth of interfaces between different geological layers in the Vuache and Jura zones is still lacking. The unavoidable limestone zones of Mandallaz have yet to be studied. Furthermore, little is known about conditions under the lake, as well as the Arve and the Rh\^{o}ne rivers. Since these conditions have a major impact on the cost and schedule of the tunnel boring, more information about the subsurface in the key areas is being gathered with geophysical and geotechnical exploration between 2024 and 2025.

\subsection{Topography, bathymetry, and other surface features}

\begin{figure}[!h]
    \centering
    \includegraphics[width=0.95\textwidth]{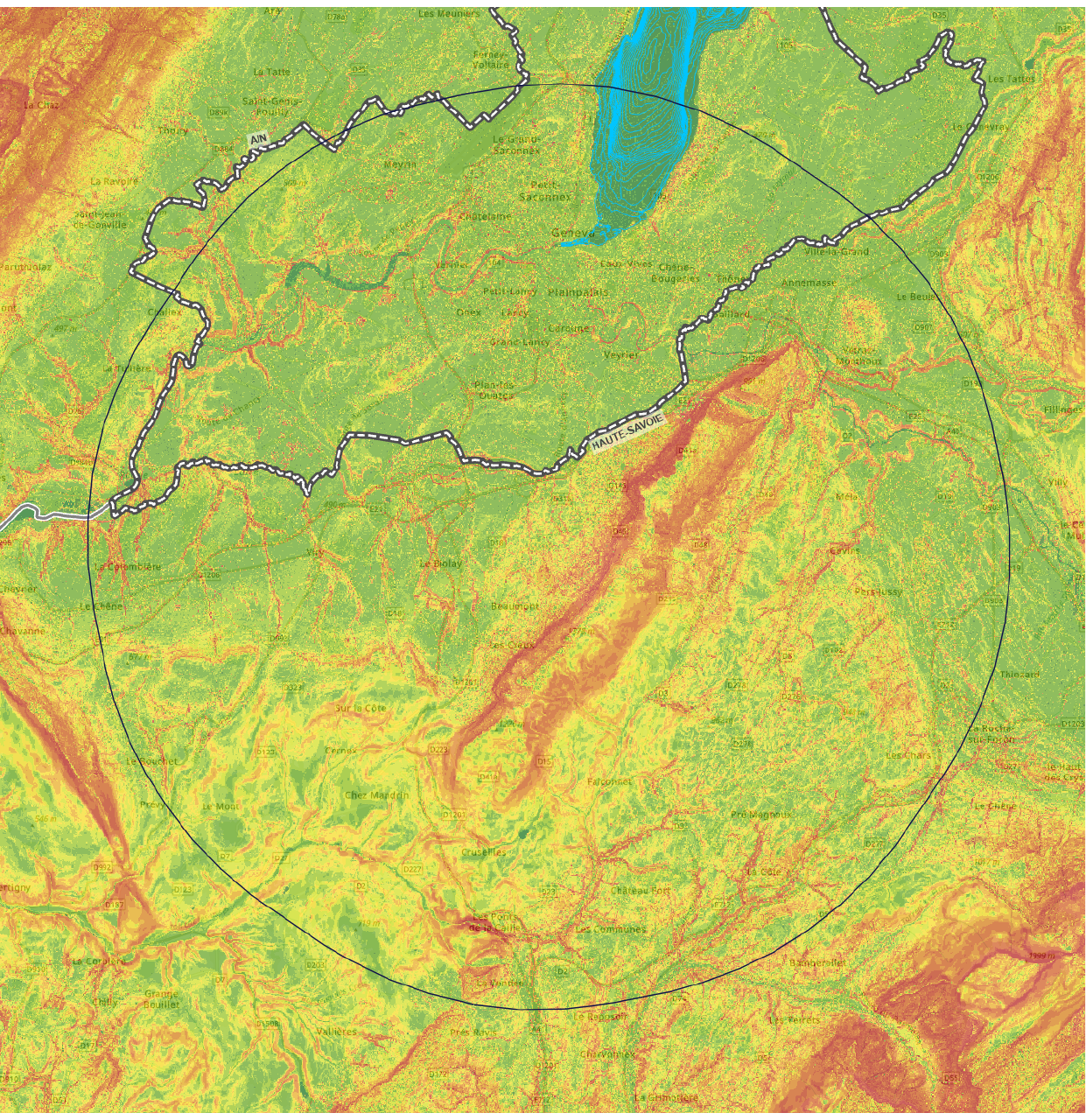}
    \caption{\label{fig:slopes}: Topography of the area. Red and orange colours indicate steep slopes.}
\end{figure}

\begin{figure}[!h]
    \centering
    \includegraphics[width=0.9\textwidth]{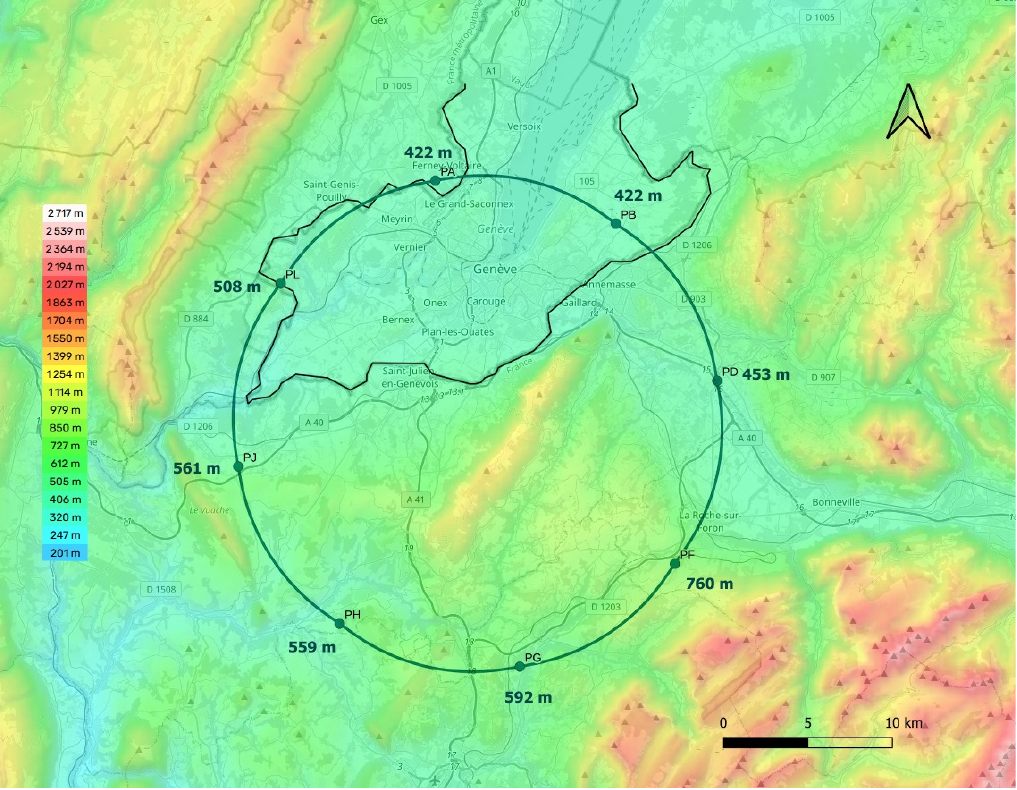}
    \caption{\label{fig:elevation}: Relief of the area. White and red colours indicate high elevations.}
\end{figure}

Information on topographical surface conditions has also been taken into account from the outset of the study. Steep slopes (Fig.~\ref{fig:slopes}) in excess of 30\% presenting risks of unstable terrain and landslides are excluded, as are cliffs, narrow valleys, and canyon-type formations. Since the tunnel must be placed at a sufficient depth below the lake bed, elevations above 750\,m are an obstacle due to the depth of the shafts and unacceptable overloading (Fig.~\ref{fig:elevation}). High elevations also lead to high overburdens, which can become challenging for tunnel boring activities, calling for more costly construction and slower progress.

\begin{figure}[!h]
    \centering
    \includegraphics[width=\textwidth]{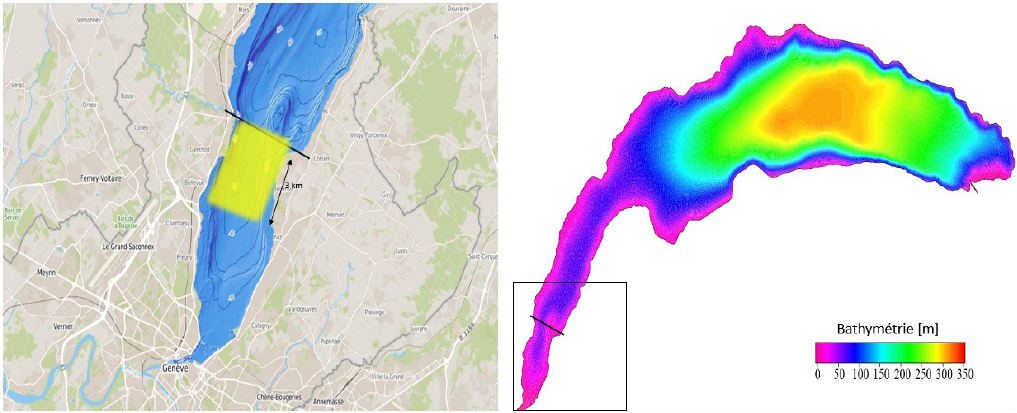}
    \caption{\label{fig:bathymetry}: Bathymetry of the Geneva lake.}
\end{figure}

The bathymetry of the lake must be taken into account to ensure that the tunnel can be placed sufficiently below the lake bed with as short a crossing as possible. Lake Geneva is more than 50\,m deep beyond the Versoix-Corsier line (see Fig.~\ref{fig:bathymetry}). To avoid the tunnel becoming too deep over its entire footprint, it is necessary to stay below this line. Crossing the lake where it is narrow avoids areas of instability and minimises the risks associated with the presence of water. At the same time, other parts of the tunnel should not be located too deep under mountainous areas.

\subsection{Scenario development flexibility}

The initial outline surface and subsurface constraints and many more constraints that are documented in detail in a specific report \cite{gutleber_2025_14773243} provided a starting point for a semi-automated search to determine approximate exclusion and candidate zones.
However, zones marked in orange (undesirable zones) and yellow (acceptable zones), which at first glance appeared compatible with the location of a surface site, proved in various cases to be unacceptable due to restrictions that are not `encoded' in the maps and publicly accessible documents which are available. This additional information was obtained in part by studying high-resolution maps, orthophotographs and data, by interviewing regional and local public administration services and local stakeholders, domain experts and by means of walking tours (visual inspection of the surrounding area).

Such additional constraints included, but were not limited to, the following: unfavourably shaped land areas (surface available too small, too narrow, non-monolithic or too irregularly shaped), presence of adjacent constraints (nature protection areas requiring buffer zones, natural hazard areas such as flood zones or unstable banks, residential areas, sensitive areas), limited access (no road or no possibility of creating one due to topographical conditions or other restrictions, rudimentary road unsuitable for regular traffic and unable to be upgraded to meet requirements), likely opposition or known conflicting development policies, known conflicting projects.

\begin{figure}[!h]
    \centering
    \includegraphics[width=\textwidth]{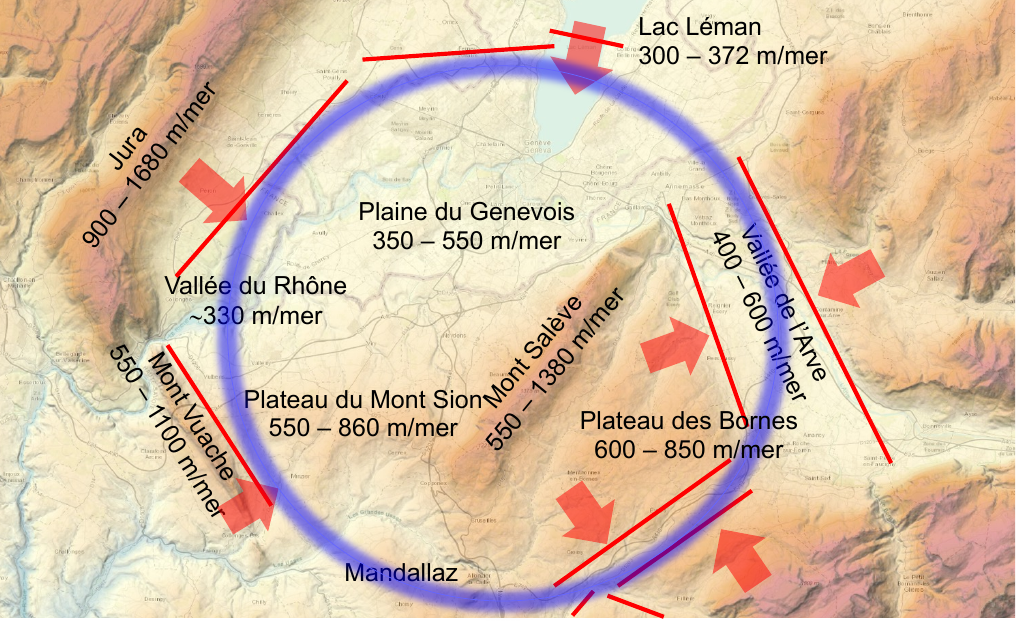}
    \caption{\label{fig:geoconstraints}: Combination of subsurface and topographic constraints that lead to a ca. 300\,m wide scenario development band and a diameter of about 28\,km.}
\end{figure}

The constraints and considering the currently known geological and topographical situation limits the circumference of a future circular collider to less than 100\,km. The layout studies conducted between 2017 and 2021 concluded that to obtain a circumference of over 90\,km that would deliver good performance and therefore yield a research infrastructure of excellence, and given the various constraints that the configuration must simultaneously meet in different locations, the available layout for the ring is limited to a strip with a width progressively reduced from 3000\,m to 300\,m depending on the area concerned (see Fig.~\ref{fig:geoconstraints}).

\section{Initial variants}
\label{imp:Initial_variants}

\subsection{Introduction}

This section provides the background which explains how the implementation scenario in the Franco-Swiss border region was developed. It explains why alternative scenarios to the west of the Jura were examined and then discarded in favour of scenarios to the east of the Jura. It reviews how the layout and placement considered during the exploratory phase of the FCC study (2014-2018) demonstrated the in-principle feasibility of a new, large-scale research infrastructure and how the specific scenario examined presented prohibitive obstacles in certain locations. Finally, it mentions the variant involving a linear collider, which was not selected for further studies.

The development of a scenario for the configuration in the Geneva region was based on the motivation of taking advantage of existing CERN infrastructures and assets built up over 70 years for a new project. The construction of a circular collider requires existing particle accelerators in operation with technical infrastructures that can serve as injectors; the availability of workspace, workshops, equipment, and skilled workforce; as well as favourable legal, administrative, and project management frameworks. It is, therefore, essential that the chosen layout remain within reasonable proximity to an existing CERN site.

In addition, the particle collider must have a circumference greater than 90\,km to accommodate two different particle colliders in the future (an electron-positron collider and a hadron collider), each of which must deliver the performance required to carry out the desired scientific research programme.
The immediate research zone would be located to the south of CERN's Meyrin and Pr\'{e}vessin sites, away from the Jura mountains. Indeed, during the Large Electron Positron collider (LEP) construction phase, major obstacles were already encountered due to the presence of karstic formations and high-pressure water penetrations due to its proximity to the Jura mountains.

Nevertheless, as the next section highlights, an unbiased examination of the situation was conducted so as not to overlook any other possible scenario.

\subsection{Scenarios west of the Jura}

\begin{figure}[!h]
    \centering
    \includegraphics[width=0.8\textwidth]{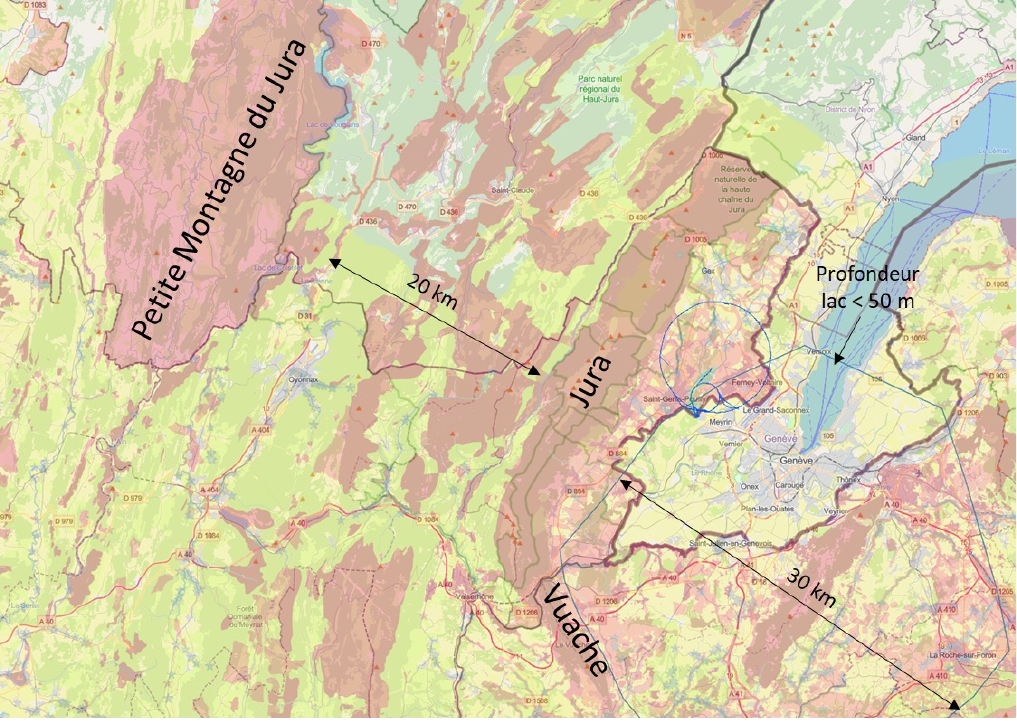}
    \caption{\label{fig:west-Jura}Comparison of land areas and protection zones to the west and east of the Jura chain.}
\end{figure}

A study\cite{tudora_2020_4545604} examined a layout in the Bresse plain (Dolois region), to the west of the Jura Massif (see Fig.~\ref{fig:west-Jura}). These options had already been studied in 1997 and 2001 and were checked again between 2014 and 2019. The ring would have been located outside the Jurassic formations, at a depth of around 40\,metres. Still, in any case, the infrastructure would have encroached on existing national and regional parks and vast nature conservation areas. In addition, the required link to CERN's particle accelerator complex would have required a long tunnel of around 60\,km for the transfer lines. This tunnel through the Jura mountain range would have passed through very unfavourable geological formations with significant overburden. Given that the distance between the Jura mountains and the Petite Montagne du Jura is less than 20\,km, it is considered unfeasible to place a sufficiently large circular particle collider in this zone. 

This variant was definitely dropped from further studies for the following reasons:
\begin{itemize}
\item A long tunnel for the beam transfer line through the Jura mountain range would pose too many difficulties, both from a civil engineering and a technical point of view. Construction times and overall costs would increase considerably, by at least 30\%.
\item This scenario would require deep access shafts in difficult-to-access mountainous regions of the Jura massif, which are largely nature conservation areas. The rock is highly unstable, and the penetration of high-pressure water is certain.
\item The loose soil in the Bresse region would require stabilisation and protection measures. Hydrophilic gypsum swells considerably (ground movements of roughly one metre recorded in the Chienberg road tunnel), which is not compatible with large-scale cavern construction nor with the requirement for a highly stable tunnel. Maintenance costs would be much higher than on stable ground.
\item The installation would most likely interfere with nature protection constraints and cause significant disturbances to local residents during the construction phase.
\item The space available between the Jura massif, its national park and the Petite Montagne du Jura area to the northwest is too small to accommodate a circular particle accelerator of sufficient size.
\item There is no reasonable way of operating and maintaining the infrastructure from CERN or any other suitable partner organisation nearby.
\end{itemize}

\subsection{Racetrack lakeside scenario}

The conceptual feasibility investigations also comprised an analysis of the opportunities and constraints related to a racetrack layout with a lakeside placement of the particle collider ring that permits connecting the infrastructure to CERN's particle accelerator complex\cite{bruening_2022}.

The concept was based on having straight sections about 11\,km long that could potentially also house a linear accelerator or collider. The motivation was to combine the possibilities of a linear electron-positron collider as a first stage with a circular hadron collider during a second stage. An alternative particle acceleration concept with linear accelerators for a first-stage circular electron-positron collider could also have been imagined. 

\begin{figure}[!ht]
  \centering
  \includegraphics[width=.5\linewidth]{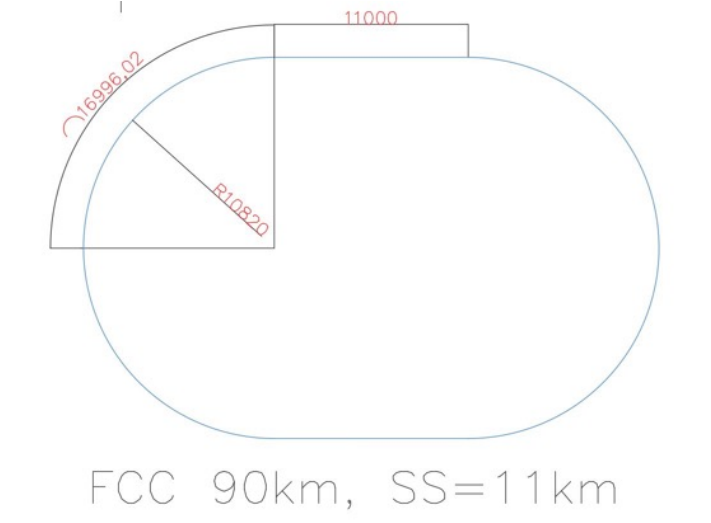}
  \caption{Racetrack layout scenario with two 11\,km long straight sections, leading to a total circumference of 90\,km.}
  \label{fig:racetrack_layout}
\end{figure}

The availability of long straight sections comes with advantages for the radiofrequency acceleration systems in terms of ample space and avoiding bending fields. A minimum crossing angle, 20\,mrad in the case of CLIC, is required, however, in order to prevent collision debris passing through the opposite linear accelerator arm. Realistically, some bended beam delivery section would be required for dispersion suppression, collimation and chromatic corrections. Such a layout would lead to a racetrack with a total circumference of 90\,km, permitting both, housing linear and circular accelerators (Fig.~\ref{fig:racetrack_layout}).

This layout comes, however, with some significant disadvantages: It only permits up to two experiments for the circular collider and also requires the inclusion of collimation, dumps, and beam transfer sections in the two straight sections.

\begin{figure}[!ht]
  \centering
  \includegraphics[width=0.8\linewidth]{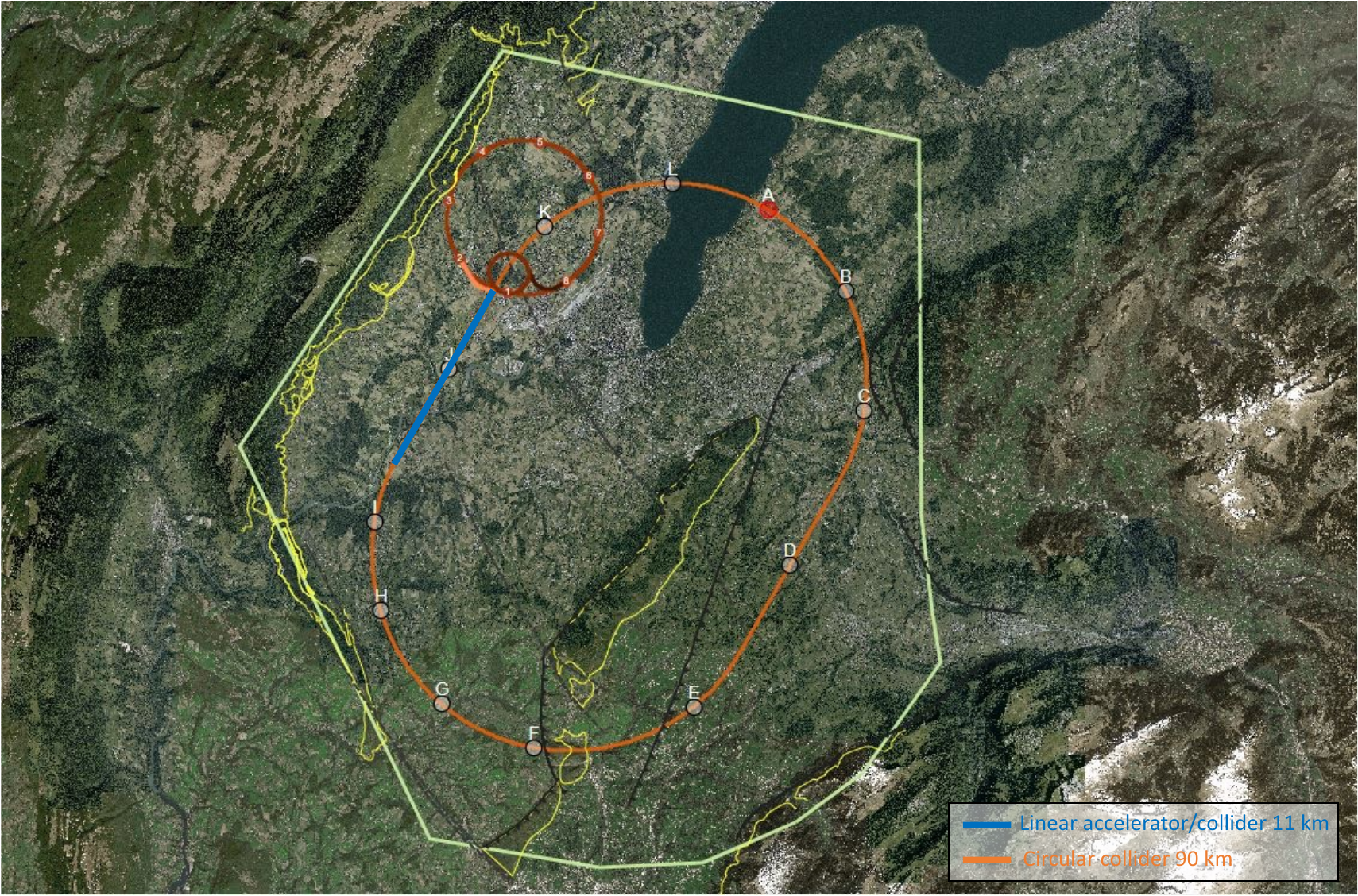}
  \caption{Placement option for a 90\,km circular / 11\,km linear racetrack scenario.}
  \label{fig:racetrack_placement}
\end{figure}

\begin{figure}[!ht]
  \centering
  \includegraphics[width=\linewidth]{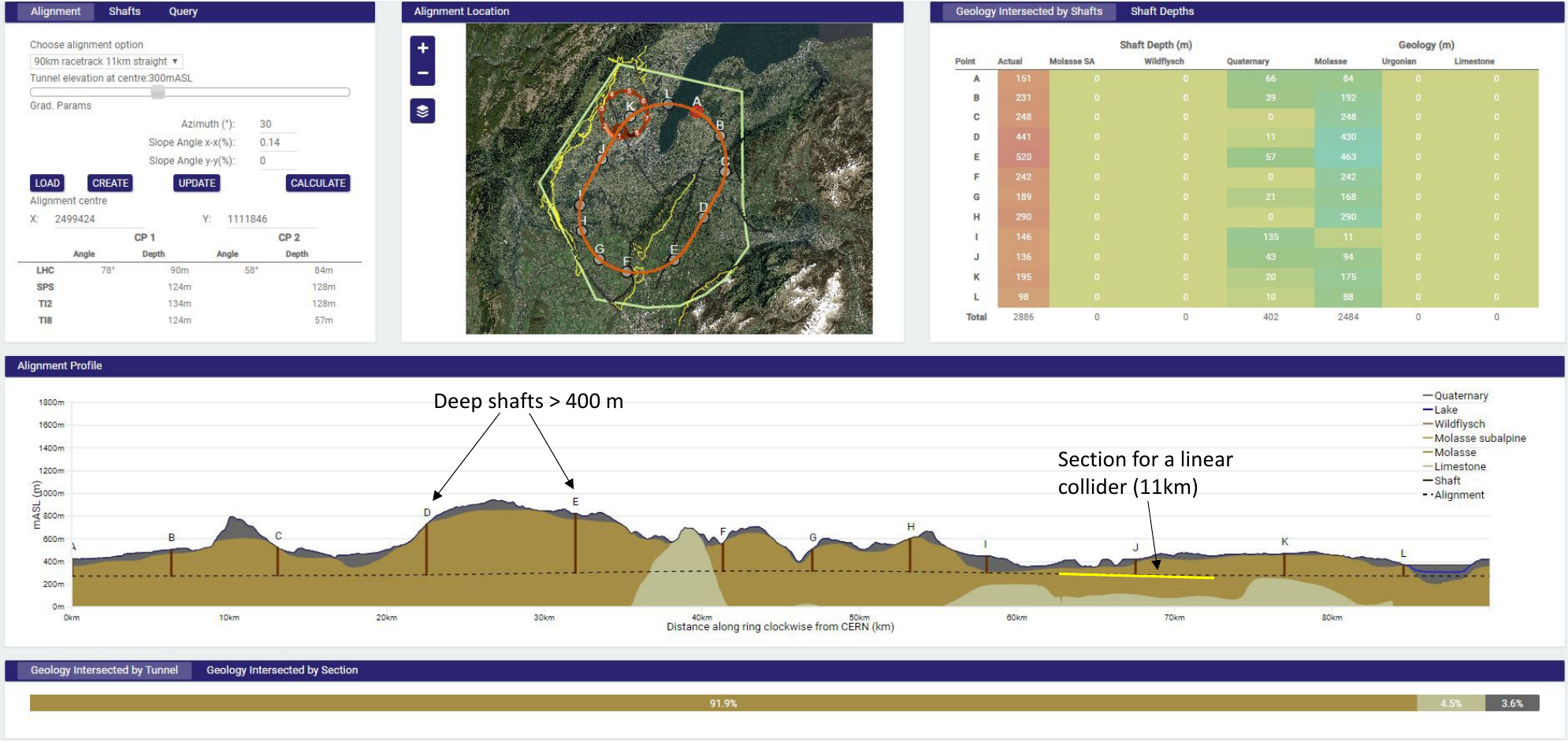}
  \caption{Example racetrack alignment scenario.}
  \label{fig:racetrack_alignment}
\end{figure}

When it comes to placement options (Fig.~\ref{fig:racetrack_placement}) for such a layout, numerous geological and environmental constraints are encountered: as with all circular scenarios, this one also needs to cross the Mandallaz limestone sector. Some shafts could be very deep in their nominal locations (Fig.~\ref{fig:racetrack_alignment}), i.e., more than 500\,m, however, with a systematic optimisation process it is likely that acceptable alternatives for displaced shafts can be found. In terms of traversal of the lake, the bathymetry starts to impose constraints with depths of 65\,m, which would need to be carefully evaluated.

\begin{figure}[!ht]
  \centering
  \includegraphics[width=0.8\linewidth]{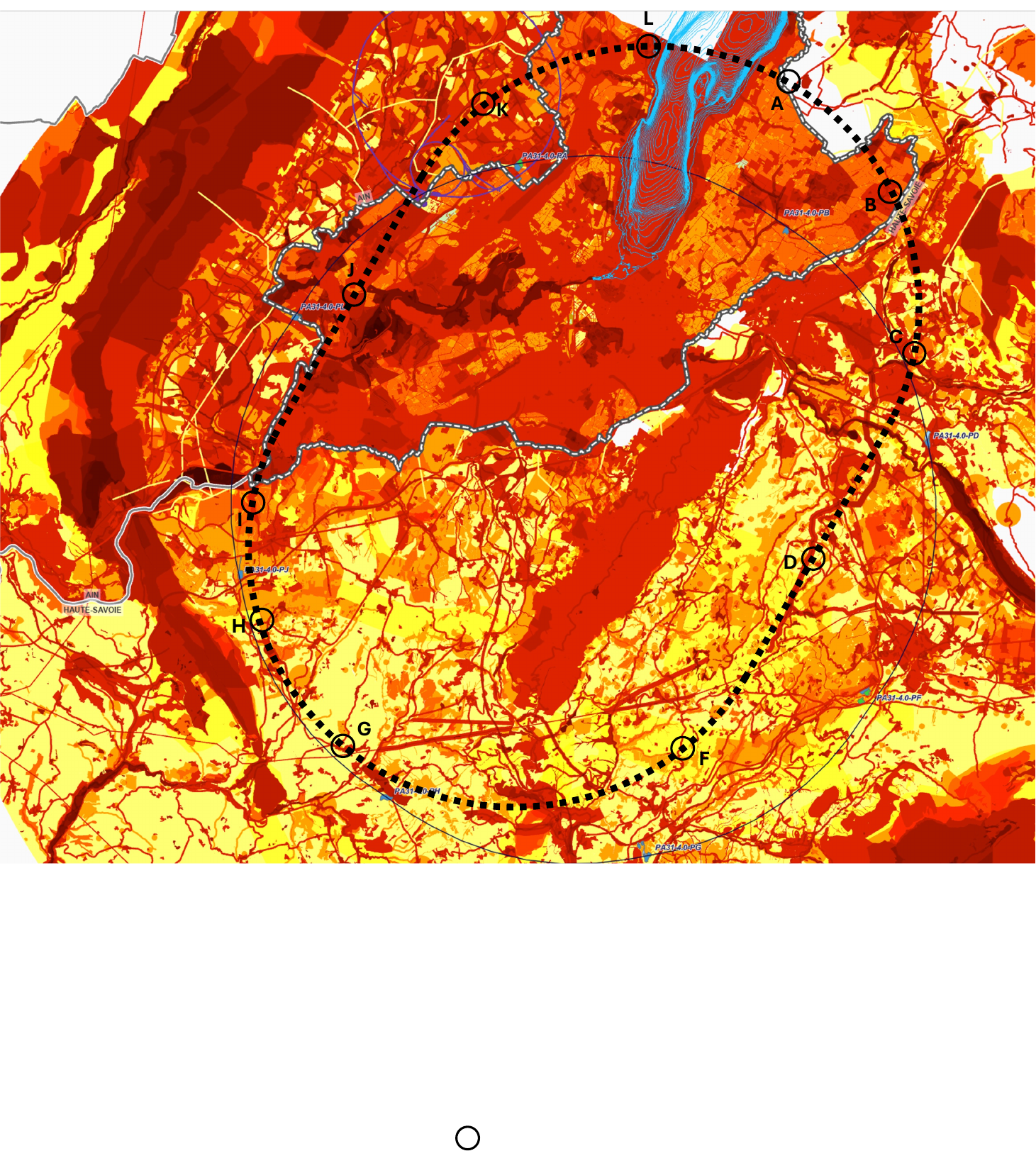}
  \caption{Example of the racetrack layout placement with the environmental constraints indicated. Site locations I, J, L and B are in highly constrained zones, rendering the identification of a feasible and societally acceptable scenario highly challenging. The lake depth at the crossing between L and A is about 65\,metres deep.}
  \label{fig:racetrack_placement_example}
\end{figure}

The strongest limitations stem from the highly challenging needs to integrate 12 surface sites in a highly urbanised area on one side and into a highly mountainous zone on the other side (Fig.~\ref{fig:racetrack_placement_example}). For instance, a site PJ in Switzerland would be in the fully protected Allondon river zone, where any construction activities on the surface and subsurface are strictly forbidden. It remains unclear where and how a feasible experiment site potentially covering up to 9 ha could be identified and rendered acceptable, considering that also major transport and electricity infrastructures would have to be created in this area which, is entirely natural.
It is a highly protected nature zone. Site locations around PI in France in the close vicinity of the river Rh\^{o}ne face similar constraints. Locations for a site PK in the Pr\'{e}vessin sector in France would fall in highly urbanised areas. Site locations in PL in Versoix in Switzerland are in fully protected lake border areas with absolute bans for subsurface activities and in highly urbanised locations. Locations for site PB in Switzerland are in a large nature preservation zone, in particular the home for amphibians which are highly protected. Potential locations could most likely be identified using the Avoid-Reduce-Compensate sequence for other sites.
In terms of cost, the scenario would lead to significant increases for the circular collider scenario, since it is not optimised for a circular collider layout. The differences are in the order of 200 to 400\,million CHF at least. Finally, the loss of at least 20\% of arc sections directly leads to a corresponding loss in the maximum achievable energy or, in the case of a smaller radius in the curved sections, to significantly higher synchrotron radiation losses.

A racetrack layout is neither optimised for a linear collider nor a circular collider. It, therefore, creates additional complications and costs and leads to lower performance. The integration into the territory is extremely challenging and significantly more complicated than the finally adopted circular layout with eight surface sites. It will lead to higher environmental impacts and increased territorial development needs (e.g., access roads) and require more and deeper shafts. Therefore, further studies on this scenario have been discontinued.

\subsection{Scenarios east of the Jura}

For the zone to the east of the Jura, an initial large search perimeter was defined in 2014 based on the following geological conditions\cite{Haas2022,Stanyard:2017wcm,fcc-ee-cdr}:

\begin{itemize}
\item To ensure stability, the tunnel must cross Lake Geneva at a sufficient depth below the lake bed. Initial ideas of placing the tunnel in unstable ground (moraines, quaternary glacial deposits) or in a construction on the lake bed had to be rejected due to a) insufficient stability and b) the impossibility of creating suitable interfaces between the tunnel on either side of the lake and the construction in the lake.
\item In order to be located at an acceptable depth below the lake bed and to remain as far as possible within the impermeable molasse layer, underground structures should be located at around 250\,m above sea level, ideally avoiding crossing the limestone/molasse interfaces.
\item A boundary line was drawn to the northwest (Jura zone) and to the west (Vuache zone) to ensure sufficient distance from high elevations leading to unacceptable overburdens, inaccessible and protected areas, unstable rocks, faults, seismically active zones, and unstable karstic formations where the penetration of high-pressure water is certain.
\item To the southwest, the Mandallaz mountainous zone forms a natural border. Its crossing cannot be avoided. Consequently, the preferred scenarios are those for which the overburden and crossing length are low.
\item To the south, a boundary line has been drawn northwest of the Montagne des Frêtes nature reserve to avoid elevations above 750\,m (unacceptable, as they would lead to excessively deep wells), inaccessible and topographically unacceptable zones, and protection zones to the south. This line crosses the Fillière valley at Thorens-Glières and spans the acceptable elevations of the Bornes plateau as far as La Roche-sur-Foron.
\item To the east, a boundary line has been drawn to avoid mountainous zones and thus remain in the Arve River valley.
\end{itemize}

\begin{figure}[!h]
    \centering
    \includegraphics[width=\textwidth]{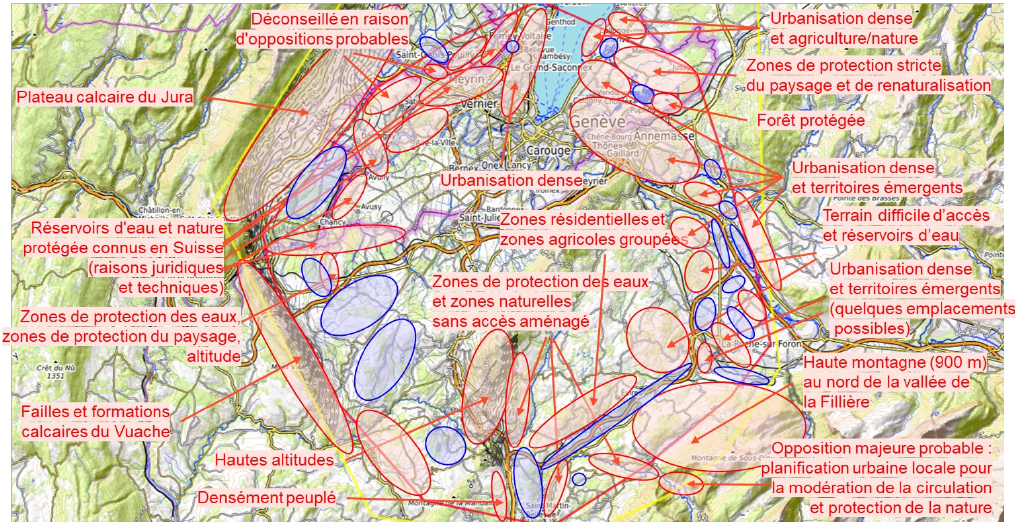}
    \caption{\label{fig:east-Jura}: Established study perimeter (red line). Exclusion zones and zones with too many difficulties for surface sites are indicated in red; favourable zones are indicated in blue.}
\end{figure}

The perimeter obtained in this initial study turned out to be compatible with a circular infrastructure with a circumference of 90 to 100\,km, and can meet geological, topographical, environmental and urban planning constraints (see Fig.~\ref{fig:east-Jura}). The maximum distance that can actually be used is 29\,km from north to south and 30\,km from west to east. The result is a zone capable of accommodating a circular collider with a circumference of up to 92\,km for a configuration comprising eight sites or up to 98\,km for a configuration comprising twelve sites.

\section{Reference scenario}
\label{imp:Reference_scenario}

\subsection{Introduction}

Of the about 100 scenarios analysed\cite{gutleber_2025_14773243}, based mainly on bibliographical data and field visits, it was the PA31 scenario that seemed the most interesting to retain for further in-depth studies and optimisation. The trace of version 1.0 (PA31-1.0) stands out from the others. It proposes a balance between territorial placement, scientific performance that the particle collider and the four interaction regions can offer and technical feasibility in terms of manageable risks and costs. This working hypothesis, developed based on the Avoid-Reduce-Compensate (ERC) process described earlier, has made it possible to gain in-depth knowledge of the region, creating the basis for continuously improving the project considering territorial requirements and constraints as well as technological advances.

\begin{figure}[!ht]
    \centering
    \includegraphics[width=\textwidth]{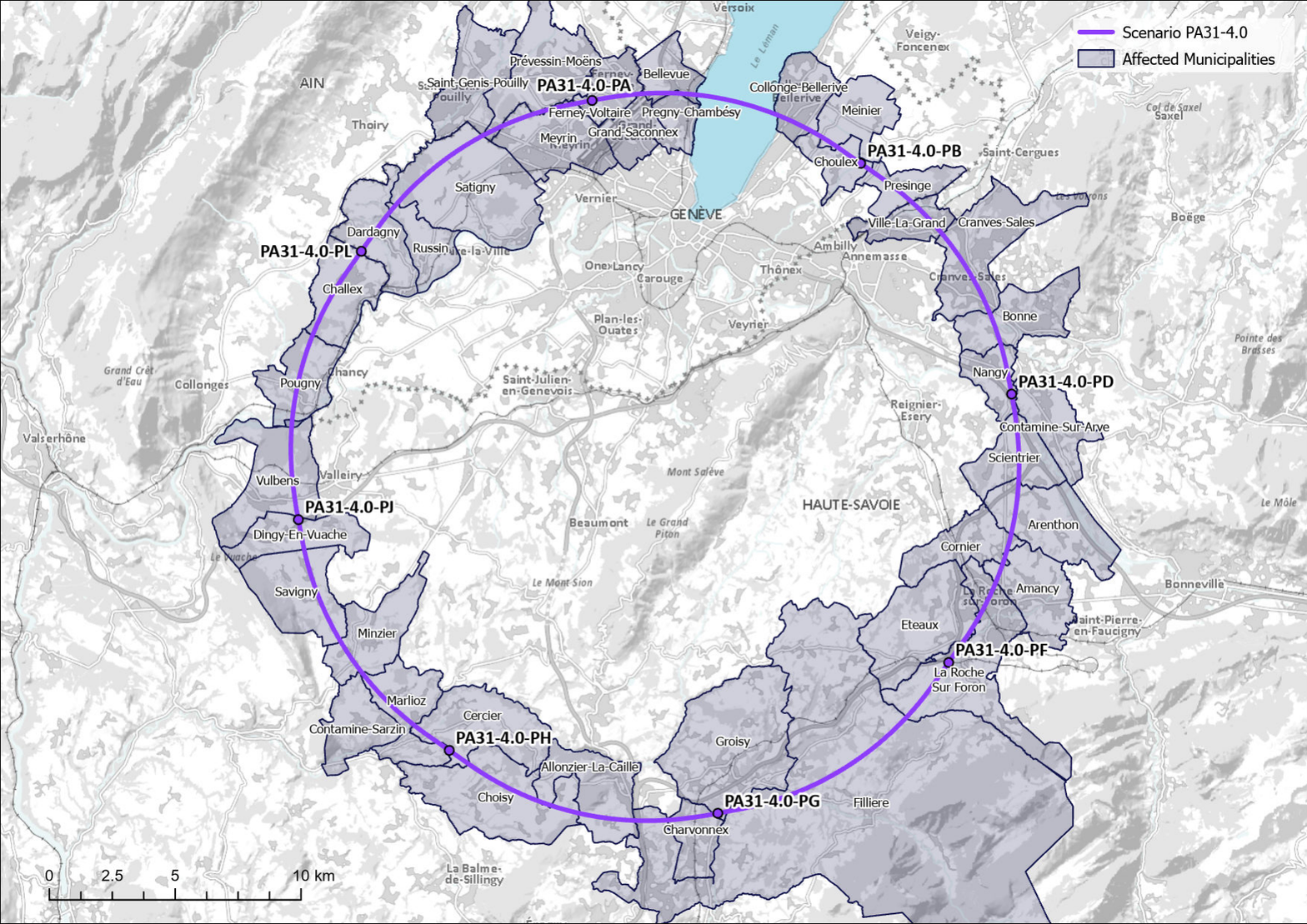}
    \caption{\label{fig:PA31-configuration}The PA31-4.0 reference scenario that served as baseline for the subsurface investigations, the analysis of the state of the environment and studies concerning connected infrastructure projects. An interactive version can be consulted at \url{https://cern.ch/fcc-overview}.}
\end{figure}

The scenario version 4.0 (PA31-4.0) shown in Fig.~\ref{fig:PA31-configuration}, presented in this chapter, is established as the reference scenario for a potential implementation project. It is the result of an optimisation process applied to the PA31-1.0 working hypothesis. It takes into account the progress made on design studies for the \mbox{FCC-ee} and FCC-hh colliders, infrastructure and civil engineering studies, as well as territorial information collected from local stakeholders (concerned municipalities, departmental directorates of the territories and the Direction R\'egionale de l'Environnement, de l'Am\'enagement et du Logement in France and the services of the Canton of Geneva). There still exists room for minor adjustments for some technical surface sites and the disposition of the experiment site shafts.

The technical feasibility of the project based on the so-called PA31-4.0 reference scenario could be confirmed with the help of a diverse set of studies. However, it should be stressed that the territory and legal frameworks in France, Switzerland and Europe are constantly evolving. Since 2014, a number of surface site location candidates and scenario traces have no longer been feasible due to these evolutions. Therefore, the situation may further evolve and the conditions for implementing the reference scenario PA31-4.0 may also change. To ease the project authorisation process and to avoid potential interference with future developments and spatial planning, it is advisable to obtain the rights on the required land plots as soon as possible.

\subsection{Scenario characteristics}

\begin{figure}[!ht]
    \centering
    \includegraphics[width=.8\textwidth]{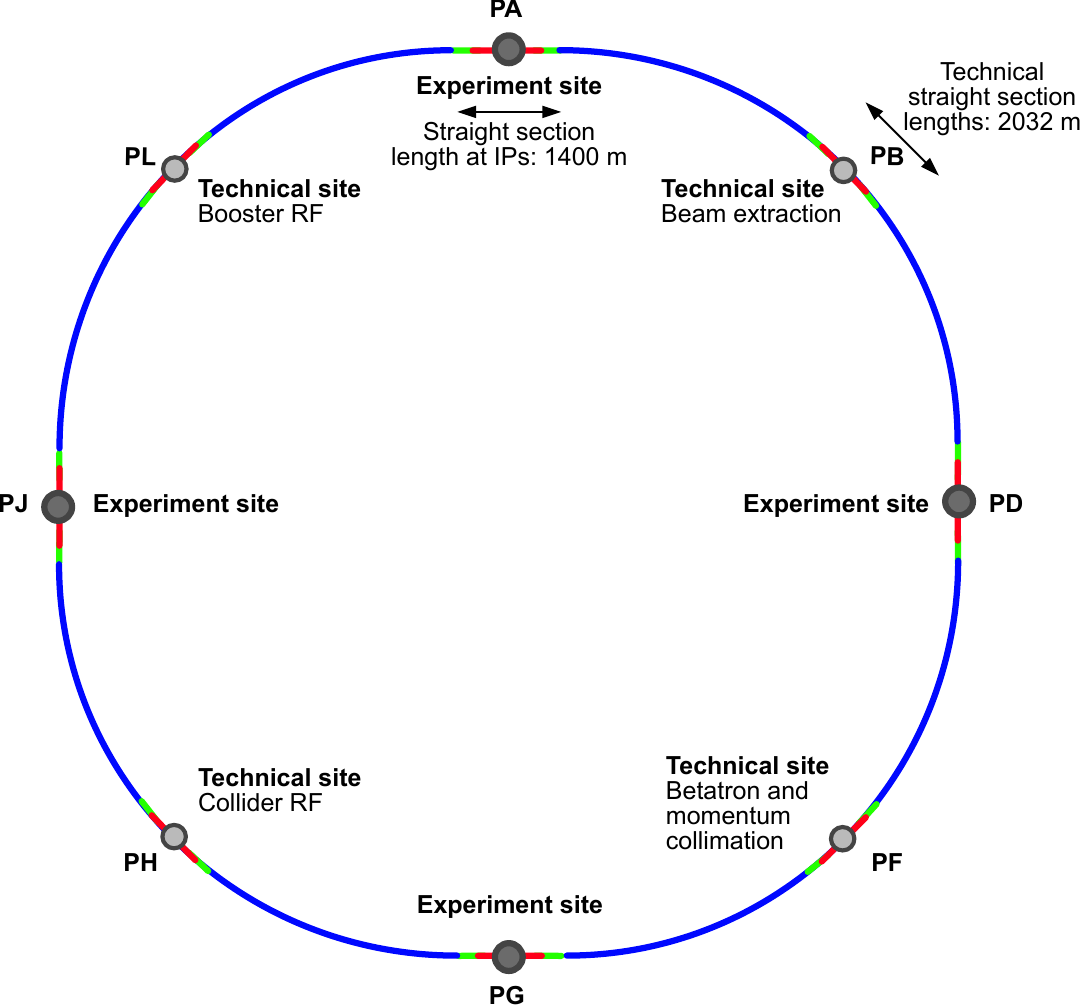}
    \caption{\label{fig:PA31-layout}The functions of the various sites in the PA31-4.0 scenario configuration.}
\end{figure}

The layout and functions of the PA31-4.0 reference scenario are shown in Fig.~\ref{fig:PA31-layout}).
The two shafts at the PA surface site (Ferney-Voltaire, France) were moved further north from the Route de Meyrin road to maintain a safe distance, to provide space for traffic on the surface site around the experiment assembly hall and shafts, and to protect against direct visibility. Moving the locations of sites PD and PJ further northward would be favourable, but is ruled out due to the presence of the departmental road at the northern limit of site PA. Moving them further eastward is also ruled out due to the presence of a compensation zone that is to be avoided and which will be turned into a fully functional wetland zone and natural habitat if the project is implemented. Such displacements would also move the surface site PG in Groisy and Charvonnex deeper into the forest and on a steep slope. Both are evolutions that are better avoided. 
The shafts at the PD surface site (Nangy, France) were rotated away from the autoroute to ensure compatibility with the RD 903 departmental road enlargement and integration with the A40 autoroute development project and to ensure the technical feasibility of creating a service shaft on the site without installing an additional cavern between the service and the experiment caverns. A move of site PA further southward is ruled out, as this would create incompatibility due to the presence of the A40 autoroute and the RD 903 road on the PD surface site, and topographical difficulties (steep slope) towards the D 1203 departmental road to Annecy on the PG site in Charvonnex.

The shifting and rotating of the scenario took into account the objective of placing the PG surface site in both the lower quality part of the forest affected and on the existing grassland plateau, where the soil turned out to be of poor quality for agriculture. In addition, the shafts should be kept away from the steep slope to the south to control construction risks and costs of building the shafts and buildings. It was not possible to move the shafts further away from the forest, towards the plateau, without creating a conflict for the PD experiment site (Nangy) with the autoroute, and for the PJ experiment site (Dingy-en-Vuache and Vulbens), due to the presence of a stream and topographical constraints.
The PJ experiment site is therefore kept at a distance from the A 40 to the south, and from the creek and topographical constraints to the west. The wildlife corridor can be maintained.
The placement of the PB technical site (Presinge, Switzerland) was optimised by taking into account maps concerning biodiversity and ecological indicators made available by cantonal services, and the results of a study of possible access routes carried out by a specialist firm in the canton of Geneva. The site would be located in Presinge, directly along the Route de Jussy road. A connection with the tunnel between the shaft and the service cavern, 98\,m long, is required.

The working hypothesis of the location for the PF technical site is in Éteaux (France), directly adjacent to the RN 203 national road. It includes a connecting tunnel between the 400\,m deep shaft and the service cavern at the collider tunnel. A variant further south is currently not considered since the creation of an inert waste deposit site (called I.S.D.I. in French) creates further constraints. However, in case the footprint of the main site is to be reduced, certain technical infrastructures that are limited to surface site constructions (e.g., ventilators, cooling towers, electrical substation) could be displaced to this area.  
The working hypothesis of the location for the PH technical site is across the border between the two communes Cercier and Marlioz (France), directly adjacent to the D 203 departmental road (Route de Choisy). Taking into account the analysis of fauna, flora and biodiversity, the state of the forest, topographical constraints and technical risks (gas pipeline north of the site), the site closely follows the Route de Choisy road to the north and stretches into the forest down a slope to the west. Should an adjustment become necessary, an alternative site location exists 900\,m in clockwise direction directly at the D2 departmental road on a set of agricultural land plots. The significant displacement from the technical straight section mid-point would require the development of a design for the fitting and installation of the radiofrequency ancillary systems (cryogenic refrigeration, powering).
The preferred location for the PL technical site (Challex, France) is on the nominal midpoint of the technical straight section to the east of the town, on a field and a plot of land containing two single-family homes. Another possible location, 600\,m west of the nominal point, has been studied and discussed with the municipality. It is slightly less preferred due to the higher visibility from residential zones of the commune. This option is also characterised by additional technical challenges that would entail additional costs due to the need to build a shaft approximately 150\,m outside the tunnel with additional civil engineering needs and more difficult access to the machine.

\subsection{Scenario parameters}

The geographical coordinates of the PA31-4.0 reference scenario resulting from this optimisation are shown in Table~\ref{tab:coordinates-IP} for the scientific sites and in Table~\ref{tab:coordinates-tech} for the technical sites.

\begin{table}[h!]
   \caption{Coordinates WGS84 of the theoretical beam interaction points for scientific sites.}
  \label{tab:coordinates-IP}
 \centering
  \begin{tabular}{|c|l|r|r|}
    \hline
    \textbf{Site} & \textbf{Location} & \textbf{Latitude} & \textbf{Longitude} \\ \hline
    \textbf{PA} & Ferney-Voltaire, Ain, France & 46.2480475$^{\circ}$
 N & 6.0986019$^{\circ}$ E \\ \hline
    \textbf{PD} & Nangy, Haute-Savoie, France & 46.1453657$^{\circ}$ N & 6.3169260$^{\circ}$ E \\ \hline
    \textbf{PG} & Charvonnex and Groisy, Haute-Savoie, France & 45.9938019$^{\circ}$ N & 6.1693009$^{\circ}$ E \\ \hline
    \textbf{PJ} & Dingy-en-Vuache and Vulbens, Haute-Savoie, France & 46.0962036$^{\circ}$ N & 5.9513024$^{\circ}$ E \\ \hline
  \end{tabular}
\end{table}

\begin{table}[h!]
  \centering
  \caption{Coordinates WGS84 of the theoretical beam interaction points for technical sites.}
  \label{tab:coordinates-tech}
  \begin{tabular}{|c|l|r|r|}
    \hline
    \textbf{Site} & \textbf{Location} & \textbf{Latitude} & \textbf{Longitude} \\ \hline
    \textbf{PB} & Presinge, Geneva, Switzerland & 46.2271027$^{\circ}$
 N & 6.2374818$^{\circ}$ E \\ \hline
    \textbf{PF} & \'Eteaux, Haute-Savoie, France & 46.0490317$^{\circ}$ N & 6.2865850$^{\circ}$ E \\ \hline
    \textbf{PH} & Cercier and Marlioz, Haute-Savoie, France & 46.0146646$^{\circ}$ N & 6.0309816$^{\circ}$ E \\ \hline
    \textbf{PL} & Challex, Ain, France & 46.1926255$^{\circ}$ N & 5.9810829$^{\circ}$ E \\ \hline
  \end{tabular}
\end{table}

Table~\ref{tab:PA31-parameters} summarises the layout parameters of the particle colliders.

\begin{table}[h!]
	\centering
	\caption{Layout parameters of the particle collider.}
	\label{tab:PA31-parameters}
	\begin{tabular}{p{0.35\textwidth} r p{0.36\textwidth}}
		\toprule
		\textbf{Parameter} & \textbf{Value} & \textbf{Comment} \\ \midrule
		Elevation of the tunnel under the PA site shaft & 202\,m above sea level & The elevation is subject to further optimisation after the availability of the geophysical and geotechnical investigations. \\ \midrule
		Length of straight sections at PA, PD, PG and PJ sites & 1400\,m & These sections feature tunnel enlargements. The ee and hh machines may have beam-optics dependent, different machine straight section lengths in these sectors. \\ \midrule
		Length of straight sections at PB, PF, PH and PL sites & 2032\,m & This is the minimum space required to ensure that all radiofrequency systems can reliably be fit. Subject to optimisation, not all available space may be used after optimisation. \\ \midrule
		East-west rotation of the collider around the PA site & 10.97 degrees & \\ \midrule
		Length of one hadron collider arc cell in the arcs & 275.792\,m & \\ \midrule
		Number of cells in a curved section (arc) per octant & 26 & \\ \midrule
		Total length of the arcs & 78\,684.476\,m & The sum of the curved machine segments of the ee and hh machine may be different \\ \midrule
		Total circumference of footprint & 90\,658.745\,m & The hh machine may be slightly shorter in the same tunnel (90,657.4\, m) \\ \bottomrule
	\end{tabular}
\end{table}
	
The functions of each surface site are shown in Table~\ref{tab:site-functions}.

\begin{table}[h!]
	\centering
	\caption{Layout parameters of the particle collider.}
	\label{tab:site-functions}
	\begin{tabular}{l p{0.7\textwidth}}
		\toprule
		\textbf{Site} & \textbf{Function} \\ \midrule
		\textbf{PA} & Scientific site with one experiment, injecting the beam line from a linear accelerator at the CERN Prevessin site into the pre-accelerator (booster) located in the same tunnel as the collider. \\ \midrule
		\textbf{PB} & Technical site with injection of booster beam line into the collider, extraction of beam line from collider. \\ \midrule
		\textbf{PD} & Scientific site with one experiment. \\  \midrule
		\textbf{PF} & Technical site with betatron and momentum collimation. \\ \midrule
		\textbf{PG} & Scientific site with one experiment. \\  \midrule
		\textbf{PH} & Technical site with radiofrequency particle acceleration system for the collider. \\ \midrule
		\textbf{PJ} & Scientific site with one experiment. \\  \midrule
		\textbf{PL} & Technical site with radiofrequency particle acceleration system for the booster pre-accelerator. \\ \bottomrule
	\end{tabular}
\end{table}

\subsection{Multi-criteria performance of the reference scenario}

\begin{figure}[!h]
    \centering
    \includegraphics[width=0.9\textwidth]{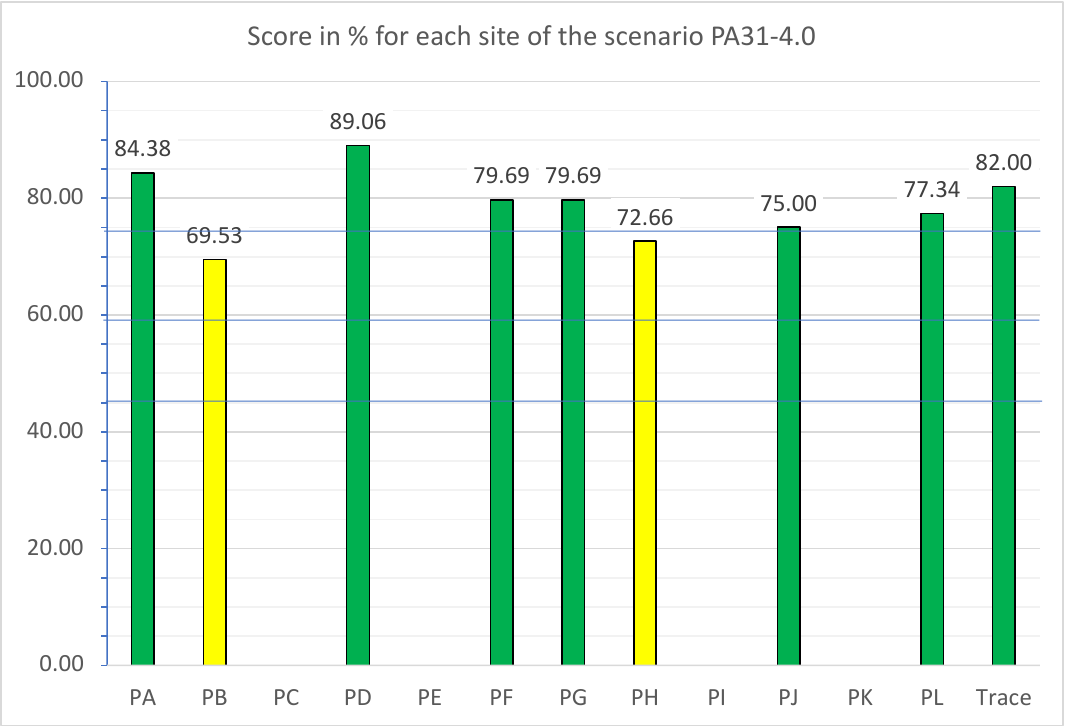}
    \caption{\label{fig:PA31-MCA}Multi-criteria analysis percentage scores of each surface site location and the entire collider trace of scenario PA31-4.0. Note that site locations PC, PE, PI and PK do not exist in 8 site scenarios. Blue lines indicate the individual performance thresholds (green, yellow, orange, red). Green indicated bars exhibit a very good performance and yellow indicated locations correspond to a good performance).}
    \label{fig:PA31-4.0-site-MCA-results}
\end{figure}

Figure~\ref{fig:VIKOR-ranking} shows the performance of a number of shortlisted scenarios that were studied in more detail before the PA31 scenario was chosen as a reference for further optimisation. Fig.~\ref{fig:PA31-4.0-site-MCA-results} shows the results of the multi-criteria analysis for this scenario in terms of territorial compatibility (T), implementation and risk management (I) and scientific value (S) for each site location and the entire trace.

\subsection{Site PA}

\subsubsection{Description of the site location}

\begin{figure}
    \centering
    \includegraphics[width=\textwidth]{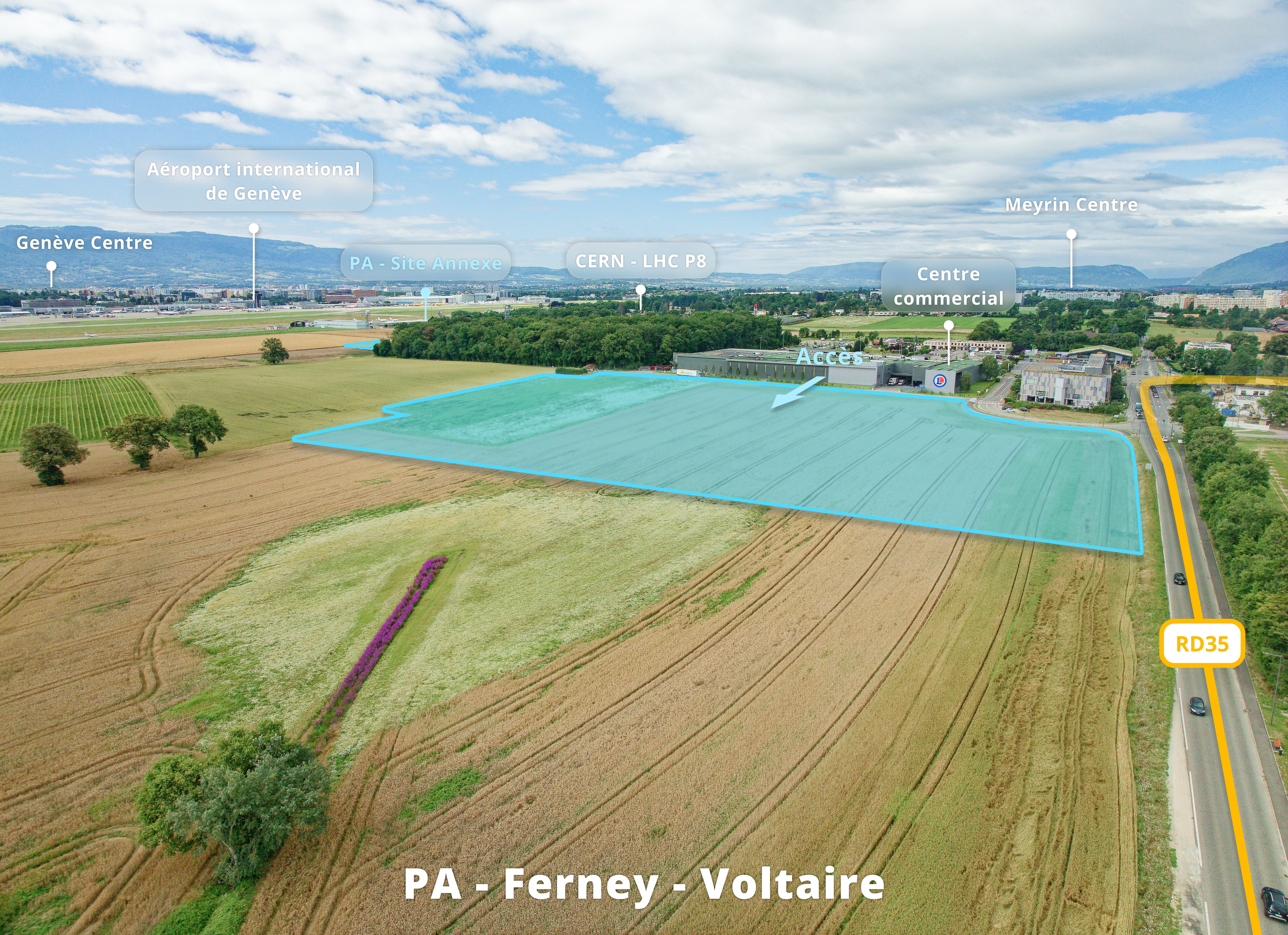}\\[1ex]
    \qrcode[height=1in]{https://fcc-eisa-media.web.cern.ch/drone_videos/PA/FCC-250121700-SEM-PA-Video-V0001.mp4}\\[1ex]
    \caption{\href{https://fcc-eisa-media.web.cern.ch/drone_videos/PA/FCC-250121700-SEM-PA-Video-V0001.mp4}{Aerial view of candidate location for surface site PA.}}
    \label{fig:drone-PA}
\end{figure}

\begin{figure}[!h]
    \centering
    \includegraphics[width=0.8\linewidth]{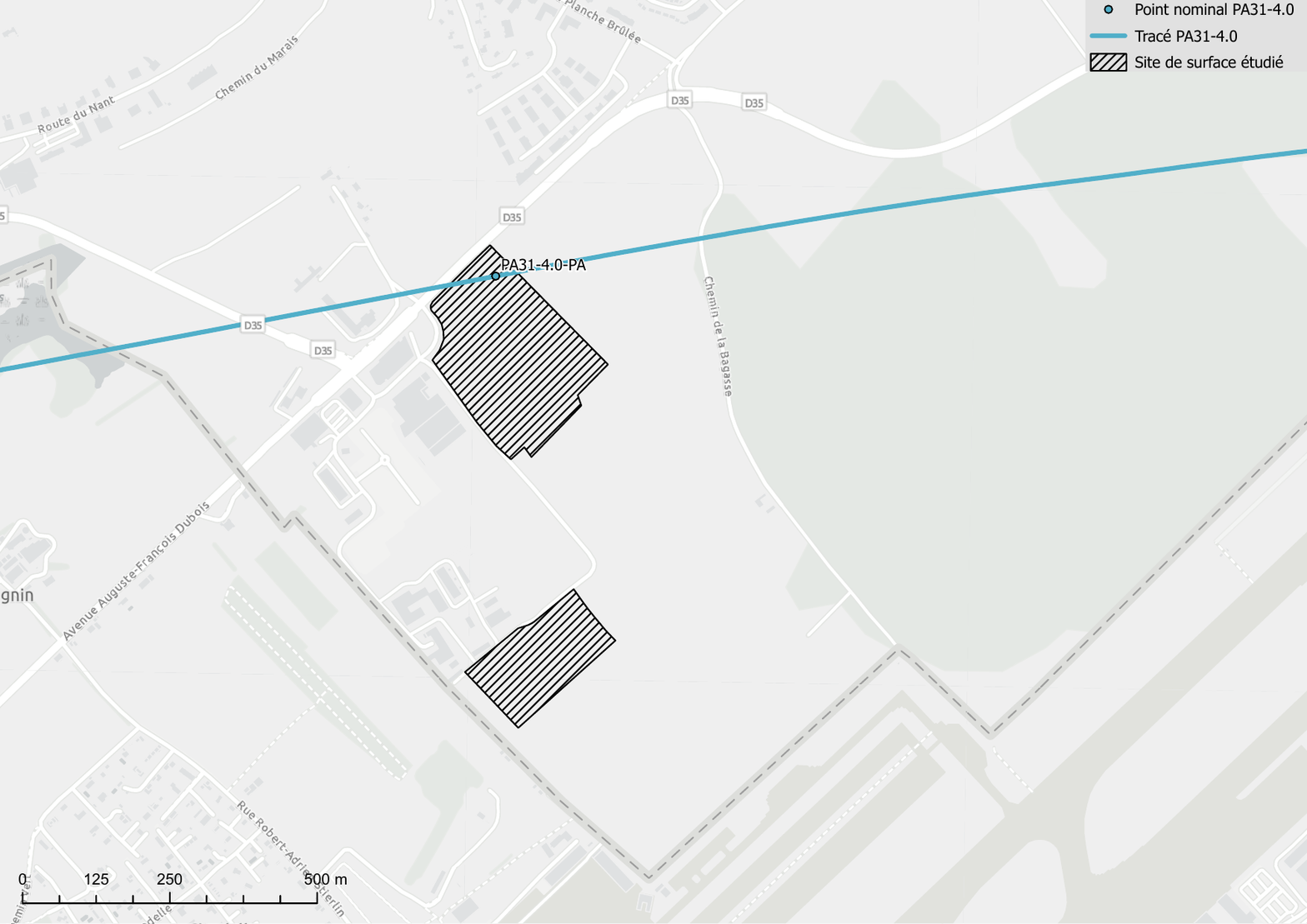}
    \caption{\label{fig:PA31-PA}PA surface site location in Ferney-Voltaire, Ain, France. The hashed space in the south represents an annex that is an extension of the existing LHC P8 site.}
\end{figure}

The main site is located in Ferney-Voltaire, Ain, France, south of the Route de Meyrin road, east of the Espace Candide shopping centre (see Fig.~\ref{fig:drone-PA} and Fig.~\ref{fig:PA31-PA}). The site would be connected to Point 8 of the LHC by the Chemin des Pr\'es Jins, a public road. An extension to the south of LHC Point 8, already planned as part of the HL-LHC project, would make it possible to accommodate technical infrastructures to reduce the land take for the main site. The surface area of the site along the Route de Meyrin is 5.2\,ha, and the surface area of the LHC Point 8 site extension is 2.7\,ha.

\subsubsection{Known constraints}

The site is located in a protected agricultural (Ap) zone, close to an environmental compensation zone classified as protected nature zone (Np). The water supply network passes close by Point 8 of the LHC.
A gas pipeline runs along the boundary of the main surface site, before crossing the border between France and Switzerland.
There is a network along the Route de Meyrin that supplies waste heat from LHC Point 8 to dwellings and to the commercial zone (ZAC, mixed development zone).
Good urban and landscape integration is necessary to preserve the view facing the Alps.
Due to heavy traffic in the area, a concept must be developed together with the municipality's technical services for efficient site access. No dedicated access road is required, though.

\subsubsection{Synergies and territorial potentials}

Waste heat from the particle collider cooling and the data centre can be recovered and supplied to the homes and business parks around the site, including the Geneva airport and industrial zone, building on the current heat supply network that has been put in place for the LHC programme.
The existing but non-functional environmental compensation area can be turned into a fully functional natural habitat and wetland, providing substantially increased environmental quality in that area. Treated residual water from the cooling system can be used for this zone (the Poirier de l'Épine zone) and potentially for agricultural purposes close to the site.
Synergy with LHC Point 8 would permit a substantial reduction of the footprint of the PA site. This requires good planning and sequencing of the new collider project with the HL-LHC programme.
Synergy with the CERN site in Pr\'evessin and the Bois-Tollot substation provides the electricity supply (existing 400 kV line to Pr\'evessin and a dedicated 63 kV link from Pr\'evessin to LHC Point 8).
A visitor centre could be created in synergy with LHC Point 8 to exploit the location fully.

\subsection{Site PB}

\subsubsection{Description of the site location}

\begin{figure}
    \centering
    \includegraphics[width=\textwidth]{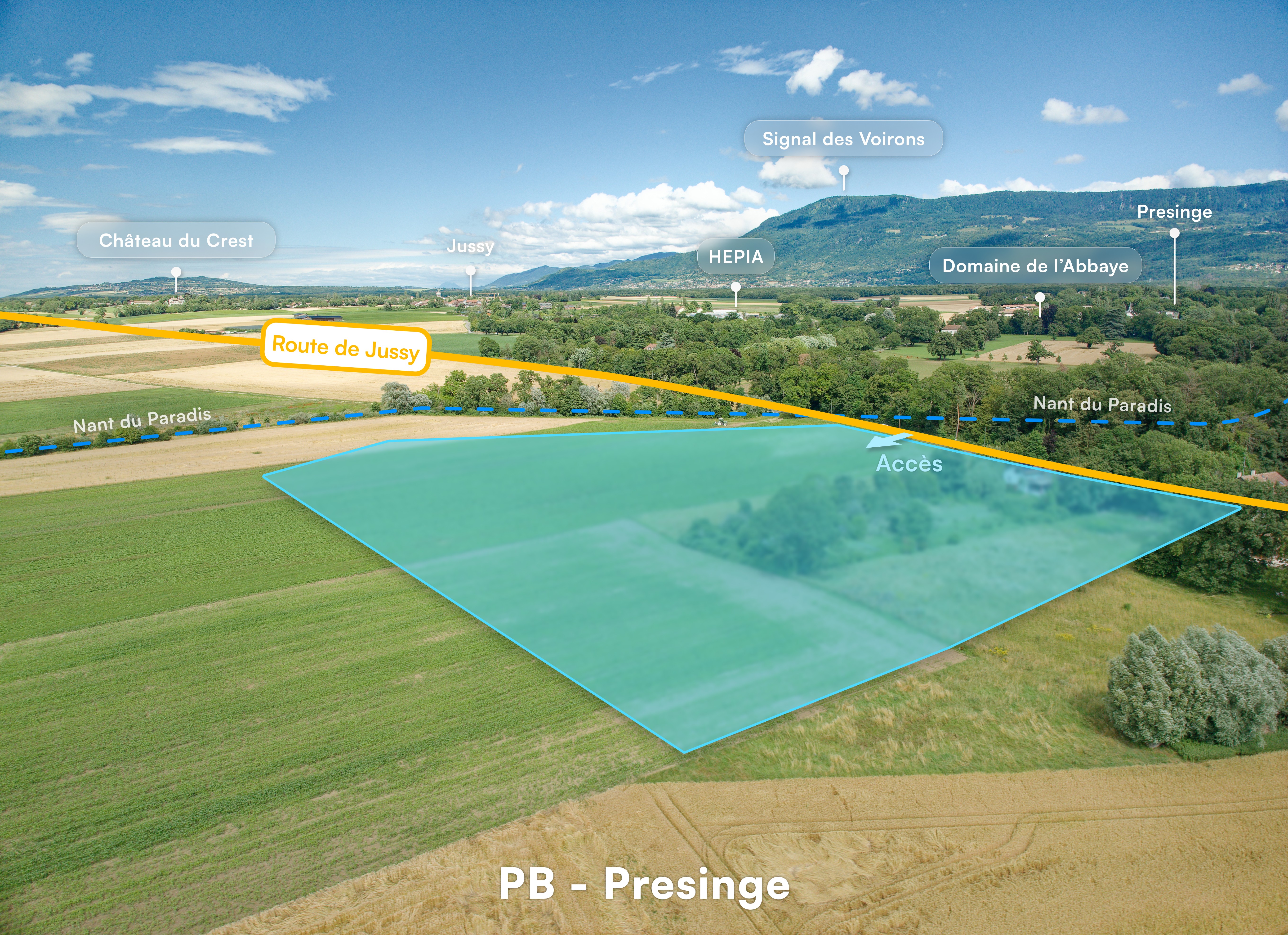}\\[1ex]
    \qrcode[height=1in]{https://fcc-eisa-media.web.cern.ch/drone_videos/PB/FCC-2501301600-SEM-PB-Video-V0001.mp4}\\[1ex]
    \caption{\href{https://fcc-eisa-media.web.cern.ch/drone_videos/PB/FCC-2501301600-SEM-PB-Video-V0001.mp4}{Aerial view of candidate location for surface site PB.}}
    \label{fig:drone-PB}
\end{figure}

\begin{figure}[!h]
    \centering
    \includegraphics[width=0.8\linewidth]{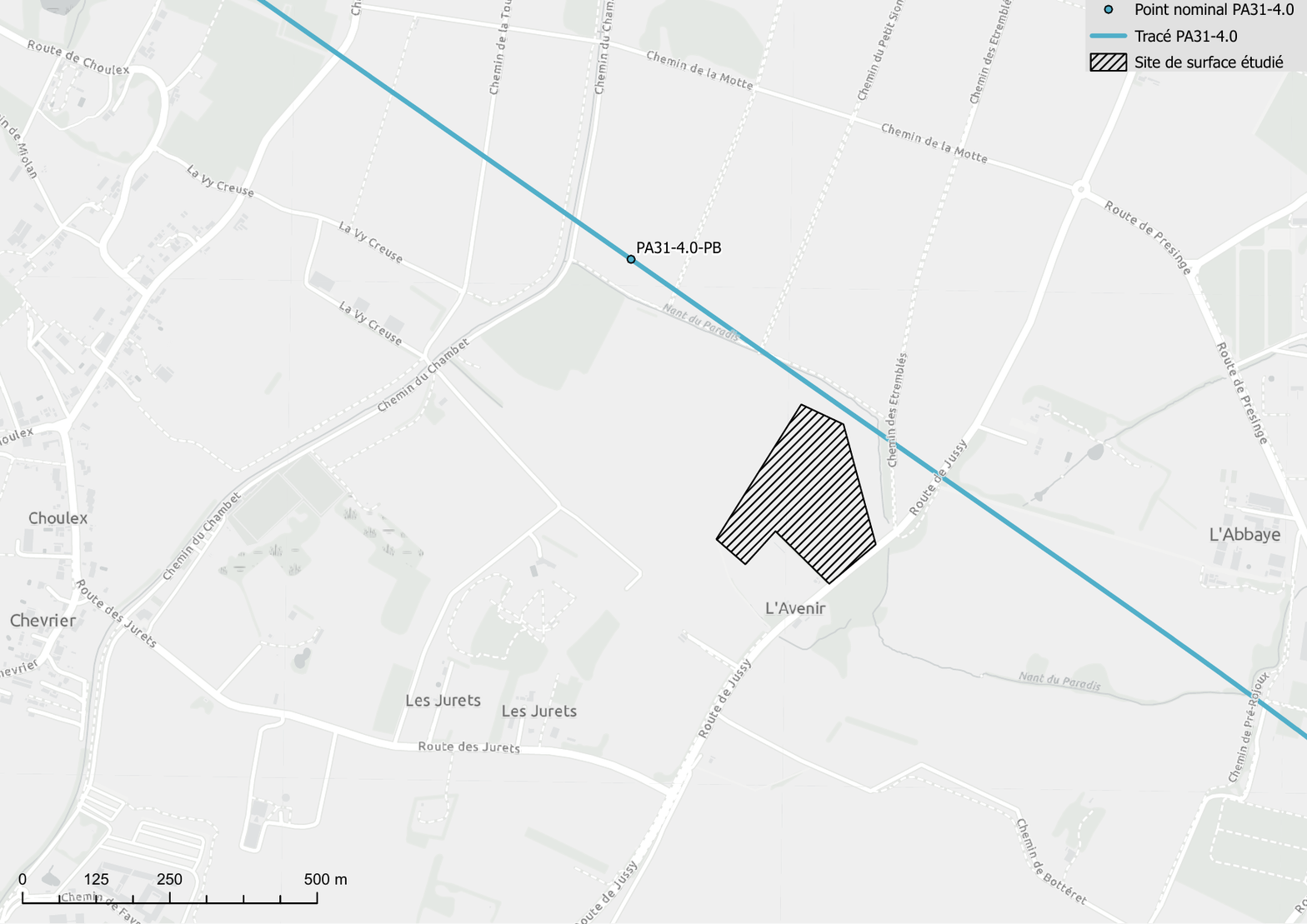}
    \caption{\label{fig:PA31-PB} PB surface site location in Presinge, canton of Geneva, Switzerland.}
\end{figure}

The site lies in Presinge, canton of Geneva, Switzerland, to the south of the Nant du Paradis stream, on a field classified as a protected crop rotation area (in French: `Surface d'Assolement' or in short SDA), bordering the Route de Jussy road  (see Fig.~\ref{fig:drone-PB} and Fig.~\ref{fig:PA31-PB}). The surface area shown on the map indicates the site area covers 4.5\,ha. The site is displaced south to the road, avoiding conflicts with nature protection zones and providing for improved access. An underground connection tunnel of about 100\,m in length is required to connect to the service cavern that is located inside the collider tunnel.

\subsubsection{Known constraints}

The prescribed protective distances between the Nant du Paradis creek that hosts an amphibian breeding area and the forest are respected. The environmental state analysis did not detect any interference with protected environments in the vicinity of the location, in particular the amphibian breeding site.
The impact on the crop rotation area (SDA) should be kept to a minimum. The space consumed by SDA has to be compensated 1:1 by transporting the topsoil to another location that needs to be identified with the help of notified cantonal services.
To blend in well with the surrounding landscape, the site needs to be as small and compact as possible and be ideally semi-underground. A study of the cooling system is necessary to reduce noise and visual disturbances.
It is also necessary to develop a concept for reducing light pollution during the construction phase.
The road access was studied by a specialised firm with in-depth knowledge of the sector. No new road needs to be constructed, but a junction would have to be created at the Route de Jussy. The detailed design and location of the recommended access option have to be validated by the notified bodies of the canton of Geneva once a construction project is proposed. A detailed technical access plan should be drawn up at a later stage for this purpose.

\subsubsection{Synergies and territorial potentials}

According to the road access study carried out by the firm, direct access to the site from the Route de Jussy is technically feasible and preferable from the project perspective. This scenario would avoid the need to create a new access road.
The green buffer zone around the site would reinforce and extend the natural environment in the vicinity of the Nant du Paradis watercourse. Surface sites that are semi-underground would not only improve the blending into the landscape but would also improve the overall ecological and biodiversity value of a non-agricultural habitat.
The reinforcement of the local power grid, necessary for the construction phase, is seen as an opportunity to improve the infrastructure for the inhabitants of the entire area.
Improved public transport, including cross-border transport as part of the FCC, also represents an opportunity for the area.
Several opportunities exist, and demand for supplying heat recovered from the particle accelerator nearby, for example, with HEPIA and a new correctional institution slated for construction by 2030 on the Champ-Dollon prison site.

\subsection{Site PD}

\subsubsection{Description of the site location}

\begin{figure}[!h]
    \centering
    \includegraphics[width=\textwidth]{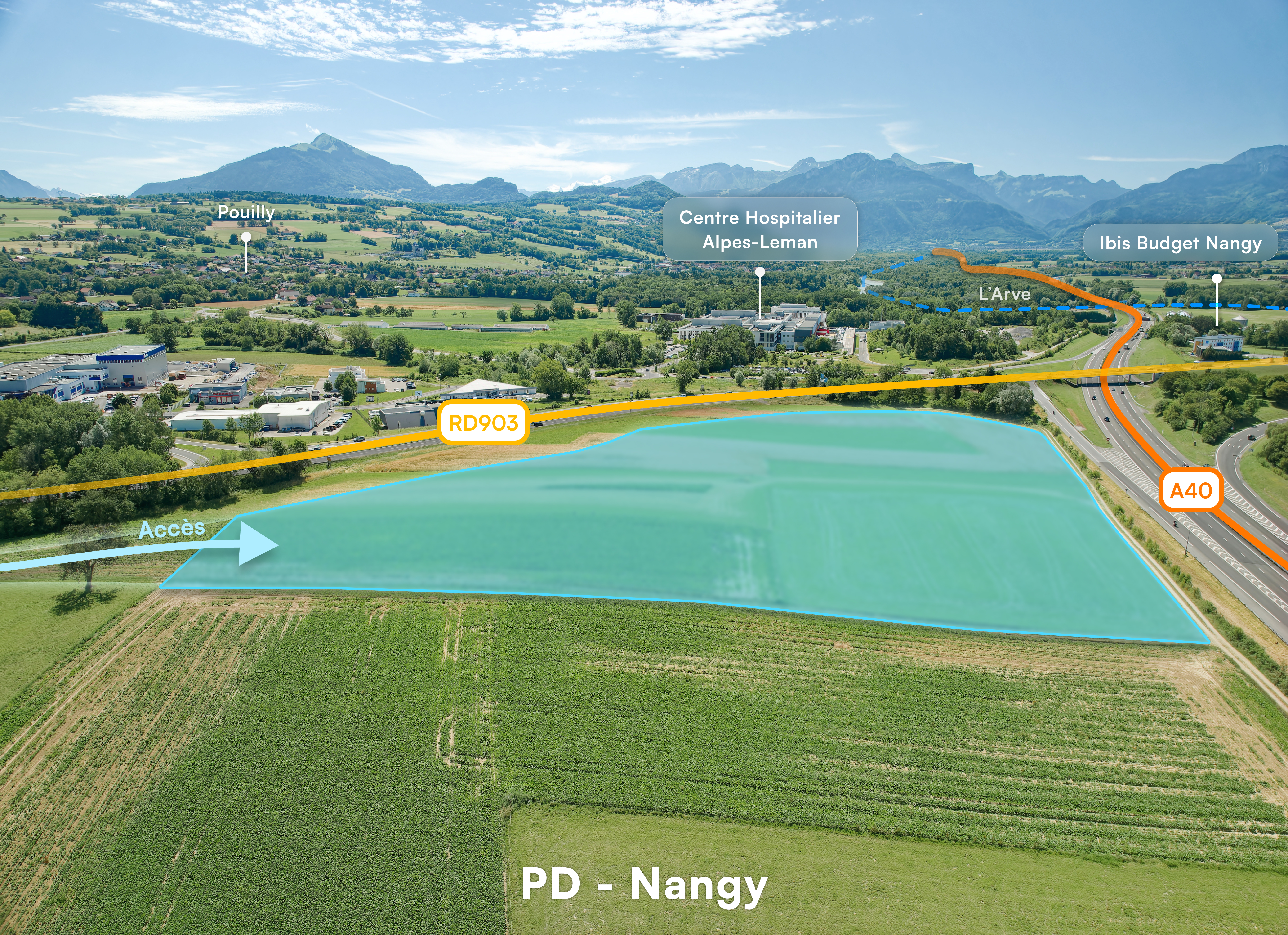}\\[1ex]
    \qrcode[height=1in]{https://fcc-eisa-media.web.cern.ch/drone_videos/PD/FCC-2412181350-SEM-PD-Video-V0002.mp4}\\[1ex]
    \caption{\href{https://fcc-eisa-media.web.cern.ch/drone_videos/PD/FCC-2412181350-SEM-PD-Video-V0002.mp4}{Aerial view of candidate location for surface site PD.}}
    \label{fig:drone-PD}
\end{figure}

\begin{figure}[!ht]
    \centering
    \includegraphics[width=0.8\linewidth]{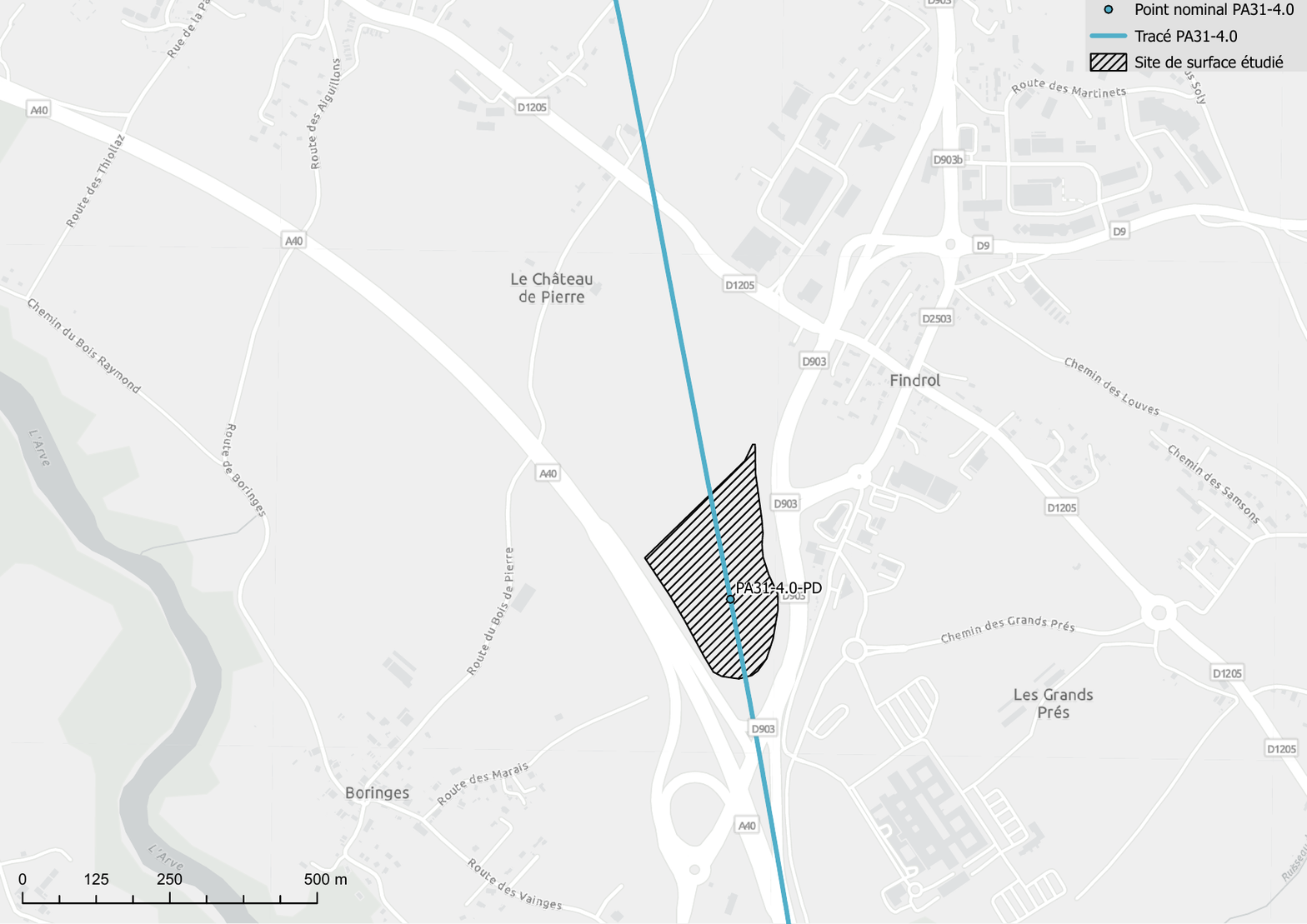}
    \caption{\label{fig:PA31-PD}PD surface site location in Nangy, Haute-Savoie, France.}
\end{figure}

The site is located in Nangy, Haute-Savoie, France (see Fig.~\ref{fig:drone-PD} and Fig.~\ref{fig:PA31-PD}) between the A40 autoroute and the RD903 departmental road on an agricultural field with zone `A' classification in the local urban plan (in French: `PLU'). The surface area of the site is approximately 4.9\,ha.

\subsubsection{Known constraints}

The site is subject to space constraints to the west, east and south due to existing roads and plans to build a new connection between the roads. Road traffic is typically high, one reason for the road development project that is expected to integrate the departmental road with the autoroute. A study for the development of different access scenarios was carried out by a specialised firm. The scenarios were discussed with the Haute-Savoie department and with the mayor of the municipality. The commonly preferred scenario with the lowest impact is one that comprises a dedicated access road to the surface site. The access to the site would be via a roundabout, which would be built to the north at the D 1205 departmental road in Fillinges. 

\subsubsection{Synergies and territorial potentials}

The site is only accessible via the dedicated, approximately 250\,m long access road.
The site is not visible from residential areas, which facilitates its integration into the landscape.
There are a number of possibilities for supplying residual heat from the cooling system, for example, to a nearby cheese factory, to an industrial and business park, to hotels, medical facilities, neighbouring communes such as Fillinges, Boringes, Scientrier, Contamine-sur-Arve and to the Alpes-Léman hospital complex. The heat can also be used for biogas production in the waste water treatment facility in Scientrier.
Currently, a study is ongoing to source water from the waste water treatment facility `SRB' in Scientrier for use in the accelerator cooling system. When the cleaned water is not used for the particle accelerator, it could be made available for agricultural and industrial purposes.
A direct connection by conveyor belt to the autoroute during the construction phase has been studied. It would facilitate the removal and possibly also the supply of materials, thus largely avoiding the need for trucks.
There are likely areas in the vicinity where excavated materials can be reused such as raised hedges, noise separation, covered trench creation, landfill and transport to quarries for rewilding purposes. To be able to develop a solid excavated materials management plan, these opportunities need to be identified with the help of local authorities and notified bodies (e.g., SAFER, DDT, DREAL).

\subsection{Site PF}

\subsubsection{Description of the site location}

\begin{figure}[!h]
    \centering
    \includegraphics[width=\textwidth]{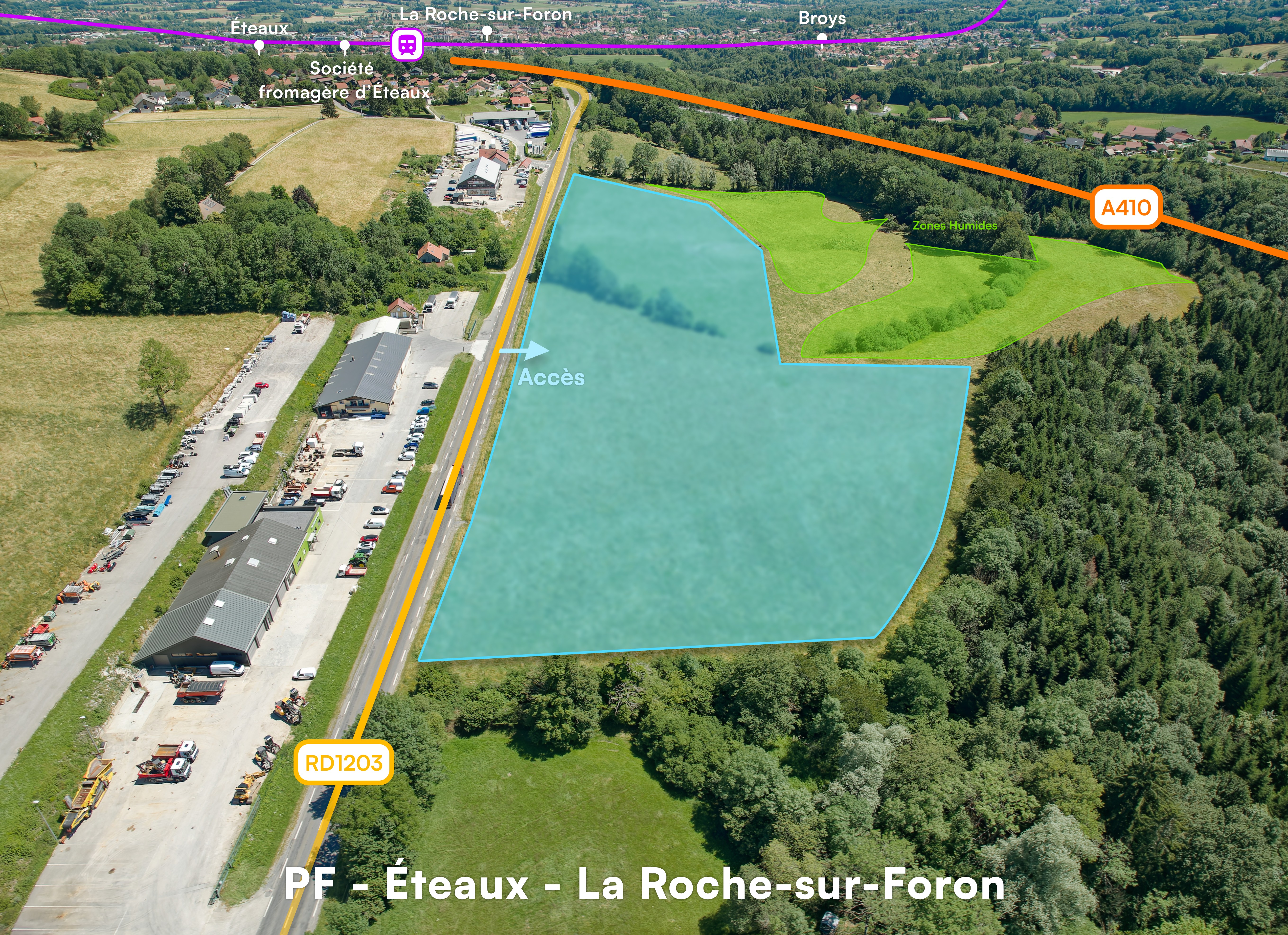}\\[1ex]
    \qrcode[height=1in]{https://fcc-eisa-media.web.cern.ch/drone_videos/PF/FCC-250121700-SEM-PF-Video-V0001.mp4}\\[1ex]
    \caption{\href{https://fcc-eisa-media.web.cern.ch/drone_videos/PF/FCC-250121700-SEM-PF-Video-V0001.mp4}{Aerial view of candidate location for surface site PF.}}
    \label{fig:drone-PF}
\end{figure}

\begin{figure}[!h]
    \centering
    \includegraphics[width=0.8\textwidth]{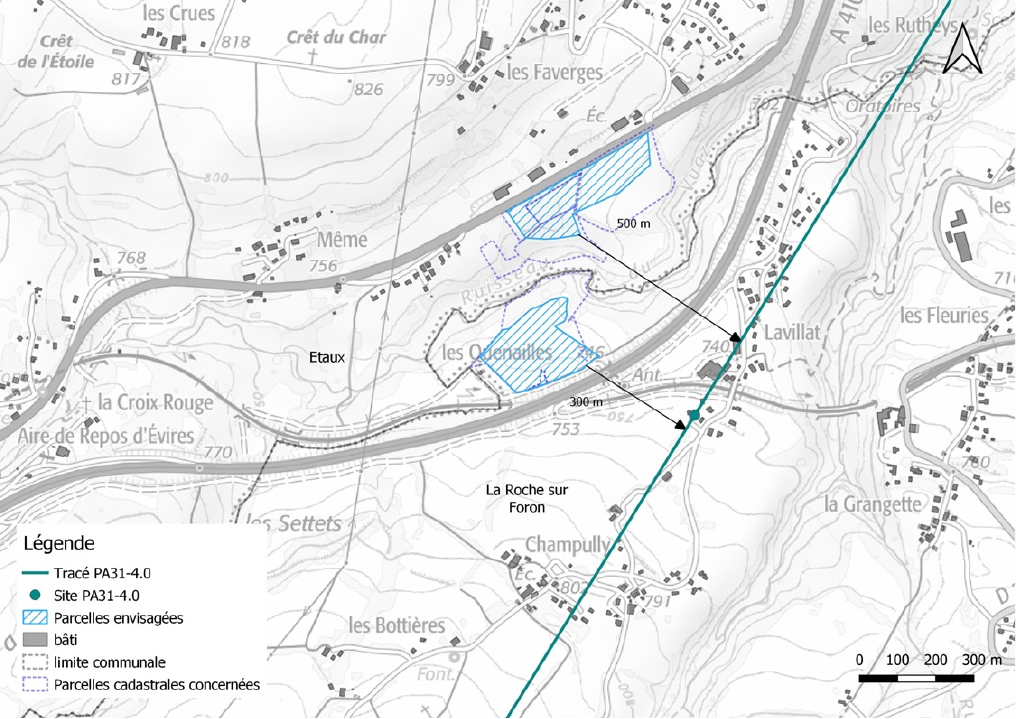}
    \caption{\label{fig:PA31-PF-nominal} Nominal location of PF indicated with a spot on the collider trace. Two surface site candidate locations requiring horizontal access tunnels are indicated with black arrows.}
\end{figure}

A technical surface site at the nominal point PF in La Roche-sur-Foron, Haute-Savoie, France (see Fig.~\ref{fig:drone-PF} and Fig.~\ref{fig:PA31-PF-nominal}) is not feasible because:
\begin{itemize}
    \item The site would be in the middle of a hamlet of single-family homes.
    \item The presence of the Lavillat road in the direction of La Roche-sur-Foron is incompatible with the construction and installation of the accelerator site and equipment.
    \item The nominal point is located on high ground (755\,m) leading to a shaft which is too deep and with an unfavourable topography.
    \item A slight movement of the ring inwards is ruled out due to the presence of steep slopes and the risk of landslides.
\end{itemize}

\begin{figure}[!h]
    \centering
    \includegraphics[width=0.8\linewidth]{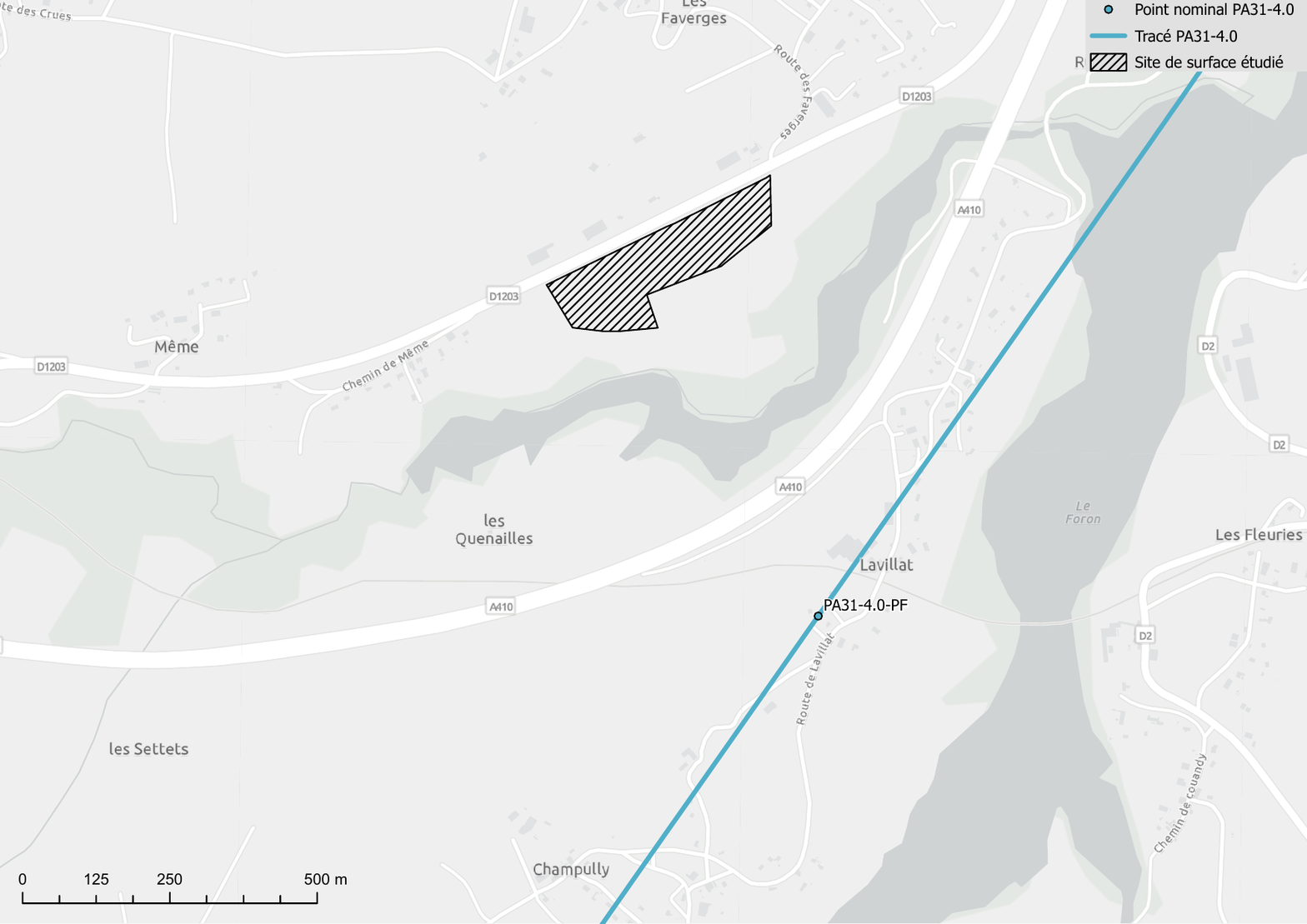}
    \caption{\label{fig:PA31-PF} PF surface site location in \'Eteaux, Haute-Savoie, France.}
\end{figure}

The preferred alternative selected for a technical surface site PF is a location in \'Eteaux, Haute-Savoie France  (see Fig.~\ref{fig:PA31-PF}), alongside the RD 1203 road, with a surface area of 4\,ha. An underground connection tunnel is required between the shaft to be located in the indicated surface site perimeter and the service cavern at the inside of the collider tunnel.

If the surface site area needs to be reduced, made more compact or certain constructions cannot be made compatible with the required landscape integration, then an annex to the south of the site, alongside the autoroute in La Roche-sur-Foron can be envisaged. The land is on the perimeter of a newly constructed inert waste storage facility (in French: `I.S.D.I.'). Once the inert waste storage is full and no longer in use, it can host constructions that are limited to the surface such as ventilation systems, cooling towers and an electrical substation.

\subsubsection{Known constraints}

\paragraph{Main site in \'Eteaux:}
The shaft of the site and the accelerator tunnel are 558.5\,m apart, depending on the position of the shaft.
The wetlands to the east of the site must be maintained and could be integrated into the site.
A certain distance must be maintained between the site and the hamlet to the east.
As a result, there is little space left for the surface site, and the size of the site should be reduced.
Work should be undertaken with the local authority and government agencies (DDT, DREAL) to define zones for recycling excavated materials in the vicinity of the site.
The area available for the surface site is very limited (4 ha). It may, therefore, be necessary to relocate certain elements, such as an electrical substation or cooling towers, to the south option for the site in La Roche-sur-Foron or to another nearby location.

\paragraph{Option of an annex in La Roche-sur-Foron:}
The technical site is located on the site of an inert waste storage facility (in France: `I.S.D.I.'), currently under construction, with the status of a facility classified for environmental protection (in France: `ICPE'), i.e., subject to particular construction, operation and monitoring constraints. Following an analysis carried out by civil engineering contractors, the location was found, in principle, to be compatible with the construction of a surface site. However, this choice would entail significant additional costs and additional technical difficulties. For example, it would be necessary to employ pillars for construction on inert waste dumps to assure the stability of the groundworks and the construction of the shaft. It would be necessary to create an access road 2.4\,km long that crosses the existing railway line, and to build an underpass under the autoroute.
Lastly, this option is located in the immediate vicinity of an environmental protection zone (in French: `ZNIEFF'), which is rich in biodiversity and includes a wildlife corridor that must be preserved.
The available surface area is roughly the same as for the location in \'Eteaux (4.1 ha, versus 4.0 ha). However, if needed, the location could accommodate certain technical infrastructures that are compatible with inert waste deposits, such as an electrical substation or cooling towers. Further analysis will be carried out during the detailed technical design phase.

\subsubsection{Synergies and territorial potentials}

For the location in \'Eteaux, it is possible to supply residual heat from the cooling system to public facilities and businesses in a radius up to 3\,km. A zone that qualifies as wetland, but which today is not well preserved, can be rewilded, protected, and be well preserved together with the habitat in the vicinity of the Vuaz stream. 
Lastly, direct access from the main road is possible and would avoid the creation of a new access road.
Electricity substations for the construction phase exist in the vicinity.
If needed, a connection to the nearby autoroute service station could be implemented to facilitate the transport of excavated materials and the supply of construction materials.

\subsection{Site PG}

\subsubsection{Description of the site location}

\begin{figure}[!h]
    \centering
    \includegraphics[width=\textwidth]{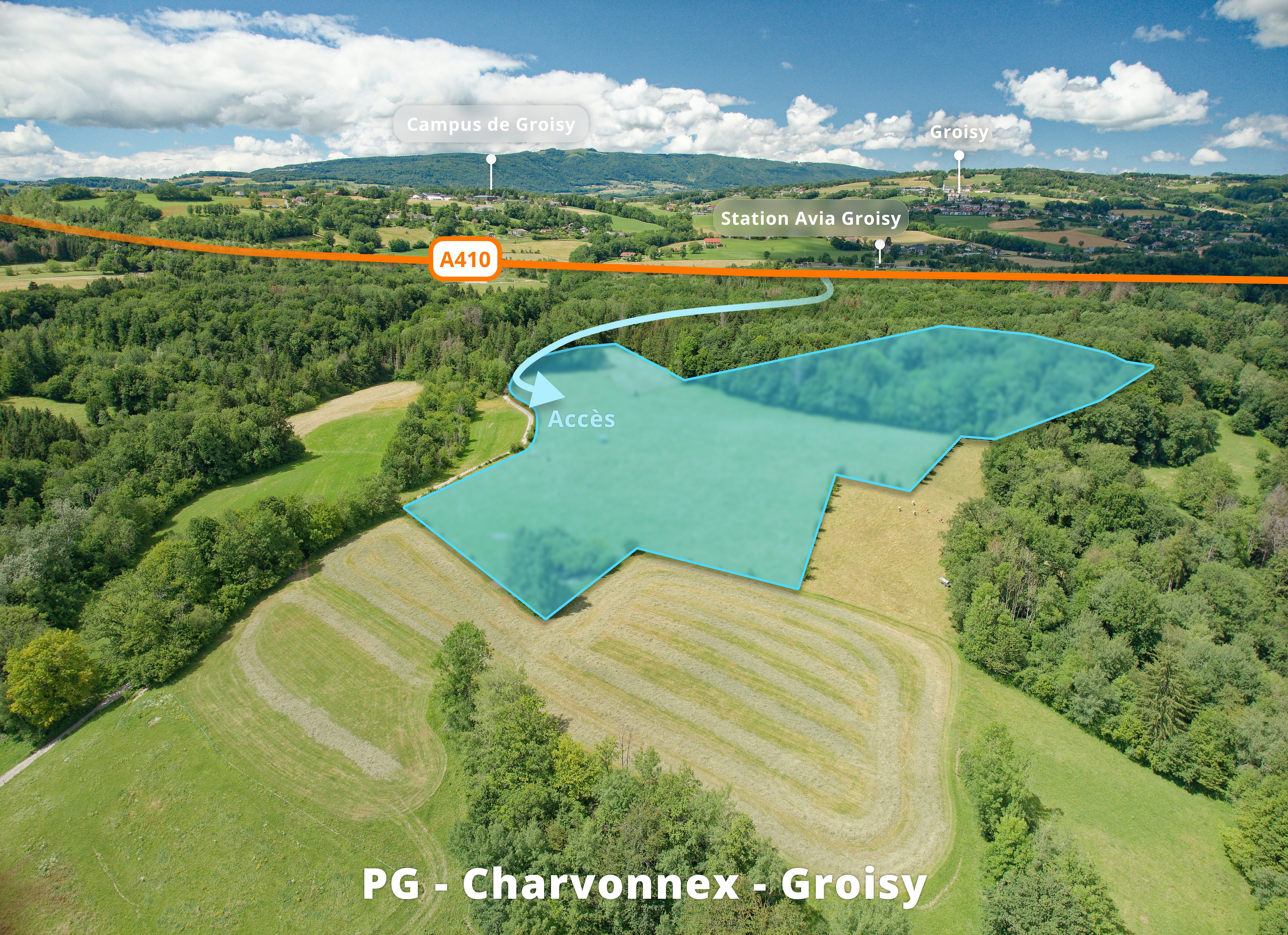}\\[1ex]
    \qrcode[height=1in]{https://fcc-eisa-media.web.cern.ch/drone_videos/PG/FCC-2501301600-SEM-PG-Video-V0001.mp4}\\[1ex]
    \caption{\href{https://fcc-eisa-media.web.cern.ch/drone_videos/PG/FCC-2501301600-SEM-PG-Video-V0001.mp4}{Aerial view of candidate location for surface site PG.}}
    \label{fig:drone-PG}
\end{figure}

\begin{figure}[!h]
    \centering
    \includegraphics[width=0.8\linewidth]{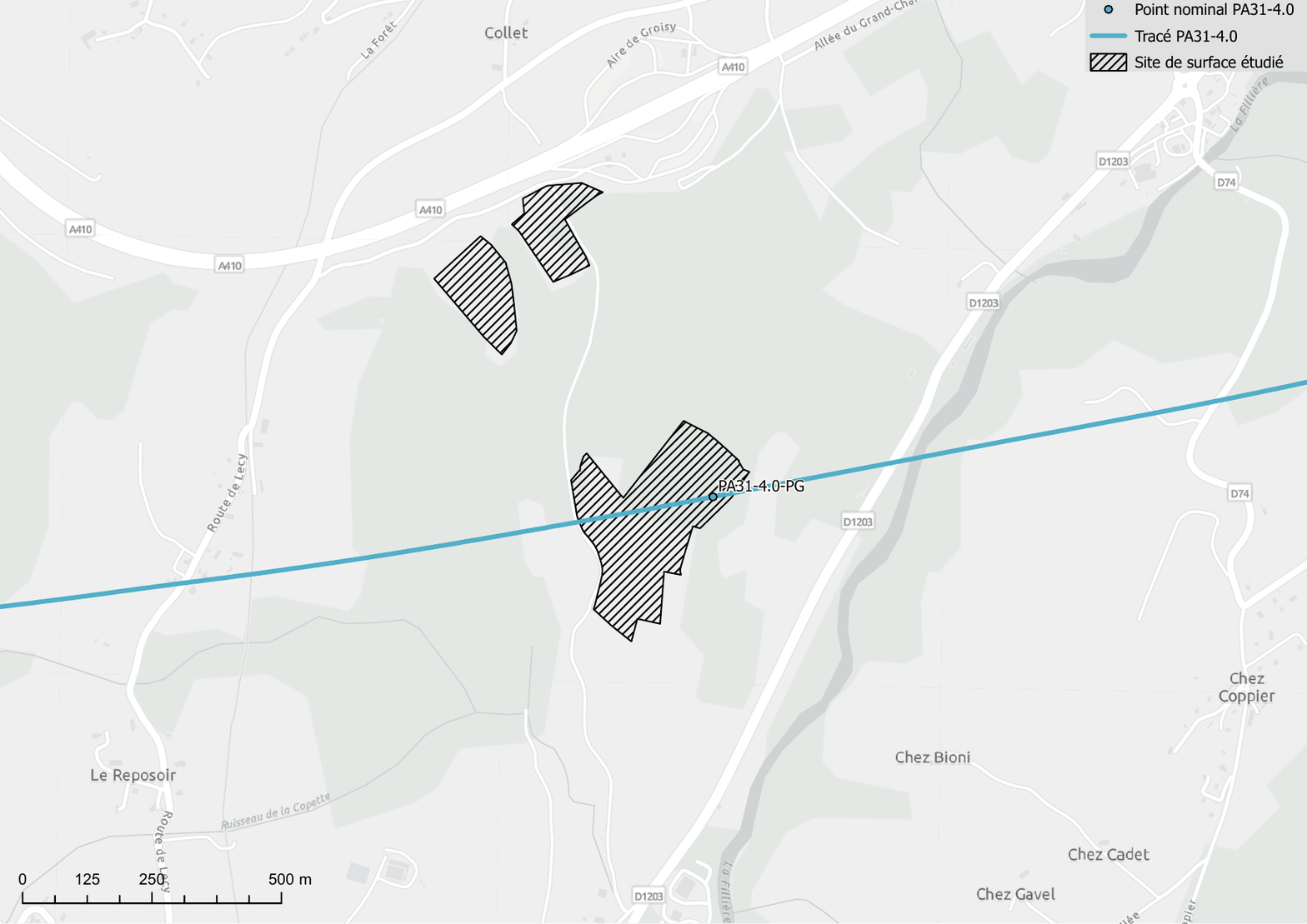}
    \caption{\label{fig:PA31-PG} PG surface site location in Charvonnex and Groisy, Haute-Savoie, France. the main site intersects the collider alignment, shown in blue. The two plots at the autoroute in the north serve for equipment that does not need to be necessarily located in the immediate proximity of the shafts. The areas also serve for handover of incoming construction materials and the evacuation of excavated materials.}
\end{figure}

The experiment site PG lies to the north of the Route d'Annecy road on a plateau, crossing the border of the communes Charvonnex and Groisy, Haute-Savoie in France (see Fig.~\ref{fig:drone-PG} and Fig.~\ref{fig:PA31-PG}). The main site is partly in a forest and partly on grasslands. It is approximately 800\,m south of the A410 autoroute area in Groisy. At this location, next to the autoroute service station, two smaller plots have been identified to host technical infrastructures that do not need to be close to the shaft. All plots are far from any dwellings.
The area marked limits of the main site is 6.9 ha. However, only a fraction of that area would be constructed. Annexes of 1.9\,ha and 1.7\,ha for equipment storage, an electrical substation and cooling towers close to the autoroute are also indicated. An existing 800\,m long, wide forest path has to be refurbished to serve as an access road to the main site.

\subsubsection{Known constraints}

The forest is a classified wooded area (in France: `EBC'), heterogeneous, valued for its biodiversity in some parts. The use of the forest for the site will be reduced and limited to the part with the lowest quality in terms of its natural environment. Clearing will be compensated for by, for example, reforesting the southern part of the plateau or existing brownfields in the forest.
The site is on the edge of a steep slope towards the Route d'Annecy to the south, and must therefore remain on the top of the plateau. It is a quiet, natural area with exceptional views towards the Aravis mountain chain to the south. The location is exceptionally well suited to host a visiting facility with recreational facilities on the site.
Measures will need to be taken to reduce noise and light pollution during the construction and operating phases.

\subsubsection{Synergies and territorial potentials}

The possibility of reforestation compensates for the clearing of approximately 1.5 ha of forest.
A forest road over flat land provides good access and makes it possible to create a route over approximately 800\,m to the Groisy autoroute area and a wide road alongside the autoroute.
Access to the autoroute during the construction phase for supplies and the removal of excavated materials has been studied and appears feasible. The use of a conveyor belt removes the need for trucks.
There are two temporary storage areas for inert waste and other materials in the immediate vicinity of the site and the forest road.
Annexes close to the autoroute could accommodate those elements most likely to generate disturbances, such as cooling towers and an electrical substation.
The supply of residual heat from the cooling system appears to be possible. Several potential consumers have been identified in the vicinity, including public facilities (elementary school, middle school, fire station), a commercial zone and a veterinary clinic. There is also the possibility of creating a district heating network in Charvonnex and Groisy. 
The fire station can serve as a base for emergency services in the immediate vicinity of the site.
Waste water from the cooling system could be used to feed the Fattes stream and wetlands in the forest.
There is enough space on the site to build a visitor centre with recreational facilities that can be reached via the Groisy autoroute area or the Route d'Annecy. The site is located near Annecy, within easy reach for CNRS/LAPP staff.
Sites for reusing excavated materials for agriculture and forestry have yet to be identified with the help of local authorities and notified government agencies (e.g., SAFER, DDT, DREAL).

\subsection{Site PH}

\subsubsection{Description of the site location}

\begin{figure}[!h]
    \centering
    \includegraphics[width=\textwidth]{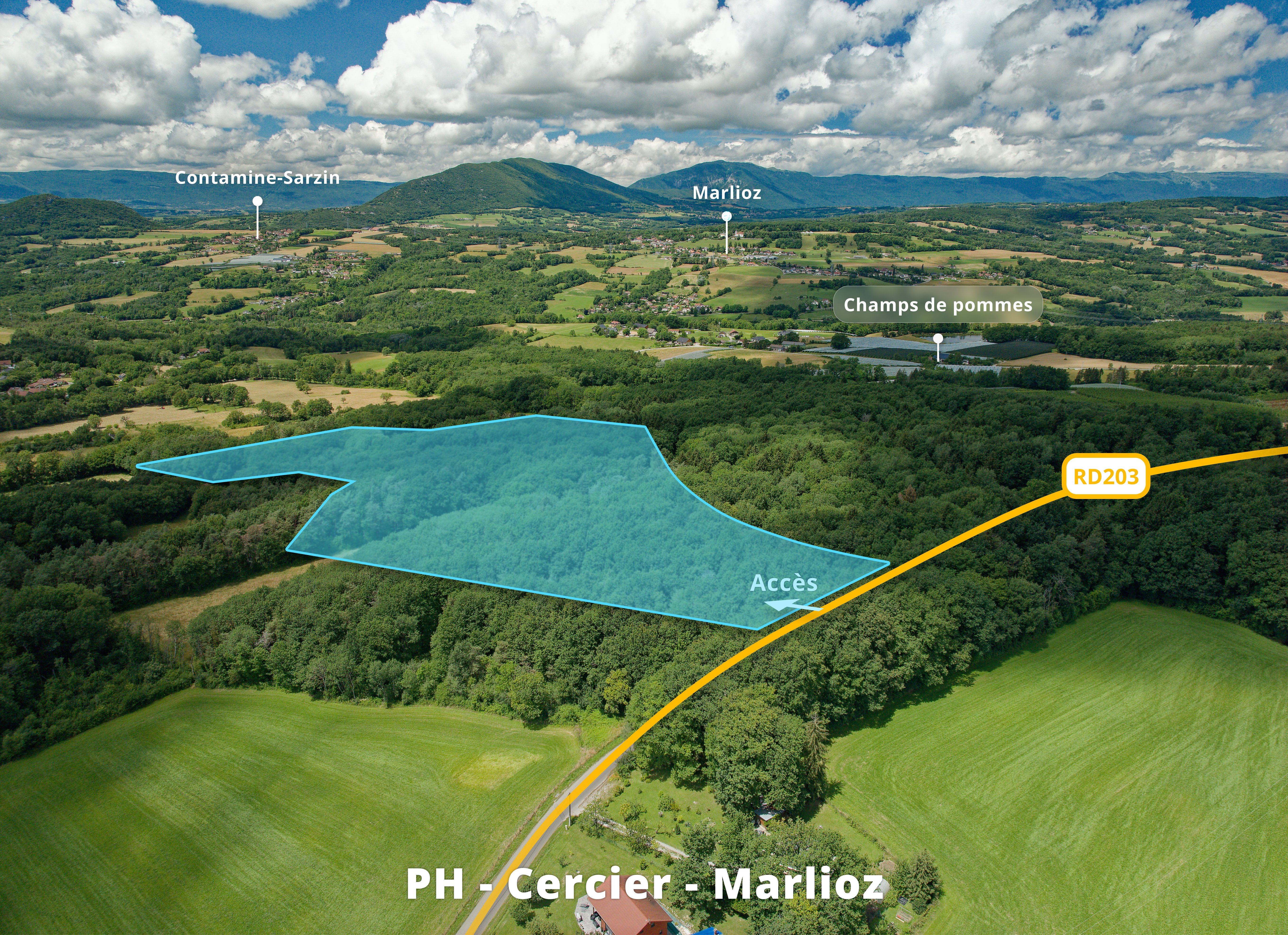}\\[1ex]
    \qrcode[height=1in]{https://fcc-eisa-media.web.cern.ch/drone_videos/PH/FCC-2501301600-SEM-PH-Video-V0001.mp4}\\[1ex]
    \caption{\href{https://fcc-eisa-media.web.cern.ch/drone_videos/PH/FCC-2501301600-SEM-PH-Video-V0001.mp4}{Aerial view of candidate location for surface site PH.}}
    \label{fig:drone-PH}
\end{figure}

\begin{figure}[!h]
    \centering
    \includegraphics[width=0.8\linewidth]{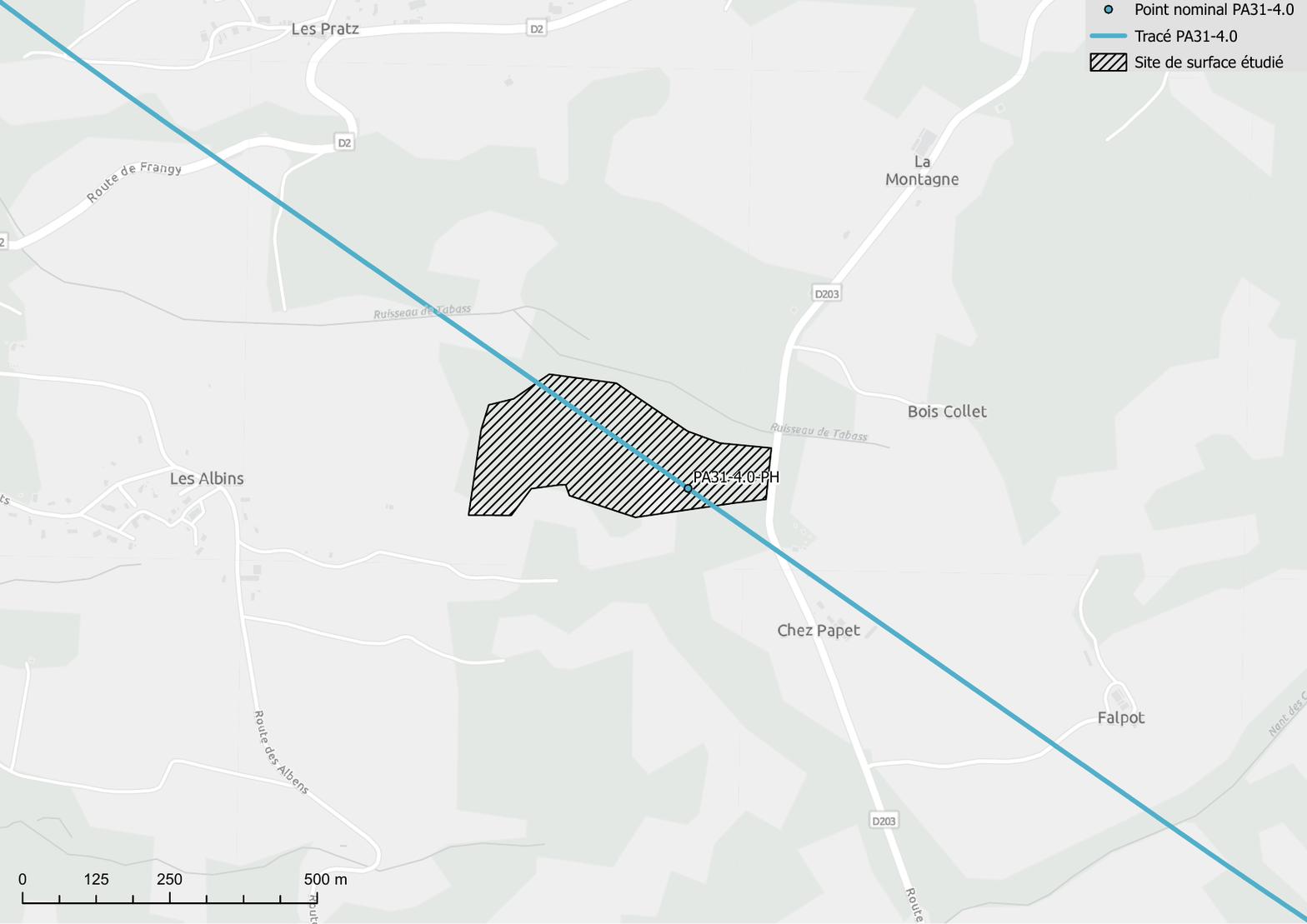}
    \caption{\label{fig:PA31-PH} PH surface site location in Cercier and Marlioz, Haute-Savoie, France.}
\end{figure}

The technical site PH is located right along the D203 road in Cercier, Haute-Savoie in France (see Fig.~\ref{fig:drone-PH} and Fig.~\ref{fig:PA31-PH}). The site is located in the forest, straddling the municipalities of Cercier and Marlioz, stretching out to the west down a slope. It is far from dwellings and is not visible. It is in a nature setting. The defined area, located in an area with less biodiversity, is 8.2 ha.

\subsubsection{Known constraints}

The environmental field investigations revealed that the part of the forested sector to the north of the site is of great value (rich biodiversity, natural environment). Therefore, use of the forest needs to be minimised and the specific locations of buildings and accesses in the forest need to be carefully developed during a subsequent design phase. The site will respect the buffer zone for the gas pipeline to the north.
Considering the quiet, natural setting, noise and light pollution must be given particular attention. A dwelling is located approximately 200\,m to the south. There is no co-visibility between the site and the dwelling.
Given the small size of the road, it is necessary to identify nearby reuse sites for excavated materials, so that they can be used in agriculture or for reforestation.
Materials would mainly be transported away to the north or the west by conveyor belt to avoid road constraints.
The location is well suited to host the radiofrequency system since a 400 kV grid line passes less than 2\,km to the north. RTE, the national grid operator will have to take into account the need to supply electricity via buried cables when carrying out the study for the creation of a substation for access to the 400\,kV line.

Should for any reason the site preferred location at the nominal point not be considered feasible, an alternative placement about 900\,m in the clockwise direction of the collider has been identified as an alternative. This site is located on agricultural fields next to the D2 departmental road. It would be suitable as the location for the electrical substation in case of placement on the nominal point or when displaced since the electrical substation does not necessarily need to be in the immediate vicinity of the shaft.

\subsubsection{Synergies and territorial potentials}

There is a great deal of interest in reusing water from the particle collider infrastructures for agricultural activities around the site (e.g., apple and pear trees). A water basin exists nearby, to the northeast. In case further, more detailed scenario developments are considered, a study will have to be carried out to find a way of integrating those opportunities with the site.
It is advisable to work with local stakeholders to identify potential consumers of the residual heat, for instance, in the fruit and vegetable sector.

\subsection{Site PJ}

\subsubsection{Description of the site location}

\begin{figure}[!h]
    \centering
    \includegraphics[width=\textwidth]{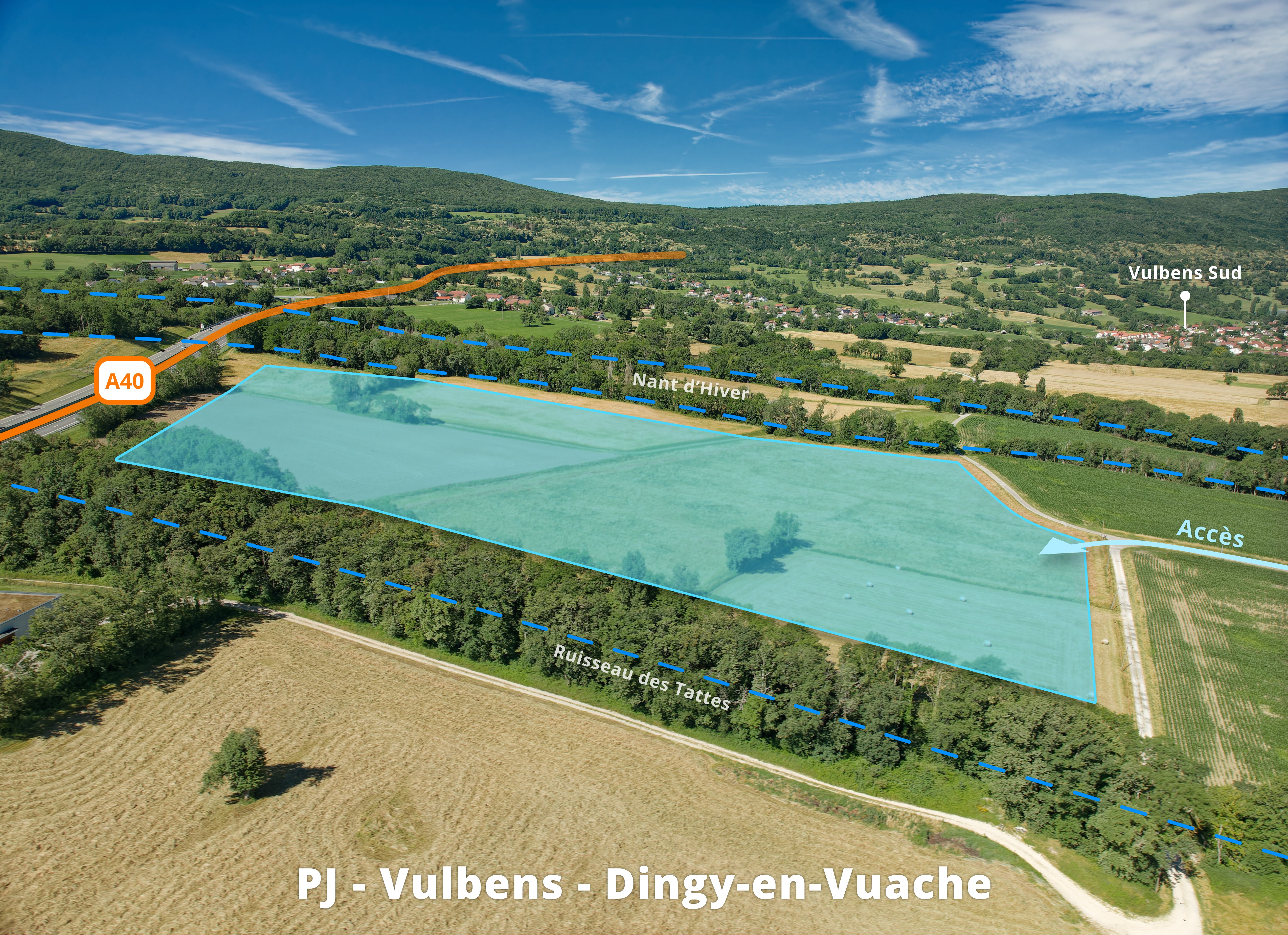}\\[1ex]
    \qrcode[height=1in]{https://fcc-eisa-media.web.cern.ch/drone_videos/PJ/FCC-250121700-SEM-PJ-Video-V0001.mp4}\\[1ex]
    \caption{\href{https://fcc-eisa-media.web.cern.ch/drone_videos/PJ/FCC-250121700-SEM-PJ-Video-V0001.mp4}{Aerial view of candidate location for surface site PJ.}}
    \label{fig:drone-PJ}
\end{figure}

\begin{figure}[!h]
    \centering
    \includegraphics[width=0.8\linewidth]{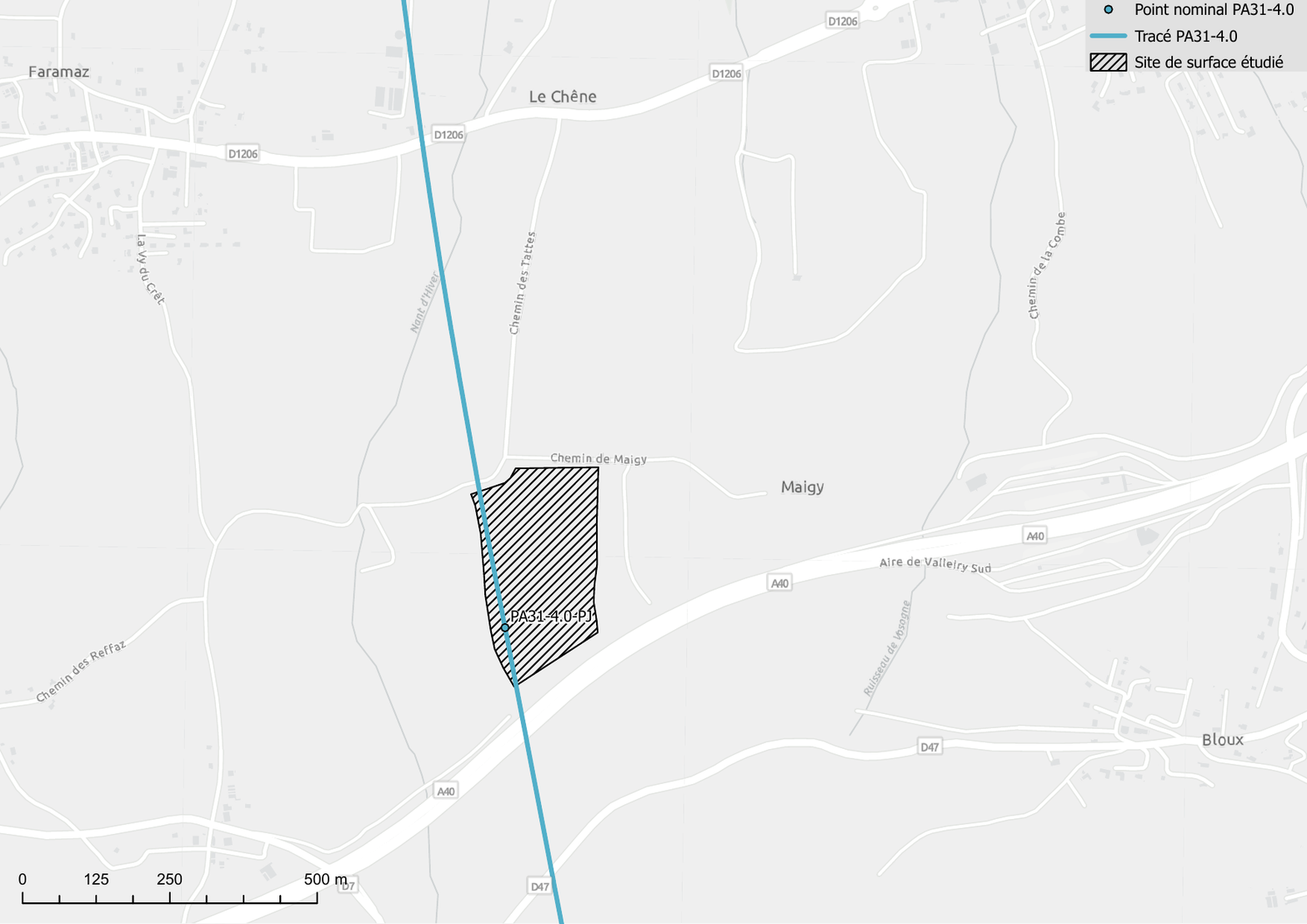}
    \caption{\label{fig:PA31-PJ}PJ surface site location in Dingy-en-Vuache and Vulbens, Haute-Savoie, France.}
\end{figure}

The experiment site PJ is located in a field on a slope to the north of the A40 autoroute and to the west of the Valleiry autoroute area, at the junction of the Chemin des Tattes and Chemin de Maigy roads across the communes of Dingy-en-Vuache and Vulbens, Haute-Savoie, France (see Fig.~\ref{fig:drone-PJ} and Fig.~\ref{fig:PA31-PJ}). To the west is a small stream. The site is far from any dwellings.
The surface area of the site is 6.1 ha.

The site is accessible by the existing paved rural path Chemin des Tattes towards the north to Vulbens. This path has to be enlarged and refurbished.

\subsubsection{Known constraints}

A wildlife corridor exists that needs to be respected in the development of the surface site. If possible, the corridor should be improved, as it currently functions poorly. An area with higher biodiversity in the middle of the site will have to be re-created at the border of the site in conjunction with the wildlife corridor. 
Site design will also have to take into account plans to improve soft mobility between Dingy-en-Vuache and Vulbens.

\subsubsection{Synergies and territorial potentials}

The site is surrounded by numerous sloping farmlands. It will be necessary to work with local authorities and government notified bodies (e.g., SAFER, the agricultural chamber, DDT, DREAL) to determine where agricultural activities can benefit from the reuse of excavated materials.
The site can also benefit from nearby developments to the south, over the same time frame (e.g., new national police station).
It seems feasible to make residual heat available to public institutions and nearby business districts in Vulbens and Valleiry.
In particular, it would be advisable to work with the department on integrating educational and training infrastructures with the surface site-related activities to generate local added value.
The creation of a visitor centre would enable the development of high-quality, sustainable tourism.
The connection to the Valleiry autoroute service area for removing materials and receiving supplies via the autoroute has been analysed and is feasible. To avoid trucks, a conveyor belt would be used to transport the materials to the autoroute area. Incoming materials could be transported via a temporary 700\,m long road to the site.
There is interest in developing synergies for the reuse of waste water to feed streams and for crop irrigation.

\subsection{Site PL}

\subsubsection{Description of the site location}

\begin{figure}[!h]
    \centering
    \includegraphics[width=\textwidth]{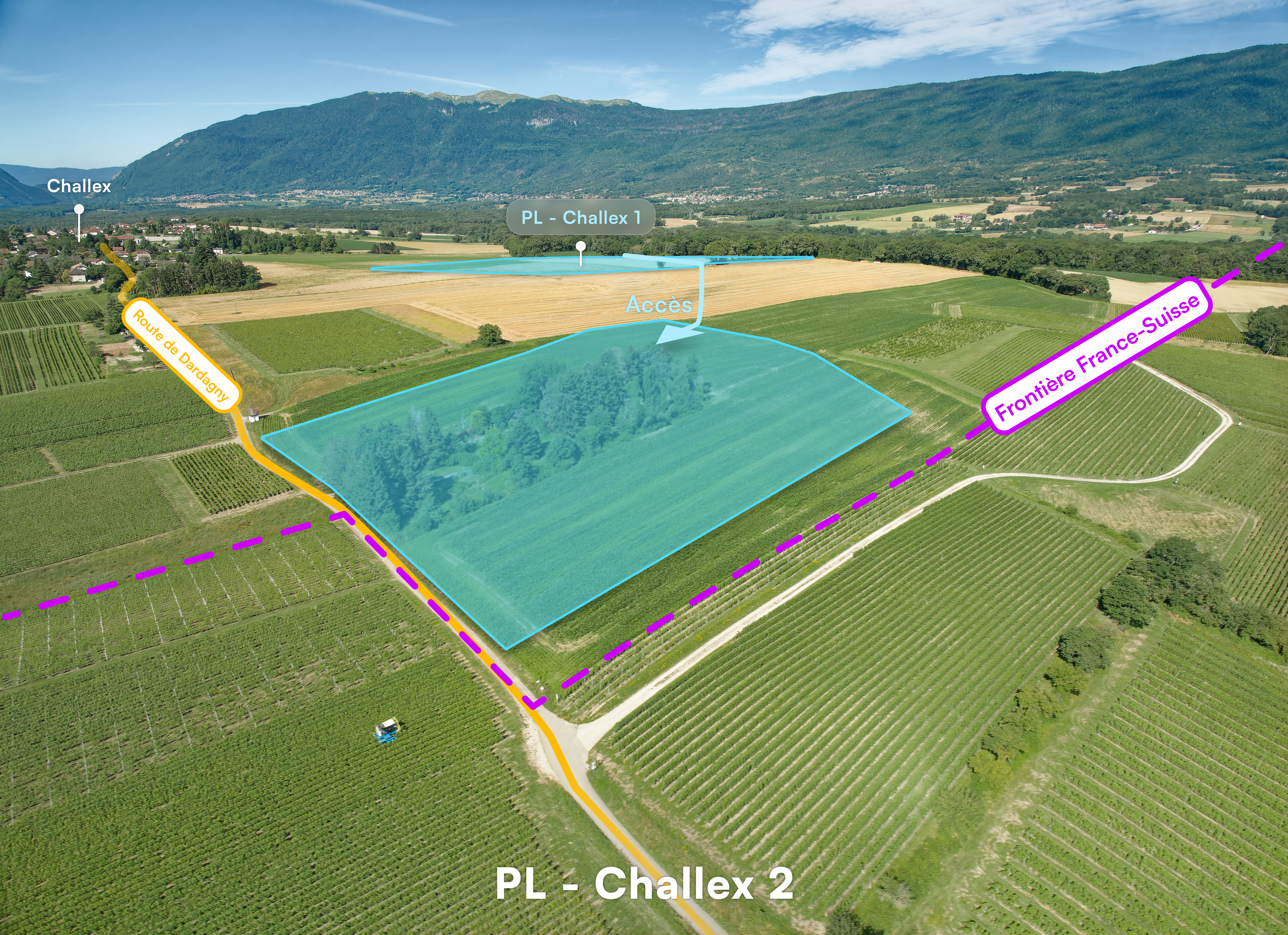}\\[1ex]
     \qrcode[height=1in]{https://fcc-eisa-media.web.cern.ch/drone_videos/PL/FCC-2501301600-SEM-PL-Video-V0002.mp4}\\[1ex]
    \caption{\href{https://fcc-eisa-media.web.cern.ch/drone_videos/PL/FCC-2501301600-SEM-PL-Video-V0002.mp4}{Aerial view of candidate location for surface site PL.}}
    \label{fig:drone-PL}
\end{figure}

\begin{figure}[!h]
    \centering
    \includegraphics[width=0.8\linewidth]{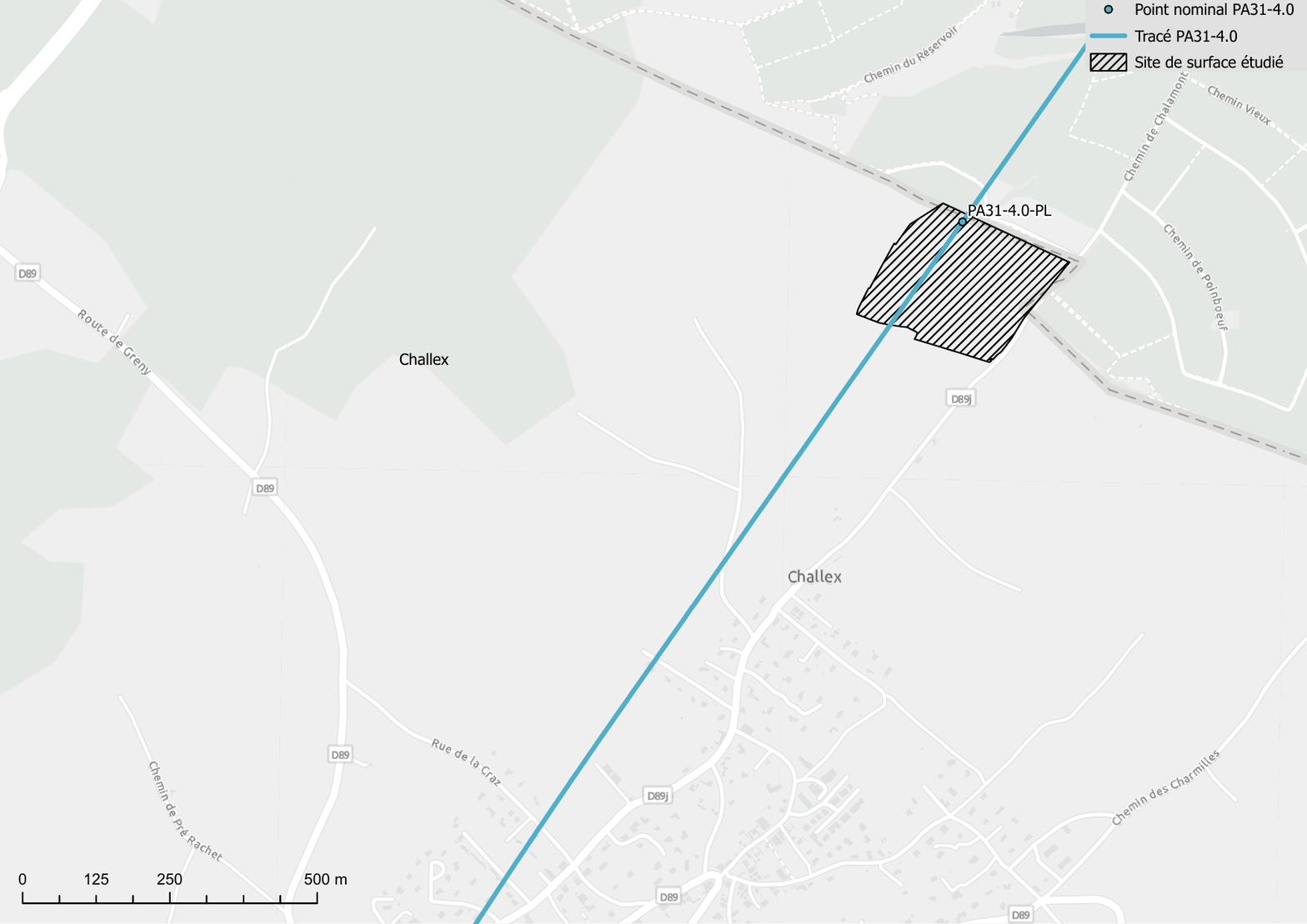}
    \caption{\label{fig:PA31-PL-option-1}PL surface site options studies in Challex, Ain, France.}
\end{figure}

The working hypothesis for the location of the technical site PL in Challex, Ain, France has been determined through various iterations with the municipality in 2023 and 2024. The reference site location at the nominal point can be seen in Fig.~\ref{fig:drone-PL} and Fig.~\ref{fig:PA31-PL-option-1}.

The site is at the nominal point in the middle of the technical straight section, near the border between France and Switzerland on an agricultural field. The perimeter includes two houses which would have to be integrated in the surface site project. The location still needs to be optimised in close cooperation with the municipality and architects. This option, located far from the village, requires the creation of an access road approximately 1.3\,km long to the D89 departmental road. The route of this road will be developed by an expert company in close cooperation with the commune, respecting the existing environmental constraints. The area of the site is 5.5 ha. A large part of this space is foreseen for rewilding.

\begin{figure}[!h]
    \centering
    \includegraphics[width=0.8\linewidth]{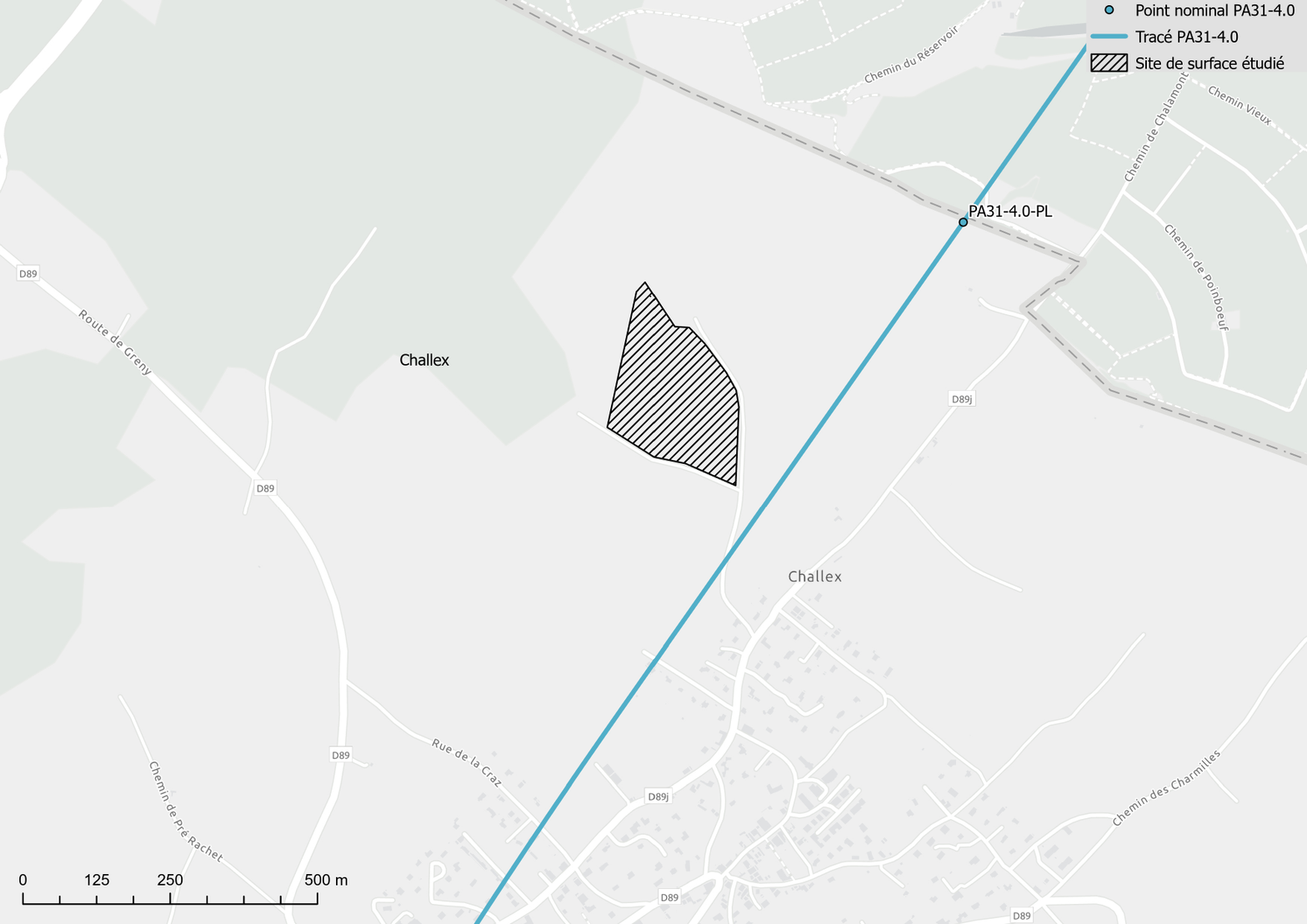}
    \caption{\label{fig:PA31-PL-2}Alternative location for site PL in Challex, Ain, France. The plot is located on the outside of the collider tunnel, which calls for an underground connection tunnel and a more complicated access to the radiofrequency gallery and the collider tunnel for equipment installation and maintenance.}
\end{figure}

Should for any reason the implementation at the nominal point turn out to be unfeasible, an alternative location has been identified at a distance of 800\,m in counter-clockwise direction along the straight section. This location would require the creation of an access shaft about 150\,m outside the line of the collider and would therefore result in significantly higher costs (Fig.~\ref{fig:PA31-PL-2}).

\subsubsection{Known constraints}

\paragraph{Option with location at nominal point}
The proximity of the Franco-Swiss border, close to vineyards, presents a challenge, as there are no roads or other infrastructure. However, one remote dwelling in the vicinity would be affected by disturbances and would therefore have to be acquired. This plot would be integrated into the surface site and serve for rewilding and visibility protection.
Access to the site through the village is ruled out. The roads are too narrow and the disturbance would be unacceptable. Building a new access route through the fields to the north of the village is complicated but feasible. The theoretical point is separated from the main roads (Route de Greny and Rue de la Craz) at the entrance to the municipality by natural areas. The forest to the north of Challex is a zone rich in biodiversity. On the Swiss side, the forest is a zone reserved for absolute protection (Ramsar site), the vineyards are protected, nature zones are protected, the village is a cultural heritage indexed zone and the slope is too steep for surface site construction.
A suspected, temporary, shallow water table was added to the Swiss maps in 2023. If this is the case, it will be confirmed through ongoing geotechnical investigations. Technically, the creation of a shaft is feasible, even if there is a temporary water table in this location. In the present, the investigations will also determine its seasonal nature and the possibility of cross-border connections. The positioning of the shaft and the entire site can be optimised according to the results of these subsurface investigations.
Following an analysis conducted with the local municipality, it has been determined that this location option is preferred.

\paragraph{Option for the location \texorpdfstring{800\,m}{800 m} east of nominal point (not selected)}
A location 800\,m east of the nominal point would not allow access inside the ring (see  Fig.~\ref{fig:PA31-PL-2}). 

The site would be approximately 150\,m from dwellings, 5 to 10\,m lower than the municipality, on a gentle slope of approximately 3\%. It would be visible from certain dwellings. The site would partially affect the protected natural area, but the impact would remain limited since it is currently a field, and the biological corridor passes further to the east of the site. It would be necessary to build an access road to the D89 (approx. 400\,m long) or to the Rue de la Craz road (approx. 300\,m long). This scenario was used as a basis for discussions with the municipality to find an alternative to the location at the nominal point. 

\begin{figure}[!ht]
    \centering
    \includegraphics[width=\linewidth]{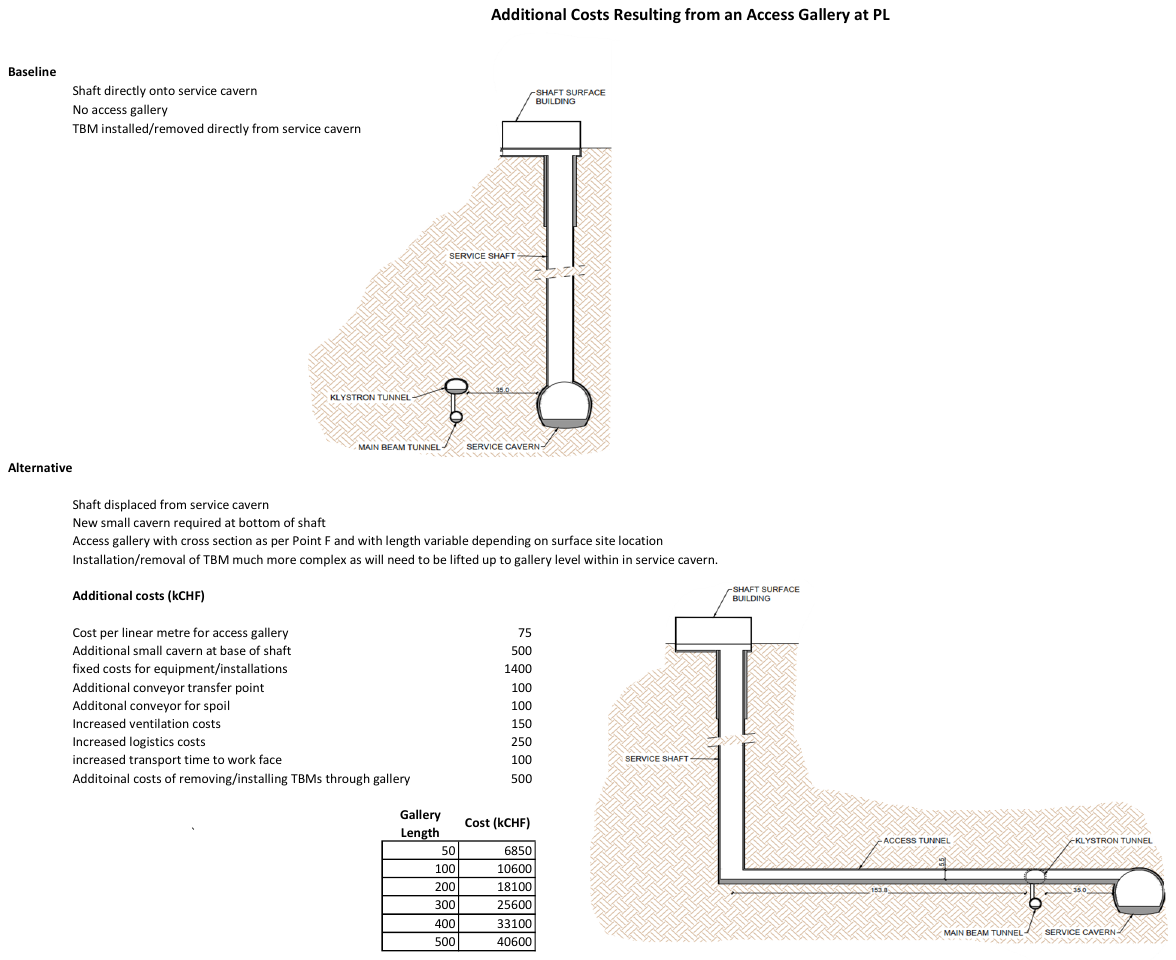}
    \caption{\label{fig:PA31-PL-outside-access}Access to the main tunnel from the exterior via an underground connection. This approach is more complicated and induces additional costs.}
\end{figure}

A technically feasible solution for access from the outside of the ring, albeit at a higher cost, has been developed. The shaft would be located about 150\,m outside the ring (see Fig.~\ref{fig:PA31-PL-outside-access}). The verification regarding its visibility from isolated dwellings on the outskirts of the municipality has shown that this location should not be prioritised. 

\subsubsection{Synergies and territorial potentials}

The proximity of the CERN sites (Pr\'evessin in France and Meyrin in Switzerland) represents an exceptional opportunity to benefit from synergies for the installation, operation, maintenance and repair of equipment in the shortest possible time.
The fields to the north of the municipality appear to be suitable for reusing excavated materials for agricultural purposes. There is no direct visibility from the village.
The proximity of the former Collonges train station at a distance of 12\,km via the de-commissioned railway tack represents an opportunity for removing materials and providing supplies, avoiding trucks as much as possible by using conveyor belts.
A connection to the La Plaine train station in Switzerland would be technically possible, but would be subject to an agreement between France and Switzerland. It could also provide an opportunity for material removal via a combination of conveyor belts and railway systems.
The improvement of the local power grid, necessary for the construction phase, would benefit the municipality and its residents. It would enable increased use of renewable energies and more robust vehicle recharging stations.
Improving the public transport network, starting at the construction phase, can also be of benefit to residents of the municipality, who currently mainly use private vehicles on the D884 in France and travel to Meyrin and La Plaine in Switzerland. Challex has traditionally been home to employees of international organisations, many of whom work at CERN. Strengthening the infrastructure would therefore benefit both the community and the project.
Residual heat from the particle collider could be used for heating individual and collective residences (e.g., the Les Cyclamens long-term care centre), as well as for the Val Thoiry commercial centre approximately 7\,km away. Other nearby communities such as Greny, Saint-Jean-de-Gonville, P\'eron and La Plaine (Switzerland) could also benefit from this heat supply. The company Firmenich (perfumes and fragrances), based in La Plaine in Switzerland, could also benefit from this heat. Heat could also be used in innovative ways for fish farming, aquaponics, market gardening, and greenhouses.

\subsection{Pr\'evessin site}

The aim of the placement studies for the injector facilities is to maximise the use of existing CERN infrastructures. Considering the strong territorial constraints that have been identified and documented throughout a 7-year period, the needs for electricity, raw water, a more than 1\,km long space on the surface for a linear accelerator and associated workshops, offices and storage zones, the existing CERN Pr\'evessin site emerged as the most suitable location for a working hypothesis. This site has the space to host the electron and positron sources, linear accelerators, damping rings and optional experimental facilities to leverage the powerful electron-positron injector as an additional scientific instrument that can be used even before the electron-positron collider enters operation. In addition, the site is at a reasonably close distance from the existing LHC surface site P8, which together with additional space in the immediate vicinity, forms the surface site PA. Hence, the implementation of the transfer line, also leveraging the existing SPS subsurface structures, leads to an advantageous configuration. Numerous technical infrastructures such as the existing 400\,kV grid connection, water supply and treatment, offices, computing facilities, room for construction activities permit the efforts that are typically linked to territorial development for a new site to be kept within limits. At the current stage of conceptual development, the injector would have a total length of approximately 1.2\,km. It could almost fit into the fenced space of the Pr\'evessin site. 

Common work with the authorities and further optimisation according to the `avoid-reduce-com\-pen\-sate' approach will be performed so that any unavoidable territorial extension will be kept as small as possible and to assure that ultimately a suitable reference scenario for the injector placement can be identified.

\begin{figure}[!ht]
  \centering
  \includegraphics[width=0.8\textwidth]{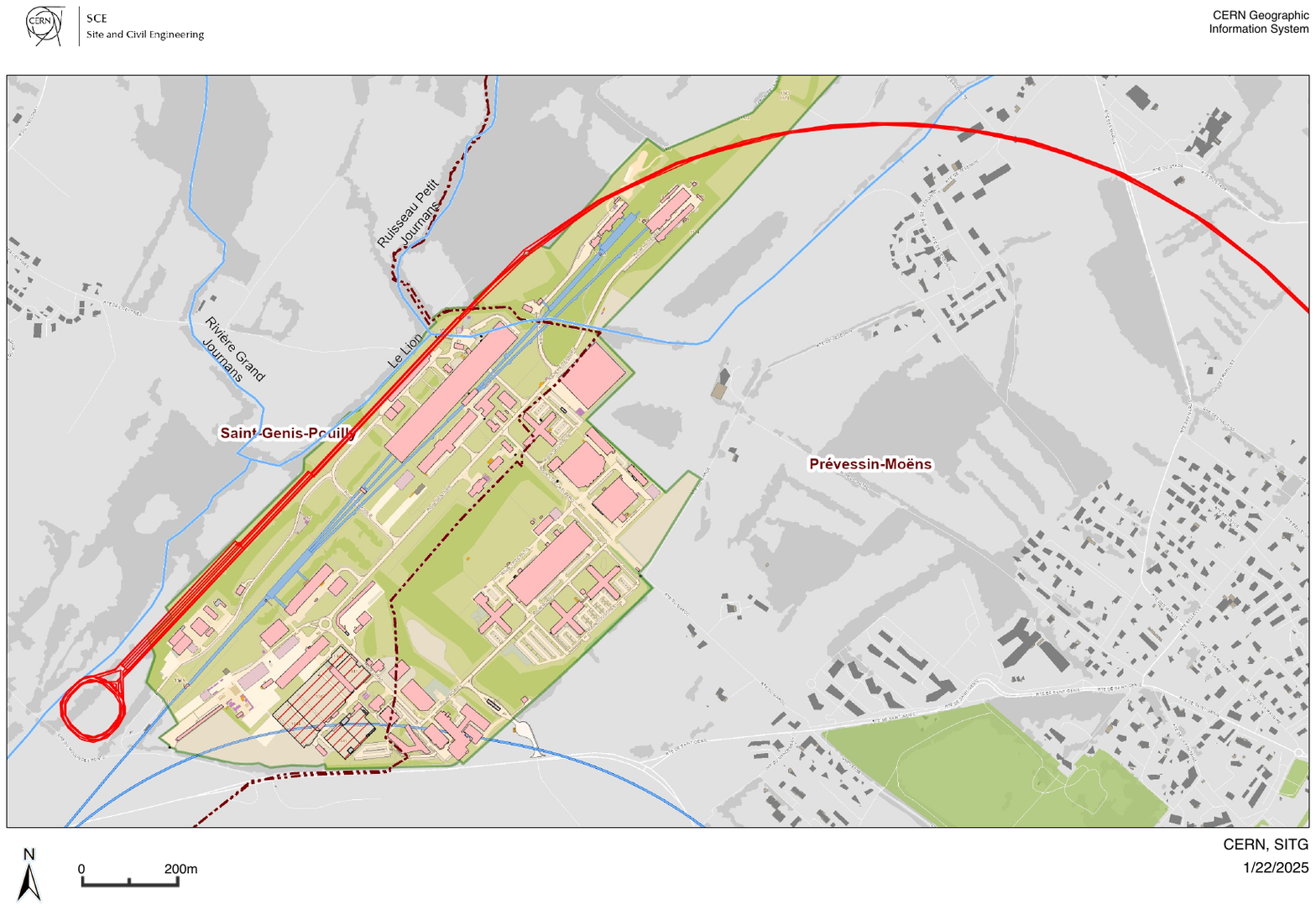}
  \caption{Sketch of the current working hypothesis for a linear-accelerator based injector that has a length of about 1.2\,km. A definitive placement remains to be developed in the frame of a design phase, considering environmental, existing experimental facility, territorial development and engineering constraints.}
  \label{fig:injector-placement}
\end{figure}

Initial explorations have excluded the possibility of extending the site towards the north and east due to numerous nature, agriculture, and visibility constraints. The current, unfinished, working hypothesis shown in Fig.~\ref{fig:injector-placement} is based on a placement between the 'Lion' creek and the North-Area beamline. The requirements and constraints of various technical concept elements, such as the exact size and shape of the damping ring, the widths and lengths of the accelerators and their integration, are today not yet at a level that permits freezing of the exact placement on the site. The analysis work carried out so far permitted, however, confirming the technical and territorial feasibility in principle and helped to identify the constraints to be considered for subsequent activities.

During a subsequent technical design phase, environmental analysis will guide the optimisation of the placement to ensure that natural constraints are respected, the required extension of the existing fenced domain is kept as small as reasonably possible, and that existing experimental facilities are not significantly impacted. A preliminary concept for the underground transfer line alignment avoids as much as possible conflicts with the projected construction areas on the surface. A definitive design will require more detailed environmental analysis, including geology and hydrogeology. The estimated efforts for these analysis and design activities are in the order of two to three years. They need to be integrated in the overall project environmental authorisation process. 

\subsection{Conclusion}

Table~\ref{tab:site-feasibility-summary} shows the status of the feasibility analysis for the surface site locations in scenario PA31-4.0. Technical feasibility was assessed using maps and data available to those working on the study and to contractors. The assessment was also based on fieldwork and environmental studies carried out by specialised companies, discussions with the relevant technical departments of the two host states, and with local elected officials (mayors, town councillors, departmental councillors, and regional councillors), who represent the citizens and act in their interests.

\begin{longtable}{c p{2cm} c p{9cm}}
 \caption{Status of technical feasibility assessment for surface sites of PA31 scenario.}
 \label{tab:site-feasibility-summary}
 \endfirsthead
 \endhead
\toprule
    \textbf{Site} & \textbf{Location} & \textbf{\makecell{Technical \\ feasibility}} & \textbf{Feasibility conditions} \\ \midrule
    \textbf{PA} & Ferney-Voltaire, Ain, France & Confirmed & Limit impact on the landscape, enhance the natural area and the existing compensation zone. Maximise synergy with LHC point 8. Compensate for the loss of agricultural space.
Develop a visitor centre, for example, on the LHC Point 8 site. Develop synergies based on heat recovery with municipalities within a 10\,km radius, including Swiss municipalities and the Geneva airport.\\ \midrule
    \textbf{PB} & Presinge, Geneva, Switzerland & Confirmed & Define the exact location of the site on the plot and initiate discussions with the municipality. Identify a location to compensate for the loss of agricultural space. Limit impact on the landscape. Take into account the sensitive location of the site in a natural setting. A detailed conceptual design of the road access is to be developed and approved by the canton of Geneva. Reach an agreement to recycle excavated materials, preferably locally. Develop synergies based on heat recovery around the site with local stakeholders and authorities. \\ \midrule
    \textbf{PD} & Nangy, Haute-Savoie, France & Confirmed & There are no specific blocking points, but the major issues in this sector call for careful joint development of the site in coordination with local stakeholders. Limit the loss of agricultural space by working on a smaller surface site. Compensate for the loss of agricultural space. Maintain compatibility with the proposed connection between the RD903 and the A40. Estimated start of construction: first quarter of 2025. Develop a transport concept for the construction phase to limit impacts on a sector that is already overburdened. Develop synergies with the communities around the nearby hospital (CHAL), the Scientrier waste water treatment plant (STEP) and the industrial zone to the north. \\ \midrule
    \textbf{PF} & \'Eteaux, Haute-Savoie, France & Confirmed & The north option along the RD1203 was confirmed, provided that nearby wetlands are avoided. Limit the loss of agricultural space. For a potential southern extension option, a decision would be required before the start of phase 2 of the ISDI (inert waste storage facility) in La Roche-sur-Foron in 2027. This would allow earthworks to be carried out on the western part for a smaller site, rather than wait for a complete development for the ISDI. Otherwise, annexes would need to be constructed on top of the inert waste. An agreement must be reached with the ISDI operator and the landowners. The access layout should be developed for this option. Any economic loss suffered by the ISDI operator must be taken into account when developing synergies between the site and the operator. Develop synergies with local authorities in terms of site development with emergency services.
\\ \midrule
    \textbf{PG} & Charvonnex and Groisy, Haute-Savoie, France & Confirmed &  The site straddles the forest and the plateau, which includes unexploited grasslands. Two annexes close to the autoroute area will accommodate certain facilities (e.g., water cooling system, electrical substation) to avoid any disturbance to the wooded area. Plan a visitor centre to develop high-quality, sustainable tourism. Develop synergies with the municipalities of Groisy and Charvonnex to develop the site with neighbourhood services for on-site researchers and emergency services. The loss of woodland can be compensated by reforestation around the site. The layout of the existing access route to the north is suitable.
\\ \midrule
    \textbf{PH} & Cercier and Marlioz, Haute-Savoie, France & Confirmed & Limit the site's footprint and remain within the wooded area to avoid impacts on the dwellings to the south of the site. Reduce the impact on natural habitats and biodiversity. Compensate for the impacts on woodland, habitat and biodiversity that cannot be avoided and reduced. Respect the easement for the gas pipeline near the site, and decide what distance to maintain between the pipeline and above-ground infrastructure. Consider a split or displaced site location along the straight section.
 \\ \midrule
    \textbf{PJ} & Dingy-en-Vuache and Vulbens, Haute-Savoie, France & Confirmed & Compensate for the loss of agricultural space. Preserve ecological corridors. Integrate the planned projects to develop soft mobility between Dingy-en-Vuache and Vulbens. Foresee a visitor centre to develop high-quality, sustainable tourism. Work with the municipalities to develop synergy for the site with regard to neighbourhood services for on-site researchers, emergency services and schools.
 \\ \midrule
    \textbf{PL} & Challex, Haute-Savoie, France & Confirmed & The location at the nominal point has been discussed with the municipality. Compensation for the loss of agricultural space. A joint optimisation of the site with the municipality is in progress.
     \\ \midrule
    \textbf{CERN} & Pr\'evessin and Saint-Genis Pouilly, Ain, France & Confirmed & The location of the injector at the existing CERN Pr\'evessin site has been verified, and its technical feasibility has been confirmed in principle. Optimisation of the placement within the site remains to be done based once detailed technical requirements and invariants are available and the state of the environment has been analysed. Territorial development outside the fenced perimeter will be kept as low as reasonably possible. The alignment of the underground transfer line to site PA will be designed considering geology, hydrogeology and the project of constructed areas on the surface.
 \\ \bottomrule

\end{longtable}

\section{Territorial infrastructure needs}
\label{imp:territorial_needs}

\subsection{Introduction}

\begin{figure}[!ht]
  \centering
  \includegraphics[width=0.85\textwidth]{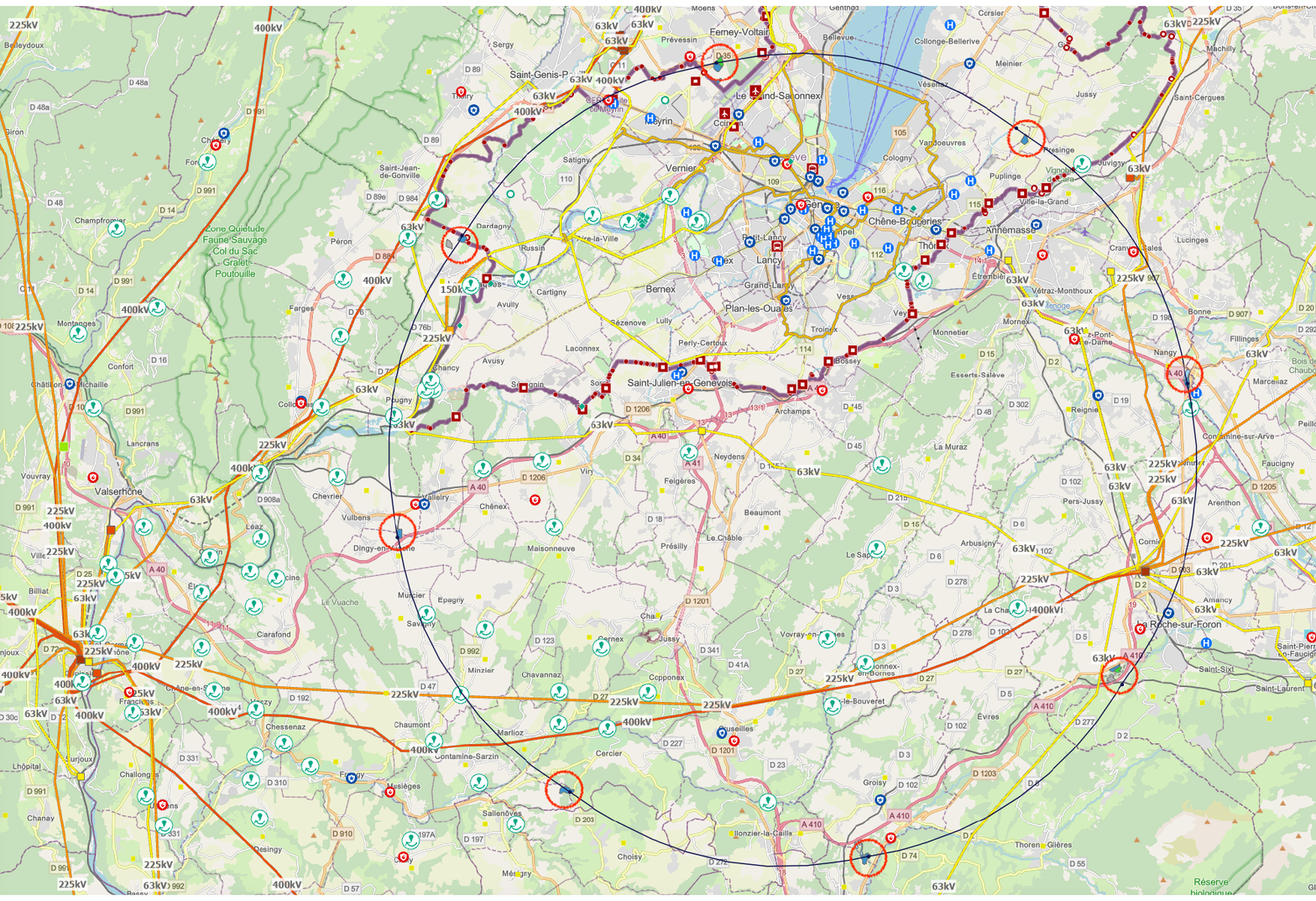}
  \caption{Noteworthy road, electricity, water treatment infrastructures, emergency services and border crossings in the vicinity of the surface sites indicated by red circles.}
  \label{fig:territory_infrastructures}
\end{figure}

To be able to construct, install and operate the particle-collider facility and its associated experiments, a number of different territorial infrastructures are required. The development of the implementation scenario put an emphasis on leveraging existing infrastructures as much as possible (Fig.~\ref{fig:territory_infrastructures}). The well-developed transport, electricity, and water networks in the region around CERN are one of the motivations to propose the facility in this region. The main infrastructures required and described in this section are: roads that provide access to the surface sites, autoroutes to evacuate excavated materials and to supply construction materials and equipment, railway lines to support possible excavated materials and construction materials transport, electricity for the construction phase, direct access to the French high-capacity power grid for the operation, and raw water for cooling purposes.

\subsection{Road access}
\label{sec:road_access}

The road network is dense along the reference scenario footprint. 
For the site access roads, in order for two heavy goods vehicles to pass each other at reduced speed, a width of 5.50\,m of roadway with two 0.50\,m shoulders is required. Subject to environmental constraints, the lanes of reinforced roads or new roads may be smaller: widths of 4.00\,m for the roadway and 0.50\,m for the shoulders are acceptable if possibilities for passage in both directions are foreseen. This reduced section will require instructions to be given to the drivers of the vehicles concerned.

\begin{figure}[h!]
    \centering
    \includegraphics[width=0.85\textwidth]{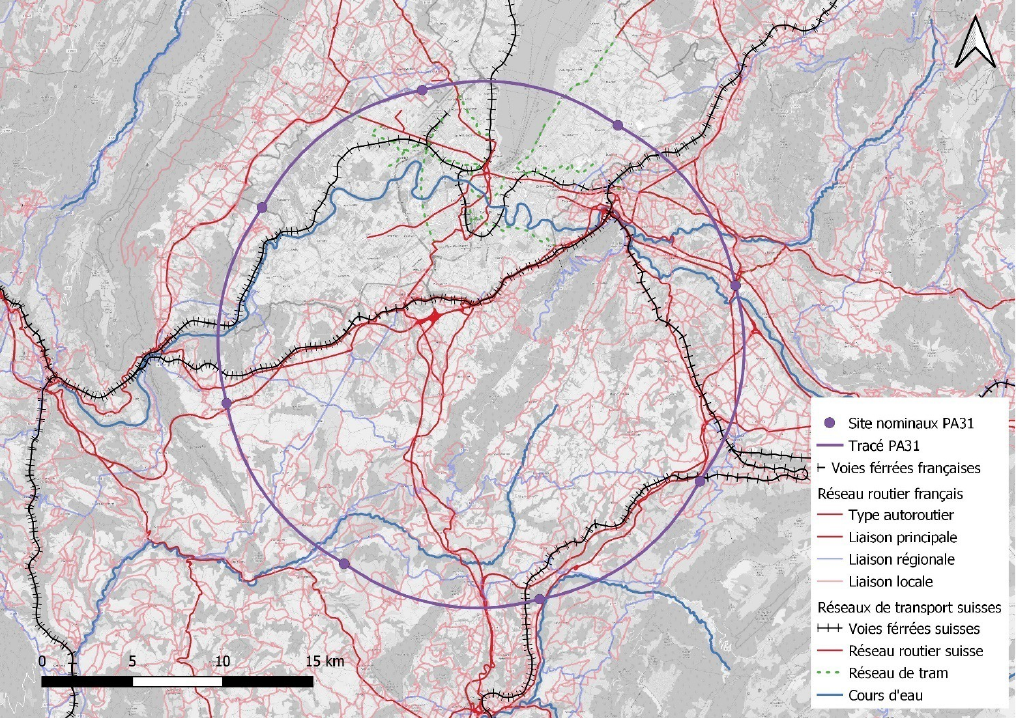}
    \caption{\label{fig:road-network} Overview of the road and railway transport network in the perimeter of the reference scenario.}
\end{figure}

Several surface sites can profit from direct road access (PA, PD, PF, PH) and some others require minor reinforcement of existing road access (800\,m of forest path to be paved for PG, 600\,m of rural path to be paved for PJ). Site PD requires the creation of a 300\,m long dedicated access. Site PL requires the creation of an approximately \SI{1300}{\metre} long dedicated access. In total, less than \SI{3}{\kilo\metre} of road have to be created.

Road traffic analysis \cite{gutleber_2025_14992820} has been carried out for all sites, and the feasibility of the construction was verified with this work.
Technical designs of road accesses have been developed for sites PB, PD and PF to ensure the feasibility in areas that are subject to particular road traffic constraints linked in PB to visibility and road safety, in PD due to a major road enlargement and autoroute development project and in PF to ensure road safety. Detailed road access designs have now to be carried out for all sites for the development of a coherent design package that is required for environmental impact assessments and project authorisation.

\clearpage

\begin{figure}[h!]
    \centering
    \includegraphics[width=\textwidth]{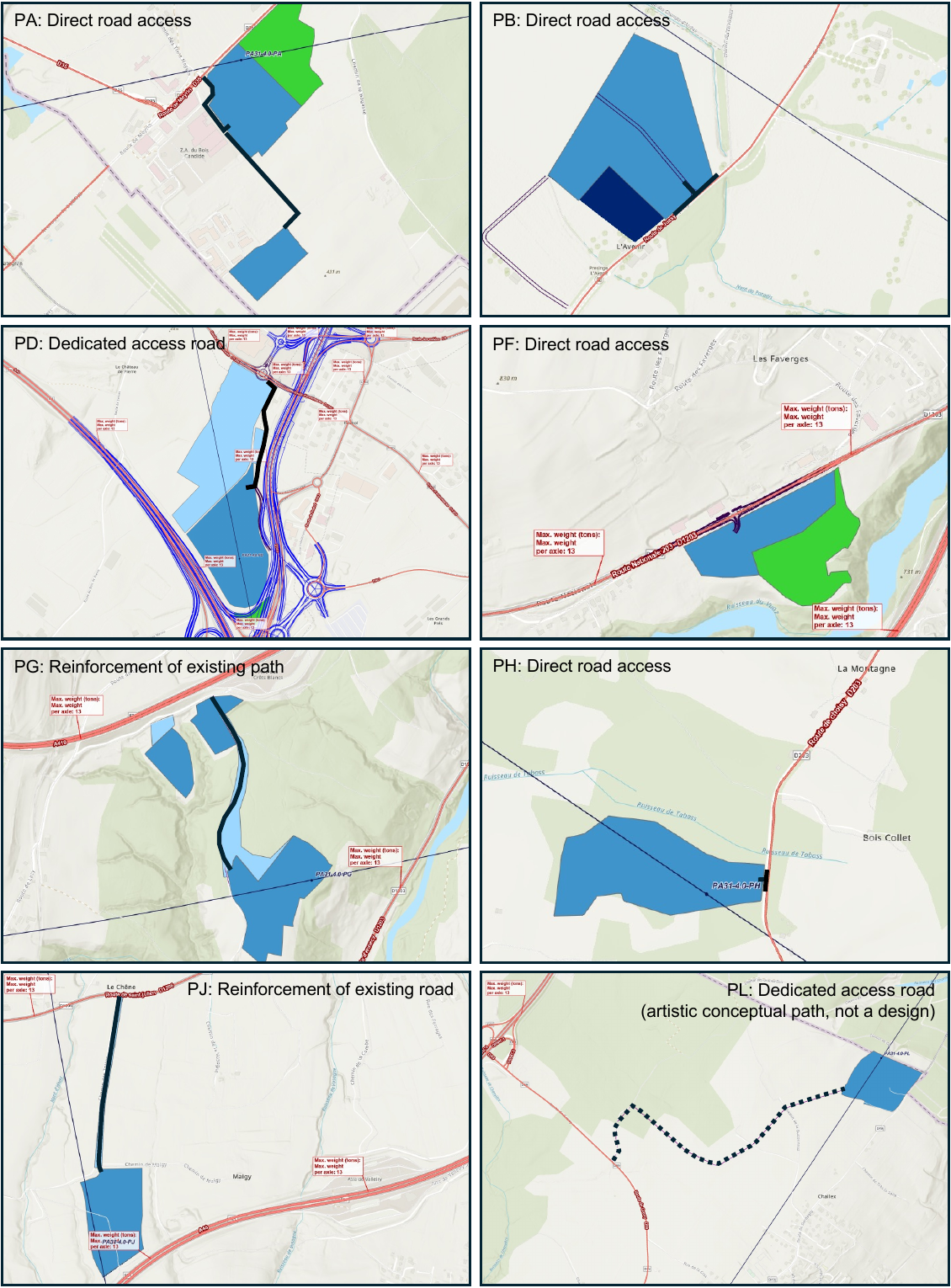}
    \caption{\label{fig:road-accesses} Overview of road access concepts for each individual surface site.}
\end{figure}

\subsection{Autoroute access}

The study examined the feasibility of connections to the autoroute network for the removal of materials and the supply of equipment during the construction phase \cite{ancelet_2022_8255742}. The possibility of obtaining autoroute connections, either directly, via conveyor belts or via temporary gravel paths during construction is a goal in the general interest, aimed at limiting the impact of the project, particularly during the construction phase. The specific technical choices will be made later, during the project development and preparation phase. To verify the technical, legal and financial feasibility of direct access to autoroutes, files with conceptual design plans and descriptions were compiled and submitted for review on 14 September, 2022 to the competent authority for granting concessions: the Direction générale des infrastructures, des transports et des mobilit\'es (DGITM),
Direction g\'en\'erale des infrastructures, des transports et des mobilit\'es / Direction des mobilit\'es routi\'eres / Sous-direction des financements innovants et du contr\^ole des concessions autorouti\`eres / Chef du Bureau des services aux usagers et de la comodalit\'e and the 
Direction g\'en\'erale des infrastructures, des transports et des mobilit\'es / Direction des mobilit\'es routières / Sous-direction des financements innovants et du contrôle des concessions autorouti\`eres / Chef du Bureau du patrimoine et de l’am\'enagement.

\begin{figure}[h!]
    \centering
    \includegraphics[width=0.8\textwidth]{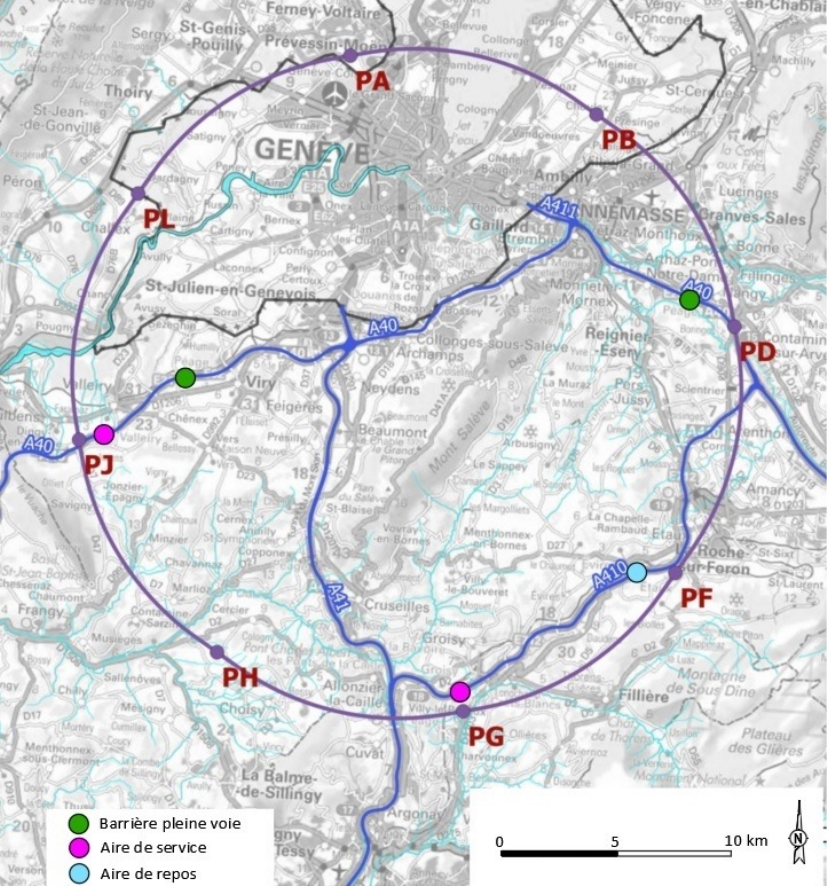}
    \caption{\label{fig:motorway-network} Location of the autoroute connections, which were studied using existing service and rest areas. Blue dots correspond to autoroute rest areas, magenta dots indicate autoroute service stations and green dots indicate toll stations.}
\end{figure}

The feasibility of four new connections to the autoroute network was analysed and confirmed to be in principle feasible (see Fig.~\ref{fig:motorway-network}) for the following sites:
\begin{enumerate}
\item PD site in Nangy (on the A40 autoroute),
\item PF site in \'Eteaux/La Roche-sur Foron (on the A410 autoroute),
\item PG site in Charvonnex/Groisy (on the A40 autoroute),
\item PJ site in Dingy-en-Vuache/Vulbens (on the A410 autoroute).
\end{enumerate}

\noindent The conclusions of the interactions with the DGITM were as follows:
\begin{itemize}
\item The access concepts and loading/unloading areas provided are in principle acceptable.
\item Adjustments will be needed, with the final decisions to be made with the autoroute operator in the project phase.
\item Detailed plans have to be developed and presented.
\item The procedure for submitting an application in the future was specified.
\item The proposed justifications, in the general interest, are acceptable;
\item Entrances and exits will need to be equipped with detection devices to manage tolls.
\end{itemize}

\subsection{Railway access}

To reduce the need for truck traffic, to generate further opportunities to supply quality construction materials from further distances and to open possibilities to transport excavated materials to appropriate deposition sites and re-use locations in an environmentally friendly and high capacity way, railway access studies have been carried out by a qualified domain-expert company \cite{egis_france_2024_10633809, egis_france_2024_10528636,egis_france_2024_10534566, ITE_PL,egis_france_2024_10533655,egis_france_2024_10534517,egis_france_2024_10533313,egis_france_2024_10532758,european_organization_for_nuclear_resear_2024_10534351}.

The current reference scenario was taken as a working hypothesis for analysing the opportunities and feasibilities concerning accessing the railway system via existing installations (goods loading and unloading facilities) and concerning the creation of new accesses (so-called ITE, `Installation Terminal Embranch\'ee' in French).

The studies used a multi-criteria analysis that considered the following indicators:
\begin{itemize}
\item Proximity of the surface site with a suitable railway track.
\item Technical and administrative compatibility for access with the French and Swiss railway infrastructure.
\item Presence of an existing service or ITE that could be leveraged.
\item Feasibility of creating a transport connection between the site and the railway track access.
\item Minimum space requirements for railway access and available space.
\item Number of convoys required for evacuating all excavated materials.
\item Capacity availability and limitations on each railway line analysed.
\end{itemize}

\begin{figure}[h!]
    \centering
    \includegraphics[width=\textwidth]{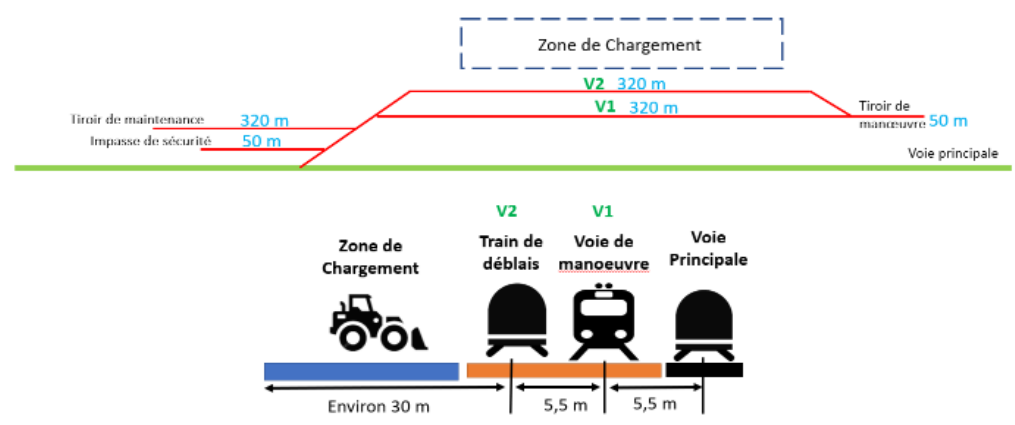}
    \caption{\label{fig:ITE} Minimum space requirements of a new railway access for goods transport.}
\end{figure}

It is important to keep in mind that each new railway access requires a space of about 400\,m by 40\,m, i.e., 1.6\,ha of land (see Fig.~\ref{fig:ITE}).

The train line 890000 through Meyrin, Collonges, today provides the necessary free capacity. Line 902000 in Vulbens and line 897000 passing in \'Eteaux and in Groisy have only limited capacities.

The analysis revealed the following `in principle' opportunities for railway access:

\begin{table}[!h]
  \centering
 \caption{\label{tab:railway-opportunities} Railway access opportunities for each surface site.}
  \begin{tabular}{clll}
        \toprule
        \textbf{Site} & \textbf{Distance} & \textbf{Feasibility} & \textbf{Description}\\ \midrule
        PA & Less than 5\,km & Unsure & Requires crossborder transport of materials \\ 
        PB & Less than 5\,km & Low & Requires crossborder transport of materials \\ 
        PD & No opportunity within 10\,km & Unfeasible & \\ 
        PF & Less than 1\,km & Medium & Space and ISDI constraints \\ 
        PG & Less than 2\,km & Medium & Implementation constraints \\ 
        PH & No opportunity within 10\,km & Unfeasible & \\ 
        PJ & Less than 5\,km & High & \\ 
        PL & Less than 10\,km & High & Via existing, unused station at Collonges \\ \bottomrule
  \end{tabular}
\end{table}

Based on the analysis, today, the most likely accesses to the railway system are the existing and unused train station in Collonges for site PL, new access north of Vulbens for site PJ and a new access west of Groisy for site PG. While the old train station of Collonges is 13\,km from the surface site PL in Challex, a connection with a conveyor belt can be a suitable approach, avoiding truck traffic.

The feasibility in principle for the creation of access in La-Roche-sur-Foron on the site of an inert waste storage facility (I.S.D.I.) exists once that facility has been filled. However, the amounts of materials produced at site PF are limited compared to other sites, and the constraints due to the status of the facility (ICPE in France) and space limitations require a careful cost/benefit and administrative feasibility analysis in addition to the technical analysis.

The technical and administrative feasibility of railway access at Geneva Airport in Switzerland for site PA in France is subject to a specific analysis that will only be carried out in 2025. If it is technically feasible, an administrative challenge is to be addressed concerning the cross-border transport of materials in both directions: France to Switzerland for excavated materials and Switzerland to France for construction materials and equipment.

The creation of a new ITE requires about 10 years of planning, detailed development of the variants to be presented for authorisation, an economic demand study concerning use beyond the FCC construction phase, environmental impact assessment and the authorisation process. The implementation for use requires about 2 years.

If train access is to be used for the evacuation of materials and the supply of construction materials and equipment, a detailed design and common project together with the French and Swiss national railway network administration services would have to be started in a forward-looking way, starting in 2025.

\subsection{Conveyor belt links}

To reduce the need for temporary paths and truck traffic due to excavated materials and construction material-related transports between sites and autoroute accesses, example studies were performed to determine the feasibility of conveyor belt links \cite{egis_france_2024_10534517, egis_france_2024_10517186} by an expert company in the domain. Two sites have been selected for the case study: PJ (Vulbens and Dingy-en-Vuache, France) and PG (Charvonnex and Groisy, France). It is worth pointing out that the findings also apply to the other sites that were examined at a high level, but no technical designs have been developed. If a preparatory project phase is launched, detailed technical design variants for traces and conveyor technologies for a construction hypothesis have to be drawn up for all eight sites and submitted for authorisation in the frame of the project environmental authorisation process.

Conveyor belts would be operated with electricity with a capacity of 120\,kW between the start and the end of the construction activities. Based on a speed of 2 to 2.5\,m/sec and a width of 650 to at most 800\,mm the following schedule has been established:
\begin{itemize}
\item 22 days per month.
\item 226 days per year.
\item Up to 8 hours per day.
\item Operation between 08h00 in the morning and 18h00 in the evening.
\item No operation during the night, weekends or holidays.
\end{itemize}

Various technologies for conveyor systems have been analysed, and their costs have been estimated for the two specific locations PG and PJ. Depending on the environmental conditions (topography, terrain, vegetation, urban constraints) different footprint and noise-limiting systems can be considered. In general, the choice is always determined by the goal to limit footprint and nuisances for the required capacities, limited by the available technical constraints. Public spaces would be leveraged whenever possible. New routes would be limited to 3\,m width. The maintenance of new routes is limited: they are not permanent and the space used will be restored after use.

The feasibility of meeting the capacity requirements associated with a construction site that operates 2 tunnel boring machines (TBM) has been confirmed from a technical perspective and an environmental perspective. Noise levels are between 65\,dB(A) directly at the conveyor and 47\,dB(A) at a distance of 64\,m with today's off-the-shelf technology. Example routes have been developed to confirm compatibility with the noise regulations in France for both study sites.

For site PJ, the conveyor can be created to the autoroute service station in Valleiry (distance \SI{700}{\metre}) and/or to new railway access north of Vulbens (\SI{1565}{\metre}) outside any residential area. It has to cross the RD 1206 departmental road. Disturbances due to noise can be avoided in both cases.

\begin{figure}[h!]
    \centering
    \includegraphics[width=\textwidth]{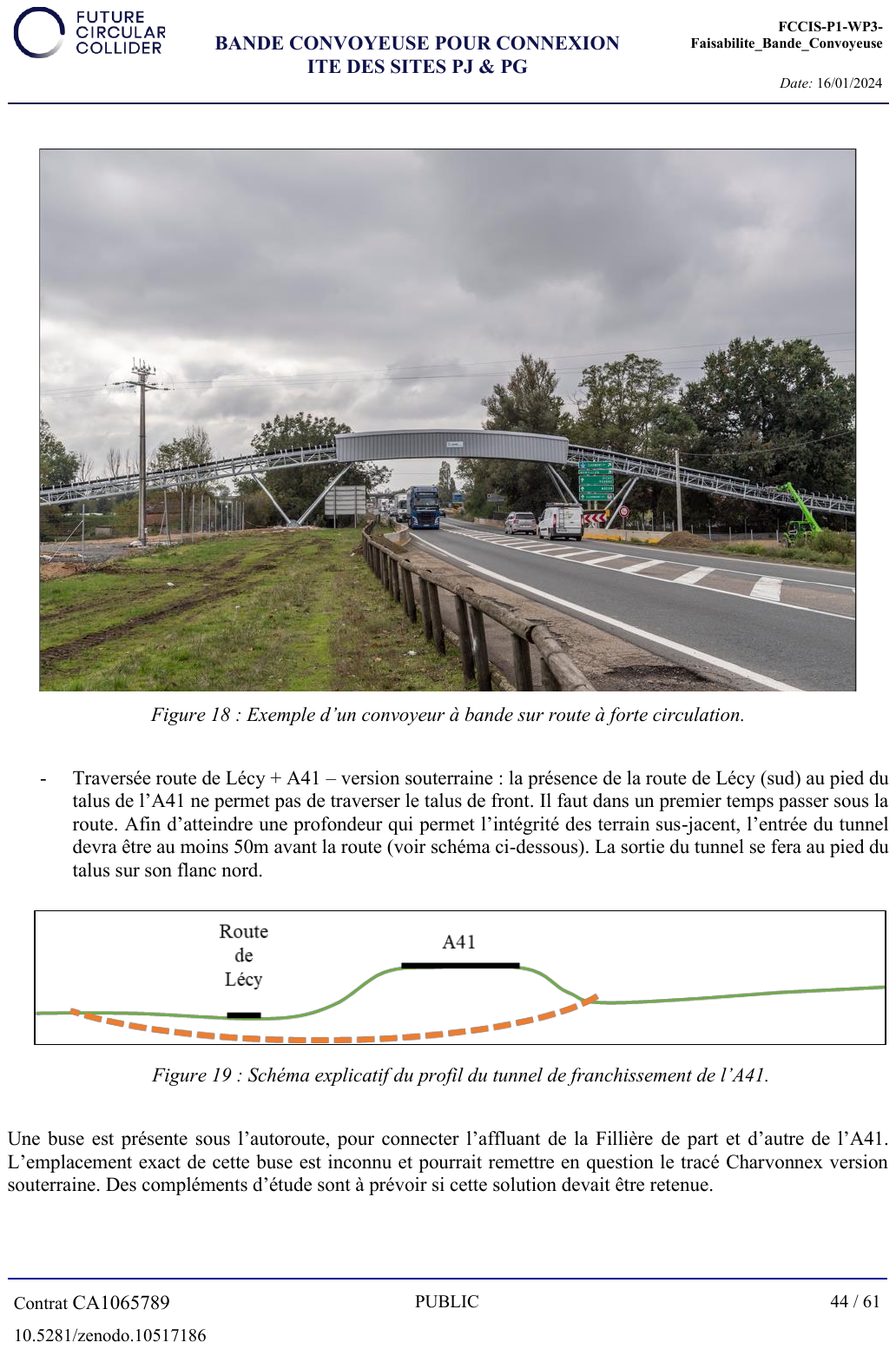}
    \caption{\label{fig:conveyor-road} Example of a conveyor crossing a major road.}
\end{figure}

For site PG, the conveyor can be created to the autoroute service area in Groisy (distance of about \SI{800}{\metre}) and/or to new railway access at the north of the autoroute (\SI{925}{\metre}). The biggest challenge, although technically feasible (see Fig.~\ref{fig:conveyor-road}) is the crossing of the A40 autoroute for a period of about 8 years. Residential areas are unlikely to be affected.

\subsection{Electricity for the construction phase}

For the supply of electricity during the construction phase, requests for hook-ups to local networks have to be made directly to the relevant national distributors (e.g., Enedis in France and SIG in Switzerland). According to information provided by company Herrenknecht, a leading TBM manufacturer, electricity requirements are around 3.7\,MVA for a construction site using a single TBM and around 7.4\,MVA for a construction site with two TBMs. The working hypothesis presented in Table~\ref{tab:electricity-construction} will be fine-tuned with the civil engineering companies during a subsequent design phase before the start of construction. This process will deliver the exact electrical power required during the construction phase. This will depend on the number and configuration of the TBMs around the ring and the specific machinery used. All that information will only be known with certainty shortly before the launch of the public works contracts. However, it has to be considered that planning, contracting, and implementing the local electricity connections for the construction phase will require several years and will need to be part of the overall environmental authorisation process.

\begin{table}[!h]
  \centering
 \caption{\label{tab:electricity-construction} Working hypothesis for connection to local electricity networks during the construction phase.}
  \begin{tabular}{cllr}
        \toprule
        \textbf{Site} & \textbf{Location} & \textbf{Operator} & \textbf{Power} \\ \midrule
        PA & Ferney-Voltaire, France & Enedis & \makecell[r]{7.0 - 13.8\,MVA\\(= 400 A at 20 kV)} \\ \midrule
        PB & Presinge, France & SIG & \makecell[r]{3.0 - 7.0\,MVA\\(= 200 A at 20 kV)} \\ \midrule
        PD & Nangy, France & Enedis & \makecell[r]{7.0 - 13.8\,MVA\\(= 400 A at 20 kV)} \\ \midrule
        PF & \'Eteaux, France & Enedis & \makecell[r]{3.0\,MVA\\(= 100 A at 20 kV)} \\ \midrule
        PG & Charvonnex, France & \'Energie et Services de Seyssel & \makecell[r]{13.8\,MVA\\(= 400 A at 20 kV)} \\ \midrule
        PH & Cercier, France & Enedis & \makecell[r]{3.0\,MVA\\(= 100 A at 20 kV)} \\ \midrule
        PJ & Vulbens, France & Enedis & \makecell[r]{13.8\,MVA\\(= 400 A at 20 kV)} \\ \midrule
        PL & Challex, France & Enedis & \makecell[r]{3.0\,MVA\\(= 200 A at 20 kV)} \\ \bottomrule       
  \end{tabular}
\end{table}

\subsection{Electricity for the operation phase}

The FCC-ee scientific research programme is based on different collider operating modes (Z, WW, ZH, $\rm t \bar{t}$ and an optional HH mode). Each mode uses a different equipment configuration. This equipment (acceleration systems by superconducting radiofrequency cavity, electrical energy conversion systems and cryogenic cooling systems) will be installed progressively during the maintenance and upgrade phases planned for these activities. One long shutdown (LS) is planned to install the radiofrequency systems for the $\rm t \bar{t}$ and optional HH operation modes. Each configuration is characterised by different electrical power needs (see Table~\ref{tab:energy_needs_per_mode}), which serve as the basis for average consumption estimates.

\begin{table}[!h]
  \centering
 \caption{\label{tab:electricity-operation} Electrical power need estimations for the operation phase. Technical feasibility studies permitted reducing power requirements by almost 45\% between 2020 (initial hypothesis) and 2024.}
  \begin{tabular}{lrrrr}
        \toprule
        \textbf{Mode} & \textbf{Beam energy} & \textbf{Operation} & \textbf{Initial capacity estimation} & \textbf{Current capacity estimation} \\ \midrule
        Z & 45.5\,GeV & 4 years & 65\,MW - 240\,MW & 30\,MW - 222\,MW \\ 
        WW & 80\,GeV & 2 years & 70\,MW - 265\,MW & 33\,MW - 247\,MW \\ 
        ZH & 120\,GeV & 3 years & 70\,MW - 294\,MW & 34\,MW - 273\,MW \\ 
        $\rm t \bar{t}$ & 175\,GeV & 5 years & 80\,MW - 350\,MW & 40\,MW - 350\,MW \\ 
        HH & 182.5\,GeV & 1 year & 50\,MW - 384\,MW & 41\,MW - 357\,MW \\ \bottomrule
  \end{tabular}
  \label{tab:energy_needs_per_mode}
\end{table}

The collider operates on a yearly schedule, with a limited period of beamline operation for physics research. The programme also includes phases for equipment testing and start-up, machine performance optimisation and routine maintenance. Depending on the operating mode, electrical energy consumption varies from year to year.

\begin{figure}[!h]
    \centering
    \includegraphics[width=\textwidth]{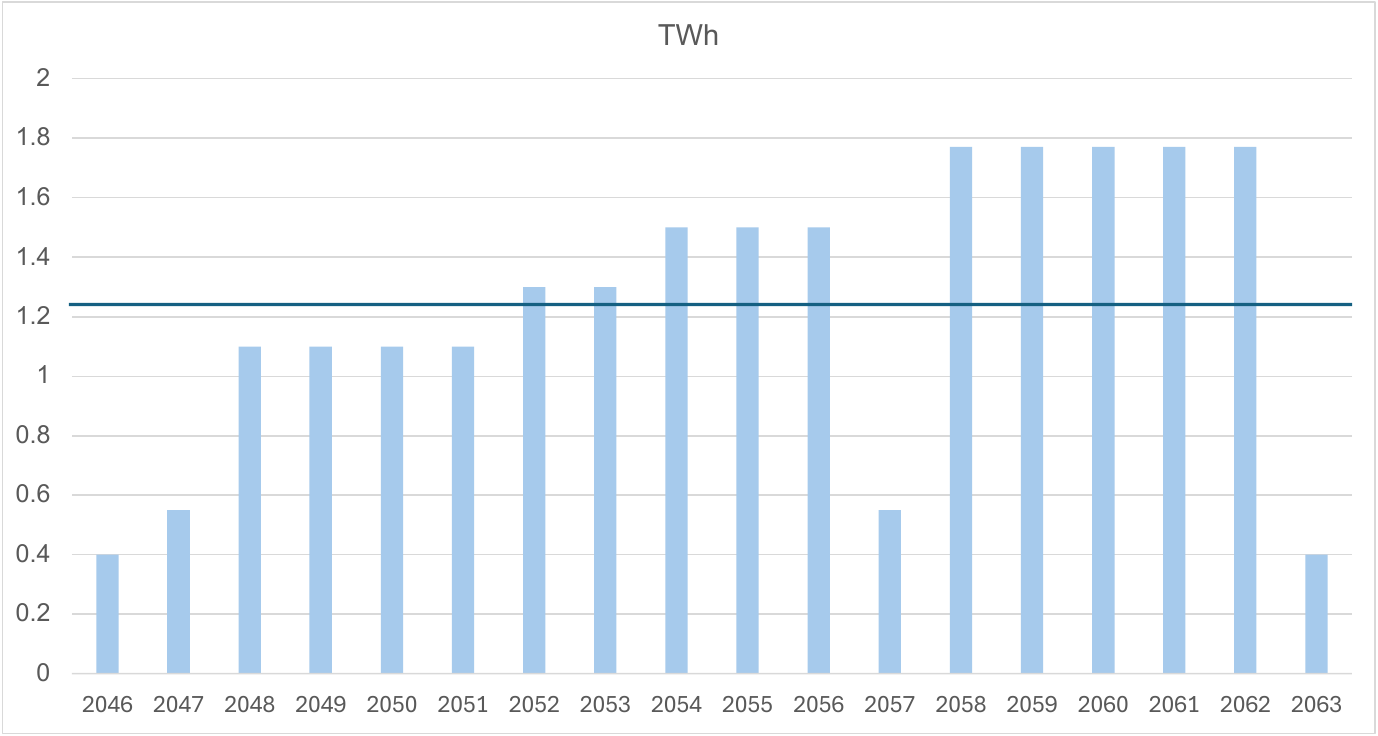}
    \caption{\label{fig:annual-electricity-need} Annual electricity requirements of the collider and its technical infrastructure. The average annual consumption over the programme is indicated by a blue horizontal line.}
\end{figure}

At this stage of the study, it is only possible to provide approximate yearly consumption figures (see Fig.~\ref{fig:annual-electricity-need}). The annual electricity consumption varies between 400\,GWh/year for basic services that are in use also without beam operation and 1770 GWh/year for the $\rm t \bar{t}$ operation phase. On average, the annual electricity consumption over the operation period is slightly above 1200 GWh/year. The capacities required to supply the particle collider with the required energy exist in France. However, a plan for a specific power purchasing agreement portfolio will have to be established, allowing ample preparation time before the energy is required to obtain favourable conditions and to contract a suitable energy mix with a low carbon footprint\cite{gutleber_2023_10023947, sophie_auchapt_2023_7781077}.

Following the preliminary and exploratory studies, more detailed concepts to optimise the energy performance of the accelerator's equipment and to improve operations based on information provided by electricity infrastructure operators and energy suppliers are required.

This approach will be based on the usual Avoid-Reduce-Compensate methodology.
\begin{itemize}
\item Avoid: the overriding goal is to limit consumption according to a cost-benefit analysis, which includes, for example: the research infrastructure layout of four experiments; limiting maximum annual power consumption, which will have impacts on luminosity and extending the duration of the research programme; and ensuring that systems do not consume energy when not in use.
\item Reduce: this means to optimise system efficiency and reduce losses. This includes, for example, improving the efficiency of electrical energy conversion for radiofrequency, reducing losses in internal distribution and equipment, energy recovery and storage, smart consumption based on needs (e.g., for ventilation and cooling), and developing systems that can switch more easily and quickly between operating modes (standby or operating mode).
\item Compensate: lastly, the goal of compensating is, on the one hand, to recover, store and supply renewable energy for society, and on the other hand, to develop synergies for the transition to energy from renewable sources, increase renewable energy capacity and cooperate internationally for the supply of renewable energy. Examples of compensation with direct economic benefits include the creation of energy communities and pooling for pre-financial-investment-decision support to build up renewable energy sources, use of waste heat in industrial processes (e.g., cheese production), for greenhouse operations, crop cultivation and the heating of public establishments such as hospitals, schools, and shopping centres.
\end{itemize}

All these measures will need to be improved over the fifteen years of the technical design and construction phases. They are an effective way of limiting electricity consumption and its impact.

With regard to the various phases of the collider maximum power requirements are only necessary during the $\rm t \bar{t}$ operation phase, and during an optional phase at the end of the programme (hh) when all the radiofrequency equipment is installed. On average, during the scientific research phase, the FCC-ee would consume approximately 1.3\,TWh per year. Over its entire lifetime, including shutdown periods and commissioning, its average electricity consumption would be around 1,3\,TWh per year. To provide a context for the impact of this electrical energy consumption, it can be compared to the electricity consumption of a state-of-the-art data centre. For example, the Altoona (IA) data centre in the USA, owned by the Meta company, most known for the Facebook, Instagram and WhatsApp applications, has an annual consumption of 1.24\,TWh\cite{MetaSustainability2024}. The carbon footprint of this data centre is 532\,158 tCO$_2$(eq) per year. This corresponds to about the carbon footprint of the entire FCC infrastructure construction. The consumption of all Meta company's data centres is 15\,TWh/year, i.e., a factor ten higher than the annual energy need of the FCC-ee. The total annual carbon footprint of all Meta's data centres totals to about 5\,million tons CO$_2$(eq). 

The specific energy needs will only be known after the detailed technical development phase, so it will be possible to take advantage of technical advances and development to improve energy efficiency.  

RTE, which was entrusted with the task of managing the electricity transmission grid in France\footnote{Decree no. 2005-1069 of 30 August 2005 approving the status of the company RTE EDF Transport: \url{https://www.legifrance.gouv.fr/loda/id/JORFTEXT000000812363}} through a public service contract dated 24 October 2005, which includes, among other things, the environmental integration of the grid (consultation, protection of landscapes and natural and urbanised environments) and safeguarding the public grid, is responsible for analysing the connection while taking into account technical choices; overseeing the hook-up process; and carrying out the necessary administrative procedures (connection agreement, grid access contract). RTE also carries out all works required to establish connections between surface site delivery points and the high capacity national electricity grid once the hook-up agreement has been signed\footnote{RTE, Instruction des demandes de raccordement, version 2, 17 Octobre 2019, \url{https://www.services-rte.com/files/live/sites/services-rte/files/documentsLibrary/DTR\%201.4.1\%20Proc\'edure\%20Racc\%20Conso\%20L342-2\%20v19\%2010\%2017\_fr}}.

According to the results of the preliminary technical feasibility study carried out by RTE \cite{billerot_2024_13364463}, which manages France's electrical grid, the connection to the high-power electrical infrastructure calls for three supply points in France at this stage. A backup power supply point in Switzerland can be considered, but if required, its technical and administrative feasibility remains to be studied.

\begin{figure}[!h]
    \centering
    \includegraphics[width=0.8\textwidth]{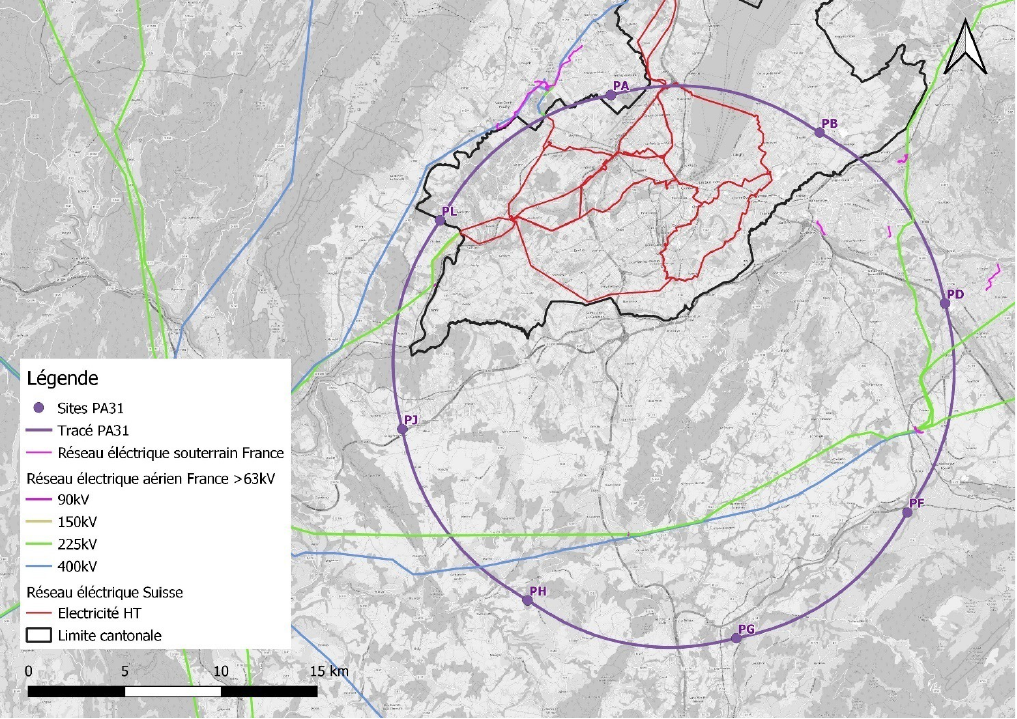}
    \caption{\label{fig:electricty-grid} Existing or planned electricity grids in France and Switzerland within the FCC perimeter. Sources: Public layer CAD\_ELEMENT\_CONDUITE from SITG 2023 for Switzerland (\url{https://sitg.ge.ch/donnees/cad-element-conduite}), available for consultation and extraction for free use; for France, Open Data R\'eseaux Énergies (ODRE, \url{https://opendata.reseaux-energies.fr)}.}
\end{figure}

At this stage, the footprint of the reference scenario is crossed by various power lines, mainly 63\,kV and 225\,kV. Two 400\,kV lines cross the PA31 footprint from west to east (see Fig.~\ref{fig:electricty-grid}). The 400\,kV lines pass close to the PL sites (Ain department in France) and the PF and PH sites (Haute-Savoie department, France). A major distribution station (Cornier) is located close to the PD and PF sites in France.
Electricity needs are higher at the PL and PH sites, as these are designed to house the radiofrequency systems that accelerate particles in the collider. 

Three supply points are currently envisaged to achieve a well-balanced electricity supply scenario: One in PL connecting to the nearby 400\,kV line. One in PD connecting to the Cornier substation. One re-using the existing CERN grid connection in Bois Tollot (Ain, France). A detailed design study is required by RTE to determine the route and specificities of the new 400\,kV connections. It may be advantageous to connect site PF instead of site PD to the 400\,kV line, leveraging the proximity of the RTE distribution point in Cornier and thus avoiding the need to cross the Arve river. It is worth noting that the planning, authorisation, contracting and creation of grid power connections require substantial lead times. Ten years should be assumed for the entire process of one connection. In addition, for the entire particle collider project's environmental authorisation process, the availability of environmental impact studies of the grid connections is required. Therefore, the detailed designs of the connections and their route variants based on the environmental constraints, technical feasibility and cost must be developed with high priority now, even if, ultimately, the connections are only required after the civil construction phase.

\subsection{Raw water supply}

The particle accelerators need raw water mainly to cool the magnets. Synchrotron radiation is the main source of heat in addition to numerous mechanical and electrical technical infrastructure components. All raw water can be taken from an existing raw water supply line provided by the local Swiss company Services Industriels de Genève (SIG)\cite{buri_2023_10014173} that sources the water from Lake Geneva, as is the case with CERN today. No raw water will be consumed from drinking water reservoirs or subsurface water layers.

The technical solutions and equipment choices will not be known until shortly before the procurement of the technical infrastructures, mid-way through the subsurface construction phase, in order to take full advantage of technical advances, including those in cooling system efficiency. Estimates of the maximum water requirements, based on consumption by CERN's existing accelerator cooling systems and on the current technical concept developments, have been established (see Table~\ref{tab:FCC-water-needs}. These values are the result of a gradual development of a concept that permitted reducing the raw water requirements from a maximum initial amount of \SI{5000000}{\cubic\metre} per year for $\rm t \bar{t}$ operation to about  \SI{3000000}{\cubic\metre} per year. The current reference water capacity needs to assume the use of closed-circuit water cooling systems with evaporation towers. The raw water consumption stems from the need to make up for the evaporated water in the secondary circuits of the cooling towers at each surface site.

\begin{table}[!h]
  \centering
 \caption{\label{tab:FCC-water-needs} Summary of the annual raw water needs.}
  \begin{tabular}{lrc}
        \toprule
      \textbf{Mode} & \makecell[c]{\textbf{Annual raw water} \\ \textbf{requirement}} & \textbf{Years} \\ \midrule
    Z & \SI{1604861}{\cubic\metre} & 4 \\ 
    WW & \SI{1928943}{\cubic\metre} & 2 \\ 
    HZ & \SI{2165458}{\cubic\metre} & 3 \\ 
    L.S. & \SI{163817}{\cubic\metre} & 1 \\ 
    $\rm t \bar{t}$ & \SI{3077591}{\cubic\metre} & 5 \\ \bottomrule
  \end{tabular}
\end{table}

For comparison, the raw water consumption at CERN in 2022 was as follows:

\begin{table}[!h]
  \centering
 \caption{\label{tab:cern-water-2022} Summary of the FCC raw water needs.}
  \begin{tabular}{lrl}
    \toprule
    \textbf{Item} & \textbf{Raw water need} & \textbf{Description} \\ \midrule
    SPS & \SI{944}{\cubic\metre} & BA2, BA4, BA5, BE2 \\ 
    LHC & \SI{795070}{\cubic\metre} & \makecell[l]{LHC complex, LHC2, LHC3.2, LHC 3.3,\\LHC4, LHC5, LHC6, LHC7, LHC8} \\ 
    \makecell[l]{Meyrin,\\Pr\'evessin} & \SI{2437988}{\cubic\metre} & \makecell[l]{Meyrin and Prévessin sites main supply, SPS BA1 and BA6,\\LHC1 safe supply, clubs, Globe} \\ \bottomrule
  \end{tabular}
\end{table}

The water supply scheme for raw water is based on the working hypothesis of using CERN's existing raw water supply, drawn from Lake Geneva by Services Industriels de Genève (SIG) in Switzerland.
The available capacity of \SI{604}{\cubic\metre/h} leading to a total capacity of more than 5\,000\,000\,m$^{3}$ per year is compatible with the needs of the FCC.
This was confirmed by an exchange between CERN and SIG in 2022. Then, in August 2023, SIG confirmed the technical feasibility of the supply within the existing contractual framework either by building a new, short connection of around 200\,m with a 500\,mm nominal diameter between the Tuileries-La Berne pipe and Point 8 of the LHC in Ferney-Voltaire (France), or by using two existing internal CERN lines which would require upgrading.

To make the distribution of water along the entire length of the accelerator technically easier and economically more advantageous, two additional water supply points can be considered in France from the Arve and/or the Rh\^one rivers. To this end, a specific territorial study would be needed, which would have to integrate the quantitative management plans available for water resources (PGRE). Such water intakes would require the creation of water filtration and treatment plants.

In an effort to reduce further the water capacity needs, initial studies have been carried out to identify promising levers (see Table~\ref{tab:water-saving}). First, the introduction of waste heat recovery and supply from the onset permits reducing the water intake needs significantly since less water needs to be evaporated if the heat is supplied to consumers.  The conservative scenario indicates the water-saving potential without adaptation of the classical operation schedule. Some commercial consumers need heat throughout the year. It has to be pointed out that the realistic waste heat reuse indicated requires adaptation of the particle collider operation to the season during which heat is required. Even higher saving potentials than the ones indicated are possible, depending on the adaptation to seasonal territorial heat needs. Second, the adaptive operation of the cooling system and the evaporation towers using advanced supervisory control and potentially artificial intelligence would permit the reduction of consumption to the strict minimum required, compatible with the actual cooling needs. Third, the creation of an additional water intake next to site PD (Nangy, France) would ease the requirements on the overall system. Given that a dedicated water filtering and treatment facility would be required, an initial study aims to verify the use of waste water from the nearby Bellecombe/Scientrier waste water treatment plant (Syndicat des Eaux de Rocailles et Bellecombe). This installation discharges treated water at an average rate of about \SI{550}{\cubic\metre/h} into the Arve River. While in principle technically feasible and economically viable (the annual operation cost per $m^3$ of water of a treatment plant required for the collider corresponds to the cost of a $m^3$ of raw water purchased from a water supplier), studies are currently ongoing to estimate the effort to reduce the residual dissolved calcium carbonate (CaCO$_3$) and germs in the water to render it compatible with use in the industrial cooling system.

\setlength{\tabcolsep}{5pt}
\begin{table}[!h]
  \centering
 \caption{\label{tab:water-saving} Water-saving potentials with the introduction of waste heat supply scenarios and use of treated waste water.}
  \begin{tabular}{crrr}
    \toprule
    \textbf{Mode} & \textbf{Conservative waste heat reuse} & \textbf{Realistic waste heat reuse} & \textbf{Treated waste water use} \\ \midrule
    Z & \SI{356800}{\cubic\metre/year} & \SI{476800}{\cubic\metre/year} & \SI{685000}-\SI{1000000}{\cubic\metre/year} \\ 
    WW & \SI{382400}{\cubic\metre/year} & \SI{523200}{\cubic\metre/year} & \SI{685000}-\SI{1000000}{\cubic\metre/year} \\ 
    HZ & \SI{409600}{\cubic\metre/year} & \SI{571200}{\cubic\metre/year} & \SI{685000}-\SI{1000000}{\cubic\metre/year}\\ 
    L.S. & \SI{96000}{\cubic\metre/year} & \SI{96000}{\cubic\metre/year} & \SI{685000}-\SI{1000000}{\cubic\metre/year}\\
    $\rm t \bar{t}$ & \SI{473600}{\cubic\metre/year} & \SI{678400}{\cubic\metre/year} & \SI{685000}-\SI{1000000}{\cubic\metre/year}\\ \bottomrule
  \end{tabular}
\end{table}
\setlength{\tabcolsep}{6pt}

At this stage, from quantitative and commercial points of view, the reference scenario for water consumption is technically, financially, and territorially feasible, since water extraction and consumption represent quantities lower than CERN's actual past consumption. The availability of a supply representing twice the total maximum requirement was confirmed in 2023.

\subsection{Waste water management}

Connections to the local sewage infrastructure are required at all sites for the management of drainage water in the subsurface structures, rainwater collected and treated water that is purged from the raw-water based cooling systems. The collected rainwater can also be reused on the sites for different project-related purposes and to maintain green spaces. Where possible, all collected water will be filtered and treated on-site before it is released into the environment. From a territorial perspective, this approach is preferred over centralized wastewater management since it is beneficial for sustaining existing creeks, wetlands, and biodiversity in rewilding projects developed in association with the surface sites. Only where the water does not qualify for free release would it be directed into the sewage system (e.g., due to a higher percentage of non-soluble residuals or during periods of heavy rain).

Alternatively, a central waste water management concept based on returning all waste water to a CERN site (e.g., LHC P8 and PA in Ferney-Voltaire) can be considered. For the management of the cooling water, a 'zero liquid discharge' (ZLD) approach can also be considered. ZLD requires the collection of solids from each surface site at regular intervals.
A definitive choice of technology has yet to be taken and calls for a comprehensive, wider Cost-Benefit Analysis over the entire project period, covering investments and operational factors. Water treatment technology is advancing rapidly due to environmental and sustainability constraints. It is, therefore, wise to consult experienced companies and consider the various options in the frame of the environmental impact assessment before finalising the technology choice.

Waste water from human activities (toilets, sinks, offices, visitor centres, caf\'es and restaurants) will be directly evacuated via connections to the public waste water system. The design for the waste water network connections is to be developed in cooperation with local public administration services in a subsequent preparatory phase, involving companies' experts in the domain. At this stage, the following configuration is envisaged for the local connections to the waste water networks:

\begin{itemize}
\item Site PA: Connection to the waste water network via the LHC Pt8 in France.
\item Site PB: Connection to the local waste water network in Presinge, Switzerland.
\item Site PD: Direct connection to the waste water treatment station (`STEP') SRB in Scientrier, France.
\item Site PF: Connection to the local waste water network in \'Eteaux, France.
\item Site PG: Connection to the local waste water network via the Groisy autoroute station. Alternatively, connection to the network in Charvonnex at the D1203 in France.
\item Site PH: Creation of a new local waste water network to the nearest connection point at 250 to 300\,m distance in France. The waste water treatment station (STEP) in the vicinity of the site is not capable of accepting all of the waste water from the site in the most demanding cases. Therefore, either the STEP needs to be reinforced for the collider project or the excess water needs to be re-directed via a waste water network in the tunnel to sites PG and PJ. Developing the strategy is part of a subsequent design phase.
\item Site PJ: Connection to the local waste water network 500\,m away in between Vulbens and Valleiry in France.
\item Site PL: Connection to the local waste water network in Challex, France.
\end{itemize}

\subsection{Emergency services}

\begin{figure}[!h]
    \centering
    \includegraphics[width=\textwidth]{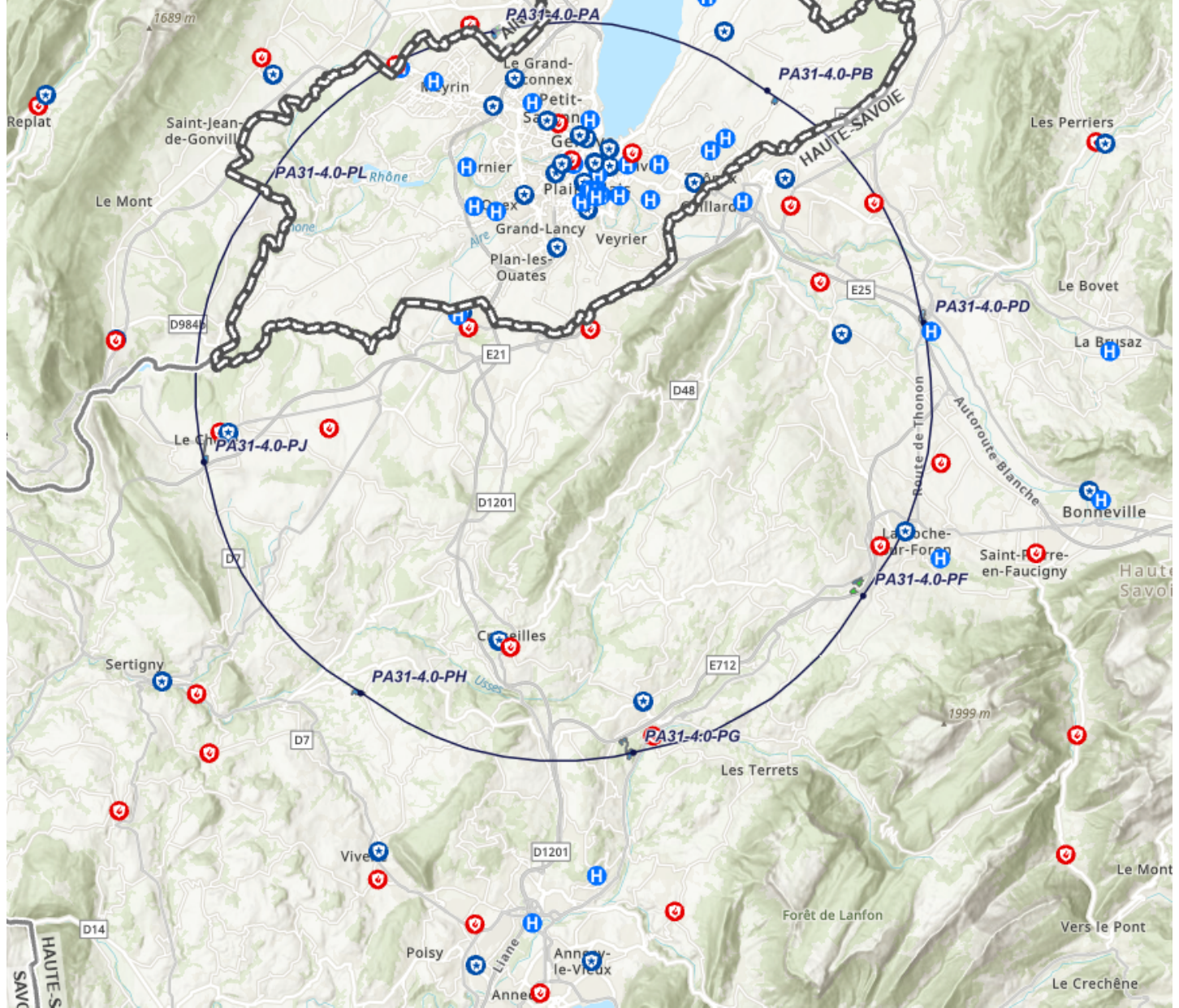}
    \caption{\label{fig:fire-emergency-services} Fire-fighting and emergency services in the perimeter of the reference scenario.}
\end{figure}

The analysis of the different implementation scenarios revealed that the continuation of the existing approach of serving surface sites of CERN's particle accelerators cannot be extended to the perimeter of the future particle collider without significant challenges. Although the surface sites are in the immediate vicinity of major road infrastructures, the distances and the traffic situation would call for intervention times, which would be too long if all sites had to be serviced from the CERN Meyrin site in Switzerland. Locating dedicated emergency service personnel on each of the surface sites is prohibitive from a financial point of view in terms of human resources and equipment, and would lead to challenges in the operational management of the facility. Subsequent analysis of a single dedicated support pole provides only limited improvements with respect to safety and emergency services. Therefore, a detailed geographical analysis was carried out to identify alternative approaches (see Fig.~\ref{fig:fire-emergency-services}).

Apart from site PH, the surface sites of the reference implementation scenario are close to firefighting and emergency service stations (see Table~\ref{tab:emergency-servicee}). Some of them have recently been constructed. Therefore, the project scenario can consider making use of support for emergency and firefighting personnel at those stations, accompanied by regular common training and an emergency guidance centre at CERN. The strengthened collaboration of CERN with those services, the contribution of equipment, materials and training also leads to socio-economic benefits. These have been analysed and are part of the wider socio-economic impact assessment.

\setlength{\tabcolsep}{3pt}

\begin{table}[!h]
  \centering
 \caption{\label{tab:emergency-servicee} Selection of emergency services in the vicinity of surface sites.}
 \small{
  \begin{tabular}{crlrl}
    \toprule
    \textbf{Site} & \textbf{Distance} & \textbf{Location} & \textbf{Capacity} & \textbf{Service} \\ \midrule
    PA & \SI{1}{\kilo\meter} & Pr\'evessin, France & 66 persons & Departmental fire fighting station CIS Est-Gessien \\ 
    PB & \SI{5.8}{\kilo\meter} & Ch\^ene-Bougeries, Switzerland & 30 persons & Cantonal fire fighting station \\ 
    PD & \SI{0.3}{\kilo\meter} & Contamine-sur-Arve, France & n/a & Hospital, Centre Hospitalier Alpes-L\'eman \\ 
    PD & \SI{9}{\kilo\meter} & Arenthon, France & 11 persons & Fire fighting and first aid station \\ 
    PD & \SI{12.1}{\kilo\meter} & \'Eteaux, France & 68 persons & Regional fire fighting station\\ 
    PF & \SI{3.2}{\kilo\meter} & \'Eteaux, France & 68 persons & Regional fire fighting station\\ 
    PG & \SI{2.3}{\kilo\meter} & Groisy, France & 50 persons & Regional fire fighting station, emergency service \\ 
    PH & \SI{10}{\kilo\meter} & Frangy, France & 37 persons &  Local fire fighting station \\ 
    PH & \SI{14}{\kilo\meter} & Epagny, France & 135 persons & Regional fire fighting station, emergency service \\ 
    PH & \SI{1.3}{\kilo\meter} & Valleiry, France & 26 persons & Regional fire fighting station, emergency service \\ 
    PL & \SI{7.4}{\kilo\meter} & Thoiry, France & 70 persons & Regional fire fighting station, emergency service \\ \bottomrule
  \end{tabular}
  }
\end{table}

\setlength{\tabcolsep}{6pt}

The build up of these resources and the transition to a new operation scheme require several years of preparation. The ten years of construction phase provide an adequate window of opportunity to launch this process, which requires agreements with the emergency and fire-fighting services.

\subsection{Territorial aspects for the management of excavated materials}

The management of the excavated materials is treated as a project management and socio-economic challenge rather than a civil engineering aspect. The availability of suitable deposits for inert waste decreases continuously, and prices for depositing materials continue to rise correspondingly. To provide an order of magnitude, the cost for the final deposit of the excavated materials can range between 385\,million euro and 825\,million euros depending on the possibilities to re-use materials or not\cite{ulrici_2025_14923266}.

While the limestone fraction of the excavated materials can be re-used in the project, the majority of the 6\,million m$^3$ of in situ volume, 95\% molasse and 3\% morraines are the focus of the re-use pathway developments. The aim is fourfold:
\begin{enumerate}
    \item Reduce the burden on the already tense situation. with respect to finding suitable deposits.
    \item Reduce the needs for long road transport.
    \item Reduce the overall project costs.
    \item Contribute to the creation of incremental socio-economic benefit generation by levering the project as a pilot for innovations in the area of excavated materials management. 
\end{enumerate}

Concerning the aim to generate socio-economic benefits beyond the project, it has to be kept in mind that the molasse basin stretches north of the Alps from the Geneva region across Switzerland, Germany, Austria up to Hungary. The use of novel methods to re-use such materials is therefore significant, and so are the socio-economic benefits potentials.

In the frame of the FCC feasibility study, leveraging the contributions of the European Commission in the H2020 co-funded FCC Innovation Study project (FCCIS), an international, challenge-based competition called `Mining the Future' was launched to explore credible, technically feasible and economically viable pathways for the development of molasse re-use on the 2030 timescale\cite{mining-future-comp-results}.

Depending on the regulatory framework conditions in France and in Switzerland, between 15\% and 30\% of the molasse materials could be considered naturally polluted (e.g., naturally present hydrocarbons, nickel, zinc and chromium). Hence, there will always be a residual quantity of materials that has to be transported to deposits that accept such materials. It is assumed that a large amount of the other materials serves the refilling of quarries and the rewilding of the filled quarries.

\begin{figure}[!ht]
  \centering
  \includegraphics[width=\textwidth]{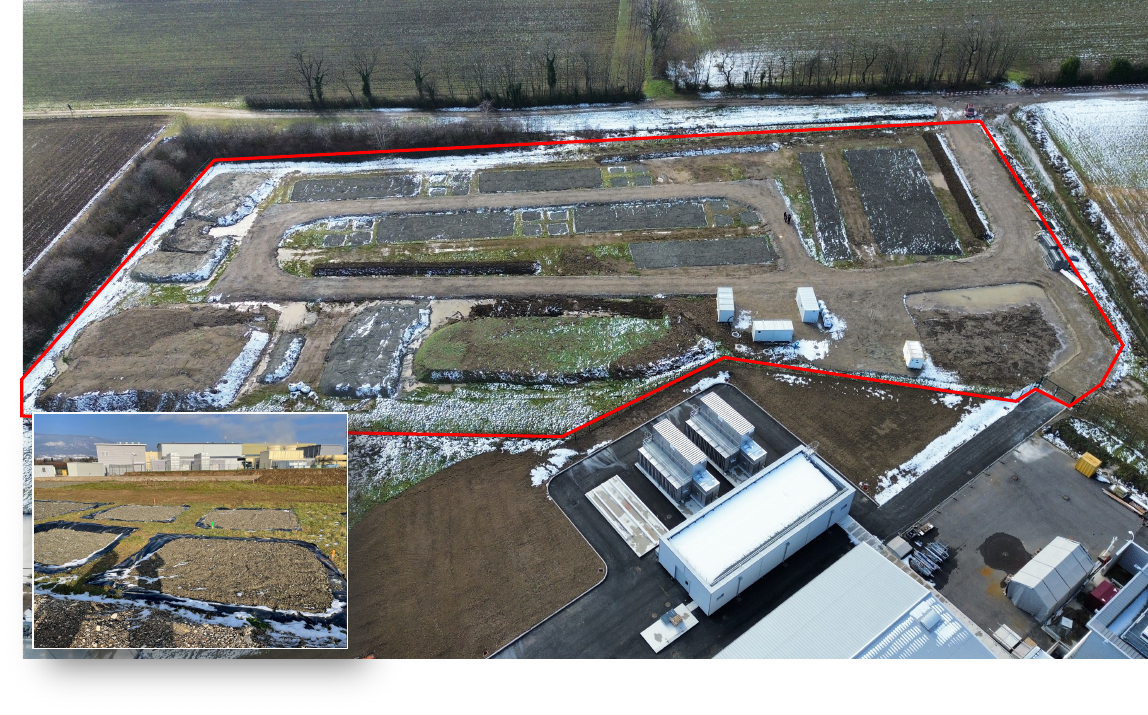}
  \caption{The OpenSkyLab field laboratory on 1\,ha of land marked with a red line, next to the CERN CMS Point 5 in Cessy, France, is developing quality managed processes for the transformation of excavated materials for use in rewilding and other societal applications. The field was prepared in the winter of 2025. The scientific development will last for at least four years.}
  \label{fig:openskylab}
\end{figure}

Further pathways are currently being conceived in the frame of a field laboratory called `OpenSyLab' (see Fig.~\ref{fig:openskylab}) on CERN premises based on scientific protocols that permit the development of quality-managed processes and products, a legal prerequisite for the re-use of materials. Since this innovation project only started in 2025 and much more extensive and detailed information is required about the characteristics of the expected excavated materials by means of geotechnical sampling (extraction of borelogs in a representative set of locations along the collider tunnel alignment), a specific plan for the management of excavated materials can only be drawn up at a later stage. It takes about 5 years to develop the processes and products to transform the molasse rock into soil that can be mixed with already fertile soil and which can be effectively used for applications in agriculture, forestry, rewilding projects, paths and roadside maintenance and further innovative applications such as thermal insulation and as construction material.

The authorisations to implement these pathways in the frame of a new construction project can only be requested from the national authorities in France and Switzerland once quality-managed processes and specific and localised descriptions of the re-use pathways exist and have been validated.

Excavated materials have waste status in national legislation due to formal criteria, independent of their origin, treatment and degree of pollution. The application of these formal criteria renders the implementation of ecological and economical meaningful principles of circular economy more difficult. On 13 April 2021, Porr Bau GmbH appealed to the Court of Justice of the European Union (CJEU) to review a preliminary ruling under Article 267 TFEU from the Landesverwaltungsgericht Steiermark (Regional Administrative Court, Styria, Austria), made by decision of 2 April 2021 in the proceedings Porr Bau GmbH vs. Bezirkshauptmannschaft Graz-Umgebung concerning the conclusion of the Austrian national court's finding that excavated materials discharged on cultivation areas constituted waste.

In case C-238/21, ECLI:EU:C:2022:885 [1], the Court (First Chamber) ruled:\\
Point 1 of Article 3 and Article 6(1) of Directive 2008/98/EC of the European Parliament and of the Council of 19 November 2008 on waste and repealing certain Directives, must be interpreted as precluding national legislation under which uncontaminated excavated materials, which, pursuant to national law, are in the highest quality class,
\begin{itemize}
\item must be classified as ‘waste’ where their holder neither intends nor is required to discard them and those materials meet the conditions laid down in Article 5(1) of that directive for being classified as ‘by-products’, 
and 
\item only loses that waste status when they are used directly as a substitute and their holder has satisfied the formal criteria which are relevant for the purposes of environmental protection, if those criteria have the effect of undermining the attainment of the objectives of that directive.
\end{itemize}

As a direct consequence of the judgement, non-contaminated excavated materials can be used for ecological and economically meaningful applications if the materials meet the applicable quality criteria. Formal criteria must not preclude and hamper the use of excavated materials and hinder the implementation of circular economy principles. The end-of-waste status can be reached by mere quality test. The pre-treatment, treatment, and transformation of the excavated materials have no impact on the possibility to use the materials, since they are integral parts of a production process described in EU Directive 2008/98/EC on waste Article 6.

Therefore, re-use of excavated materials is possible in an EU country if one knows the quality of the excavated materials to a sufficient extent to organise the re-use before extraction and concludes agreements with customers for the materials for applications that are compatible with the materials' quality that can be delivered including any potential processing before delivery and the re-use pathway is ecologically justified and economically viable.

To support this process, the project needs to implement a comprehensive monitoring system for the analysis of the materials excavated, their treatment on the excavation site, their pre-processing and processing, the production of the end product, the transport to the customers (audit and traceability) and the monitoring of the re-used materials after the customers have accepted them.

Finally, the transnational transport must be carefully planned. Since the fair share principle yields an amount of materials under Swiss territory that is larger than the volume excavated on this territory, a solution for either the repatriation of excavated materials from France to Switzerland or a form of compensation needs to be developed before the excavation process can start.

\begin{figure}[!ht]
  \centering
  \includegraphics[width=\textwidth]{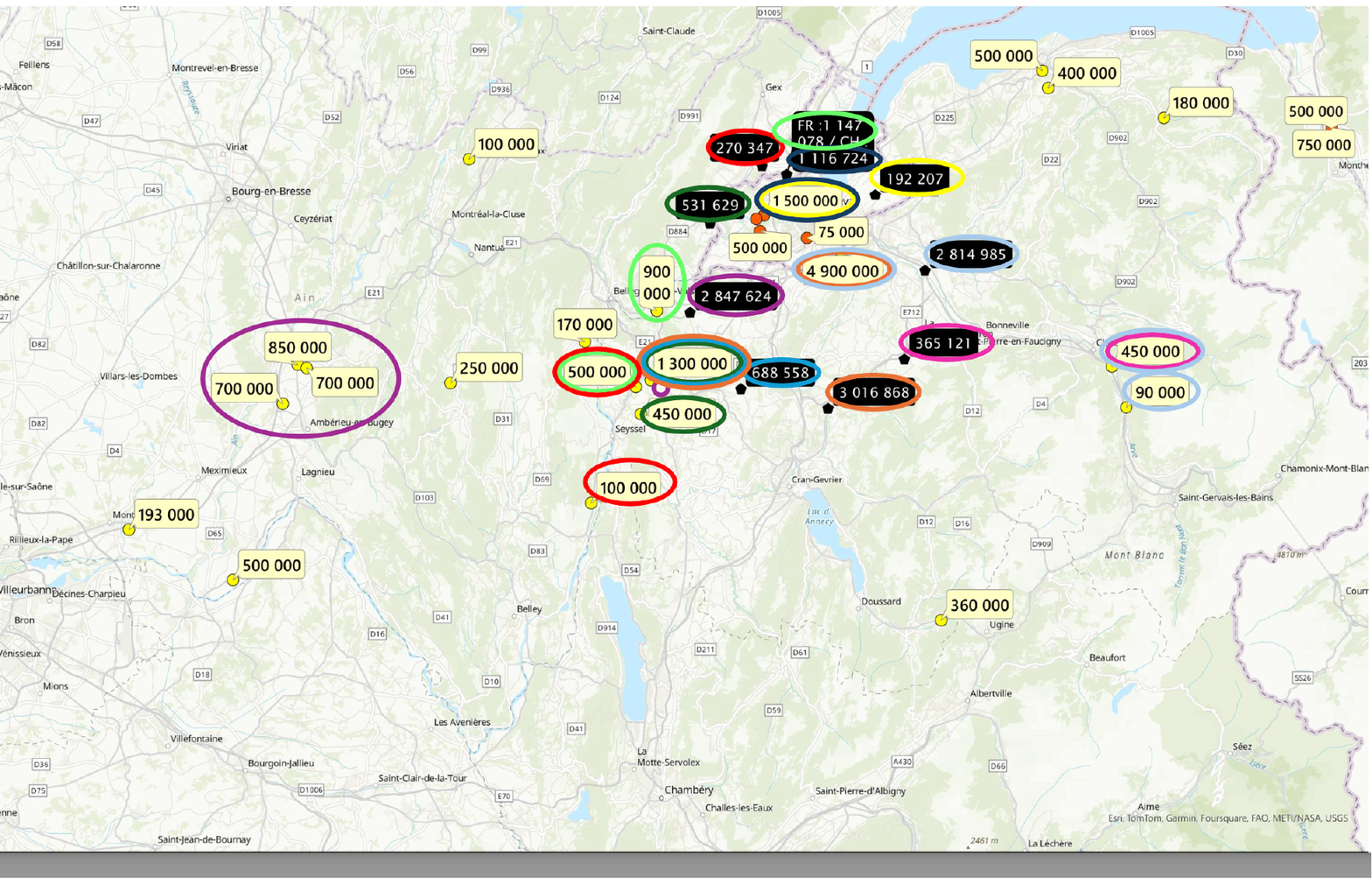}
  \caption{Capacities of quarries that could be re-filled in the vicinity of the project on French territory and in the canton of Geneva in Switzerland. Capacities are indicated in tonnes.}
  \label{fig:quarries}
\end{figure}

For the refilling of quarries, the feasibility study has identified suitable locations on the 2030 time horizon. However, it must be kept in mind that these locations are rapidly being consumed by other construction processes, and therefore, capacities must be reserved in good time via agreements between CERN and the quarries. While the study in France could be exhaustively carried out for the region concerned (Fig.~\ref{fig:quarries}), data for potential quarries and mines that can be re-filled remain incomplete for Swiss territory today.

While transport of the materials in the vicinity of the project by truck along major transport axes remains the primary assumption, transport between the excavation sites and the transport axes is preferably carried out with alternative means such as conveyor belts. Transport to re-use locations at further distances is preferably carried out by train.

As a first step, a strategy for the management of excavated materials has been established as a joint effort of technical domain experts and organisations in host states that regularly accompany large-scale construction projects. Currently, processes and products are being developed to demonstrate the reuse potentials of molasse materials. At the same time, a baseline plan for the management of the expected materials has to be drawn up that confirms that the quarries and mines and the deposit availabilities can serve a conservative excavated materials scenario in both host countries. Work with host state authorities is needed to present the quality managed re-use processes and to obtain the authorisations for the application of these processes. Agreements with customers need to be established in order to be able to implement the re-use pathways. Significantly more samples and detailed analysis of geological samples need to be taken along the tunnel alignment to come to a more precise estimate of a re-use scenario. To technically support the strategy, the online materials analysis during tunnelling and the on-site treatment and processing of the materials in a modular plant need to be demonstrated and brought to the industrial application level (TRL level 9) within at most five years.
The open question of railway transport, crossborder traffic and repatriation of excavated materials needs to be resolved. Finally, alternative re-use possibilities at larger distances should be investigated, since once materials are on a railway track they can be delivered with only a little additional carbon footprint. This opens up possibilities for re-uses that are currently not considered.

\subsection{Transport and mobility}
\label{sec:transport_and_mobility}

Concerning transport of materials and equipment and mobility of persons, the following topics are identified in the frame of the territorial implementation project development:

\begin{enumerate}
\item Transport of construction materials to the construction sites.
\item Evacuation of excavated materials from the construction sites.
\item Commute of construction workers to and from construction sites.
\item Transport of technical infrastructure and particle accelerator equipment to the surface sites.
\item Transport of goods and consumables to the surface sites for operation, maintenance, and repair.
\item Commute of personnel to and from surface sites for operation, maintenance, and repair.
\item Visitor-induced traffic at experiment sites.
\end{enumerate}

The most important contributor to additionally induced traffic is the evacuation of excavated materials. The strategy for a future project is to limit the use of trucks for transport from the sites and to rely on alternative approaches such as conveyor belts and ropeways to create links to nearby major transport axes (e.g., autoroute service stations, railway terminals, multi-lane departmental roads). Such modalities are also suited for bringing in certain construction materials from the major transport routes to the sites. Where this is not possible, trucks bring in equipment from the major transport axes via temporary routes, e.g., for pre-cast concrete elements and bulky particle accelerator and technical infrastructure equipment. Only where such an approach cannot be avoided local roads will be used for limited construction and installation-related activities. Special transports will be unavoidable, but occur only very rarely and according to planned and authorised schedules and conditions. Construction materials-related traffic is very limited, typically less than 10 deliveries per site that features a tunnel boring machine. The traffic during the installation phase is of the order of 9 deliveries for a technical site and 18 deliveries for an experiment site. The limiting factor is the amount of materials and equipment that can be transferred from the surface to the subsurface, as well as the limited transport and installation capacities in the underground structure.

\begin{figure}[!ht]
  \centering
  \includegraphics[width=\textwidth]{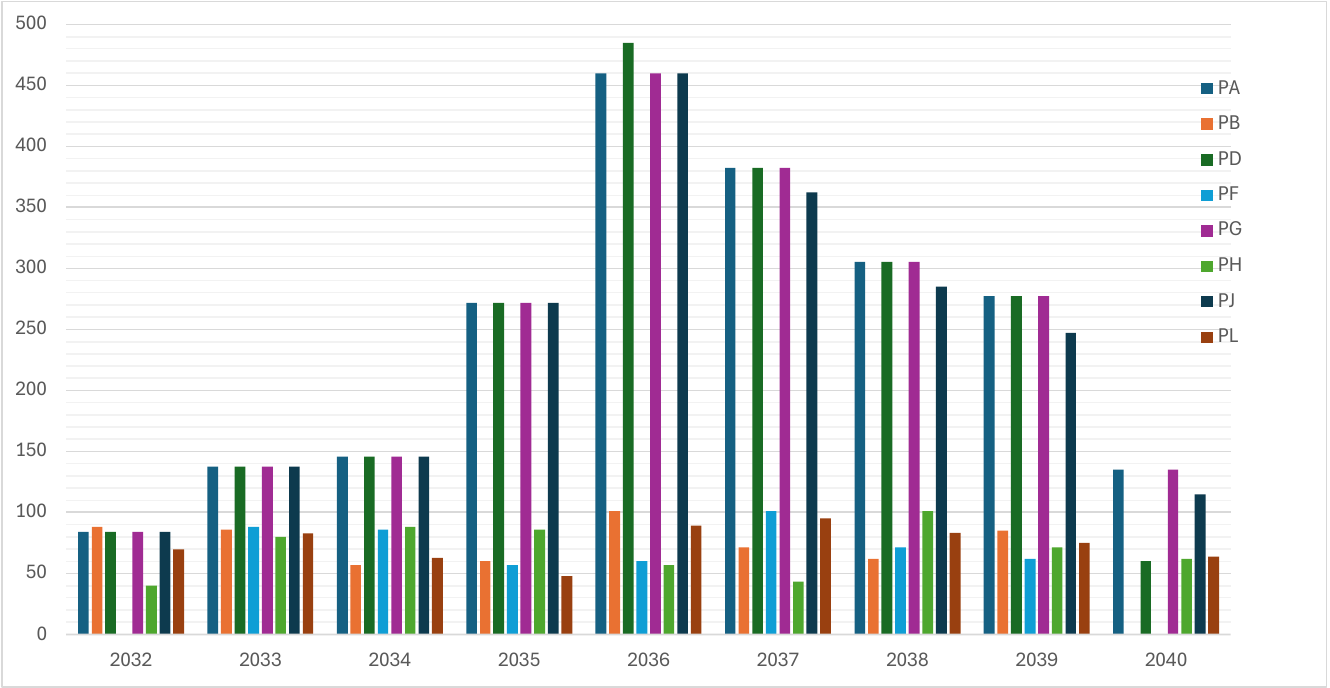}
  \caption{Example scenario of construction workers per construction site and year. A specific personnel and mobility plan has to be developed in the frame of a preparatory phase project when construction activities are defined at a more detailed level.}
  \label{fig:workers}
\end{figure}

Today, a preliminary estimate of the number of construction workers per site shows that the presence varies between approximately 50 and 450 throughout the multi-year construction phase (Fig.~\ref{fig:workers}). Significant differences in workers on construction sites occur between sites with tunnel boring machines (PA, PD, PG, PJ) and sites without tunnel boring machines (PB, PF, PH, PL). Sites without tunnel boring machines do not see the presence of more than 100 persons at a time. The amount of workers exceeds 250 people only during the peak activity period between 2035 and 2039. A specific personnel mobility plan can only be developed once the construction activities are defined in greater detail during a preparatory phase project to prepare market surveys and tenders for the construction. To avoid that workers commute individually between the construction sites and their residences, the construction plan will have organised shuttle transfers as is best practice in construction and industrial operation. The same approach will be taken for the installation phase during which about 200 to 300 people would be active on experiment sites and about 100 people on technical sites. A small fraction of people will have to rely on individual car transport. However, it is foreseen to keep such traffic limited to cases where such transport is needed.

Personnel commuting during operation is negligible. For maintenance and repair periods, shuttle transfer can again be foreseen. This also supports the participation of members of the international collaboration who do not necessarily have individual transport means. No more than 10 to 30 people are expected to be present on a surface site, depending on the type: a technical site or an experiment site.

Visitor traffic to and from experiment sites is typically centrally organised with shuttles. If a site features a visitor facility or additional cultural, educational, and leisure infrastructure, capacities for individual traffic have to be planned. Assuming an expected presence of 25,000 visitors per year and per site and 50\% of individual visitors with 2 persons per vehicle, the individual traffic is in the order of 5 to 10 vehicle round trips per site and day. It needs to be noted that visits only take place during about 250 days per year and during about 8 hours per day.

\subsection{Landscape integration and architectural strategy}

\subsubsection{Context}

Generic technical designs of surface site buildings have been developed in the frame of the feasibility study to understand the space needs better, the compatibility of the required technical infrastructures with the numerous surface site location constraints (e.g., topography, relief, visibility, nature, technological risks, access, and many others) and the costs. These conceptual developments are not to be confused with specific surface site designs that are eventually turned into plans to be included in the environmental authorisation process. This activity also required and engagement of local stakeholders in participative workshops, compliance with numerous applicable requirements and constraints such as functional and structural performance requirements, zoning limitations (permitted land use), soil and subsurface protection, urban planning documents and regulations, national and international norms, building codes and standards, energy efficiency regulations, safety, emergency response, noise protection, containment of artificial light pollution, accessibility, habitat and biodiversity protection (e.g., corridors and continuities), cultural and heritage considerations, landscape integration and ultimately, a societal licence to operate.

For these reasons and to prepare a potential subsequent project preparatory phase that needs to include the developments of the specific surface sites and the constructions on those sites, the feasibility study launched a first exploratory analysis to identify suitable components for an architectural toolkit that can guide this subsequent work. The architecture concepts have been developed by an expert company with experience in the field of innovative architecture, building design and urban development and planning. A multi-sectoral team of architects, planners, designers and urban planners came together to draw up the seeds that need to be further developed into a comprehensive architectural toolkit for the design phase.

\begin{figure}[!ht]
  \centering
  \includegraphics[width=\linewidth]{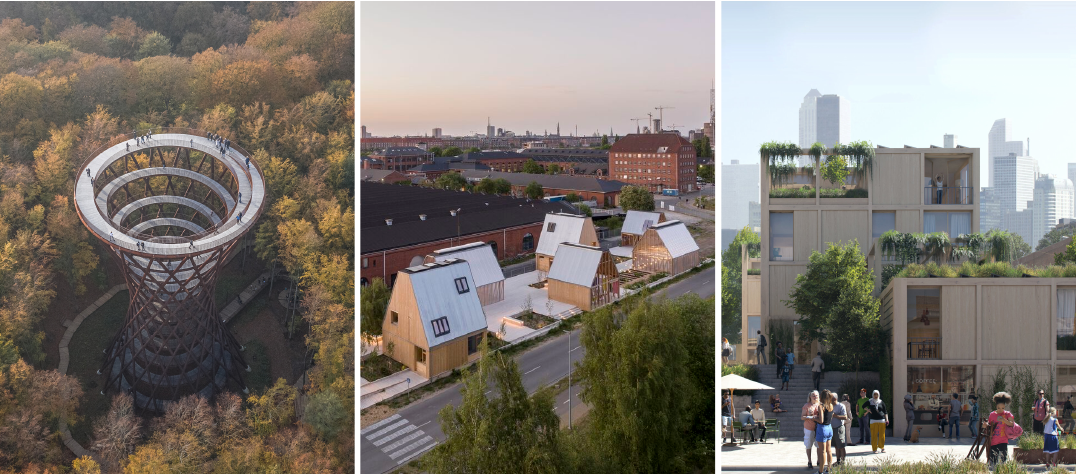}
  \caption{Three examples of innovative projects designed by the company that was engaged in the feasibility study to develop and architecture guidance toolkit. From left to right: forest tower, living places, urban village.}
  \label{fig:effekt-projects}
\end{figure}

The elements of the architecture toolkit are all based on actually implemented projects Three noteworthy examples (Fig.~\ref{fig:effekt-projects} developed by the contract company include:

\begin{enumerate}
\item The Forest Tower\footnote{\url{https://www.campadventure.dk/en/skovtaarnet/}} at Camp Adventure Park in Gisselfeld Klosters Skove, Denmark. This project features a 900-metre boardwalk connected to a 45-metre-tall observation tower, allowing visitors to experience the forest from a unique vantage point. The continuous ramp design ensures accessibility for all visitors, regardless of physical condition.
\item The Living Places Copenhagen, which demonstrates a new way of building homes with a carbon footprint of 3.8 kg/CO$_2$/m$^2$/year, three times lower than the current average. This project showcases working sustainability and innovative designs.
\item The Urban Village project develops a model for developing affordable and liveable homes. It comprises the designs of modular, affordable and low-carbon footprint buildings that can be easily assembled in constrained, urban environments.
\end{enumerate}

Another example is the project CO-EVOLUTION, a Danish/Chinese collaboration on sustainable urban development in China, which was awarded the Golden Lion in 2006 at the Venice Biennale of Architecture.

\subsubsection{Architectural toolkit}

The motivation for the development of an architectural toolkit is to support the integration of surface sites as early as possible into their territorial contexts which differ from site to site. An additional question that the toolkit aims to address is how new surface sites can create benefits to the territory and the local communities around the sites.

The toolkit builds on three pillars:
\begin{enumerate}
\item Architecture concepts and elements,
\item Landscape and
\item Community 
\end{enumerate}

\begin{figure}[!ht]
  \centering
  \includegraphics[width=\linewidth]{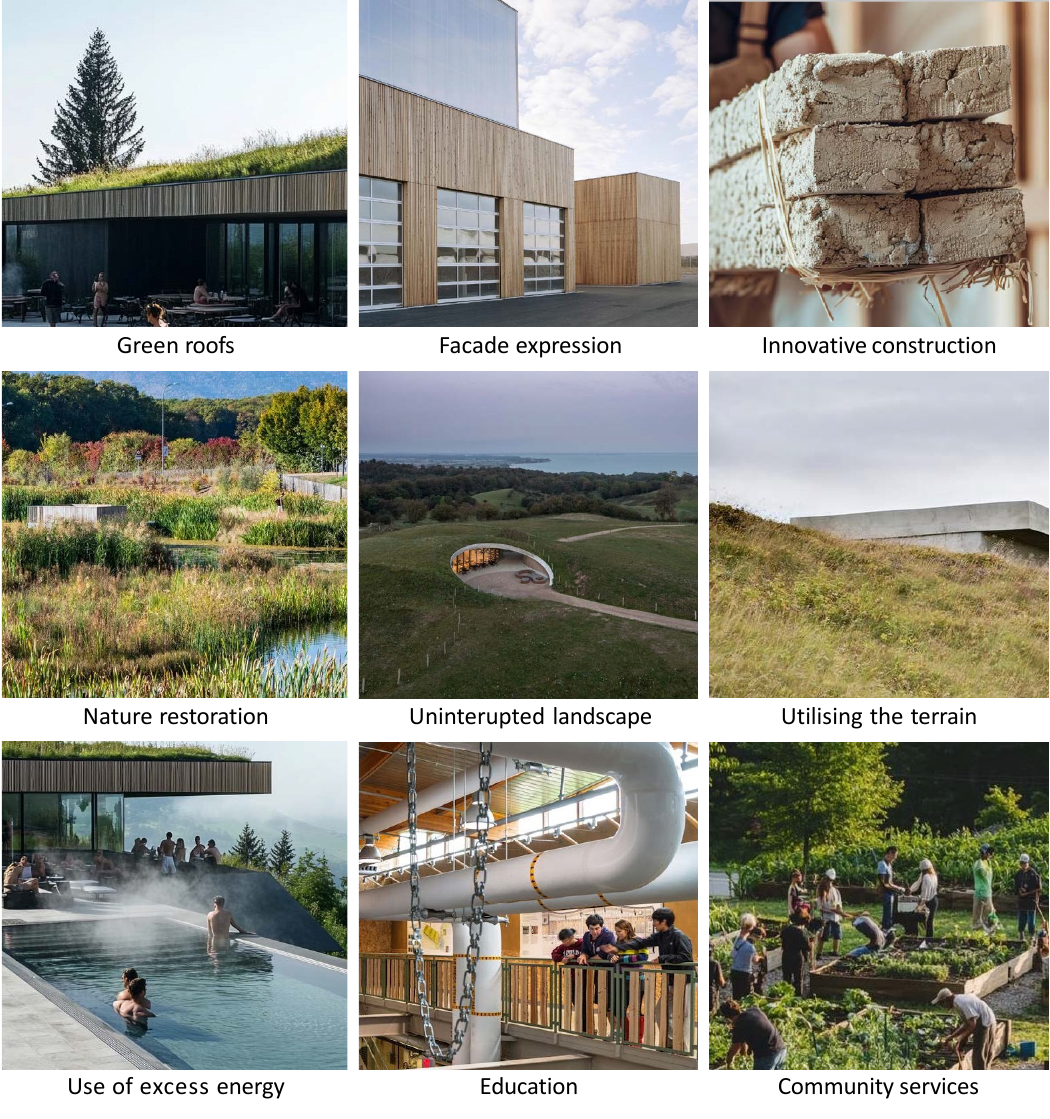}
  \caption{Examples of elements from the architectural toolkit conceived as a basis to further develop architectural guidelines for surface site developments.}
  \label{fig:architectural-toolkit}
\end{figure}

The three building blocks, architecture, landscape, and community (Fig.~\ref{fig:architectural-toolkit})) are the focus of the integration strategies for the surface sites. The strategies aim to balance technical requirements with environmental sustainability and community engagement. The incorporation of architectural innovation, landscape restoration, and public amenities seeks to minimise the ecological and visual footprint while fostering biodiversity and local engagement.

\paragraph{Architecture}

\begin{figure}[!ht]
  \centering
  \includegraphics[width=\textwidth]{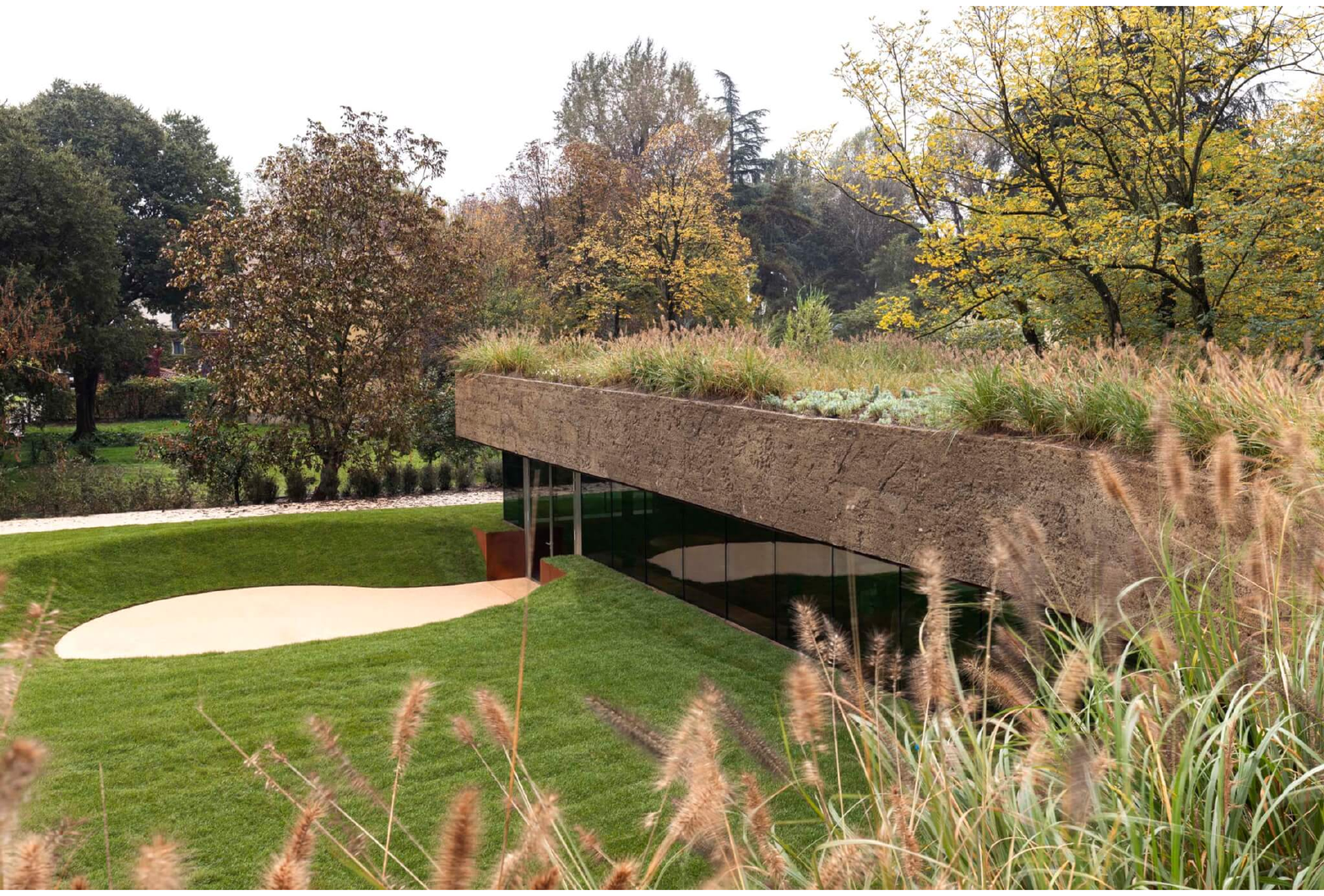}
  \caption{The green roof created from the excavated materials of the construction site allows the building to blend with its natural surroundings. (Agnese Bedini and Melania Della Grave/DSL Studio)}
  \label{fig:mutti}
\end{figure}

The architectural design concepts emphasise sustainable and aesthetic integration into the environment. Designers will adopt modern approaches such for instance green roofs, innovative construction techniques, and facade expressions to achieve these goals. These methods reduce the industrial structures' visual dominance. They facilitate a seamless transition between buildings and the surrounding landscape. Designers are encouraged to explore novel materials and construction techniques that align with both environmental and economic objectives. The incorporation of green roofs not only mitigates environmental impacts but also enhances the visual cohesion between built and natural environments. An example for a recent implementation of this concept is the one of Carlo Ratti for the Mutti food processing company in Italy \cite{Roche2024} (See Fig.~\ref{fig:mutti}).

\paragraph{Landscape design}
The project plans to employ strategies that prioritise ecological restoration and visual harmony. Key components include:

\begin{itemize}
\item Nature restoration: Restoration efforts enhance local ecosystems by promoting biodiversity and increasing resilience against environmental changes.
Utilisation of natural terrain: By embedding structures into the existing landscape, designers minimise visual disruptions and maintain natural terrain continuity.
\item Utilising the terrain: Working with the natural terrain minimises the facilities’ visual presence in the surroundings.
\item Green buffers: The implementation of vegetative barriers reduces the visual and environmental impact of facilities, supporting local flora and fauna.
The landscape design ensures that facilities are integrated subtly into the environment, allowing nature to thrive while masking infrastructural elements from view.
\item Uninterrupted landscape: By embedding constructions into the relief and the landscape, nature can flow uninterrupted above, while the facilities can be partially hidden from view.
\end{itemize}

\paragraph{Community engagement}

The project is committed to actively involve local communities surrounding the sites by creating shared spaces, educational opportunities, cultural spaces and visit facilities. Depending on the community support, professional and/or leisure services can be foreseen in the designs. Examples of specific actions include, for instance, the establishment of recreational areas such as community gardens, which utilise unused spaces while benefiting local populations. Providing educational platforms and visiting facilities allows the public to learn about the project’s scientific objectives and technical achievements. They permit opportunities for direct engagements with scientists and the international engineering community. Leveraging excess energy for community purposes, such as heating public swimming pools, health and relaxation facilities and supporting other leisure activities, are further possibilities that can be taken into consideration.
Through these initiatives, the project fosters a sense of inclusion and provides value to surrounding communities beyond its core scientific mission.

The integration strategies for the FCC surface sites reflect a comprehensive approach to addressing environmental, technical, and social dimensions. By combining architectural innovation, ecological restoration, and community-centred design, the project ensures a sustainable and harmonious coexistence with its environment. These strategies not only aim at mitigating potential negative impacts but also provide tangible and concrete actions to enhance local ecosystems and engage the public effectively.

\subsubsection{Site PA}

\begin{figure}[!ht]
  \centering
  \includegraphics[width=\linewidth]{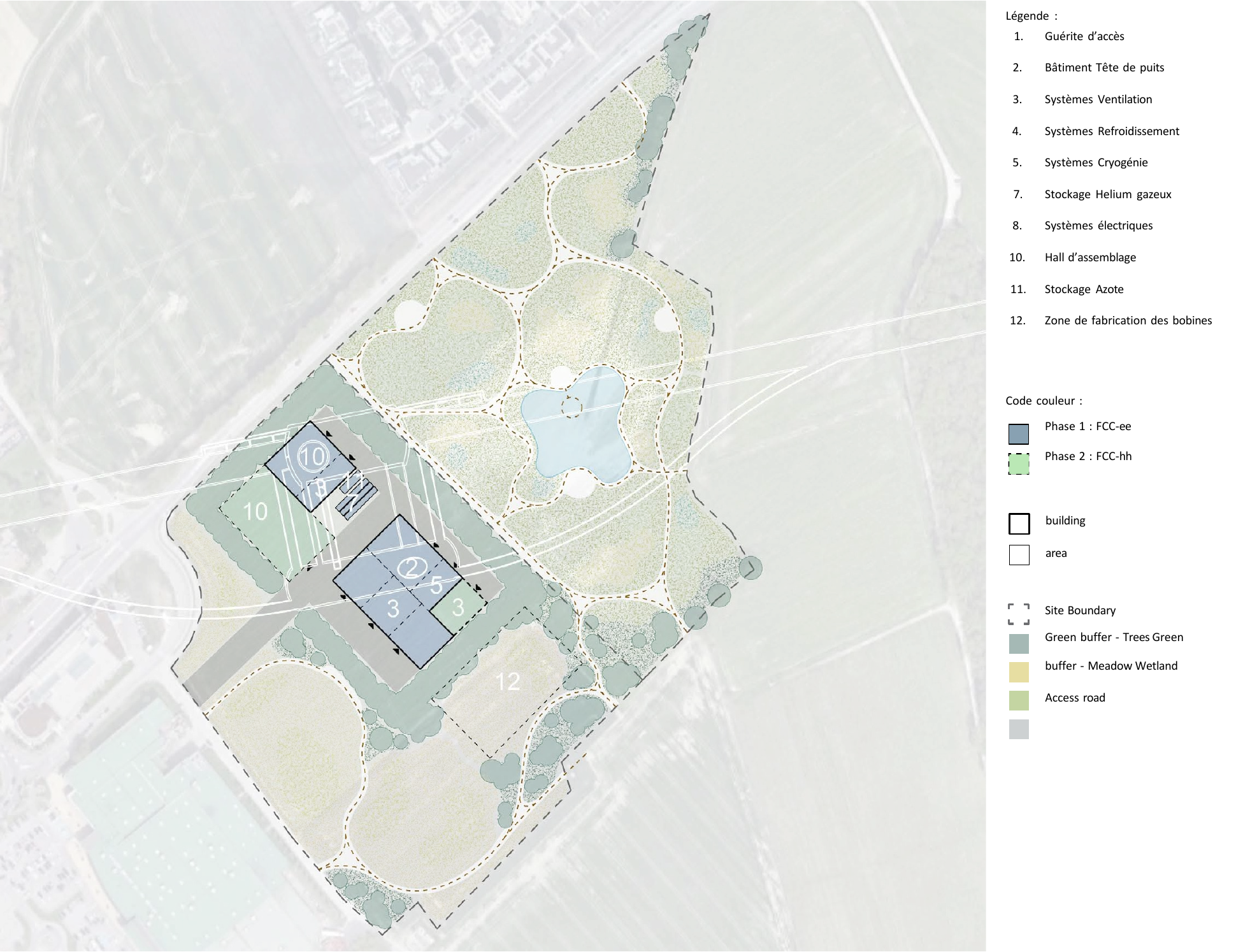}
  \caption{Space requirements for a site and landscape integration of experiment site PA in Ferney-Voltaire, France.}
  \label{fig:PA-2D}
\end{figure}

Experiment Site PA in Ferney-Voltaire (Fig.~\ref{fig:PA-2D}) is located in an already urban environment that continues to see further constructions such as a commercial and high-tech innovation quarter, health-care providers, residential buildings and commercial facilities. Located in a crossborder context with the Geneva airport  and a border crossing in the immediate vicinity, the remaining open space would be partially occupied by the surface site. Therefore, a good integration and preservation of the view of the Mont Blanc mountain chain need to be ensured. The surroundings of the site can be restored to create added value for an existing habitat and nature corridor. The image below gives an impression of how the site could embed into the open land, exploiting the immediate vicinity of the existing LHC point 8 (not shown) for any constructions and infrastructures that are not necessarily required to be in the immediate proximity of the two shafts.

\subsubsection{Site PB}

\begin{figure}[!ht]
  \centering
  \includegraphics[width=\linewidth]{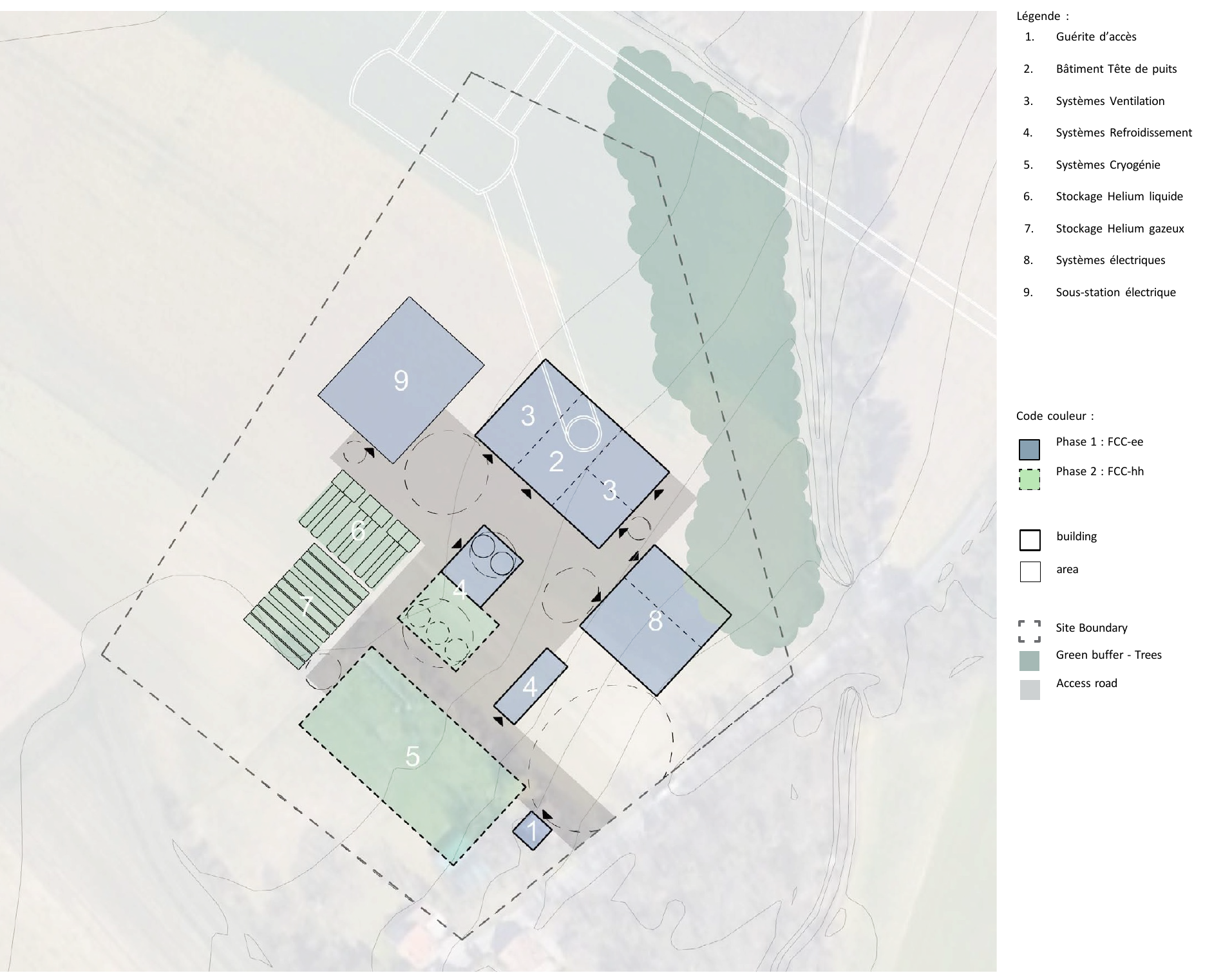}
  \caption{Space requirements for a site and landscape integration of experiment site PB in Presinge, Switzerland.}
  \label{fig:PB-2D}
\end{figure}

Technical site PB in Presinge, Switzerland, is located in an open agricultural landscape away from villages. The area is, however, frequently used for leisure activities by residents, such as walking and running. Together with the protected nature spaces in the immediate vicinity, the context calls for good landscape integration. Conventional industrial buildings are not an option for this location. The architectural toolkit aims to provide guiding principles that need to be applied together with the local stakeholders to yield an acceptable site that provides the technical functionalities required for the science project. Figure~\ref{fig:PB-2D} outlines the total space requirements for the FCC-ee and the FCC-hh phases.
The envelope indicated can be kept as small as possible but as large as needed to develop appropriate landscape integration based on an uninterrupted landscape and nature restoration. The aim is to create additional value by extending the nature preservation site and providing additional habitat to create improved conditions for biodiversity growth and human leisure activities.

\subsubsection{Site PD}

Experiment site PD in Nangy, France is located between the A40 autoroute and the newly developed RD903 multi-lane departmental road that passses right outside the southern end of the site. The surroundings are dominated by the large regional hospital `CHAL', an industrial zone in the north with a large milk processing facility and a mixed commercial and residential quarter on the opposite side of the multi-lane departmental road. Although the area is large and open, the site is not highly visible, since it would be developed on a slope. The topographic conditions call for a terracing approach from north to south. Although buildings and equipment for the hadron collider phase do not need to be constructed for the first, lepton collider phase, the site development and landscape integration call for a planned site development of the terraces from the onset. Green spaces will, therefore, dominate the site during the first operation phase. Access to the site from the north is exclusive, so no additional traffic is generated.

\begin{figure}[!ht]
  \centering
  \includegraphics[width=\linewidth]{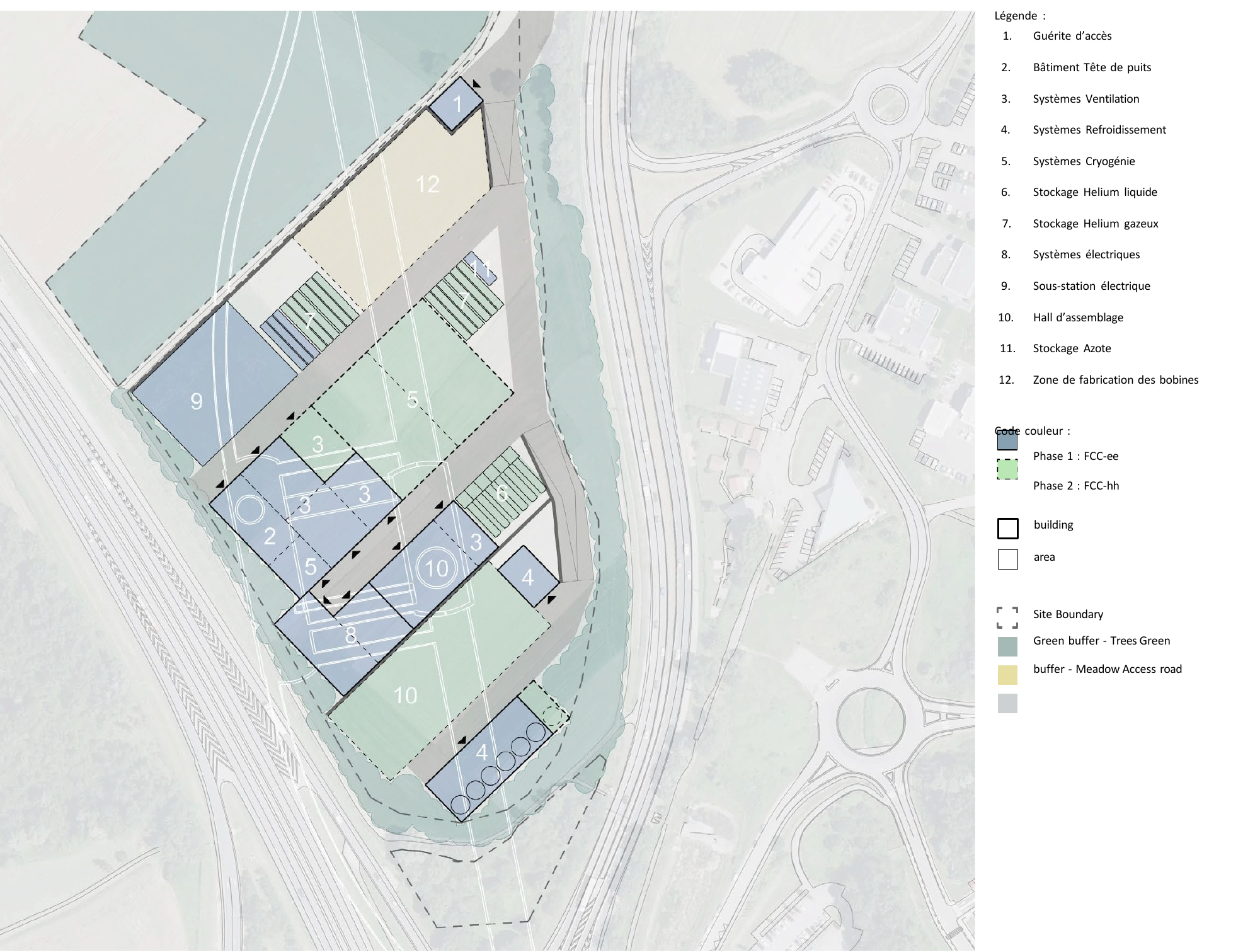}
  \caption{Space requirements for a site and landscape integration of experiment site PD in Nangy, France.}
  \label{fig:PD-2D}
\end{figure}

\subsubsection{Site PF}

\begin{figure}[!ht]
  \centering
  \includegraphics[width=\linewidth]{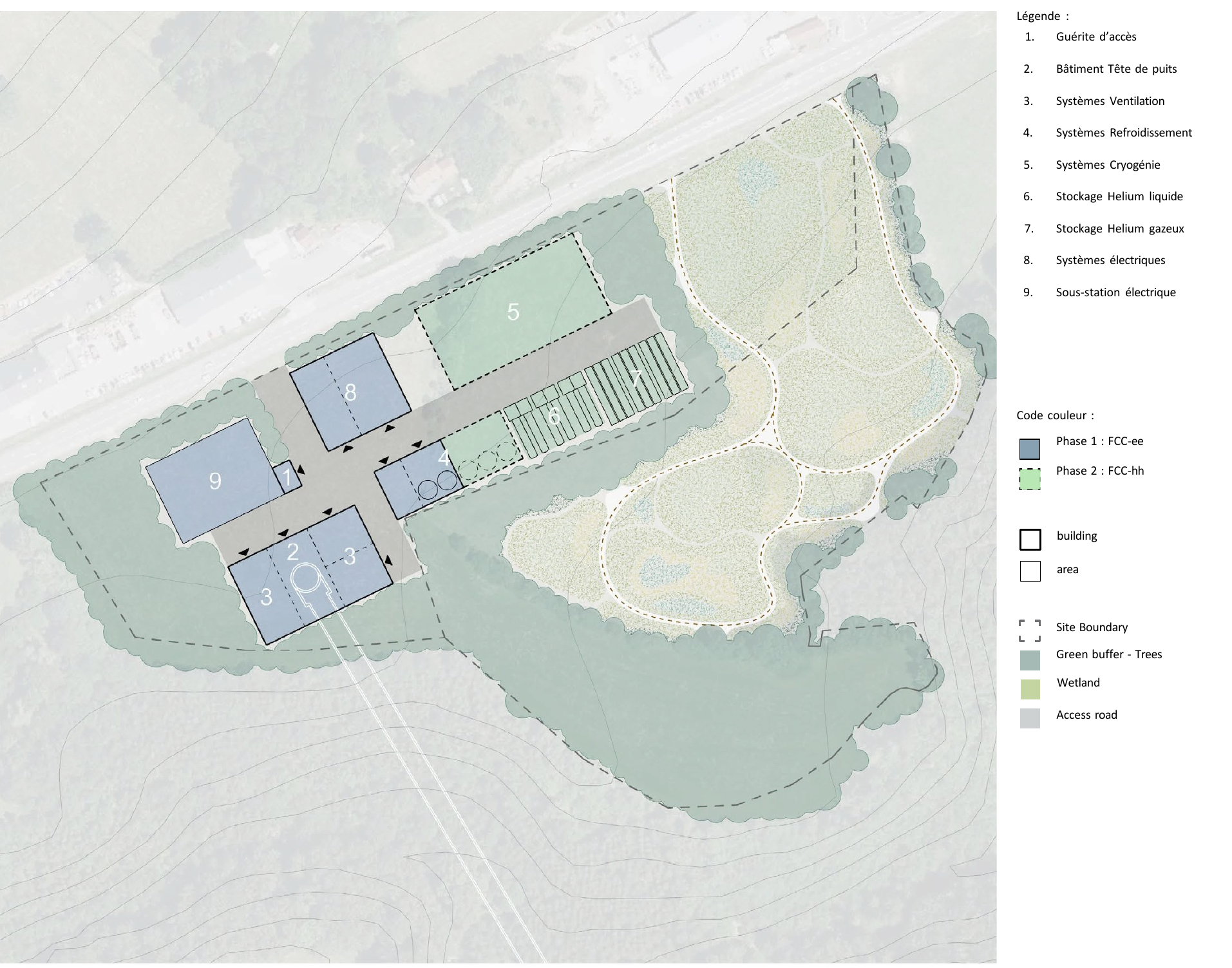}
  \caption{Space requirements for a site and landscape integration of technical site PF in \'Eteaux, France.}
  \label{fig:PF-2D}
\end{figure}

Technical site PF in \'Eteaux, France is directly located on a national road with heavy traffic. A public works company is established on the opposite side of the road. The area is characterised by an open view towards the Pre-Alp mountains and the slope falling off to a forest and small creek in the south at the A410 autoroute making the site  partially visible from the national road. The architectural toolkit provides means to foresee the restoration of nature in this location, thus creating added value despite the consumption of agricultural space. Wetlands in the immediate vicinity of the site that are not adequately preserved today can be improved to create habitats and catalyse the lasting growth of biodiversity in this area. Figure~\ref{fig:PF-2D} shows the space requirements for the technical site and the opportunities for the wetland restoration in the west. The restored space is approximately as large as the constructed space.

\subsubsection{Site PG}

\begin{figure}[!ht]
  \centering
  \includegraphics[width=\linewidth]{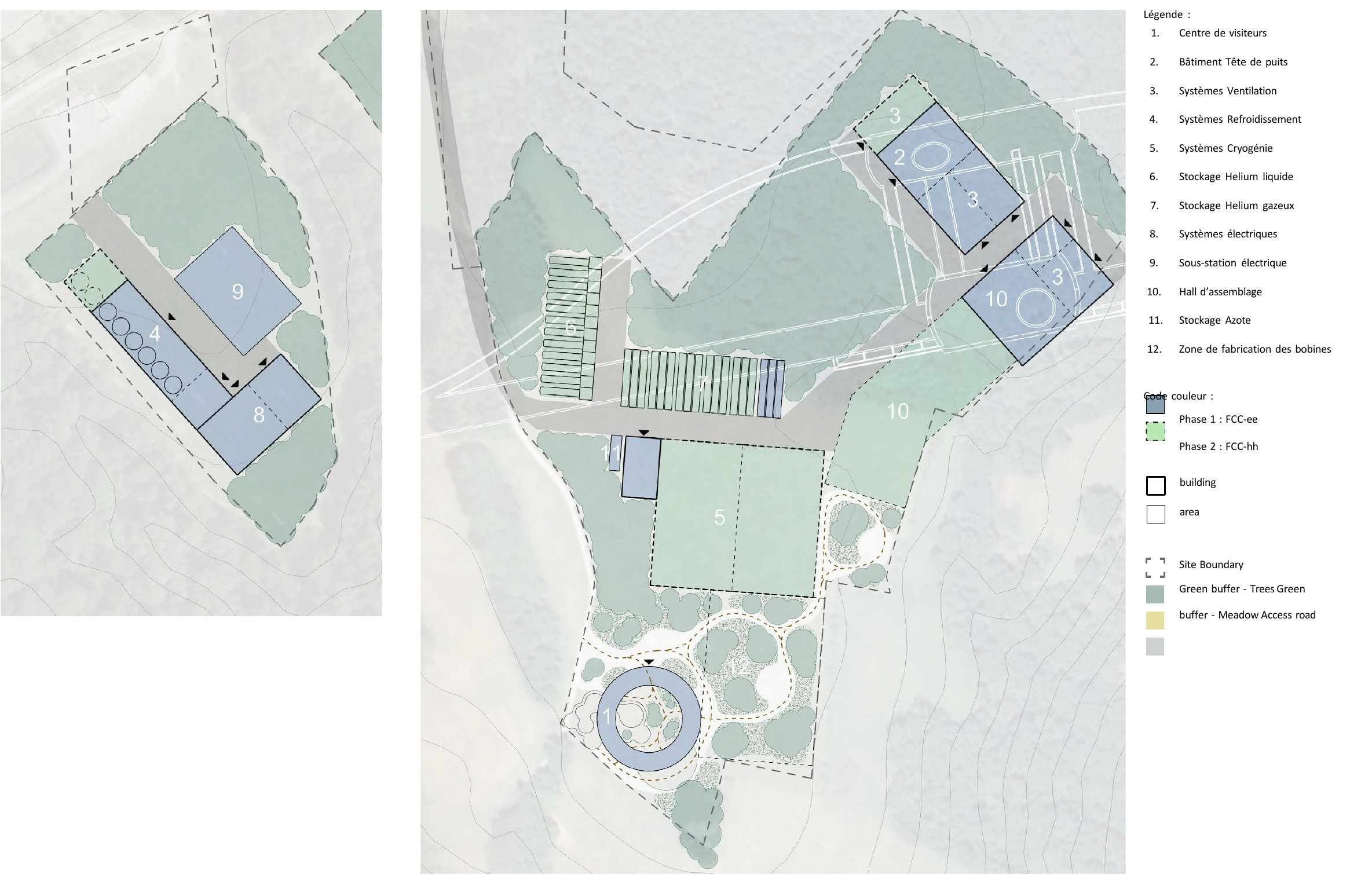}
  \caption{Space requirements for a site and landscape integration of experiment site PG in Charvonnex and Groisy, France. The right image concerns the main site around the shafts and the left image depicts the annex close to the autoroute that helps to reduce the impacts on nature, habitat, biodiversity, visibility and noise.}
  \label{fig:PG-2D}
\end{figure}

Experiment site PG spanning an area across the borders of Groisy and Charvonnex in France is located in a mixed natural environment, a forest, and a pasture used for cattle. Avoiding the forest is not possible, since the main shaft to the experiment cavern cannot be displaced and only little flexibility exists for adjusting the location of the shaft to the service cavern. The habitat and biodiversity constraints generated by the forest call for a reduction of woodland consumption as much as possible. Consequently, a split site location is envisaged that places visually and noise-impacting infrastructures close to the autoroute in the north, facilities that can be displaced to the open plateau outside the forest and keep only strictly needed equipment such as lifts and ventilation close to the shafts. The result is a site that covers a larger area but which impacts nature much less than a monolithic site. The configuration permits the creation of added value through a visiting facility. The first phase dedicated to the lepton collider would keep the spaces largely free of construction and green. Only the second phase, dedicated to the hadron collider, calls for temporary constructions for winding the coil of the detectors and permanent cryogenic refrigeration systems. An area and trees equivalent to the cleared area and trees can be re-created on the site area that is currently open pasture and on clearings in the forest that have been created historically. This means that the site development will eventually aim at an overall neutral balance with respect to habitat and biodiversity preservation and it can create added value through visitor facilities. Figure~\ref{fig:PG-2D} shows the space requirements for constructions at the main site location (right) and the annex close to the autoroute (left). The green spaces for nature restoration and an example of a visitor facility can be seen in the bottom part of the main site.

\subsubsection{Site PH}

\begin{figure}[!ht]
  \centering
  \includegraphics[width=\linewidth]{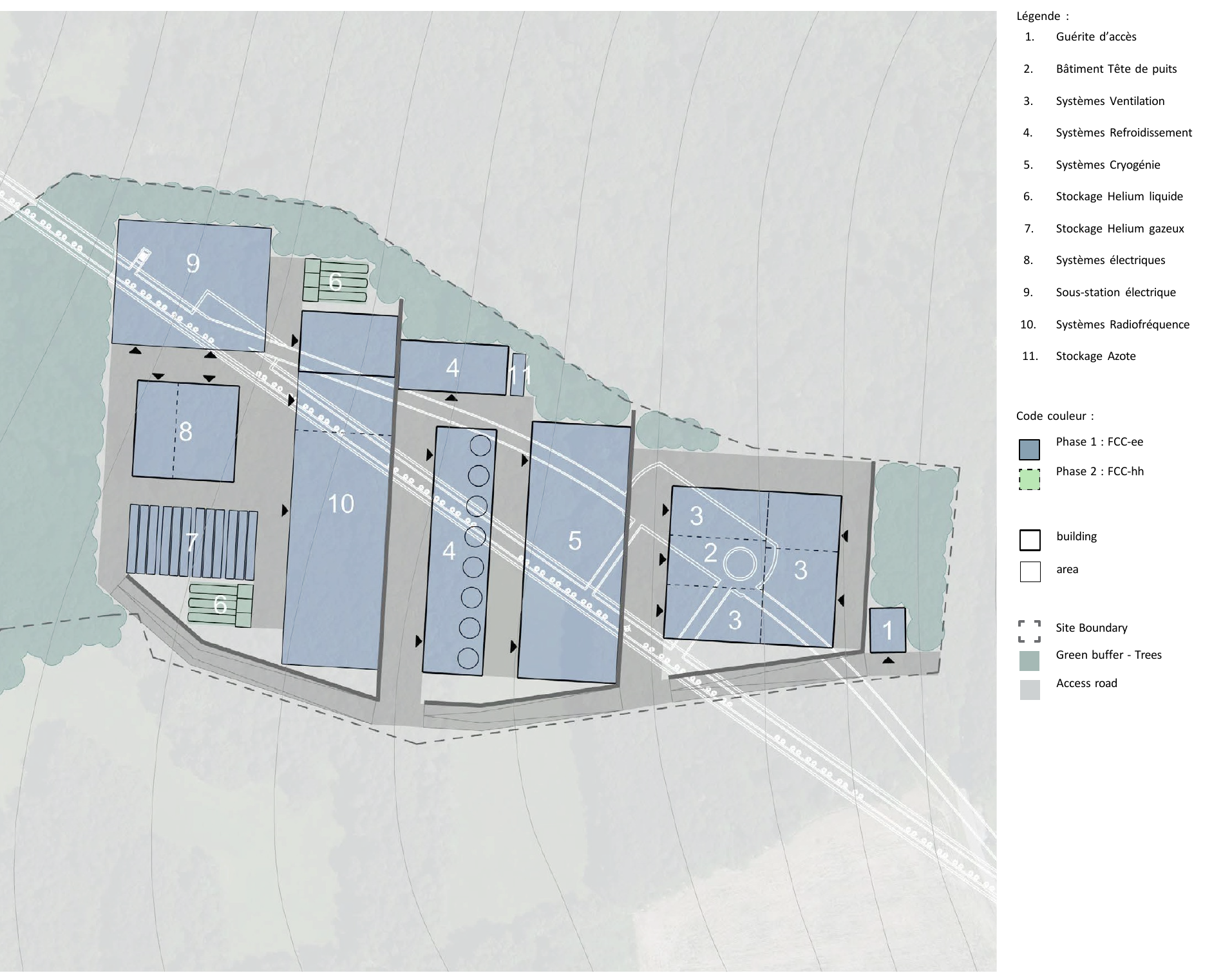}
  \caption{Space requirements for a site integration of technical site PH in Cercier and Marlioz, France.}
  \label{fig:PH-2D}
\end{figure}

Technical site PH in Cercier and Marlioz in France would  be entirely located in a woodland on a rather steep slope at the nominal location. In terms of integration, the main technical challenge is the slope that calls for a terracing approach. The site would be entirely hidden in the forest. The technical requirements for hosting all equipment to operate the radiofrequency system require, however, substantial space (Fig.~\ref{fig:PH-2D}). The major construction elements are the 400\,kV electrical substation, the power converters, and the cryogenic refrigeration system.

Although the location is technically feasible, environmental impact analysis and engagement with local stakeholders following the avoid-reduce-compensate scheme may still call for further reduction of space consumption. In this case, splitting of the site into an electrical part that can be further displaced and elements that need to be close to the shaft (cryogenics, cooling, ventilation) can be re-considered.

From an architectural perspective, the site buildings are less demanding than the others since the site is invisible from the outside.

\subsubsection{Site PJ}

\begin{figure}[!ht]
  \centering
  \includegraphics[width=\linewidth]{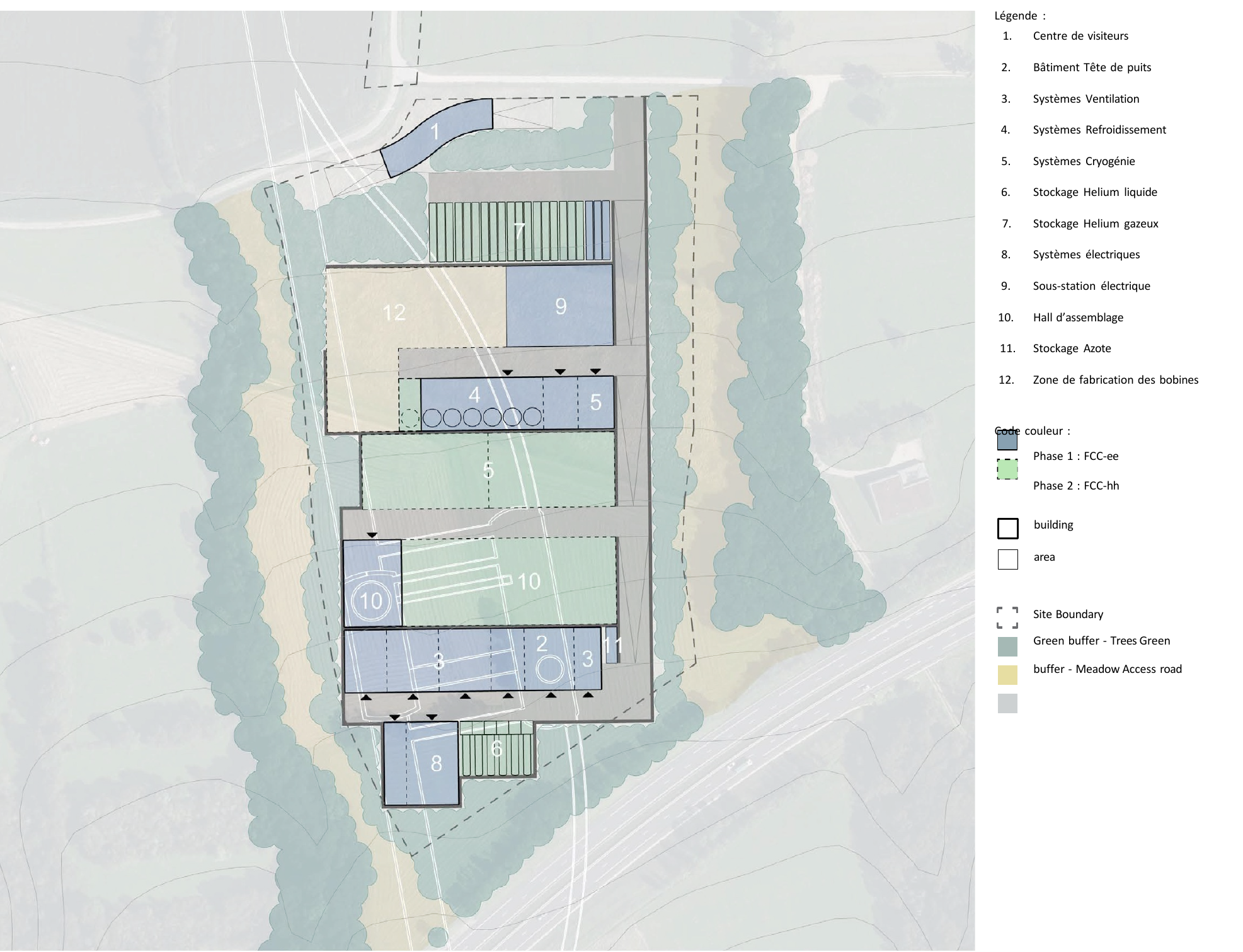}
  \caption{Space requirements for a site integration of experiment site PJ in Dingy-en-Vuache and Vulbens, France. The architecture toolkit suggests centring the site between existing tree lines to ensure that functional fauna corridors are created. Compared to today, they can lead to an improvement of the habitat conditions.}
  \label{fig:PJ-2D}
\end{figure}

Experiment site PJ in Dingy-en-Vuache and Vulbens in France is located in an agricultural area just south of the A40 autoroute, with existing road access to Vulbens. The existence of a fauna corridor will need to be considered when developing the site integration. The site is not visible from the autoroute or the communes. It is remote from any hamlets or individual houses. The topographic and relief constraints call again for a terracing approach that makes it necessary to prepare the entire site for both collider phases (ee and hh) from the onset. The terrain provides a good means to integrate the site visually and to foresee a visit facility. The existence of developing soft mobility connections between the communes in the vicinity can also be integrated in the site planning. Most of the surface would be kept as grassland until the second particle collider is installed. Several spaces that would be needed for this phase would also be created already during the first phase, since the terracing approach permits the creation of covered and half-buried volumes that can be used for different purposes at different times. As at other experiment sites, coil winding and detector assembly facilities would only be temporary. These areas would be restored to green fields afterwards (Fig.~\ref{fig:PJ-2D}).

\subsubsection{Site PL}

Technical site PL in Challex, Ain department, France would be located on an open, flat, agricultural space and the location of two individual houses. Vineyards in the vicinity on Swiss territory exist on slopes that fall steeply off to the Rh\^{o}ne river and to the Allondon river zone. The vicinity of the site location outside the village is used for leisure activities such as walking, hiking, running and visiting the vineyards by local residents and by tourists. Full exploitation of the architecture guidance toolkit for very good integration into the landscape is therefore a primary goal. Since the very even terrain cannot be effectively utilized to integrate the site into the landscape, lowering the site a little to avoid disrupting the landscape can mitigate some of the visibility challenges. Facade expressions, natural construction materials, green facades and roofs, and visibility screens will be essential elements for the site architecture and design. The surface area foreseen for this site includes an additional green buffer on space that is unusable for agricultural exploitation to be able to ensure that the visibility of the site can be kept low and that the integration can be very well planned and implemented. Such integration also helps improve the habitat value, supporting the thriving of biodiversity and making the area more attractive for leisure activities. Only a small difference between the lepton and the hadron collider phase constructions is planned, ensuring that the site is developed definitely as much as possible from the onset.

\begin{figure}[!ht]
  \centering
  \includegraphics[width=\linewidth]{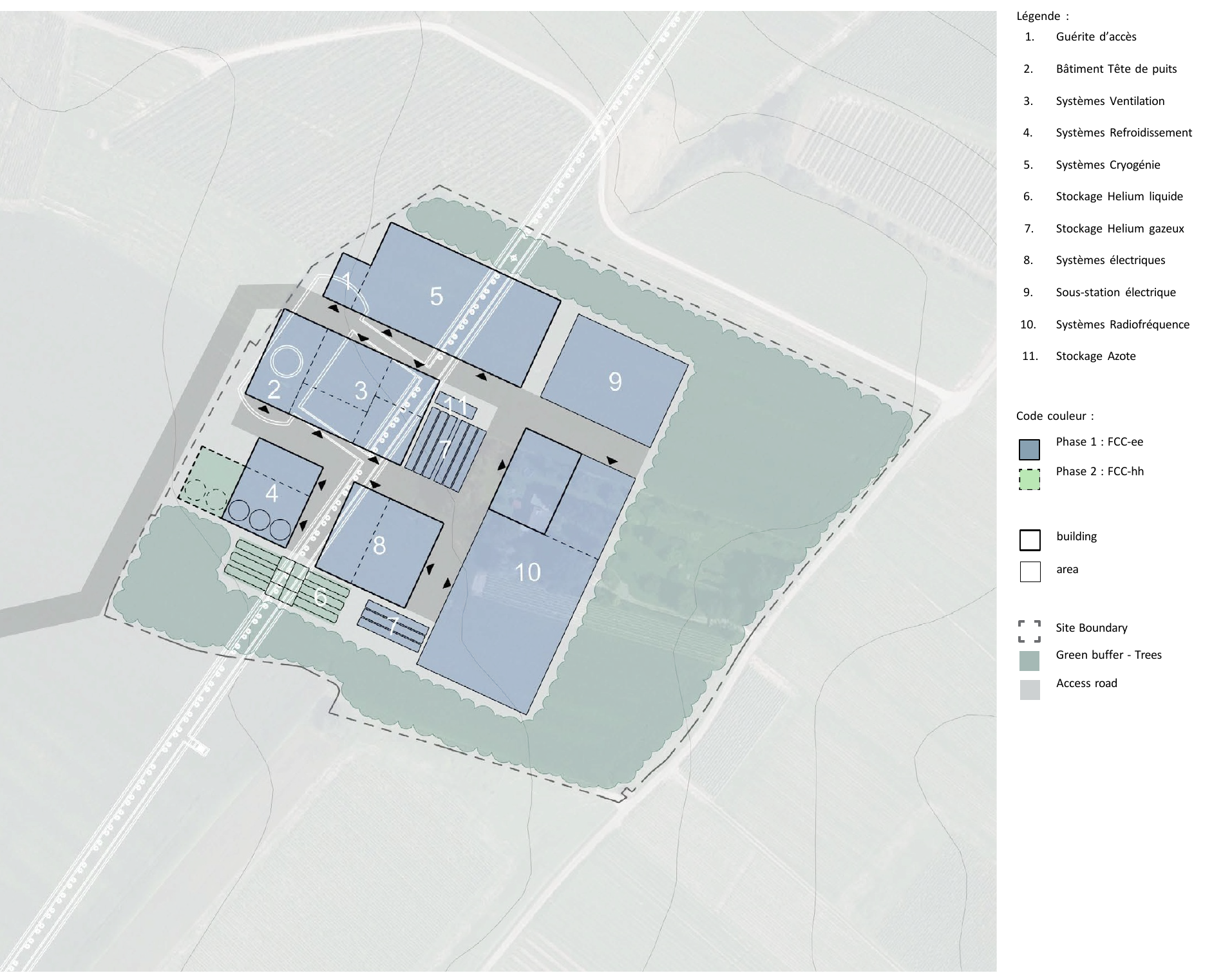}
  \caption{Space requirements for a site integration of technical site PL in Challex, France.}
  \label{fig:PL-2D}
\end{figure}

\subsubsection{Conclusions}

\begin{figure}[!ht]
  \centering
  \includegraphics[width=\linewidth]{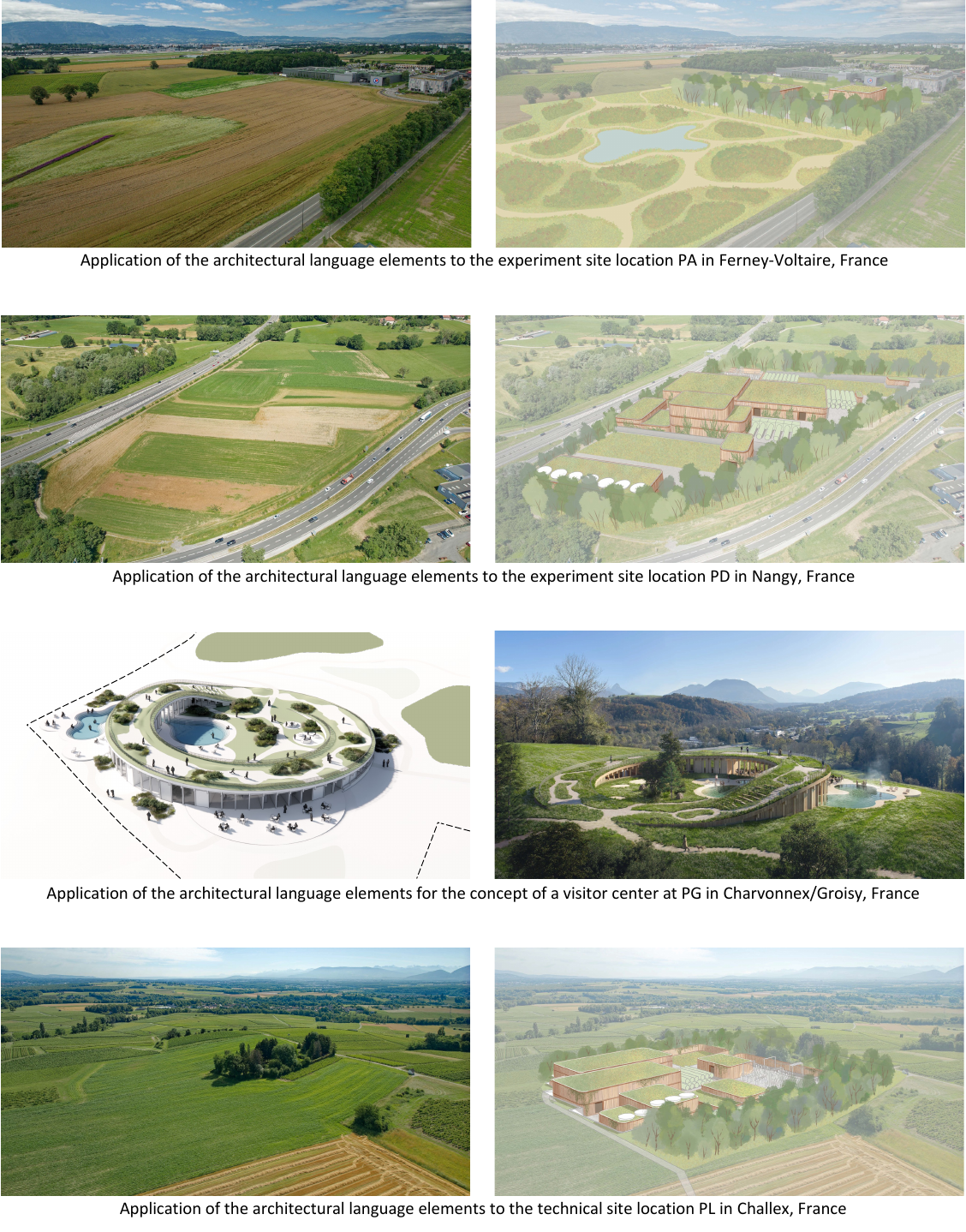}
  \caption{A collection of some artist concepts that result from the application of the architecture toolkit to individual site locations. Landscape integrated designs call for early preparation of the plots, and green buffers will have to be prepared during the subsurface construction phase to be able to screen the sites from the beginning.}
  \label{fig:architecture-concepts}
\end{figure}

Figure~\ref{fig:architecture-concepts} depicts the results of the application of the architectural toolkit on some of the surface site locations. These works do not represent designs or specific architectural choices for the implementation. They serve as a tool to engage with the local stakeholders to start a dialogue on the needs and constraints in each location. They permit developing architectural concepts and designs as a cooperative effort, federating the scientists and engineers designing the research infrastructure, the architects and landscape planners developing the sites, and the local stakeholders. The engagement first takes the form of collaborative workshops with  representatives of the population and interested local stakeholders in each location. This phase also includes common field visits to get a better understanding of what can realistically be envisaged and what is needed both from technical and stakeholder sides. Requirements and constraints from the technical project will be integrated in that activity. Once the boundary conditions are agreed upon and documented, a further, more detailed phase can start in which architect-engineers develop specific scenarios for the sites. The goal is to come to well-balanced and adequately integrated site designs in a subsequent project preparatory phase. The results in the form of design plans, descriptions, and construction prescriptions will be included in the environmental authorisation process in each host state.

It should be noted that the architectural concept and design developments of a public project of this scale with eight new surface sites occupying several ha of space each call for a timely market survey and tendering process for a qualified partner. The availability of such an architectural company must be ensured in terms of minimum personnel and uninterrupted time to work on the project. The entire development process is estimated to span several years.

\chapter{Environment}

\section{Context}
\label{env:environmental_context}

The international, collaborative FCC feasibility study created a basis for the subsequent environmental authorisation processes of the project in the two Host States, aiming at a single permit for the project in each country. The term \textit{project}, in agreement with the national and internationally applicable regulatory frameworks, refers to \textit{any intervention in the natural space including the soil}. It requires three essential parts:

\begin{enumerate}
\item An intent to construct that is communicated to authorities.
\item A documented definition of the scope of the project, comprising indirect and induced connected projects.
\item A documentation of the project boundaries and interfaces.
\end{enumerate}

The term \textit{project} is to be understood in a broad sense. It embraces all elements that are required to construct and operate the particle-collider-based research infrastructure. It includes, therefore, (1) the research infrastructure and (2) all territorial enabling developments and elements in France and in Switzerland that are required to construct and operate the research infrastructure. 

The research infrastructure, in turn, is composed of the subsurface and surface structures, all technical infrastructure, the particle accelerators and the experiment detectors. The territorial development elements include, but may not be limited to: raw and drinking water supply; non-public sewage and water treatment; low, medium and high voltage supplies; access roads; optional autoroute and railway accesses; facilities for waste treatment and management; excavated materials storage; final depot and re-use sites; compensation sites. 

Territorial annex projects will need to be considered and developed collaboratively with the Host States. For instance, these may include, but are not limited to, emergency and safety services, as well as temporary housing related to the construction phase, reinforcement of education facilities for workers and project participants (e.g., schools), health services for workers and project participants, district heating networks to supply the waste heat to consumers, water intakes that are shared with public clients, a regional geodetic network. Consequently, the project consists of segments that can ultimately be undertaken by different organisations and may involve multiple actors with distinct responsibilities. Defining the scope, boundaries, and interfaces will therefore be of utmost importance for a construction preparatory phase. This study was limited to the identification of the most relevant project segments and the launch of the process for gathering their requirements and key characteristics.

\begin{figure}[h]
  \centering
  \includegraphics[scale=0.8]{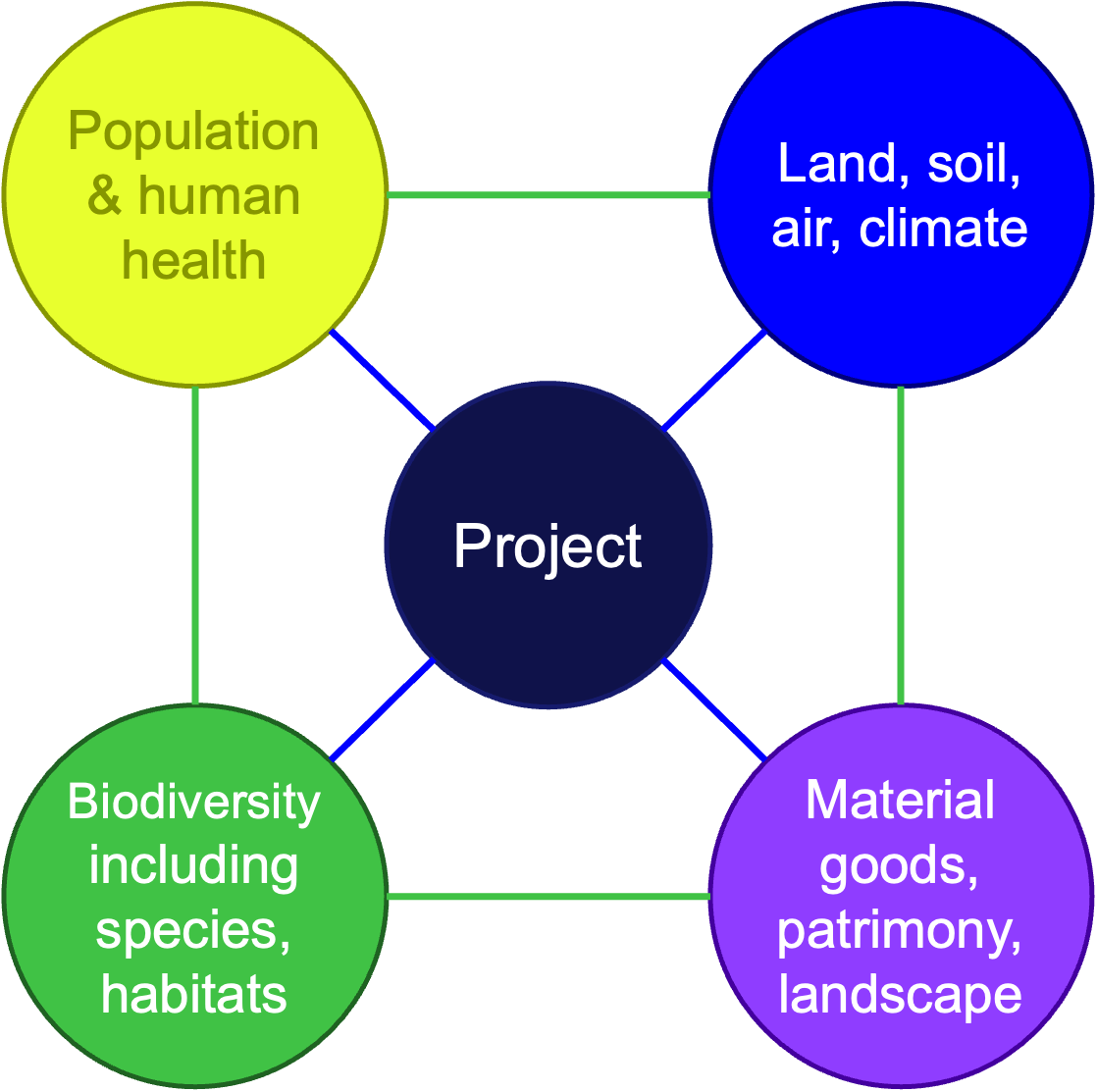}
  \caption{The environment refers to all elements that surround a project and their interactions with the project and among them. It is to be understood in the largest sense.}
  \label{fig:environment}
\end{figure}

The term \textit{environment} refers to all elements that surround the project (Fig.~\ref{fig:environment}). According to the EU regulation EC 2011/92/EU and the French `Code de l'environnement', articles L.122-1, R.122-2 and the European Norm EN 14001 section 3.2.1 the environment is to be understood in the largest sense. Specifically, the French \textit{Code de l'environnement} (article L110-1) includes aspects such as the spatial context, natural resources, habitats, noise, odour, sites, landscape, air quality, water quality, all living species, biodiversity, soil, geodiversity (subsurface conditions), fauna, flora, climate, social coherence, economy and the well-being of human beings. In Switzerland, the \textit{Manuel de l'Étude d'Impact sur l'Environnement} (Manuel EIE) guides environmental impact assessments under the \textit{Ordonnance sur l'étude d'impact sur l'environnement} (OEIE), derived from the \textit{Loi sur la protection de l’environnement} (LPE) at federal level. It covers climate, sites, historical monuments, archaeology, natural dangers, territorial and spatial development, air, noise, vibrations, energy, non-ionising and ionising radiation, subsurface water, surface water, aquatic ecosystems, water management, soil, polluted sites, waste, dangerous substances, dangerous organisms, prevention of major accidents, forests, flora, fauna, biotopes, landscape, light immission, paths for pedestrians and hiking, cross-border effects, as well as cantonal environmental protection regulations as long as they do not disproportionately limit the implementation. 

Following the requirement to identify and understand the direct and indirect interactions of the segments and the potential challenges that emerge from them, a variety of topics need to be considered. These include, for instance, but are not limited to the geology, hydrogeology, urbanism, health and safety of people, the well-being of all living species concerned, traffic and mobility, services and infrastructures, natural and cultural heritage, traded goods such as waste of all types, synergies, and conflicts with other planned projects, material goods such as physical items that are privately or publicly owned, technical risks, and many more.

An environmental \textit{aspect} refers to any project element or process that interacts with the environment. Identifying and prioritising these aspects hierarchically is a prerequisite for environmental impact assessment. Several can be identified and quantified early; though eventually, they depend on requirements and design choices. The challenge in large-scale particle collider projects is their iterative, decades-long design process. While subsurface structures and construction are planned first, technical infrastructures follow accelerator and experiment designs, with surface structures finalized later to incorporate advanced techniques and stakeholder-driven landscape integration. This evolving process, guided by a plan-do-act-check cycle, generates a challenge for a timely impact assessment for long-term projects with territorial implications.

\begin{figure}[h]
  \centering
  \includegraphics[width=\textwidth]{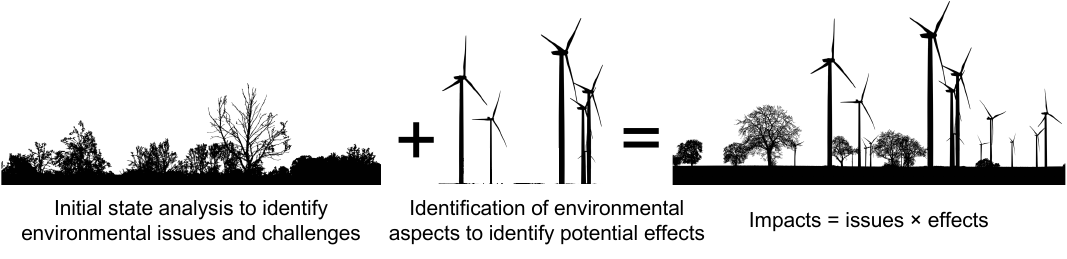}
  \caption{Environmental impacts are a consequence of the environmental sensitivity at a certain location (issues and challenges) and the potential effects of environmental aspects of a project at that location.}
  \label{fig:environmental-aspects-effects-impacts}
\end{figure}

An environmental \textit{impact} is any change to the project environment resulting from an identified, relevant environmental aspect. 
It emerges from the intersection of the environmental sensitivity in a certain location (environmental issues and challenges) and the potential effects of an environmental aspect of a project element (Fig.~\ref{fig:environmental-aspects-effects-impacts}). The impacts can, therefore, only be analysed and assessed for a specific, localised project scenario. The impact assessment also requires a certain stability in the designs and a sufficient level of design detail. In practice, the following elements are required for the impact assessment:

\begin{enumerate}
\item A specific project scenario (description, location, scope, boundaries, interfaces).
\item Specific construction plans.
\item Specific designs of the technical infrastructures, machine and experiment elements.
\item Descriptions of the construction processes.
\item Detailed operation concepts.
\item Prescribed procedures for operation, including the handling of failure cases and accidents.
\item An analysis of the current (initial) state of the project environment (at least over four seasons).
\item As a result of the initial state analysis, the prioritised issues and challenges of the locations in which the project will be embedded.
\item The descriptions of the goods, products and resources (and their origins) used during all project phases, e.g., a procurement and transport scenario hypothesis.
\item Relevant management concepts for all waste during all project phases.
\end{enumerate}

Assessing the environmental compatibility of the project involves an environmental impact study. It is part of the single environmental authorisation process in each Host State. It concerns the analysis and assessment of the relevant environmental effects of aspects of the project by crossing them with the environmental issues and challenges (also called `stakes') that have been identified and recorded in the frame of the initial state analysis in project-specific locations. This legally required process concerns the verification of the project's compliance with the regulations. It includes an iterative improvement of the project scenario following the Avoid-Reduce-Compensate scheme that has already been adopted and used during the development of the layout and placement scenario. Public participation is a key element of this process, in line with the Aarhus agreement \cite{Aarhus-ratification}, a UN convention that has been ratified by France in 2002 and Switzerland in 2014 to ensure access to information and public participation in decision-making and access to justice in environmental matters. Additionally, the Espoo agreement \cite{Espoo},
signed by France in 2001 and by Switzerland in 1996, regulates the environmental impact assessment requirements in a cross-border context, applicable as well to this project. Given that the project cannot be subdivided and the authorisation applicants cannot be separated, a single unified project will be presented in both Host States during the authorisation process.

\section{Environmental aspects}
\label{env:environmental_aspects}

\subsection{Methodology}

Identifying and gathering the environmental aspects, i.e., any element of the project that can interact with the environment, is a pre-requisite to be able to plan the environmental impact assessment during a subsequent project preparatory phase. This work started with the availability of technical design concepts in 2024 and will continue after the end of the feasibility study. It permits the establishment of a baseline of potential environmental effects assuming current state-of-the-art technical choices, i.e., without optimisation and not foreseeing technical evolutions and improvements beyond what industrial partners expect that they can deliver today.

The identification of environmental aspects and the assessment of potential environmental effects are conducted systematically using configuration-managed system description sheets. These sheets are jointly developed by environmental engineers and subsystem engineers for all project elements.

However, technical infrastructures located outside the FCC site boundaries — such as access roads, highway connections, potential railway links, public electricity grids, and sewage and water networks — are not included in this approach. These elements are addressed in separate, dedicated environmental impact studies, instead.

Information systematically gathered and compiled in an environmental aspect report \cite{EA_2025} serves the following purposes:

\begin{itemize}
\item Identify companies suitable to produce environmental impact studies;
\item Inform Host State environmental notified bodies about the project scope, contents, and interfaces in order to engage them effectively;
\item Inform the public about noteworthy project elements and their environmentally relevant elements;
\item Provides a baseline for the project engineers for the implementation of the eco-design approach;
\item Guides the environmental impact assessment;
\end{itemize}

Consequently, all information gathered is documented in French and in a language that permits non-experts to understand the project, the project's technical equipment, and the potential environmental effects linked to those elements.

Depending on the potential environmental effects, the documentation concerns different project segments at system or subsystem levels. Irrespective of the granularity of the element concerned, the following information is gathered and documented:

\begin{itemize}
\item Name of the system or subsystem;
\item Maturity level of the requirements and designs;
\item Linkages and connections to other systems such as a particle accelerator, a technical infrastructure or an experiment;
\item A short (around 80 words or one paragraph) non-technical description of the system's purpose: What does the system do and why does it exist;
\item A compact (up to one page) functional description of the system: what functions does the system implement, how is it composed and how does it accomplish its purpose;
\item The system's capacity and performance including the following information: capacities in different operation modes with a description, number of units deployed, energy needs, resource and raw materials needs, if and how the system transforms energy, types and quantities of emissions together with an estimate of the environmental relevance (including a description of waste production related to the system), known losses and inefficiencies in supplying the intended function, known space requirements. For all system capacities and performance characteristics, an indication of the confidence level and stability of the information is provided.
\item Locations at which the system is deployed.
\end{itemize}

This description is complemented by a structured summary of noteworthy environmental aspects and effects during normal operation, in degraded functioning, or in case of an accident. The following types of aspects are considered: 
\begin{itemize}
\item Release into water, soil, subsurface, air (including odour)~;
\item Waste production in terms of conventional and radioactive waste~;
\item Use of electrical and other sources of energy~;
\item Consumption of raw materials~;
\item Consumption of natural resources: water, land, natural habitats, aquatic habitats, protected zones, protected subsurface, agricultural space, protected agricultural space and others~ ;
\item Use of consumables (chemical products, for example) ;
\item Energy emissions such as non-ionising and ionising radiation, noise, vibration, heat, light or other forms of energy.
\item Effects on human environment, material goods and heritage such as traffic and transport ways, landscape, visibility and co-visibility, health, increase of technological risks, agriculture and forestry, constructed spaces, technical infrastructures, demographic development, territorial developments, urbanism, tourism and leisure activities.
\end{itemize}

The aspects are reported per project phase (construction, installation, operation, maintenance and repair, decommissioning). For each aspect, the likelihood of occurrence is indicated and the potential effect in case it materialises. Its relevance is also indicated in order to be able to create a hierarchy of environmental aspects. Finally, the level of confidence is provided for that piece of information.

The descriptions are complemented with high-level functional diagrams and drawings and references to supplementary information that provide evidence for the source of information and to permit technically interested readers to obtain further information.

\subsection{Scope}

The systematic methodological approach to establishing a first inventory of environmental aspects of the entire project comprises a large variety of project elements with the aim of being as exhaustive as possible given the constraint that the level of requirements and designs at this preliminary stage is still low.

The following topics and elements are included in the inventory and report:

\begin{enumerate}
\item Lepton injector: sources, positron production, linear accelerator and transfer lines to the booster.
\item Lepton collider: the collider as a whole and indivisible entity, magnet circuits, magnet power supply, vacuum system, radiofrequency system, beam transfer systems, beam interception equipment. The systems and equipment dealt with are also applicable to the environmental aspects of the booster.
\item Geodetic network and installations.
\item Data communication networks in the entire perimeter of the project, including safety-related communication equipment.
\item Experiments: the research programme and its duration up to the end of the century, a generic detector and a working assumption for its subdetectors, the technical infrastructure systems required to operate the detector, the data acquisition and online computing infrastructure.
\item Technical infrastructures: ventilation systems, compressed air production, connections to the French national high capacity electricity grid for operation, local electricity distribution networks primarily intended for construction, short-term energy buffering systems to ensure the stability of the electricity supply, emergency energy supply, general communication networks, cryogenic refrigeration system, cryogen storage, drinking water supply, raw water supply, water cooling systems, demineralised water production, chilled water production, cooling tower water management, underground water drainage recovery and evacuation, management of used water, clear water management (including rainwater), and personal safety systems.
\item Hadron collider: the research programme, the injectors, the collider, the experiments and the technical infrastructures at a high level, with a baseline of today's technologies.
\item Subsurface structures with a focus on the construction: shafts, transfer line tunnels, caverns, accesses, elevators, and subsurface transport systems.
\item Surface sites with a focus on the construction: Site locations and functions, each individual net site (PA, PB, PD, PF, PG, PH, PJ, PL), the re-use of LHC point 8 for site PA, the Prévessin and Meyrin sites at a high level.
\item Territorial developments in France: electricity supply for the construction, electricity supply for operation, drinking water, sewage, used water treatment, options for additional supply of cooling water, access roads, access to autoroutes, options for railway accesses, emergency services, communication network needs, needs concerning the management of excavated materials, requirements for the supply of waste heat, local transport needs for the construction, installation, and operation phases.
\item Territorial developments in Switzerland: electricity supply for construction, electricity supply for operation, drinking water, sewage, used water treatment, raw water supply for cooling, site access, options for railway access, emergency services, communication network needs, needs concerning the management of excavated materials, requirements for the supply of waste heat, local transport needs for the construction, installation, and operation phases.
\item Construction phase: a dedicated description of the environmental aspects linked to the subsurface construction activities and the development of the surface sites, a preliminary schedule, descriptions of the construction sites, preparation of the sites before construction starts, management of the construction sites, management of the personnel engaged in the construction, the supply of construction materials, resources required for the construction, shaft and cavern construction processes, tunnel construction processes, surface site civil structure construction approach, principles for the management of the excavated materials, site restoration after construction.
\item Installation phase: Installation of technical infrastructures, particle accelerators, the detectors, supply of the cryogens, transition of the technical infrastructures, the detectors, and the accelerators into operation.
\end{enumerate}

\subsection{Status and conclusions}

The establishment of the comprehensive environmental aspects report \cite{EA_2025} is currently in progress. It is an iterative work that advances with the gathering of requirements and suitable technologies to establish a baseline and credible outlooks for future technologies and approaches.

The civil engineering construction phase (underground structures and surface sites), the technical infrastructure systems, and the lepton collider systems are described according to the currently established baseline, along with the corresponding preliminary confidence and detail level. This preliminary analysis will evolve as more detailed construction details and plans become available. 

Requirements for materials and resources that form the basis for an iteration of the environmental aspects analysis must be understood to be approximate and linked to certain uncertainties. In particular, only a high-level description of a generic experiment detector is available today as a baseline for further studies. This description is based on current state-of-the-art technologies for detectors, typical detector technical infrastructures and data acquisition and processing. The subsequent design phase must at least include individual descriptions of four specific experiment detectors with more tangible and technical design assumptions based on a reasonable technological outlook for the coming years.

Conducting an environmental impact assessment requires detailed plans. If technical designs for particle accelerators, experiment detectors, and their associated infrastructures are unavailable, voluntary commitments on performance characteristics must be established.

The lepton injector and its systems and the installation phase, including commissioning, are currently under pre-study. Descriptions are, therefore, not included at this stage.
The territorial developments in France and Switzerland have not yet been precisely defined. However, the minimum needs are known and have been included. It should be noted that some adjustments of site locations may still be required, depending on the progress of the dialogue with local stakeholders in the vicinity of such sites.

At this stage, the main environmental aspects discussed in this chapter are linked to the construction activities: the territorial development of some 50\,ha of land spread over eight new surface sites, the generation of $\sim$6\,million\,m$^3$ (in situ) of excavated materials and the nuisances that are linked to any construction activity (e.g., noise, vibrations, dust, artificial light pollution, additional traffic).

Consequently, the study anticipated a life-cycle analysis of the construction activities to gain a better understanding of other potential environmental effects on climate and the depletion of natural resources~\cite{mauree_2024_13899160}. Contrary to conventional subsurface constructions, the creation of civil engineering structures for a research infrastructure is less impactful. The LCA was based on actual available low-carbon footprint products backed by EN 15804 certified Environmental Product Declarations (EPDs) and construction processes that foster short supply chains and construction prescriptions that consider responsible use of resources. The resulting estimated carbon footprint covering the A1 to A5 lifecycle phases concerning the construction phase is about 0.53\,million\,tCO2(eq). This value can be compared to CERN's current carbon footprint of 0.36\,million\,tCO2(eq) in 2022, covering Scope 1, 2 and 3 emissions. It must be noted that for carbon accounting the emissions must be distributed over the potentially contributing countries; consequently, the resulting annual impact per capita is about 0.11\,kg, limited to the construction period. For comparison, the Paris Agreement defines a carbon budget of 2\,000\,kg per capita per year. The effects are mainly induced by two types of construction materials: ready-made concrete and steel. Design optimisations and technological progress will permit further lowering the footprint. An update of the lifecycle analysis can be carried out during the technical design phase to quantitatively report on the reduction effects. A dedicated environmental impact assessment will consider the directly related construction effects and their management. The planning of the construction process will have to include an environmental impact management plan that describes the responsible use of resources (e.g., water), the management of the construction activity-related waste production and the management of the construction-related traffic.

Concerning the management of the excavated materials, the initial inventory of locations to permit refilling quarries and deposit of non-reusable materials revealed the technical and managerial feasibility of the construction. However, in an effort to increase the reuse, a dedicated multi-year study has been launched in 2024 on CERN terrain with excavated molasse materials from the HL-LHC construction to develop quality-managed processes to transform the molasse into reusable materials. The pathways include rewilding projects (e.g., restoration of wasteland and preparation of the lower soil layers for new agricultural areas with fertile topsoil), the creation of raised hedges, the stabilisation of roadsides, the treatment of rural and forest paths, the creation of parks and mini forests and the improvement of poor agricultural soil. Further applications, such as the production of insulation materials and non-structural construction materials, are being developed with industrial partners. The goal is to limit the pure deposit of non-reusable materials to 30\%.
It is important to recognize that, at this stage, no definite statements about the re-use fraction can be provided due to the need to get an understanding of the excavated material’s characteristics with dedicated subsurface investigations. However, FCC collaboration is actively investing in research and development of different innovative solutions to maximise material reuse and minimise environmental impact.

Although the additional traffic induced by the construction activities is limited with respect to the existing road traffic (between 2 and 15 transports per hour at the 8 sites during working hours, representing an addition of 0.15 to 1.8\% of the total traffic, depending on the site) in the immediate vicinity of the construction sites, care will have to be taken to develop a solid plan of the transport activities. This will include the preference to transport excavated materials to the nearest major transport axe by conveyor belts and to optimise the supply of materials from the transport axes to the sites (e.g., conveyor belts and optimised logistics). The goal is to avoid transit through residential neighbourhoods and small roads as much as possible. The transport concept is part of the construction planning, the associated authorisation process, and the subsequent requirements that are to be included in the tendering procedures.

The artificialisation\footnote{May be defined as: the transformation of a soil of agricultural, natural or forestry character by management actions, which may result in its total or partial waterproofing.} of land and the linked loss of habitat and biodiversity has been quantified and included in the socio-economic impact study. This effect is intended to be largely mitigated with rewilding projects around the sites and the re-creation of agricultural spaces. The environmental impact assessment will need to develop these measures in detail.

During construction and exploitation, noise is a topic that can lead to notable effects in a few locations where residential houses are within a perimeter of between 100 and 300\,m around the sites. The effects have been quantified and are included in the socio-economic impact study. The environmental impact assessment will have to include the development of adequate protection measures following the avoid-reduce and compensate approach. So far, the potential effects revealed are limited to less than 10 households in total for the entire project.

The effects of ionising radiation have also been quantified and have been found to be insignificant (actual annual dose in the vicinity of surface sites below 0.01\,mSv/year well within the natural background radiation of about 0.8\,mSv/year). This is because, based on CERN's multi-decade experience of operation, the scientific research installations generate additional ionising radiation far below the permitted thresholds with no health effects. Nevertheless, the applicable socio-economic quantification methods have been used to cost the effects and report them. The environmental impact assessment will need to include all the required monitoring and protection measures to ensure that the potential effects are maintained at an insignificant level.

Noteworthy effects are expected to be primarily indirect, arising from the supply of raw and construction materials, as well as off-site emissions. These include Scope 2 emissions, which result from the electricity, heating, or cooling purchased to power project-related operations, and Scope 3 emissions, which encompass the broader environmental impact across the value chain, such as the production and transportation of materials, construction activities, and waste management. To address both aspects linked to the construction of the facility and the scientific instruments and their operation, a procurement plan that includes the environmental aspects will have to be developed and implemented. A first analysis carried out with experts in the domain of energy procurement revealed that a large coverage of supply from renewable energy sources keeps the indirect emissions low. Privileging electrified construction methods on a time horizon of 10 years can be foreseen. Entering operation on a 2050-time horizon, in turn, offers the possibility for substantial energy supply coverage from renewable energy sources. Supply of waste heat, in particular when the research infrastructure operation is adapted to the local and seasonal needs, can largely help to reduce the carbon footprint by replacing conventional, fossil heat sources. The supply of raw and construction materials for the infrastructures and instruments will need to be well-planned and included in the procurement and in-kind supply requirements. The market is currently evolving fast in Europe towards fostering recycled materials (e.g., steel) and low-carbon production (e.g., concrete). The application and requirement of European Norms such as product comparison based on EPD (Environmental Product Declarations) are needed to ensure a low environmental footprint. Further lifecycle analysis for technical infrastructures, particle accelerator components and experiment detectors are needed in the subsequent design phase to develop these requirements. Gases that are harmful to the environment will need to be avoided if they have not already been taken off the market.

The surface sites lead to noteworthy environmental effects that are mainly linked to visibility. These must be addressed in the design phase through the engagement of the public who is directly affected or concerned in cooperation with architects who have experience in landscape integration of industrial and functional buildings. Preliminary studies have been carried out, and they confirm the existence of suitable approaches and technologies today.

Each site presents unique requirements and constraints that must be carefully considered. A key challenge for the integrated project is the evolving definition of the space required to accommodate technical equipment for the lepton collider, along with the long-term needs for a potential future hadron collider. In locations where seamless landscape integration is a priority, constraints will play a determining role in decision-making. A balance must be struck between minimising the site footprint for efficient land use and ensuring that the infrastructure can accommodate both the initial and potential second collider phases.

The consumption of agricultural space and forest not only leads to a reduction of habitat and biodiversity but also has tangible economic impacts. The loss of income (direct, upstream and downstream) has been analysed and quantified and is included in the socio-economic study. Compensation proposals need to be developed. While in Switzerland, only a one-to-one compensation of the space is acceptable, in France, collective compensation measures can also be jointly developed with affected stakeholders during the environmental impact evaluation phase. The direct compensation for the loss of protected agricultural space involves the transfer of the high-quality topsoil to locations with either poorer soil quality or wasteland that will need to be refurbished. The loss of forests limited to France can for example be partially compensated through re-forestation in the vicinity of the surface sites and at specific locations. The notified bodies will be consulted for recommendations and validation for compensatory measures.

Water consumption remains within current levels, as the requirements for the FCC-ee are similar to CERN’s current needs. Recognising the importance of responsible resource management, efforts will be made to limit water use. The FCC-ee replaces the Large Hadron Collider’s water demand.
Considering the continuous depletion of water resources, protection of water resources and reduction of water consumption are part of the eco-design goals. Therefore, studies are underway to understand if and how treated wastewater can be used for cooling purposes. The initial results confirm, in principle, its technical and financial feasibility. Local effects may remain with respect to water needed during the construction phase. The subsequent environmental impact study must therefore include a water sourcing and management plan to ensure that there are no noteworthy local impacts due to the construction activities which affect the population and the existing economic activities.

The construction and operation of the research infrastructure entail generating significant amounts of conventional waste. The subsequent environmental impact study will detail the types and quantities. The effect stems mainly from procured goods and therefore to manage this issue, waste avoidance and management must be included in the procurement process. Since the operation of a new research infrastructure replaces the operation of CERN's current flagship project and the number of persons involved is expected to be stable over time, the daily waste production is not expected to grow. Continuous efforts to reduce waste generation, fostering reuse first and recycling in line with the Host State regulations is expected to lead to a continuation of the waste reduction already engaged today.

Finally, the aim to keep environmental impacts within acceptable limits calls for an adaptive operation concept. Considering climate change effects and the need to generate environmental benefits to balance negative externalities requires the inclusion of external constraints into the operation. For instance, the optimisation of the waste heat supply may call for a shift of operation into colder seasons. The operation of the particle accelerators could be linked to the availability of abundant renewable energy via online notifications received by a link with the energy supplier. The operation may also need to include consideration of the changes in secondary circuit cooling water temperatures due to the long-term evolution of the climate. The presence of personnel also means having to consider seasonally changing working conditions. For instance, it can be more sustainable to work during colder seasons than in hotter ones since heating can be achieved by waste-heat recovery, but cooling requires additional energy.

The preliminary inventory and analysis of environmental aspects have led the environment consortium, composed of a number of companies, to develop a first version of an environmental strategy that is described in Section~\ref{sec:environmental_strategy}. 
The guidelines that this strategy includes are iteratively provided to the designers of the facility and are included in new procurement actions relating to studies and technical designs. In this way, a culture of integrating the environmental impact into the design process is gradually being established. This eco-design approach is expected to play a fundamental role in the subsequent design, construction, and operation of the FCC.

\subsection{Environmental strategy}
\label{sec:environmental_strategy}

\subsubsection{Purpose and scope}

The purpose of the strategic vision \cite{fcc-ecodesign-guideline-2024} for the environment, eco-design, and sustainable development for the project is to define and set guidelines in order to:
\begin{itemize}
\item Set out the project owner's ambitions in relation to the environment, eco-design and sustainable development.
\item Steer and design the project to consider and integrate environmental aspects.
\item Embed the project in an overarching sustainable development approach based on the three sustainability pillars: environmental, social and economic.
\item Establish a cross-disciplinary framework for the project, as well as for the international partners and collaborators, that can be applied to all phases of the project’s development and operation.
\end{itemize}
The strategic vision applies to all phases of the project’s development: planning, design, construction, operation, maintenance, and future development. It, therefore, concerns the construction of the infrastructure, the particle accelerators, and the experiments that will be carried out there.
From a geographical point of view, the strategic vision is transnational in scope and incorporates the national and specific regulations applicable at every stage of the project.

\subsubsection{Organisation and steering}

The environmental strategy and its organisation and steering are based on the principles \cite{CERN-Enviro} 
already being followed at CERN as the project host. The strategy also incorporates the recommendations issued in the context of the environmental standards that serve as guidelines~\cite{ISO-Enviro, HQCert}, including the recommendations relating to the key issues to be addressed. 
CERN’s main objectives for the period 2021–2025~\cite{CERN-objectives-2025} and beyond place the environment at the top of the agenda.
This 2E+SD (Environment, eco-design and Sustainable Development) strategic vision reflects a strong commitment to design a new project that encompasses local and regional developments.
To date, CERN’s environmental objectives have been focused on existing activities. An evolution is needed to meet the environmental challenges of building and then operating a new large-scale research infrastructure while taking into account the protection and development of the region. The project needs to be conceived considering the environment in which it will be embedded from the outset.
To develop a scenario for a project, the project owner should set up an appropriate structure that is responsible for the integration of the 2E+SD strategy.
This structure will make it possible for the strategy and its implementation to be steered independently of the entities responsible for the study, design, and implementation. It must report directly to the project management and have clearly identified reference people in all the entities involved in the various phases of the project.
It will help to define the project management procedures and will set requirements for the operational management (organisational issues) and/or for the steering.

\subsubsection{Collective and individual responsibilities}

The project management team develops the strategy, communicates it clearly, and ensures all international collaborators adhere to it. They establish an organisational structure and implement all necessary measures to achieve the objectives outlined in the 2E+SD strategic vision. To accomplish this, the following actions will be taken:

\paragraph{Short term (2024-2030)}
\begin{itemize}
\item Understand the environmental and regulatory context in which the research infrastructure project and its related projects unfold.
\item Draw up an `Environment, eco-design and Sustainable Development' roadmap to structure the way in which the environment is taken into account throughout the project and to identify and involve all stakeholders.
\item Set targets to be reached and continuous improvement objectives that are applicable to all stakeholders and to all phases of the project based on the UN’s Sustainable Development Goals.
\item Engage in a systematic process of avoiding, minimising and offsetting the project’s impacts through follow-up and improvement measures.
\item Monitor regulatory and technical developments to anticipate changes in the fields of the environment, eco-design and sustainable development.
\item Integrate environmental and sustainable development objectives into all activities relating to the project, in particular during the project design phase.
\end{itemize}

\paragraph{Medium term (2030-2040+)}
\begin{itemize}
\item Steer the integration of environmental and sustainable development objectives into the construction of infrastructure, technical equipment (particle accelerators, scientific experiments and all the technical infrastructure needed to operate them) and territorial development projects.
\item Develop partnerships with local stakeholders planned during the previous period (2024–2030).
\item Update the strategic vision and adapt it to changing circumstances based on environmental monitoring.
\end{itemize}

\paragraph{Long term (2040+)}
\begin{itemize}
\item Continue to integrate the sustainability enablers, in particular solutions with the lowest energy consumption and greenhouse gas emissions, provided that they are compatible with the scientific operation of the project and are economically viable.
\end{itemize}

All those involved in project-related activities actively contribute to the implementation of the 2E+SD strategy through exemplary conduct and by keeping in mind the associated objectives and adopting best market practices on the basis of assessment of emerging technologies.

At every stage of the project, while pursuing the aim of achieving its scientific objectives, everyone must actively seek out information enabling them to minimise the project’s impact on the environment and must fulfil the environment, eco-design, and sustainable development responsibilities entrusted to them. 

Decisions to adopt measures must be based on a cost–benefit assessment of the various options, including lifecycle analysis where appropriate.

It is understood that actions to avoid and reduce impacts must be technically feasible, economically viable, and compatible with the need to deliver sustained scientific excellence on a global scale and with the scientific goals of the new research infrastructure.

\subsubsection{Priority environmental themes}

Among the many environmental aspects, the following eight topics have been assigned a priority for the subsequent design phase:

\paragraph{Water}

\begin{itemize}
    \item Limit water consumption, avoid sensitive periods and do not consume water from sensitive sites.
    \item Monitor and control the quality of released water and how it is managed.
    \item Optimise water use, recycling and reuse.
\end{itemize}

\paragraph{Waste}

\begin{itemize}
    \item Channel all waste to appropriate sorting facilities.
    \item Repurpose to avoid and reduce final waste.
    \item Use recycled and organic materials where possible.
\end{itemize}

\paragraph{Energy}

\begin{itemize}
    \item Limit consumption and increase system efficiency.
    \item Favour renewable energy sources.
    \item Recover and store energy; supply residual heat.
\end{itemize}

\paragraph{Ionising radiation}

\begin{itemize}
    \item Apply the standards and the agreement of CERN with the two Host States~\cite{CERN_radiation_2010}.
    \item Study solutions to reduce and limit the generation of ionising radiation.
    \item Identify and quantify local risks.
\end{itemize}

\paragraph{Biodiversity}

\begin{itemize}
    \item Identify and preserve sensitive sites.
    \item Identify local protected and heritage species.
    \item Take appropriate environmental offsetting measures.
\end{itemize}

\paragraph{Emissions}

\begin{itemize}
    \item Quantify emissions, including in terms of carbon footprint.
    \item Develop recovery and management systems.
    \item Monitor and control the quality of emissions and what happens to them.
\end{itemize}

\paragraph{Placement}

\begin{itemize}
    \item Avoid sites that are sensitive or are subject to strong constraints.
    \item Reduce surface areas and optimise the site layout.
    \item Blend the infrastructure and activities into the environment.
\end{itemize}

\paragraph{Societal dimensions}

\begin{itemize}
    \item Limit activities to necessary locations and periods.
    \item Limit disruption and preserve the comfort of local residents.
    \item Plan the creation of added value for society as part of the project.
\end{itemize}

\subsubsection{Specific guidelines}
As a result of the development of the strategic vision, the following specific guidelines have been formulated. They are to be integrated in the subsequent design phases by all infrastructure and equipment developers.

\paragraph{Water}

\begin{itemize}
    \item Avoid using drinking water for non-sanitary purposes.
    \item Avoid extracting groundwater.
    \item From the planning stage onwards, preserve the quality and quantity of groundwater.
    \item Do not create interaction between layers of groundwater or mix their waters.
    \item Put rainwater collection systems in place for purposes that do not require drinking water.
    \item Limit evaporation by using closed circuits and, where possible, dry systems.
    \item Optimise water use, in particular by recycling and reusing it.
    \item Minimise direct extraction from watercourses.
    \item Ensure that water released into the natural environment is of a quality at least equivalent to that of the water quality of the environment.
    \item Develop synergies for sharing water and reusing waste water for the benefit of other consumers, subject to technical and regulatory feasibility.
    \item Evaluate the use of waste water for industrial processes.
\end{itemize}

\paragraph{Waste}

\begin{itemize}
    \item Where possible, choose materials with a low impact on the environment (e.g., recycled, organic).
    \item Include a requirement in tendering processes for companies to avoid packaging and to remove their waste.
    \item Channel all waste to the appropriate sorting facility.
    \item Repurpose excavated materials, if possible, locally.
    \item Avoid single-use and limited-use equipment and containers.
    \item Share space and equipment.
\end{itemize}

\paragraph{Energy}

\begin{itemize}
    \item Take the Host States’ energy and climate goals into account when planning, building and operating future facilities.
    \item Limit energy consumption to the quantities and durations required.
    \item Recover and reuse as much energy as possible in the research infrastructure.
    \item Take energy recovery measures.
    \item Limit the use of fossil fuels.
    \item Plan for the timely implementation of energy supply contracts or power purchasing agreements for the supply of renewable energies.
    \item Plan to recover, store and supply energy, including waste heat, first for use within the project, with the possibility of subsequently making it available outside the project.
    \item Avoid unnecessary consumption (e.g., by operating systems only where and when necessary, avoiding systems that run in standby mode and assessing the possibility of switching systems off).
    \item Increase the temperature of the cooling water and ambient air in areas requiring air handling.
    \item Limit the areas in which air handling is required and the duration thereof.
    \item Avoid unnecessary lighting and operation of machinery.
    \item Optimise schedules by striking a balance between different criteria, in particular by gearing consumption towards periods when cheap renewable energy is available and by making the best possible use of recovered waste heat.
    \item Optimise and pool means of transport (e.g., conveyors, trains, electric vehicles) and commuting between sites.
    \item Adapt IT and data processing systems to requirements.
    \item Design and construct buildings with ambitious energy performance targets.
\end{itemize}

\paragraph{Ionising radiation}

\begin{itemize}
    \item Limit activities that produce ionising radiation.
    \item Locate hazardous activities in non-sensitive areas.
    \item Identify, quantify and control all radioactive emissions and immission in the environment and waste.
    \item Keep track of the locally applicable regulations.
    \item Draw up risk management protocols.
\end{itemize}

\paragraph{Biodiversity}

\begin{itemize}
    \item Identify local biodiversity issues to avoid destroying sensitive sites, habitats and ecological corridors. Where avoidance is not possible, reduce and compensate.
    \item Limit indirect impacts that disturb the environment (e.g., noise, light, vibration).
    \item Contain unavoidable impacts (e.g., lighting intensity and colour, operating periods).
    \item Plan measures to preserve flora and fauna and where not possible, develop mitigation measures to address residual impacts.
    \item Enhance biodiversity on and around the sites.
    \item Plant only local species and avoid planting any undesirable ornamental species.
\end{itemize}

\paragraph{Emissions}

\begin{itemize}
    \item Take into account the climate goals set by Switzerland and France in line with the IPCC recommendations.
    \item Identify all sources of emissions.
    \item Treat emissions and immissions in the environment before release.
    \item Determine the carbon footprint of emissions and immissions in the environment.
    \item Keep all activities on a virtuous emission pathway.
    \item Monitor and measure activities that are likely to pollute the air, soil and water.
    \item Avoid gases that have a significant greenhouse effect (e.g., SF$_6$) and use alternatives.
    \item Opt for electricity and hydrogen (fuel cells) from renewable sources for the powering of machinery and the transport of equipment.
    \item Encourage soft mobility and electric transport.
    \item Remain below the thresholds in force for the immission of suspended dust (PM10) and nitrogen dioxide (NO$_2$) around the perimeter of the future extensions.
    \item Identify and limit greenhouse gas emissions in the life cycle of the research infrastructure, taking into account the scientific objectives, technical feasibility and economic sustainability, including during equipment construction and procurement activities.
\end{itemize}

\paragraph{Placement}

\begin{itemize}
    \item Avoid locations that present major concerns.
    \item Reduce the footprint of surface sites to the absolute minimum, compliant with the requirements.
    \item Optimise the use of space by increasing the density of sites and buildings.
    \item Ensure that the infrastructure blends well into the landscape, the local environment and the urban setting.
    \item Avoid ground sealing and manage rainwater.
    \item Incorporate islands of freshness.
    \item Limit the impact of activities within the site.
    \item Plan for the resilience of the sites.
\end{itemize}

\paragraph{Societal dimensions}

\begin{itemize}
    \item Identify local sensitivities and issues.
    \item Limit noise, odour and light pollution for local residents.
    \item Integrate activities into the economic and social environment and develop synergies with local services, such as the fire brigades and other emergency and security services.
    \item As far as is compatible with the project plan, limit disturbance during rest periods (nights, weekends).
    \item Minimise road traffic and put in place alternatives (e.g., pooled transport, public transport, rail transport, conveyors).
    \item Provide parking spaces and loading and unloading areas during the construction phase, based on requirements identified and quantified in advance.
\end{itemize}

\paragraph{Raw materials}

\begin{itemize}
    \item Analyse the lifecycle of materials and infrastructure to anticipate the end of their life.
    \item Favour the use of low-carbon construction materials (e.g., low-carbon concrete), where feasible.
    \item Favour the use of recycled materials (e.g., steel, copper), where feasible.
    \item Prioritise local sourcing and regional production.
    \item Favour organic resources (e.g., wood), where feasible.
\end{itemize}

\subsection{Energy use}

The project is committed to promoting and continuously increasing the use of renewable energy sources \cite{gutleber_johannes_2023_8074977}. Fossil fuel-based energy will generally be avoided, and where its use is unavoidable, it will be minimised as much as reasonably possible. The internal heating of the particle accelerator infrastructure will reuse waste heat and renewable energy sources whenever feasible and economically viable in the long term. The use of fossil energy sources for backup electricity supply should be avoided. Instead, hydrogen to power a fuel-cell based system can be a valid alternative in addition to the short-term electricity supply via battery-based energy storage systems (BESS).

\begin{figure}[h]
  \centering
  \includegraphics[width=\linewidth]{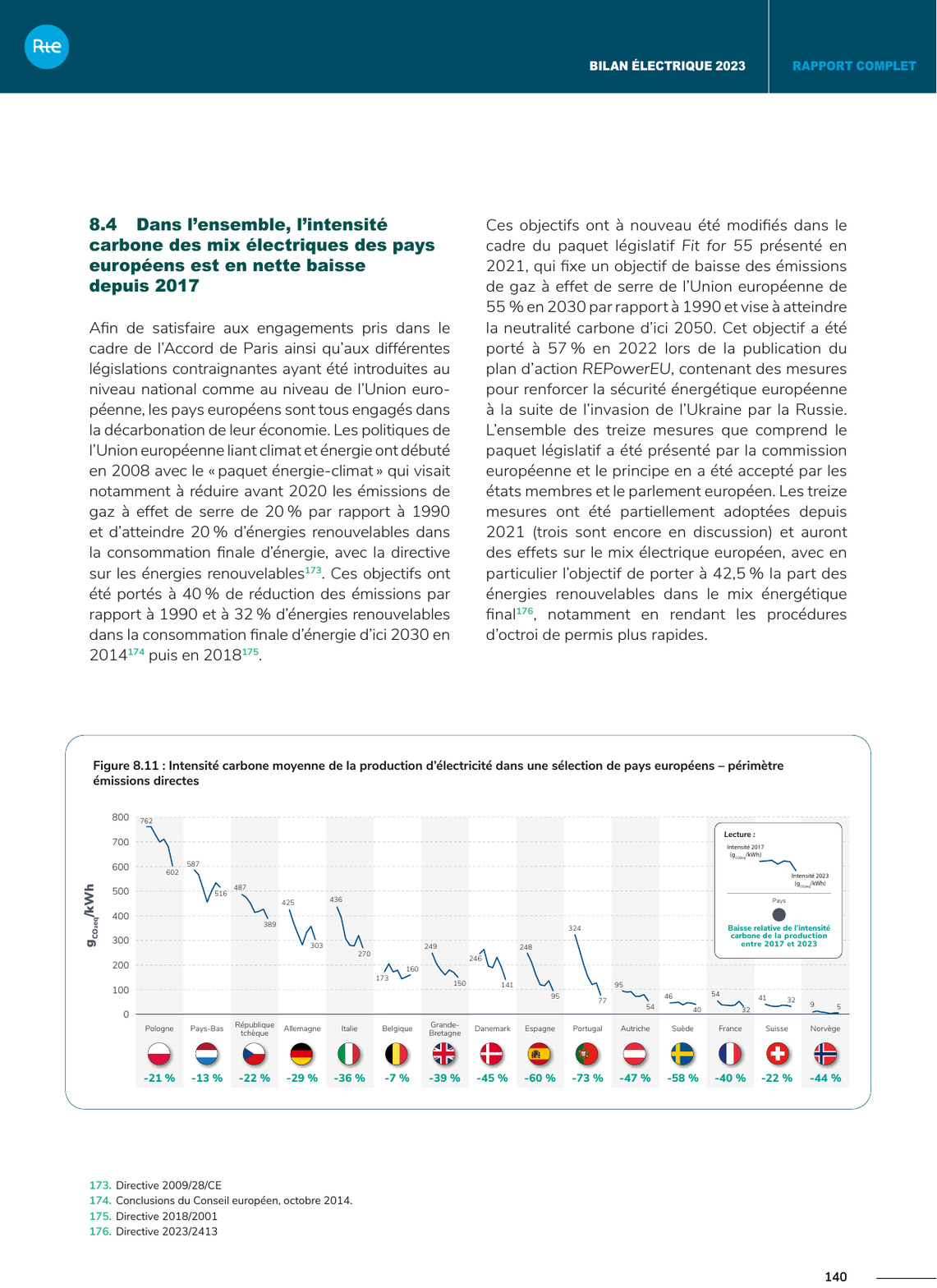}
  \caption{Evolutions of the carbon intensities of national grids in Europe. Source: RTE\cite{rte2023}.}
  \label{fig:carbon-intensity-europe}
\end{figure}

Studies carried out with external consultants to understand how far renewable energy sources can be leveraged, show that a strategy based on sourcing the majority of electricity from renewable energy sources at the national level can satisfy the project needs\cite{gutleber_2023_10023947}. Continuous increase of renewable energy sources and cross-border electricity transfer capacities pledged by the French government\cite{PPE_France_2024} support the validity of this concept \cite{RationalisinRTE}. In line with other European countries, the French electricity grid has experienced a continuous reduction of its carbon intensity \cite{rte2023} and is today amongst the countries with the lowest location-based electricity carbon intensity (Fig.~\ref{fig:carbon-intensity-europe}, Sweden, France, Switzerland, and Norway). The long-term planning \cite{RTE_SDDR_2040} published in early 2025 confirms the feasibility of the supply of the required energy from the grid and via renewable energy sources. 

The studies permit expecting that by 2035 the energy needs relating to the construction activities can be entirely satisfied by renewable energy sources. Implementing the approach via the local electricity supplies requires, however, that dedicated renewable energy supply contracts with certificates of origin are settled. This in turn calls for a sufficiently exhaustive analysis of the energy needs at the construction sites over the construction years and a reliable construction planning that is sufficiently stable (planning security). 

To prepare the energy supply for the operation phase, up to ten years of preparatory time may be required. This time frame serves well for defining the energy needs and consumption patterns, the invariant energy needs, the development of concepts and techniques that permit adapting the research infrastructure to supply possibilities in terms of changing availability and price. Furthermore, it allows time to develop the portfolio of power purchasing agreements in terms of functions and capacities, the contractual conditions and the tender processes. Eventually, the portfolio will lead to a mix of energy production technologies that match the needs of the research infrastructure, the ability of suppliers to grant flexibility, the carbon footprint, and general contract-related conditions. While the studies with external consultants show that on the time scale of 2035 covering more than 1\,TWh with renewable energy in the region is feasible at the 60\% level, significantly higher coverage will be feasible in the 2050 time frame.

It is worth keeping in mind that the supply and consumption of energy from renewable sources do not rely on a physical connection between the producer and the consumer. In an interconnected electricity grid, electrons are fungible, i.e., indistinguishable and interchangeable. Therefore, any volume of purchased energy from renewable sources is accounted in the financial, societal and environmental accounting scheme that determines the sustainability level of the project. It is not relevant where the energy has been physically produced and when it has been produced in the frame of a power purchasing agreement. It is, therefore, important to distinguish the carbon footprint of the power grid (location-based carbon footprint of consumption) from the certificate of electricity origin carbon footprint of a project (market-based carbon footprint of consumption). Renewable energy certificates and guarantees of origin that are part of power purchasing agreements play a crucial role in the carbon accounting of a project. The Greenhouse Gas Protocol Scope 2 Guidance explicitly states the requirement to report both figures~\cite{GHG:Scope2Guidance}. To achieve good financial performance, a more detailed forecast of the volumes required over an operating year and the possibility to adapt to the availability of renewable energy and its changing rewards are desirable. Consequently, a portfolio of renewable energy power purchasing agreements must be developed with adequate preparation time. The portfolio will evolve during the operation phase since the energy consumption needs to evolve according to different operation phases. 

\begin{figure}[h]
  \centering
  \includegraphics[width=\linewidth]{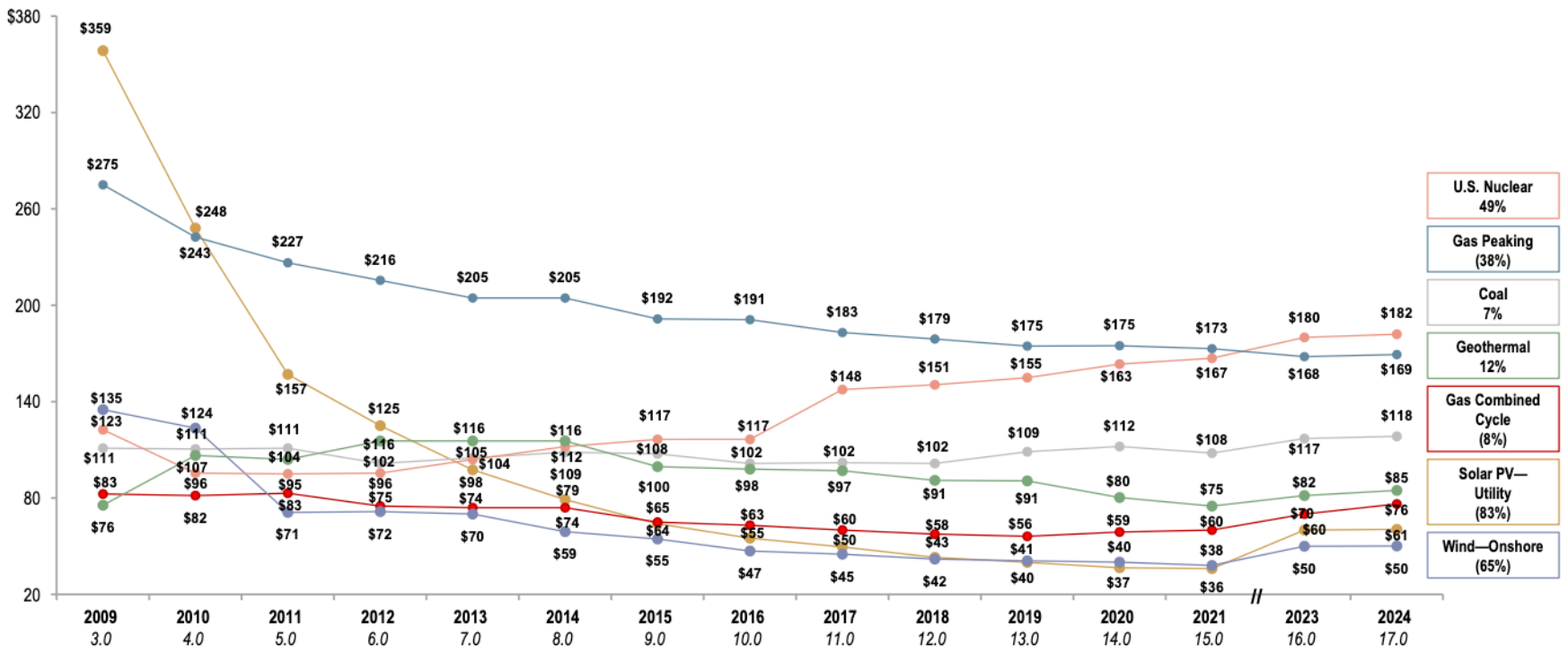}
  \caption{Evolution of the levelized cost of electricity (LCOE, Source: Lazard's Levelized Cost of Energy Version 17.0).}
  \label{fig:LCOE}
\end{figure}

\begin{figure}[h]
  \centering
  \includegraphics[width=.8\linewidth]{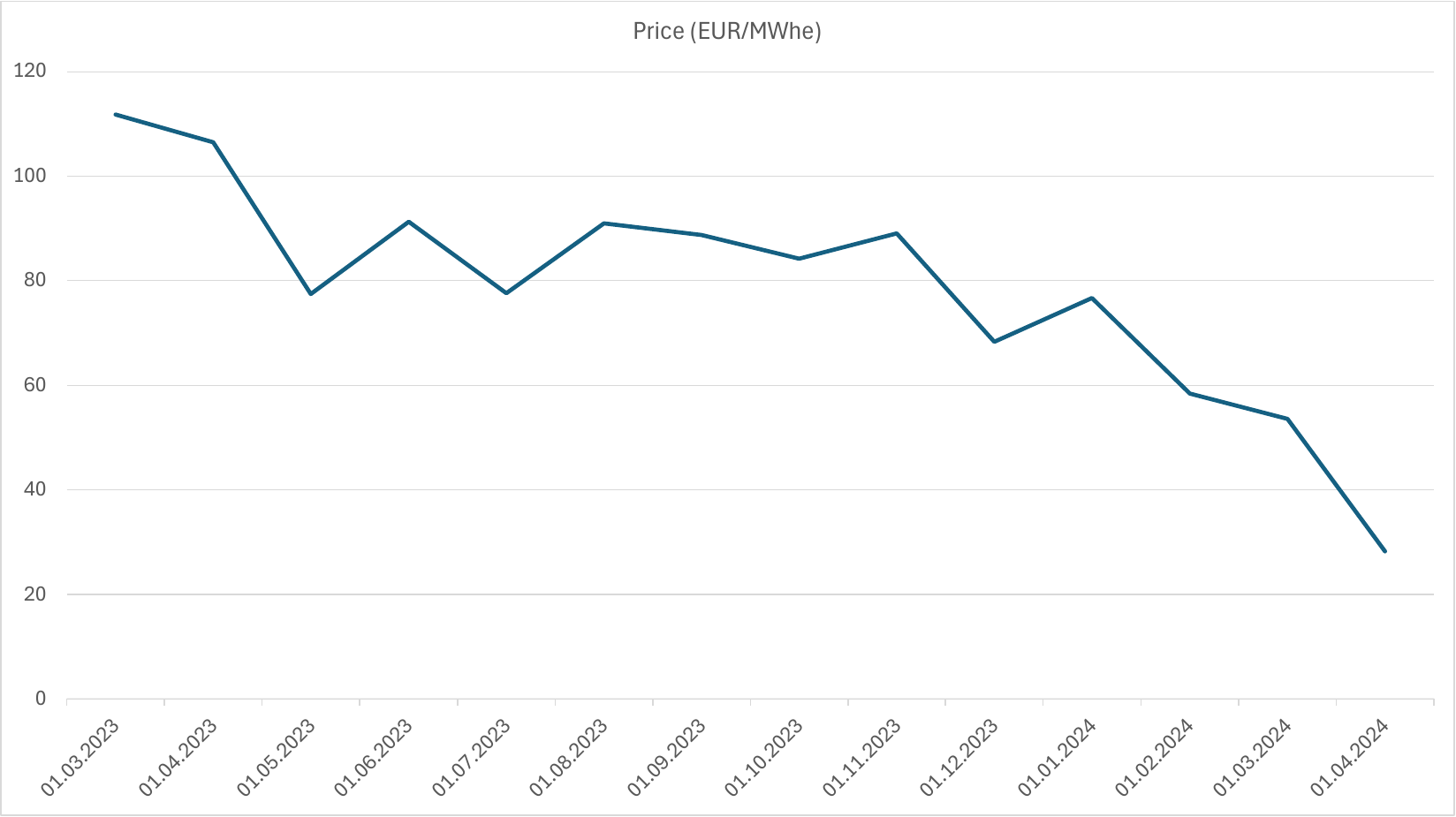}
  \caption{Example of monthly electricity spot price in France between 2023 and 2024 (Source: Ember Energy \cite{ember}
  ).}
  \label{fig:monthly_spot_prize}
\end{figure}

By 2023, the so-called levelized electricity costs (LCOE)\footnote{LCOE (Levelized Cost of Energy) is a metric used to assess the cost of generating electricity from different energy sources over their lifetime. It represents the per-unit cost (e.g., USD/MWh) of building and operating a power plant, considering all costs and revenues.} of all renewable energy sources are below the production costs of electricity production from fossil fuels\cite{lazard2023} (Fig.~\ref{fig:LCOE}). Seasonal fluctuations in the daily electricity prices are a result of a complex interplay of production capacity (supply), demand and non-technical factors. With respect to the pre-pandemic, electricity price evolution remains volatile. The production location can be different from the consumption location in the interconnected grid. Therefore, no general statement can be given, and the observation of large-volume electricity prices over multiple seasons did not permit an unambiguous seasonal cost variation pattern to be derived (Fig.~\ref{fig:monthly_spot_prize}). In particular, the production costs of electricity from renewable energy sources are diverse: while wind power is 10\% cheaper in winter and 20\% cheaper in spring than in summer, the situation for production from photo-voltaic sources is, in general, the inverse. In continental climate zones, a quasi-equilibrium of supply and demand exists\cite{iea:RES-variability}.

So far, it has not been possible to identify a significant difference in the cost of energy from renewable energy sources between seasons.

To achieve good coverage with low-carbon electricity sources, off-site energy storage could be foreseen in the power purchasing strategy. For example, hybrid PPAs are rapidly entering the market and, by the 2040 to 2050 period, will be an integral part of large-volume power purchasing. Missing volumes of renewable energy sources can be complemented with nuclear energy, which has a higher LCOE than renewable energy sources but is characterised by a very low carbon footprint.

\begin{figure}[h]
  \centering
  \includegraphics[width=.6\linewidth]{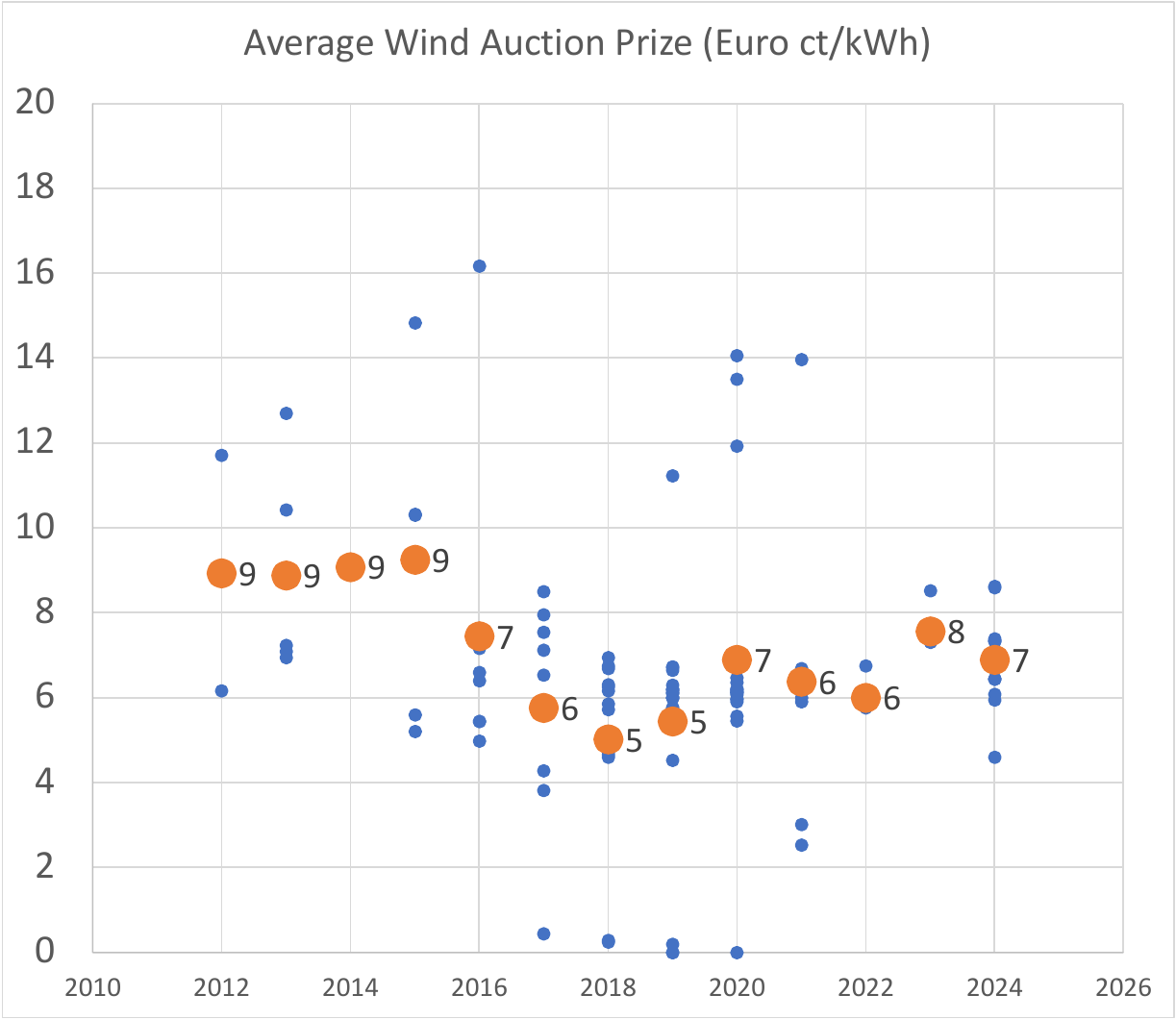}
  \caption{Historic contract market prices of electricity from wind power in Europe in cents/kWh up to the end of 2024 (Source: graph derived from public European wind auctions).}
  \label{fig:wind_auctions}
\end{figure}

By way of comparison, medium-sized offshore wind farms with a capacity of 600\,MW, such as Kriegers Flak (Denmark) or Dunkerque, produce around 2.5\,TWh of electricity per year at a cost of 44\,euro/MWh at 2021 prices. Due to periods of economic crisis that caused an energy price increase between 2020 and 2023, the auction prices for offshore renewable wind energy in France are decreasing again. In France, auction prices for renewable wind power sources are currently, on average, 69\,euro/MWh. These auction prices are a suitable indicator for setting the electricity price to be assumed for the annual energy costs in the overall sustainability assessment. Based on today's prices (Fig.~\ref{fig:wind_auctions}) and environmental performances of energy production, an average price range of 70 to 80 euro/MWh and a market-based carbon intensity of 15 to 25 gCO$_2$(eq)/kWh of electricity is very conservatively assumed today. 

\begin{figure}[h]
  \centering
  \includegraphics[width=\linewidth]{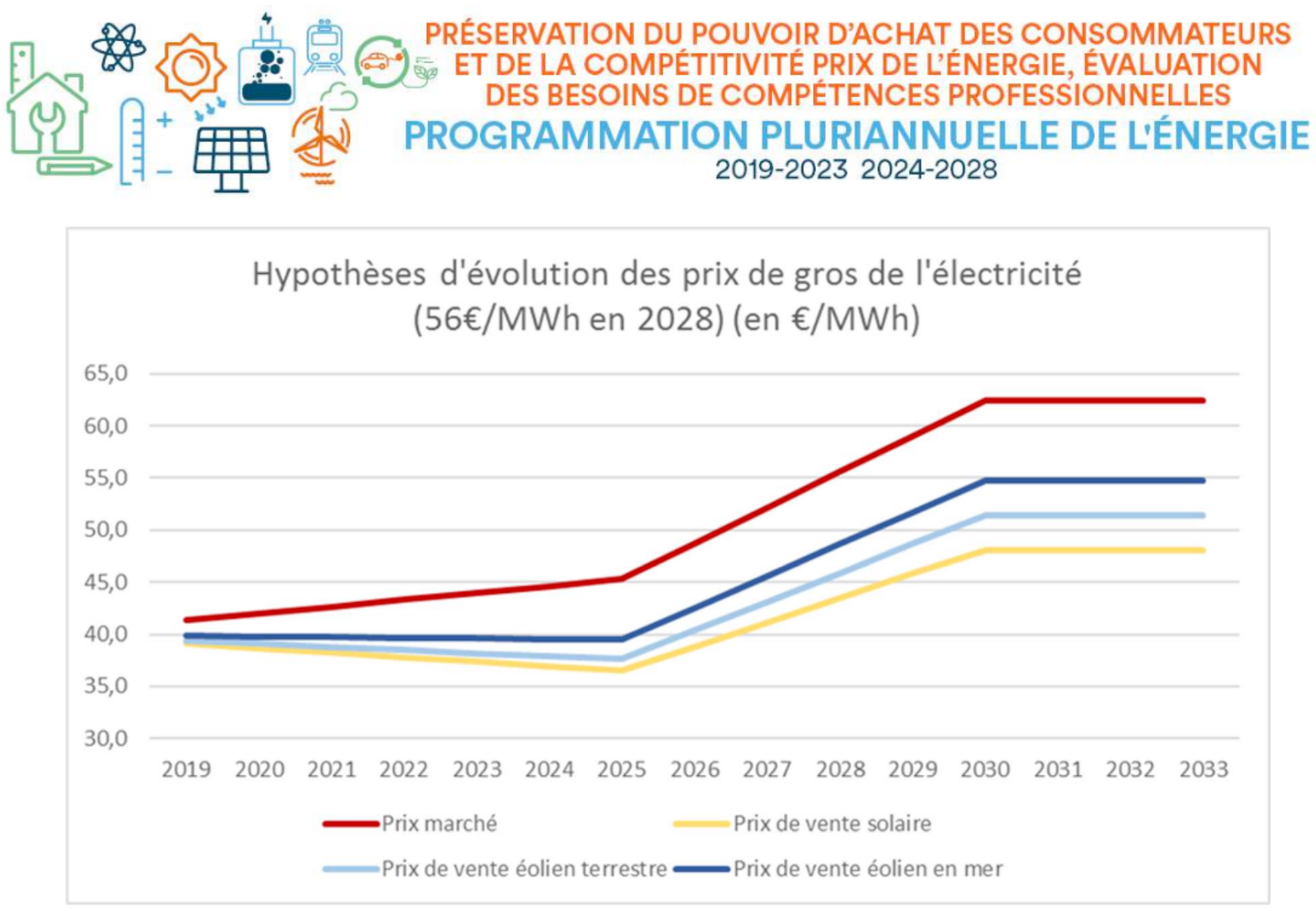}
  \caption{Stated electricity market price targets for 2030 published by the French government in the pluri-annual energy programme\cite{PPE_France_2024}).}
  \label{fig:electricity_cost_France}
\end{figure}

This value is significantly higher than the price goal of 56\,euro/MWh in 2030 in France, communicated in the public multi-annual energy planning of the government\cite{PPE_France_2024} (Fig.~\ref{fig:electricity_cost_France}).

To maximise waste heat supply potential while maintaining reasonably-sized heat buffering systems, operational flexibility and shifting operations to cooler seasons are preferred. Given the regional climate patterns, working conditions are generally more favorable in autumn and spring compared to summer.

The subsequent project preparatory phase must include the preparation of the power purchasing strategy and a plan for the construction phase. This will serve as a learning experience to develop and negotiate the power purchasing agreement portfolio for the operation phase, which is assumed to start after 2045.

\subsection{Waste heat use}

Almost all of the energy used to operate the technical infrastructures and subsystems of a particle accelerator is eventually converted into heat. Energy used to operate accelerator magnets, amplify radiofrequency energy, absorb synchrotron radiation, air management systems, operate electronics and data processing equipment is almost entirely converted into low-grade heat, typically below \SI{45}{\celsius}. Temperatures above \SI{50}{\celsius}, but still below \SI{70}{\celsius}, can rarely be reached, for instance, when cooling cryogenic refrigeration system equipment and electrical transformers and substations. This heat is typically dissipated in the ambient air via water-cooling and free-to-air cooling systems and is thus lost.
Given the amount of heat that particle-accelerator based research infrastructures generate, there is an interest to explore ways to recover that heat and convert it into a valuable resource. The use of the heat for other purposes inside and outside the project boundaries has the following socio-economic benefit potentials:

\begin{itemize}
\item Reduction of electrical energy consumption and associated costs due to reduced cooling system operation requirements.
\item Reduction of raw water consumption and associated costs due to reduced cooling system operation requirements.
\item Increased cooling system lifetime and reduction of associated maintenance and repair costs due to lower operational load on equipment.
\item Reduction of heat-generation-related carbon emissions due to the avoidance of dedicated heat production.
\item Lower heat costs for consumers.
\item Opportunities for creating new economic activities in the vicinity of the heat source.
\end{itemize}

\noindent However, heat recovery and supply also require additional efforts and costs:

\begin{itemize}
\item Additional components to recover the heat.
\item A dedicated network to transport the heat to where it is needed. 
\item Potential additional components to raise the temperature of low-grade heat supplied for specific needs.
\item Short-term and long-term heat storage systems to ensure supply stability and provide heat when needed.
\item The need to refurbish existing buildings and the need for new buildings that are equipped with heating and cooling that function with the low-grade heat.
\item The need for a heat supply operator if the heat is used outside the research infrastructure boundaries.
\end{itemize}

A technical-economic study has been carried out by an expert engineering company to create an inventory of the heat sources in the research infrastructure, to confirm the technical and financial feasibility and, most importantly, to draw up a detailed cartography of the heat demand within the surface sites perimeters. The study also included an assessment of how far heat recovery and supply is technically feasible and economically viable \cite{ginger_burgeap_2024_11192180, ginger_burgeap_2025_14719832, ginger_burgeap_2024_11192000}.

The information gathered was also integrated in the comprehensive, wider socio-economic impact analysis, estimating the contribution to the net present value of the project and subsequently reporting on the overall net benefit.
This approach ensures that an informed decision-making process is implemented.
If the system is put in place, it is strongly advised that the heat supply is continuously monitored to report on the efficiency of the approach and to be able to further optimise the waste heat recovery and supply process.

Whilst retrofitting heat recovery and supply to existing cooling systems is technically possible, it can be more costly than planning the concept from the onset. Depending on the existing equipment, infrastructures and environment around the particle accelerator facility, retrofitting may be less efficient since the operating temperature of the equipment supplying the heat may not be matched to the consumer needs e.g., magnet water cooling circuit temperatures too low, lack of data centre rack cooling infrastructure, mismatch of the waste heat characteristics with existing district heating networks or missing low-temperature district heating networks and finally lack of space and missing agreements with consumers. It is, however, preferred over no heat recovery. Therefore, from the outset, the FCC project has adopted an eco-design approach that integrates heat recovery and supply into the research infrastructure while embedding it within its broader socio-economic and environmental context.

Waste-heat recovery and supply are already implemented at CERN in the frame of the LHC project. One installation supplies waste heat from the cryogenic refrigeration system at LHC Pt8 in Ferney-Voltaire. A district heating network developed by the company Dalkia for the municipality, supported by the state, the region, and the French environmental organisation ADEME, connects to this surface site. The surface site PA in Ferney-Voltaire is envisaged as an immediate extension of this LHC site, leveraging not only the existing district heating network but also opening a window of opportunity to connect to structured heating networks~\cite{ge2024rts} on the nearby Swiss territory that also supplies Geneva airport and major industrial and commercial facilities as well as residential buildings in the vicinity of the surface site. The second example is the newly constructed data centre at the CERN Pr\'evessin site. Its recovered waste heat will largely cover CERN's campus heating needs. A third example is the heat recovery project at the CERN LHC surface site point 1 (ATLAS experiment) that will supply heat to the CERN Meyrin campus.

The following are examples of equipment that can serve as a starting point for the study of the functionality of integrating heat recovery and supply.

\begin{itemize}
\item Normal conducting magnets (recovery of cooling water at temperatures between \SI{25}-\SI{45}{\celsius}).
\item Normal conducting radiofrequency cavities (cooling water temperature between \SI{25}-\SI{45}{\celsius}).
\item Synchrotron radiation absorbers (cooling water temperature between \SI{25}-\SI{45}{\celsius}).
\item Rack mounted electronics (water cooled with a \textdelta T of 20\,K and a temperature range between \SI{27}-\SI{49}{\celsius} on the outer circuit with a temperature range between \SI{39}-\SI{60}{\celsius} on the rack cooling circuit).
\item Radiofrequency amplifiers of different types e.g., solid state, klystrons, IOTs with a cooling water circuit temperature range between \SI{25}-\SI{35}{\celsius} with high stability and low-temperature fluctuation constraint -- depending on the case, as tight as \SI{0.1}{\celsius}. By design, the maximum water temperature at the klystron's collector may be allowed to reach \SI{63}{\celsius}, but so far, applications operating in this regime are not known.
\item Cryogenic refrigeration plants for superconducting components (e.g., magnets, radiofrequency cavities) with equipment cooling circuit water temperatures in the range of \SI{50}{\celsius} (e.g., compressors) to \SI{75}{\celsius} (e.g., oil separator).
\item Power electronics and converters (temperature range of circuits between \SI{30}-\SI{60}{\celsius} for directly water-cooled IGBT systems, for example).
\item Electrical transformer stations with water cooling-based systems in the range of \SI{20}-\SI{70}{\celsius} (e.g., oil-based transformers).
\item Data centres~\cite{YUAN2023113777} (recovery of \SI{15}-\SI{20}{\celsius} air, \SI{40}-\SI{50}{\celsius} heat from the CRAHs and \SI{50}-\SI{60}{\celsius} from liquid cooling systems).
\item Ventilation and air management systems (e.g., heat from motors and air-to-air transfer) from \SI{25} up to about \SI{40}{\celsius}.
\end{itemize}

The initial equipment and heat load analysis will inform the designers which components are the ones that produce most of the heat and with which characteristics (stability, temperature).
This permits the creation of a hierarchy of heat-producing components that can guide the development of the heat recovery and supply concept.
A multi-criteria analysis with different weights for the individual aspects is a suitable approach for this first step.
The analysis process should at least include the following non-exhaustive list of aspects:

\begin{itemize}
\item Operational temperature requirements and constraints from the equipment components to be cooled and their temperature variation tolerances.
\item Temperatures of the heat recovery potential for the different equipment to be cooled.
\item Variability and stability of the heat generation (hourly, daily, weekly, monthly).
\item Climatic and weather conditions in the environment of the research infrastructure (for instance, a particle accelerator facility and a data centre operated in the north of Europe permit the use of different heat recovery and supply technologies than in a southern European region).
\item Use cases for the recovered heat inside the research infrastructure (e.g., pre-warming of water, offices, workshops, assembly halls, guest houses).
\item Demand of industrial heat consumers outside the research infrastructure (e.g., food production and processing industries, offices, hotels, airports, shopping malls, theatres, cinemas, congress centres).
\item Demand for heating public spaces and institutions (e.g., schools and universities, hospitals, prisons, train stations)
\item Demand for heating of private spaces (e.g., apartment buildings and individual houses).
\item Demand for hot water production (the required temperature is above \SI{55}{\celsius} for sanitary reasons and therefore priming with water boilers and heat pumps may be needed on an individual basis). 
\item Distances between heat production and consumers (note that distances up to \SI{10}{\kilo\meter} are feasible with modern pipe technology for low-grade district heating systems in the \SI{50}{\celsius} range).
\item Gap analysis concerning the need for heat buffering (e.g., capacity, space, duration, technology, investment costs, operation costs).
\item Investment costs for heat recovery, buffering and supply.
\item Operation costs for heat recovery, buffering and supply.
\item Capital and operation expenditures outside the system boundary (e.g., district heating network operation host, private heat pumps and priming equipment).
\item Public co-financing possibilities.
\item Duration and observation period envisaged for the heat recovery and supply.
\item Baseline for avoiding fossil energy sources for heating and the avoidance of primary energy for heating. Only the energy for stepping up the temperature for specific end-use cases needs to be considered.
\item Definition of the interface between the heat recovery and supply system that is part of the research infrastructure and the segment that is outside the responsibility of the research infrastructure.
\item Conditions of the heat supply operator which provides the infrastructure up to the consumer and that supplies the heat with guarantees or with contractual conditions that require the consumer to generate or obtain the gap between supplied and required heat.
\item The proposed heat supply technology (e.g., direct, indirect via a loop, indirect via heating the soil or other approaches).
\end{itemize}

Based on technical designs in the subsequent development phase, all data need to be collected and the most promising heat sources that qualify for a heat recovery case have to be identified. The viable heat consumers have to be confirmed in the frame of a territorial co-development activity. Eventually, the following non-exhaustive list of aspects to tune the heat recovery and supply scenario should be considered:

\begin{itemize}
\item Increase of the heat supplied by relaxing the equipment cooling requirements (e.g., water-based magnet cooling up to \SI{50}{\celsius}, increase of ambient air temperature inside the facility up to \SI{40}{\celsius} and possibly beyond).
\item Validation that mission-critical systems remain within their required operation margins (e.g., increasing the cooling water temperature of klystrons or relaxing their temperature stability may lead to unacceptable performance or render operation unfeasible).
\item Total amount of CO$_2$ emission reduction potential as a result of avoidance of fossil fuel and any primary energy, based on a credible estimate for the energy required to prime the heat for the end-use applications.
\item Optimisation of the heat supply by adjusting the operation schedule and introducing the possibility of reacting dynamically to heat needs.
\item Adaptation of the particle accelerator operation schedule to the actual societal heat demand to increase the overall socio-economic performance.
\item The potentially different energy costs for the research infrastructure operator when adjusting the operation schedule of the particle accelerator or when introducing the capability to dynamically react to both electricity supply and heat demand constraints.
\item Additional societal and economic benefits that can be generated by creating new heat consumers in the vicinity of the supplied heat (e.g., food processing industries, agricultural producers, biogas production, thermal baths and recreational installations).
\item Introduction of temporary heat buffers (daily, weekly, monthly).
\item Availability of specific public co-financing facilities and loans with specific conditions.
\item Optimisation of the interface between research infrastructure, operator and heat consumers.
\end{itemize}

The recovered heat will not be consumed at all times and the consumption pattern will change. Therefore, care must be taken not to under-dimension the cooling, ventilation and evaporation systems for the research infrastructure. If the heat is not consumed or cannot be delivered, it must be possible to cool all components reliably to ensure operation for scientific research purposes.

The techno-economic analysis permitted establishing the demand-based waste-heat supply scenarios based on the three different assumptions shown in Table~\ref{tab:waste-heat-scenarios}.

\begin{table}[h]
  \centering
  \caption{Waste heat supply potential according to local demand, supply scenario and operation mode.}
  \label{tab:waste-heat-scenarios}
  \begin{tabular}{lrcc}
    \toprule
    \textbf{Mode} & \textbf{\makecell[c]{Minimum\\supply potential}} & \textbf{\makecell[c]{Adaptation of operation\\schedule to demand}} & \textbf{\makecell[c]{Adaptation to demand\\and redistribution between sites}} \\ \midrule
    Z & 223 GWh/year & 308 GWh/year & 414 GWh/year  \\ 
    WW & 239 GWh/year & 339 GWh/year & 471 GWh/year \\ 
    HZ & 256 GWh/year & 371 GWh/year & 529 GWh/year \\ 
    L.S. & 60 GWh/year & 60 GWh/year & 60 GWh/year  \\ 
    $\rm t \bar{t}$ & 296 GWh/year & 441 GWh/year & 710 GWh/year \\ \bottomrule
  \end{tabular}
\end{table}

Based on this scenario, the potential of avoiding carbon emissions in the region by substituting conventionally created heat with waste heat that is largely produced from renewable energy sources can be estimated. Table~\ref{tab:avoided-emissions} shows the estimates of carbon emissions avoided, based on the following assumptions: an average market-based carbon footprint of 25\,tCO$_2$/GWh of electricity supplied to the FCC. A weighted average of about 190\,tCO$_2$ per GWh of conventional heat produced that can be substituted with waste heat\footnote{Regional heat production mix: 36\% electricity at 147\,gCO$_2$/kWh, 34\% gas at 227\,gCO$_2$/kWh, 16\% oil at 324\,gCO$_2$/kWh, 11\% wood at 30\,gCO$_2$/kWh, 0.3\% heat at 49\,gCO$_2$/kWh leads to a total footprint of 186.7\,gCO$_2$/kWh = 186.7\,tCO$_2$/GWh}. The resulting net carbon footprint avoided by supplying waste heat is \mbox{190-25\,=\,165\,tCO$_2$/GWh.} Comparing the avoidable carbon emissions by supplying waste heat with the range of total Scope 2 related carbon emissions of the collider operation between 305\,000 and 509\,000\,tCO$_2$ shows that the supply of waste heat can be partially substitute conventional heat sources at the same level, thus generating substantial positive environmental externalities that are made visible in the comprehensive socio-economic impact assessment.

\begin{table}[h]
  \centering
  \caption{One scenario outlining the potential for avoiding carbon emissions by supplying waste heat based on an average of 165 t avoided CO$_2$ per GWh of heat supplied.}
  \label{tab:avoided-emissions}
  \begin{tabular}{lcrrrr}
    \toprule
    \textbf{Mode} & \textbf{Years} & \textbf{Heat supplied/year} & \textbf{Heat supplied} & \textbf{CO$_2$ avoided/year} &\textbf{CO$_2$ avoided} \\ \midrule
    Z & 4 & 308\,GWh/year & 1232\,GWh & 50\,820 tCO$_2$/year & 203\,280 tCO$_2$ \\ 
    WW & 2 & 339\,GWh/year & 678\,GWh & 55\,935 tCO$_2$/year & 111\,870  tCO$_2$ \\
    HZ & 3 & 371\,GWh/year & 1113\,GWh & 61\,215 tCO$_2$/year & 183\,645 tCO$_2$ \\ 
    L.S. & 1 & 60\,GWh/year & 60\,GWh & 9900 tCO$_2$/year & 9900 tCO$_2$ \\ 
    $\rm t \bar{t}$ & 5 & 441\,GWh/year & 2205\,GWh & 72 765 CO$_2$/year & 363 825 CO$_2$\\ \midrule
    \textbf{Total} & & & \textbf{5288\,GWh} & & \textbf{872\,520 tCO$_2$} \\ 
    \multicolumn{5}{l}{\textbf{Operation Scope 2 emissions} (for comparison, 20\,350 GWh at 25 tCO$_2$/GWh)} & \textbf{508\,750 tCO$_2$} \\ \bottomrule
  \end{tabular}
\end{table}

\begin{figure}[h]
  \centering
  \includegraphics[width=\linewidth]{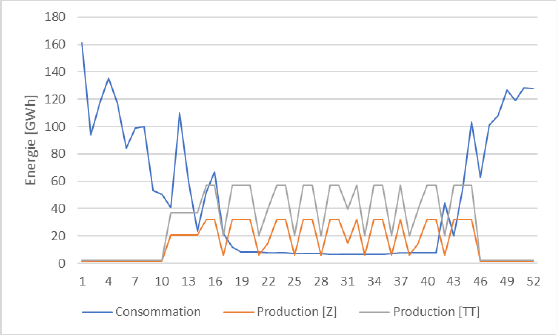}
  \caption{Weekly heat demand and supply for an example schedule of the Z operation mode.}
  \label{fig:Z-heat-demand-offer}
\end{figure}

The `minimum' scenario is based on the hypothesis that the particle collider starts operating in March and ends at the latest in November. Fig.~\ref{fig:Z-heat-demand-offer} shows as an example the weekly overview of the cumulative heat demand around a perimeter of 5\,km around each surface site.

\begin{figure}[h]
  \centering
  \includegraphics[width=\linewidth]{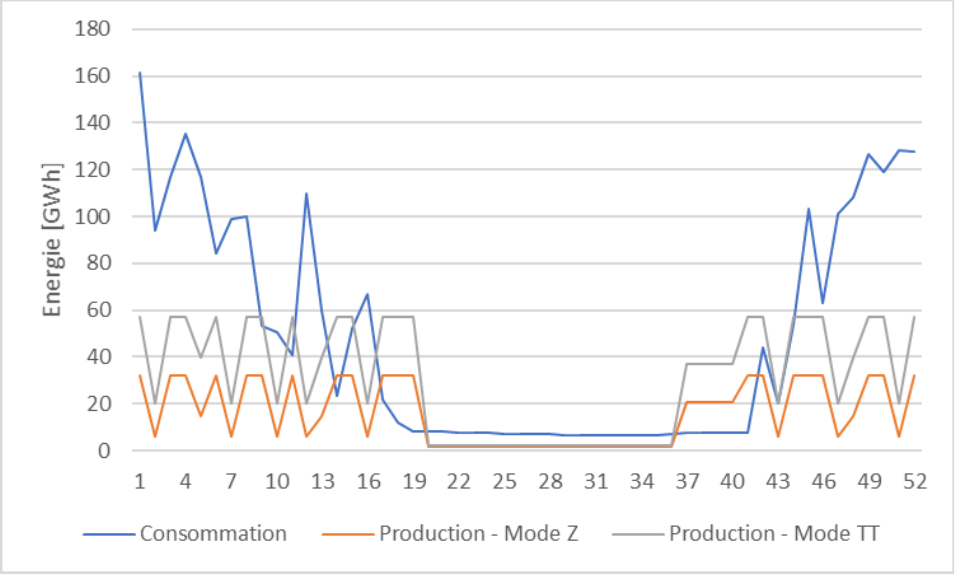}
  \caption{Weekly heat demand and supply for an adapted schedule of the Z operation mode.}
  \label{fig:Z-heat-demand-offer-adapted}
\end{figure}

Figure~\ref{fig:Z-heat-demand-offer-adapted} shows the heat demand and supply during the Z mode throughout the year, with a schedule that is better adapted to the heat demand.

\begin{figure}[h]
  \centering
  \includegraphics[width=\linewidth]{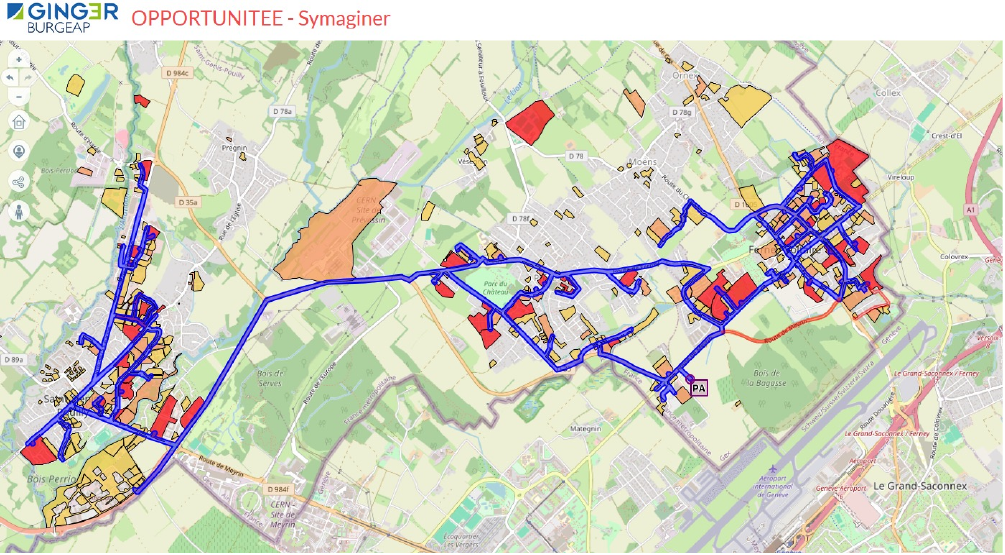}
  \caption{Example from the heat demand study at site PA in Ferney-Voltaire, within a perimeter of $\sim$5\,km around the site.}
  \label{fig:waste-heat-PA}
\end{figure}

Figure~\ref{fig:waste-heat-PA} shows the site PA in Ferney-Voltaire as an example for the heat demand study that permitted the development of the concept for the heat supply.

\begin{figure}[h]
  \centering
  \includegraphics[width=\linewidth]{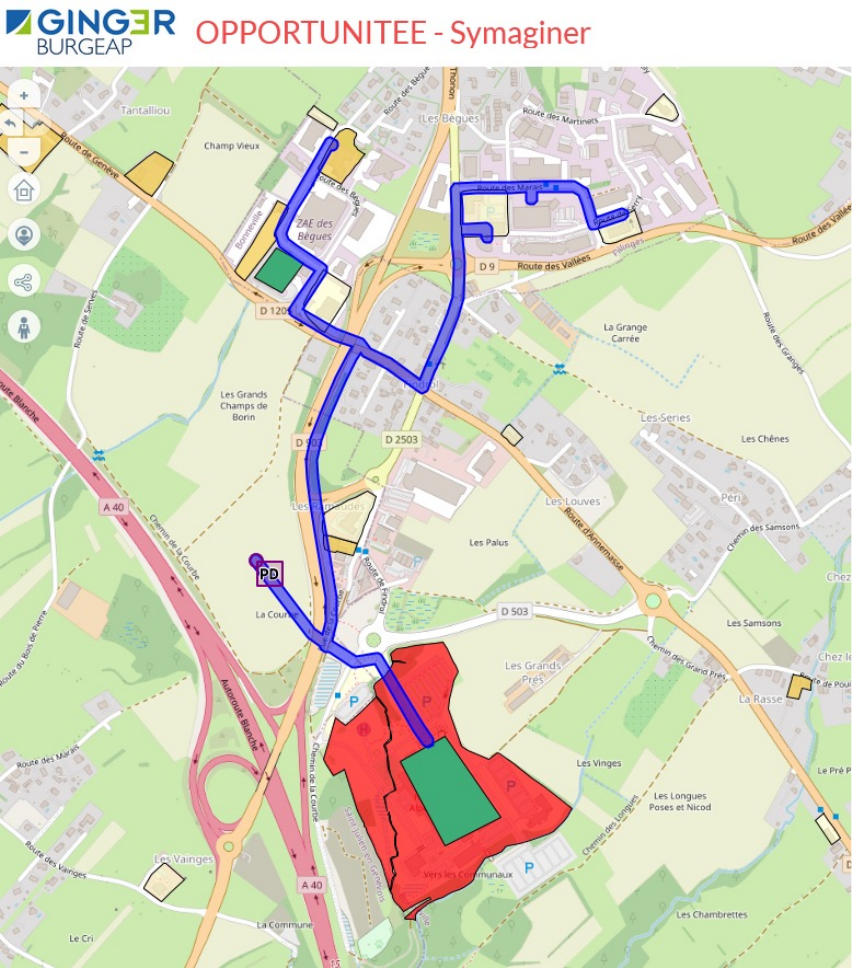}
  \caption{Example from the heat demand study at site PD in Nangy, concerning a nearby hospital (south), a cheese producer (north) and a mixed industrial/residential zone (north-east).}
  \label{fig:waste-heat-PD}
\end{figure}

Several industrial and public heat consumers were identified in the vicinity of several surface sites. They include, for example, a hospital, a school, cheese production facilities, an airport, commercial zones and public housing. Such heat consumers are preferred over supply to individual houses that are more difficult to connect. Public, industrial and commercial consumers typically have a higher and more stable heat demand. Site PD in Nangy (see Fig.~\ref{fig:waste-heat-PD}) is one example where significant amounts of heat in the 10\,GWh/year range can be supplied in the close vicinity of the surface site.

The techno-economical study revealed that waste heat supply is challenging at site PH (Cercier and Marlioz) and would be modest at sites PL (Challex) and site PB (Presinge). Therefore, a redistribution of the heat from PH to PG, from PL to PA and between PB and PD could be considered to improve the yield.

Table~\ref{tab:waste-heat-potentials} gives an overview of the potential total waste heat demand that exists today in the perimeters studied around each site. Waste heat is best re-used with the creation of new consumers, such as healthcare facilities, thermal baths, greenhouses, industrial facilities, and residential buildings that are connected to the new network from the outset.

\begin{table}[h]
  \centering
  \caption{Overview of the total potential waste heat demand that exists today in the perimeters studied around each site}
  \label{tab:waste-heat-potentials}
  \begin{tabular}{ccrl}
    \toprule
    \textbf{Site} & \textbf{Potential} & \textbf{Demand} & \textbf{Consumers} \\ \midrule
  PA & High & 200 GWh/year & \makecell[l]{Schools, commercial zones, residential} \\ 
  PA Extended & High & 1700 GWh/year & \makecell[l]{Extension to Switzerland: airport, commercial and\\industrial activities, residential, hospitals} \\ 
  PB & Medium & 30 - 200 GWh/year & \makecell[l]{School, greenhouses, penitentiary,\\hospital, commercial activities, housing} \\ 
  PB Extended & High & > 200 GWh/year & \makecell[l]{Commercial and residential demands\\in nearby France, sector Annemasse} \\ 
  PD & Low & 14 GWh/year & \makecell[l]{Hospital, industrial} \\ 
  PD Extended & Medium & 50 GWh/year & \makecell[l]{Schools, healthcare, residential} \\ 
  PF & Medium & 60 GWh/year & \makecell[l]{Sector La Roche-sur-Foron: industrial\\expo center, schools} \\ 
  PG & Low & 16 GWh/year & \makecell[l]{Schools, residential} \\ 
  PG Extended & Medium & 35 GWh/year & \makecell[l]{Schools, residential} \\ 
  PG Annecy & High & 860 GWh/year & \makecell[l]{Annecy at distance of 7\,km:\\Industrial, commercial, residential} \\  
  PH & Very low & 14 GWh/year & \makecell[l]{At 8\,km distance: Retirement home, residential} \\ 
  PJ & Low & 20 GWh/year & \makecell[l]{Schools, commercial, residential} \\ 
  PL & Low & 30 GWh/year & \makecell[l]{Spread over 5\,km: school, commercial, residential} \\ \bottomrule
  \end{tabular}
\end{table}

The supply of residual heat (or waste heat) from the FCC creates windfalls in three ways:
\begin{enumerate}
    \item  Balance the non-avoidable and non-reducible residual carbon footprint of electrical energy:
supplying the FCC with electricity (the working hypothesis is based on using electricity partly from renewable sources) would represent an average carbon footprint of around 40\,000\,tCO$_2$(eq)/year. The yearly supply potential from residual waste heat is around 320\,GWh of energy per year. The total maximum residual heat capacity is around 1\,600\,GWh per year. Consequently the supply of waste heat can substitute for the carbon footprint of the electrical energy consumed, provided that this heat supply can be implemented and that the demand can be satisfied via a heat distribution infrastructure. The technical-economic study~\cite{ginger_burgeap_2024_11192180,ginger_burgeap_2025_14719832,ginger_burgeap_2024_11192000} showed that 220\,GWh to 300\,GWh of heat could be consumed within a radius of approximately 5\,km around the surface sites. However, to increase the efficacy of waste heat supply, the particle collider operation schedule needs to adapt within acceptable limits to the heat needs.

\item Reduction of the carbon footprint of heating and cooling in the region:
the supply of energy by a heating network using a high proportion of renewable energies (including residual heat) avoids the need for other sources of heat. Based on the minimum reuse hypothesis (220 - 300 GWh/year), the production of 27\,500 to 38\,500 tCO$_2$(eq) could be avoided each year. Reasonable adaptation of the operating schedule throughout the year would be required to make this approach an effective lever.

\item Increasing the purchasing power of the local population:
The organisation operating the FCC is not a profit-making organisation. Energy can, therefore, be supplied by network operators at very competitive prices. According to the French multi-annual energy programme (PPE), waste-to-energy plants (which recover waste heat) sell this heat at a very competitive price, between 10 and €25/MWh. As a result, the heat supplied can then be the preferred choice for heating, hot water, and cooling. In the Anergie network (Ferney-Voltaire), the price of waste heat supplied by CERN equipment is even lower. According to a study by AMORCE and ADEME, the average selling price for networks supplied mainly by renewable and recovered energies (“EnR\&R”) was €78.2\,/MWh, incl. tax, in 2020 (these prices do not take into account the initial investment costs of the networks). According to the waste heat supply study carried out for this project by the engineering firm Ginger BURGEAP, the average price was at the same level more recently. Based on this model, the potential savings are currently estimated at €50/MWh  compared with gas heating and €140/MWh  compared with electricity, using energy prices of 1st November 2023 as a reference.

\item Limiting water consumption: 
reusing residual heat would also reduce the need for cooling water. The potentials are described in Section~\ref{sec:water-saving}.

\end{enumerate}

\subsection{Construction related carbon footprint}

\begin{figure}[h]
  \centering
  \includegraphics[width=\linewidth]{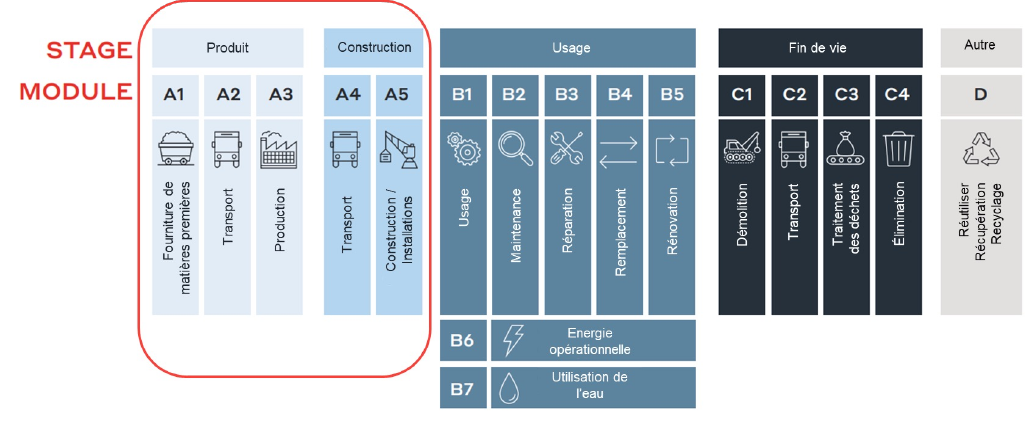}
  \caption{Stages considered for the LCA of the infrastructure construction.}
  \label{fig:LCA-construction-stages}
\end{figure}

As the global focus on combating climate change intensifies, reducing greenhouse gas (GHG) emissions has become a top priority. Infrastructures—spanning transport, construction, and scientific instruments — play a critical role in this transition. Carbon budget analysis, as part of a more comprehensive Lifecycle Analysis (LCA) (see Fig.~\ref{fig:LCA-construction-stages}), has emerged as an essential tool for measuring and managing these emissions effectively.

A comprehensive lifecycle analysis (LCA) conforming to the applicable ISO standards and European Norms, EN 14040 and EN 14044, has been carried out\cite{mauree_2024_13899160}. For the construction sector, the EN 17472 norm has been followed. In addition to the use of generic databases, a specific procurement scenario has been analysed, based on currently available Environmental Product Declarations (EPD) conforming to the European Norm, EN 15804+2, and the French `Fiche de Déclaration Environnementale et Sanitaire' (FDES).

The goal of the work was firstly to identify the key drivers for quantitatively estimated environmental impacts of the construction and, secondly, to establish a credible reference scenario for the carbon budget as a baseline for further designs and optimisations. It is important to note that a generic LCA cannot provide adequate indications for the carbon footprint, but is limited to the capability of identifying drivers. A specific and geo-localised project scenario with a particular sourcing scenario is necessary to be able to report absolute and credible estimates.
The methodology adopted comprised the following steps:
\begin{enumerate}
\item Identification of components : a detailed inventory of materials was compiled based on the bill of quantities for the subsurface construction, the 4 experiment sites and the technical sites. The products concerned are used throughout the infrastructure's lifecycle.
\item EPDs acquisition: EPDs were sourced for each material and product identified, ensuring compliance with EN 15804+A2, the foundation of EN 17472. The materials were selected based on expert knowledge of the local environment and state-of-the-art products. The availability of the products was confirmed by the suppliers.
\item Software Tool: A certified tool compatible with French and Swiss Environmental Product Declaration, \textsc{One Click LCA}, was chosen, ensuring robust and accurate calculations.
\item Data Entry: data was imported into the tool and project-specific data was entered, including material quantities and lifecycle phases, transportation for excavated and construction material.
\item Calculations: the tool was used to estimate the carbon budget, drawing on EPDs' data to evaluate GHG emissions for each lifecycle phase.
\item Result Analysis: the results were analysed to pinpoint major emission sources, comparing them against benchmarks and reduction targets.
\item Guidance: the results were used to sensitise engineers and scientists to include a carbon budget as a further input to the subsequent design phase and to ensure that requirements are well formulated and justified so that the infrastructure corresponds to what is required, thus limiting the environmental impacts.
\end{enumerate}

The analysis provided a breakdown of GHG emissions and additional quantitative potential environmental impacts across the various lifecycle phases, products and materials. The key emission sources are reinforced steel (14\%), precast concrete (49\%) and concrete (23\%). This highlights opportunities for emission reduction by designing the infrastructure with carbon reduction in mind, making careful material selection, optimising the construction process and making energy efficiency improvements. The most effective environmental impact management approach is a combination of the establishment of technical requirements for the infrastructures with rationales for the requirement, design of the infrastructure based on the requirements with carbon reduction in mind, careful material selection in agreement with the requirements, construction process optimisation, and energy efficiency improvements.  The GHG impacts of the initial and benchmark scenarios are given in Table~\ref{tab:carbon-reference}.

\begin{table}[h]
  \centering
  \caption{Reference carbon footprint of the FCC infrastructure construction for a period of 10\,years based on specific sourcing and procurement scenarios with EPDs.}
  \label{tab:carbon-reference}
  \begin{tabular}{lr}
    \toprule
    \textbf{Item} &  \textbf{Carbon footprint} \\ \midrule
   Subsurface & 477\,388 tCO$_2$(eq) \\ 
   4 technical sites & 17\,600 tCO$_2$(eq) \\
   4 experiment sites & 31\,200 tCO$_2$(eq) \\ 
   Total & 526\,188 tCO$_2$(eq) \\ \bottomrule
      \end{tabular}
\end{table}

The 526\,188 tCO$_{2}$(eq) carbon footprint of the construction over a period of 10\,years can be compared to CERN's current annual carbon footprint (184\,173\,tCO$_{2}$(eq) \cite{CERN-Enviro-2021}.
Thus, the construction corresponds to about 3\,years of CERN's annual footprint or about 30\% of CERN's annual carbon footprint per construction year. The FCC construction carbon footprint can also be compared to that of the Olympic Games in Paris which had an estimated carbon budget of 1\,580\,000\,tCO$_{2}$(eq) \cite{Paris2024}.

\begin{figure}[h]
  \centering
  \includegraphics[width=.9\linewidth]{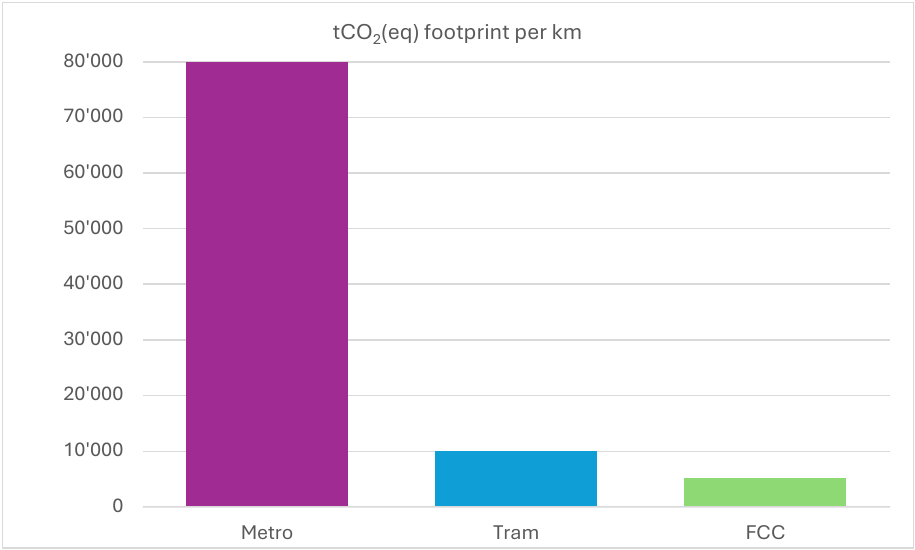}
  \caption{The construction-related carbon footprint per km of the FCC tunnel compared to typical public transport linear structures~\cite{mauree_2024_13899160}}
  \label{fig:linear-carbon-footprint}
\end{figure}

With respect to conventional construction projects, the carbon footprint of a research infrastructure is significantly lower (see Fig.~\ref{fig:linear-carbon-footprint}) than a small-scale metro line or a tramway line. For instance, the construction of the U5 Metro line in Berlin, Germany had a carbon footprint of 98\,000 tCO$_{2}$(eq) per km \cite{Berlin-Ubahn}.
On average, per km of underground transport line construction, the carbon footprint is 80\,000 tCO$_{2}$(eq). The construction of a tram line has a carbon footprint between, 7\,600 and 10\,850 tCO$_{2}$(eq) per km.

Based on the analysis, a number of recommendations were developed with the help of the expert company that carried out the LCA.

Although CERN's annual carbon footprint will gradually be reduced, in line with the established environmental goals,  national climate protection plans and IPCC recommendations, it would be advisable to limit adding construction-related climate impacts to the operation-related carbon budget. Hence, while the construction of a new facility ramps up, it would be prudent to reduce the activities related to other carbon-intense activities.
With respect to the construction, the following strategies can help to reduce the potential impacts further:
\begin{itemize}
\item Implementation of a thorough systems engineering methodology to develop well-justified and fully documented technical requirements for both subsurface and surface structural elements. These requirements will represent the absolute minimum necessary to accomplish the scientific research programme effectively (for example, determining the precise dimensions and volumes of structures that meet the essential technical and physical requirements).
\item Integration of an eco-design in the comprehensive systems engineering process with carbon reduction in mind that matches the established requirements (e.g., appropriate sizing of the caverns, shafts, alcoves).
\item Structural modification of the scenario by reducing the inner line thickness of subsurface structures by at least 5\,cm, leading to a reduction of 16\% for precast concrete and rebar steel. 
\item Material substitution, considering low-impact materials which meet the established requirements and working with industrial partners to innovate and produce locally wherever possible. 
\item Construction process optimisation to minimise emissions.
\item Reuse of excavated materials, for instance, in concrete production. 
\end{itemize}

With respect to the initial baseline scenario, the reduction of the circumference to about 91\,km and the suppression of 4 shafts have led to a significant reduction of the construction-related carbon footprint during the scenario development phase. Furthermore, the carbon footprint must be seen in the context of establishing an infrastructure that serves a worldwide community of about 15\,000 scientists for two subsequent particle colliders until the end of the century. Although according to international norms the quantities reported by the LCA must be entirely accounted for the first project phase only (FCC-ee), it can be seen as an investment that benefits the second phase, the high-energy hadron collider phase (FCC-hh).

\subsection{Operation related carbon footprint}

\subsubsection{Detectors}
While the carbon footprint of today's scientific research facilities, such as the LHC and its experiments, is largely caused by gases used in the detectors, this contribution will be only marginally relevant for the FCC era: gases with significant climate effects are already being banned, and the list of such products is growing rapidly. 

Today's working gases will largely be unavailable by the year 2050 when the first collider enters its operation phase. 
The performance and long-term operation of gaseous detectors rely primarily on the use of the optimal gas mixture, which is the active medium where the primary ionisation happens in the detector. Several gaseous detector technologies make use of gas mixtures containing expensive or greenhouse gases, which have specific properties that allow optimal detector performance and avoid ageing effects. The GHGs most in use are the $\rm C_2H_2F_4$ (known as R134a, GWP of 1430) and SF$_6$ (GWP of 23,900) for the Resistive Plate Chambers, the $\rm C_4F_{10}$ (GWP of 8860) for Cherenkov detectors and the CF$_4$ (GWP of 7390) for wire chambers, Cherenkov detectors and micro pattern gaseous detectors (MPGDs). These gases are necessary to mitigate ageing phenomena, to act as a Cherenkov radiator and to contain charge development (thanks to their electronegative properties) or to improve time resolution.

The detector volumes range from a few m$^3$ to hundreds of m$^3$ making the use of gas recirculation systems compulsory to reduce operational costs and GHG emissions. Even with the implementation of these systems, in some cases, emissions can be present mostly due to detector requirements or the presence of leaks. Residual leaks are mainly due to failures of plastic pipes and connectors that break due to built-in fragility and mechanical stress. These leaks are not accessible during run periods and, in some cases, during regular technical stops. Big leak search and repair campaigns usually take place during long shutdown periods, but leaks typically keep developing.

To reduce emissions, today's strategy~\cite{CERN-Environment-2023-003} is based on three lines of action:
\begin{itemize}
    \item \textbf{Gas Recirculation.} The gas mixture is taken at the output of the detectors, purified and sent back to the detectors. It is technically possible to recycle 100\% of the gas mixture.
    \item \textbf{Gas Recuperation.} The gas mixture is sent to a recuperation plant where the GHG is extracted, stored and re-used. This system is always used in combination with a gas recirculation system to allow a further GHG reduction.
    \item \textbf{Alternative gases.} Search for alternative gas mixtures suitable for particle detectors that do not contain or have limited use of GHGs.
\end{itemize}

The substitution of fluorinated gases (F-gases) is fundamental because of the implementation in Europe of the F-gas regulation that will render such substances unavailable by 2050. Also in 2023  the European Chemicals Agency (ECHA) released a proposal regarding restriction on PFAS, i.e., per- and polyfluoroalkyl substances, which contain at least one fully fluorinated methyl (CF$_3$-) or methylene (-CF$_2$-) carbon atom (without any H/Cl/Br/I attached to it). The proposal covers over 10\,000 different PFAS, which are considered environmental pollutants with links to harmful health effects. Most of the so-called `eco-friendly' gases belong to the PFAS family.

In addition, devices and circuits using gases either for cooling or for particle detection purposes must be designed to be leak-tight and use a re-circulation principle.

\subsubsection{Scope 2}

The principal cause of carbon footprint will be the use of electrical energy. 

The same approach will be used for the optimisation of all other resource use, such as electrical energy. The adoption of the hierarchical `Avoid-Reduce-Compensate' principle guides the iterative planning, implementation, checking and taking action cycle that leads to continuous optimisation.

A comprehensive technical requirements gathering process has to be established to document and scrutinise where and when electrical energy is needed. Following a baseline scenario, the eco-design approach helps to conceive designs and choose products that lead to reduced energy consumption. A review of the operation model that will be guided by overall sustainability goals integrating economic, ecological and societal aspects will guide the development of different scenarios. Some measures require another iteration of the eco-design cycle by introducing new requirements that foster sustainability, such as the integration of waste-heat recovery and supply functionality. It permits, for instance, substituting fossil energy used inside and outside the project, but it comes with constraints such as additional investment and operation costs, the necessity to buffer energy, the need to dynamically adapt operation, requirements on financial and ecological accounting and the requirement to establish administrative and commercial frameworks for successful implementation. All relevant ESG parameters need to be taken into consideration, and it is therefore recommended that experienced companies be employed for the overall energy optimisation of the infrastructures.

Official national values for the selected base year of the sustainability analysis must be used to determine the carbon footprint of the electricity consumed during the design and planning phases. Consequently, the carbon footprint depends on the technology (offshore wind, onshore wind, photovoltaic, hydro, nuclear) and the geographical location of the energy production. Even if all electricity is transported and supplied to the FCC by the French electricity grid, operated by RTE, this does not mean that the electrical energy needs to be sourced solely from French territory. Studies carried out with experts in the domain \cite{sophie_auchapt_2023_7781077} showed that a comprehensive portfolio of energy supply contracts or power purchasing agreements (PPA) would be best established with sufficient lead time (order of 10 years). It must be able to respond to the evolving needs of the commissioning and operation phases. In order to foster the use of renewable energy sources and exploit the waste heat supply functionality, the particle collider would need to adapt to the energy supply contract conditions that eventually are determined by commercial conditions on the one hand and by the availability of renewable energy capacities on the other.

Depending on the country of origin, the carbon intensity of electricity is made available from different sources. Table~\ref{tab:ElectricityCarbonFootprint} provides an illustration.

\begin{table}
\centering
\caption{Examples of some official electrical energy carbon footprint sources.}
\label{tab:ElectricityCarbonFootprint}
\begin{tabular}{lllrc}
\toprule
\textbf{Country} & \textbf{Source} & \textbf{Energy} & \textbf{\makecell[r]{Carbon footprint\\gCO$_2$/kWh\\tCO$_2$/GWh}} & \textbf{Year} \\ \midrule
France & Ademe Base Empreinte & Offshore wind & 15.6 & 2023 \\
France & Ademe Base Empreinte & Unqualified energy mix & 52.0 & 2022 \\
Germany & Umweltbundesamt (UBA)\footnote{https://www.umweltbundesamt.de/publikationen/entwicklung-der-spezifischen-treibhausgas-10} & Unqualified energy mix & 380.0 & 2023 \\
Italy & ISPRA & Unqualified energy mix & 257.2 & 2022 \\
Switzerland & BAFU/OFEV\footnote{https://www.bafu.admin.ch/bafu/de/home/themen/klima/fragen-antworten.html} & Unqualified energy mix & 54.7 & 2018 \\
Switzerland & BAFU/OFEV & Renewable energy mix & 15.7 & 2018 \\
\bottomrule
\end{tabular}
\end{table}

Once a specific energy supply contract is active, emission accounting will be carried out using supplier-provided emission information. For instance, a typical consumer-oriented electricity contract based entirely on renewable energy sources   (62\% hydro, 31\% wind and 7\% PV, 100\% are certified to be of renewable energy sources) in France today has an actual carbon footprint of 34\,kgCO$_2$(eq)/MWh\cite{Engie}.
This value is higher than the ADEME indicated emission factor for renewable energy sources to be used for the planning and design phases since the consumption of electricity in the frame of a live contract includes all emissions along the value chain and not only the production-related emissions. In addition, renewable energy that is potentially sourced from physical PPAs is not accounted for in the carbon footprint.

Different assumptions were taken to estimate the Scope 2 carbon footprint related to the operation. One assumption is based on today's energy mix, including nuclear energy and renewable energy sources with a varying contribution of up to 75\% and a carbon footprint of 24 gCO$_2$/kWh. This configuration leads to an average annual indirect carbon footprint of about 500\,000\,tCO$_2$ or approximately 35\,000\,tCO$_2$ per year depending on the renewable PPA and energy supply contract portfolio\footnote{20\,900\,GWh * 24 tCO$_2$(eq)/Gwh = 501\,600\,tCO$_2$(eq).}.

\begin{table}[h]
  \centering
  \caption{Lowest carbon footprint energy mix assumption used for the  estimate of the carbon footprint for operation on a 2050 time horizon. The carbon intensity figures were obtained from the Ademe Base Empreinte database, 2023.}
  \label{tab:energy-mix-2050}
  \begin{tabular}{lrr}
    \toprule
    \textbf{Energy Source} & \textbf{Contribution} & \textbf{Carbon intensity} \\ \midrule
    Nuclear & 10\% & 3.7 gCO$_2$(eq)/kWh  \\ 
    Offshore wind & 55\% & 15.6 gCO$_2$(eq)/kWh \\ 
    Onshore wind & 10\% & 14.1 gCO$_2$(eq)/kWh \\ 
    Photovoltaic & 15\% & 25.2 gCO$_2$(eq)/kWh \\ 
    Hydro power & 10\% & 6.0 gCO$_2$(eq)/kWh \\ \midrule
    \textbf{Total} & 100\% & \textbf{14.74 gCO$_2$(eq)/kWh} \\ \bottomrule 
  \end{tabular}
\end{table}

For an optimistic reference estimate, it was assumed that an energy mix based on a portfolio of physical and non-physical PPAs and energy supply contracts on the 2050 time horizon, entirely sourced from French territory via the national grid operated by RTE (see Table~\ref{tab:energy-mix-2050}). Based on this mix, the total carbon footprint of about 300\,000 tCO$_2$(eq) was estimated\footnote{20\,900\,GWh * 14.74 tCO$_2$(eq)/GWh = 308\,066 tCO$_2$(eq).} for the entire scientific research period over 15 years. This corresponds to an annual average of about 20\,500 tCO$_2$(eq). Using a pessimistic estimate of 25 tCO$_2$(eq)/GWh leads to a carbon footprint of 522\,500 tCO$_2$(eq) of an annual average of about 34\,800 tCO$_2$(eq).

The residual carbon footprint can in principle also be offset by the supply of waste heat. Depending on the amount of waste heat re-used, this opens an opportunity to evolve towards net-zero operation scenario after 2050. Such a scenario requires the careful development of energy supply contracts and/or long-term PPAs and the creation of district heating and industrial heat supply networks, with an adaptation of the accelerator operation to the energy supply and the heat demand. Introducing temporary heat buffering will help achieve the goal.

\subsection{Water use and saving}
\label{sec:water-saving}

Drinking water from the existing local distribution networks will only be used for drinking and sanitary purposes. All raw water required for industrial cooling systems will be sourced from CERN's existing water supply (SIG) in Switzerland, which sources the water from Lake Geneva.

From a quantitative point of view, the reference scenario for water consumption is technically, financially, and territorially feasible since water extraction and consumption represent quantities lower than CERN's actual consumption in 2022. The availability of a supply representing twice the total needs was confirmed by SIG in 2023.

During the subsequent design phase, it will be necessary to verify aspects relating to the sharing of water with the other local stakeholders who use the same water sources, particularly with regard to related catchment areas which currently experience chronic deficits (Pays de Gex, La Roche-sur-Foron and possibly the Usses). Numerous synergies can be considered with local authorities in the vicinity of surface sites in terms of the reuse of residual heat and released water.

Therefore, a study has been launched recently to determine the feasibility and the conditions to also source water from a water treatment station near the site PD in Nangy (STEP SRB in Scientrier, see Table~\ref{tab:water-needs}). Initial results are promising, requiring, however, a reduction of the non-soluble content in the water and an effective treatment of bacteria. A demonstration will be required for such an installation. If considered viable, it can lead to substantial raw water reduction and can help supply treated water for other industrial purposes when the particle accelerator does not require it.

\setlength{\tabcolsep}{5pt}

\begin{table}[h]
  \centering
  \caption{Overview of the raw water needs for cooling for each operation mode, waste water re-use potentials for sites PD, PF, PG and the residual treated water that can be made publicly available.}
  \label{tab:water-needs}
  \begin{tabular}{ccrrrr}
    \toprule
    \textbf{Mode} & \textbf{Years} & \textbf{\makecell[r]{Initial FCC\\water need}} & \textbf{\makecell[r]{Waste water used\\at PD, PF, PG}} & \textbf{\makecell[r]{Residual\\water needs}} & \textbf{\makecell[r]{Treated waste water\\available to society}}\\ \midrule
    Z & 4 & 1\,604\,861\,m$^3$/year & 560\,304\,m$^3$/year & 1\,044\,557\,m$^3$/year & 2\,049\,792\,m$^3$/year  \\ 
    WW & 2 & 1\,928\,943\,m$^3$/year & 619\,017\,m$^3$/year & 1\,309\,927\,m$^3$/year & 1\,991\,080\,m$^3$/year  \\ 
    ZH & 3 & 2\,165\,458\,m$^3$/year & 705\,173\,m$^3$/year & 1\,460\,285\,m$^3$/year & 1\,904\,924\,m$^3$/year  \\ 
    L.S. & 1 & 163\,817\,m$^3$/year & 78\,840\,m$^3$/year & \,977\,m$^3$/year & 2 531 256\,m$^3$/year  \\
    $\rm t \bar{t}$ & 5 & 3\,077\,591\,m$^3$/year & 897\,334\,m$^3$/year & 2\,180\,258\,m$^3$/year & 1\,712\,763\,m$^3$/year  \\ \bottomrule
  \end{tabular}
\end{table}

\setlength{\tabcolsep}{6pt}

\begin{figure}[h]
  \centering
  \includegraphics[width=\textwidth]{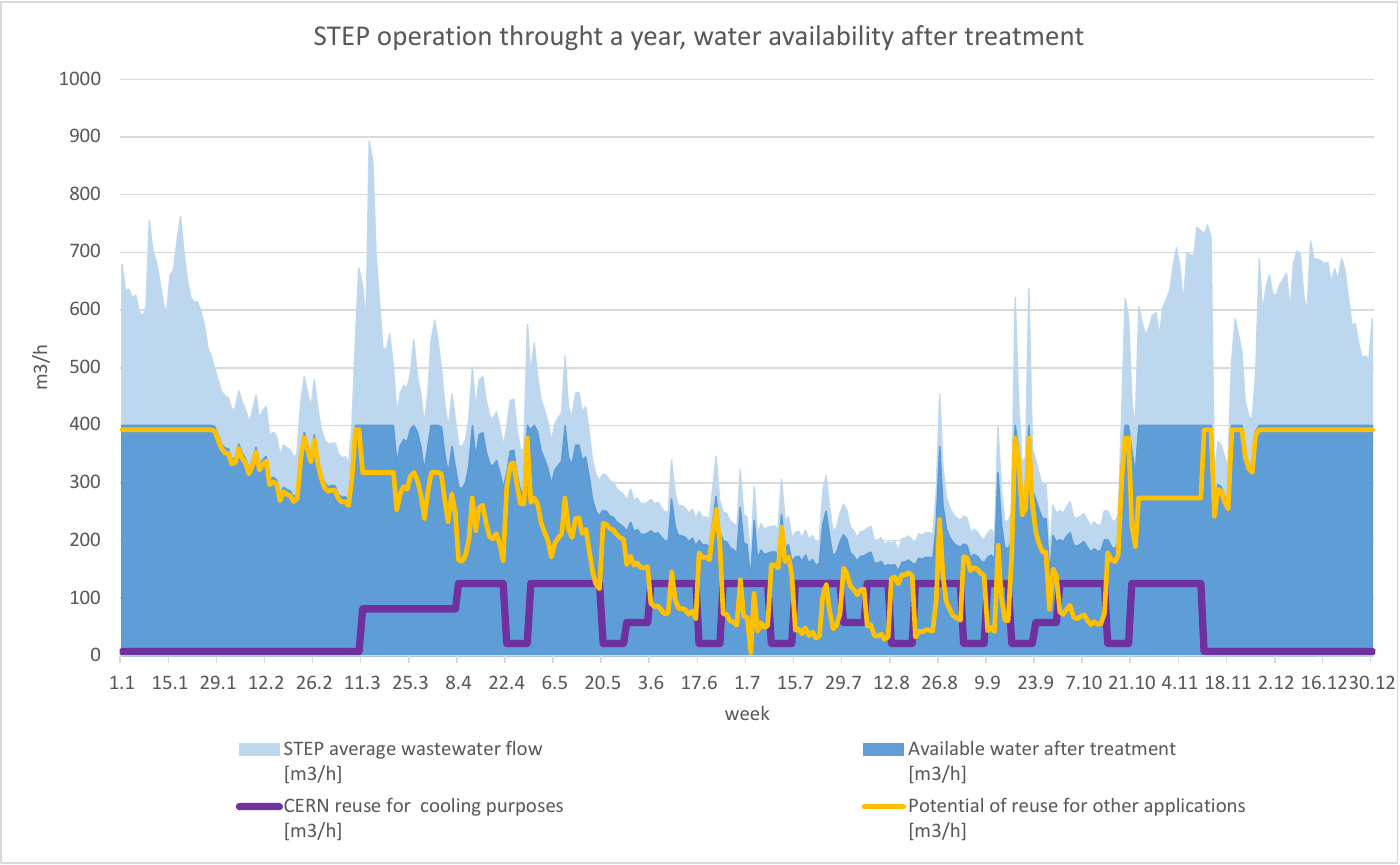}
  \caption{Operation of the STEP in Scientrier throughout a typical year, matching with FCC Z mode water cooling needs and the capacities made available by water treatment to the FCC and society.}
  \label{fig:STEP-capacity}
\end{figure}

The preliminary analysis has been carried out based on the water supplied to the treatment plant during a typical year and an assumption of a treatment capacity of up to 400\,m$^3$/hour. Although a water treatment plant has, in principle, the capacity to provide more water than is needed for the cooling of the particle collider, it is prudent to assume that the water is mixed with ordinary raw water, that the treated water is not always compliant with the needs, and that the treatment is not always fully efficient or available. Figure~\ref{fig:STEP-capacity} gives an impression of the typical annual operation of the water treatment plant in Scientrier close to site PD and how wastewater recovery would map to the collider cooling needs for the Z mode as an example.

It should be noted that ultimately the feasibility of this approach and the actual capacities depend on the technical designs and the possibility of implementing the concept in cooperation with the national, regional and local stakeholders in France.

Furthermore, water reduction can be achieved with the recovery and supply of waste heat, since less heat needs to be dissipated with evaporation towers. The reduction potential estimated in the dedicated engineering study was approximately 1.6\,m$^3$ of water per MWh of waste heat supplied. 
Table~\ref{tab:water-saving-heat-supply} outlines the potential amount of cooling water savings for each operation mode and for one of three different waste heat supply scenarios:

\begin{itemize}
\item[] Scenario 1: the heat is distributed in a 5\,km radius with no adaptation of the schedule. This means that the FCC would be operated from April to September, when the heating needs are low.
\item[] Scenario 2: the heat is distributed in a 5\,km radius with an adapted schedule. This means that the FCC would be operated from October to March, when the heating needs are high.
\item[] Scenario 3: the heat is distributed in a 5\,km radius with an adapted schedule and distribution. This means that the infrastructure would operate from October to March, when heating demand is highest, and the heat would be distributed to areas where there is higher demand than in scenario\,2.
\end{itemize}

\begin{table}[h]
  \centering
  \caption{Water saving potential related to waste heat supply (scenario 2, adapting the collider operation to seasonal demand).}
  \label{tab:water-saving-heat-supply}
  \begin{tabular}{lcrr}
    \toprule
    \textbf{Mode} & \textbf{Years} & \textbf{Heat supplied} & \textbf{Water saved} \\ \midrule
    Z & 4 & 308\,GWh/year & 477\,000\,m$^3$/year \\ 
    WW & 2 & 339\,GWh/year & 523\,000\,m$^3$/year \\ 
    HZ & 3 & 371\,GWh/year & 571\,000\,m$^3$/year \\ 
    L.S. & 1 & 60\,GWh/year & 96\,000\,m$^3$/year \\ 
    $\rm t \bar{t}$ & 5 & 441\,GWh/year & 678\,000\,m$^3$/year \\ \midrule
    \textbf{Total} & & \textbf{5288\,GWh} & \textbf{8\,155\,200\,m$^3$} \\ \bottomrule
  \end{tabular}
\end{table}

The water savings also translate into a modest reduction of operating cost of about 6\,million\,CHF. Whilst this is not a noteworthy financial sustainability contribution, together with some annual electricity savings of 4\, GWh per year, in total about 60\,GWh which is worth another 5\,million\,CHF, financially and economically compensating the additional effort to operate the waste heat recovery system. 

\subsection{Induced road traffic}
\label{sec:road_traffic}

\begin{figure}[h]
  \centering
  \includegraphics[width=\linewidth]{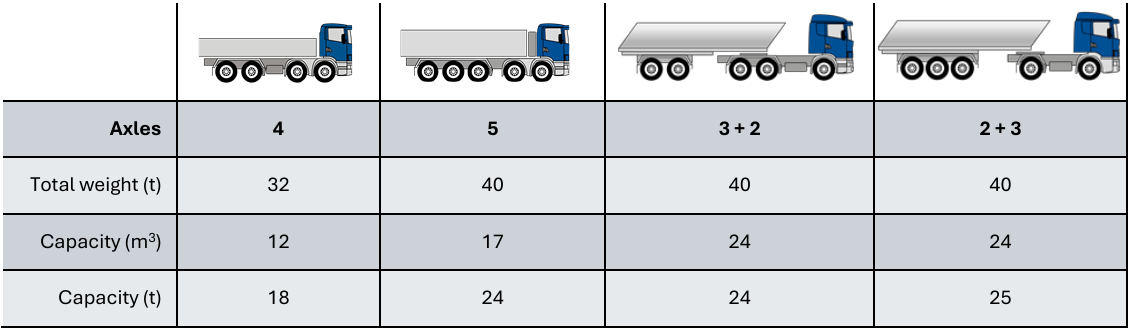}
  \caption{Capacities of different standard trucks used to analyse the construction site induced additional road traffic.}
  \label{fig:truck_capacities}
\end{figure}

\begin{figure}[h]
  \centering
  \includegraphics[width=\linewidth]{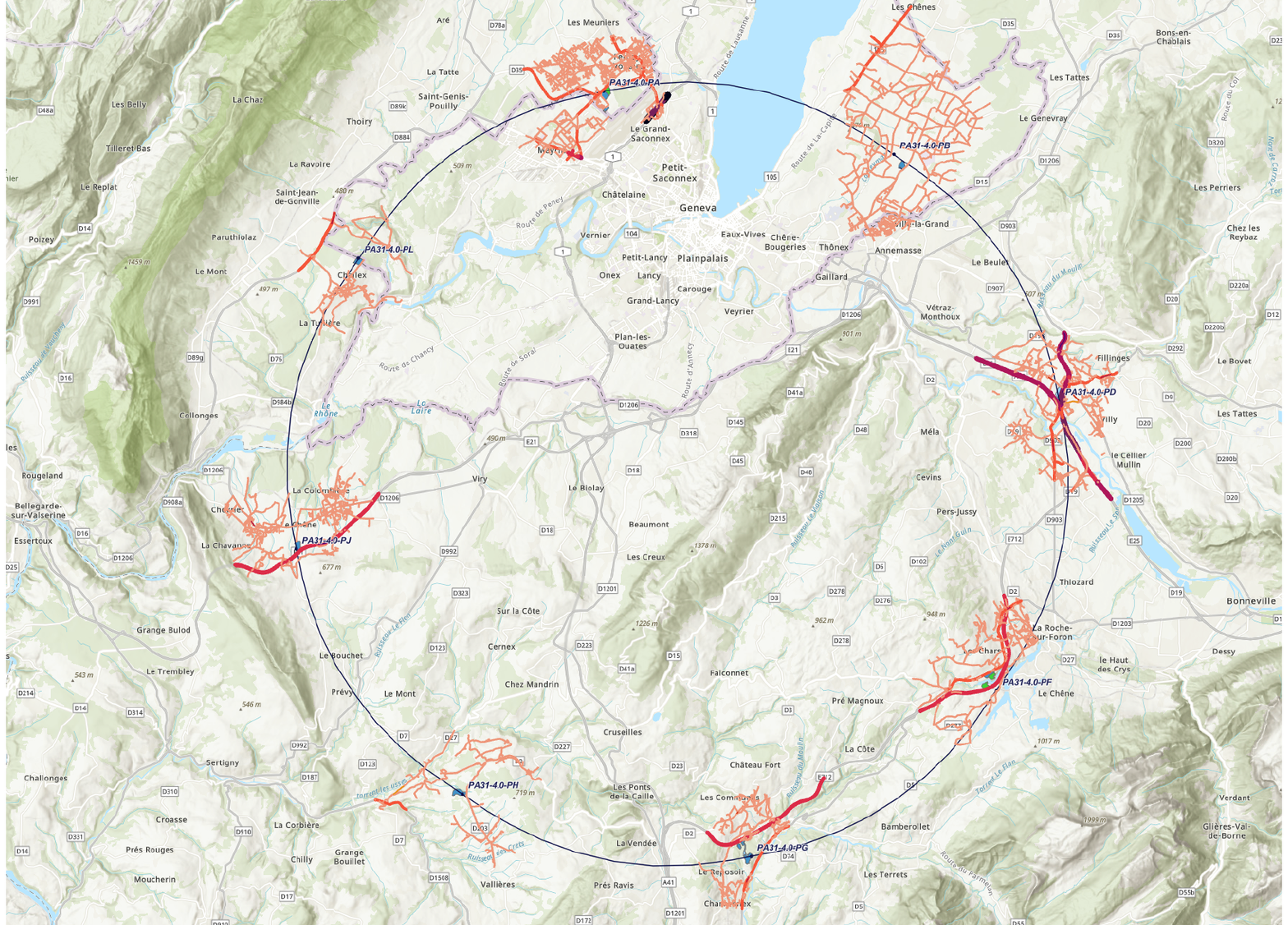}
  \caption{Perimeters of the traffic studies for each construction site. The bolder lines indicate the traffic recorded on major transport routes.}
  \label{fig:traffic_study}
\end{figure}

To estimate the additional traffic induced by the construction activities, a traffic analysis has been carried out using up-to-date traffic data in approximately 5\,km wide perimeters around the surface site locations and comparing different standard trucks used for construction (Fig.~\ref{fig:truck_capacities}). Based on this traffic analysis (Fig.~\ref{fig:traffic_study}) and the assumption of a worst-case scenario in which all materials need to be transported by trucks on roads, it can be shown that the additional traffic induced by the project, distributed over nine construction sites and more than eight years is only marginal (Table~\ref{tab:matex-traffic}). Traffic is assumed to be limited to working days and regular working hours, avoiding morning and evening peak hours. 

\begin{table}[h]
  \centering
  \caption{Worst case scenario during the construction period for excavated material transport using trucks only. The traffic can vary with different excavation scenarios, leading to less traffic at PL and PB and more traffic at PA.}
  \label{tab:matex-traffic}
  \begin{tabular}{crrrrr}
    \toprule
    \textbf{Site} & \textbf{\makecell[r]{Traffic\\per day}} & \textbf{\makecell[r]{Additional 5-axle\\trucks per day}} & \textbf{\makecell[r]{Additional traffic\\in percent}} & \textbf{\makecell[r]{Additional 5-axle\\trucks per hour}}  \\ \midrule
    PA & 7803 & 46 & 0.5 \% & 4  \\ 
    PB & 5918 & 37 & 0.6 \% & 3 \\ 
    PD & 20,475 & 93 & 0.4 \% & 8 \\ 
    PF & 11,331 & 12 & 0.1 \% & 1 \\ 
    PG & 13,681 & 100 & 0.7 \% & 8 \\ 
    PH & 1709 & 23 & 1.3 \% & 2 \\ 
    PJ & 13,954 & 95 & 0.7 \% & 8 \\ 
    PL & 3380 & 46 & 1.4 \% & 4 \\ 
    Injector & 7803 & 9 & 0.1\% &  1 \\ \bottomrule
      \end{tabular}
\end{table}

The subsequent project design will emphasise reducing road-based transport requirements. The traffic can be further reduced by using larger trucks (2+3 axles or 3+2 axles). Conveyor belts are the preferred means to bring excavated materials to major road and rail transport networks so that no local traffic is induced. Where possible, construction materials are brought in via autoroutes to site connections or via major roads to avoid passing through residential areas. 

The traffic for the five years of installation of the particle accelerator equipment is small. It ranges between 9 and 18 trucks per site and day, i.e., between 1 and 2 trucks per working hour. No traffic is foreseen during the night and non-working days. The limitation to traffic during the installation is due to the limited capacity of transferring equipment through the access shafts. The handling speed between surface and underground and in the constrained underground environment determines the amount of equipment that can be brought in. Between about 200 and 300 people per day are expected to be present on experiment sites and about 100 people on technical sites (see Table~\ref{tab:persons-on-site}). This leads to daily work-related traffic of about 100 to 200 cars. This traffic can be significantly reduced by organising the work and providing the possibility of carpools and bus transport.

During operational periods, technical sites are expected to have minimal personnel presence, based on system requirements, resulting in minimal traffic impact. For experiment sites, the highest projected staffing scenario involves small teams of up to 20 people working across three shifts. Transportation should be organized to limit daily vehicle movement to several dozen cars entering and exiting each site. However, visitor traffic at sites PA, PD, PG, and PJ may be substantially higher.

\begin{table}[h]
  \centering
  \caption{Number of people on-site during various different activity periods.}
  \label{tab:persons-on-site}
  \begin{tabular}{cllll}
    \toprule
    \textbf{Site} & \textbf{\makecell[l]{Installation\\5 years}} & \textbf{\makecell[l]{Operation\\10\,months per year}} & \textbf{\makecell[l]{Maintenance\\2\,months per year}} & \textbf{\makecell[l]{Shutdown\\every few years}} \\ \midrule
   PA, PD, PG, PJ & 200 - 300 people & 15 to 20 people & 100 people & 200 - 300 people \\ 
   PB, PD, PH, PL & 100 people & 0 to 10 people & 15 - 30 people & Up to 100 people \\ \bottomrule
      \end{tabular}
\end{table}

\subsection{Environmental monitoring}

An environmental monitoring system will be put in place once the new research infrastructure is completely constructed. Such a system will help track compliance with the initially set goals and support safe operation. Such a system typically comprises the following functionalities:
\begin{itemize}
\item Clear water monitoring: Measurement stations for effluent water integrate continuous monitoring of temperature, pH, hydrocarbons, foam, turbidity, conductivity and flow rate. Alarms are triggered based on threshold and trend conditions. Where a surface site is equipped with water retention facilities, additional monitors will be installed to activate retention when needed.
\item Sewage water monitoring: Measurement stations integrate continuous monitoring of temperature, pH, conductivity and flow rate for water released into the public sewage network. Periodic sampling is also implemented.
\item Process water sampling: Such stations serve periodic sampling of residual effluents from the processes such as for instance water recycling and treatment.
\item Air quality monitoring: Such stations continuously monitor the air, including oxides (nitrogen oxides and ozone) that may be byproducts of the synchrotron operation. Those systems are coupled to air recycling functions. Continuous comparison with the existing background air conditions will be implemented.
\item Noise monitoring: Such stations measure and record continuously the noise levels at the surface site locations and in sensitive areas in the vicinity.
\item Meteorological monitoring: These stations are equipped with anemometers and pluviometers for assessment of hazards due to potentially radioactive substances and fumes in case of fire. Cooperation with the national metereological services will be considered for data exchange.
\item Radiological monitoring: Monitoring of radiological parame-
ters during and after the operation to provide evidence for compliance with the dose constraints and limits. The system comprises equipment for on-site monitoring for the safety of workers and on and off-site monitoring for the environment.
\end{itemize}

In addition to the stationary environmental monitoring facilities, additional portable devices will be used for periodic in-field monitoring. Samples such as water, soil and plants will be analysed regularly by environmental laboratories.

\section{Current state of the environment}
\label{env:state_of_the_environment}

\subsection{Methodology}

The current state of the environment in the perimeter of the reference implementation scenario and at the surface site candidates has been analysed following the national regulations in the two Host States, France and Switzerland. The resulting single, integrated report \cite{EISA_2025}, complemented by interactive maps, audio, image and video materials as well as by an Environmental Information System based on the ESRI geographical information system, is an essential preparation work for the subsequent environmental authorisation process that has to include an environmental impact assessment in a transnational context. This work has been carried out between 2023 and 2025 with a consortium of expert companies. The following section gives a glimpse of the scope of the work carried out and provides the basic conclusions. The initial state of the environment is only valid for a limited period of time since the environment is constantly evolving. Therefore, further complementary studies (e.g., hydrogeological investigations, further studies on fauna and flora, more comprehensive environmental measurements) and studies for potential alternative site locations have to be engaged during a subsequent environmental impact assessment phase.

\subsection{Air and climate}

\subsubsection{Climate}
The climatic conditions and their foreseeable evolution have to be considered for a long-term programme like the FCC that will extend until the end of the century. The environmental state analysis includes the collection of climate data from bibliographical sources and climate observation facilities in Geneva (Switzerland) and Annecy (France). Despite the limited distance of about 30\,km between the extreme sites of the FCC, the climatic conditions are notably different. The north is characterised by a subcontinental climate with hot summers, but moderate weather conditions due to the protective effects of the mountains and the moderating effect of Lake Geneva water mass. The south experiences a mountain continental climate with greater weather and temperature differences between the summer and winter seasons. Winds are rather constant in all areas. The climate evolution shows a sustained and evolving temperature anomaly between +1.7\textdegree{} and +2.5\textdegree C compared to the pre-industrial period. The number of frosty days has decreased significantly by 20\% since 1950. Precipitation is highly variable, without noteworthy changes over the time period. Extreme heat conditions during summer periods are expected to evolve further until the end of the century. They are typically accompanied by dry periods with effects on the soil and the amount of precipitation is expected to decrease. Rain is expected to intensify during the wet periods. 

Lake temperatures have risen in recent years due to climate change. Since 1980, the surface water in most lakes has warmed by about 0.4\textdegree C per decade. The warming of the deep water is more variable and mostly ranges between 0.0 and 0.2\textdegree C per decade in the deep lakes. A further increase in the temperature of the surface water layer down to 1\,m deep is expected in all Swiss lakes: for a scenario without climate change mitigation of between 3 and 4\textdegree C in most lakes towards the end of the century. The water temperature at depths where raw water intake occurs will only slightly increase.

The evolution of climatic conditions is, however,  not expected to lead to noteworthy effects on the particle collider, for instance the water cooling systems. Nevertheless, the FCC must account for climate evolution in its design and the development of an operational concept. In the presence of an average envisaged temperature increase of around 4\textdegree C until the end of the century, these steps concern in particular, the adoption of operation to conditions that permit efficient work (e.g., avoiding periods that are too hot) and increasing the benefits of waste-heat reuse project, both internally and in the region (e.g., shift operation to a colder season).

\subsubsection{Air}

\begin{figure}[h]
  \centering
  \includegraphics[width=0.7\textwidth]{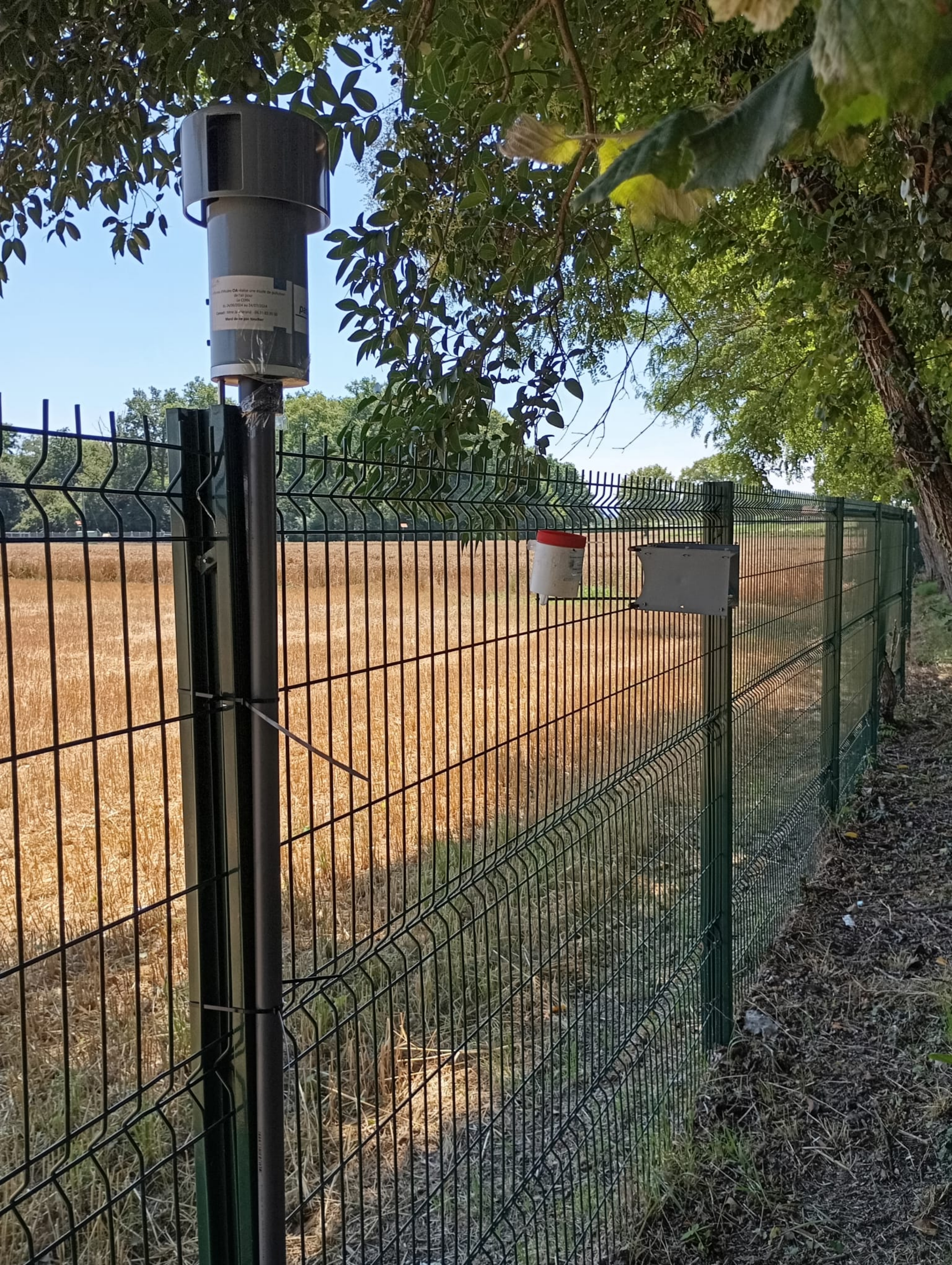}
  \caption{Air quality measurements carried out using monitoring equipment to assess atmospheric pollutants in the vicinity of the surface site.}
\label{fig:Photo_AirMeasurements_Point11pm.pdf}
\end{figure}

The air condition within the perimeter of the FCC is generally good, and the air quality is improving constantly. The main cause of air pollution is road traffic, mainly diesel-powered trucks. Fine dust particles have their origin mainly in the agricultural sector and industry, in particular construction and quarries. All typical air pollutants have been studied to document baselines, and air quality measurements have been carried out to establish references in the vicinity of the surface site. These references serve as valuable input for the development of the construction process and the design of the infrastructures in order to be able to properly and adequately ensure the protection of the air quality and meet the EU zero pollution vision for 2050. If it is decided to go ahead with the project, continuous air monitoring has to be implemented at the surface sites to monitor the evolution and to be able to control the impacts on the air quality during the construction and operation phases.

\subsubsection{Climate protection plans}

France has reduced its emissions by 27\% with respect to 1990, in line with an average reduction of the emissions in Europe by about 31\%. The per capita footprint in 2022 was about 6.5\,tCO2(eq) per year~\cite{EUClimateAction2023}, a value that is below the European Union average of 8 metric tons per capita. While the energy and industry sectors were able to reduce their footprints considerably, the transport sector did not yet see the same improvements. Emissions due to transport even slightly increased. This situation is also reflected at a regional scale, with the transport sector remaining the main producer of greenhouse gas emissions. France has adopted the 2015 Paris Agreement, aiming for a reduction of the net emissions by 55\% with respect to 1990 by the year 2030 and climate effect neutrality by 2050. Locally, the climate protection goals can differ. For instance, for the Pays de Gex around CERN today, the goals established are -66\% by 2050 with respect to the year 2015. For `Grand Annecy' the goals are -55\% by 2030 and -87\% by 2050. 

In Switzerland, greenhouse gas emissions have decreased continuously since 2010. Transport, agriculture, and industry remain the main emission contributors. As of 2025, two laws are in place that impose requirements: the law on CO$_2$ and the law on climate and innovation.
The established goals are a reduction of 60\% of the greenhouse gas emissions by 2030 with respect to 1990. The long-term goal for the country is to achieve net carbon neutrality by 2050. Also, at the cantonal level, an operational climate protection plan is in place. It prescribes goals and 41 specific measures to be implemented by 2030. The emission reduction objectives are in line with values established at the federal level.

\subsubsection{Summary and conclusions}

\begin{figure}[h]
  \centering
  \includegraphics[width=0.5\textwidth]{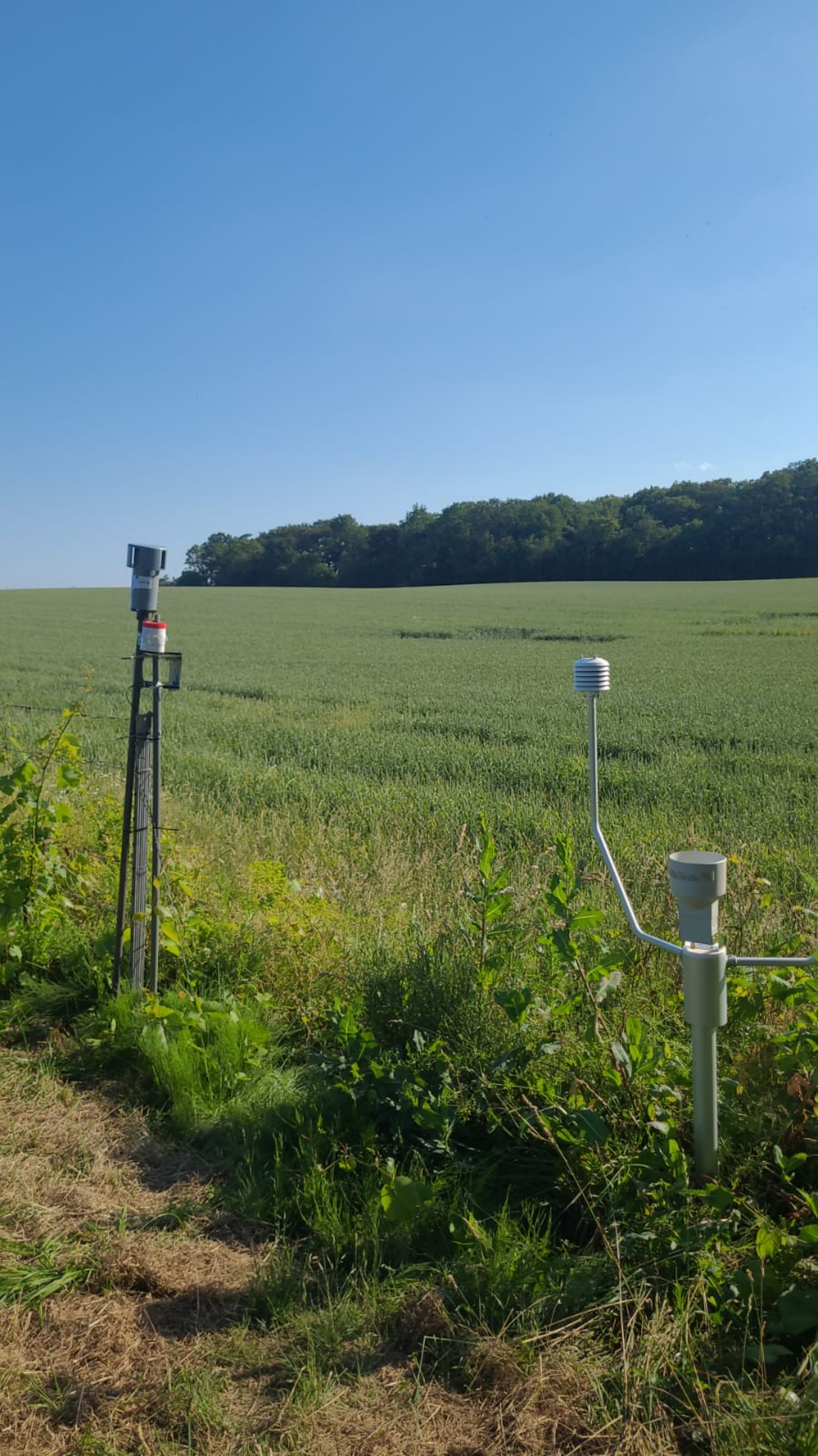}
  \caption{Air quality measurements conducted to assess atmospheric pollutants in the vicinity of the surface site.}
\label{fig:Photo_AirMeasurements_Point10.pdf}
\end{figure}

The planning, construction and operation of a future research infrastructure at CERN takes into account the plans and regulatory requirements that have been established at the national levels to achieve the goals of the Paris Agreement. Aspects related to energy and greenhouse gas emissions are considered according to the current policies established by CERN, aiming at keeping the energy required for its activity as low as possible, ensuring that the established research programme goals can be achieved. This includes continuous improvement of its energy efficiency, recovering and supplying waste heat and reducing greenhouse gas emissions.
CERN is committed to demonstrating that appropriate measures are taken with respect to energy and greenhouse gas emissions during all phases of a future research infrastructure, in line with the plans and regulatory requirements established by the Host States.

At the end of 2019, CERN established a target for the reduction of its direct CO$_2$ emissions by 28\% with respect to 2018 (baseline year) by the end of the Large Hadron Collider Operation Run 3 (around 2026). This target has been highlighted since September 2020 in the CERN public environment reports that are published biennially. Recently, CERN has adopted a target for the reduction of its direct CO$_2$ emissions by 50\% with respect to the baseline year 2018 for the year 2030 (\cite{CERN:environmental_goals_2030}. Indirect emissions related to electricity consumption and other emission categories, such as those related to procurement, are also reported and published. Associated reduction targets are under investigation.

If a new research infrastructure is approved and included in CERN's overall environmental performance management, emission reduction plans and goals, including the new research infrastructure, will be set with respect to the corresponding updated envelope of policies.

The integration of a new research infrastructure into CERN's environment will generate additional emissions that the organisation will strive to minimise within the established goals to be achieved for the scientific research performance. A rigorous eco-design approach applied to the construction activities, the technical infrastructures, the particle accelerators, and the detectors will be introduced at the level of the organisation and will be required from the entire international collaboration contributing to the project, ensuring that in-kind contributions will comply with the relevant environmental rules and regulations. 

The proposed particle collider will not operate concurrently with the Large Hadron Collider. Therefore, the emissions generated by the operation of the new collider and its experiments will replace those linked to the LHC operation. The new collider and its experiments will technologically be significantly more advanced, and the eco-design will make it a research infrastructure with lower emissions than the LHC today. 

Together with continued efforts to technologically upgrade other CERN research facilities and activities, effective support of the Host States efforts to meet their climate protection goals remains achievable.

In the context of the environmental authorisation process, a climate protection plan integrating the new project is expected to be included as part of the environmental impact assessment.

\subsection{Water}

\subsubsection{Context}
The situation of surface and subsurface water within the perimeter of the scenario has been analysed based on bibliographical data, databases, maps, and reports. The legal frameworks in France and Switzerland are very different with respect to this topic, which makes the comparison and integration of data challenging. Whilst in France, the European Union definitions and directives are applied, different water protection frameworks exist in Switzerland at federal and cantonal levels. For this reason, the current state of the water has to be looked at separately for each of the two countries. In addition, subsurface water tables have to be distinguished from surface water. However, both water masses, subsurface and surface, extend across the national borders and lead to cross-border aspects with respect to subsequent analysis of potential environmental aspects: effects and impacts that will need to be considered during a project preparatory phase.

\subsubsection{Subsurface}

\paragraph{France}

Concerning subsurface water, five distinct relevant water masses have been identified. FRDG517 extends from Gex across the Geneva basin to the Grande C\^ote de Bonmont and the Usses sector. FRDG208 lies below the FRDG517, extending from the Jura to the lake and towards the Usses sector in a calcareous geology. FRDG231 is located north of the Genevois zone in France, partially under the FRDG517, partially limited like a river between the Rh\^one and Divonne-les-Bains, touching the Swiss border at the height of Versoix. FRDG511 is located in the Savoie alpine region, south of FRDG517. FRDG364, corresponding to the Arve valley, passes at about 100\,m of the site PD in Nangy. 

With respect to the European legislation, a goal has been set to maintain the so-called `good state' level quality attained in 2015 of all water masses. The standard needs to be maintained both in qualitative and quantitative terms. 

Water-bearing layers exist in relation to these water masses. All apart from in the implementation perimeter are `free', i.e., there are no water-tight layers towards the surface. None of the hydrogeological entities is characterised by an aquifer.

Because of the need to maintain the quality levels and the potential interactions between water masses, particular protection measures are in place that will need to be observed with respect to subsurface works.

\paragraph{Switzerland}

In Switzerland, various types of water-bearing layers are distinguishable based on their water capacity and depth. Deep layers can also exist in the molasse layer which is preferred for tunnel construction. Today, four water-bearing layers are identified in the canton of Geneva, and two of them are used for providing drinking water: Allondon and Genevois. Montfleury and Rh\^one are currently being studied in terms of capacities and quality with respect to serving as a drinking water supply. The deep layers are today not mapped and therefore dedicated subsurface investigations are required for a future particle collider project. A multitude of shallow and temporary layers that are between 2 and 10\,m below the surface are spread over the entire canton. They need to be analysed where shafts and surface sites are planned. One of them is at a distance of about 140\,m from site PB in Presinge at a depth of 2\,m. No potential mutual effects could be determined.

Another water-bearing layer is at about 90\,m from site PA in France. The Montfleury layer at a depth of about 45\,m is located 800\,m south of the main site. One shallow layer is directly at the surface site location PL in Challex in France. It spans the border, and no information is available on this layer. Sites PL and PA require particular analysis to take these layers into account for the choice of the shaft location and the shaft construction technique. All three sites, PL, PA and PB, need to consider the potential water layers in the vicinity of their surface site and consider cross-border aspects.

\subsubsection{Surface}

\paragraph{France}

Four main surface water bodies are within a radius of 1\,km of the surface sites. Three of them are in good state (FRDR11960, FRDR559, FRDR537) and one has moderate quality (FRDR555c) with a goal to achieve good quality by 2033. The site PF in \'Eteaux is located in the vicinity of a stream (50 to 100\,m). Also, site PH is in the vicinity of a small creek (30\,m). Other sites such as PD (600\,m from the Arve) and PG (300\,m from the Filli\`ere) are further from rivers.

\paragraph{Switzerland}

Lake Geneva in Switzerland is the most important water body at the surface. It is 4\,km from the PA and PB sites. Site PB is located in the vicinity of a small stream (30\,m). 

\subsubsection{Summary}

\begin{figure}
    \centering
    \includegraphics[width=\textwidth]{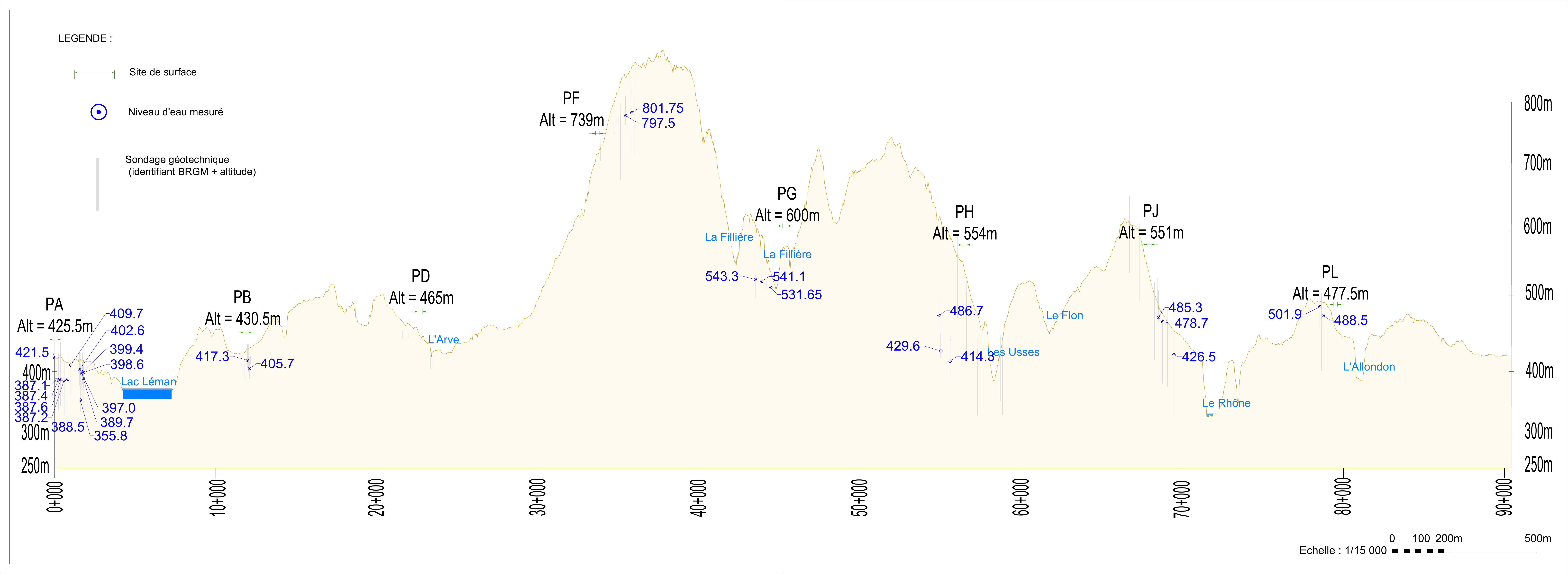}
    \caption{\label{fig:maquette-hydro} Overview of surface and subsurface water bodies in the scenario perimeter. In this conceptual drawing, the subsurface structures are assumed to be at an elevation of 250\,m. The uncertainty of the water table heights is indicated in grey. The surface site locations and elevations are indicated.}
\end{figure}

The projection from the line of the implementation scenario to the subsurface intersects with the geographical location of numerous subsurface water bodies. However, the tunnel will be located significantly below them (see Fig.~\ref{fig:maquette-hydro}). No intake from any subsurface water-bearing layer is expected for the construction and operation of the project. Water for drinking will be consumed from connections to the existing drinking water network. Raw water for cooling purposes will be taken from the existing water supply network for CERN in Switzerland, which takes water from Lake Geneva. Drinking water protection zones are largely avoided for the entire project so that any potential adverse effects can be excluded. 

The majority of subsurface layers are not located under a watertight layer. Thus, they are potentially subject to pollution. Also molasse layers may include water-bearing volumes.
Therefore, dedicated hydrogeological investigations will be carried out to optimise the location of the shafts to avoid potential adverse effects with water-bearing layers in PA (Ferney-Voltaire, France) at a distance of 35\,m from a drilling ban zone (sector B in Switzerland) and PL (Challex, France). Particular attention will be devoted to avoid affecting nearby creeks and the biodiversity around sites PB (Presinge, Switzerland) and PH (Cercier and Marlioz, France). PH is, in addition, in an area with chronic lack of water and therefore subsurface works such as the creation of a shaft will have adequate protection measures in their design in case there is a risk of affecting any water bearing layers.

Concerning the release of water into the environment and particularly into nearby creeks, the design of the infrastructure will include filtering and cleaning before release. Connections to wastewater networks will ensure that all other water is released via the existing water treatment infrastructure at all times, ensuring compliance with the legal and regulatory frameworks in the two Host States.

\subsection{Subsurface}

\subsubsection{Relief}

\begin{figure}
    \centering
    \includegraphics[width=\textwidth]{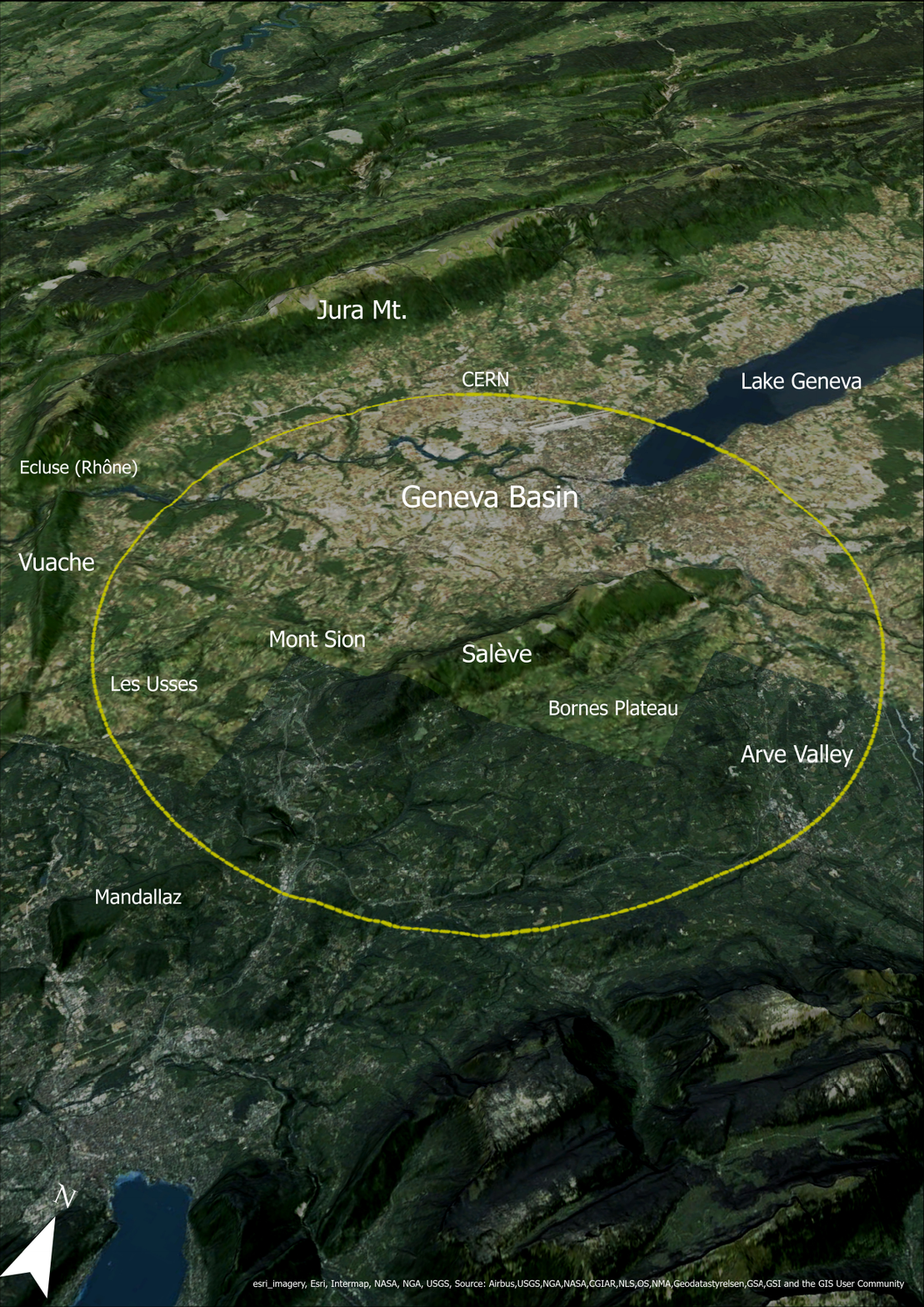}
    \caption{\label{fig:relief} Relief in the area of the FCC reference scenario.}
\end{figure}

The FCC perimeter is constrained by noteworthy mountains (see Fig.~\ref{fig:relief}). The Jura is located north and west of Lake Geneva with peaks of 1720\,m. The south-east is characterised by the Bornes plateau evolving into the pre-alps that lead to Mont Blanc which has an elevation of 4806\,m. The west is constrained by the Vuache mountain reaching 1112\,m. The Sal\`{e}ve (1379\,m) is located in the centre of the FCC circular alignment. The Aravis mountain range is located just south of sites PD, PF and PG. Mont Sion (78t\,m)  is located between the Vuache range and the Sal\'{e}ve . In the south, the tunnel crosses under the Mandallaz range (923\,m), also called Balme, that is part of the pre-Alps. The Massif du Chablais is northeast of Lake Geneva and east of the PB and PD sites. Despite the presence of the numerous mountain chains and peaks, all surface sites are located on lower ground in flat areas at between 400 and 700\,m elevation.

\subsubsection{Topography}

Site PA is located at an elevation of approximately 425\,m in an open area with a very slight slope of about 2 to 5\%. Site PB is at an elevation of about 430\,m on an even and open area. There are hills only a few kilometres to the north. Site PD at an elevation of approximately 460\,m is located on a slope of about 5\% towards the north. The zone is hilly in all directions, but the terrain is cut by major transport routes (A40 autoroute, D903 departmental road). The absence of bushes and hedges makes the terrain visible. Site PF reaches elevations of between 730 and 745\,m on a slope of about 7\% from west to east. This makes the area very visible towards the mountains and from the RD1203 road. However, it is not visible from the A410 autoroute. Site PG is separated from the A410 autoroute in the north by a forest on a slight slope of 5\% in north-south and west-east directions at about 600\,m elevation. At the southern boundary, the slope starts to become steep -- from 25\% to 40\% towards the RD1203,:Annecy road. this area is avoided. Site PH is located on a slope of 10 to 20\% at an elevation of between 517 and 591\,m, falling off from the RD203 road. The entire area is located in a forest that covers the zone towards the Usses Valley in the west. The topography imposes a terracing approach for the site.
The PJ site is located on a wide and long slope of 6\% at an elevation of between 496 and 532\,m that falls off from the A40 autoroute to the Rh\^{o}ne valley. This location also requires terracing. Tree lines that break the even space and slope exist in the vicinity. The PL site is located at 500\,m elevation on a rather flat area. The absence of noteworthy trees or hedges in the vicinity makes the terrain highly visible. The land falls off steeply towards the Rh\^{o}ne valley on the opposite side of the nearby pathway towards Switzerland.

\subsubsection{Geology}

The geology is covered in greater detail in the sections on civil engineering since it determines the placement of the subsurface works. Here, a general overview provides the context.
The geological landscape of the Genève Basin spans between the Jura Mountains in the northwest and the Prealps in the southeast, with a Tertiary molasse basin overlaying Cretaceous formations. Shaped by Quaternary glaciations, the basin features deep valleys and sedimentary deposits up to 400\,m thick, especially in the Grand-Lac, with Holocene sediments varying by river currents. The Rh\^one Delta sees Holocene sediments exceeding 100\,m in thickness. The basin's geology includes three main units: Jura sedimentary rocks in the northwest, central Tertiary sandstone (molasse), and thrusted molasse with Prealpine units in the southeast.

\begin{figure}
    \centering
    \includegraphics[width=\textwidth]{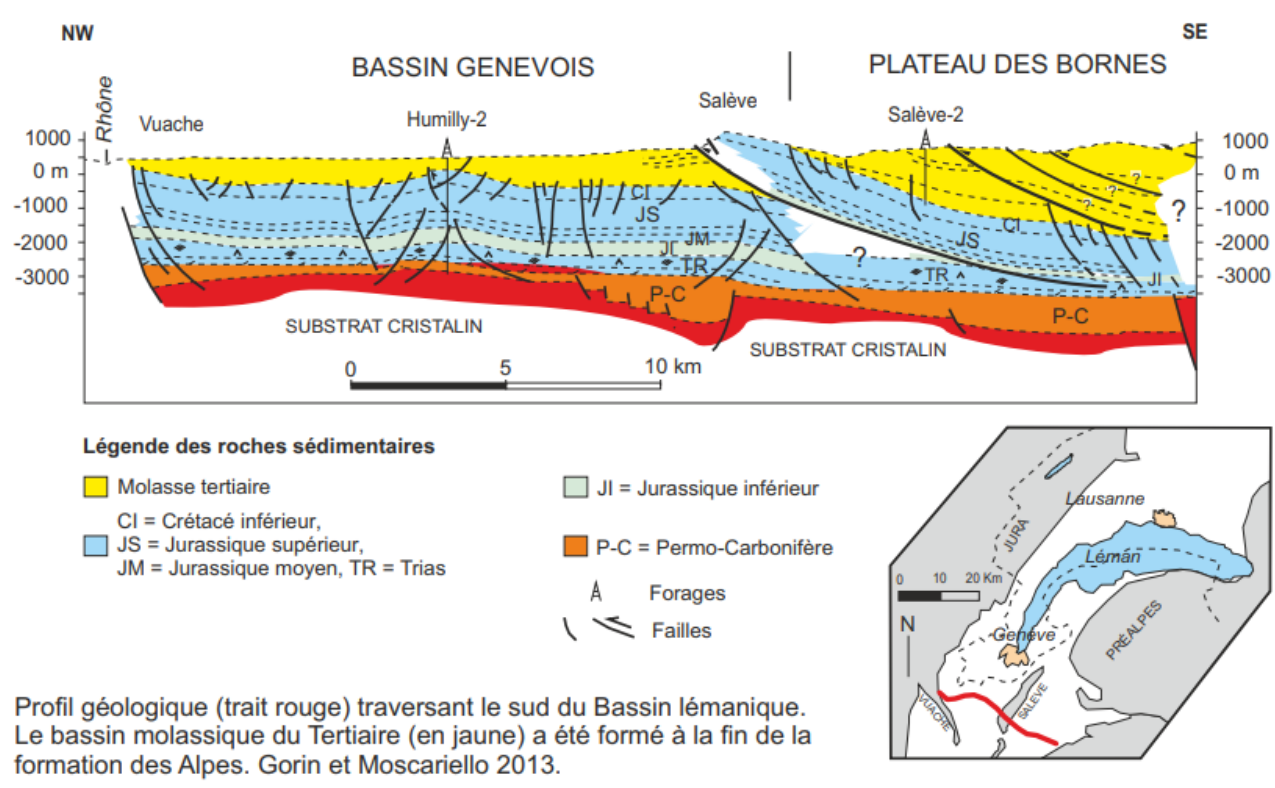}
 \caption{\label{fig:geologybassin} Geological profile of the Geneva Basin, illustrating stratigraphic units, tectonic features, and sedimentary formations shaping the region.}
\end{figure}

The Ain department is located on two very different geographical and geological domains with, to the west, the large plains of Bresse and Dombes the western plains of Bresse and Dombes, a tectonic rift filled with Tertiary deposits, and to the east, Jura mountains, marking the southwestern edge of the Swiss molasse plain. In addition, the department is covered by three large geological units, Bresse and Dombes including C\^oti\`ere and part of the Val de Sa\^one; Bugey and the southern part of Revermont and finally the Pays de Gex. The Pays de Gex, where the PA and PL sites are located, features mountainous limestone formations (Middle and Upper Jurassic, Lower Cretaceous) with karst systems, bordered by faults and glacial deposits. Similarly, the surface sites PD, PF, PG, PH, PJ in Haute-Savoie in France and PB in Switzerland are part of the molasse basin and showcase diverse geological features from crystalline Alpine massifs to sedimentary molasse basins like the Plateau des Bornes. These molasse deposits, formed from Alpine erosion during the Oligocene to Miocene, vary in thickness and are tectonically influenced by Alpine uplift.
The geological formations of the different surface sites are presented in Table~\ref{tab:subsurface_geologicalformation} 
\begin{table}[h]
	\centering
	\caption{Geological formations present at surface sites.}
	\label{tab:subsurface_geologicalformation}
	\begin{tabular}{ccc}
		\toprule
		\textbf{Site} & \textbf{Deposits} & \textbf{ Composition of materials} \\ \midrule
		PA & Würmian glacio-lacustrine & Layered clays and silts \\ \midrule
		PB & Morainic & Clays, silts, and sands \\ \midrule
		PD & Glacio-lacustrine & Clays and silts \\ \midrule
		PF & Würmian to post-Würmian morainic & \makecell[c]{Silts, sands, pebbles, gravels, with\\ localised presence of clays} \\ \midrule
		PG & Würmian morainic & \makecell[c]{Clays, sands, pebbles,\\  stones,and boulders }\\ \midrule
		PH & Würmian to post-Würmian morainic & \makecell[c]{Clays, sands, pebbles,\\ stones, and boulders} \\ \midrule
		PJ &\makecell[c]{ Würmian to post-Würmian morainic\\ (or colluvium)} & \makecell[c]{Silts, sands, pebbles, gravels, with\\ localised presence of clays} \\ \midrule
		PL & Würmian morainic & \makecell[c]{Pebbles, gravels, sands, and limestone\\ with localised presence of clays }\\ \bottomrule
	\end{tabular}
\end{table}.

\subsubsection{Soil}

\begin{figure}[h]
  \centering
  \includegraphics[width=0.6\textwidth]{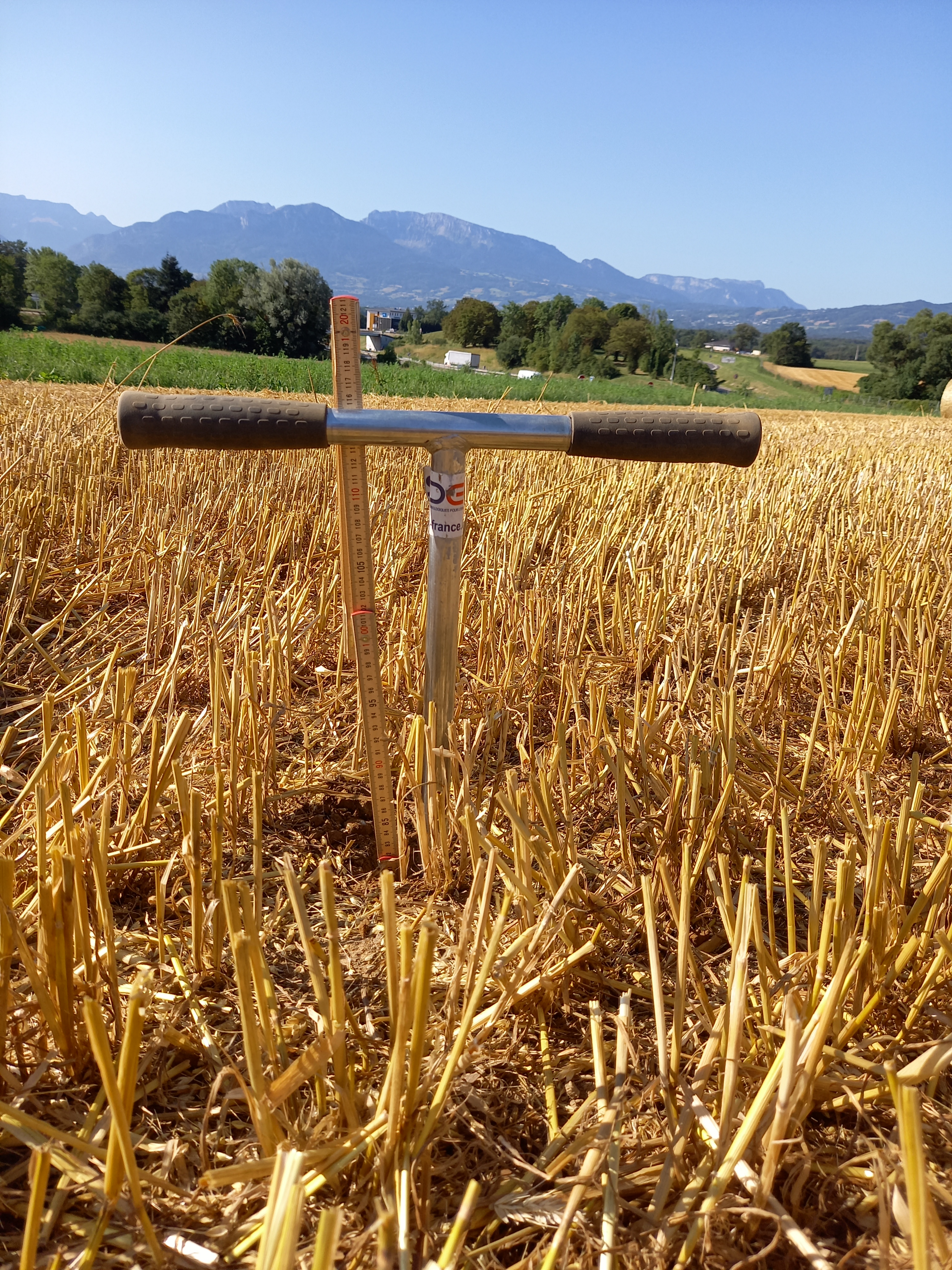}
  \caption{The auger boring method used to obtain soil samples from different depths.}
\label{fig:Photo_SoilInvestigations_Auger_PD.pdf}
\end{figure}

Field soil investigations were carried out in July 2023 for site PD and in April 2024 for the sites PA, PB, PF, PG, PH, PJ and PL. The primary objective was to analyse the pedological characteristics of the soil of these sites to determine their suitability for agricultural activities and other potential uses.

The work consisted in the identification and description of existing vegetation and crops, as well as the assessment of factors impacting agricultural productivity, such as accessibility, topography, non-cultivable areas, wet zones, and irrigation infrastructure. Soil samples were collected using hand augers up to a depth of 120\,cm to examine soil layers, coarse elements, and traces of waterlogging.
The number of samples per site was determined based on plot size and soil diversity in order to ensure adequate representation. A total of 48 samples were collected across the eight sites.

The PA site is dominated by deep loamy soils (neoluvisols) with a low content of coarse elements, although areas near roads feature shallower and rockier soil. 

At the PB site, the surface horizon consists of clay-loamy soils with low amounts of coarse elements, but signs of pseudo-gleying appear in deeper layers. 

PD site is characterised by deep loamy soils, healthy and well-drained, with low, coarse elements throughout the soil profile and limited hydromorphy. The soil is highly suitable for agriculture due to its deep, fertile, and easily mechanisable soils. 

At the PF site, the surface layer consists of loamy neoluvisols with low rocky content, while clay accumulation begins at around 70\,cm depth. These parameters are suitable for permanent grasslands. 

\begin{figure}[h]
  \centering
  \includegraphics[width=0.7\textwidth]{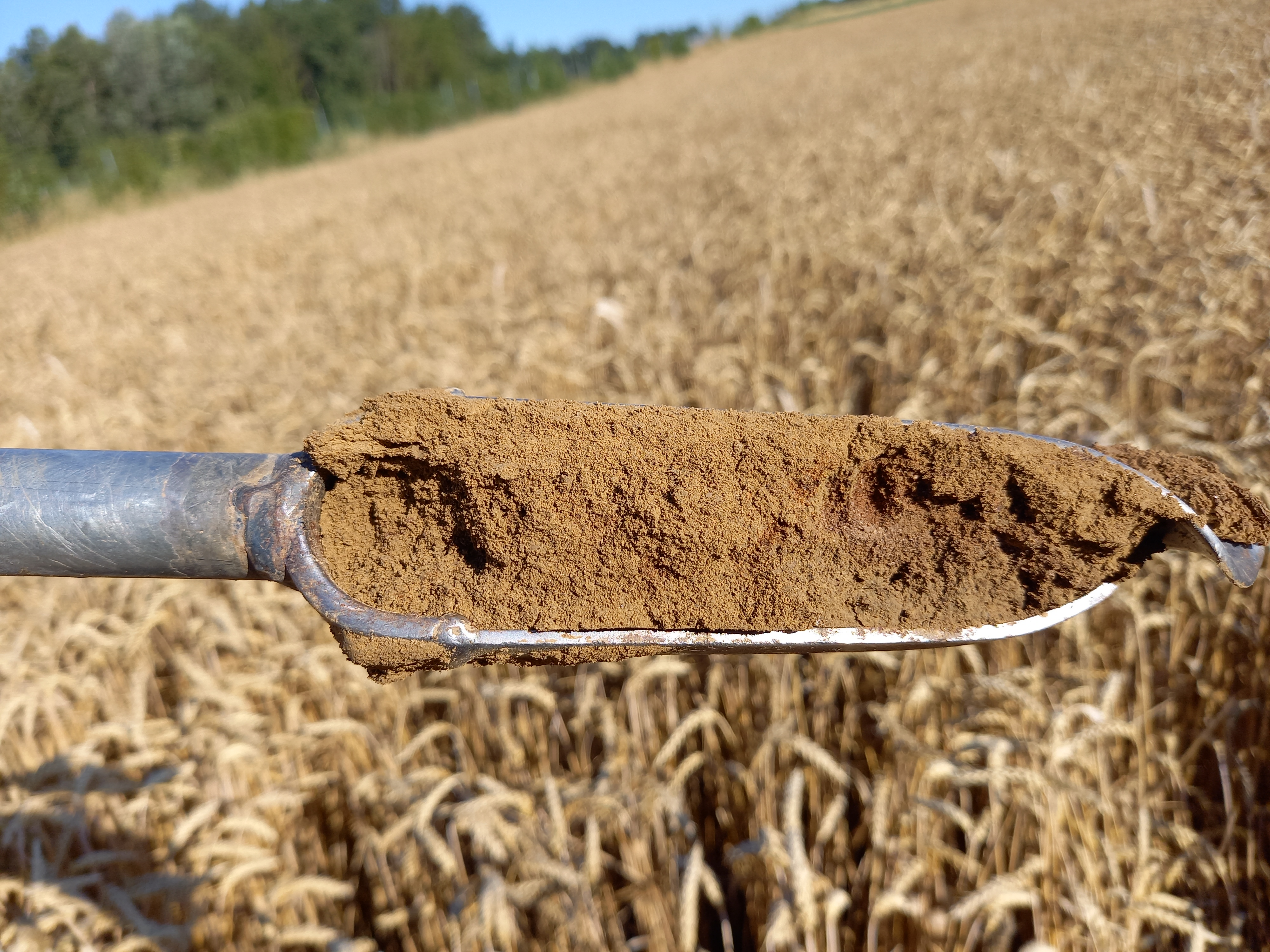}
  \caption{Soil sample obtained by using auger boring method.}
\label{fig:Photo_SoilInvestigations_soil_PD.pdf}
\end{figure}

The surveys carried out at the PG site and its annexes show that the present soil does not match the soil recorded in the French GIS-Sol database. The soil of PG main site is deep, homogeneous, and range from clayey to clay-loamy textures, with low amounts of coarse elements and temporary hydromorphy traces. Annex sites have lighter, powdery soils, moderately deep with a clay-loamy texture. 

The PH site is covered by deep Brunisol-type soils with Redoxisol characteristics, where strong hydromorphic features are already apparent near the surface, indicating limited drainage. The survey confirms that the permanent grassland at a small part of the PH surface site corresponds to the soil cover listed in the database. 

PJ site is characterised by deep argilo-limoneux soils with minor coarse elements. Moderate hydromorphy is visible near the surface, with traces of altered limestone beyond 70\,cm. The area is suitable for both permanent and temporary grasslands. 

PL site at the border with Switzerland is predominated by calcosols with areas of colluviosols in southern regions. Soils vary from superficial to deep, with coarse elements more common in superficial areas. Hydromorphic features are less pronounced, and soils are generally suitable for agriculture.

Overall, most surface sites have deep soils with limited coarse elements, ensuring good agricultural potential, although hydromorphic tendencies and limited drainage at certain locations limits their agricultural use.
Additional laboratory analyses of the soil are currently being conducted to assess its quality in terms of the presence of various mineral elements and their availability to plants, and potential contaminants. 

\subsubsection{Summary}

Concerning relief and topography only, site PH in Cercier and Marlioz exhibits significant issues due to the slope in the forest. PJ in Dingy-en-Vuache and Vulbens is also located on a slope, but it is in an open area and significantly less steep with good access. Sites PA, PB and PL on very open and flat areas call for particular integration with respect to visibility and co-visibility. Site PD and PF do not exhibit any particular challenges, although good landscape integration is advisable to reduce visibility and to blend into the terrain. Care needs to be taken at site PG to stay clear of the steep slope towards the Annecy road. 

The geological context of the study area highlights the diversity of formations across the Geneva Basin and the Ain  and Haute-Savoie departments. The basin features sedimentary deposits influenced by quaternary glaciations, with thick holocene sediments in the Rh\^one Delta and Grand-Lac. The sites span different geological units: Pays de Gex (limestone formations with karst systems) and molasse basins (formed by Alpine erosion), showcasing a variety of materials, including glacio-lacustrine clays, morainic sands, gravels, and silts. These variations underline the region's complex geological history, which is crucial for subsurface work placement and excavated material management.

The pedological surveys confirmed diverse soil characteristics, ranging from deep, fertile, and well-drained loamy soils to hydromorphic and clayey profiles with drainage limitations. Sites such as PD and PL are highly suitable for agriculture due to their deep, fertile, and easily mechanisable soils. PA, PB, PF, and PJ show varying degrees of hydromorphy, particularly in deeper layers, making them more suitable for grasslands or selective agricultural practices. PH and PG present significant hydromorphic constraints, with PH showing pronounced surface water retention issues due to its altered limestone subsurface.
Variations in texture, depth, and coarse element content reflect the influence of local geology and historical land use.

\subsection{Biodiversity}

\begin{figure}[h]
  \centering
  \includegraphics[width=0.7
  \textwidth]{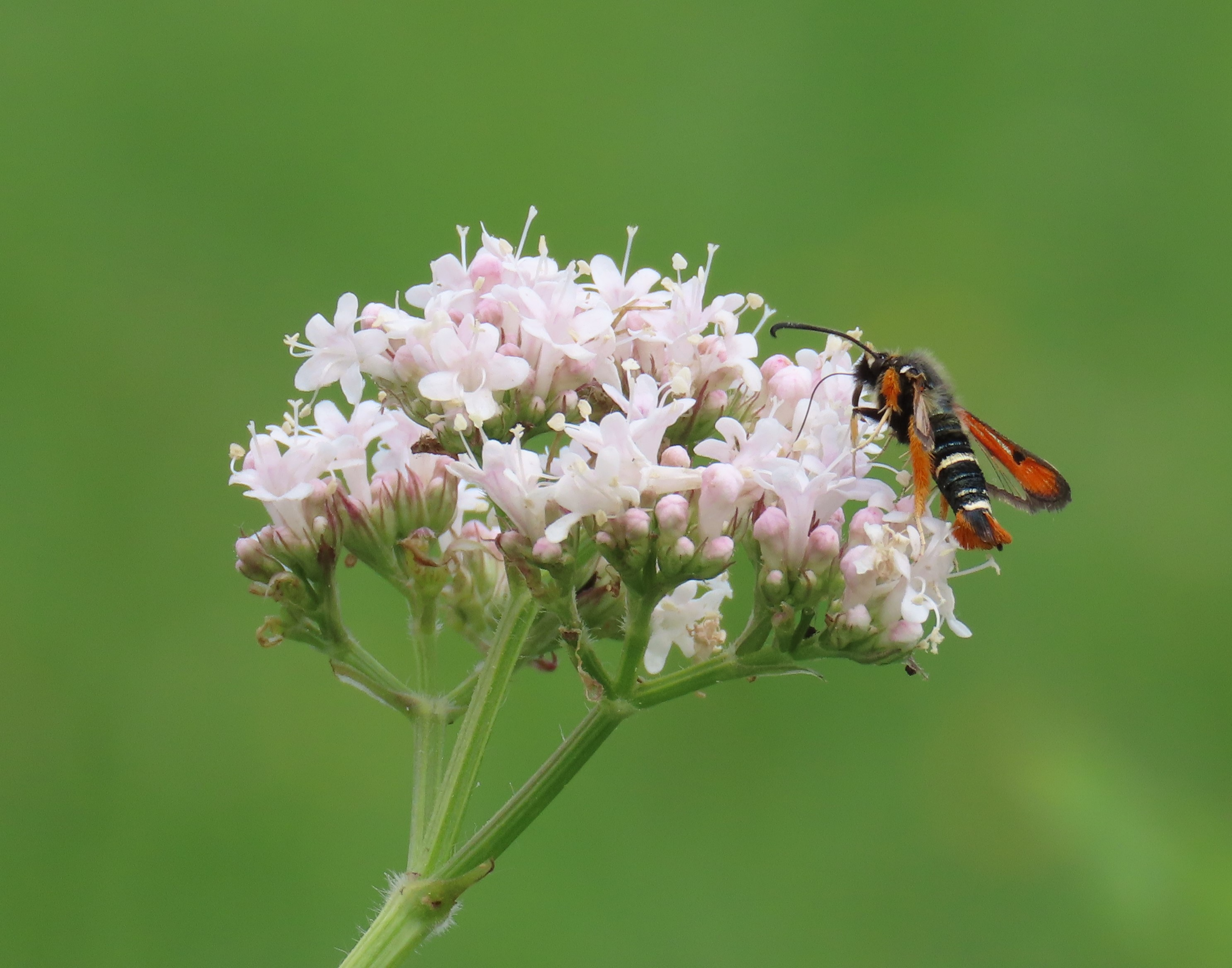}
  \caption{Fiery Clearwing Moth {\textit{(Pyropteron chrysidiformis)} feeding on a plant.}}
\label{fig:Photo_Biodiversity_FieryClearwingMoth_Plant_PD.pdf}
\end{figure}

Biodiversity refers to the diversity of species, ecosystems, habitats and ecological processes. It includes both natural and human-modified environments, which together create the conditions for the existence and functioning of various organisms. 
A variety of species in the natural spaces are not locally confined. They extend across fauna corridors and ecologically coherent zones that comprise the terrestrial and aquatic species, namely plants, insects, mammals, amphibians and reptiles, which are considered as an integrated whole.

To assess potential impacts on biodiversity, the current situation was analysed for all candidate surface site locations, first using bibliographical information. This includes, for example, the study of protection zones at regional, national and international levels, inventories of natural heritage, ecological corridors and their continuations. Then, field investigations were carried out by numerous experts over a time frame of more than one year, covering the four seasons. They served not only to confirm and complement the existing bibliographic information but also as necessary input to establish a baseline for the avoid-reduce-compensation approach, the optimisation of the sites and the subsequent environmental impact assessment. The investigations were not restricted to the area within the perimeters of the surface site locations. They were extended to a perimeter of up to several hundred metres larger, depending on the topography, but the extension was investigated in less detail. An extended perimeter covering up to 5\,km was analysed at a high level with the help of existing databases and cartographic materials. 

Site PA in Ferney-Voltaire is in the vicinity of nature protection and humid zones as well as forests that serve as cross-border corridors for animals. Amphibian breeding zones are noted in the enlarged perimeter on the Swiss side. The forests also serve as retreat areas for migrating birds. There are nature protection and wetland zones, including amphibian reproduction zones, in the vicinity of the PB site in Presinge. There is a protection zone for migrating birds at some distance. For site PD in the Arve area there is a nature protection zone in the vicinity. Nature protection zones and ecological corridors have been identified in the vicinity of site PF in \'Eteaux. Several nature protection zones also exist in the vicinity of site PG in Groisy and Charvonnex. Further away, there is a biotope protection zone. Nature protection zones also exist in the vicinity of site PH in Cercier and Marlioz. At some distance, there is a biotope protection zone. Rare species are found in the immediate vicinity of the site and partially at the limit of the site, mainly close to a creek that passes at the site boundary to the north. There are nature protection zones in the vicinity of site PJ in Dingy-en-Vuache and Vulbens. There are nature protection zones and cross-border ecological corridors, linking forest spaces and the Allondon zone with the Rh\^one river zone in the immediate vicinity of site PL in Challex. Strict bird protection zones on the Swiss side of the border extend into the forest on the French side. There, biotope protection zones are known.

None of the sites is directly affected by a nature protection restriction, a national park or international biodiversity protection regulations (e.g., Natura 2000, RAMSAR, UNESCO or national parks).

\subsubsection{Natural habitats}

\begin{figure}[h]
  \centering
  \includegraphics[width=0.8\textwidth]{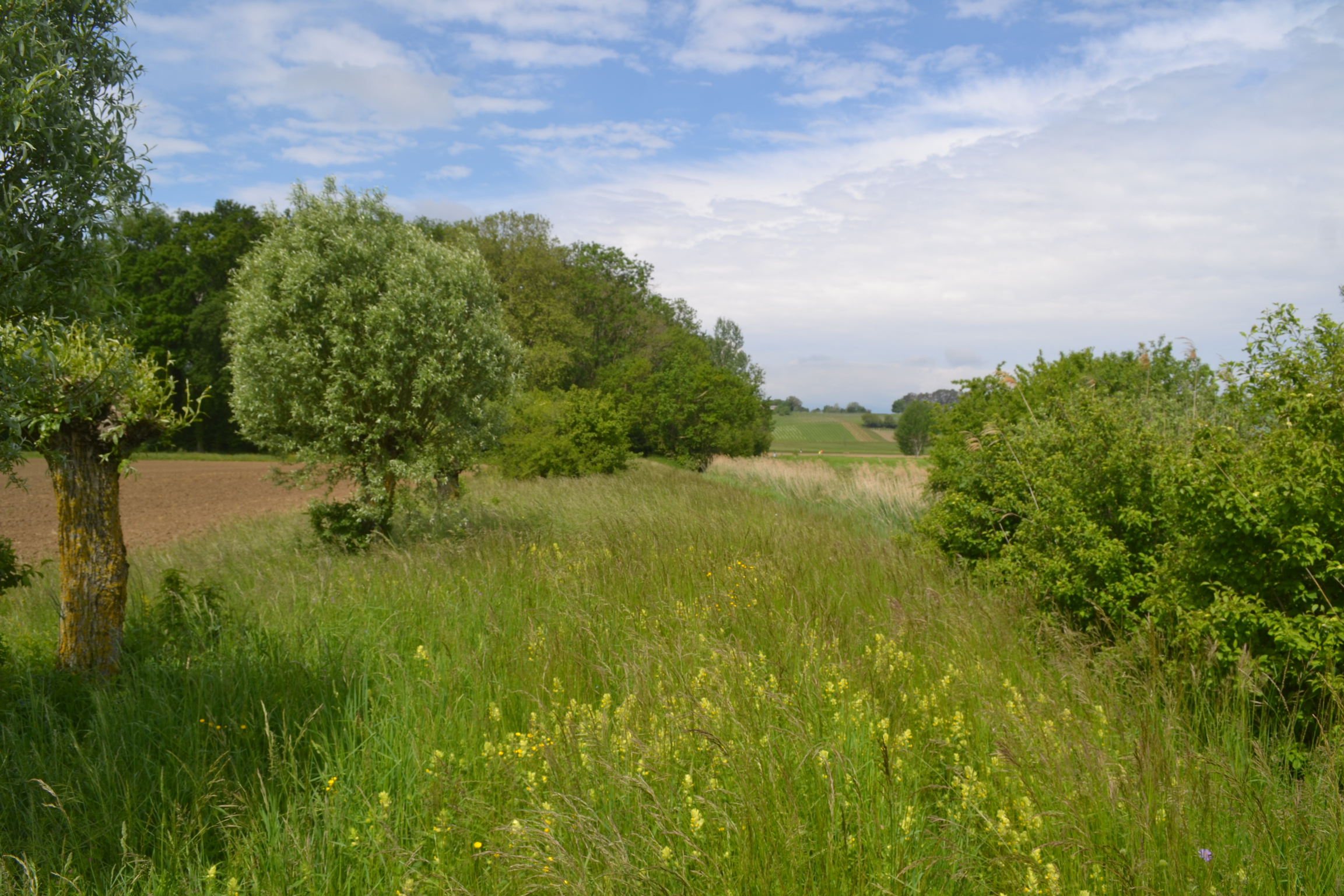}
  \caption{Bocage hay meadow photographed during the field investigations.}
\label{fig:Photo_Habitats_BocageMeadow_SitePB.pdf}
\end{figure}

In the context of environmental analysis, a habitat refers to the natural environment or ecosystem where a particular species, community, or group of living organisms lives, grows, and thrives. It includes the physical, chemical, and biological components that support life, such as soil, water, air, terrain, other plants and animals and the interaction among all those components. The studies carried out comprised establishing an inventory of the existing habits at the surface sites, investigating the perimeters around them, and evaluating their characteristics and qualities. This work helped to determine the sensitivity levels associated with the larger zones concerned by the surface sites in order to apply the avoid-reduce-compensate scheme for further optimisation of the project scenario.

Site PA in Ferney-Voltaire, France is predominantly a peri-urban, open agricultural habitat of about 70\,ha that is also closely linked to Geneva airport, agricultural spaces in Switzerland and forest spaces between Switzerland and France. Its mainly formed by prairie, bushes, trees, monocultures and artificial habitats (paths, roads, industrial and commercial buildings). The groves in the vicinity of the site exhibit high sensitivity. Overall, the sensitivity on the site is average to low, often only partially fulfilling the requirements of a habitat.

The site PB in Presinge, Switzerland, is also an open agricultural space dominated by monocultures with some nearby woodlands and isolated hedges. Only a thin strip in the close vicinity of the nearby creek exhibits a moderate to average habitat quality. 

Site PD in Nangy, France is located in a rural context, constrained at the side by an autoroute and a departmental road and one side is bordered by a hamlet. Some isolated bushes and trees can be found in the vicinity. The habitat quality is very low to low. Only in a small patch at the southern end at the surface site border, the bushes represent a high value in this much-constrained space.

Site PF in \'Eteaux, France is in a mixed agricultural and prairie zone. Woodland starts to appear at the borders. The north is limited by a heavily used national road. The enlarged investigation zone includes a wetland. At the location of the surface site the habitat quality is low to average. The neighbouring wetlands and forest lands have a strong sensitivity.

Site PG in Groisy and Charvonnex in France is located in a mixed environment consisting of prairie, woodland, grassland and rural/forest paths. The enlarged space close to the autoroute is dominated by artificialised spaces such as retention basins, temporary inert waste buffering, roads, paths and constructed areas. The zone can, in general, be characterised as predominantly a forest habitat. On average, the whole zone studied exhibits a low to average level of habitat quality. Where areas can be considered wetlands, both inside and outside woodlands, the habitat quality can be considered average. Two forest areas that are outside the site perimeter, but in the immediate vicinity, have been identified to have a high sensitivity due to the quality of the trees.

Site PH in Cercier and Marlioz in France is almost entirely located in a woodland. There exist some cleared spots that host prairies and grassland. Overall, the habitat quality is low and where monocultures exist, it is very low. A zone exists that can be considered a wetland due to its characteristics south of the site. Its effects extend to a 0.4\,ha large part of the site, making this area a high-quality habitat zone. 

Site PJ in Dingy-en-Vuache and Vulbens in France is located in an open rural, agricultural space that is limited at two sides by treelines, a temporary creek, by an autoroute in the north and a rural path and further agricultural areas in the south. The extended study perimeter also includes zones that can be characterised as humid. The habitat quality is heterogeneous, ranging from very low to high. The space occupied by the surface site can be split in two halves: an area of low and an area of high habitat quality. The half that is closer to the autoroute exhibits the higher value. 

Site PL in Challex in France is located in a rural context. The site and the larger study perimeter spanning 100\,ha include agricultural fields, wasteland, bushes, and trees around houses that are on the surface site. The herbaceous spaces carry characteristics of wetlands. The agricultural constructed spaces with trees and gardens currently have a low to very low habitat quality. The humid prairies in the vicinity, but not on the surface site, have an average habitat quality.

\subsubsection{Wetlands}

\begin{figure}[h]
  \centering
  \includegraphics[width=0.6\textwidth]{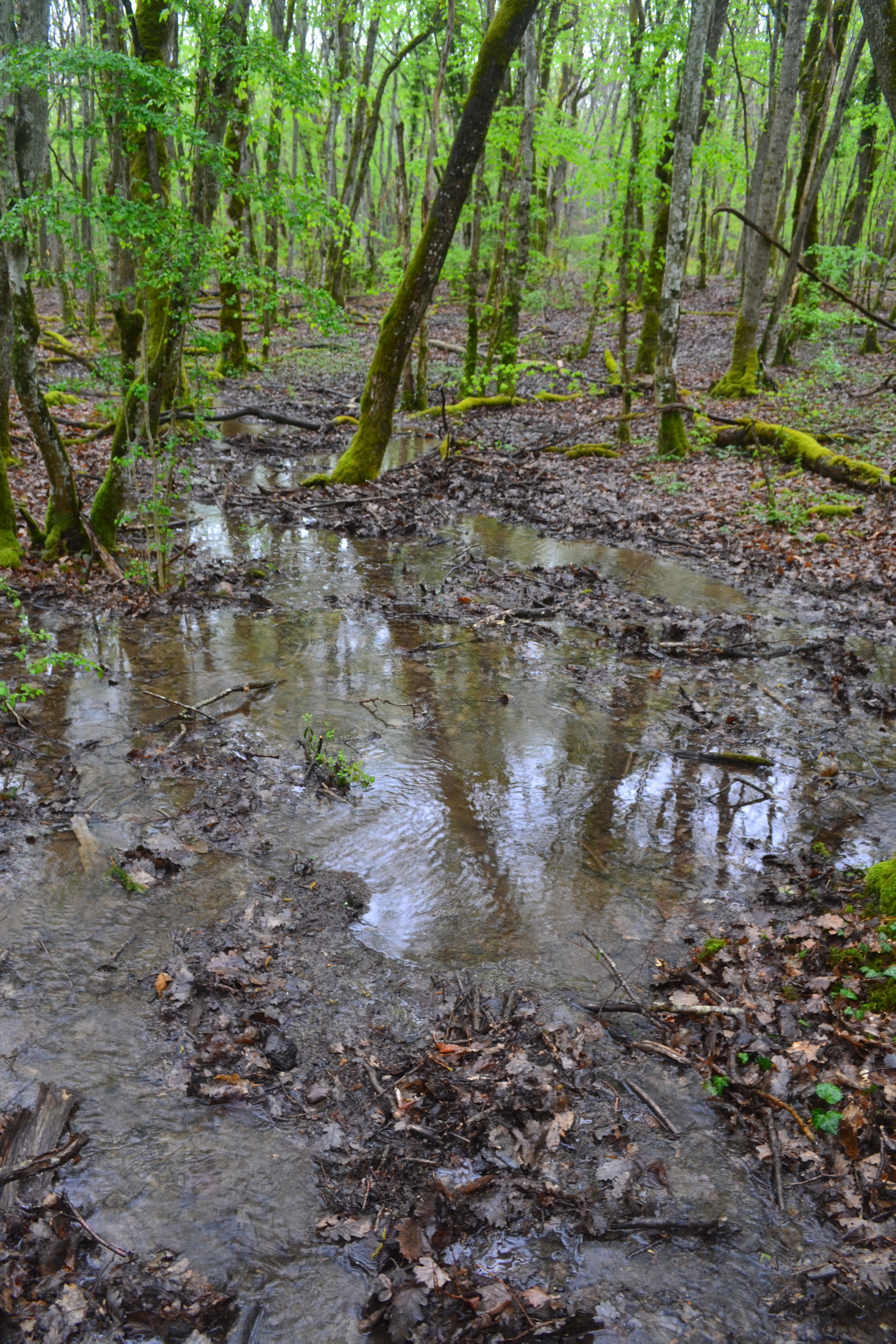}
  \caption{Wetland observed during the field investigations.}
\label{fig:Photo_Wetland_Mandallaz.pdf}
\end{figure}

Wetlands provide numerous services for ecosystems, particularly in terms of regulation, water storage, and  conservation of biodiversity. In the context of this study, investigations were carried out on the presence of wetlands in areas potentially affected by the surface sites. The study of wetlands is important environmentally and also a regulatory requirement under French law. This study aims to assist in decision-making for the location of sites and their optimisation according to the avoid-reduce-compensate approach. In France, in the case of the destruction of a wetland, the regulations impose compensatory measures with a ratio that can reach 1.5 to 2 times the wetland area impacted. Measures can include the improvement of partially degraded wetland functions and monitoring over a defined period to evaluate their effectiveness.

The study of wetlands was carried out in the same way for Swiss and French territories to obtain comparable and coherent data. In Switzerland, only wetlands listed in the federal inventories of low marshes, riparian zones, OROEM, RAMSAR sites, and amphibian breeding sites are potentially protected. In the absence of a direct equivalent of wetlands in Swiss legislation, this study was inspired by the definition of wetlands given in French legislation. According to Article L211-1 of the French Environment Code \cite{Gouv_FR_L211-1}, \textit{``Wetlands are understood as areas, used or not, usually flooded or saturated with freshwater, saltwater, or brackish water, permanently or temporarily; vegetation, when present, is dominated by hygrophilous plants for at least part of the year''}. Article R211-108 of the Environment Code specifies that: \textit{``The criteria to be retained for the definition of wetlands [...] relate to the soil morphology linked to the prolonged presence of water of natural origin and the possible presence of hygrophilous plants. These are defined based on lists established by biogeographical regions. In the absence of hygrophilous vegetation, the morphology of the soils is enough to define a wetland.''} Thus, French legislation defines wetlands based on floristic and/or pedological criteria.

The delineation of wetlands on surfaces concerned by the potential site locations was based on departmental/cantonal, Swiss, and French inventories, floristic criteria, and pedological inventories. Government inventories of the departments of Ain and Haute-Savoie allowed a first delineation of the known wetlands currently present on surface sites. Inventories of flora allowing a second delineation of wetlands were carried out during the flowering period in spring 2023 and by an expert company to identify and delimit the types of habitats potentially present on the immediate and extended perimeters around each site. These inventories also highlighted `pro parte' surfaces, i.e., surfaces where the habitat identified was not systematically or entirely characteristic of wetlands. Finally, pedological inventories were conducted by another expert company in 2023 and 2024. These inventories identified soils characteristic of wetlands for surfaces where the presence of wetland characteristics was eventually determined to be actually present or not in 2024 using shallow subsurface investigations (12 to 90\,cm deep).

At the border of site PA in Ferney-Voltaire, two wetlands are known with a total size of 6.3\,ha. The site does not directly impact the zone. The concept for the site has a rewilding project to improve the quality of this area and to make it a permanent and protected natural habitat with recreational characteristics. There are zones that are comparable to the French definition of wetlands in the vicinity of site PB in Presinge. None of these is in the immediate perimeter of the site, and the site will not impact any of these protection zones. The conceptual plan for the surface site includes the integration of one part of the nearby creek's area to rewild the space used for agricultural purposes today and to make it a fully protected habitat. 

Three wetlands zones can be found in the extended perimeter of site PD in Nangy, separated from the site by an autoroute. The site does not affect any of these wetlands. A number of wetland zones can be found in the immediate vicinity of site PF in \'Eteaux. The field investigations revealed that the zone is larger than registered in the regional inventory. The site does not impact these zones. However, the conceptual plan for the site concept includes the creation of a green buffer that includes one of the zones currently used for agricultural purposes to rewild it and to make it a protected habitat. There are also several wetland zones in the forest in the vicinity of site PG in Charvonnex and Groisy. The shape of the site has been adapted to avoid potential negative effects on these zones. A wetland zone exists in the forest at the northern border of site PH in Cercier and Marlioz and cuts through the site towards the south. Out of 16\,ha about 0.8\,ha are in the perimeter of the currently indicated site boundary. Consequently, the site will be further optimised during a subsequent design phase to either exclude significant effects on the wetland zone or to develop appropriate compensatory measures where the effect cannot be avoided. For example, a nearby area of land which has very poor biodiversity and habitat value has been identified and it can serve as an optional space in case the surface site equipment does not fit within the reduced surface site geometry. 

The vicinity of site PJ, in particular close to the autoroute, is characterised by extended wetland zones, covering 13.6\,ha. 1.6\,ha are on an agriculturally exploited area of the surface site. The subsequent site design phase will take in account the presence of this zone to establish avoidance and reduction measures, including the possibility of creating an annex further in the north on land that is not affected. Compensation measures may have to be developed for the part that cannot be entirely avoided. Wetland zones also exist at some distance from site PL in Challex, in the forest. The site does not affect them. 

Summing up, none of the sites is directly affected by wetland induced constraints. The restoration of an ecological compensation zone in the immediate vicinity of site PA provides an opportunity to increase the value of the zone and thus compensate for the loss of space by fostering the development of a natural habitat and the increase of biodiversity opportunities. Sites PH and PJ deserve particular attention during the further development of surface site designs due to zones that have characteristics of wetlands entering partially the site perimeters. Care also needs to be taken during the optimisation of the perimeter of site PG to ensure the avoidance of potential wetland-like areas outside the site limits.

\subsubsection{Flora}
\label{sec:flora}

\begin{figure}[h]
  \centering
  \includegraphics[width=0.5\textwidth]{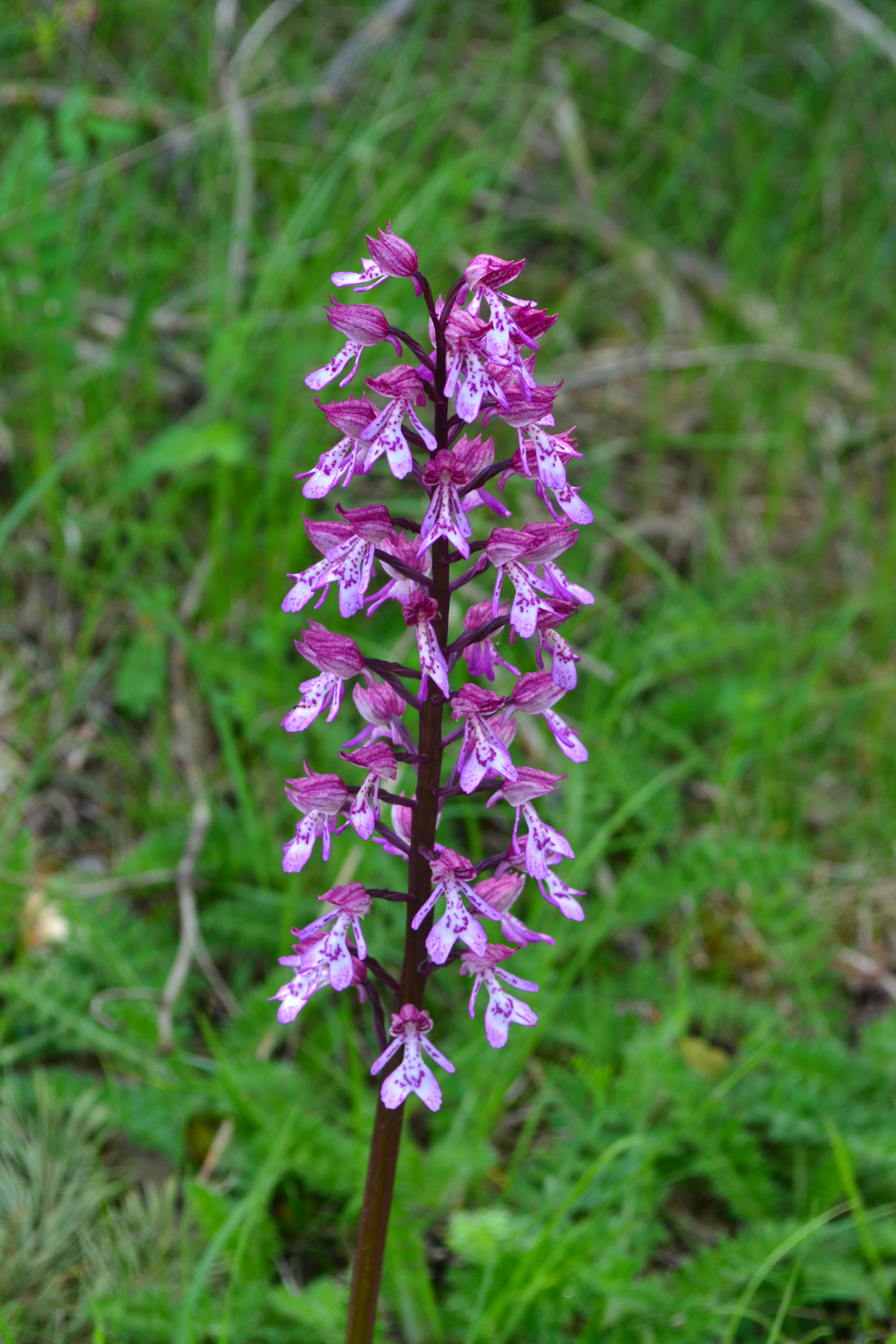}
  \caption{Military orchid {\textit{(Orchis Militaris)} observed during field investigations, providing valuable data on local wildlife presence and ecosystem dynamics.}}
\label{fig:Photo_Flora_Orchid_OrchisMilitaris_PG.pdf}
\end{figure}

On-site field visits were conducted in 2023 to validate bibliographic data and inventory the species present at the surface sites. These investigations aimed to confirm the presence of remarkable flora and invasive species and provide insights into local ecological contexts. ``Remarkable flora'' refers to plants that are notable or extraordinary due to their unique characteristics, ecological importance, rarity, or cultural significance. These plants often stand out because of their striking appearance, unusual adaptations, or the critical roles they play in their ecosystems. The term can apply to native, endemic, or even cultivated plants. Invasive plants are non-native species that are introduced to a particular ecosystem, where they spread rapidly and often outcompete native plants. These plants typically lack natural predators, diseases, or other controls in their new environment, allowing them to thrive unchecked. As a result, they can disrupt local ecosystems, reduce biodiversity, and cause environmental and economic harm.

For this study, remarkable species are those with legal protection status (national, regional, or departmental) in one of the two Host States and those listed as ``near-threatened'' (NT) and those that appear in regional and national red lists.

Special attention was given to invasive species, since soil that hosts invasive species must not be simply transferred to land compensation areas in order to avoid spreading of such species. Risk levels established at regional and national criteria were considered to determine the sensitivity of areas with respect to invasive species.

The habitats and floral compositions in the study area of site PA in Ferney-Voltaire, France are dominated by the agricultural space. During the field inventory, no remarkable species were observed. However, the municipal species list for Ferney-Voltaire identifies 3 noteworthy species that could in principle be encountered in the enlarged study area around the site. Complementary studies will be required during a project preparatory phase to confirm or exclude the presence in all areas that would potentially be affected by the surface site. Several species of invasive plants were observed in the PA study area that need to be considered when developing agricultural space compensation plans.

The habitats and composition in the Presinge study area in Switzerland concerning site PB are mainly agricultural. Woodland exists at a further distance and close to the Nant de Paradis creek, plants typical of wetlands (wetlands, riverside vegetation) are found. During the field visits, two remarkable species, protected at the cantonal and national level, were observed. Apart from these exceptions, no other important species listed in the bibliography were recorded. Several invasive species exist in the surroundings, and there are even some exotic ones. The agricultural space is free from remarkable and invasive species.

\begin{figure}[h]
  \centering
  \includegraphics[width=0.7\textwidth]{Enviro/Environment-Figs/Photo_Flora_SalviaGlutinosa_PD.pdf}
  \caption{Glutinous sage {\textit{(Salvia glutinosa)} observed during field investigations, providing valuable data on local wildlife presence and ecosystem dynamics.}}
\label{fig:Photo_Flora_SalviaGlutinosa_PD.pdf}
\end{figure}

Applying the municipal catalogue of remarkable species for site PD in Nangy, France does not lead to any significant sensitivity of the site. During the field inventory, no noteworthy species were found. However, several invasive species were observed that may negatively affect agricultural cultivation. Therefore, it remains to be studied if the soil can be transported for compensation purposes as is or if particular measures to eliminate those species will be required.

No remarkable species were recorded during the field inventory of the site PF in \'Eteaux in France. Three noteworthy species listed in the bibliography may, in principle, occur in the enlarged study area including wet meadows and wooded areas, outside of the surface site perimeter.

The municipal species catalogue of Groisy and Charvonnex applicable to site PG in France mentions three remarkable species that could be present in the study area. None of them were identified during the field inventory and no other remarkable species from the bibliography were observed either. A few invasive species have been recorded in the study area and topsoil that is removed should be cleared of those species.

The list of flora species of the communes of Cercier and Marlioz applicable to site PH in France includes a considerable number of remarkable species, but none of them is present in the study area, and no remarkable species were observed during the field visit.

None of the remarkable species listed in the bibliography for Dingy-en-Vuache and Vulbens for site PJ in France occur in the study area and no remarkable species were observed during the field visits. Some remarkable species listed in the communal lists could, however, be potentially present in the area that exhibit sufficient characteristics to support those species. They concern mainly humid zones at the edges of the surface site candidate perimeter.

For site PL in Challex, France, four species are highlighted as noteworthy in the bibliography. Tow out of them may potentially occur in the area of the PL site. However, none of them were observed during the field visits. On the contrary, two invasive species were inventoried in the study area, which should be taken into account in the case of reuse of agricultural topsoil.

Summing up, no particular sensitivity of any surface site candidate could be established with respect to remarkable flora. However, some sites will require attention with respect to the treatment of invasive species before the topsoil can be re-used for compensation measures. The PB surface site design requires attention due to two remarkable species observed during the field visit and the large number of potentially present species listed in the bibliography in the larger area. This site is, therefore, still considered to have a strong sensitivity at its edges. No outstanding flora species were inventoried at the PA, PG, PH, PJ, PL and PF sites, but these areas require complementary studies during a preparatory project phase to confirm the state and to plan for the construction site activities. So in general, additional field visits are required to reliably assess the environmental impacts. The sensitivity of site PF is also considered strong due to potential remarkable flora at the very edges of the site. Attention will need to be paid to invasive species present in several locations, and this will need to be taken into account when considering the reuse of agricultural soil in other locations.

\subsubsection{Fauna}

\paragraph{Amphibians}

\begin{figure}[h]
  \centering
  \includegraphics[width=0.8\textwidth]{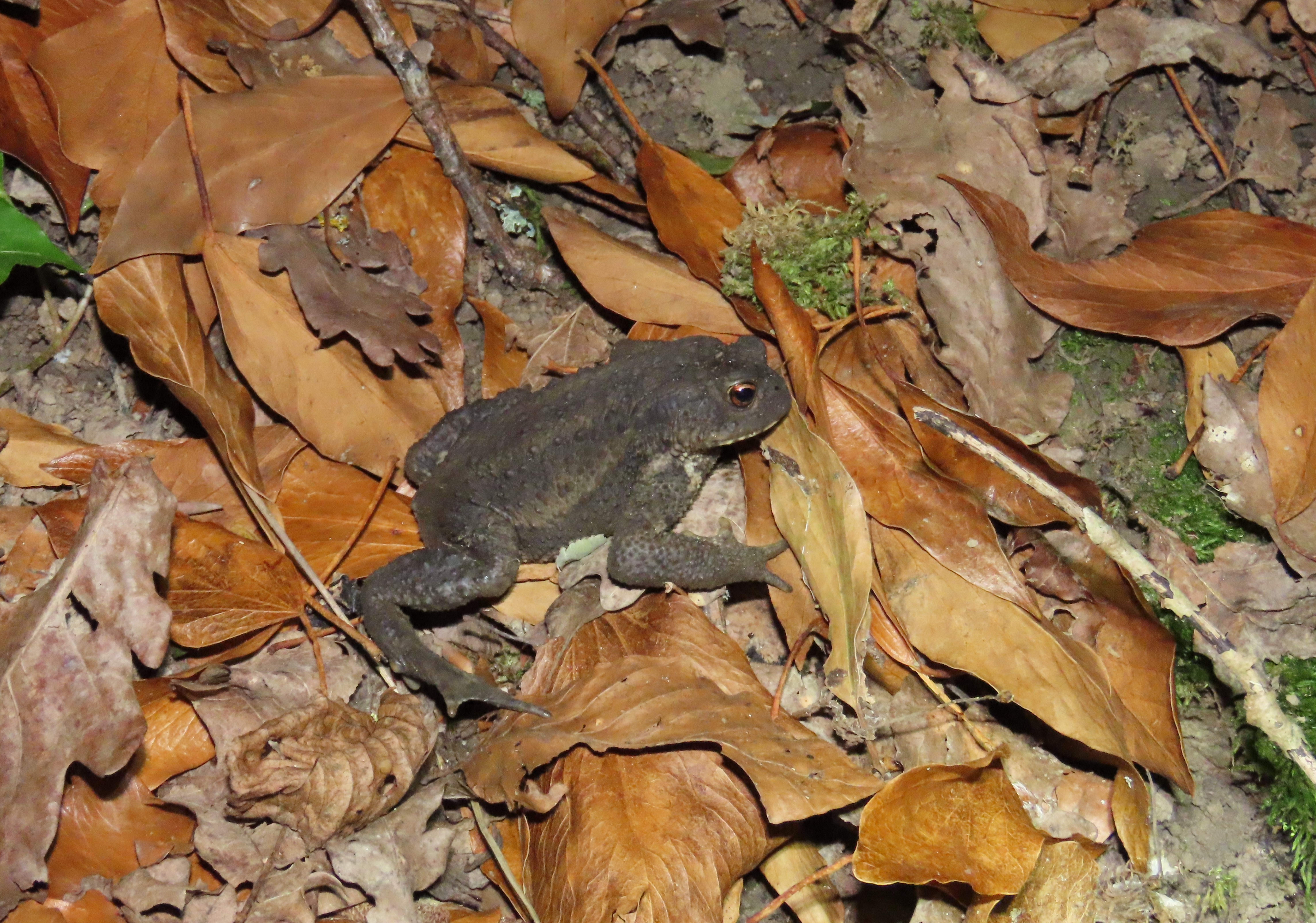}
  \caption{Common toad {\textit{(Bufo bufo)} observed during field investigations, providing valuable data on local wildlife presence and ecosystem dynamics.}}
\label{fig:Photo_Amphibians_BufoBufo_PF.pdf}
\end{figure}

For some groups of amphibians, identification down to the species level cannot be carried out without genetic analysis due to the strong hybridisation within these groups. However, an inventory based on bibliographical data and field investigations was established not only on the sites, but also for the perimeters in the vicinity of the site. Investigations were done during the day and the night for several months. Amphibians are frequently found in the vicinity of water spaces, creeks, rivers, and wetland zones.

Amphibians were observed in the vicinity of site PA in Ferney-Voltaire across the French and Swiss territories. Some of them are considered endangered in Switzerland. No amphibians are present on the surface site location. In the immediate vicinity of site PB in Presinge the bibliographic inventory was confirmed by the field investigations and the issue\footnote{Environmental issues mentioned here and in the following sections may also be referred to as `stakes' - an alternative translation of the French `enjeux'.} is big, and therefore the river zone will be entirely avoided by the project. The integration of a rewilding project in the site will help to raise the protection level of that zone further. The area is currently partially used as agricultural space. Site PD in Nangy and its vicinity have a low stake concerning amphibians. Although the site PF in \'Eteaux is not directly concerned, the immediate vicinity revealed species with a high stake. At site PG in Charvonnex and Groisy several amphibians are confirmed in the vicinity and few on the site. The stake concerning amphibians is also high in the vicinity of site PF. On the site, no issues could be identified. Some amphibians were observed in the vicinity of site PJ, but the site did not reveal particular issues. Some amphibians were observed in the larger area around site PL in Challex, including Switzerland. Some of them are listed as endangered and protected. The site does not host amphibians.

Summing up, the zones in the vicinity of sites PA and PD feature a few habitats that would give them a high value. For sites PF and PJ, the availability of habitats also seems relatively limited, but the forest/hedge environment and the presence of streams give them a higher value. The vicinity of site PB, meanwhile, has the particularity of being partly located on an amphibian reproduction site, that increases its value level, despite the lack of direct observation of high-valued species. The sites PG, PH, and PL have high values due to a high density of aquatic habitats and the observation of a large number of high-value species in their vicinities.

\paragraph{Birds}

\begin{figure}[h]
  \centering
  \includegraphics[width=0.7\textwidth]{Enviro/Environment-Figs/Photo_Fauna_Birds_AcrocephalusScirpaceus_PB.pdf}
  \caption{Reed warbler {\textit{(Acrocephalus scirpaceus)} observed during field investigations, providing valuable data on local wildlife presence and ecosystem dynamics.}}
\label{fig:Photo_Fauna_Birds_AcrocephalusScirpaceus_PB.pdf}
\end{figure}

Ornithological surveys were conducted using sound recorders and visually with binoculars and cameras. The survey period spanned 2023 and 2024. An inventory was carried out for the 8 surface site locations over several seasons. Five ornithologists took part in the surveys. The inventories were conducted under suitable weather conditions (no rain and little/no wind). Field observations were recorded on-site using computer equipment. Observation intensity varied depending on the seasons. Increased observation intensity was applied to the spring inventories due to the sensitive period for breeding birds. A matrix was established using European guidelines for endangered species that is only applicable to the French territory. For Switzerland, the cantonal and federal protection lists were used to analyse the risks.

Birds find living space in the vicinity of site PA in Ferney-Voltaire, mainly in trees, bushes, and hedges. Some noteworthy species have been found in the vicinity. Birds are also prevalent in the vicinity of the agricultural space in site PB in Presinge and next to the small creeks and rivers. Protected and endangered species were found in the extended zone around the site. Areas and trees in the vicinity of site PD in Nangy provide protection spaces for birds. Some noteworthy species have been found on the surface site location. The surface site PF in \'Eteaux today includes several bushes that serve birds to find food and rest during the migration. Some noteworthy species have been found on the site. The forest on site PG in Charvonnex and Groisy provides living space for a variety of birds, and the bushes around serve as a retreat and reproduction space. Some noteworthy species have been found in the vicinity of the site. Also the forest on site PH in Cercier and Marlioz provides space for birds. Some heritage species were found on the site. Site PJ is surrounded by bushes and hedges that serve birds as living and reproduction space and for rest during migration. Some heritage species were found on the site and in the immediate vicinity. Also the hedges nearby site PL provide protection for birds and noteworthy species were found there.

\begin{figure}[h]
  \centering
  \includegraphics[width=0.7\textwidth]{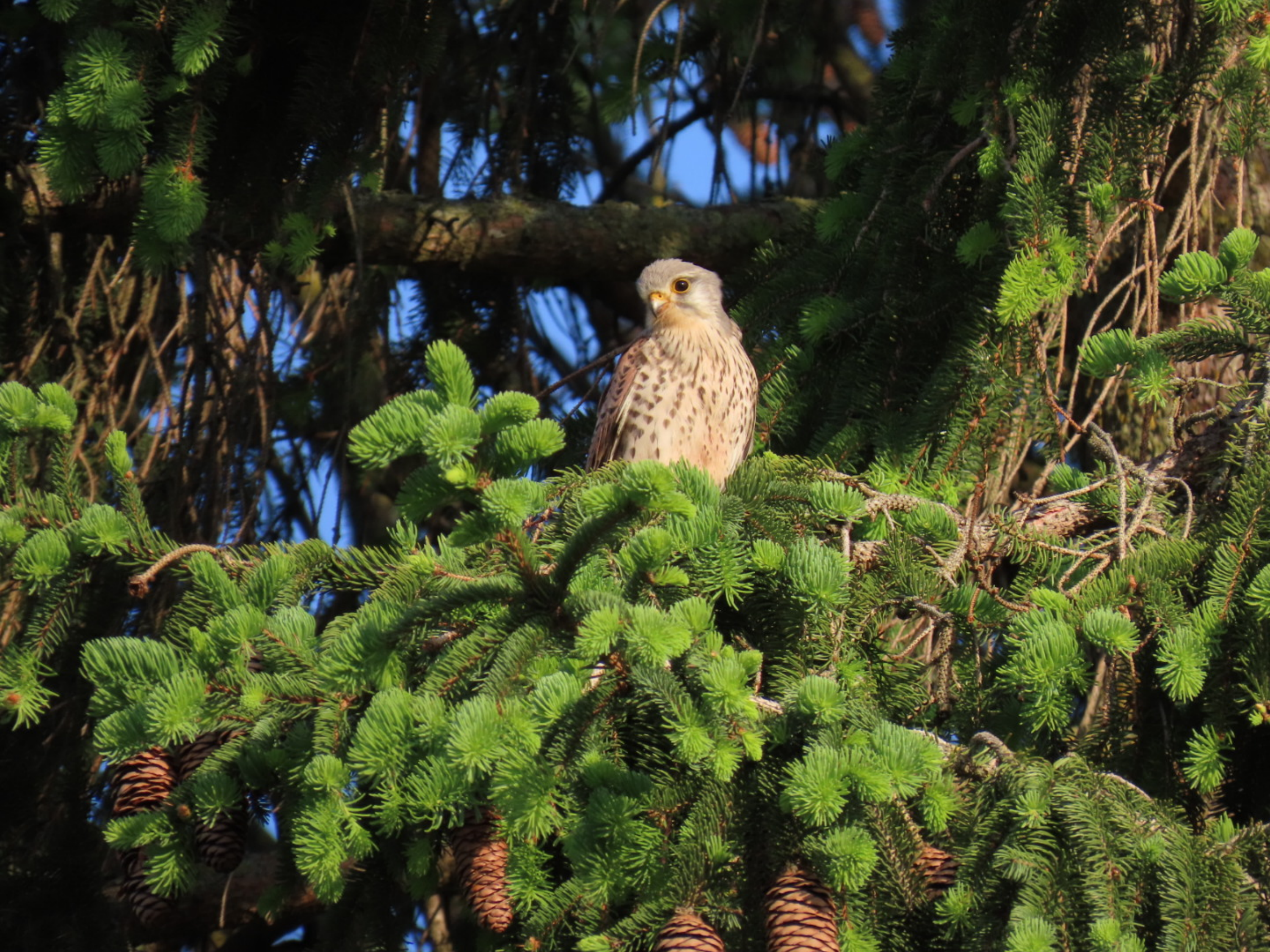}
  \caption{European kestrel  {\textit{(Falco tinnunculus)} observed during field investigations, providing valuable data on local wildlife presence and ecosystem dynamics.}}
\label{fig:Photo_Fauna_Birds_FalcoTinnunculus_Arve.pdf}
\end{figure}

A full database of species observed was established, and it has been integrated into a project-wide geographical information system. The bird observations have limited validity and must be continued if a decision to advance with a project design is taken.

In summary, the sites PD, PL, and PJ in France, as well as PB in Switzerland and their surroundings, contain hedgerows and open areas that support nesting species of high conservation value. The sites PF, PG, and PH in France feature hedges and forests adjacent to extensive pastures, which are crucial habitats for uncommon species throughout the year. Site PA in France consists primarily of conventional agricultural land that becomes waterlogged in winter, serving as an important stopover site for waders.

To mitigate the environmental impact, dedicated measures will be implemented during the design phase to recreate bird nesting and living areas comparable to those lost, either within the green buffers of the sites or in their immediate vicinity. These efforts will be integrated into rewilding projects associated with the development.

\paragraph{Mammals}

\begin{figure}[h]
  \centering
  \includegraphics[width=0.7\textwidth]{Enviro/Environment-Figs/Photo_Fauna_Mammals_Vulpes_SitePF.pdf}
  \caption{Red fox {\textit{(Vulpes vulpes)} observed during field investigations, providing valuable data on local wildlife presence and ecosystem dynamics.}}
\label{fig:Photo_Fauna_Mammals_Vulpes_SitePF.pdf}
\end{figure}

Data on terrestrial mammals was collected during all naturalist assessments of other groups on the surface sites. Therefore, data on terrestrial mammals were collected both day and night at all surface sites throughout four seasons. Terrestrial mammal species were recorded visually or through indicators such as tracks, faeces, burrows, etc. The particular attention given to weather conditions for other groups was also applied during the assessments of terrestrial mammals. All observations were geolocalised and recorded in the project-wide geographical information system.

The PA site in Ferney-Voltaire has been the subject of infrequent observations of terrestrial mammals during naturalist investigations. They include the European roe deer, wild boar, European badger, European hare, brown rat, and greater white-toothed shrew. In the wider surrounding bibliographic data also revealed the presence of species of medium concern, such as for instance the hedgehog and beaver.

In the vicinity of site PB in Presinge, two species with a high conservation priority in Switzerland were observed: the beaver and the hare. Other species with a low priority that were encountered are the roe deer, badger and fox. Bibliographic data also provided information about species with a very high priority: the harvest mouse, polecat, weasel, wolf and the dormouse. Two species with a medium priority are the hedgehog and the stoat. However, these species were not observed.

The observations on site PD in Nangy of medium importance are the beaver and hedgehog. A species with lower importance, the fox was also observed. Bibliographic data also reports on the wildcat, the polecat, the black rat, but they were not observed.

Several observations have been recorded at the PF site at \'Eteaux: the hedgehog, with medium importance, and the fox and the badger, with very low importance. Bibliographic data details the presence of four species of medium conservation concern, including the Alpine Ibex, European beaver, European rabbit and the wildcat.

Observations at the PG site in Charvonnex and Groisy ranging from very low to low importance are: European roe deer, fox, squirrel, wild boar, European hare, chamois. In the surroundings, the bibliography also reports species with high importance: European otter, polecat. The following species with medium importance are reported: hedgehog and wildcat. However, these species were not observed.

At the PH site in Cercier and Marlioz the presence of species with medium importance such as the beaver and the squirrel were recorded. Species of lower importance concern the red deer, roe deer, wild boar and fox. The bibliography also mentions the presence of species with very high concern such as the grey wolf,  otter and medium concern such as the wildcat and the hedgehog. These species were, however, not confirmed.

At the PJ site in Dingy-en-Vuache and Vulbens several observations of mammals with medium importance were made, such as the European beaver and the wildcat. Species with lower importance are the squirrel and with even lower importance the red deer, roe deer, hare, wild boar, badger, fox and the greater white-toothed shrew. The bibliography also mentions the presence of the lynx and the otter with very high importance and the rabbit with medium importance, but they were not observed.

Observations on site PL in Challex with a very low importance include the badger, fox, roe deer and hare. The analysis of the bibliography highlights four species with medium concern: wildcat, beaver, rabbit, and hedgehog. The polecat is cited with high importance and the lynx with very high importance. These species have, however, not been observed.

Summing up the bibliographic and field investigations of mammals carried out on all surface site candidates and in extended perimeter around the sites shows that overall a strong sensitivity exists for all sites although on none of the sites direct observations of relevant species were confirmed. The reason is that mammals were observed in the vicinities and extended surroundings of the sites and they can traverse the sites today, but they will face limitations during construction periods and when the sites are constructed. However, no protected or endangered species would be affected directly by the surface sites. A subsequent preparatory project phase needs to consider the mammals in the surroundings of the site in the design of the construction sites and the surface sites. Ecological corridors need to be considered and preserved. Where possible, green buffers and sites that can be traversed by mammals can preserve their current behaviour. Embedding the presence of animals in general in the concepts of surface sites can also help to improve their habitats and eventually even support the increase of biodiversity and conditions for them.

\paragraph{Chiropters}

\begin{figure}[h]
  \centering
  \includegraphics[width=0.7\textwidth]{Enviro/Environment-Figs/Photo_Fauna_Bats_detector.pdf}
  \caption{Bat detectors used to detect the presence of bats by converting their echolocation ultrasound signals.}
\label{fig:Photo_Fauna_Bats_detector.pdf}
\end{figure}

Chiropters (bats) are mammals. The name chiroptera comes from the Greek words cheir (hand) and pteron (wing), meaning `hand-wing'. This reflects the unique structure of their wings, where elongated fingers are covered by a thin membrane of skin that allows them to fly. Chiropters are in the focus of environmental studies due to their ecological importance, vulnerability to habitat changes, and the legal protections they enjoy in France and Switzerland through national laws, as well as in the frame of the EU habitats directive, the `EUROBATS' agreement, the `Bern Convention' and `Natura 2000 site'. Violating protections of bats in France can result in significant penalties consisting of high fines and prison sentences.

This study comprised dedicated field investigations applying different techniques to establish an inventory of bats on and in the vicinity of candidate surface site locations to permit developing avoidance, reduction, compensation, and accompanying measures during a subsequent project design and preparatory phase.

Bibliographic analysis could unfortunately not be carried out since it requires a detailed description of the areas to be investigated, and the surface site locations were not sufficiently defined at the time of making such inquiries for data, which are time-consuming processes with different data owners. Therefore, indirect (e.g., passive and active acoustic searches with ultrasound detectors and microphones, search for traces) searches were immediately carried out at the larger candidate surface site zones.

On site PA in Ferney-Voltaire, France, the existence of an ecological corridor renders the passing of chiropters likely in a band at the southern end of the surface site and outside that zone. 12 species were recorded in the larger area around the surface site location, in particular in the neighbouring woodlands and groves. There were no sightings on the site itself, probably due to it being an open space currently subject to light pollution during nighttime.

On site PB in Presinge, Switzerland, the zones at the borders of the surface site towards the Nant de Paradis creek represent an area of interest for chiropters for hunting. The same applies for individual bushes, the Seymaz zone and woodlands in the vicinity and garden areas in the Avenir hamlet. 2 species have been observed at houses that would be removed for the construction of the site and 18 species were found in the larger area. 7 of them have a preservation status. This makes the area sensitive and calls for measures to preserve the habitats and avoid and reduce light pollution as far as reasonably possible. 

On site PD in Nangy, France, the larger zone is of interest for Chiropters for hunting where bushes, trees and houses exist. 3 species were observed on the site and 8 in the extended investigation area. They have protection status. As with site PB, the larger area is sensitive and calls for measures to preserve the habitats and avoid and reduce additional light pollution as far as reasonably possible.

On site PF in \'Eteaux, France, the neighbouring areas are of interest for chiropters, in particular zones with hedges and woodland. This concerns for instance the boundaries of the surface site, houses and gardens along the national road and the wetland zones. 13 species were observed on the larger area and some of them on the site close to trees. 3 enjoy a protection status. The larger area represents an interest for this species and calls for measures to preserve the habitats and avoid and reduce additional light pollution as far as reasonably possible.

On site PG in Groisy and Charvonnex, France, the woodland is of interest for chiropters. It extends to the autoroute zone in the north. 18 species have been observed in the larger area and 13 species were found in some locations on the surface site, primarily in the forest. 8 enjoy a protection status. The strong presence of chiropters in the forest calls for a minimisation of the impact on those zones during the design phase and envisages compensation and accompanying measures.

On site PH in Cercier and Marlioz, France, chiropters find hunting areas and a corridor in and between the woodlands. 13 species have been observed within the site perimeter in the woodland. 5 species enjoy a very high protection status and two are highly protected. The density and diversity diminishes as one approaches the road on the eastern side of the site. These findings call for further optimisation of the site, reduction, compensation and accompanying measures in the subsequent phase.

On site PJ in Dingy-en-Vuache and Vulbens, France, the woodlands at the creeks and the hedges at the limits of the surface site are interesting spaces for chiropters. Ten species were found on the border of the site and 16 in the larger area around the site. Some enjoy a particular protection status. The density and diversity are higher around hedges and trees close to the motorway. This location will need some attention during the subsequent site design with respect to avoidance, reduction, compensation and accompanying. Artificial light pollution is one issue to be considered in this area.

On site PL in Challex, France, chiropters find suitable locations in the woods, hedges, hamlets and individual houses, their gardens and the nearby vineyards. Two species with medium to low protection status were recorded at the houses that would have to be removed to construct the surface site.

Summing up, the stakes with respect to chiropters directly on the surface sites are low in PA, PB, PD and PL. PJ has no particular sensitivity on the entire site, but a small sector that is close to the woodlands requires particular attention. The stakes are average for PF, high for PG and very high for PH. The limitations of impacts on chiropters, the preservation and, where possible, the improvement of their habitats have to be included in the eco-design approach to be implemented in the subsequent phase.

\paragraph{Reptiles}

\begin{figure}[h]
  \centering
  \includegraphics[width=0.7\textwidth]{Enviro/Environment-Figs/Photo_Fauna_Reptiles_PodarcisMuralis.pdf}
  \caption{Common wall lizard {\textit{(Podarcis muralis)} observed during field investigations, providing valuable data on local wildlife presence and ecosystem dynamics.}}
\label{fig:Photo_Fauna_Reptiles_PodarcisMuralis.pdf}
\end{figure}

Field investigations on the candidate surface site locations focused on identifying reptile populations in environments that are known to be favourable habitats for them. They include for example semi-open areas, forest boundary zones, cavities and stone or woodpiles as well as constructed areas. These inventories were carried out during the reptile's primary activity periods from May to June and from September to October. Each observation was geolocated. No intrusive methods were used to avoid disturbance to the species. Several methodological limitations, an unfavourable rainy spring and an exceptionally hot summer as well as access limitations set limits on the quality and reliability of the results. This concerns mainly the data for snakes that are more difficult to observe than other reptiles. For a preparatory phase project and a comprehensive environmental impact assessment, the initial state of reptiles has to be updated with complementary studies.

The natural habitats investigated for reptiles varied across the sites, leading to different levels of suitability for living spaces and ecological corridors.

At site PA in Ferney-Voltaire, France, the forest edges and wood strips were identified as high-value habitats. Isolated trees and cultivated areas had a lower interest. Only one species with a low conservation value, the common wall lizard, was observed on the annex site south of LHC point 8. Bibliographic data do not indicate the presence of species with high stakes.

At site PB in Presinge, Switzerland, the zones with vegetation at the nearby creek Nant de Paradis, gardens in the Avenir hamlet and small woods in the vicinity are of interest to reptiles, but not the surface site location. Only a single species was found in the larger area around the site. No species were found on the perimeter of the site.

At site PD in Nangy, France, interesting locations for reptiles are at the southern end of the site in the hedges and the borders of the autoroute, but not the surface site. Three species were found in the larger area around the site and at the extreme edge in the south, outside the site.

At site PF in \'Eteaux, France, the hedges at the border of the surface site are of interest for reptiles. Only a single species was found in the larger environment around the site. No observations could be confirmed on the site or in the immediate vicinity.

At site PG in Groisy and Charvonnex, France, the forest zone and the limits of the woodlands are of interest for reptiles. Also, areas close to the highway can be relevant spots. 2 species were found in the larger area around the site, but no observations could be confirmed on the site directly.

At site PH in Cercier and Marlioz, France, the entire forest occupied by the surface site location is of interest for reptiles. 2 species were found in the area around the site, sometimes entering the perimeter of the site. However, no clear pattern of presence or movement could be determined that would permit drawing sound conclusions on the permanent presence of reptiles on the site.

At site PJ in Dingy-en-Vuache and Vulbens in France, areas in the vicinity of the creeks, bushes, groves and trees including the zone close to the motorway are of interest for reptiles. 7 species were found in the larger area around the site, leading also to the conclusion that at the edges of the site, species could be present.

At site PL in Challex, France, gardens, bushes, and trees are of interest for reptiles but not the majority of the surface site. Only one species could be found at a distance of the site, close to the forest in the north. No observations on the site could be confirmed.

Summing up, only very low sensitivity with respect to reptiles exists for sites PA, PB, PD and PL. In the vicinity of PF some care needs to be taken to preserve the living spaces of reptiles. For PG and PH sensitivity may exist in some parts of the forest and the project design needs to respect this condition. An update of the initial state with complementary field investigations is required to optimise the integration of the surface site, applying avoidance and reduction measures. The architectural designs of the surface site constructions should integrate concepts that favour the creation of habitats and thus help to increase the presence of reptiles. 

\paragraph{Insects}

\begin{figure}[h]
  \centering
  \includegraphics[width=0.7\textwidth]{Enviro/Environment-Figs/Photo_Fauna_Insects_Butterfly_AporiaCrataegi.pdf}
  \caption{Butterfly black-veined white {\textit{(Aporia crataegi)} observed during field investigations, providing valuable data on local wildlife presence and ecosystem dynamics.}}
\label{fig:Photo_Fauna_Insects_Butterfly_AporiaCrataegi.pdf}
\end{figure}

Bibliographic research and field investigations of the sites and their vicinities by specialised companies were used to identify the issues with respect to insects. Field investigations were carried out for several months during good weather conditions, low wind and during both day and night.

Despite the highly urbanised environment, a strong population of insects was observed in the surroundings of site PA in Ferney-Voltaire due to the favourable living and breeding spaces. The site itself, an agricultural area is not subject to a strong presence of insects and thus, therefore, has only a very low sensitivity. The vicinity of the border between France and Switzerland towards the Geneva airport may require the consideration of a cross-border impact since some species observed in this zone outside the site perimeter are protected in Switzerland. Complementary field investigations are required during a project preparatory phase to reveal any such potential case.

The surroundings of site PB in Presinge feature an important population of insects, mainly those relating to aquatic habitats. Some species observed in this area outside the site perimeter enjoy particular protection status in Switzerland. The site itself is an agricultural space with low stakes apart from the areas that are close to the nearby creek.

\begin{figure}[h]
  \centering
  \includegraphics[width=0.7\textwidth]{Enviro/Environment-Figs/Photo_Fauna_Insects_Dragonfly_PB.pdf}
  \caption{Golden-ringed dragonfly {\textit{(Cordulegaster boltonii)} observed during field investigations, providing valuable data on local wildlife presence and ecosystem dynamics.}}
\label{fig:Photo_Fauna_Insects_Dragonfly_PB.pdf}
\end{figure}

The surroundings of site PD in Nangy are characterised by a few insect species. The site is an agricultural space and has low stakes with respect to insects. No species with protection status were observed.

The extended surroundings of site PF in \'Eteaux are attractive for insects. However, on the site and in the immediate vicinity, no relevant species were identified, and the sensitivity is low.

The forest at and around site PG in Charvonnex and Groisy is moderately attractive for insects, as are the surrounding hedges. The site itself does not show particular issues with respect to insects. Also, the woodland spaces did not reveal the presence of relevant insects.

The forest spaces at and around site PH in Cercier and Marlioz provide, in principle, a favourable habitat for insects. However, only little presence of insects was revealed within the candidate site perimeter. Due to some observations at the border of the site, the site has a medium level sensitivity.

Site PJ in Dingy-en-Vuache and Vulbens, in principle, offers favourable conditions due to the trees and hedges around it. The stakes are high in the extended surroundings, but they are low on the site itself.

The hedges and bushes around site PL in Challex also, in principle, provide a favourable habitat for insects. However, the site shows only low-level issues with respect to insects, though a presence cannot be entirely excluded. Some species are protected in the nearby Swiss territory. A preparatory project phase will have to make complementary field investigations to consider potential cross-border impacts.

Summing up, none of the surface site candidates exhibits particular sensitivity with respect to insect presence. The surroundings of site PB present a habitat with high sensitivity. The wider surroundings of site PF present, in principle, some sensitivity. For sites PG and PH, despite being woodland, the favourable habitat for insects turns out to be less favourable than expected. The presence of some relevant species in the surroundings raises the sensitivity of the areas that are in the vicinity of the site. There were observations of some species with medium sensitivity in the surroundings of site PJ, suggesting that they might also be present on the site.

\subsubsection{Aquatic}

\begin{figure}[h]
  \centering
  \includegraphics[width=0.5\textwidth]{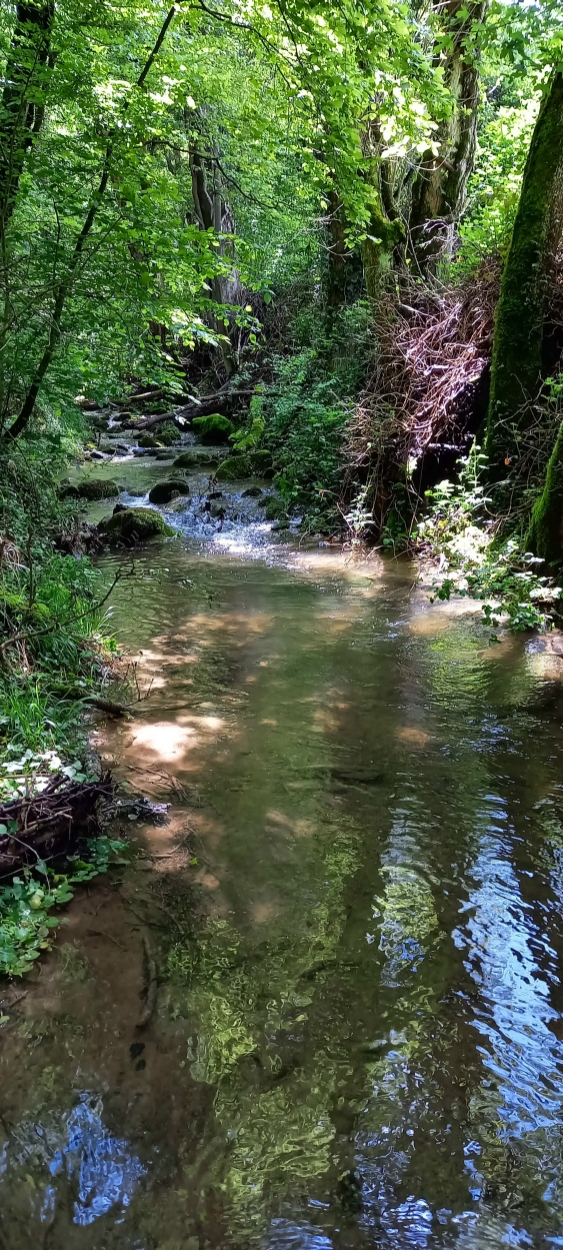}
  \caption{Stream photographed during the field investigations.}
\label{fig:Photo_Aquatic_NantDHiver_PJ.pdf}
\end{figure}

Fauna in the aquatic context has been analysed based on cartographic and orthophoto information. Systematic investigations and water analysis for all aquatic aspects were not carried out at this stage. They are planned to be done if a design phase is launched. However, nearby surface water in the vicinity of sites PF, PG, PH and PJ was analysed. The stake concerning macroinvertebrates is often high in the extended surroundings around surface sites. Aquatic benthic macroinvertebrates are insects in their nymph and larval stages, snails, worms, crayfish, and clams that spend at least part of their lives in water. Fireflies are another important species that are sometimes encountered in the wider perimeters of the sites.

Although the site PA and its immediate surroundings are not subject to aquatic fauna presence, the wider environment is known to be a habitat for vulnerable and endangered species. Protected and endangered species are close to site PB, although the site does not affect their habitats. Sites PD and PF and their surroundings are not concerned by aquatic fauna. The vicinity of site PG is characterised by important aquatic habitats with vulnerable and endangered species that the project will avoid and aim not to affect. The site perimeter has already been adjusted as a result of various stakes identified during the initial state analysis. Although site PH lies in the forest and close to a creek the presence of noteworthy aquatic species is low and the site itself is not affected. Sites PJ and PL and their surroundings also have no issues with respect to aquatic fauna.

Summing up, the PJ site does not present any observations of benthic macroinvertebrates. The wider surroundings of the PD, PF, and PL sites show a few habitats favourable to the development of benthic macroinvertebrates, but only a few taxa are present. Also, the PH site does not present any taxa with significance; however, the habitats are varied and the taxonomic diversity is high near the site, although with very low significance. The PG site hosts taxa with `medium' significance, but the habitats are varied and taxonomic diversity is high near the site border. The wider PA site surroundings are subject to an observation with `high' significance, but the habitats are poor and have little interest for the benthic macrofauna. The PB wider site surroundings have several observations of taxa with `very high' significance. No surface site has any relation with fis,h although some sites (PB, PF, PG, PH, PJ) are in the vicinity of small creeks.

\subsubsection{Forest}

\begin{figure}[h]
  \centering
  \includegraphics[width=0.7\textwidth]{Enviro/Environment-Figs/Photo_Forest_SitePA.pdf}
  \caption{Woodland photographed during the field investigations.}
\label{fig:Photo_Forest_SitePA.pdf}
\end{figure}

During this study, expert companies and forest evaluation consultants have made a comprehensive and detailed forest quality and value analysis. This included the potential loss of biodiversity, habitat and economic income over a sustained period of several decades. The results are also integrated into the comprehensive, wider socio-economic assessment. The French `Indice de biodiversité potentielle' (IBP) methodology was applied \cite{Bouget2014, IBP_online}. The project will not affect any existing forest spaces in Switzerland. Forests are in the vicinities of sites PA, PB, PD, PF, PG, PJ and PL. Only sites PG and PH will require clearings. 

The forest that would be affected by site PG has quite a diverse character. The xeric woodlands located in the southeast mainly consist of small-diameter oak woods, showing a strong to weak stake depending on the area. The ravine woodlands in the west feature a composition of large and very large trees rich in dendro-microhabitats, potentially hosting a rich and diverse biodiversity. The central area of the forest presents a character more easily exploitable for forest owners. This leads to more or less diverse stands depending on the plots. The presence of large dead wood is less important overall. The challenge in terms of biodiversity varies from strong to weak depending on the age of the trees. Consequently, the surface site has been adapted to reduce the affected forest as much as possible and to select an area for the access shaft locations that has a lower quality than the surroundings. In total about 2.4\,ha of forest may be affected by the surface site development.

The forest at PH features relatively young woodlands resulting from agricultural abandonment, thereby having few dendro-microhabitats. The woodlands located on the northern fringe of the site are the oldest and there is presence of a temporary watercourse. Medium wood is less conducive to supporting forest biodiversity than large and very large wood. In total, up to 10\,ha of forest may be affected in the communes of Cercier and Marlioz. Further surface site designs are needed to determine the exact surface requirements, taking into consideration all the environmental issues that have been identified.

\subsubsection{Summary}
Detailed data are included in specific paragraphs of the biodiversity section.
\begin{figure}[h]
  \centering
  \includegraphics[width=0.7\textwidth]{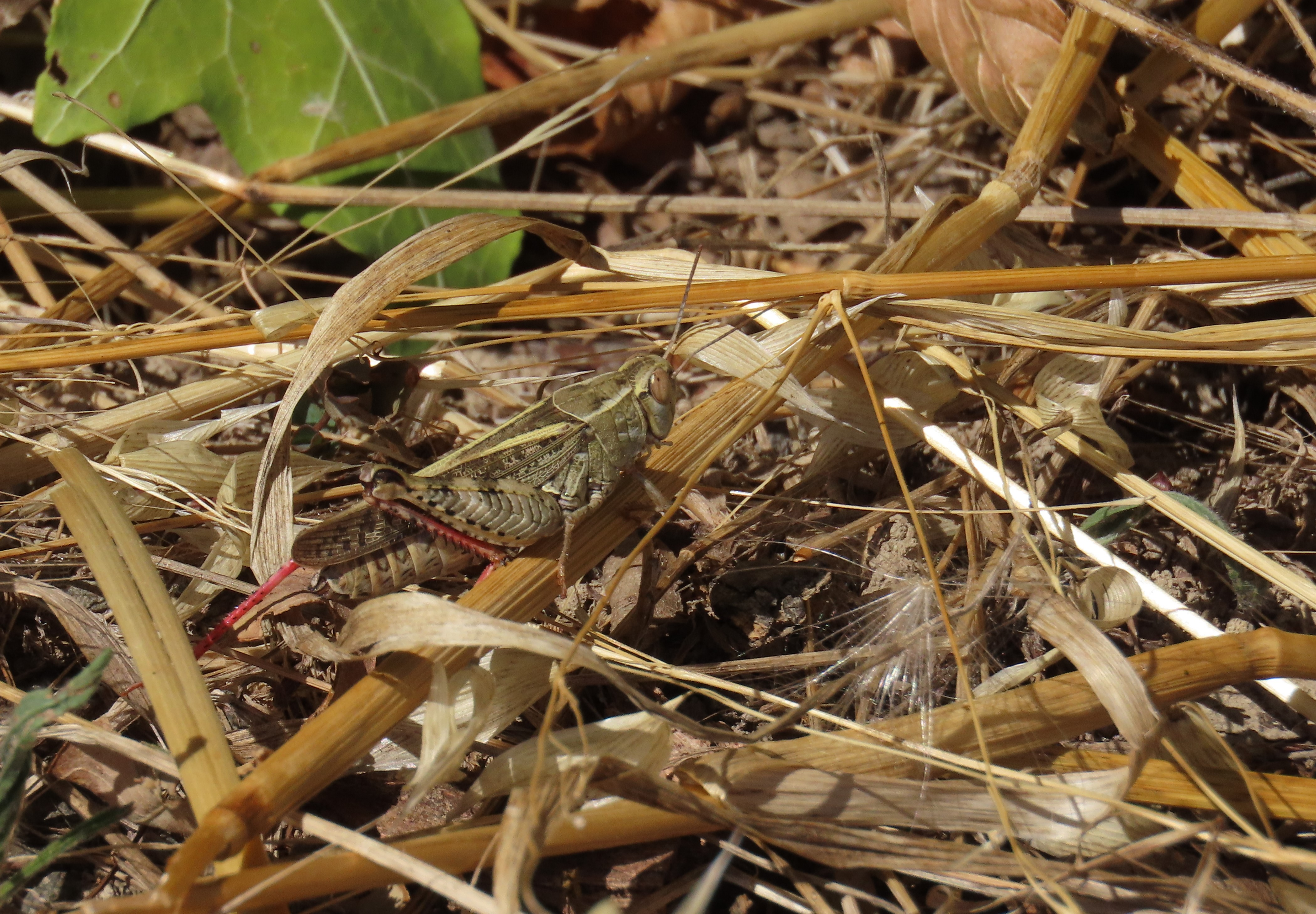}
  \caption{Italian locust {\textit{(Calliptamus italicus)} observed during field investigations, providing valuable data on local wildlife presence and ecosystem dynamics.}}
\label{fig:Photo_Fauna_Insects_Grasshopper.pdf}
\end{figure}
To the east of the PA are wetlands that serve as a migratory stopover for bird species, and to the south, near LHCb Pt8, is a forest of ecological value due to the presence of bats and insects. Some agricultural patches near the site with isolated trees also play a valuable role for birds. The PA surface sites have low ecological stakes, however the future construction must carefully consider the surrounding elements to maintain the key habitats and species.

The PB surface site presents low to moderate ecological stakes, however the presence of diverse bird species, aquatic insects and protected plants along the Nant du Paradis stream highlights the ecological value of the site's nearby areas. The future construction must take into account the ecological sensitivity of these places to ensure the preservation of biodiversity and habitats.

The site PD is entirely covered by agricultural land which represents low-level stakes. Small areas of high ecological importance are located further north of the site. These include hedges and old trees that provide habitat for birds and bats of low or moderate concern. Future construction in this area has to take into consideration small fauna species that might be present on or crossing the site.

The PF site is used as an agricultural meadow and presents moderate ecological states. However, the site is located within the ecological corridor for wildlife movement, near the wetland area and forest with the stream providing habitats for a range of species, including birds, bats, and amphibians. 
The site layout must be carefully planned to avoid disrupting the sensitive ecosystems and ensure good functionality of the fauna corridor.

The main PG surface site is partly in a forest area, which represents high stakes, and partly in pasture land with low stakes. The annex site to the north, located near the highway, shows low ecological stakes. The main constraint rPG site is the valuable forests that hosts fauna and flora species, therefore the effort has to be put into limiting deforestation and maintaining the current habitats.

The PH surface site is located in the forest with small clearings and wetland area. Part of the PF surface site shows very high stakes, mainly in the north, due to the presence of bird species, including those of high conservation concern, bat species with their ecological corridor as well as small fauna. Stakes of strong character, with parts of medium and low are located in the southwestern part of the site. Limiting the use of areas with high stakes will be necessary to preserve its biodiversity and ecological value.

Most of the PJ surface site presents low stakes, excluding an inventoried small wetland and hedgerow that are considered as strong stake. The current layout of the site already foresees the space for an ecological corridor and grassy meadows in order to maintain connectivity between natural areas for wildlife and to keep areas for birds to hunt.

The major part of the PL surface site is located on the agricultural land presenting low stakes. Some hedges within the site and in the vicinity of the site play an important role for bird species and must be preserved whenever possible.

To ensure responsible development of the areas, site construction must be carefully planned to minimise habitat disturbance, maintain ecological corridors, and preserve biodiversity. By integrating these considerations into project design plans at the early stage, it is possible to balance infrastructure needs with the preservation of local biodiversity and ecological integrity. 

\subsection{Ecological functionality}

\begin{figure}[h]
  \centering
  \includegraphics[width=0.7\textwidth]{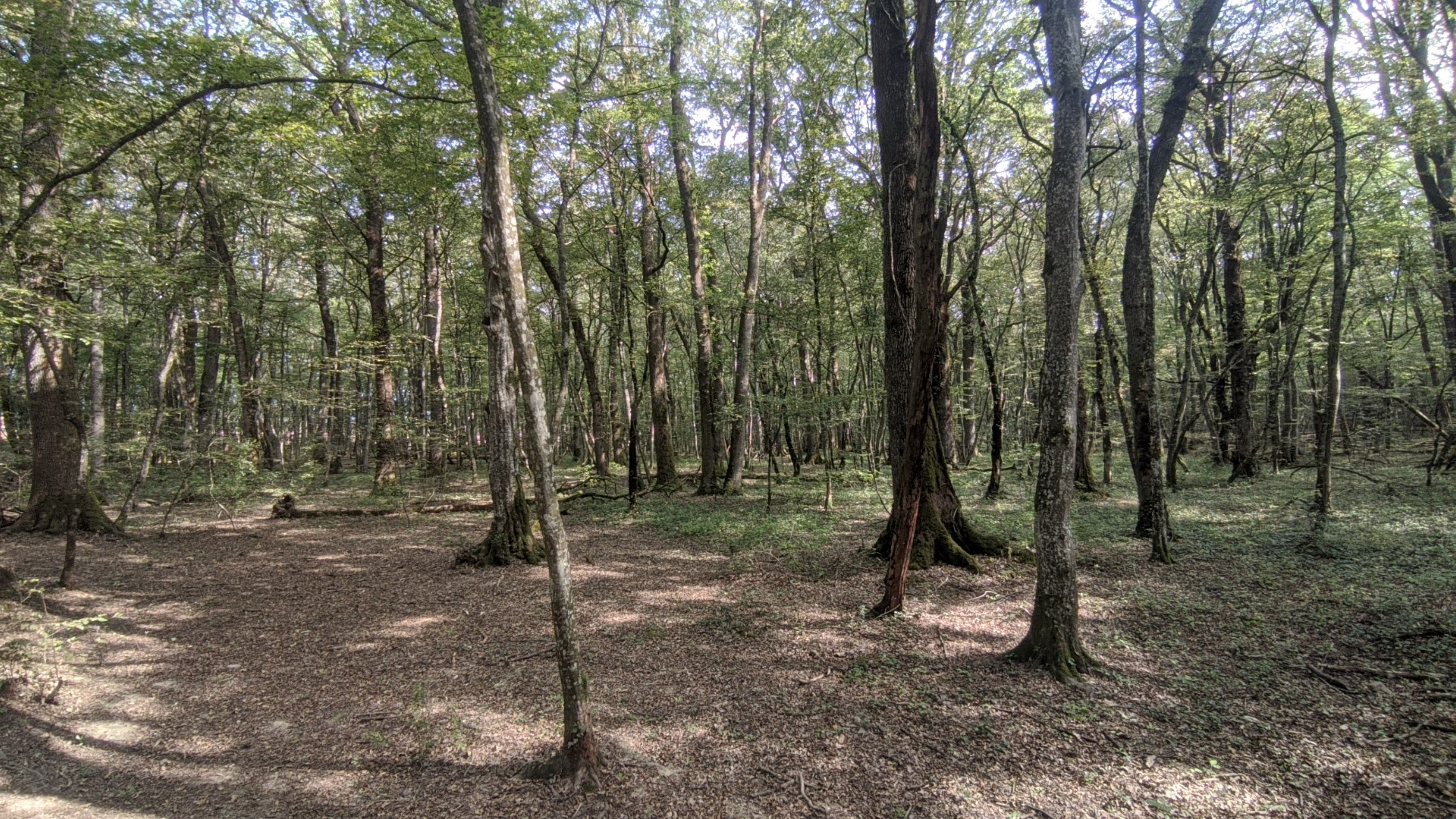}
  \caption{Trees in the woodland photographed during the field investigations.}
\label{fig:Photo_Forest_Mature_SitePA.pdf}
\end{figure}

In France, the ecological network, known as the `Trame verte et bleue' (Green and Blue Framework), is a component of the national and regional planning framework. It integrates terrestrial and aquatic ecological continuity. This network comprises biodiversity reservoirs, ecological corridors, permeable terrestrial and aquatic spaces, and large agricultural areas. These elements collectively support species movement, lifecycle completion, and biodiversity preservation.
In Switzerland, the ecological network identifies essential zones for nature and their connections. The REN's (R\'eseau \'ecologique national) framework includes nodal zones (vital habitats for species lifecycle completion), extension zones (lower-quality or smaller zones), continua (interconnected areas such as forests and wetlands), and development zones (partial habitats). The network also features ecological corridors linking key areas.

The difference in both countries lies in implementation and structure. In France, the ecological network emphasises integrating green and blue corridors at a national scale under regional governance. In contrast, Switzerland’s REN adopts a more localised approach, focusing on specific ecological zones and their physical connectivity. Additionally, the French framework includes significant agricultural spaces and functions of water bodies while the Swiss REN categorises its zones with a greater focus on inter-zonal ecological dynamics.

The site PA and nearby area are permeable agricultural spaces that represent only limited interest in terms of ecological functionality. A regionally important ecological corridor was identified south of the surface site. Wetlands in the vicinity of the surface site contribute to the ecological functionality in the context of the blue network. According to the local urban development plan, three isolated trees located to the east of the main surface site and trees to the south-east are considered a landscape element to be preserved. The layout and design of the surface site constructions will need to ensure that the ecological continuity is maintained.

Site PB is located on large agricultural land with some extensive areas and a continuum of dry grasslands. However, these areas which are conducive to biodiversity are not located within the surface site. There are two locally important movement corridors for large fauna, one of which passes west to the surface site without crossing it and the other through the eastern part of the surface site area, near the hedgerows, which should be taken into account in the design phase. The PB surface site lies in the immediate vicinity of the blue continuum, which follows the hedgerow line to the north-east. However, the location of the site does not directly interrupt the functionality of this network. 

The area of PD surface site is located on the permeable relay agricultural land. South of the PD surface, under RD903, a secondary corridor of medium fauna that is to be considered in the design phase was identified. The surface site does not encroach on any element related to wetlands or water areas and therefore does not affect the functionality of the region's blue network. According to the Nangy local urban plan, no element of biodiversity reservoirs or ecological corridors is present in the study area.

The PF surface site is located within permeable agricultural spaces and a forest ecological corridor connecting the areas located to the north-west and south-west of the site. It is a remote area used by large fauna. The surface site is close to two wetland areas of the blue network, the functionality of which has not yet been analysed. This aspect must be taken into account in the design phase of the surface site. It will also be necessary to ensure the functionality of ecological corridors and large fauna movement routes between the forest areas to the north and south of the site surface, as required by the \'Eteaux commune's local urban plan.

To the north of the PG surface site, across the highway, there are two forested corridors, while to the south there is a linear corridor, also forested. The PG area with its annex is located in permeable spaces, partly on agricultural and forest land, however the PG surface area is not located in any ecological corridor and no biodiversity reservoir has been inventoried. The surface site does not concern any element of the blue network. According to the Groisy local urban plan, the surface site and the possible access road are located on classified forested areas - Espaces Bois\'es Class\'es - which results in preservation or appropriate reforestation in the case of felling, depending on the agreed compensation method, therefore construction in the forested area will be limited to the minimum necessary. 

The PH surface site is concerned only with permeable relay spaces and does not cross any ecological corridors or biodiversity reservoirs. The surface site's proximity to the Tabassé stream, which joins the Usses watercourse, is one of the main elements of the blue network of the area. During the field visits, the wetland was inventoried on the perimeter of the surface site, but it does not seem to represent a major functional role in the blue structure. The local urban development plan of Val des Usses and Cercier also mentions a sector of ecological interest related to the watercourse to the north, which will be taken into account when designing the layout of the infrastructure.

The PJ surface site is located on agricultural land with permeable relay spaces, as well as an ecological corridor connecting forested biodiversity reservoirs between the north and the south, which connects with another ecological corridor in the further part. There are two small streams, west and east of the PJ site, and a wetland has been inventoried on the site itself, but its functionality has not yet been analysed. In accordance with the local urban plans of two municipalities concerned, the surface site is located between wooded area ensuring the ecological continuity, and contains two lines of protected hedges. The preservation of ecological continuity and the ecological corridor will be taken into account during the design phase.

The PL surface site is located on the relay permeable spaces and agricultural spaces. A regional wetland from departmental inventories and a corridor for the movement of large fauna are located to the north of the site, however, these structures do not cross the site. According to the local urban plan of the municipality of Challex, small hedges that are within the surface site perimeter and in close vicinity are considered natural structures and are to be preserved for ecological reasons.

In summary, the PA site includes a regional ecological corridor with wetlands and hedges that need preservation to maintain ecological continuity. The PB site is near two fauna movement corridors and the blue network along the eastern side. Though the site does not directly impact the blue structure, the surface site layout has to take it into consideration. The PD site consists of agricultural spaces with a secondary fauna corridor in the south, near the RD903 but does not encroach on wetlands or biodiversity reservoirs. The PF site is located near two wetlands and within an ecological corridor, which will have to be taken into account in the design phase. Part of the PG site includes a classified forest area, and wood cutting should be minimised. The design of the site will have to ensure that the functionality of nearby ecological corridor and biodiversity reservoir is maintained. The PH site is close to the stream, and a larger watercourse is to be considered during the layout design. It also has some wetlands present but these play a minor functional role. The PJ site lies in a large ecological corridor, including an area of wetlands and hedgerows of ecological value. Lastly, the PL site consists of agricultural land with hedges that need preservation, while a regional wetland and fauna corridor further to the north remains unaffected. Overall, ecological continuity and protected elements must be carefully considered during the design phase to ensure their functionality.

\subsection{Urbanism}

The urbanism aspects comprise all local and regional policies and plans for territorial developments.
All land plots are subject to such plans, and they are regularly reviewed and updated at the municipality level and at the local and
regional public administration levels.

The urban aspects of the perimeters of the surface sites, enlarged and extended zones of several kilometres around the surface sites have been analysed based on the relevant regional and local urban plans in France and in Switzerland. In addition, field visits permitted the publicly available information to be complemented and enriched  with the up-to-date situation. This survey permits all regulatory constraints with respect to the territorial development to be compiled and anticipating the planned evolution of the territory from an environmental point of view, considering all of the applicable laws and regulations. The relevant documents and associated geographical information systems and maps comprise PLU, PLUi, PLUIh, PADD, SCoT in France and PDCn and PDCm of the canton in Geneva in Switzerland.

Table~\ref{table:urbanism_stakes} shows that all extended or enlarged perimeters around the surface sites are subject to strong or other urbanistic and territorial development issues. These issues have to be taken into account during the detailed project plan development.

\begin{table}[!h]
\centering
\caption{Summary of urbanism issues.}
\label{table:urbanism_stakes}
\small{
\begin{tabular}{lll} 
\toprule  
\textbf{Topic} & \textbf{Extended perimeter around site} & \textbf{Description of stakes} \\
\midrule
Strong urbanism stakes & \makecell[l]{PA, PB, PD, PF, PG, PH, PJ, PL} &
\makecell[l]{Nature and agriculture protection zones,\\humid zones.} \\ \midrule
Other urbanism stakes & \makecell[l]{PA, PB, PD, PF, PG, PJ, PL} &
\makecell[l]{Protected or valuable architecture in the\\vicinity, sport facilities, agriculture,\\roads, highways.} \\ \midrule 
Public utility servitude & \makecell[l]{PB, PF, PH} &
\makecell[l]{Pipelines, electricity lines and protection\\buffers around those infrastructures.} \\ \midrule
\makecell[l]{Urban environment and\\applicable regulations} &
\makecell[l]{PA, PB, PD, PG, PF, PJ, PL} &
\makecell[l]{Co-visibility, requirement for integration\\in the urban context, topographic constraints.} \\ \bottomrule
\end{tabular}
}
\end{table}

Table~\ref{table:urbanism_sites} shows that half of the sites are directly affected by urban constraints that need to be considered during the site design development and that require particular attention during the project authorisation process.

\begin{table}[!h]
\centering
\caption{Summary of urbanism topics by site.}
\label{table:urbanism_sites}
\begin{tabular}{ccl} \toprule
\textbf{Site} & \textbf{Stakes} & \textbf{Urbanism topics} \\
\midrule
PA & High & Protected agriculture zone, gas pipeline at the site border \\ 
PB & High & Protected agriculture zone (SDA), ecological corridor, landscape integration \\ 
PD & Low & Agricultural space, road development project \\ 
PF & Low & Agricultural space \\ 
PG & Low & Agricultural space, nature zone (forest) \\ 
PH & High & Agricultural space, nature zone (forest), pipeline at the northern site limit \\ 
PJ & High & Agricultural zone, wetland zone, protected agricultural space, ecological corridor \\ 
PL & Medium & Protected agricultural zone, nature protection zone \\ \bottomrule
\end{tabular}
\end{table}

\subsection{Mobility}

Data about the public transport infrastructures and road traffic as well as soft mobility and multi-modal mobility within the perimeter of the project have been collected and analysed.

\subsubsection{Public transport}
 The PA Site is well served by the Geneva public transport system (TPG) which operates across the Swiss/French border and serves the entire zone. The bus connection is direct, and there is a tramway connection within a reasonable distance. Site PB in Switzerland is also served by the TPG network with a bus station in the immediate vicinity and a tramway connection at a reasonable distance. The PD Site in France is well served by regional French bus lines, including a park+ride facility connecting to Geneva. The presence of the large hospital (CHAL) ensures that the connections are maintained and potentially further developed. The location of site PF is poorly served by public transport. However, 2\,km away, La Roche-sur-Foron train station is an important multi-modal transport pole including a connection to Geneva via the L\'eman Express. The PG site is not served by public transport. However, the Groisy train station is 2\,km away and provides regular connections to Annecy and Geneva via the L\'eman Express line. The PH site is not served by public transport and no public transport exists in the vicinity. Site PJ is not served by public transport, but 2\,km away, a Swiss TPG bus line connects to Vulbens and in Valleiry, there is a train station on the line between Bellegarde and Evian via Annemasse. Although there are Swiss TPG bus stops in Challex in the vicinity of site PL, the frequency is modest. Better connections exist in nearby Dardagny in Switzerland. At a distance of 4\,km in La Pleine, there are very good connections to bus and train lines.

Summing up, the experiment sites PA and PD are well served by public transport and the experiment sites PG and PJ are reasonably connected at some distance. The technical sites PB, PF, PH and PL are not well served by public transport. An analysis of the demand from these sites would be required if further development of the public transport system is considered in relation to the FCC project. It could be in the mutual interest of the project and the local stakeholders to develop public transport around the PG and PJ experiment sites at least. 

\subsubsection{Road network}

\begin{figure}
    \centering
    \includegraphics[width=\textwidth]{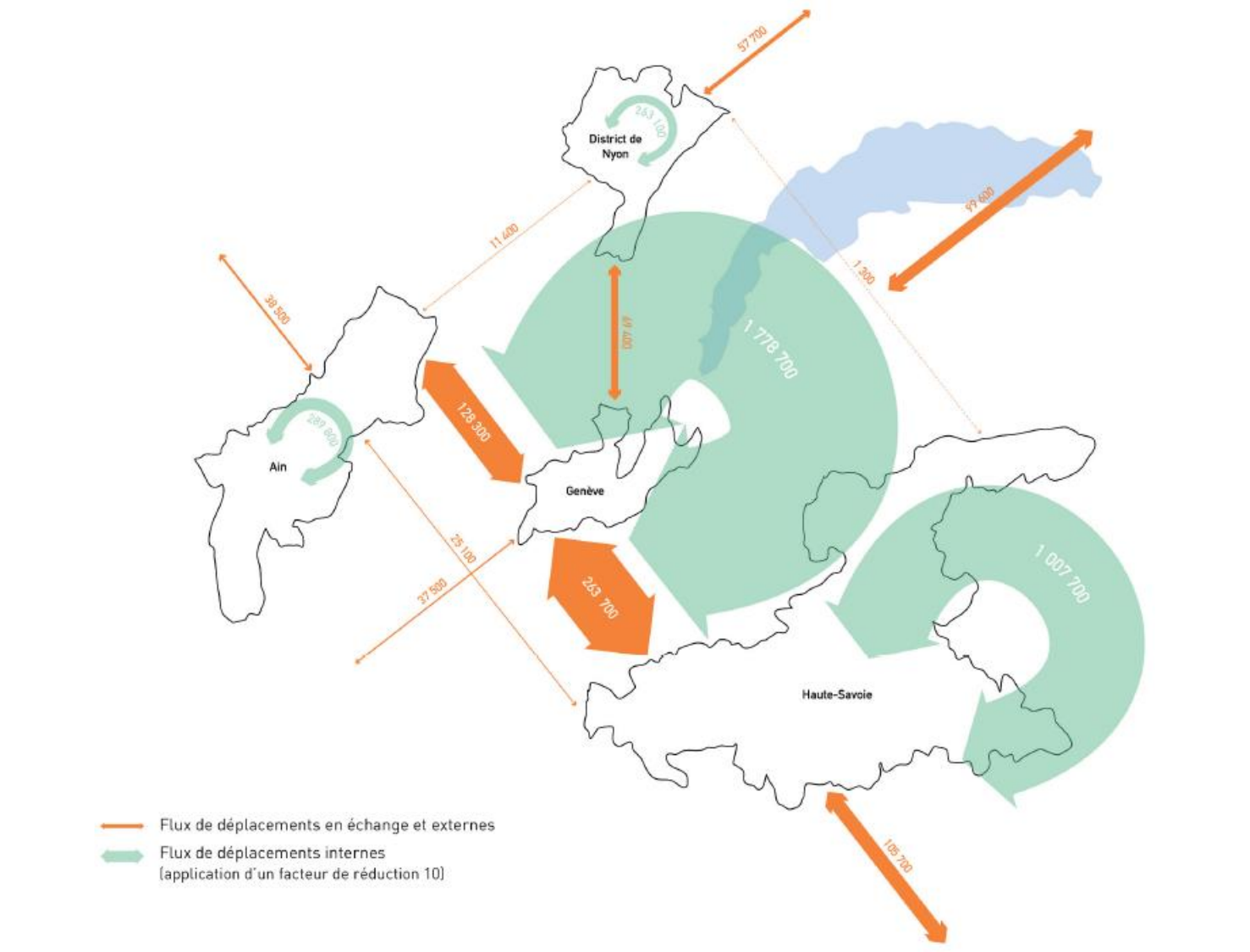}
    \caption {Daily travel flows (all modes combined) in Grand Geneve (Source Grand-Genève \cite{GrandGenevePA4}, based on MRMT – EDGT 2015 - 2016)
    Although Geneva remains a central hub, the analysis of flows reveals a more multipolar organisation, with a majority of internal travel within the large territories (Geneva, French Geneva, Nyon district).}
    {\label{fig:mobility_allreasons}}
\end{figure}

The experiment sites PA, PD, PG and PJ are in the immediate vicinity of major roads and would profit from direct access to the autoroute infrastructure for construction and installation purposes. This would eliminate any potential residual local traffic challenges around the sites during these phases. The PA site is also well-connected to the CERN Pr\'evessin and Meyrin sites via a major departmental road. In addition, the PL technical site is in the vicinity of a major departmental road and is also well-connected to the existing CERN sites. The PB Site is directly on a good road, but the traffic situation through Geneva and to nearby Annemasse in France is challenging. Technical site PF is also well-connected via a major departmental road. Although site PH is directly located on a departmental road, it is isolated and is distant from major transport routes. The closest autoroute access is at a distance of 10\,km in Allonzier-la-Caille. Installation of bulky equipment at this site needs to be carefully studied, developed and planned.

\subsubsection{Other transport modes}
The study also included the analysis of dedicated bicycle tracks in France and Switzerland. Dedicated lanes and tracks have recently been constructed in the vicinity of site PA and these are expected to be further developed. These span the Franco-Swiss border. A dedicated bicycle track is also being constructed directly at the PB site. Sites PD and PF are not equipped for soft mobility. No dedicated bicycle lanes exist around site PG, but the road to Groisy is well-adapted for cyclists. Site PH has no soft mobility infrastructures. The surroundings of site PJ are being developed with a view to strengthening soft mobility, aiming to link the nearby municipalities Vulbens and Valleiry. The creation of access to site PJ also permits the connection of Dingy-en-Vuache to this system. PL is not particularly equipped for soft mobility. Although walking and biking are easy in the commune, there are no dedicated links to other municipalities in France and Switzerland.

\subsubsection{Foreseeable evolution}

The Grand Gen\`eve area is developing a multi-modal transport plan that aims to improve further the transport infrastructure across the Franco/Swiss border. The continued housing development and demographic evolution of about +1.2\% per year in the neighbouring French departments of Ain and Haute-Savoie, which is unrelated to CERN's activities, calls for such developments. There are about 1.2\,million journeys today within the perimeter of the project, but about 4.2\,million per day are expected by 2040. As will be described later, the activities relating to the construction, installation, and operation of the project are insignificant compared to the existing and future mobility in the region.

\begin{figure}
    \centering
    \includegraphics[width=0.7\textwidth]{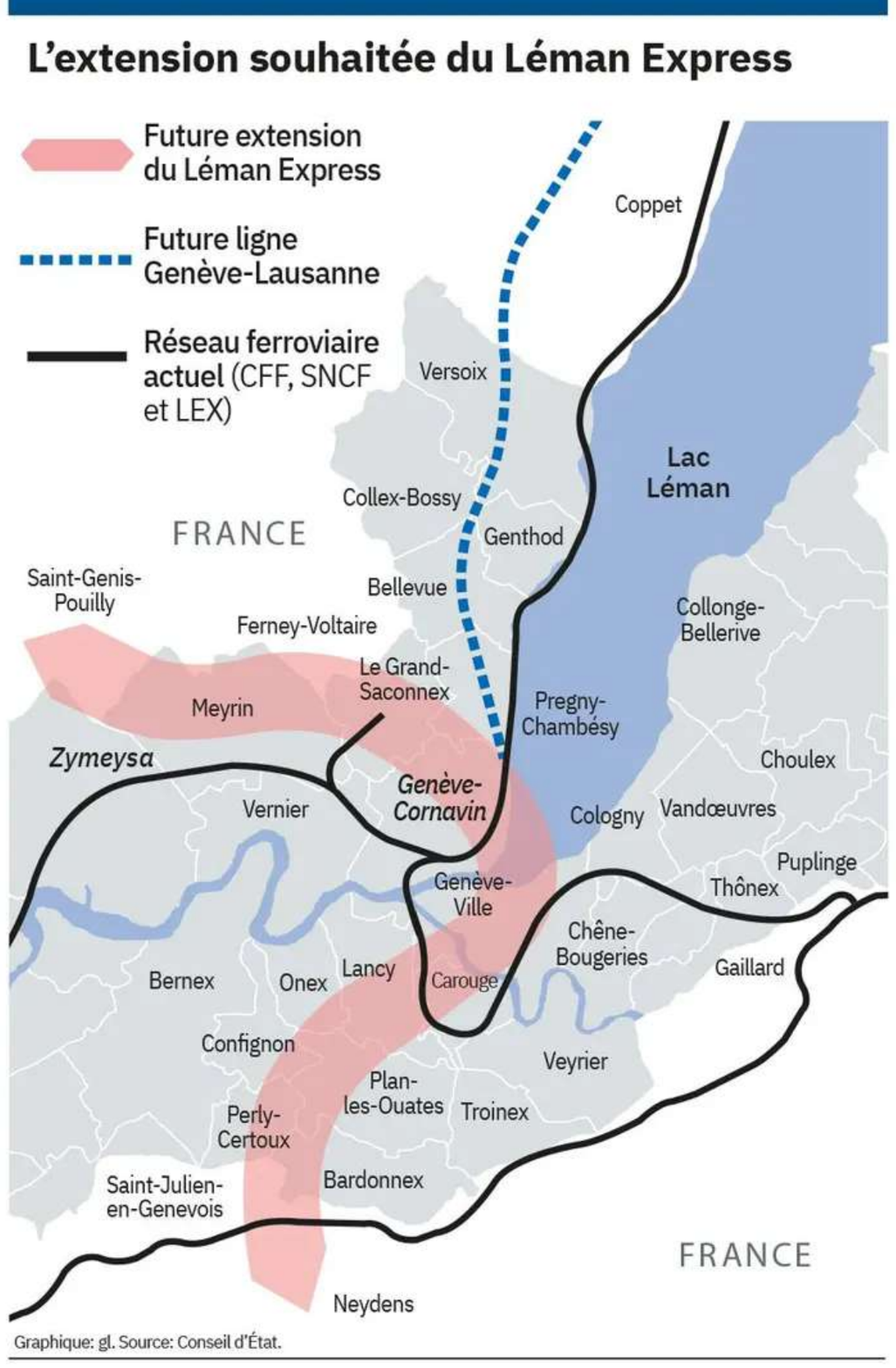}
    \caption{\label{fig:LemanExpress} Current and potential future extension of the Leman Express}
\end{figure}

The extension of the Swiss railway system continues to increase its daily train capacity, mainly due to the L\'eman Express lines (see Fig.~\ref{fig:LemanExpress}). This also includes an improvement of the services to the Arve valley (La Roche-sur-Forton), Groisy and Annecy for completion in 2030. In the longer term, developments are planned for increasing the service to La Plaine and beyond to Bellegarde. Autoroute extension projects in Switzerland have been planned, but were recently put on hold. A project to connect both sides of Lake Geneva sides by a tunnel under the lake has been studied, but any potential implementation before 2050 is unlikely and therefore the project is not included in specific plans. In France, an autoroute is planned to connect the A40 (Arve valley) to the A412(Th\^onon) via a wide departmental road (RD903) connecting to the A40 in Nangy. Bus line developments between major agglomerations and the hospital next to the PD site are likely.

\subsubsection{Project induced traffic}

Project induced traffic refers to the following types:
\begin{enumerate}
\item Workers commuting to and from the construction sites during an approximately ten-year-long construction phase.
\item Evacuation of excavated materials from construction sites. While all sites require the evacuation of materials during the first two years, only four sites will see relevant transport of excavated materials during another 6 to 8 years due to the deployment of tunnel boring machines.
\item Transport construction materials to the construction sites. All sites will see a moderate inflow of construction materials during the first two years, but only four sites will continue to have relevant construction materials inflow during the period when the tunnel boring machines are deployed. Construction materials also need to be brought in when the surface site buildings are constructed.
\item Transport of accelerator equipment during the six to eight-year installation phase that overlaps partially with the civil construction activities.
\item Transport of the experiment detector equipment during experiment installation that overlaps with accelerator installation and testing phases.
\item Commute of engineers and scientists during the installation and testing phases.
\item Commute of engineers and scientists during the operation phase.
\item Commute of engineers and scientists during regular maintenance periods and in the frame of repair activities.
\item Traffic induced by visitors to experiment sites during the operation and shutdown phases.
\end{enumerate}

The main project-induced traffic is linked to the evacuation of the excavated materials. It will be confined to major transport routes. The inflow of construction materials represents a minor additional contribution to this traffic. Commute of workers is intended to be organised centrally, as is best practice for construction sites. The installation of equipment for the particle accelerator represents a minimal contribution to the traffic. The same is true for the transport of the experiment detector equipment. For this case, some isolated, exceptional loads may be required due to the size of pieces manufactured off-site. Once operational, only a few scientists and engineers commute daily to experiment sites. Technical sites are predominantly operated remotely and will see very little traffic for maintenance and repair. The traffic of an estimated total number of 25\,000 visitors per year to an experiment site during the operation phase can be compared to the traffic induced by a typical museum or archaeological site such as the Grottes de Cerdon in the region (about 50\,000 visitors per year\footnote{\url{https://www.ain.cci.fr/sites/g/files/mwbcuj1466/files/2024-02/Chiffres\%20cles\%202024\%20AIN_.pdf}})
or the Ch\^ateau Voltaire in Ferney-Voltare (about 50\,000 visitors per year \cite{lemanbleu_voltaire}\footnote{\url{https://www.lemanbleu.ch/fr/Actualite/Archives/Le-chateau-de-Voltaire-entierement-renove.html}}
). This traffic of individual visitors can be managed with appropriate directions to the site and through support by public transport. Visitors in groups arrive in buses, presenting a minimal amount of additional traffic.

More information about the quantities and the additional traffic induced can be found in sections Section~\ref{sec:road_access} and Section~\ref{sec:transport_and_mobility}. A first quantitative traffic analysis has been carried out to confirm that the traffic is manageable and that with respect to the nearby major transport routes, it represents only a minor addition.

\subsection{Human activities}

\paragraph{Human activities}
The analysis of the environmental state established the situation concerning human activities around the surface sites. The environment around site PA in Ferney-Voltaire is dominated by commercial activities, the LHC Pt8 surface sites and the airport. The presence of a surface site does not add to these activities. The environment around PB in Presinge is rural and characterised by small hamlets. The Geneva Landscape and Architecture School (HEPIA) is an academic activity zone in the vicinity. Towards the west, the zone starts to be dominated by urban activities. The environment of site PD in Nangy is characterised by a mix of agricultural, commercial and industrial activities, residential zones and a major hospital (CHAL). Due to this mixed environment and the major transport routes available, the location is advantageous for an experiment site with a permanent presence of scientists and engineers. PF in \'Eteaux is located on a main transport route with dispersed residential areas, small businesses and artisans and a major public works company opposite the site. PG in Groisy and Charvonnex does not have major human activities. The vicinity of Groisy, with schools and potential for local development, is an opportunity to develop a main experiment site with the presence of scientists and engineers. Site PH in Cercier is very rural, with agricultural activities (fruit production) and some small hamlets. No major human activities are carried out in this area. The environment around the PJ site in Dingy-en-Vuache and Vulbens is agricultural with commercial activities to the north-east in Valleiry. PL in Challex is in an agricultural zone with low-density residential areas. Commercial activities take place at a distance in La Plaine, Switzerland, which is not directly linked to the site.

\begin{figure}[h]
  \centering
  \includegraphics[width=0.7\textwidth]{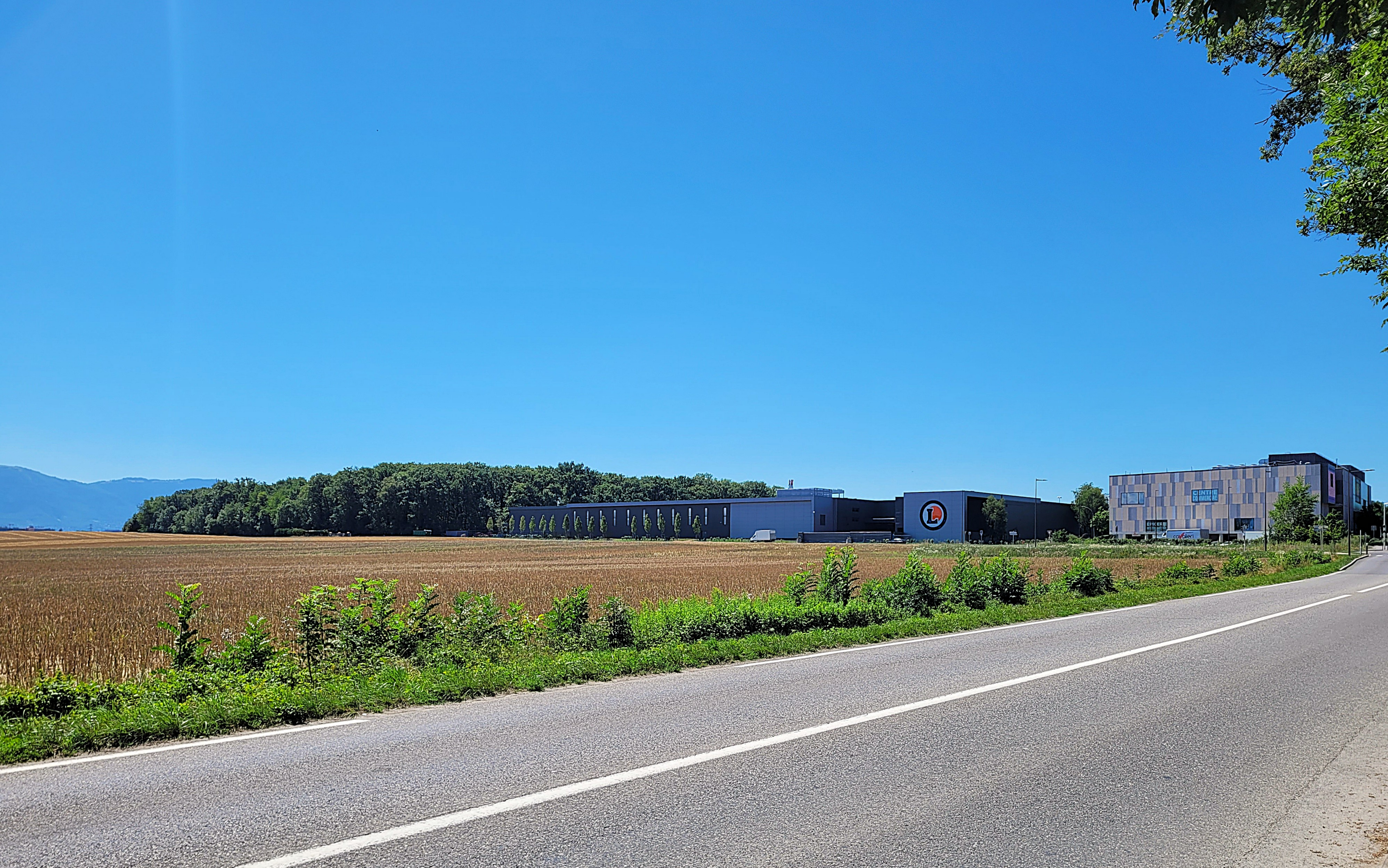}
  \caption{Commercial district near the PA site photographed during the field investigations.}
\label{fig:Photo_Activite_Commerce_PA.pdf}
\end{figure}

\paragraph{Poplulation}
The population density around the eight surface sites varies significantly. Around PA, the area is moderately populated with about 34\,000 inhabitants within a perimeter of 2\,km. The major communes are Ferney-Voltaire, Pr\'evessin-Moens in France, Meyrin and Grand-Saconnex in Switzerland. The population growth in this area is 2.4\% per year but this is unrelated to CERN's activities. Around PB the environment is sparsely populated, at most 6000 inhabitants can be counted in a perimeter of 2\,km. The main communes are Presinge, Pupling, Choulex, Meinier, Jussy in Switzerland and Ville-La-Grand in France. There is no annual population growth in Switzerland (between -0.3 and +0.3\%). In neighbouring France, the annual growth is modest (+1.1\%), comparable to the average in Haute-Savoie. Site PD is located in a moderately dense environment with about 4000 inhabitants in a perimeter of 2\,km, the majority located in Nangy and Contamine-sur-Arve and Fillinges. The annual population growths in these communes are highly diverse, ranging from negative 0.5\% in Nangy to +3\% in Contamine-sur-Arve. The PF site is located in a moderately populated zone with about 4000 inhabitants in a 2\,km radius with the majority of people in \'Eteaux and La Roche-sur-Foron. The population evolution is stable, ranging between -0.6\% and +1.0\% per year. PG is located in a sparsely populated area with around 4000 inhabitants in a perimeter of 2\,km, mainly in Charvonnex and Groisy. However, the two communes see an annual population growth of between 2.2 and 3.2\%. The environment around the PH site is very weakly populated. There are only 1500 inhabitants in a perimeter of 2\,km in Cercier, Choisy and Marlioz. The population growth is around 1.5\% per year. The area around PJ is also sparsely populated with about 6000 inhabitants in a perimeter of 2\,km in Vulbens and Dingy-en-Vuache. Valleiry is a more densely populated commune nearby. The annual population growth is 2 to 3.5\%, higher than the average in Haute-Savoie. PL in Challex is in a sparsely populated area with only about 3000 inhabitants in a perimeter of 2\,km including Challex in France and Dardagny in Switzerland. Both communes experience an annual population growth of about 3\% per year.

\paragraph{Housing}

\begin{figure}[h]
  \centering
  \includegraphics[width=0.7\textwidth]{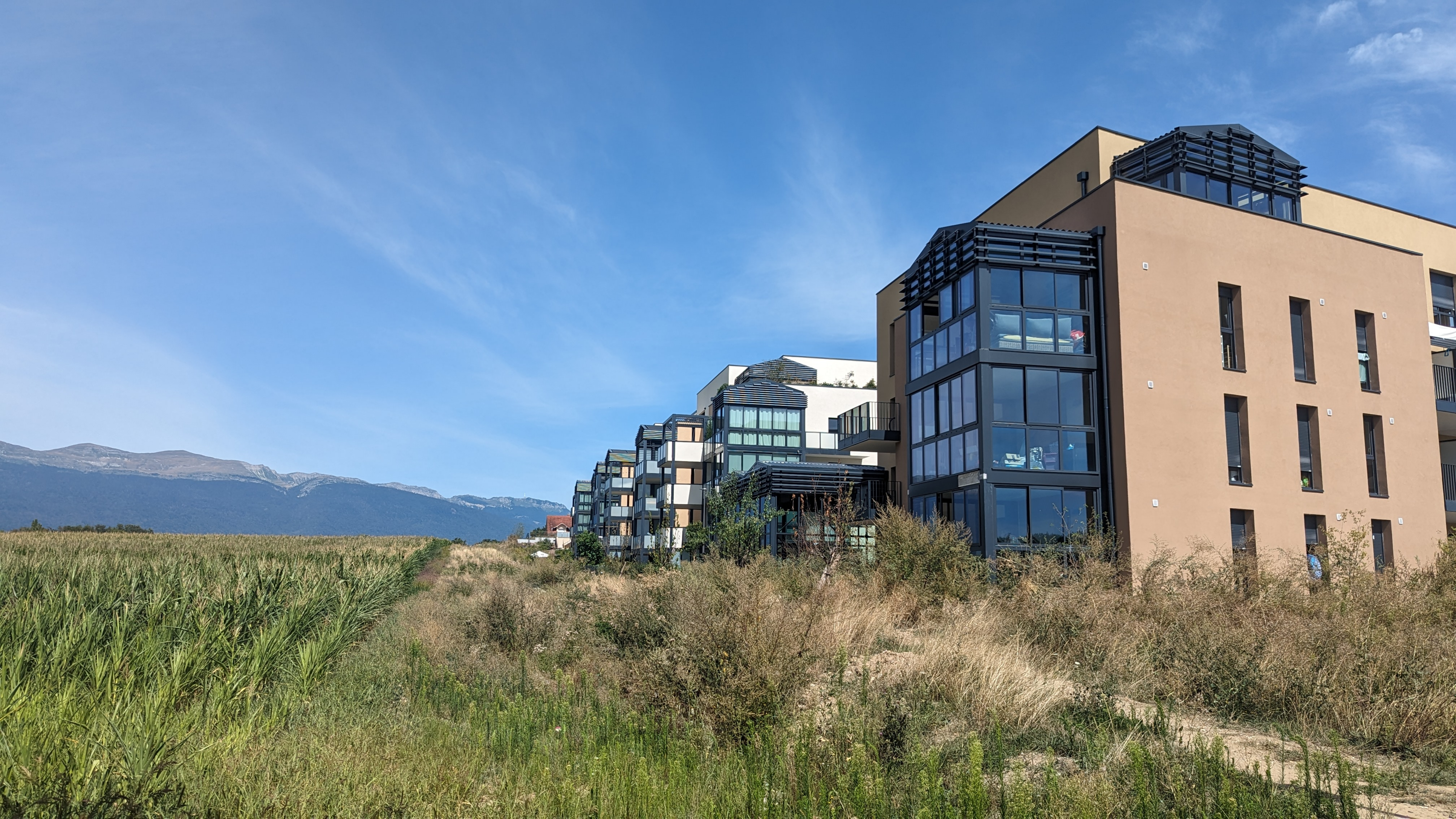}
  \caption{Residential district photographed during the field investigations.}
\label{fig:Photo_Urbanism_Housing_GardenParl_SitePA.pdf}
\end{figure}

As with the population, the housing situation also varies significantly amongst the surface site areas. In 
the immediate environment around PA, the housing sector is dominated by apartment buildings. Pr\'evessin is dominated by individual houses. The housing sector around PB is mixed with individual houses in the closer vicinity and residential buildings further away. The communes in the vicinity of the site are not dominated by individual houses, but see a mix of houses and residential buildings. The area around PD is dominated by individual houses. Individual houses are also predominantly found around site PF. The entire area around PG is dominated by individual houses, as is the case with sites PH and PJ. Only Valleiry is an exception with a significant number of residential buildings and apartments. The surroundings of site PL have a highly contrasting housing sector. Mainly individual houses are found in the French communes of Challex, P\'eron, Saint-Jean-de-Gonville. Residential buildings are predominant in the Swiss communes of Avully and Dardagny.

\begin{figure}[!h]
    \centering
    \includegraphics[width=0.8\textwidth]{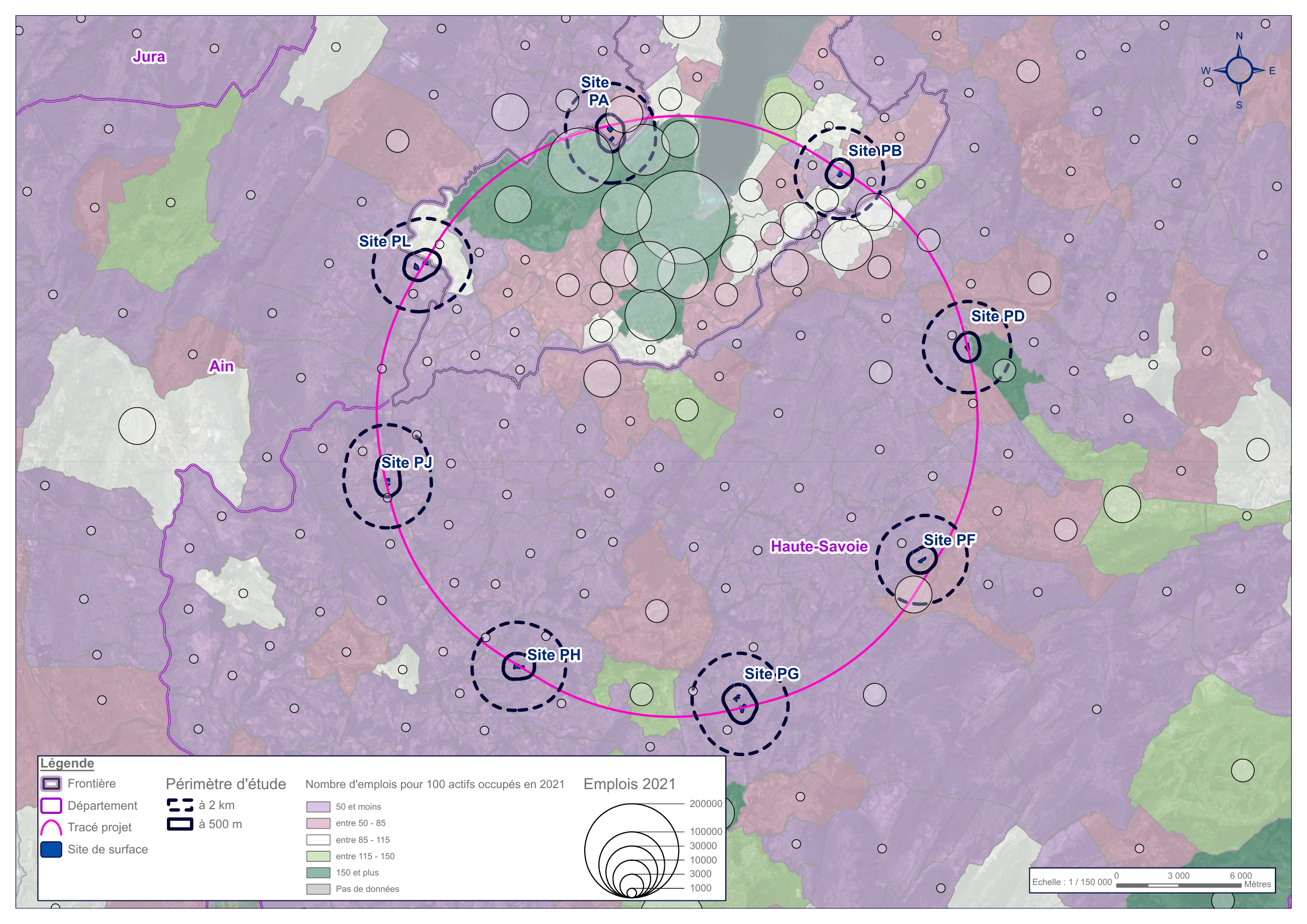}
    \caption{\label{fig:employement} Overview of the employment opportunities in the perimeter of the FCC reference scenario.}
\end{figure}

\paragraph{Employment}
At a general level, the employment sector in the northern sectors of the FCC is dominated by Geneva and its surroundings on French territory (see also Fig.~\ref{fig:employement}). The southern zone in Haute-Savoie is characterised by significantly fewer job opportunities and an overcapacity of available workforce. Depending on the region, the tertiary sector is only weakly developed. The secondary sector (industrial activities) offers opportunities in selected locations. Like Geneva, the Department of Ain in France has many more employment opportunities than the Haute-Savoie department.

Around PA, employment opportunities are mainly found in the tertiary sector. In the Swiss territory, the secondary (industrial) sector is also a major source of employment opportunities. Employment around PB is dominated by the primary sector (agriculture) and there is a slightly smaller number of job opportunities than people seeking work. PD has employment opportunities in the industrial sector and the health sector (CHAL hospital), although the job opportunities are limited. The area around PF offers few employment opportunities and those that exist are distributed between the secondary and tertiary sectors. The primary sector (agriculture) is rather weak in this area. There are fewer job offers than the available workforce. Also around PG there are fewer job opportunities than available workforce. In particular,r the tertiary sector is only weakly developed. The same is true for site PH in which the primary sector (agriculture) is the main employer for almost one-third of the population. Also the area around PJ sees more job seekers than job offers. The main employment sector is agriculture. Around PL the situation is mixed: residents find their employment in France and in Switzerland.

\paragraph{Agriculture}

\begin{figure}[h]
  \centering
  \includegraphics[width=0.7\textwidth]{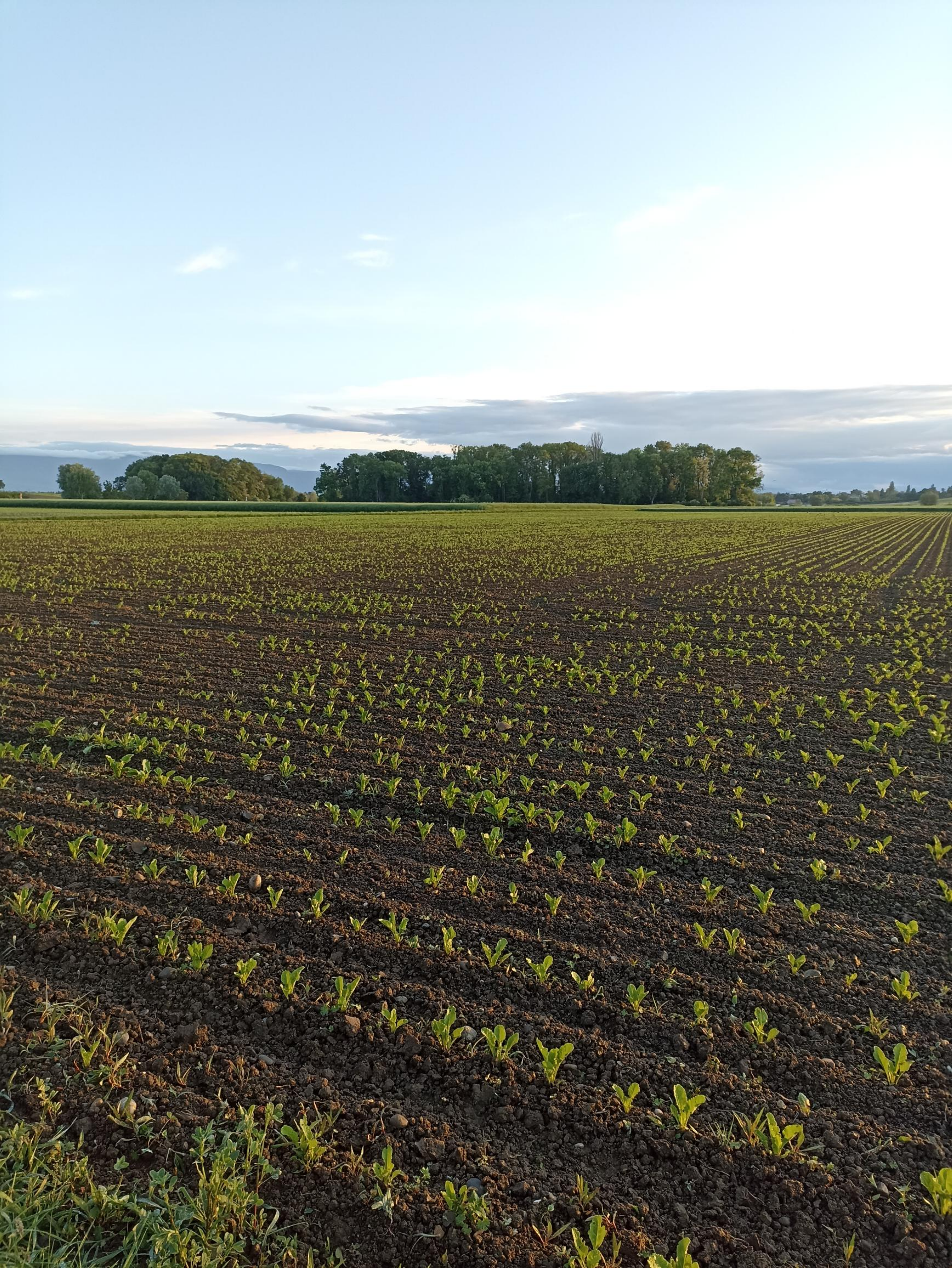}
  \caption{Agricultural land photographed during the field investigations.}
\label{fig:Photo_Photo_Agriculture_SitePB.pdf}
\end{figure}

The area used for agriculture in the Pays de Gex between the site PL in Challex and PA in Ferney-Voltaire has been stable over the last 30 years. The evolution of plot ownership and agricultural exploitations in the 1990s led to fewer but larger farms. The typical size of a single exploitation today is about 90 ha per farm, covering a total of about 3600\,ha. The main product remains milk, although this sector is decreasing. Cereals represent a minor contribution. Free trade zones between France and Switzerland in this sector facilitate goods exchange, from which the sector for milk and milk-derived products profits mainly. Out of the 23\,million litres produced in the Pays de Gex, 14\, million are sold to cooperatives in Geneva. The volume of the milk production in the area is larger than the entire volume in Switzerland. The cereals produced by about 90 producers are almost entirely sold in Switzerland. The production of cereals saw a constant increase with a growth of more than 60\% between 2000 and 2010. 

Milk production is also an important agricultural sector in Haute-Savoie. Major, globally acting firms and groups (e.g., Soci\'et\'e Laiti\`ere des Hauts de Savoie / Lactalis) are important economic players in the region who also produce milk derived products, in particular Reblochon and Abondance cheese. Green spaces for feeding the cattle are important for this industry. Growing urbanisation puts this branch of the economy under pressure, although the economic potential is high. Production and sales of hay as cattle food is a related main economic branch. 

The agriculture sector around PB in Switzerland is dominated by the production of cereals. Recently, following the construction of glass houses, different types of agriculture are fostering the local production of various products. Local agriculture is one of the economic pillars of the canton, although the sizes of the individual operators are much smaller than in France (typically less than 50\,ha). Vineyards are another relevant part of the cantonal agricultural sector. Land consumption creates pressure on the agricultural sector: about 20\,ha of agricultural space is lost every year. 

The sector around PD in Haute-Savoie experiences different climatic conditions than the northern sector. The countryside is more mountainous and the agricultural activities are more diversified, although the main product is hay and grass related to milk production. Almost 5000\,ha are exploited in this area for milk and cheese production. 17\,million litres of milk are processed by a firm in Fillinges alone. 

Moving further south (sites PF and PG) shows that the importance of agricultural activities has decreased over recent decades and small farms dominate. The activities are diverse, including cereals, cattle farming, meat production, hay production, vegetables, poultry and some local wine production. Dairy production remains the dominant branch.

Towards the south-east (PH) the activities are still milk and cheese production with a total area of more than 12\,000\,ha devoted to it. Other important activities are fruit (apples and pears) and cereals (including maize) production.  The entire zone processes per year about 45\,million litres of milk. The eastern sector (PD) processes about 12\,million litres of milk yearly and also sees dairy product industries (e.g., Baiko). Cattle farming for meat production is a related activity in the sector.

\paragraph{Forestry}

Forestry is mainly important in the French Haute-Savoie department. About 150\,000\,ha are utilised in the area. 70\% of the forests are privately owned. Wood production comprises a variety of different tree types, both deciduous and conifer trees. Forestry in the French department of Ain is also well-developed, mainly supplying wood for carpentry and construction works.

The project scenario only affects forestry in a very limited way in PG (Groisy) and PH (Cercier and Marlioz) in the Haute-Savoie department in France. The economic loss has been quantified, and the potential effects on habitat and biodiversity have been analysed. Mitigation measures can be developed in a subsequent design phase based on this analysis.

\paragraph{Viticulture}

\begin{figure}[h]
  \centering
  \includegraphics[width=0.7\textwidth]{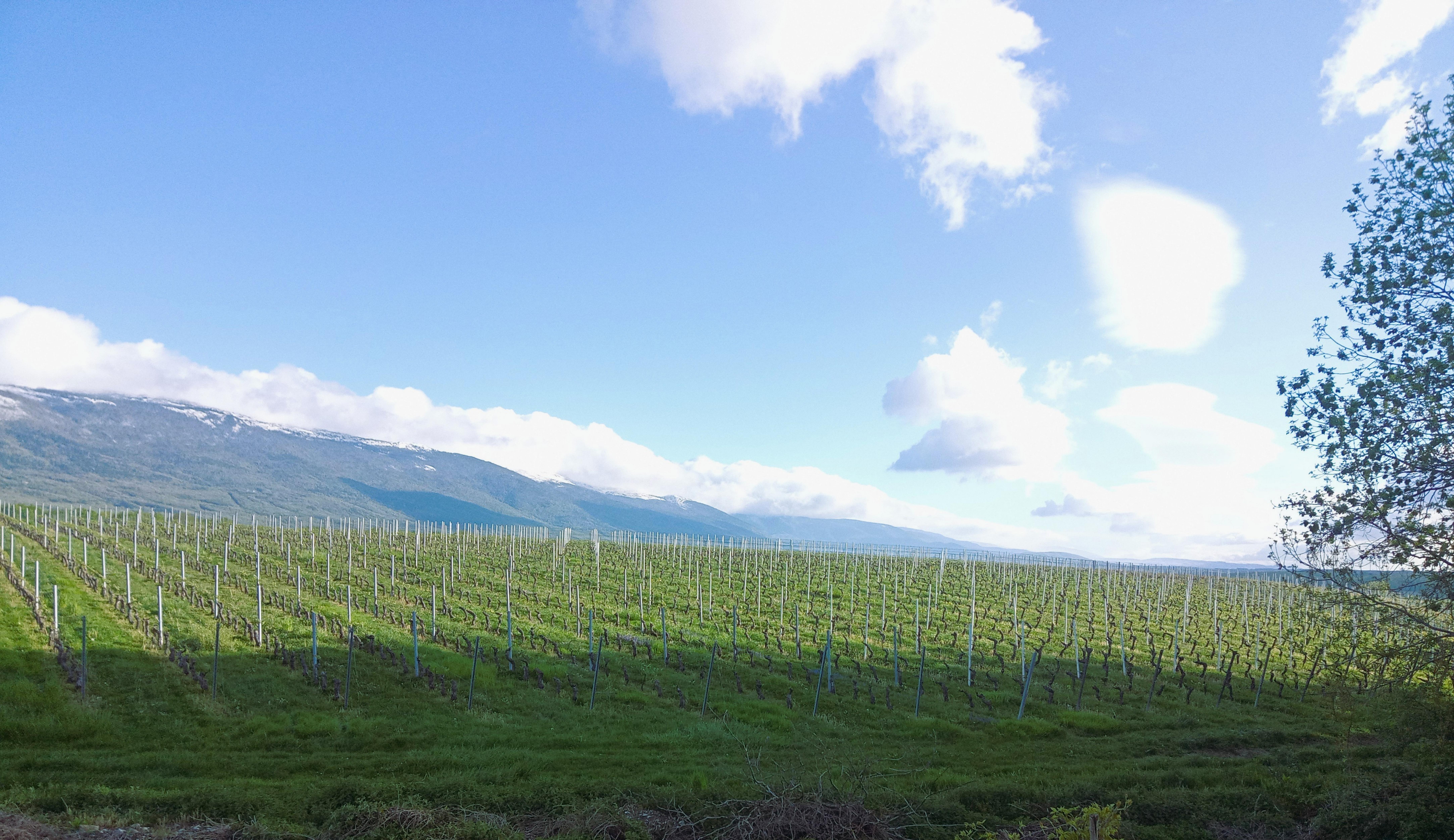}
  \caption{Vineyards photographed during the field investigations.}
\label{fig:Photo_Activite_Vineyards_SitePL.pdf}
\end{figure}

Vineyards are present in the Ain and Haute-Savoie departments in France, as well as in the canton of Geneva. However, the implementation scenario does not affect any of them. Wine production exists in the vicinity of the PL site in Challex, on both the French and Swiss sides. Wine production also exists at some distance from the surface site PB in Switzerland.

\paragraph{Tourism}

Tourism is already a major economic factor in Geneva and the immediate vicinity of CERN because of its Science Gateway visitor centre and the numerous exhibitions and guided tours that CERN offers. Every year more than 400\,000 people visit CERN either individually or as part of groups. The economic impact due to local spending that creates indirect, direct and induced jobs is significant, since people typically combine their visit with other activities in the region, on average for a stay of four days. For instance Ferney-Voltaire (PA site) is known for the Ch\^ateau Voltiare, its typical weekly market and numerous tourist attractions in the immediate vicinity  (e.g., the Jura mountains in France and Coppet and Nyon in Switzerland). Geneva and its numerous tourist attractions are not exhaustively listed here. However, the United Nations, the International Red Cross Organisation, the old town, the Jet d'Eau, the watch museums, the cathedral are some examples.
The economic benefits in this area today have been recorded and estimations for the effects to be expected with a future collider projects have been estimated and are reported in the section on socio-economic impact (see Section~{sustainability:results}). With the increase of CERN's activities in the region, high-quality science tourism is expected to expand into the Haute-Savoie region, contributing to the tourism development there as well.

Concerning the new sites, PB in Presinge features a number of small, but relevant, tourist attractions that invite pedestrian and bicycle tourists, extending their excursions through the vineyards to the lake or towards France to the Arve valley.

The area around the PD site in Nangy in France is mostly known for its bicycle tracks that will be significantly extended by 2030. A track from the lake to Mont-Blanc is planned.

The surroundings of the PF site in \'Eteaux do not yet feature dedicated tourist attractions. However, several opportunities exist for discovering local cheese products between sites PD and PF. Together with a visitor centre at PG in Charvonnex and Groisy the entire zone could profit from a well-planned development of high-quality tourism that links the lake through the Arve valley with the Annecy area, which is highly developed in terms of tourism. Numerous mountain walking tours can be included in this programme.

The area around site PH in Cercier and Marlioz is not currently developed touristically and does not have noteworthy infrastructures.

The zone close to the Vuache next to site PD in Dingy-en-Vuache and Vulbens, on the contrary features, numerous bicycle paths (including ViaRh\^ona) that are inviting for tourists interested in nature.

The site PL in Challex is embedded in a regional nature and bicycle tourism area, linking the Jura mountains with the Rh\^one valley. It features hiking paths and trails through the vineyards. The area around the Swiss commune Dardagny is known as one of the most beautiful in Switzerland and is therefore particularly protected along with the landscape that surrounds it.

\paragraph{Economic development projects}

Development projects potentially relevant for the FCC scenario are included in the Geneva cantonal master plan (Plan Directeur Cantonal, PDCn) and concern the period to 2030. A project for new apartments on about 180\,ha in the vicinity of the airport, involving major construction works and the employment of about 5700 workers,  is potentially relevant for site PA in France. Also, about 1300 apartments are planned to be constructed in nearby Grand-Saconnex, involving up to 2400 workers. No development projects are registered in the vicinity of sites PL in Challex and PB in Presinge on the Swiss side.

In France, the potential main developments are in the vicinity of site PA in Ferney-Voltaire. A future commercial activity zone (ZAC) on about 65\,ha and the creation of an additional 2500 apartments and educational facilities are planned for the period up to 2030.

In the vicinity of site PD in Nangy a new road widening project (RD903) and the intersection with the A40 autoroute are scheduled to be implemented before 2030.

\paragraph{Scientific activities}

The main scientific project in the region is the upgrade of the Large Hadron Collider, known as the High-Luminosity Large Hadron Collider (HL-LHC). It guarantees the continuation of the presence of scientists and engineers at today's level until the 2040s. The FCC implementation scenario builds on this activity, leveraging the existing LHC Pt8 for the creation of surface site PA and hosting the injector on the CERN Pr\'evessin site. Also site PL in Challex profits from being in the vicinity of CERN Meyrin.

For site at PB, the presence of the Geneva Architecture and Landscape School (HEPIA) in Lullier may be relevant. This facility is close to the site and potentially provides services that the surface site can leverage to reduce its own requirements. HEPIA could also potentially profit from common developments in different areas around the FCC, including agricultural studies and technical infrastructures. A collaboration already exists between the study, CERN and HEPIA to work on the re-use of excavated materials and agricultural aspects. 

The national milk and meat industry school can be found in the vicinity of the PF site in \'Eteaux in La Roche-sur-Foron. The infrastructure also hosts a technical and scientific high-school. The school is active in scientific partnerships with INRAE, CNRS, universities and national research centres. Potential synergies can, for example, be found in the development of water and heat re-use.

\subsection{Heritage}

\begin{figure}[h]
  \centering
  \includegraphics[width=0.8\textwidth]{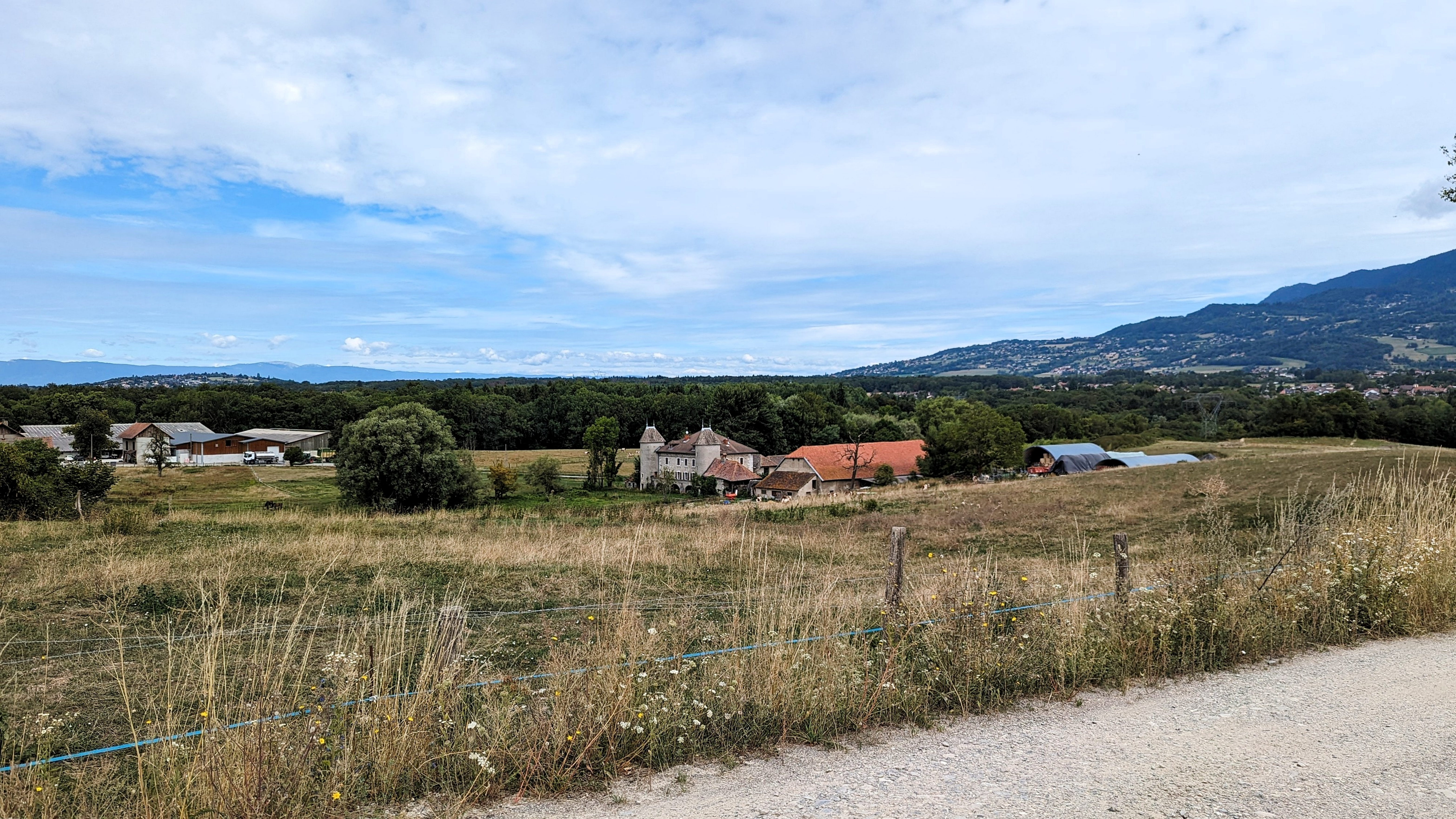}
  \caption{Architectural heritage photographed during the field investigations.}
\label{fig:Photo_Patrimony_Nangy.pdf}
\end{figure}

The analyses carried out in the frame of this study related to cultural, architectural and archaeological heritage first took into account all noteworthy elements that could be identified in a perimeter of 500\,m around the surface sites. A subsequent, more detailed assessment focused on the elements in the immediate vicinity of the surface site candidate locations. This choice of perimeter was established in order to understand the richness of each surface site which, subsequently, would need to be considered in the designs and territorial integration of the sites.

Bibliographic analyses did not reveal any registered archaeological sites or interest in archaeological relevance at and in close proximity to the surface sites. There are no buildings or monuments of historical importance on the surface sites or directly affected by potential surface site constructions due to visibility or co-visibility. There are a few buildings of historical significance in the wider perimeters of some of the surface sites.

The sensitivity of the area of PA in Ferney-Voltaire, France with respect to heritage is low, particularly at the surface site itself. The site would be located on protected agricultural land. To the west of the study area, there is the protected woodland, Bois de la Mouille, also classified as wetland. To the northeast, in the vicinity of the site, there are a few isolated protected trees and a wetland with low functionality. Nevertheless, the implementation of a surface site site avoids these classified and protected areas, despite the fact that some of them are on agricultural land.

The area around the site PB in Presinge, Switzerland has a strong heritage, architectural, cultural and landscape character. The surface site is relatively far from heritage sites but it remains partially visible from the hamlets to the north as well as those to the south. The existing wooded strips partially protect the views towards the villages of Choulex, Puplinge and Presinge, but remain open towards the communes of Jussy and Meinier. In order to preserve the rural character of the area, it is advisable to consider good landscape integration, such as a semi-buried surface site so that visibility from nearby hamlets is limited.

The heritage sensitivity within the study area of the PD site in Nangy, France, is considered medium to very high depending on the part of the site. The presence of the Ch\^{a}teau de Pierres and a classified wood located to the north and west, 180 to 300\,m from the surface site induced this rating. In the area directly concerned by the site, the sensitivity is very low. The site is in the immediate vicinity of an autoroute and a departmental road which will be reorganised to improve traffic flow. No relevant visibility issues are recorded in an area that is dominated by high traffic, a hospital, or industrial and commercial facilities.

Sensitivity to heritage in the area of the PF surface site in \'{E}teaux in France is low. Very high issues are noted in the wider vicinity of the site to the south and the southeast due to a woodland around a creek that is classified and protected. To the north-west, at a distance of about 400\,m, there is a notable building, constructed in 1923, which formerly housed a cheese factory. There are also several groups of isolated trees that are witnesses to the history of agriculture. However, the implementation of the surface site avoids all sensitive spaces and stays close to the departmental road, directly opposite a public works construction company.

The sensitivity related to heritage at site PG in Groisy and Charvonnex in France is considered to be low. Some areas in the south of the main site are remarkable due to the presence of areas with wetland characteristics and unclassified but high-quality woodlands. There are a few old buildings linked to the networks of old mills located in the surrounding larger study area. The surface site does not directly affect any heritage elements.

Heritage sensitivity in the immediate area of the PF surface site in Cercier and Marlioz, France is considered very low. This assessment is linked to the presence of the large forest (Grand Bois) and the Tabass\'{e} creek to the north of the surface site. Both present ecological interests, but are not considered heritage sites.

The heritage sensitivity of the PJ surface site in Dingy-en-Vuache and Vulbens in France is considered medium to high due to the location of the site in an area classified as a protected agricultural area and an ecological corridor protection zone. The entire zone is of agricultural character. In the south-west, in the immediate vicinity, some areas exhibit very high sensitivity due to the presence of classified woodlands along small creeks. The current shape of the surface site respects the integrity of the constraints. Still, particular attention will have to be paid for the specific design and integration of the surface site to maintain the registered ecological corridor, ensuring the free movement of fauna in the area.

The heritage sensitivity for the PL surface site in Challex, France is considered to be medium to high. The area around the site consists largely of protected agricultural and, further away, natural lands. There are several groups of isolated trees and alignments that bear witness to the history of agriculture. The site itself is located in a protected agricultural area and borders a protected ecological corridor registered in the municipal urban plan. The main issue is the visibility of the site from several locations in nearby villages. Therefore, care has to be taken for the specific design and landscape integration of the surface site.

In summary, the main heritage constraints are mostly related to the rural and agricultural character of the areas, with the presence of historical buildings at some distance from the sites. Particular attention related to heritage has be paid during site design and landscape integration at the site locations PB, PJ and PL. 

\subsection{Landscape}

\begin{figure}[h]
  \centering
  \includegraphics[width=0.7\textwidth]{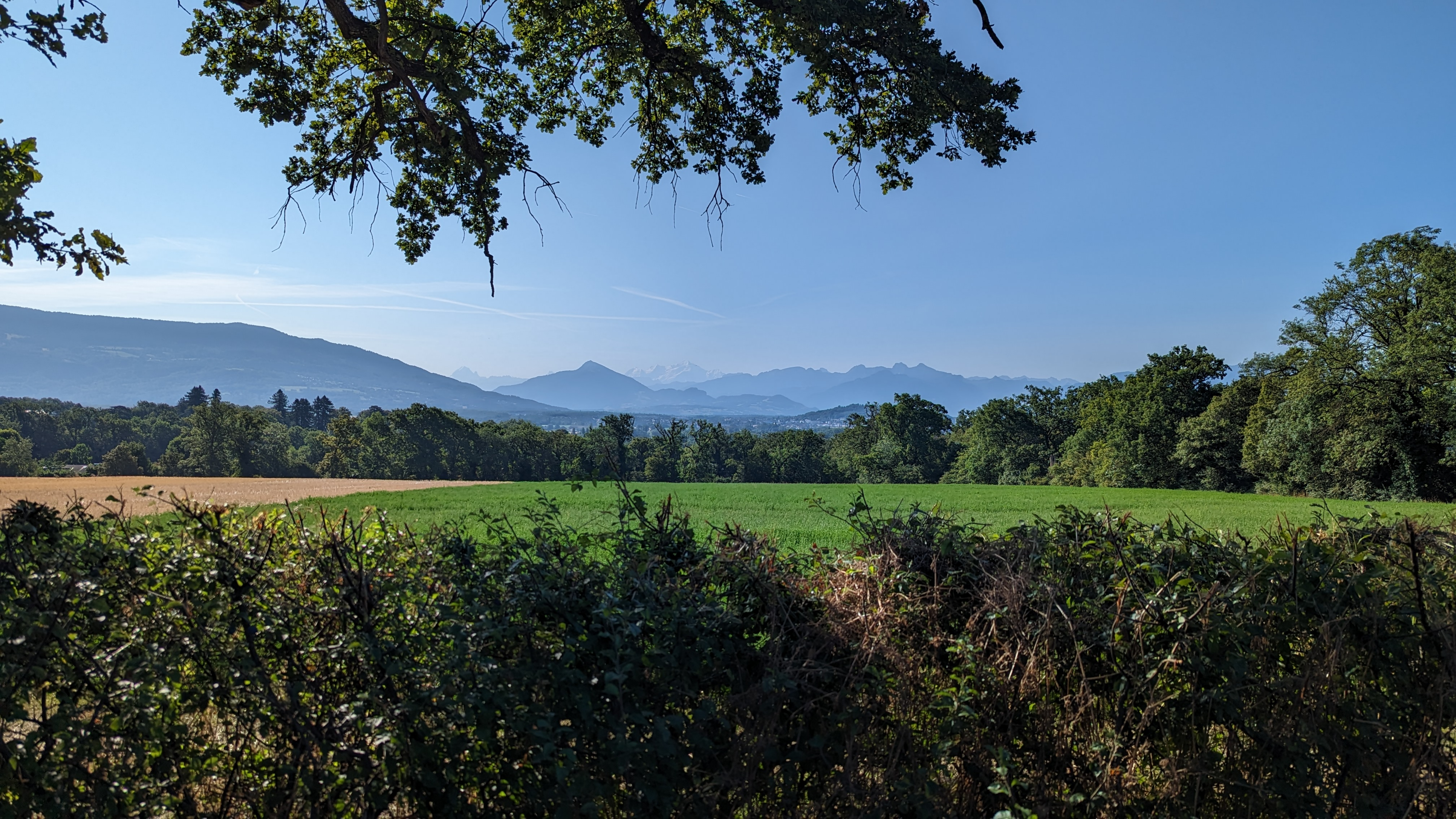}
  \caption{Mountainous landscape with agricultural fields documented during field investigations, highlighting the interaction between natural and cultivated environments.}
\label{fig:Photo_Landscape_mountains_Chem.DesPrinces_PB.pdf}
\end{figure}

The initial state analysis included a detailed landscape analysis that was carried out with expert companies during field visits over an entire year. The landscape analysis extended up to 5\,km around the surface sites, since the visibility of the sites depends significantly on the topography, urban and vegetation environment. Detailed landscape analysis and maps were developed that show the topography, the views from and to the sites and the visibility issues. They will serve as input for subsequent architectural design works and landscape integration concepts.

The site PA and its surroundings in Ferney-Voltaire (France) are dominated by agricultural areas, Geneva airport, highly urbanised commercial surroundings and an ecological corridor between the residual forests that occupy the buffer along the Franco-Swiss border. The area is in an open landscape with a view towards the Alps and Mont Blanc that has to be considered during the architectural design of the surface site. The nearby LHC site Pt8 (LHCb) and its planned extension provide an excellent opportunity to reduce the PA site footprint as much as possible to preserve the view and the ecological corridor that also stretches across the border.

The site PB in Presinge (Switzerland) is located in open countryside with valuable views towards the Alps and the Sal\`eve mountain. The landscape continuity is remarkable and the views from nearby hills over the landscape and its cultural and landscape heritage are highly valued. Much care must be taken in the architectural designs of the site, ideally integrating the site in the landscape as much as possible, taking care that the view from the nearby villages towards the mountains remains unobstructed.

The PD site in Nangy (France) is located in a mixed urban, agricultural and industrial environment that is dominated by major transport routes. Direct views to the sites are limited.

The PF site in \'Eteaux (France) is located in a natural environment, directly on highly frequented transport routes and commercial premises. No direct visibility from La-Roche-sur-Foron exists. However, the mountain views towards the Alps need to be considered when developing the architectural design of the site.

The site PG in Charvonnex and Groisy (France) is located in a highly natural environment, partially forested and with grass fields. The mountain views from the site are highly valued. A direct, but limited view of the site only exists from the Oli\`eres plateau.

The PH site in Cercier and Marlioz (France) is located in a rural environment with a forest. Some hamlets exist in the vicinity, but due to the forest and the topography, no direct view of the site exists.

The PJ site in Dingy-en-Vuache and in Vulbens (France) is located in a mixed natural and agricultural environment. The landscape is open and the view from Vulbens must be taken into consideration during the architectural design.

The site PL in Challex (France) is located in an agricultural/vineyard environment with views to the Jura mountain, the Rh\^one valley and the Alps. The open views require careful landscape integration, considering in particular the site's visibility from the nearby hamlets.

\subsection{Noise}

\begin{figure}[h]
  \centering
  \includegraphics[width=0.7\textwidth]{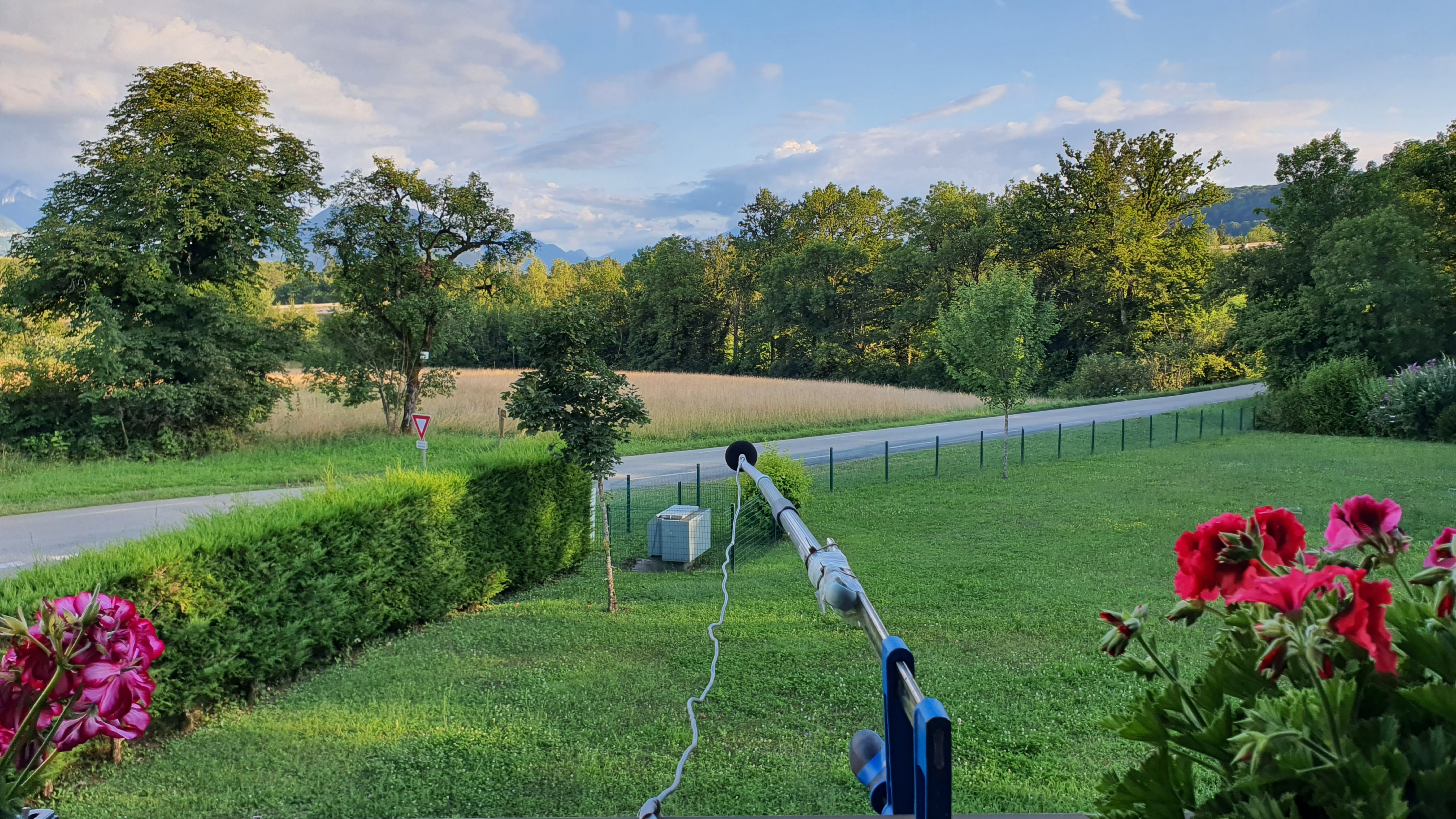}
  \caption{Field-based noise measurements conducted to assess ambient sound levels.}
\label{fig:Photo_NoiseMeasurements_Garden.pdf}
\end{figure}

\begin{figure}[h]
  \centering
  \includegraphics[width=0.7\textwidth]{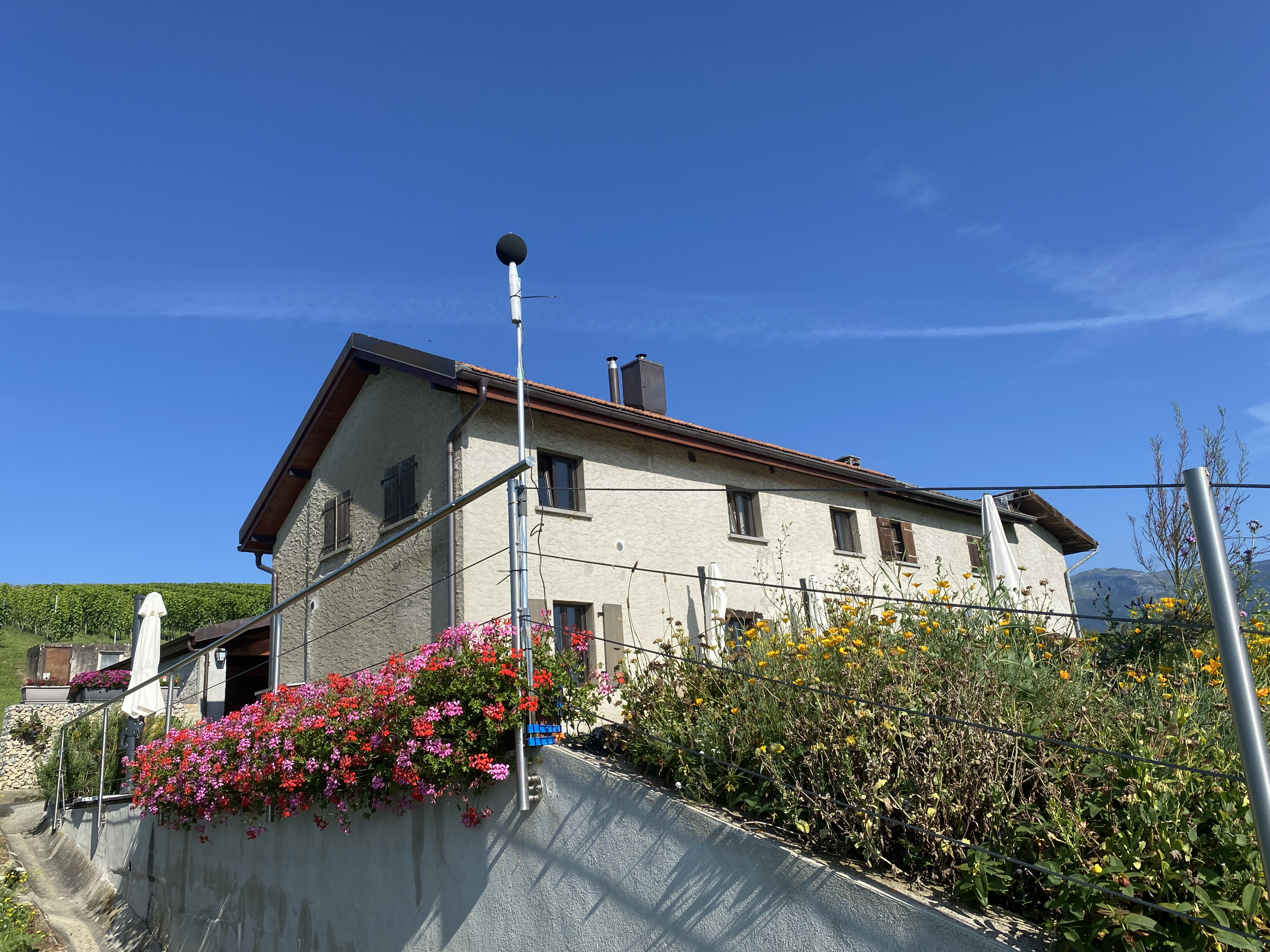}
  \caption{Noise measurements taken near a residential area to understand the current noise level.}
\label{fig:Photo_NoiseMeasurements_House.pdf}
\end{figure}

Noise, is considered to be unpleasant or annoying and is an environmental aspect that is relevant during the construction of the surface and subsurface structures and during the operation of the particle collider. The regulations concerning noise protection differ significantly between France and Switzerland. However, in all cases the impact due to noise depends on the presence of people and animals that would be affected by the noise. The relevant indicators are not only frequency and amplitude, but also the time during which an exposure to noise occurs and the duration of the exposure: during ordinary working hours, during the night, on non-working days, the typical presence of people in their homes, the age, health and other social conditions of the potentially affected people (e.g., noise in the vicinity of a hospital, a school or a retirement home). Typically, noise generated by machinery is considered to have noteworthy health effects from 40\,dB(A) and higher. Noise up to 30\,dB(A) is a level that is generally experienced by persons in a typical living environment (see Table~\ref{tab:noise-level-examples}).

\begin{table}[ht!]
	\label{tab:echelle_des_bruits}
	\centering
	\caption{Typical noise levels with practical examples.}
	\label{tab:noise-level-examples}
	\begin{tabular}{lccc}
		\toprule
		\textbf{Noise} & \textbf{dB(A)} & \textbf{Level} & \textbf{Conversation} \\ \midrule
		Jet plane takeoff & < \textbf{130} & Above pain threshold & \multirow{3}{*}{\textbf{Impossible}}\\ \cmidrule{1-3}
		Jackhammer at 1\,m & < \textbf{110} & \makecell[c]{Supportable for a\\short moment} & \\ \midrule
		Motorcycle at 2\,m & < \textbf{90} & Annoying & \makecell[c]{\textbf{Possible when shouting}}\\ \midrule
		Heavy road traffic & < \textbf{80} & Very noisy & \textbf{Difficult without shouting}\\ \midrule
		\makecell[l]{Dwelling close\\to an autoroute} & <\textbf{70} & Noisy & \textbf{\makecell[c]{Possible when\\ speaking loud}}
		\\ \midrule
		Working on a computer & < \textbf{60} & Moderate & \multirow{3}{*}{\textbf{\makecell{Possible with\\ normal voice}}}
		\\ \cmidrule{1-3} 
		\makecell[l]{Noise level in a\\ city during the day} & < \textbf{50} & Rather quiet & \\ \midrule
		\makecell[l]{Construction noise at\\ 100\,to\,200\,m distance\\ from a surface site} & \textbf{40 - 45} & - & \multirow{11}{*}{\textbf{Possible with quiet voice}}\\ \cmidrule{1-3} 
		\makecell[l]{Noise level in the\\ countryside during\\ the day} & < \textbf{40} & Quiet & \\
		\cmidrule{1-3} 
		\makecell[l]{Noise level in the\\ countryside during the\\ night without wind} & < \textbf{30} & Very quiet & \\ \cmidrule{1-3} 
		\makecell[l]{Snow falling in the\\ mountains, recording studio} & < \textbf{15} & Silent & \\ \bottomrule
	\end{tabular}
\end{table}

Therefore, the first step required is to establish the existing background noise in the vicinity of the candidate surface sites. This activity is followed by estimations of noise generation without protection measures. This identifies sensitive areas. An eco-design is subsequently applied to reduce the noise where potential impacts on the environment are expected. The approach is to first work on low-noise designs. This also includes the relocation of noise-generating devices on the surface sites. Where this process turns out to be insufficient, noise protection measures to reduce the impacts are studied. Relocation of noise-generating equipment or adjustment of an entire surface site is an option that is considered if no adequate mitigation measures can be identified.

The background noise was established by first consulting national noise emission databases and then completing the data with field measurements. The measurements were carried out by expert companies on various days in 2024 at a selection of different locations and at different distances from the surface sites. Detailed analysis of the frequencies, amplitudes, and directions of origins have been made to permit mixing potential noise sources at the surface sites with the ambient noise at a later stage. The number of potential residents that could be affected by noise has been obtained using an analysis of the data collected with the geographical information system and statistical information (see Table~\ref{tab:background-noise}).

\begin{table}[h]
  \centering
	\caption{Typical background noise measured in the vicinity of surface sites and the numbers of people potentially affected within the distances indicated from the site.}
\label{tab:background-noise}
  \begin{tabular}{crrrr}
    \toprule
    \textbf{Site} & \textbf{\makecell[r]{Background noise dB(A)\\7h-22h}} & \textbf{\makecell[r]{Background noise dB(A)\\22h-7h}} & \textbf{\makecell[r]{Potentially\\affected people}} & \textbf{Distance to site} \\ \midrule
    PA & 55 & 45 & 0 & 100 - 200\,m \\ 
    PB & 36 & 36 & 0$^\dagger$ (< 10$^\ddagger$) & 100\,m \\ 
    PD & 48 & 44 & 0 & 100 - 200\,m \\ 
    PF & 48 & 40 & \textasciitilde 10 & 100 - 200\,m \\ 
    PG & 36 - 48 & 37 - 45 & 0 & 100 - 300\,m \\ 
    PH & 35 & 29 & < 10 & 100\,m \\ 
    PJ & 39 - 47 & 35 - 44 & 0 & 300\,m \\ 
    PL & 40 & 30 & < 5 & 200\,m \\ \bottomrule
	\multicolumn{5}{l}{~$\dagger$ \footnotesize People exposed according to Swiss limits, DS III}\\
	\multicolumn{5}{l}{~$\ddagger$ \footnotesize People exposed according to permitted noise emissions in France}\\	  
  \end{tabular}
\end{table}

The measurements indicate that the majority of sites are affected by significant background noise with no people present in a perimeter of 200 to 300\,m. Therefore, neither the noise from construction nor operation, that will be kept as low as possible with noise protection measures, are expected to impact the population at sites PA (Ferney-Voltaire), PD (Nangy), PF (\'Eteaux), PG (Charvonnex and Groisy), PJ (Dingy-en-Vuache and Vulbens). At the PL site (Challex), construction activities during the night could lead to an excess of background noise, and therefore, the construction schedule would have to be adapted if a detailed technical design of the construction site in view of noise protection turns out to be insufficient. At sites PB (Presinge) and PH (Cercier and Marlioz) the background noise is particularly low with about 35\,dB(A) compared to all other locations. In all cases, operation related noise can be kept below the threshold for residents potentially affected in the vicinity, but for the construction planning, care must be taken to respect the times during which noise exposure is more impactful (e.g., during the night, weekends and non-working days). For site PH, 3 to 4 houses at a distance of 100\,m would be affected, and at site PB, two houses at a distance of 200\,m would be affected. The construction-induced noise should, therefore, not significantly exceed the background noise during resting times in these locations.

\subsection{Vibrations}

Vibrations are rapid oscillating movements that propagate through solid paths and can be transmitted to the human body, especially through direct contact with the ground or the structure concerned. They are thus physical phenomena characterised by a wave with its amplitude and frequency.

The risks associated with vibrations are mainly damage to buildings and disturbance of people. Vibrations can occur due to seismic activities, ground movements, meteorological conditions and artificial sources such as construction activities and machinery. The environmental analysis revealed that in the entire perimeter of the project, seismic activities have a very weak to weak potential to create relevant vibrations, despite different occurrence probabilities for small-scale seismic events. Concerning vibrations caused by landslides surface site locations PA, PB, PD, PH, PJ and PL are at significant distances from risk zones and no subsurface cavities exist within a relevant distance of surface sites. A potential landslide risk zone exists only  at a distance of 500\,m from the PF site, but the likelihood of an event is very small. A cavity that is part of a defence infrastructure exists at a distance of 250\,m. It is unlikely that this cavity would collapse. Weather phenomena such as violent winds and lightning may cause vibrations that affect trees and building structures. However, such events are infrequent in the perimeter of the scenario. Various human-induced vibration sources over the coming ten years were identified in the vicinity of surface sites PA, PD and PF. Road traffic is another source of ongoing vibration. It needs to be considered for sites PA, PD, PF, PG and PJ. No other noteworthy sources of vibrations could be identified in the vicinity of surface sites.

\subsection{Light}

Light pollution refers to the excess or poor management of artificial light in the nighttime environment. This phenomenon is increasingly concerning, both for its effects on biodiversity and human health. There are two types of light pollution. First, direct light pollution relates to the impact of light directly produced. Secondly, indirect light pollution results from an accumulation of light, creating a halo that obscures the stars and degrades the quality of the night sky. With respect to the surface sites, light pollution is a relevant environmental topic to be assessed during the detailed design phase concerning the construction activities and the subsequent operation phase. Therefore, at this stage, the existing light pollution in the vicinity of the candidate locations for surface sites has been analysed and documented to serve as a baseline for the development of the detailed design.

The main source of light pollution in the areas concerned by the project is public lightning along roads in residential areas. In addition, commercial zones are a source of artificial light, in particular when they use large-scale public screens. Frequently, lights are not controlled. Some outdoor lights of private houses contribute to light pollution. Finally, industrial installations for operation during nighttime and to ensure safety are a strong source of light pollution.

For the current state, data have been purchased from various sources, and maps have been developed using these data for the areas around the surface sites. The zone around site PA is strongly illuminated during the night. One source is the Geneva airport and another one is the commercial zone in the immediate vicinity. Road lighting and newly built residential houses also contribute. The Grand Gen\`eve plan foresees that the entire area is gradually restored to protect the ecological corridors and the fauna. However, no protection measures are currently planned at local French urban planning level. Site PB in Switzerland is located in the countryside and only few buildings are found in the vicinity. In the Grand Gen\`eve plan, the perimeter is classified in order to protect the `night'. In particular, the zone close to the road is subject to refurbishment and the zone along the creek has a stake. However, no protection measures are mentioned at local urban levels. Site PD is in the immediate vicinity of major transport routes and industrial buildings on a field. No particular protection measures are planned at local urban planning level. Site PG is in a natural environment and in the forest with only little light pollution. Although no particular protection measures for the location exist, it can be assumed that preserving the current situation is a priority. Also, site PH in the forest is in an area with very low light pollution. 

Although no particular protection levels are indicated it can be assumed that the current situation is to be preserved as much as possible.  Site PJ in the countryside experiences small light sources in the vicinity from some distant urban constructions. The zone is classified to be preserved, and the ecological corridors constitute a stake. General protection measures apply to forbid artificial light along the water courses although at local urban planning level, no protection zones are indicated. Site PL is located in an agricultural zone and rather isolated next to fields, forests and vineyards. In Switzerland, nature protection zones forbid artificial light. At the local urban level, no protections apply in France.

Summing up, the site PD and its surroundings are located in an area where the night should be preserved with priority. However, no noteworthy species are affected and therefore the stake is average. The PA site is, however, in an area with high light pollution and there are ecological stakes in the surrounding area. Therefore, there is a will to restore the night environment, and the stake is high. The stake is also high for the PB site in Switzerland, since the night is intended to be preserved and light pollution should even be further reduced. Also, the stake for site PF is high since the site is located in an area with almost no light pollution today. Similar conditions with high stakes apply to sites PJ and PL. The wider surroundings of sites PG and PH are in a forest with low light pollution and nearby noteworthy biodiversity reservoirs. Therefore, the stakes with respect to light pollution are very high. These findings have to be considered and integrated in the design and planning of the construction sites and a plan to preserve the night has to be developed for all sites concerning the operation phase.

\subsection{Radiation}

\begin{figure}[h]
  \centering
  \includegraphics[width=0.7\textwidth]{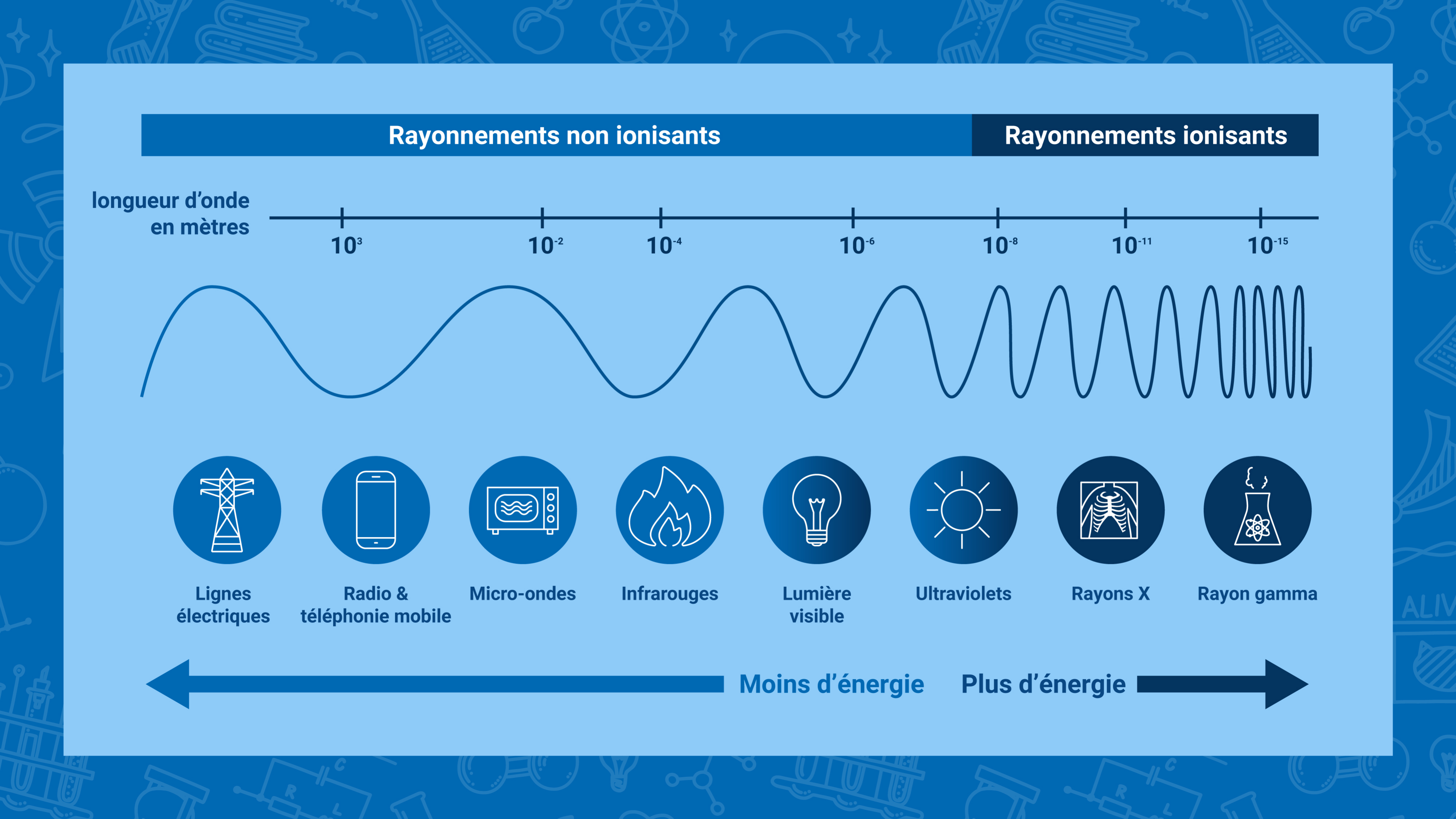}
  \caption{Types of radiation (Source: IAEA
  \cite{radiation_introduction_IAEA}).}
\label{fig:Radiation}
\end{figure}

Possible risks for the population and environment emerging from non-ionising and ionising radiation exposure at different locations on the surface were analysed based on available data in both Host States. Figure~\ref{fig:Radiation} shows the different types of radiation, non-ionising and ionising, that exist. 

The Host States notified services in charge of carrying out measurements to establish national databases will require updated data before a particle accelerator is put into operation to establish a baseline for the ionising radiation.

\subsubsection{Non-ionising radiation}

\begin{figure}[h]
  \centering
  \includegraphics[width=0.7\textwidth]{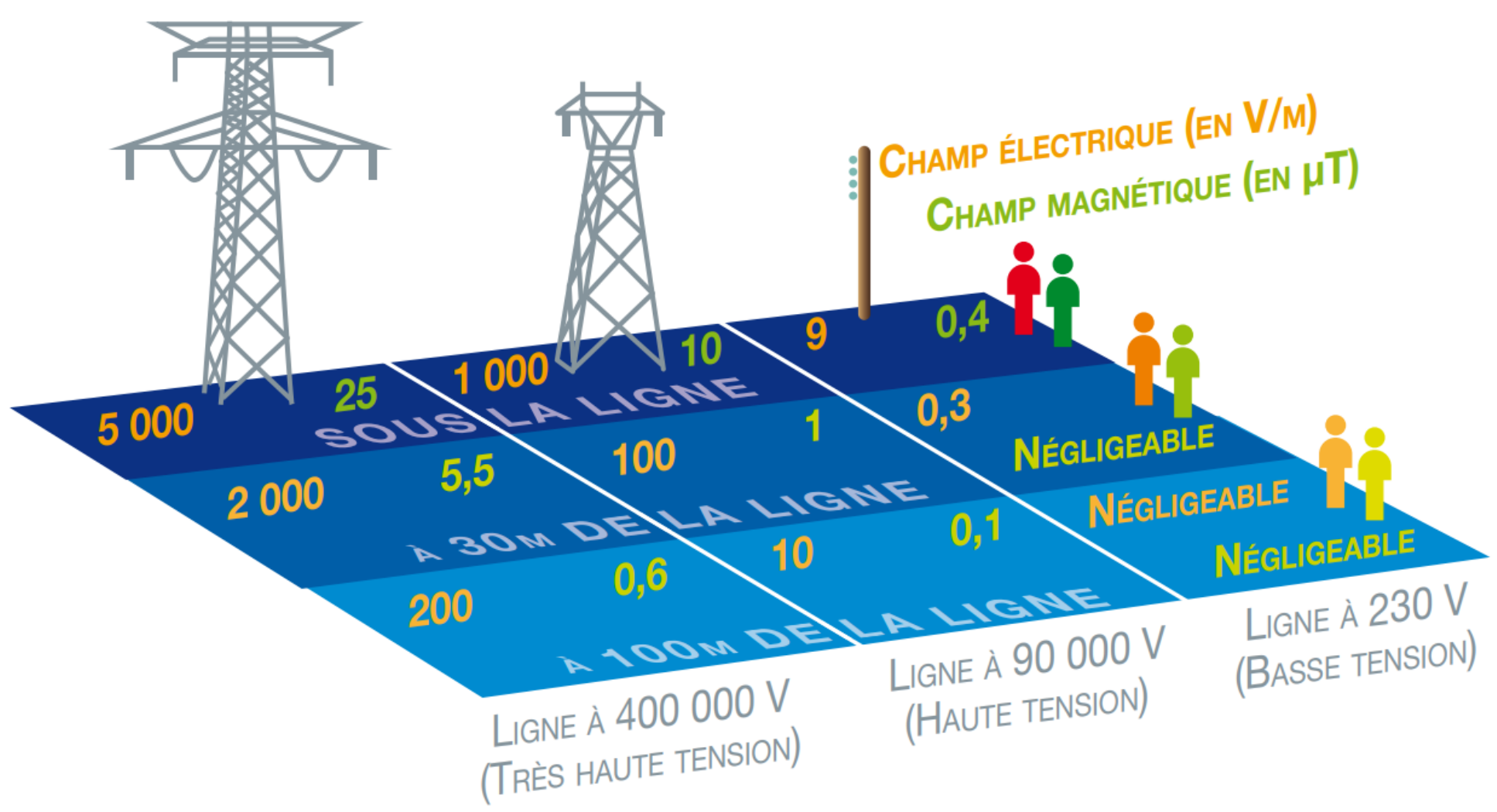}
  \caption{Average values of the electric and magnetic fields around overhead power transmission lines at 50 Hz (Source: DGS 
  \cite{electric_magnetic_field_law_freq}).}
\label{fig:Lines_electromagnetic_field.pdf}
\end{figure}

In France, exposure limits vary depending on the frequency range used, typically between 28 V/m and 87 V/m for the electric field. The Swiss regulations distinguish between public areas in which the persons stay only for brief intervals of time, such as roads or sports facilities, and sensitive areas, in which the persons may stay for a certain limited period of time including houses and apartments, schools, and hospitals. Switzerland imposes stricter installation limits for radio equipment than France, with maximum electric field values set at 4 V/m for 900\,MHz, 6 V/m for 1800\,MHz, and 5\,V/m for installations operating across multiple frequencies. These Swiss regulations are designed to provide enhanced protection for sensitive locations compared to the International Commission on Non-Ionizing Radiation Protection (ICNIRP) guidelines. Swiss limits with respect to immission, which represent the cumulative exposure from all emitters, align with European recommendations and are identical to those in France.
Figure~\ref{fig:Lines_electromagnetic_field.pdf} shows the values of the electric and magnetic fields measured under the line, as well as at 30\,m and 100\,m from the line, for very high voltage, high voltage and low voltage overhead lines.

No electricity lines are emitting low-frequency radiation in the vicinity of sites PB in Switzerland, PD, PG, PH, PJ and PL in France. A \num{63} \unit{kV} line runs \num{250}\,m in the vicinity of the PF site in \'{E}teaux in France, which emits a negligible electromagnetic field onto the candidate surface site. The \num{66} \unit{kV} underground electricity line supplying LHC Point 8 in Ferney-Voltaire in France is just next to the PA site. It induces a very weak electromagnetic field onto the site.

Some radio stations that are sources of radio waves are present near the surface locations, but the recorded exposure measurements are well below the reference levels. The data collected from measurement stations near PA, PB, PD, PG, PJ show that all exposure measurements are significantly below the reference levels. No data was found for measurements within a 5\,km radius of the PF, PH and PL locations in France.
Detailed measurements will have to be carried out before the installation is put into operation, but the risk of exceeding the limits for ionising and non-ionising radiation specified in the regulations is considered to be very low.

\subsubsection{Ionising radiation}

Ionising radiation carries sufficient energy to ionise atoms or molecules, leading to molecular changes. High doses can cause cellular damage and mortality, whereas controlled applications are pivotal in industries, scientific research, and medicine. The sources of exposure are of natural and artificial origins:

\begin{itemize}
\item Natural cosmic radiation: energetic particles from space contribute to exposure, influenced by altitude and geographic location.
Terrestrial radiation: radioactive elements in the Earth's crust (e.g., uranium, thorium) emit radiation, varying regionally due to geological factors.
\item Natural radon gas: a significant contributor, radon accumulates in poorly ventilated indoor spaces. It constitutes 33\% of annual exposure in France.
Incorporated radionuclides: naturally occurring radionuclides in food and water contribute 12\% to annual exposure.
\item Artificial sources, nuclear accidents and testing: fallout from historical nuclear tests and accidents like Chernobyl contributes marginally to current exposure.
\item Nuclear facilities: emissions from civilian and military installations are tightly regulated, with negligible contributions (~0.001–0.01\,mSv/year near facilities).
\item Artificial sources, medical applications: diagnostic and therapeutic uses dominate artificial exposure, contributing 34\% of total exposure in France.
\item Artificial sources, scientific research activities: the particle accelerators and colliders operated by CERN contributing less than 0.01\,mSv/year.
\end{itemize}

Table~\ref{tab:radiation_exposure} gives an overview of the current ionising-radiation background conditions at the surface site locations.

\begin{table}[!h]
\caption{Level of exposure to ionising radiation at the surface site locations today (Source : Autorité de sûreté nucléaire et de radioprotection (ASNR))}
\label{tab:radiation_exposure}
\centering
\begin{tabular}{lcl}
\toprule
\textbf{Nearest data entry} & \textbf{Annual exposure}  & \textbf{Surface site}\\
\midrule
Ferney-Voltaire &  5.2\,mSv/year & PA (Ferney-Voltaire, France) \\ 
Ville-la-Grand & 4.0\,mSv/year & PB (Presinge, Switzerland)  \\ 
Nangy & 4.3\,mSv/year & PD (Nangy, France)  \\ 
Éteaux & 4.4\,mSv/year & PF (Éteaux, France)  \\ 
Groisy & 4.4\,mSv/year & PG (Charvonnex and Groisy, France)  \\ 
Charvonnex & 4.9\,mSv/year & PG (Charvonnex and Groisy, France) \\ 
Marlioz & 4.9\,mSv/year & PH (Cercier and Marlioz, France)  \\ 
Cercier & 5.0\,mSv/year & PH (Cercier and Marlioz, France) \\ 
Dingy-en-Vuache & 5.0\,mSv/year & PJ (Dingy-en-Vuache and Vulbens, France) \\ 
Vulbens & 5.8\,mSv/year & PJ (Dingy-en-Vuache and Vulbens, France) \\ 
Challex & 5.5\,mSv/year & PL (Challex, France) \\ \bottomrule
\end{tabular}
\end{table}

Exposure to cosmic radiation, originating from high-energy particles such as protons and heavy ions from space, poses hazards due to its ionising nature, which can damage living tissues. The intensity of exposure increases with altitude, being roughly twice as high at 1500\,metres compared to sea level and further elevated during air travel. While the Earth's magnetic field reduces exposure near the equator, higher doses are received near the poles. In France, cosmic radiation accounts for about 7\% of the average annual exposure to ionising radiation, with an effective dose at the surface sites between 0.32 and 0.70\,mSv/year. In Switzerland, the dose associated to cosmic rays in the vicinity of the site is about 0.35\,mSv/year.

Telluric radiation refers to ionising radiation emitted by naturally radioactive elements like uranium, thorium, and potassium-40 found in the Earth's crust. Exposure levels vary depending on regional geology, with areas rich in granitic or volcanic rock typically exhibiting higher radiation levels. In France, telluric radiation contributes approximately 14\% of the average annual exposure to ionising radiation, with an effective dose of around 0.51 to 0.55\,mSv/year in the areas of the surface sites in France and 0.33\, mSv/year in the vicinity of the surface site in Switzerland.

Exposure to natural radionuclides occurs through the ingestion of radioactive elements, such as polonium-210, present in food, water, and air, which originate from terrestrial rocks, soils, or cosmic interactions. Foods like seafood are particularly rich in these radionuclides, and tobacco inhalation can also contribute significantly. In France, this accounts for approximately 12\% of the average annual exposure to ionising radiation, with an effective dose of about 0.55\,mSv/year in France, varying between 0.4 and 3.1\,mSv/year depending on dietary habits. In Switzerland, the exposure is about 0.40\,mSv/year. Tobacco consumption accounts for about 0.04\,mSv/year.

Exposure to radon, a naturally radioactive gas, varies between France and Switzerland due to differences in geology and measurement methodologies. In France, radon contributes 33\% of the average annual ionising radiation exposure, with a mean dose of 1.5\,mSv/year, ranging between 0.54 and 3.2\,mSv/year. Specific areas concerned by the surface sites in Ain and Haute-Savoie report slightly lower exposures, around 1.47 and 1.14\,mSv/year, respectively. In Switzerland, radon is the largest natural contributor to radiation, with an average dose of 3.3\,mSv/year, reflecting a higher impact. This discrepancy partly arises from different dose conversion factors used; France employs the UNSCEAR coefficient, while Switzerland uses the CIPR's updated factors, which approximately double the estimated risks. Consequently, radon exposure is relatively more significant in Switzerland.

Exposure to ionising radiation from artificial sources in France and Switzerland includes contributions from nuclear accidents, medical applications, nuclear installations and scientific research facilities.

Medical applications in both France and Switzerland, are the largest contributors to artificial ionising radiation exposure. In France, they account for approximately 1.5\,mSv/year per capita, representing 34\% of the total radiation exposure. Similarly, in Switzerland, medical applications contribute about 1.49\,mSv/year per capita, highlighting the widespread use of diagnostic and therapeutic procedures involving ionising radiation in both countries.

Radiation exposure due to past nuclear accidents and fallout, such as Chernobyl, and atmospheric nuclear tests is now minimal in both countries. In France, the average annual dose is 0.012\,mSv/year, with slightly higher doses in areas with significant fallout. In Switzerland, this exposure is even lower, contributing only a few hundredths of a mSv annually, reflecting the diminished impact of residual fallout over time.

Exposure from nuclear facilities is negligible in both France and Switzerland due to strict regulatory measures. In France, people living within 10\,km of nuclear installations receive an annual dose of 0.001 to 0.01\,mSv under normal operational conditions. In Switzerland, exposure near nuclear facilities and the scientific facilities of CERN, is similarly low, with annual doses generally below 0.004\,mSv, demonstrating effective safety and monitoring measures.

Summing up, the ionising radiation context at the surface sites is significantly larger than the residual ionising radiation that is generated by CERN's scientific particle accelerators.

\subsection{Technical risks}

\subsubsection{Natural Hazards}

As part of this study, several types of natural hazards that can potentially lead to interactions with the project have been identified. These will be analysed in further detail during a subsequent environmental impact assessment. The topics concern flooding, ground movement, seismic activities, avalanches, forest fires, cavities, radon, clay swelling and technological hazards. These natural hazards may also be caused by climate change effects. The evolution of the climate has been taken into consideration in the study of the current state of the environment, as well as the evolution of the environment in which the project would be embedded.

\paragraph{Flood and ground movement} hazards that can be caused by numerous factors, such as rain, groundwater and rivers, are considered to have low probability for all the locations. Effects due to unstable ground, such as landslides, mudflows and erosion have also been taken into account. Overall, no significant hazards or potential effects leading to relevant risks were identified in the concerned locations. No surface site on French territory is subject to a ground movement. For the site PB in Presinge in Switzerland, the closest potentially unstable ground area is located \num{700} \unit{m} east of the site, however the landslide risk is superficial. The candidate surface site location is not subject to the hazards linked to flooding or earth movement. 

\paragraph{Seismic} hazards vary depending on the location. France is divided into five seismic activity zones. The sites studied are located in zones of moderate or medium seismic activity. In particular, sites PA, PL and PJ are located in zones of moderate (level 3) activity, while sites PG, PF and PD are located in zones of medium (level 4) activity. This requires the application of earthquake-resistant construction principles during the surface site design development. In addition, some of the sites studied are located near active tectonic faults. For example, site PA in Ferney-Voltiare is located about \num{400} \unit{m} from a fault. In Switzerland, the seismic hazard model classifies locations into five zones, of which site PB in Presinge is located in zone Z1b (level 2), characterised by low seismic activity. Further field studies are needed, and cooperation with scientific institutions to assess the local seismic risk in detail and develop specific engineering models will be undertaken. 

\paragraph{Forest fires} turn out to be a very low hazard in the zones concerned by the surface sites, although two locations are in forest zones. In France, the Forest Fire Risk Prevention Plan (PPRIF) established at the municipal or inter-municipal level, targets areas exposed to significant risk levels and strong land pressure. In addition, the PPRIF may also impose clearing of areas in order to isolate the buildings. It may also require that access roads be sized to allow fire trucks to pass and for people to evacuate in the event of a fire. No French municipality affected by the surface sites is subject to a forest fire risk prevention plan and none of the areas surrounding the surface sites in France are subject to the legal obligation of clearing. Although the site location PH in Cercier and Marlioz is not exposed to the natural forest fire hazard according to various regulatory and management tools, there always remains a residual and low hazard of forest fires in this area due to the presence of timber. In Switzerland and more specifically in the canton of Geneva, there is no law, prevention plan or specific forest fire risk management plan. The site PB in Presinge is not in the vicinity of a woodland that would generate a forest fire hazard to the site.

\paragraph{Avalanches} are not considered a relevant hazard for the surface sites due to relatively flat terrain and low forest density in the vicinity of the locations analysed.

\paragraph{Cavities} natural or man-made voids can be considered hazards for construction activities. The analysis addressed their presence and potential effects with respect to the implementation of the surface sites. No cavities are located within a radius of \num{500}~\unit{m} around the French surface sites, except for the PG Charvonnex site where several cavities linked to military works are located within a radius of more than \num{250} \unit{m} around this site. Apart from their existence and location, no additional information is available. Thus, the potential hazard from underground cavities in this area requires additional data gathering.

\paragraph{Radon} is a colourless, odourless radioactive gas that originates from the decay chains of uranium and thorium, both naturally present in the Earth's rocks. A key factor influencing radon concentration levels in buildings is the geology, in particular, the uranium content in the subsurface layers.
In some areas, specific underground features (e.g., faults, mining works, hydrothermal sources) can exacerbate radon transfer to the surface, locally increasing its potential. For new constructions, radon transfer can be limited by enhancing the building's seal between the ground and the structure. In France, the mapping method estimates the radon potential of geological formations by considering factors that influence both radon production in the subsurface and its transport to the surface. Zones are classified as having low, moderate, or high potential.  All French surface sites studied are located in low radon potential zones.

In Switzerland, the radon map shows the probability in \% of exceeding \num{300} \unit {\becquerel\per\cubic\m} in buildings, divided into four categories: $\leq$1\%, \num{2} and \num{10}\%, \num{11} and \num{20}\%, and >\num{20}\%. Unlike France, there is no national campaign that estimates radon thresholds in buildings. The PB site in Presinge lies in a low-risk area, in the range of \num{2} to \num{10}\% probability of exceeding \num{300} \unit {\becquerel\per\cubic\m}.

\begin{figure}
    \centering
    \includegraphics[width=0.7\textwidth]{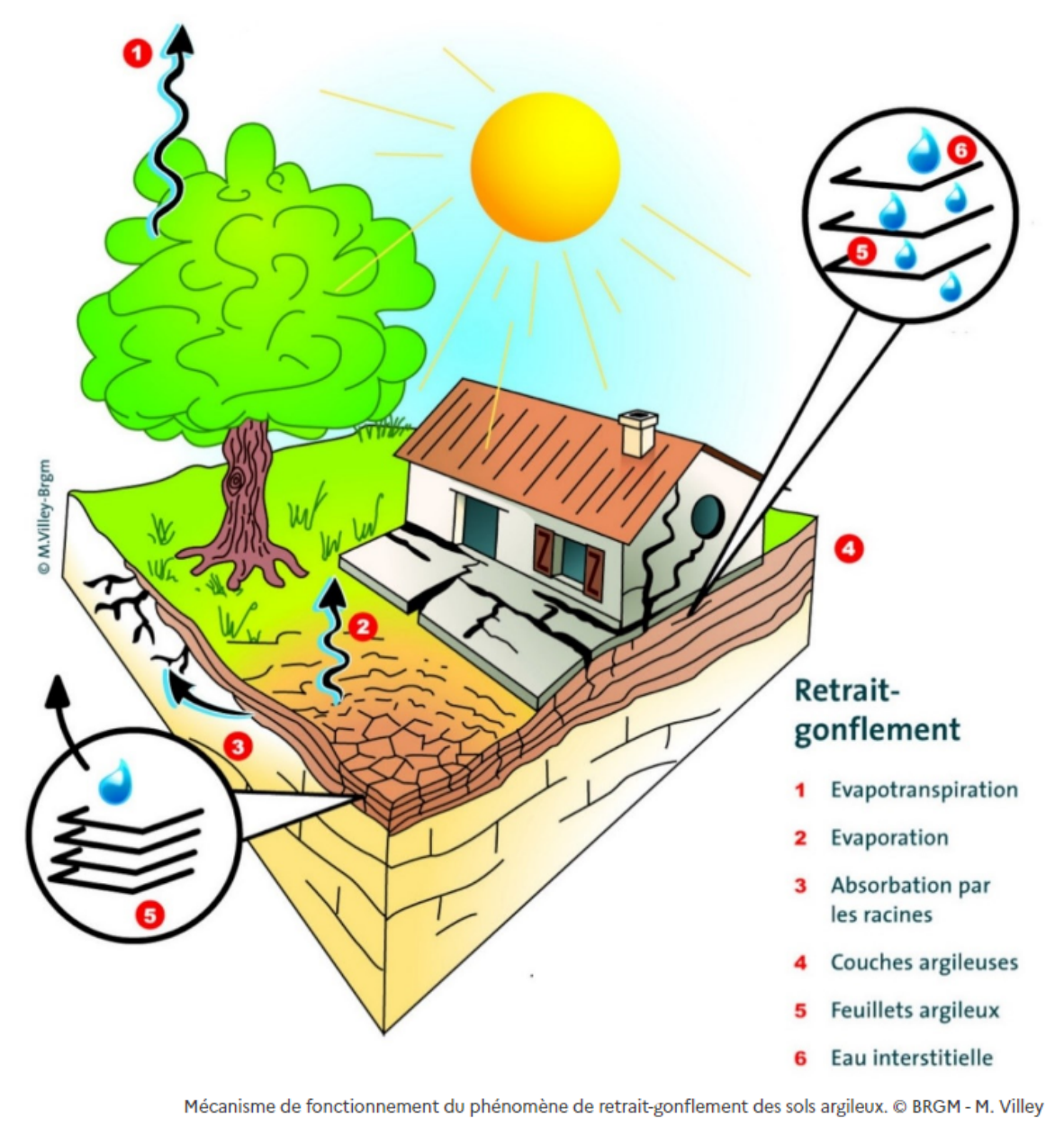}
    \caption{\label{fig:Schema_clay} Expansion and shrinkage of clays. Damage to the building is caused by the rain induced expansion on one side and shrinkage caused by removal of water by the tree on the other.}
\end{figure}

\paragraph{Shrinking and swelling of clays} is a natural hazard that has been assessed for all surface site locations. Superficial clay soils have the ability to change consistency depending on their water content. When the water content of clay soil increases, its volume expands, a phenomenon referred to as the swelling of clays. Conversely, a reduction in water content causes the opposite effect, known as shrinkage of clays.  Figure~\ref{fig:Schema_clay} explains the mechanism of the shrink-swell phenomenon. This shrinkage or swelling can cause structural damage to buildings.

From January 1, 2024, in France enforcement the of construction regulations has been tightened, and a new certificate for clay shrink-swell behaviour is now mandatory in zones subject to medium or high exposure. 
The surface sites located in France are subject to low to medium exposure to shrink-swell risks.
Site PA in Ferney-Voltaire is mostly in a low hazard area, except for a small part of the site near the LHC site point 8 (LHCb) where the exposure level is medium. Sites PD, PF, PG and PF are rates as low exposure. PH and PJ sites are located in zone with medium exposure. The PL surface site in Challex is mainly located in the zone with low exposure, except for a small fragment in the southern part of the site with medium exposure, which is not foreseen to host constructions, but a green buffer.
In Switzerland, a lithological map that provides an overview of the subsoil classified according to lithological and petrographic criteria indicates that the site PB in Presinge is located in soils containing clays. However, the concentration and depth of this clay element in the soil are not precisely described. Detailed soil investigations are required during a subsequent environmental impact and surface site design phase to clarify the conditions and, if needed, to take them into account during the design of the surface site.

\subsubsection{Technological hazards} 

Technological hazards linked to industrial, environmental, and infrastructural activities could potentially have effects on human health, safety and the surrounding environment. Risks emerging from these hazards and effects arise from industrial installations, the transport of hazardous materials and other activities involving dangerous substances. Therefore, their presence near candidate surface site locations has been analysed. 

In France, two categories of industrial facilities are considered to generate risk. The first group concerns establishments reporting emissions and transfers of pollutants into the air, water or soil and the production of hazardous waste. This covers establishments such as large industrial installations, large municipal sewage treatment plants and certain livestock farms. Second, installations classified for environmental protection (called ICPE) that are industrial or agricultural facilities subject to  regulations for the prevention of environmental risks and are classified according to regimes ranging from simple declarations to authorisations, depending on the risk. In addition, the SEVESO regulations (EU Directive 2012/18/EU) impose strict controls on facilities processing hazardous chemicals in order to prevent serious accidents.

In Switzerland, the Ordinance on Major Accident Prevention (so called OPAM) seeks to protect people and the environment from severe damage caused by extraordinary events in installations or during the transport of hazardous materials. OPAM applies to businesses exceeding specific hazardous material thresholds, genetic organisms under strict confinement, and transport infrastructures like pipelines, railways, and highways used for dangerous goods. The planning authorities (cantons and municipalities) are responsible for integrating aspects of major accident prevention into their plan for land use. Although Switzerland does not apply the SEVESO directive, OPAM adopts even stricter thresholds for hazardous substances and includes transport routes, pipelines, and dangerous microorganisms within its scope.

A gas pipeline passes through the PA annex surface site near the LHCb, extending from Switzerland and running along the edge of the main PA surface site towards the north. This installation must be taken into account when designing the surface infrastructure; however, there are no hazardous industrial installations located near the PA site, and the main threats might come from the proximity of major communication arteries (airport and departmental routes) that can transport dangerous goods. 

There is a gas pipeline approximately \num{200} \unit{m} south-west of the PB surface site which poses a potential risk of an accident involving hazardous materials. The Route de Jussy, classified as a major transit route, runs alongside the eastern boundary of the site. This road is subject to regulations concerning the transport of dangerous goods under the Swiss OPAM ordinance. Apart from these, there are no other industrial installations, facilities declaring pollutant emissions, or safety perimeters within the vicinity of Site PB.

The PD, PF and PG surface sites are located near major transport routes (highways and departmental roads) that are used for regional and national goods transport, including hazardous materials, creating potential risks of accidents or spillages. A railway line currently used for passenger transport passes near the PG site. Neither the PD or PG sites have other classified industrial facilities or pollutant-emitting sources within a \num{500} \unit{m} radius, ensuring a relatively low environmental risk. The classified installation for environmental protection SARL Luc Maulet, which handles inert waste, is located \num{150} \unit{m} from the PF surface site near the A410 but poses no risk as its activities are related to the storage of non-recoverable inert waste. 

There are no major roads or highways within a \num{500} \unit{m} radius of PH surface site, there are just local roads and paths, which significantly reduces transport-related risks involving hazardous materials. There are also no classified installations, pollutant-emitting facilities, or other industrial sites in proximity. However, there is a gas pipeline near the PH site and special attention should be paid to heavy transport that would cross the gas pipeline route.

Apart from the presence of the A40 autoroute located \num{50} \unit{m} to the south of the PJ surface site and the associated potential exposure to risks associated with the transport of dangerous goods, there are no classified installations or establishments reporting releases and transfers of pollutants in the vicinity.

Also, for the PL surface site, apart from the proximity of the departmental road, no classified installations or establishments reporting releases and transfers of pollutants were identified within a radius of \num{500} \unit{m}, including the Swiss territory.

The risk of the dam bursting has also been analysed. Such an event could cause water to leak from the reservoir and flood the surrounding areas to a greater or lesser extent. None of the communes in France where the surface sites are located are at risk of dam failure. In the canton of Geneva, there are two dams, each downstream of the PB surface site thus, the site is not subject to the risk of dam failure.

There is no nuclear installation present near the surface sites, in France and Switzerland. 
Furthermore, there is no facility subject to the ICPE nomenclature for activities related to pyrotechnic risk, within a radius of \num{500} \unit{m} around the sites, on French territory.
In Lake Geneva, the entire geographical area of the Petit Lac is polluted by the presence of munitions dumped between 1948 and 1979. These are munitions such as shells, rifle cartridges, aircraft bombs and other explosive residues. The PA and PB sites are located \num{4} \unit{km} from the shores of the lake. Thus, these surface sites are not subject to the pyrotechnic risk associated with un-decommissioned munitions in the lake.

In summary, some sites are located near transport routes such as autoroutes, departmental roads and railways that carry or may carry dangerous goods. There is only a gas pipeline in close proximity to the PA, PB and PH sites, which needs to be considered when designing the surface site to minimise the risk of an environmental accident. However, there are no nearby nuclear facilities, polluting facilities or classified industrial facilities within \num{500}\unit{m} of any surface site, and the risk of a dam failure or pyrotechnic hazards from munitions in Lake Geneva is minimal. Overall, these sites pose a relatively low environmental and safety risk, with special consideration being given to infrastructure near transport routes.

\subsubsection{Polluted sites}

In France, polluted sites and soils are described as locations that, due to past waste disposal or pollution infiltration, pose ongoing risks to people or the environment. These issues often stem from outdated waste disposal practices, chemical leaks, or accidents. Some areas also experience pollution from atmospheric fallout, which has accumulated over many years. It is different from diffuse pollution, like that from agricultural practices or car emissions near major roads. Industrial or agricultural activities that could cause pollution or risks to the local population are considered classified installations and are subject to regulations. The national policy aims to prevent future pollution while managing existing sites and ensuring they are safe for their intended use. Three complementary databases (BASIAS, BASOL, SIS) provide comprehensive pollution diagnostics, with detailed inventories of polluted sites and their risk levels.

In Switzerland, polluted sites refer to locations where waste has been permanently stored, such as landfills, as well as areas where waste has been stored or infiltrated. Contaminated sites are those that cause harmful or disruptive effects on the environment or have the potential to do so in the future. 
Thanks to waste regulations established in the 1990s, Switzerland has prevented the creation of new contaminated sites, by prohibiting hazardous waste landfills and untreated urban waste. It has put the infrastructure for proper waste treatment in place, ensuring that any waste is handled responsibly. 

For surface sites, PA, PD, PG, PH and PB there are sites listed in BASIAS, BASOL, SIS or Swiss database located within a radius of \num{500}\unit{m}, however, none of the areas of PA, PB, PD, PF, PG, PH, PJ and PL have contaminated sites in immediate proximity.

\subsection{Other projects}

\subsubsection{Introduction}
In an environment that is subject to continuous development, urbanisation and demographic development in a cross-border context, the constraints and opportunities that can emerge from other projects must be considered when conceiving a new research infrastructure with significant territorial development needs. Therefore, the analysis of the initial state of the environment included the establishment of an inventory of projects that were planned and constructed and that were potentially relevant to the FCC. Continuous monitoring needs to be implemented to keep this inventory up to date and to act rapidly in case a new project concept is developed, in order to understand if synergies are possible and to avoid potential conflicts emerging.

Today, the most relevant other projects to be considered are:
\begin{itemize}
\item Geothermal exploitation
\item Lake crossing
\item Grand Gen\`{e}ve development project
\item Grand Annecy development project
\item Enlargement of the departmental road D903 and integration with the A40 autoroute in France
\item District heat network in Switzerland
\item Heat networks in France
\item Water networks in Switzerland
\item Water networks in France
\item Railway network development Nord Gen\`{e}ve
\item Railway network developments in France
\item Territorial development in the Dingy-en-Vuache, Vulbens, Valleiry sector
\item Development of Groisy
\item Development of Ferney-Voltaire and Geneva international sector
\end{itemize}

\subsubsection{Geothermal exploitation}

The energy provisioning strategy of the canton of Geneva foresees extensive further development of heat recovery from geothermal sources that range from about 100\,m depth to several hundred metres. The investigations carried out in the frame of the feasibility study reveal no incompatibilities between the FCC and existing geothermal installations  today. Given the depth of the FCC subsurface structures, it cannot be excluded that further geothermal probes lead to potential conflicts. It is, therefore, of utmost importance to implement an early warning system based on the current reference trace that permits CERN, as the project owner, to engage with the persons who plan to create such installations. A minor displacement of the probe can frequently resolve any incompatibility. Should a probe be built without knowledge or until the subsurface volumes are properly protected from other projects, the conflicting geothermal probe would have to be removed, and the loss would need to be replaced. In France, no geothermal probes are known in the vicinity of the current reference scenario and no knowledge of a department or region-wide geothermal campaign exists. However, the need for reserving the subsurface volumes is equally important on French territory.

\subsubsection{Lake crossing}

Various subsurface lake-crossing projects have been conceived on Swiss territory during recent decades in the sector of the FCC, but they are not in its immediate vicinity. Most recently, a metro railway tunnel project has been proposed. Such a project can lead to potential synergies as well as to potential conflicts. Synergies would exist around subsurface investigations, tunnelling technologies, and excavated materials management. Conflicts could arise from overlapping construction schedules that could lead to increased nuisances, increased difficulties for the management of excavated materials, availability issues for engineering companies and workers, and organisation of construction sites and activities. The convolution of the authorisation processes of two almost concurrent projects could eventually generate societal acceptance issues for both projects. Care must also be taken to ensure appropriate designs for both projects in case the two tunnelling projects overlap, although the depths differ. 

\subsubsection{Grand Gen\`{e}ve development project}

Territorial planning documents in both host countries integrate the continuous development of a `Grand Gen\`{e}ve' in their planning documents. In addition, it is a cross-border project that involves substantial coordination. The project involves numerous improvements, such as road and railway mobility, urbanism, economic development, environmental sustainability, housing, social cohesion and cross-border governance. The project facilitates the extension of CERN's scientific activities in a larger zone. It also creates constraints, since the urban development implies stricter preservation of natural and agricultural zones. This rapid evolution needs to be considered in the further developments and authorisation processes of the research infrastructure. The existing planning documents span a time horizon until 2030. Updates that are imminent need to include a future particle collider project in the region. Therefore, the timeliness of deciding for an intent to advance with a construction project or not is important. As the example of the departmental road enlargement in the Nangy sector has shown, the improvement of mobility can be an opportunity. However, it can also rapidly jeopardise the feasibility of the current scenario when surface site candidate locations are concerned.

\subsubsection{Grand Annecy development project}

The regional development around Annecy is captured in numerous planning documents in France with a long-term horizon of 2050. It concerns mobility, management of natural resources, common use of infrastructures and services, preservation of nature and agricultural spaces, support for innovation and energy transition, development of eco-responsible tourism, development of education and training offers, social cohesion and solidarity, and the development of improved joint governance among municipalities, public and private actors, associations, and the public. This development permits the development of synergies with the FCC that for instance is also aiming at the development of innovation, education and training as well as high-quality tourism in the region, in particular in the sectors Groisy and Charvonnex that would host an experiment site. Conflicts could potentially emerge from the increased protection measures.

\subsubsection{RD903 enlargement and A40 integration}

The project to link the A40 autoroute with the 903 departmental road and to enlarge this road significantly in the immediate vicinity of the experiment site PD in Nangy was a potential risk for the feasibility. Timely interaction with the Host State administration permitted the development of a suitable solution to adapt the surface site to the road project. Depending on the start date and duration of the road construction works, very good coordination between the two projects is required.

\begin{figure}[h!]
    \centering
    \includegraphics[width=0.6\textwidth]{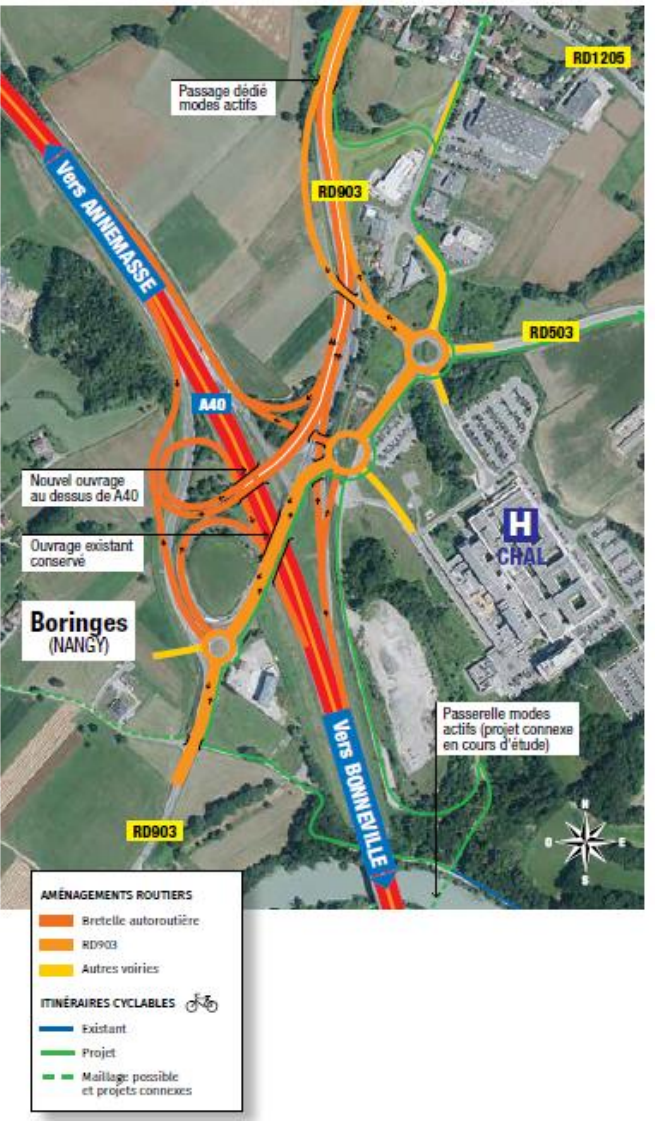}
    \caption {Overview of the RD903 redevelopment project between the A40 Findrol interchange and the Chasseurs crossroads.}
    {\label{fig:RD903_findrol.pdf}}
\end{figure}

\subsubsection{District heat network in Switzerland}

The canton of Geneva has committed\footnote{\url{https://www.ge.ch/installer-remplacer-chauffage/reseaux-thermiques-structurants-rts-0}} to substantially enlarge its heat distribution networks, in particular with major heat transport axes. The integration of geothermal heat, the GeniLac and the GeniTerre systems ensure heating and cooling capabilities in view of supporting the achievement of climate protection goals. This project leads to concrete synergies with the future research infrastructure that will be able to supply significant amounts of waste heat that can be injected into this network. No conflicts are identified.

\subsubsection{Heat networks in France}

Several district heat networks are starting to be conceived and developed in the perimeters of various candidate surface site locations in France. This concerns, but is not limited to, the PA site in Ferney-Voltaire, for example. This development does not generate conflicts, but as the case of Ferney-Voltaire shows, it creates concrete synergy potentials. The network in Ferney-Voltaire today profits from the heat supplied by the Large Hadron Collider. A future particle collider would not only ensure continuity, but would permit a significant increase in the heat supplied. Specific studies of the demand have been carried out in the frame of the feasibility study and these have confirmed this potential synergy.

\subsubsection{Water networks in Switzerland}

The regional plans for water treatment in the Geneva region are currently under revision, but details are not available. Although conflicts with the FCC project are unlikely, it is prudent to monitor the evolution of the regional water management plan developments. The raw water for cooling the particle collider may have a high fraction of undissolved residuals after multiple recirculation. It is important to ensure that the residual materials, including the total dissolved solids (TDS) in the cooling water, can and will be accepted by the regional water treatment stations.

\subsubsection{Water networks in France}

Similar to the situation in Switzerland, the knowledge of the planning and evolution of the local water networks and water treatment facilities is not centrally available. Although conflicts with the FCC project are unlikely, it is prudent to monitor the evolution of the regional water management plan developments. As mentioned above, it is important to ensure that the cooling water can and will be accepted by the regional water treatment stations.

The water treatment facility in the vicinity of site PD represents an opportunity for the project. The study included the verification of the feasibility of extending the water treatment facility with infrastructures to accept the residual cooling water from the particle collider. In addition, the study revealed that it is, in principle, technically and economically feasible to treat the waste water from the treatment plant and to use it for water cooling systems. This generates a potential to reduce raw water consumption and creates a socio-economic benefit potential for using treated waste water when the particle collider does not need it.

\subsubsection{Railway network development Nord Gen\`{e}ve}

The existing Léman Express is already reaching its maximum capacity and road traffic saturation calls for further development of the railway infrastructure.
The development of a north-south railway transport axis is a priority of the Geneva canton. A conflict with the FCC is unlikely and the additional transport may, on the other hand, lead to potential synergies for the operation period of the new particle collider infrastructure. 

\subsubsection{Railway network developments in France}

The continuous increase of road traffic, accompanied by traffic saturation and air quality impacts also calls for a development of the railway infrastructure in France in the region concerned by the FCC. This mainly involves the Léman Express lines that pass in the vicinity of some of the FCC surface sites. Therefore, the development of the railway system presents opportunities for the project and potentially for the territory.

\subsubsection{Territorial development in the Dingy-en-Vuache, Vulbens, Valleiry sector}

The modernisation of the A40 autoroute in the Dingy-en-Vuache, Vulbens and Valleiry sectors is planned by the operating company ATMB for between 2023 and 2028. It is important that the construction of the FCC starts after the tunnel through the Vuache has been refurbished to avoid limitations. If this is the case, the planned autoroute works do not lead to a conflict with the FCC. However, timely design work must be foreseen concerning the autoroute access for site PD for the FCC construction period. It would take about 10 years to put such access in place.

The territorial developments around local mobility (e.g., bike paths), education (e.g., high school) and emergency services (e.g., fire brigade) represent attractive opportunities to develop synergies with the FCC. Timely engagement with the local, departmental and regional stakeholders is required to leverage the FCC project, support these territorial development projects and integrate them into the surface site activities at PJ.

\subsubsection{Development of Groisy}

A school development project was validated by the department of Haute Savoie in 2023. There is no conflict with the FCC project. However, this and related further development projects in Groisy and Charvonnex represent attractive opportunities to generate synergies such as the supply of waste heat, high-quality tourism and the creation of apprenticeship programmes and other cooperation in the education domain.

\subsubsection{Development of Ferney-Voltaire and Geneva international sector}

The commune of Ferney-Voltaire has recently launched a large-scale redevelopment and modernisation programme around a new commercial activity zone (ZAC) in the vicinity of the PA surface site. The supply of waste heat from the existing CERN LHC point 8 (LHCb) site is also integrated in the development of a local district heating network. Developments that create synergies with the Geneva airport are also envisaged in this sector. The plan also foresees the creation of high-tech and innovative facilities and a significant increase in housing capacity. At the same time, on the Swiss side of the border, significant development activities have started in Grand Saconnex. The development of the local tramway system connecting the Geneva international sector (United Nations) with Ferney-Voltaire has been planned.

\begin{figure}
    \centering
    \includegraphics[width=0.8\textwidth]{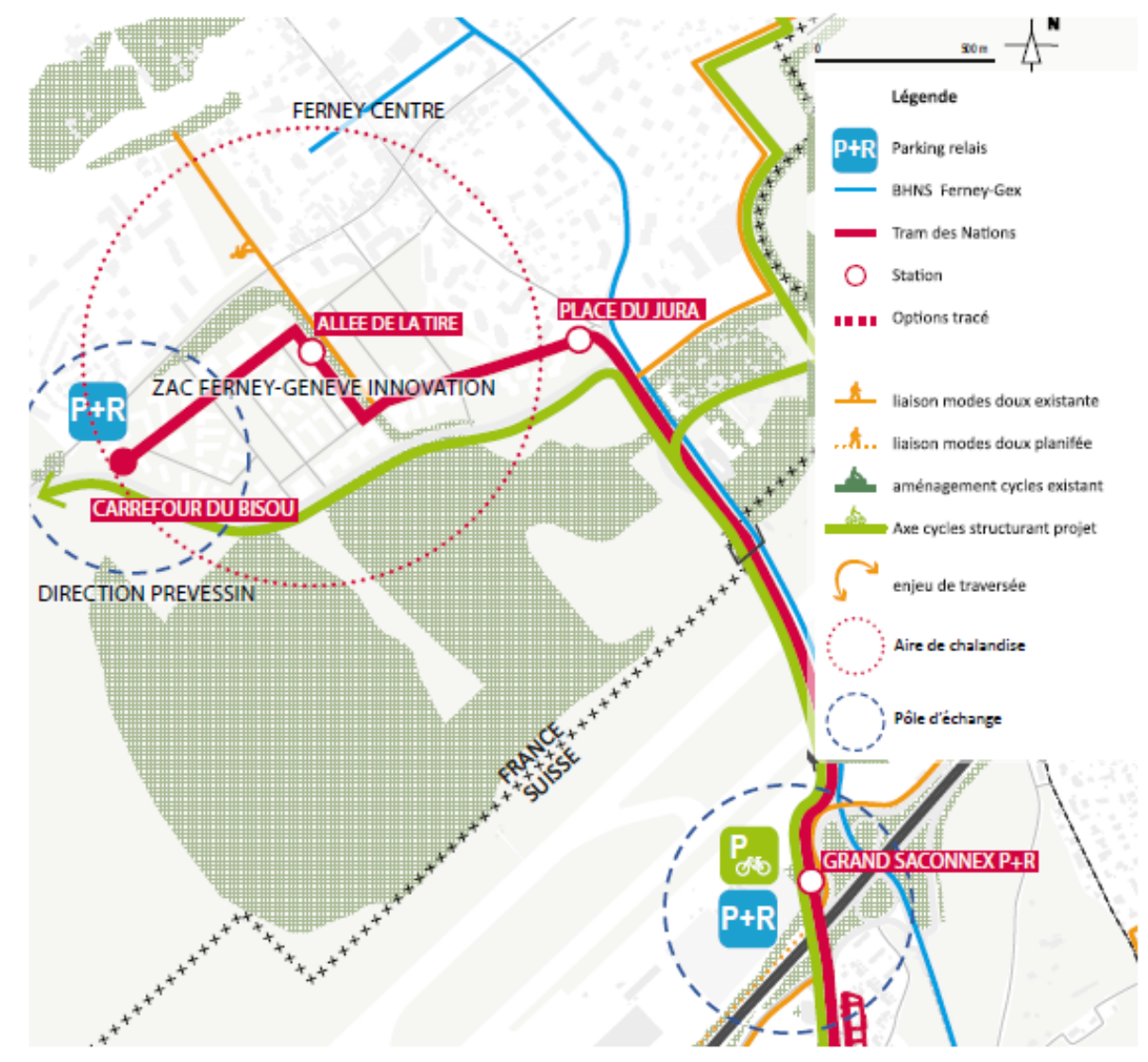}
    \caption {The future Ferney Genève ZAC project with an extension of the tram line and terminus located less than a kilometre from the PA site.}
    {\label{fig:Tram_nations_ferney.pdf}}
\end{figure}

All these local economic development activities can potentially create conflicts with the FCC construction activities due to a local concentration of construction sites and construction-related traffic. A timely coordination with the local actors on the planning of an FCC construction is therefore needed.

On the other hand, these developments can lead to significant cross-fertilisation in terms of waste heat supply, mobility, education and economic developments.

\section{Conclusion}

The feasibility study phase anticipated studies and field investigations that are part of the environmental authorisation process in both countries. Those include, for example, the establishment of an initial state of the environment, the identification and prioritisation of noteworthy environmental aspects, the identification of indirect and induced connected enabling projects (road accesses, electricity, water and water treatment) as well as informal engagements with the public and their representatives. The scenario development process has been documented together with variants, and evolution of versions, the reference scenario and project invariants.

The studies permitted the identification of relevant issues and points requiring attention. The critical subsurface zones were located and the need for a continuation of subsurface investigations in order to establish a comprehensive and detailed 3D model was identified. The latter will serve as input to the construction process. The results of the preliminary subsurface investigations will provide information about an improved depth and inclination of the tunnel alignment. The initial state analysis revealed that different surface sites are subject to very different environmental constraints and present different opportunities for the creation of synergies. The work on the description of the project elements and the environmental aspects showed that an iterative approach is required due to the long and iterative development of the technical designs. A detailed description of the subsurface construction works is needed as a first step to assess the environmental impacts of the construction phase. 
A plan for the adequate management of the excavated materials according to industrial best practices and suitable for presentation to notified bodies in both Host States needs to be established as soon as the preliminary subsurface investigation results are available. The particle accelerator requirements need to be formally documented to permit a sufficiently detailed development of the technical infrastructure designs that are needed to assess the environmental impacts. An eco-design strategy and guidelines have been developed that need to be integrated in subsequent project development by putting a project-wide, transverse systems engineering approach in place.

Although challenges and environmental sensitivities were revealed and documented, no fundamental showstopper with respect to technical, engineering and environmental feasibility could be identified.

The reference locations of the surface sites will require very different optimisation, reduction and integration approaches. Where residual impacts cannot be mitigated with these approaches, compensatory measures may be required. Such measures can differ substantially in terms of type, size and approach among the site locations. For the loss of protected agricultural spaces, a 1-to-1 compensation by re-creating the lost area by transplanting the topsoil to wastelands or areas with poorer quality is envisaged. Rewilding is another way of compensating for the loss of habitats and biodiversity. 

Reforestation around surface sites where forests have been removed is an option, along with new forest establishment in previously non-forested areas. Landscape integration and preservation of visibility can be achieved by half-buried site elements, the separation of sites into different segments and by terracing. 

The use of decarbonised, including renewable energies, aims at keeping Scope 2 emissions low. Incremental socio-economic benefits can, for example, be generated by entering energy supply contracts and agreements that include the creation of new renewable energy sources and by integrating waste heat recovery and supply from the onset. Location-specific innovation projects can lead to further reductions of the environmental footprint and incremental socio-economic benefits such as the use of treated waste water, the creation of soft mobility infrastructures, the development of high-quality tourism and local services around some of the surface sites and the creation of training and education opportunities in the project as a whole and at individual surface sites.

\chapter{Sustainability}

\section{Context}
\label{sustainability:context}

CERN’s longstanding commitment to sustainability, integrating scientific, environmental, societal goals is a guiding principle in the development of a future particle collider infrastructure. The results presented in this chapter show the integration of strategies in the scenario development process that align with the Organization’s guiding principles and policies, ensuring that the FCC is conceived as a model of responsible large-scale scientific infrastructure development.

From the outset, the studies and developments for the FCC integrated CERN's existing guiding principles for the protection of the environment. It includes topics that extend beyond the biosphere, covering several aspects of the three sustainability dimensions. Examples include optimising resource use, limiting greenhouse gas emissions, and prioritising sustainable excavation and material reuse. The feasibility study also considered CERN’s ongoing biodiversity and land-use management efforts, ensuring that construction and operational activities respect local ecosystems. It builds on the results and experience of CERN’s past and present energy efficiency and waste heat recovery initiatives, aiming to reduce overall energy consumption, make responsible use of energy and contribute to regional energy networks. Water conservation, noise management, and emissions mitigation are further key elements of the study, reinforcing CERN’s goal of balancing cutting-edge scientific progress with responsible environmental stewardship.

Achieving these objectives requires close cooperation with the authorities in both Host States, as well as meaningful engagement with the public. The feasibility study provides the foundation for the dialogue with national and regional regulatory bodies, ensuring that the FCC project aligns with environmental legislation and sustainability goals in both France and Switzerland. Furthermore, open and continued discussions with local communities will be essential to address concerns, share best practices, and build a collaborative approach to come to an environmental, social, and governance (ESG) framework that can ensure the long-term sustainability of such a new research infrastructure.

By incorporating lessons from past and existing projects at CERN and by leveraging the innovations adopted by other large-scale infrastructure projects in science and beyond, the FCC feasibility study seeks to set new standards for sustainability in large-scale scientific projects. 
It builds on the principles laid out in international and national laws and regulations, best practices and CERN’s published policies and strategies, integrating best practices in impact avoidance, reduction and mitigation, circular economy principles, and stakeholder engagement. As such, this chapter lays the groundwork for a potential FCC implementation project that not only advances knowledge creation through fundamental physics research but does so with a firm commitment to long-term sustainability that embraces scientific excellence, the environment and society.

\section{Introduction}
\label{sustainability:introduction}

\subsection{Sustainable research infrastructures}

Research infrastructures (RI) are facilities that provide resources and services for research communities to conduct research and foster innovation~\cite{EURIsite}. An RI can be a single facility, such as, for example, the European Spallation Source (ESS) or the European XFEL, or it can be a facility that is part of an organisation that hosts multiple facilities, such as the Large Hadron Collider (LHC) and other particle accelerators at CERN (e.g., Antiproton Decelerator, ELENA, PS, SPS) that offer dedicated scientific research opportunities. The Future Circular Collider will be a research infrastructure that is hosted by CERN, conceived and constructed by an international collaboration, providing open access to a worldwide community of scientists.

\begin{figure}[h]
    \centering
    \includegraphics[width=0.4\textwidth]{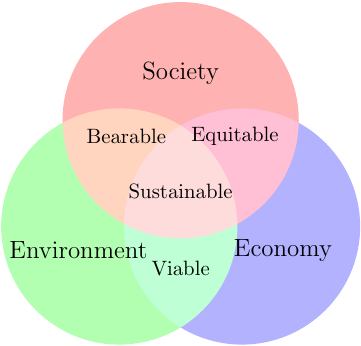}
    \caption{Sustainability dimensions.}
    \label{fig:sustainability_dimensions}
\end{figure}

Sustainability refers to the ability to maintain an activity at a certain rate or level. It integrates three stakes: Society, Economy and Environment(Fig.~\ref{fig:sustainability_dimensions}). A science programme or project can be considered sustainable if it is able to successfully address and complete its scientific core mission satisfying the requirements of three stakes: it obtains a `social license' to operate~\cite{social_license}, it maintains an ecological balance applying an avoid-reduce-compensate sequence~\cite{Cerema2021}, and it is affordable in the long term with well-understood and managed risks.

\begin{figure}[h]
    \centering
    \includegraphics[width=0.8\textwidth]{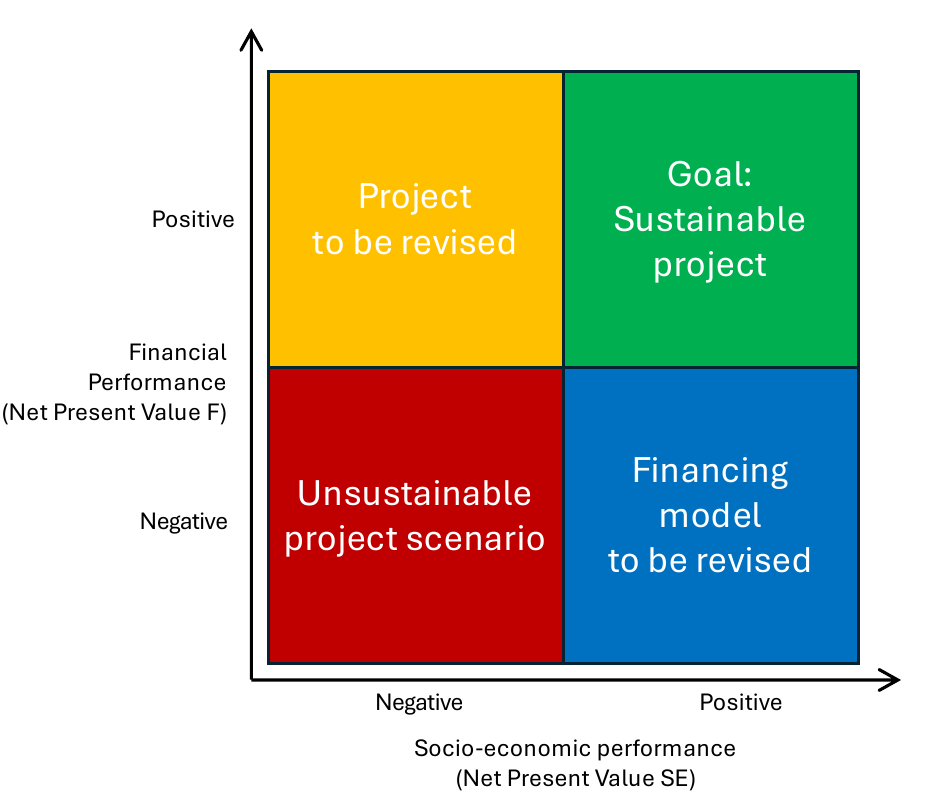}
    \caption{Guidance for the sustainability of public investment projects. From Ref.~\cite{SGPI2018}.}
    \label{fig:guide_sustainability}
\end{figure}

An appraisal process helps understanding if a programme or project is sustainable and provides means to identify pathways to make them more sustainable. This process integrates financial and socio-economic aspects. The latter comprise social and environmental benefits as well as negative effects including environmental costs. The ISO standard for monetisation of environmental impacts (ISO 14008\cite{iso14008}) and the guidelines for determining environmental costs and benefits (ISO 14007\cite{iso14007}) are evidence for the importance of applying such environmental unit costs in environmental and sustainability reporting at the organisation and project levels.
Science projects can achieve a positive socio-economic net present value. If their financial performance is too low, the funding model needs to be revised. Projects that are financially viable but have no or little socio-economic benefits need to be revised. Projects that are financially and socio-economically positive can be sustainable (Fig.~\ref{fig:guide_sustainability}).

Early and accompanying socio-economic analysis that integrates all sustainability dimensions helps to benchmark variants and versions of project scenarios and permits planning for long-term sustainability. An RI needs to periodically monitor and track the social, environmental, and economic performance against the initial estimations in order to implement a continuous improvement process.

In line with the `Ecodesign' EU Directive 2009/125/EC~\cite{ECecodesign}, sustainable development requires proper consideration of all three sustainability elements, including the identification and accounting of positive and negative impacts throughout the entire lifecycle of a programme or project. For RIs carrying out science missions, at least the following stakes should be considered:
\begin{itemize}
    \item Economy
        \begin{itemize}
            \item Scientific excellence
            \item Total costs (capital and operation expenditures)
            \item Risks and residual risks after mitigation
            \item Direct, indirect and induced `value added' and employment
            \item Quantified incremental economic benefit potentials
        \end{itemize}
    \item Society
        \begin{itemize}
            \item Quantified incremental social benefit potentials
            \item Common good value (the value of the science mission as perceived by people)
            \item Territorial compatibility
            \item Social license
        \end{itemize}
    \item Environment
        \begin{itemize}
            \item Quantified negative externalities
            \item Quantified incremental environmental benefit potentials
        \end{itemize}
\end{itemize}

\subsection{Approach}
\label{par:RISustainability}

A new particle collider and its enabling technical infrastructure represent significant long-term investments for participating countries and funding partners. An appraisal process is necessary to gain
an understanding of the cost drivers and benefit levers. Reducing the former, developing the latter and conceiving a long-term financing model that relies on international collaboration are required to ensure long-term sustainability from the conceptual phase onwards. Funding agencies, science strategy bodies (e.g., ESFRI), national notified bodies in charge of issuing authorisations and investment banks who potentially grant long-term loans for publicly funded projects (e.g., EIB), request evidence for the project's viability. The structured process of assessing the case for proceeding with an implementation preparation project and validating the project's viability is called `project appraisal'.

This chapter also compiles a set of regulatory frameworks and appraisal guidelines that exist at the European level, in France and in Switzerland to provide a landscape of the sustainability requirements that govern the planning and implementation of new research infrastructure projects with strategic importance and territorial development needs at CERN. In the countries that are part of the European Research Area (ERA), legal frameworks govern the compliance with sustainability aspects. Other countries such as Australia, the UK and the USA are still largely applying panel-based project decision taking, although the UK does carry out ex-post evaluations on a case-by-case basis and Australia is evaluating the sustainability of research infrastructures periodically. Sustainability and environmental criteria such as climate considerations are increasingly being incorporated into proposal design and assessment. Switzerland has recently made sustainability aspects explicit for CERN's strategically important projects with territorial development in the update of the law~\cite{Swiss_parliament_20240214,Swiss_parliament_20240927} for the encouragement of research and innovation that introduces an authorisation process at federal level. The `Ordonnance concernant l’approbation des plans des constructions et installations du CERN (OCIC)'~\cite{fedlex_2025_02_06}, the prescription on how this plan for the authorisation process is to be implemented, has been issued for public review in February 2025. In addition to the commonly known environmental topics (e.g., air, water and soil) it includes the need for the assessment of the impacts on the climate and specific measures to comply with climate protection laws.

Consideration of quantitative environmental externalities\footnote{An environmental externality is a cost or benefit that affects a third party who didn’t choose to incur that cost or benefit, specifically in relation to the environment.}, both positive and negative, is still an emerging approach in the domain of research infrastructures unless it is explicitly required by national legislations and governed by national guidelines. Such a `wider' socio-economic analysis is, however, already common practice in the general infrastructure project appraisal in numerous European countries such as France~\cite{CGEDD:2017}, Germany~\cite{BMBF:2024, LeNa-framework:2023}, Italy and Switzerland and at EU level through EC funding conditions for numerous sectors, e.g., Connecting Europe~\cite{CINEA_CBA:2022}, transport~\cite{UNECE_CBA:2003}, energy~\cite{ENTSOE_CBA:2024}, regional investment projects~\cite{ECCBAGuide2014} as well as nature preservation and restoration~\cite{Natura200:2013}.

Methodologies for assessing public investment and assuring that a project contributes to the increase of public welfare\footnote{Public welfare refers to the collective well-being of society that is promoted through public services and institutions, aiming to ensure equity, social cohesion, and access to essential goods and services for all.
Paraphrased summary from Florio, M. (2019). Public Enterprises: Resurgence and the Future of the Public Sector. Springer.} exist across different policy sectors and institutions~\cite{CSILFlorio2023,10143094060} including environmental impacts\cite{OECD_CBA_2018} are already used for research infrastructure projects including particle accelerators like the ALBA light source~\cite{ALBA2023}, the SOLEIL light source~\cite{SOLEIL2020}, the DESY/PETRA III synchrotron\cite{DESY2023PetraIIIImpact}, the CNAO hadron therapy facility~\cite{BATTISTONI201679}, the LHC~\cite{Florio:2036479}, High Luminosity LHC (HL-LHC)~\cite{Bastianin2021,Bastianin:2319300, doi:10.1080/00036846.2022.2140763} and the Compact Linear Collider (CLIC) study~\cite{Magazinik:2836841}. The Future Circular Collider (FCC)~\cite{csil_2024_10653396} study has, in particular, devoted significant resources to this topic over a time frame of almost ten years and has contributed to the advancement of the research infrastructure appraisal process at an international scale through scientific contributions to the field and participation in collaborative impact assessment projects at EU level. The approach is also used for other science facilities such as the Einstein Telescope~\cite{EinsteinTelescope2021}, the Paris Saclay heat supply facility~\cite{Saclay_2015}, the Nantes University hospital~\cite{Nantes_2020}, and by the Commonwealth Scientific and Industrial Research Organisation (CSIRO)~\cite{CSIRO:2023} to evaluate a number of different science programmes. Ever more RIs, such as the Square Kilometre Array, are considering the approach in view of planning for sustainability~\cite{SKA:2023}. Environmental factors are increasingly being integrated into these assessments~\cite{FS2022,SGPI2018,Enerland2023,STM2022}. A comprehensive socio-economic impact assessment is needed when a new research infrastructure project proposal requests being entered into the European Strategy Forum for Research Infrastructures (ESFRI) roadmap \cite{ESFRI-roadmap-form} as indicated by the 2026 roadmap guide (see page 25 of Ref.~\cite{esfri2026guide}.

\subsection{The context in Europe}

Research infrastructure long-term sustainability at European Research Area (ERA \cite{ERA-strategy}) level is primarily guided by the European Strategy Forum for Research Infrastructures (ESFRI) \cite{ESFRI, ERIC} \footnote{ESFRI (\url{https://www.esfri.eu}) was established in 2002 with the purpose of developing a European approach to Research Infrastructure policy as a key element of the emerging European Research Area (ERA). Further Research Infrastructures contribute to this effort in the frame of the EIROforum and the European Research Infrastructure Consortium (ERIC).} 

EU regulation 2021/695 \cite{HorizonEurope} defining the Horizon Europe Framework Programme for Research and Innovation explicitly includes requirements to address global challenges, including climate change and the United Nations Sustainable Development Goals (SDGs).

EU member states typically translate and integrate the strategies, policies, and guidelines into their national roadmaps and plans in addition to their existing national policies.

A robust long-term vision is a prerequisite to successfully and sustainably operate a research infrastructure. Therefore, ESFRI issued a number of recommendations and actions\cite{ESFRI:2017}.

Scientific excellence is the condition sine qua non. However, sufficient funding and sustainable funding models, required across the entire life cycle, are indispensable for a successful strategy for a new research infrastructure. Together with adequate human resources, it is crucial for the operational phase. Effective governance is another key element for ensuring long-term sustainability. Moreover, RIs should contribute to their sustainability by contributing to their carbon neutrality in order to support the carbon neutrality goals at the national level. It is worth mentioning that carbon neutrality accounting is taking place at a national level.

While the current recommendations place an explicit focus on scientific excellence, financial sustainability and societal acceptance, the call for comprehensive sustainability and impact assessment implicitly includes all environmental aspects. The recommendations also spell out that RIs should dedicate sufficient resources to periodically evaluate and communicate their socio-economic performance to various audiences. CERN's pioneering activities in this domain are explicitly mentioned by the ESFRI guide\cite{esfri2020cost}, encouraging national authorities to support the approach in cooperation with experts in the field.

The ESFRI policy brief \cite{ESFRI:2023} details the need to tailor the impact assessment to the specific project and requests the inclusion in the roadmap of project proposals that aim for implementation on a ten-year time frame that the wider socio-economic impact assessment is provided. Following the OECD definition \cite{OECD-criteria} these impacts comprise \textit{``the extent to which the intervention has generated or is expected to generate positive or negative, intended or unintended, higher-level effects''}. The European Commission \cite{europa-guidelines} states that ``\textit{The term impact describes all the changes which are expected to happen due to the implementation and application of a given policy option/intervention [such as investment in a Research Infrastructure and its activities]. Such impacts may occur over different timescales, affect different actors and be relevant at different scales (local, regional, national and EU)}''. Ex-post assessment is needed to determine whether the intended objectives and the ex-ante estimations have actually been achieved. 

EU Regulation 2021/1060 of the European Parliament and of the Council of 24 June 2021 lay down common provisions on the European Regional Development Fund, the European Social Fund Plus, the Cohesion Fund, the Just Transition Fund and the European Maritime, Fisheries and Aquaculture Fund and financial rules for those and for the Asylum, Migration and Integration Fund, the Internal Security Fund and the Instrument for Financial Support for Border Management and Visa Policy. According to Article 100 of Regulation (EU) No 1303/2013 \cite{eu1303_2013}, a major project is an investment operation comprising “a series of works, activities or services intended in itself to accomplish an indivisible task of a precise economic or technical nature which has clearly identified goals and for which the total eligible cost exceeds EUR 50 000 000 […]”. A socio-economic impact assessment of such major projects is recommended. The European Commission Economic Appraisal Vademecum 2021-2027 \cite{ECVademecum} captures the general principles and provides sector application examples for impact assessment, including research and innovation in Annex I. It extends and complements the common provisions regulation that recommends cost-benefit analysis in line with EU regulation (EU) No 207/2015 \cite{ECCBA2015}. The methodology is explained in detail in the `European Commission Guide to Cost-Benefit Analysis of Investment Projects' \cite{ECCBAGuide2014}. 

Climate change adaptation and mitigation and disaster resilience are covered by this regulation. For instance, the volume of greenhouse gas (GHG) externality \footnote{An externality is a cost or benefit that is caused by one party but financially incurred or received by another. Externalities can be negative or positive. A negative externality is the indirect imposition of a cost by one party onto another. A positive externality, on the other hand, is when one party receives an indirect benefit as a result of actions taken by another.}  and the external cost of carbon is explicitly included, and alignment with the EU 2050 decarbonisation objectives is required. Concerning climate adaptation, the costs of measures aiming at enhancing the resilience of the project to climate change impacts that are duly justified in feasibility studies should be included in the economic analysis. The benefits of these measures, e.g., measures taken to limit the emissions of GHG or enhance the resilience to climate change, weather extremes and other natural disasters, should also be assessed and included in the economic analysis, if possible quantified; otherwise, they should be properly described. 

An analysis of sustainable development (environmental protection, resources efficiency, climate change mitigation and adaptation, biodiversity and risk prevention) should be covered. An analysis of the options considering technical, operational, economic, environmental and social criteria for the location of the infrastructure is requested to be in a feasibility study. This also describes the project's consistency with the applicable environmental policy. Considerations should include resource efficiency, preservation of biodiversity and ecosystems, reduction of GHG emissions, and resilience to climate change impacts. The process needs to fulfil the Directive 2011/92/EU, that defines the environmental impact assessment (EIA) process, which ensures that projects likely to have significant effects on the environment are made subject to an assessment prior to their authorisation. Therefore, the total costs of the negative environmental impacts and their compensation have to be included. Environmental benefits can be assessed and added as well. They include, for instance, contributions to improve water supply and sanitation, waste management, energy capacity and stability, transport, ports (airports, seaports, inter-modal), research and innovation and broadband communication.

The appraisal process aims to assess if a project will contribute to overall social welfare and economic growth, taking into account benefits and costs to society. The EC \cite{ECCBAGuide2014} and UNIDO \cite{UNIDO_handbook} handbooks focus on economic and societal topics, although aspects such as assessment of the environmental externalities are typically part of project appraisal as required per EU regulation. For instance, the shadow cost of carbon and the GHG emissions are shown in project appraisal examples. Eventually, the requirements for appraisal are defined for each project, specifically by the notified body for that project. For instance, to obtain funds from the European Investment Bank (EIB), a comprehensive appraisal study including the positive and negative environmental externalities is required. The EIB published a guide \cite{EIBCBAGuide2023} dedicated to this topic that includes references for the shadow cost of carbon and also comprehensive calculation examples and guidelines for capturing environmental externalities.

Other requirements emerge, for instance, from applying to the Connecting Europe Facility (CEF). The InvestEU regulation introduces climate, environmental and social sustainability as elements in the decision-making process when applying for the InvestEU Fund. The process is also a requirement in the framework of the preparatory phase of ESFRI projects. The European Bank for Reconstruction and Development (EBRD) requires project assessment with potentially relevant greenhouse gas (GHG) emissions of more than 25\,000 tonnes of CO$_2$(eq) per year with respect to a baseline of 100\,000 tonnes of CO$_2$(eq) emissions per year. National requirements differ substantially from each other, requiring assessments for investment projects with public funding as low as for instance 300\,000\,euros in Lithuania.

\subsection{The context in France}

In France, the `Code de l'environnement' \cite{CodeEnvironnementFrance} guides the approach to develop a sustainable project scenario in the frame of an environmental evaluation process that is part of the authorisation of the project. The term `environment' is to be understood in its original meaning and in a wide sense: the surroundings and conditions in which the project will be placed. 

\begin{figure}[h]
    \centering
    \includegraphics[width=\textwidth]{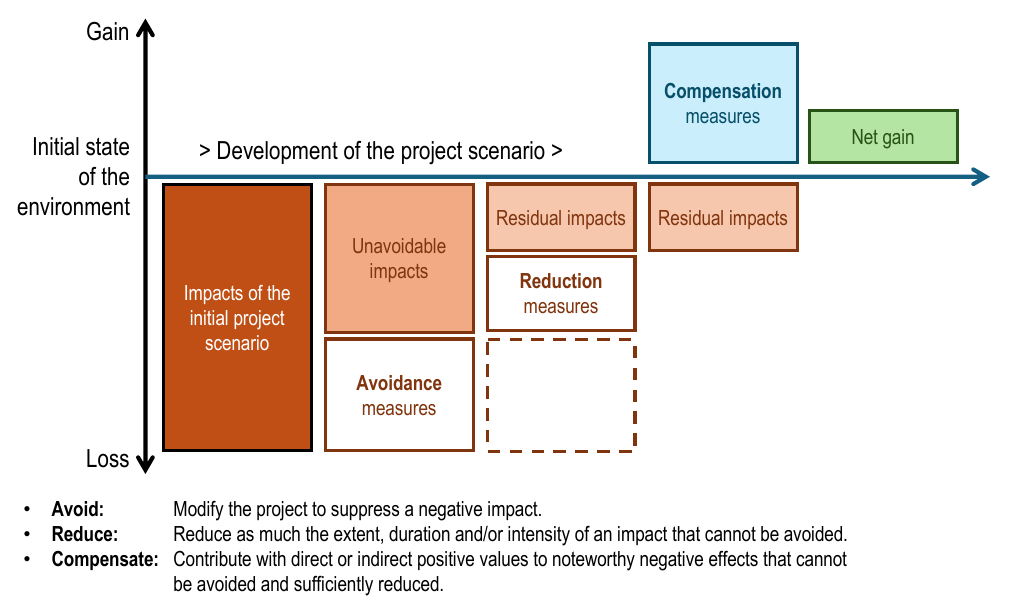}
    \caption{`Avoid-Reduce-Compensate' approach for iterative development of a sustainable project scenario to achieve an ecological balance and, as far as possible, a net gain\cite{ERC:2021}.}
    \label{fig:ERC2}
\end{figure}

The goal of the process is to develop a feasible and sustainable project scenario following the `avoid-reduce-compensate' method (French: éviter-réduire-compenser, ERC), which is anchored in European regulations and which is implemented in French law \cite{EU-reg-38247366}.
As a consequence, the project scenario to be authorised aims at a net positive value (see Fig.~\ref{fig:ERC2}).

This approach of iterative development of variants and versions that are continuously improved is also documented by the international standard ISO 14001 concerning environmental management. 
It describes the Plan-Do-Check-Act (PDCA) model as an iterative process to ensure continuous improvement (ISO 14001, Step by Step, Chapter 1). 

Considering that programme and project evaluation in France is compulsory for public investments that exceed 20\,million euros and a second assessment is required for investments that exceed 100\,million euros, the national assembly tracks and reports on such projects every year using a standard information sheet approach that includes the net present value as an overall sustainability indicator and the greenhouse gas emissions avoided. 
Such information is published as an annex of the annual budget  law \cite{ProjetBudget2024France}. In 2023, 13 research investments were subject to comprehensive socio-economic evaluations. 

The evaluation concerns the entire project throughout all life cycle phases. It is understood that the level of description detail for individual phases and project segments can initially differ. The process accompanies the project throughout its entire lifetime. Consequently, the documentation needs to be regularly updated. The process records the initial state of the environment, including nature and a variety of other topics such as urbanism, public health, population safety and economic impacts. 

The process actively identifies and assesses relevant effects on the project environment (environmental impact analysis) and gives input to the design process following the avoid, reduce and compensate steps. 
The process includes the involvement of the population via, for instance, an informal dialogue and participation phase and a formal public consultation processes (`d\'{e}bat public') and public inquiry (`enqu\^{e}te publique'). 
The latter processes is carried out with national notified bodies.

To capture the sustainability performance properly, the French government requires a socio-economic evaluation for each project that is funded via public investments of more than 20 million euros. The evaluation must be carried out according to the guidelines of the `Secr\'{e}tariat g\'{e}n\'{e}ral pour l'investissement' (SGPI)\cite{SGPI:2023}. 
Projects with a total public investment of more than 100 million euros are subject to a second expert assessment, on which the SGPI provides an assessment to the French parliament and the Prime Minister as well as to the minister under whose responsibility the project falls.

The socio-economic evaluation comprises a detailed description of the project, its variants and alternatives, the key characteristics, the implementation schedule, a list of relevant socio-economic indicators, a list of indicator values showing how the project performs with respect to public policies (e.g., climate impact reduction objectives), the financial plan, compliance with laws and regulations and a risk registry.

Until 2020, socio-economic evaluations of programmes and projects included only limited environmental aspects such as habitats and green spaces. The ever-growing importance of more global topics led to a broadening of the positive impacts and negative externalities covered. 
The 2023 impact assessment guide in France~\cite{SGPI:2023} explicitly includes all environmental aspects that are project relevant such as CO$_2$ equivalent emissions, noise, air pollution, water use and pollution, soil use and pollution and requires them to be associated with monetary values and to quantify the positive contributions in monetary terms. 
For instance, the legal value of one tonne of CO$_2$ for the socio-economic evaluation has been set to \euro{32} (base year 2010) in the 2023 guideline, which corresponds to about \euro{40} in 2024. 
Despite this value, the annex that indicates the guidelines of converting non-tradeable goods and externalities into monetary terms indicates a progression of the shadow cost of carbon \euro{250} per CO$_2$(eq) to 2030 and \euro{775} per tCO$_2$(eq) in the year 2050. This progression is in line with the EIB recommendations of the shadow cost of carbon.

The socio-economic evaluation of the FCC is based on a standard cost-benefit analysis, reporting a net present value and a benefit/cost ratio based on the comparison of discounted total costs and benefits over a defined observation period. Costs and benefits were quantified with respect to a "counterfactual scenario" in which CERN operates the LHC until its expected end of life and continues to act as a scientific research platform with the existing particle accelerator complex. Noteworthy positive and negative direct effects and externalities including environmental effects were taken into consideration for the entire life cycle of the project.

For the social discount rate (SDR), the 2023 guidelines for socio-economic impact assessment in France \cite{FranceStrategie2023} recommend using a value of 3.2\% for projects with a lifetime up to the year 2070. However, for very long-term and low-risk projects such as research infrastructures, a lower social discount rate in the range of 2.5\% can be justified. For the FCC CBA, economists selected a SDR of 2.8\%, based on the weighting of the SDRs in each individual CERN Member State considering its annual financial contribution to CERN.

If there is a `d\'{e}claration d'utilit\'{e} publique' (DUP), which governs the authorisation process for public investment projects, the socio-economic evaluation is part of the legally required `enquête publique' files to obtain project authorisation.

The socio-economic evaluation has to be updated regularly, following the iterative approach of refining the project scenario and comparing expected effects to the actual effects once the project is implemented (`ex-post' analysis).

A project can be considered sustainable if both the socio-economic and the financial net present value are positive. 
In addition, the social discount rate applied should be larger than the average weighted cost of the capital needed to finance the project. This can be verified by determining the so-called Internal Return Rate (IRR), which is closely linked to the calculation of the net present value. It expresses at which value of the social discount rate the net present value of the project would become zero. For instance, if the loans for the project are granted at 3\% but the net present value becomes zero at only 5\%, the condition would be satisfied.

If the project does not generate revenues, i.e., in the case of a research infrastructure for fundamental scientific research, no financial net present value can be provided, a positive socio-economic net present value and an internal return rate higher than the cost of the capital support the sustainability condition.
If the socio-economic net present value is negative and the financial net present value is positive, the investment may be financially viable, but it is not recommended from a societal point of view.

\subsection{The context in Switzerland}

The authorisation process in Switzerland is based on guiding principles that require the demonstration of compliance of the scenario with all applicable, individual laws and regulations. All laws and regulations that would apply to the project need to be identified and explicitly listed. It is considered best practice to follow the Avoid-Reduce-Compensate approach as required in France and Europe with continuous involvement of notified bodies, public administration services, and the population. It is to be accompanied by monitoring and continuous improvement measures.

With the recent update of the law for the encouragement of research and innovation (LERI)~\cite{LERI:2023} and the development of a dedicated plan for new constructions and installations of CERN~\cite{ParlamentSchweiz:2024}, guidelines with respect to the sustainability of research infrastructures become more specific.

The consumption of land, in particular, high-quality crop-rotation arable land that is regarded as a precious resource, needs to be avoided. Where this cannot be achieved, the residual consumption of such land that is registered in a national inventory needs, in principle, to be compensated. Therefore, alternatives and variants of the project need to be documented and assessed. Compensation can, for instance, be achieved by stripping the topsoil of the original land affected and using that soil for creating high-quality arable land within an agreed geographical perimeter, for instance on wastelands or degraded soil. The authorities facilitate the process of identifying suitable compensation areas and administrative processes.

The sector plan for CERN's constructions and installations with territorial needs (PS CERN)~\cite{PlanSectorielCERN2025} highlights a number of topics that need to be considered when developing new projects:
\begin{itemize}
\item Water-bearing layers need to be protected and should not be used to provide raw water for cooling purposes. Care must be taken not to pollute water-bearing layers or to mix the water of superimposed water-bearing layers.

\item Exclusion buffers next to watercourses protect the habitats.

\item Forest clearings are generally forbidden in Switzerland. An exemption can only be granted in exceptional cases for installations that are in the public interest.

\item Measures need to be taken to manage rainwater and used water to avoid flooding and polluting the environment. The same holds for the management of water used to extinguish fires.

\item Limitations for dust and particle emissions need to be respected.

\item Limitations for noise emissions need to be respected. Noise emission limits in Switzerland are absolute and defined for different sensitivity zones as opposed to France, which considers noise limitations in excess of existing background noise.

\item Measures need to be taken to manage waste of any kind during all phases of the project.

\item The national and cantonal energy and climate protection goals are to be taken into consideration during the planning, construction and operation of future research installations. Energy consumption is to be reduced to a minimum compatible with the capacity to fulfil the scientific research goals.

\item The emission of greenhouse gases is to be minimised within the boundaries needed to meet the scientific goals and requirements of the project.

\item The design and implementation of technical systems have to integrate the optimisation of energy efficiency and the recovery and use of waste heat. The capacities and characteristics available for internal and external use need to be identified and documented.

\item Buildings must comply with the national energy efficiency regulations.

\item In general, any optimisation and minimisation effort is understood to be conditioned by the technical feasibility and economic viability. These efforts must always be compatible with the feasibility of achieving the specified scientific research goals.

\item A mobility plan that supports the national and cantonal climate plans is to be developed. Parking spaces are to be optimised, in line with the applicable regulations.
\end{itemize}

\section{Methodology}

The study adopted the approach and methodology to carry out a comprehensive, wider quantitative sustainability assessment based on the well-established cost-benefit analysis methodology~\cite{RICBA:2016} defined by the European Commission, the European Investment Bank and as described by national guidelines such as the one in France~\cite{SGPI:2023}. Such an assessment includes the identification and quantification of costs, benefits, and positive and negative externalities.

The goal is to present an implementation scenario that is likely to be long-term sustainable and for which sustainability level can be continuously improved with respect to an initial forecast that serves as a baseline. The approach permits the analysis of variants and sustainability boundary conditions to be established.

Following a lifecycle approach, it is a best practice to develop an initial cost-benefit assessment as early as possible at the pre-feasibility stage and to update it regularly as long as the key project features continue to be adjusted during the design phase and even during the implementation. This approach helps to deal with the challenge that exhaustive coverage of costs and benefits cannot be achieved and to focus on gaining a good understanding of the key sustainability enablers and risks.

Sustainability analysis is an ingredient of informed decision-making. Project and sustainability appraisal of the scientific research project does not, however, capture the opportunity and value of the underlying scientific mission. The indicators must therefore not be used to compare research projects or to make an investment choice between several projects.

To be useful during project authorisation, implementation and operation, the project must plan periodic sustainability monitoring and tracking throughout its entire lifetime and implement an iterative improvement process following the standard `Plan-Do-Check-Act' principle either at the infrastructure level or for new programmes and projects.

Life Cycle Assessment (or analysis) (LCA)~\cite{iso14040:2006,ILCD:2010a} is a methodology that is suitable for individual project segments. It allows assessment of a set of environmental aspects that are relevant for achieving sustainability. Its goal and scope must be well-defined, and its results must be integrated in the overall project and sustainability appraisal. When carrying out an LCA, care must be taken to be as specific as possible and to exhaustively document the scenario variant and version assessed, the assumptions, the input parameters, the data quality and the allocation procedure (algorithms and tools) to ensure that the results have a meaningful value for appraisal. LCA is not able to capture all relevant environmental aspects and does not therefore replace an environmental impact evaluation. It can therefore not be used to report on the overall environmental performance of a project.

For a comprehensive sustainability assessment, a quantitative Cost-Benefit Analysis (CBA) method based on state-of-the-art economic knowledge that integrates total costs, negative externalities (including environmental ones), industrial, social and environmental benefits is the preferred and adopted approach. A more complete description of the approach is provided in Section~\ref{annex:CBA}.

Project developers have to keep up to date with national and international regulations and legal frameworks in the subsequent design and preparatory phase.
They must continuously improve their knowledge about state-of-the-art project appraisal methods and the development of approaches to convert positive and negative impacts and externalities into quantitative terms suitable to report an overall sustainability indicator in the form of a project's net present value and benefit-cost ratio.
For this reason, as long as additional knowledge and assumptions about the project scenario and the project's key features are developed, the Benefit-Cost Ratio (BCR) and the Internal Return Rate (IRR) evolve throughout the lifecycle phases.

\section{Socio-economic sustainability enablers}

\subsection{Impact pathways}

\begin{figure}[ht]
    \centering
    \includegraphics[width=0.5\textwidth]{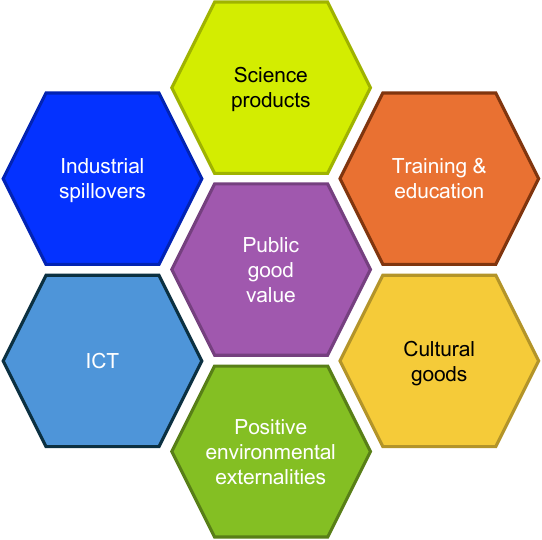}
    \caption{Main socio-economic impact pathways that are sustainability enablers of particle accelerator and particle physics research infrastructures.}
    \label{fig:main_impact_pathways}
\end{figure}

Socio-economic impact pathways are a valuable inventory of sustainability enablers. Based on the concept of `Theory of Change' \cite{vogel2012review}, the RI-PATHS EU-funded project has developed a toolkit~\cite{impact_toolkit}, federating research infrastructures across various domains, including particle accelerator facilities such as ALBA, CERN and DESY. The main pathways typical for particle accelerator-based facilities are shown in Fig.~\ref{fig:main_impact_pathways}. This section outlines briefly how such pathways can lead directly to benefits and positive externalities, and gives some specific examples. 

Future projects are highly encouraged to carry out a comprehensive charting of the impact pathway potentials and develop sustainability-enabling measures around such a systematic exploration and design activity.

\subsubsection{Fundamental Physics Knowledge}

Science is formalised knowledge that is rationally explicable and tested against reality, logic and the scrutiny of peers, is a global public good~\cite{CouncilScience:2021}. Ultimately it is the main output of research infrastructures. In our transforming society, knowledge is about to become the basic economic resource, gradually replacing capital, land and labour. There is evidence that the public sector has and is continuously funding much of the innovative science that stimulates private sector responses~\cite{mazzucato2013entrepreneurial}.
But what is the value of that knowledge that fundamental science projects generate? Are continued investments in new particle accelerator projects for scientific research sustainable in the perception of those people who do not directly use those instruments and who do not directly profit from the generated knowledge? Which means exist to capture the value of such investments?

Particle accelerators for fundamental science may or may not generate knowledge that can be directly used by society. Other particle accelerators, such as synchrotron light sources, typically serve applied research. This means that a lower bound for the value of the knowledge generated can be quantified in terms of the cost of the publications, the fees charged for the use of the facility and the resources that the users invest in this research. The approach is based on the assumption that the research results eventually end up in products and services that society uses. An example is the characterisation of the COVID-19 virus structure at the BESSY-II synchrotron~\cite{Scudellari:2020}. Particle accelerators and colliders that serve curiosity-driven research, such as the LHC or the Electron-Ion Collider, cannot use this model since it is not obvious how the knowledge gain leads to direct societal applications. There exists, however, no doubt that the scientific outcomes advance society. This view is shared by the majority of people, even if the asset is not directly used.\footnote{The notion for this concept is the `public good' as opposed to the `common good' that is jointly used. Examples of common goods are fish stocks in the ocean and public roads. Examples of public goods are knowledge, natural and cultural heritage, open software and datasets and even tangible assets such as street lighting.}

Evidence for the validity of the approach is available through the award of several Nobel Prizes in economics for advances in this domain. It is proof that this value generation process is of fundamental importance for society and welfare economics \cite{public-choice, growth-theory, spence-stiglitz, ostrom-pool, tirole, milgrom}. 

To capture the value that science missions and research infrastructures represent to laypeople who can not directly use them, methods that originally have been developed to elucidate the public good value of outdoor recreational spaces~\cite{davis1963value}, to identify suitable value levels for environmental protection measures~\cite{NOAA:1993}, to determine adequate levels of environmental incident mitigation measures~\cite{ExxonValdez:2003} and to determine investment levels to protect natural and cultural heritage can be used. 
This approach has first been taken to estimate the value of the LHC project~\cite{RePEc:mil:wpdepa:2016-03}, then for the HL-LHC project~\cite{CERN_Courier:2018} and now for the FCC~\cite{GIFFONI2023104627}.
The approach has recently also been adopted by an atmospheric research infrastructure~\cite{Actris:2023}. Briefly, the value that people associate with a science project can be captured by estimating their willingness to participate financially using well-established approaches in economics, such as contingent valuation, also referred to as ‘stated preference’.

The valuation of non-market impacts is challenging but could be undertaken wherever possible, and the `EU better regulation toolkit' indicates the instruments for it \cite{EC_Better_Regulation_Toolbox}.
This will enable a better understanding of the public perception towards the science vision and test the validity of the public funding sustainability hypothesis, i.e., what level of periodic investment is justified in the public perception. To design, plan and carry out a public good value estimation, experts in the domain and qualified companies need to be employed.
It is important to ensure that such a willingness to participate in analysis, typically based on surveys, is carried out independently, without the influence of the project owner and according to the internationally established guidelines on ethics and the quality standards required of survey-based analytics. Examples of this approach can be found in Refs.~\cite{secci_2023_7766949, GIFFONI2023104627}.

Depending on the number of countries or funding agencies that would carry out the project, such a survey can require substantial resources. Economists, therefore, use techniques based on the `Benefit Transfer Method'~\cite{Johnston:2015} to estimate the perceived value of a large population based on the identification of only a few significant parameters that can be derived from a limited number of samples. These parameters may be different for different projects. Therefore, such a study requires a pilot phase to determine the significant parameters, followed by a mass survey to determine the value that the public associates with the investment.

If the estimated willingness to financially participate (WTP) is larger than the actual or expected contribution to the planned new project, the public good value of the funded infrastructure can be considered consistent with the financial participation and, therefore, can also be considered justified. If the WTP is less than the actual or planned contribution, there is a discrepancy between the value perceived by the people and the contribution supplied. In this case, the project should be revised.

\subsubsection{Sustainability through education and training}

Particle accelerator and experimental physics research infrastructures offer the possibility to engage people at all education and training levels, from apprentices to post-doctoral researchers. If people have the opportunity to actively participate in the design, construction and operation of such projects and programmes, they enjoy benefits that translate directly into a \SI{2} to \SI{10}{\percent} lifetime salary premium~\cite{csil_2024_10653396, Camporesi_2017} compared to their peers who were enrolled in conventional training programmes at schools and universities. A dedicated programme to integrate people after their active time in a research facility into the labour market in their country or region of origin can significantly improve the return of investment for the participating country or region~\cite{crescenzi_2024_13166167}.

\subsubsection{Sustainability through participation of international science communities}
Involving a significantly large and continuously growing community of scientists and engineers in the form of collaborative research projects over long periods of time ensures the long-term sustainability of a science project or programme. The reasons are 
\begin{itemize}
\item The personnel costs are distributed over many contributing organisations and countries.
\item A stable generation of scientific products~\cite{csil_2024_13920183, MORRETTA2022121730} (i.e., books, peer-reviewed articles, pre-prints and technical reports, proceedings and presentations) can be guaranteed.
\item The knowledge can be effectively transferred into lasting education curricula through publications and direct training over generations.
\item Impact can be generated through the training of early career professionals (see previous section). 
\end{itemize}
High value is mainly achieved by ensuring that the science products generated directly by the research project or programme are taken up effectively and cited by an even larger second-tier community. Open Access publication of trusted (peer-reviewed) publications is a pre-requisite to generating research infrastructure sustainability through scientific product generation. Finally, the concept of Open Innovation that involves a large number of knowledge domains around a scientific core mission will ensure that the probability of knowledge generation that is stimulated through the challenges of the science mission will eventually spill over to society in domains that are immediately relevant to them~\cite{Gutleber2025}.

\subsubsection{Sustainability through industrial spillovers}

Industrial spillovers from science projects generate benefits directly for industry and society~\cite{Griniece:2020, Castelnovo:2018, Florio:2016}. This impact pathway is most effective if science projects construct their instruments and infrastructures in collaboration with industrial partners~\cite{Florio:2818, Autio:2004, Nordberg:2003}. It works best when companies work closely with the research project to develop technologically intensive solutions and deliver non-standard services~\cite{Sirtori:2019}. This approach is more laborious than a conventional client/supplier relationship since it requires exchanges of ideas and knowledge, the development of integrated processes, the mutual adaptation of working methodologies, and risk sharing. This process, however, leads to the creation of lasting processes, products, and services that the industrial partners can leverage in other markets with earnings multipliers above 3 and effects that can last between five and eight years. Lasting territorial effects have also been reliably documented~\cite{Crescenzi2025, Moretti:2010}.

Sustainability also depends on the design of the science mission: short-lived and non-upgradable programmes will lead to lower and shorter-lasting industrial spillovers than long-lasting programmes that are characterised by periodic upgrades and operational efficiency improvement measures involving industrial partners in continuous challenge-based activities.

Industrial spillovers can also lead to lasting positive environmental externalities. The challenge of reducing the carbon footprint and environmental impact of the construction of a new particle accelerator facility creates an opportunity for industrial innovations in numerous construction-related domains. For instance, it creates a potential of developing or putting in place low-carbon concrete and other construction materials production facilities. It serves as a pilot platform for improved construction techniques, such as using natural resources such as wood and compressed earth. Showcasing the application creates market interest. Developing processes and products for the use of excavated materials can generate benefits significantly beyond the needs of the research infrastructure project since the management of construction waste is a challenge that society faces and for which conventional construction projects typically have insufficient time and budget. Eco-design based industrial architecture is another emerging discipline for which particle accelerator facilities are suitable early adopters.
Environmental benefits can also be generated in the area of technical infrastructures that are developed with industrial partners. They include:
\begin{itemize}
    \item More efficient refrigeration systems which find their application, for example, in gas liquefaction and transport.
    \item More efficient water cooling systems (adaptive water intake, use of waste water).
    \item Improved electrical systems (loss reduction via DC-based systems.
    \item High-speed power control management.
    \item Short and medium-term energy storage.
    \item Adaptive and machine learning infrastructure operations.
    \item Waste heat buffering and supply. 
\end{itemize}

\subsubsection{Sustainability through open information and computing technologies}

The development of Information and Communication Technology (ICT) and the generation of widely available data not limited to scientific results (e.g., engineering test data, operation monitoring, system tests, etc.) are essential outputs of particle accelerator-based research infrastructures~\cite{FLORIO201638}. Putting a global data sharing- and processing infrastructure in place has already led to the creation of numerous, openly accessible software packages, platforms and online services, which are also used in environments outside high-energy physics. They range from scalable data storage and distribution middleware through data analysis and visualisation to data management and workflow systems. Also, openly accessible software has been developed with a value that extends beyond scientific collaborations. Examples include innovative Cloud computing services (e.g., Helix Nebula Science Cloud serving five scientific research domains), meeting and event management software (e.g., Indico), particle/matter interaction modelling and analysis (e.g., \textsc{Geant4, FLUKA} and \textsc{ActiWiz}) and electronic library and information access software (e.g., Invenio and Zenodo). Long-term data preservation is another technological domain with high societal relevance that is gaining importance and is primarily driven by fundamental science research infrastructures.

Particle accelerator and high-energy physics research projects are strongly encouraged to create an inventory of potential ICT tools that can be made available openly and free-of-charge for societal use. The feasibility of quantifying the societal impact of these solutions relies strongly on accounting for the software uptake and use (e.g., number of installations, number of times it has been integrated in commercial and other open software packages) over sustained periods of time. Today, there is a lack of such accounting, which makes quantitative estimation of the societal impact challenging and labour intensive. However, it is important establish systematic monitoring and evaluation since ICT tools developed in the science environment have a major impact factor with significant and tangible value potentials~\cite{CrespoGarrido:2025a, CrespoGarrido:2025b}.

Open and freely available software developed and maintained by science projects and programmes are sustainability enablers for research infrastructures. Their investment and continuous development and maintenance costs are marginal with respect to the societal benefits they can generate over decades. Future particle accelerator projects and programmes should focus on properly managing and promoting ICT developments and make sure they are taken up by society through a technological competence leveraging process and sustained online presence in potential user domains.

\subsubsection{Sustainability through cultural goods}

Creating an interest in science among all citizens is part of the mission of any research infrastructure. 
Projects and organisations can develop a broad variety of activities to attract lay people, for instance, through permanent and travelling exhibitions, open days, guided tours, engagement with schools and teachers in joint workshops, citizen science projects, websites, social media, engagement with video bloggers, online and TV documentaries, art internships, common art projects, feature movies, books, science fairs, presence in radio and TV shows and much more. The limit of cultural good creation is the limit of the imagination of an ever-changing and diverse group of creative people that are best employed over sustained periods of time. Each of these cultural goods has the potential to generate value for society. Creating a sustained interest in the science project breaks down barriers and fear of people who are outside the science community.

It is important for particle accelerator and physics projects to engage lay people in playful and entertaining ways rather than aiming at education and teaching, which is a different socio-economic impact pathway. The two should be kept separate, although effective cultural good creation and engagement raise the possibilities for explaining the underlying science in a second step.

The identification of value for cultural goods can be very different for each good. Therefore, future projects and programmes will need to focus initially on a few cultural pathways. If the estimated value of a pathway turns out to have sufficient potential, it should be further developed with a view to sustainable engagement. Continuous monitoring of the value should be carried out.

The value of on-site visits represents a cultural good of science infrastructures that can generate substantial tangible economic value in a sustainable way~\cite{CrespoGarrido:2025c, NASA_economic_snapshot:2024}. Such visits include the discovery of the environment and nature~\cite{Weaver:2011, Wyk-Jacobs:2018}. It should, therefore, be the first to be considered for development.

Social media presence has the most impact on online cultural goods today and should therefore also be considered with priority~\cite{Bastianin2021}. It is particularly important to create a sustained interest in science and to explain in which ways science impacts the environment and people's everyday lives. Different means of communication are needed for different generations and socio-economic groups~\cite{blekman2024crowdsourcedparticlephysicsstories}.

Finally, citizen science projects on the periphery of the physics science mission are effective tools to engage lay people and to reinforce the environmental and territorial compatibility of the research infrastructure. An example is the initiatives to create biodiversity inventories of the surface sites, to improve the quality of habitats on and around sites.

\subsubsection{Sustainability through positive environmental externalities}

Any future particle-accelerator-based project has the potential to create positive environmental externalities that can compensate for the residual, unavoidable negative environmental effects that cannot be reduced further. Before developing compensation and accompanying measures, negative effects on the environment have to be avoided. If they cannot be avoided, they should be reduced as far as is compatible with achieving the goals and objectives of the infrastructure within accepted cost and schedule constraints.

National and international legal and regulatory frameworks define boundary conditions for the avoidance, reduction, and compensation approach. The recently adopted update of the law for research and innovation in Switzerland, for instance, explicitly requires the consideration of the national and regional climate protection plans and energy-related aspects~\cite{fedlex:24.029}. Other countries, such as France, have already encoded the fight against climate change, resource and biodiversity protection, circular economy, and sustainable territorial development in the environmental protection laws (L110-1 of~\cite{CodeEnvironnementFrance} and~\cite{Loi_2021_1104}) that govern the authorisation of new projects~\cite{Loi_2021_1104}. Hence, `environmental' is always to be interpreted in a wide sense. The greenhouse gas emission reduction goals defined by the Paris Agreement are to be integrated in new projects at EU level and following its translation into the national laws~\cite{EU:2018_1999}.

If properly planned, evaluated and monitored, collective compensation can even lead to a net positive effect~\cite{Cerema:2018}. Switzerland does not foresee such an approach and requires compensation based on equivalent surface and quality in principle within the affected canton~\cite{ARE2020, BAFU:2022, Fedlex:451}. Despite this constraint, the overall environmental performance may still be neutral or positive from a socio-economic assessment point of view if the positive environmental externalities are properly evaluated and integrated in the net present value of the project.

Some typical examples of positive environmental externalities that particle-accelerator facilities can integrate in their designs from the onset are:

\begin{itemize}
    \item Re-creation of agricultural spaces by transferring top soil to low quality land plots and wasteland.
    \item Creation of green spaces with covered roads, fertilising wastelands and backfilled quarries, creation of parks, greening of roofs, reforestation around sites, creation and quality improvement of natural habitats and wetlands.
    \item Increase of biodiversity by the creation of new habitats on the research infrastructure's domain.
    \item Improvement or creation of new ecological corridors, green and blue continuities.
    \item Creation of forests with trees and plants  adapted for climate change.
    \item Introduction of forest management in view of protection from wildfires.
    \item Improved water management of existing, but low-quality water courses.
    \item Creation of water reservoirs.
    \item Creation of raised hedges to fight soil erosion and create new habitats.
    \item Creation of soft or multimodal mobility concepts for use beyond the research facility.
    \item Creation and improvement of natural habitats and nature protection zones on the research infrastructure sites.
    \item Carbon footprint reduction by helping to avoid fossil fuel use through waste heat supply.
    \item Supply of raw water for non-drinking purposes by supplying purified waste-water when not used by the research facility.
    \item Increase of renewable energy resource capacities through long-term power purchase agreements and energy purchase communities.
    \item Development of products and processes that leverage circular economy principles and low environmental impact technologies that spill over into the industrial domain (e.g., in the areas of electrical substations, construction materials, architecture solutions, power transmission and buffering, industrial cooling).
    \item Dismantling of infrastructures which are no longer used to create environmental and societal value.
\end{itemize}

\subsection{Innovation and R\&D}
Particle accelerators and experiment detectors at the collision points of particle colliders are projects which use cutting-edge technologies over periods spanning several decades. As a consequence, they tend to develop and use technologies that do not exist at the time of design. This process continues for the duration of the operation as detectors are upgraded. The required technical advances may be designed in-house and/or in partner institutions (such as universities) in conjunction with external companies. The development can take the form of joint design or procurement of technologies to be developed by the company.  
In both cases, industries profit from a technological knowledge transfer which may lead to technological breakthroughs, patents, or new business opportunities~\cite{bastianin2019technologicallearninginnovationgestation,Sirtori:2670056}. Such processes are encouraged and accompanied by dedicated knowledge transfer (KT) centres at the participating organisations.\cite{CERN-KT, desy-kt, infn-kt, fnal-kt} As enabler of the science mission, such technological advancement is a valuable output, which significantly affects the social sustainability of a project.

\subsubsection{Other sustainability potentials}

International science projects such as the FCC also provide possibilities to develop soft skills and international cooperation. Such projects bring nations closer together as they aim to pursue a common, peaceful goal. As recently seen with the ceremonies of CERN's 70th anniversary, heads of state and ministers come to visit the site and have bilateral discussions that may or may not be related to the scientific mission of the project or even to science. National news outlets have reported on the event, focusing on the common project and these side aspects. Often, who discusses with whom is seen as more important than the event itself. Collaboration leading to more collaboration, there is a societal interest to bring decision makers together to discuss issues of the world in an informal setting. Spinning off from particle accelerator scientific research, the SESAME project in Jordan~\cite{sesame-proj}
was set up with international collaboration in mind, as much or even more than the proposed scientific project. The Heidelberg, CNAO and MedAustron light-ion cancer therapy facilities are concrete examples of the direct transfer of knowledge, expertise, skills, and technology from fundamental particle physics and particle accelerator research to meaningful societal application. A future circular collider can achieve all this for Europe and beyond. 

\section{Comprehensive sustainability performance assessment based on Cost-Benefit Analysis}
\label{annex:CBA}

\begin{figure}[ht]
    \centering
    \includegraphics[width=\textwidth]{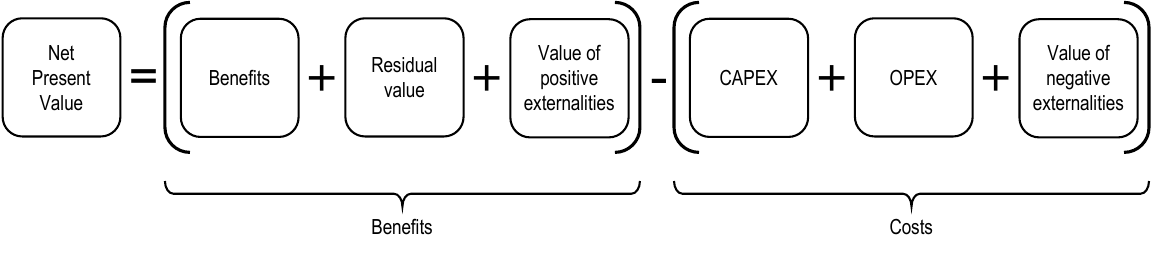}
    \caption{Expression to determine the net present Value (NPV) of an investment project considering economic, societal and environmental aspects.}
    \label{fig:NPV_formula}
\end{figure}

The approach is to create an inventory of negative cost and positive benefit items including externalities and convert them into monetary terms that are combined to determine a net present value (NPV) of the investment at the end of a chosen observation period (see Fig.~\ref{fig:NPV_formula}).
The ratio of the benefits of a project relative to its costs (BCR) including externalities and the so-called internal return rate (IRR) are two measures that can be derived from this analysis to provide valuable insight in the sustainability of the undertaking.
The IRR is the discount rate that would bring the NPV of the entire project over the observation period to zero.
An investment can be considered sustainable if it is financially feasible (net cash flow is non-zero) and profitable from a socio-econonomic perspective.
A positive BCR has been determined through a comprehensive socio-economic impact assessment under the most conservative assumptions, including total costs, externalities and benefits and an IRR that is greater than the cost of obtaining the capital required. These results support the overall sustainability of the project.

In line with the guidelines on cost estimation of research infrastructures \cite{StrESFRI}, the following items need to be established:
\begin{enumerate}
\item Unit of analysis with clear scope and boundary descriptions.
\item Reference period with a start and an end date of the observation based on the expected useful life (note that this period may be different from the physical life of the infrastructure and may lead to residual asset values).
\item Base year, i.e., the point in time when the quantitative estimation is made that does not necessarily coincide with the start date of the investment or the project operation. Past values are capitalised and future values are discounted with respect to the base year using the same mathematical formula. Conversion rates of the unit of measure with respect to other relevant currencies are to be recorded for the base year.
\item Unit of measure for costs and benefits in a specific monetary currency.
\item Approach for converting in-kind contributions in monetary terms.
\item Definition and description of the counterfactual scenario.
\item Project-specific Social Discount Rate that is justified by an economist expert advisory committee.
\item Structure of capital and operation expenditures.
\item Structure of negative externalities considered.
\item Structure of positive impacts (benefits) considered.
\end{enumerate}

In addition to full financial costs, negative externalities must be integrated in the assessment, as far as they can be reasonably identified, accounted and converted in monetary terms.

Concerning potential impacts on the climate, the methodology includes the accounting of project relevant emission factors, the estimation of the net greenhouse gas (GHG) emissions (a negative externality) and avoided emissions (positive impact or benefit) compared with a counterfactual baseline scenario.
The resulting amount of generated and avoided GHG emissions in tonnes of carbon dioxide equivalent (tCO$_2$(eq)) has to be converted into monetary terms using a shadow price of carbon.
In line with the EC technical guidance on the climate proofing of infrastructure, \cite{ECClimateProofing}, for the shadow cost of carbon, it is recommended to use the values established by the European Investment Bank (EIB) as the best available evidence on the cost of meeting the goals of the Paris Agreement \cite{ParisTreaty}.\footnote{The Paris Agreement is a legally binding international treaty on climate change. It was adopted by 196 Parties at the UN Climate Change Conference (COP21) in Paris, France, on 12 December 2015. It entered into force on 4 November 2016.}

Input-Output analysis \cite{IOModels} is a methodology that only captures economic linkages and determines the value-added of an investment and can be used to estimate the effect on the job market and in which domains and geographical regions the economic activations take place.
This tool is regularly used by governments and at the international level (e.g., EU, OECD) for empirical economic research and structural analysis and is therefore widely known.
However, it is purely an economic instrument that should only be used as follows:
\begin{itemize}
    \item As a complementary tool to document economic linkages, permitting the development of international in-cash and in-kind contributions.
    \item To identify project-relevant industrial sectors.
    \item To develop targeted common activities and synergies.
    \item To understand the job market implications. 
    \item To develop regional specialisation policies.
    \item To develop focused training and skilled labour mobility plans.
\end{itemize}

Recently, the FCC study implemented the economists' recommendation to carry out a complementary analysis of the public perception \cite{secci_2023_7766949} using a stated preference and Willingness-To-Financially-Participate (WTP) approach \cite{RePEc:ucp:jaerec:doi:10.1086/691697}. Such a survey can help to reveal if the full costs and cumulative impacts of a research infrastructure or scientific investment are at least justified with respect to the perception of the public.

The establishment of a risk registry, a risk assessment, and an evaluation of the residual risks is important for new science missions and research infrastructure projects and programmes. It should include economic, social, and environmental domains in addition to standard topics such as financial and project management related matters.
Selected chapters are to be analysed using a simplified appraisal process with a limited and self-contained CBA or a multi-criteria analysis (MCA).
Presenting several variants and versions of the project and individual segments together with their full financial costs, economic, social, and environmental performance levels is considered good practice in project appraisal and has been implemented by this study.

The results of the integrating and wider sustainability analysis are reported in a condensed summary form using the following key performance indicators:

\textbf{Net Present Value (NPV)} is the discounted sum of all future benefits less the discounted sum of all future costs over the appraisal period as a whole.
To properly estimate the NPV, realistic estimates are required of the streams of benefits and costs over the appraisal period that can reasonably comprise around 30 years. Beyond this time frame, quantitative socio-economic estimates become challenging.
The key to determining both these streams is knowledge of the times at which the various elements would come into play.
Investment costs will typically be incurred prior to the date of opening, whilst operating costs (for example, personnel and resources for operation and maintenance) and user benefits would arise after the year of opening.
User benefits, operating costs and revenues can be estimated from model runs for two or more years and the stream of benefits can be derived by interpolation and extrapolation between the benefits for the modelled years.

\textbf{Benefit/Cost Ratio (BCR)} is given by the ratio of the discounted sum of all future costs and benefits. The BCR is, therefore, a value-for-money measure, which indicates how much net social benefit could be obtained in return for each unit of investment. Although the values are monetised, i.e., converted into monetary units in this approach, they do not necessarily represent financial investments (e.g., the monetary shadow cost of carbon is not a paid amount of money. It represents a monetary cost to restore the effects linked to the quantified, potentially emitted carbon dioxide equivalent).
Formulae for the NPV and BCR will be found in CBA textbooks, a good example of which is Pearce and Nash (1981).

\textbf{Internal Rate of Return (IRR)}. Whereas the NPV and BCR measures require a test discount rate to be specified, the IRR reports the average rate of return on investment costs over the appraisal period. This can be compared with the test discount rate to see whether the project yields a higher or lower return than is required to break even in social terms. Calculation of the IRR and issues surrounding it are discussed in Pearce and Nash, Chapter 4 and in other cost-benefit textbooks.

\section{Limitations}
\label{sustainability:limitations}

Particle accelerator facilities are characterised by a diverse set of investment and operation cost items, negative externalities, benefits and positive externalities. Research infrastructures dedicated to fundamental science generally do not generate financial revenue. Unlike commercial ventures, research facilities exploring fundamental questions about nature (like the origin of matter or the structure of the universe) don’t sell a product or service that generates income. Consequently, they cannot generate profit\footnote{Financial profit is the next income that is earned after deducing all explicit costs from the total revenue}. This makes it difficult to evaluate their performance using standard business metrics like return on investment, earnings or profit.

This situation presents challenges for conducting comprehensive project appraisals and assessing long-term sustainability through traditional financial metrics. In the absence of a direct positive financial return by definition, the value of these projects is primarily reflected in their broader socio-economic impact. Thus, the assessment of such infrastructures relies on identifying and, where possible, quantifying the societal benefits in relation to the associated costs and externalities.

The public good value refers to the importance or benefit of a public good provided without profit to all members of a society. A science infrastructure such as particle collider are an example for a public good.
Elucidating its value as perceived by the public is an essential complementary ingredient to understand if society considers it worth investing in such scientific research activity.
The total public good value in the countries that would potentially financially contribute to a future particle collider-based research infrastructure was found to be larger than the sum of all known, identified, and quantified costs. This result supports, at least from a societal point of view, the intent.
Establishing the hypothesis that only CERN Member States contribute to the project implementation is not a likely scenario.
An infrastructure of the scale of the FCC will attract participation and funding from beyond the European Research Area, as was the case with the Large Hadron Collider.
Therefore, to estimate the public good value, those countries have been selected that historically financially participated in CERN's international research programmes.
An extremely conservative scenario can be built that includes only CERN Members and Associated Member States. However, no less than those countries should be considered.

\begin{figure}[h]
    \centering
    \includegraphics[width=.6\textwidth]{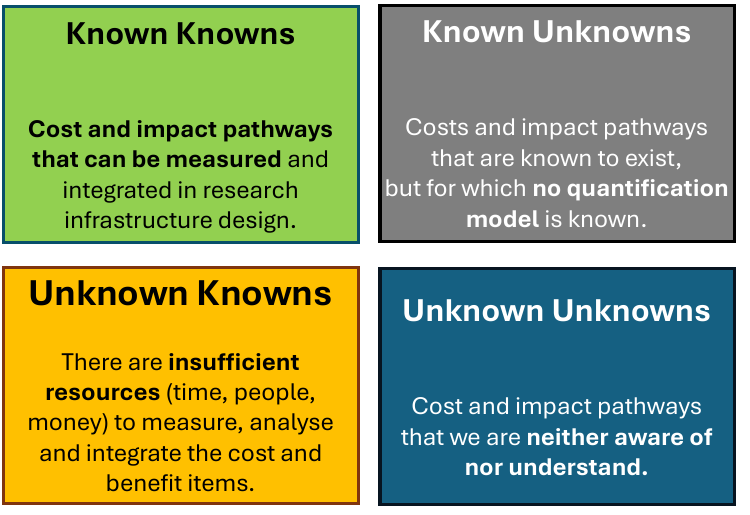}
    \caption{The Rumsfeld matrix~\cite{Krogerus_2012} applied to the identification of costs and benefits of research infrastructures.}
    \label{fig:rumsfeld}
\end{figure}

\begin{figure}[h]
    \centering
    \includegraphics[width=.5\textwidth]{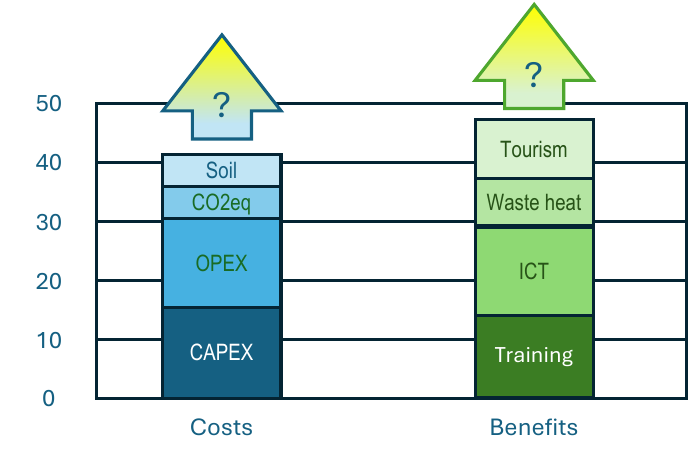}
    \caption{Examples for the accounting of full costs and benefits for an integrated sustainability analysis. Only the identified cost and benefit items that can be converted to monetary terms were considered in this specific assessment. Exhaustive coverage of all negative and positive externalities is challenging and and needs to be accepted in all socio-economic impact assessments.}
    \label{fig:fullcoverage}
\end{figure}

As recommended by the OECD~\cite{pearce2006cost, oecd2018cost}, the French government~\cite{quinet2014evaluation}, the UK government~\cite{UK_greenbook:2022}, and the European Commission~\cite{sartori2014guide, ECVademecum}, a comprehensive sustainability assessment should consider all of the above elements, i.e., total costs, total benefits, negative externalities, positive impact, and environmental benefit potentials.
In practice, not all contributing elements are relevant for a specific investment project, not all elements can  be reliably quantified, and not all elements are known or understood (Fig.~\ref{fig:rumsfeld}).
Both cost and benefit items are affected by uncertainties (Fig.~\ref{fig:fullcoverage}).
The results presented in this chapter must be interpreted in this light. An uncertainty analysis will eventually shed light on the ranges. The EC CBA Guide \cite{ECCBAGuide2014} recommends a Monte Carlo based approach. However, this method also relies on the knowledge of the probability distributions of the individual cost and benefit components, which are in turn challenging to obtain with a high degree of confidence. 

As far as possible, the comprehensive socio-economic study report describes the scope of the study as precisely as possible, and the aspects considered are described together with the assumptions used for the quantification of the aspects and their conversion into monetary terms.

Rationalising comprehensive sustainability assessment using standard cost-benefit analysis~\cite{OMAHONY2021106587} helps to identify the sustainability limiting and enabling aspects and can guide the design and iterative evolution of the research infrastructure to increase long-term sustainability.
In this approach, time is accounted in a rigorous way, which is essential when considering effects on the environment at large.
While methods to quantify cost items, including negative externalities, are typically well defined in the existing guidelines, the approaches to quantifying and monetising benefits and positive externalities are not exhaustively captured by a general catalogue or taxonomy.
They are project-specific and need to be identified and captured on a case-by-case basis using different methods and models.
Still, not all cost items and negative externalities are known at all times, and even if they are known, they cannot always be reliably quantified, monetised, and assigned with uncertainty ranges. Much depends on the evolution of the economy and society, such as the availability and demand for excavated material deposits, the valuation of climate impacts, the evolution of water supply and management, and the local and regional demands for goods and services. An ex-ante analysis is, therefore, always only valid in the current context, projecting all potential future costs and benefits into today's socio-economic environment using the currently known uncertainties.
All assumptions underlying the cost and benefit quantification, the models used, and the monetisation methods need to be documented, together with the analysis results.

\begin{figure}[h]
    \centering
    \includegraphics[width=0.8\textwidth]{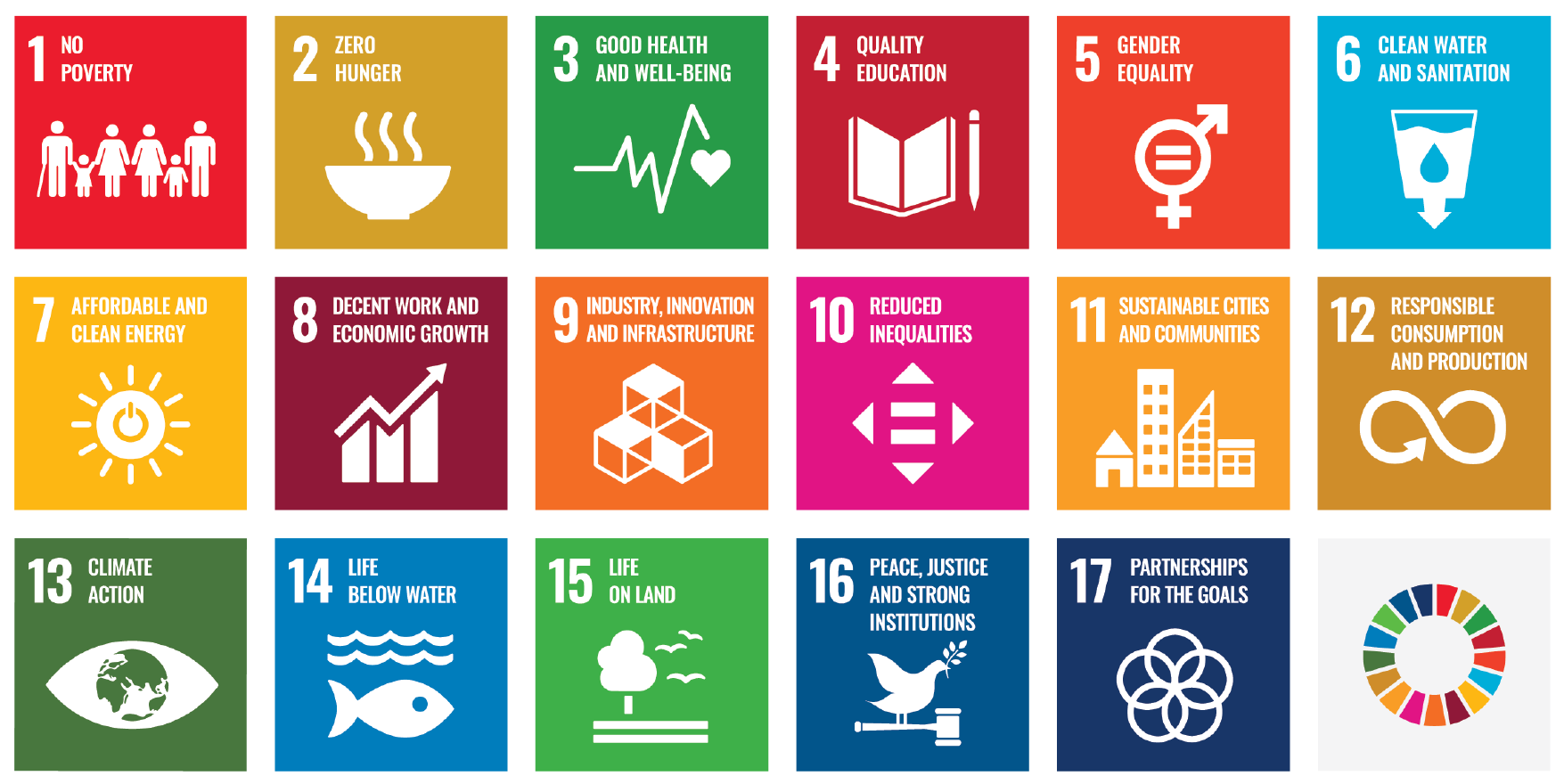}
    \caption{The UN sustainability development goals (SDG).}
    \label{fig:UN_SDG_Goals}
\end{figure}

Sustainability might refer narrowly to the internal sustainability of the project. In reality, it refers more broadly to a whole range of external economic, social or environmental factors  which can be influenced by the project.
The 17 high-level development objectives (with more than 160 sub-objectives) of the 2015 UN Sustainability Development Goals (SDGs) shown in Fig.~\ref{fig:UN_SDG_Goals} illustrate the breadth of these factors.
The ultimate goal is to identify and document quantified positive and negative impacts considering and leaning on the UN Sustainability Goals as a guiding principle for the topics~\cite{un-sdg-com} as a reference matrix.
Since quantitative sustainability analysis cannot provide full coverage, the UN SDGs can be regarded as a complementary catalogue of potential negative and positive impacts that can be looked at in terms of causal relationships with the project.
The global indicators and monitoring framework for the SDGs~\cite{sdg_transformation_center} have been designed to help countries, not research organisations, develop strategies~\cite{UNSDGMonitoringFramework}.
They are therefore not directly usable for reporting on scientific programmes and projects. A research infrastructure can, however, report relevant positive and negative impacts in quantitative terms as far as they have been assessed for each UN goal activity.
This approach has for example, been implemented with CERN's periodic environment report that relies on the GRI framework~\cite{CERN_env_report_GRI_SDG}.
In the absence of established guidelines, the SDG tracker~\cite{owid-sdgs} and the European Commission SDG information site~\cite{ec-sdg} provide a good starting point for indicator recommendations based on the UN SDG goal specification.
A full exploration of the FCC's contribution to the UN SDGs has not been done at this point.
It should, however, be developed during the subsequent design phase that includes the project authorisation phase.

Such a qualitative analysis is not expected to give evidence of compliance with specific national or international targets.
Rather, the goal is to show how the research infrastructure affects the environment at large in negative ways and how it contributes in very tangible ways to each of the goals.

The choice of the Social Discount Rate (SDR) affects the integration of cost and benefit items. Using different SDR rates can be used to perform a sensitivity analysis and to show the robustness of the overall net present value. For this study only one SDR value has been used.

\section{Lifecycle analysis}

\subsection{Context}

As the global focus on combating climate change intensifies, reducing greenhouse gas (GHG) emissions has become a top priority. Infrastructures spanning transport, construction and scientific instruments play a critical role in this transition. A lifecycle analysis is an essential tool to analyse the sources of emissions that affect the climate, to understand the drivers and to optimise the project in view of improved environmental compatibility.

The topic is comprehensive and complex, determined by many uncertainties and by the fact that the different segments of the project are only gradually known as the concept development and design advance. A long-term project like the future circular collider does not deliver detailed designs of the technical infrastructures and particle accelerator components until shortly before procurement. Detectors that constitute the scientific experiments will only be built and procured as a collaborative effort at a much later stage. To be able to leverage the technical advances, detailed designs are only developed when the particle collider equipment is well known, procured and potentially already being installed.

In addition to this time dependency, LCA and the associated carbon budget assessment strongly depend on the procurement scenario and the scenario of specific technologies. Therefore, the FCC study has taken the approach of focusing first on the best-known project segment, the environmental footprint of civil construction based on the current concept.

The study also included the estimate of the Scope 2 footprint for operation, i.e., the carbon emissions that would be associated with the consumption of energy. Also, a specific procurement hypothesis had to be assumed for this analysis.

It must be understood that the results presented herein are subject to further evolution throughout the lifecycle phases of the project, should a decision to pursue it be taken.
In particular, they do not consider certain negative externalities related to the construction of the particle accelerators and experiments, optimisations, which are possible with the use of further advanced technologies, materials, products, local production, responsible sourcing and procurement and the outcomes of commercial negotiations during the procurement processes.
Consequently, continuous socio-economic performance monitoring, incorporating eco-design based on an iterative Plan-Do-Check-Act cycle, is important and should be established before entering a preparatory project phase.

The analysis carried out using the European LCA Norms ISO/EN 14040 and ISO/EN 14044 aimed to estimate the carbon budget of the infrastructure construction that would serve two subsequently installed particle colliders.
It aims to establish a credible benchmark, leveraging the use of European norm EN 15804+A2 and French FDES norm standardised Environmental Product Declarations (EPDs) of materials and products available today.
This approach ensures a comprehensive assessment in line with the European standard and norm EN 17472, which provides requirements and guidelines for calculating and reporting GHG emissions associated with infrastructure projects.
The work also included the recommendations of Cerema for the evaluation of greenhouse gas emissions for road projects, the Swiss `Koordinationskonferenz der Bau- und Liegenschaftsorgane der öffentlichen Bauherren' (KBOB) and French `Les données environnementales et sanitaires de référence pour le bâtiment' (INIES) databases for ecological product footprints, and the ecoinvent (ecoinvent is an internationally active, mission-driven organisation devoted to supporting high-quality, science-based environmental assessments) database for products and materials where no EPDs were available.
The work revealed some realistic pathways to further reduce the consequences of the construction and gave some recommendations on future aspects that need to be considered during the subsequent design phase.

The results are integrated in the integrating, wider socio-economic impact assessment by converting the results of the LCA into monetary terms using the recommended approach of the EIB and the European Norms EN 14007 and EN 14008.

\subsection{Methodology}

The work for the estimation of the construction-related greenhouse warming potential included the following steps:
\begin{enumerate}
\item Component identification: Compilation of the detailed inventory of materials based on the bill of quantities for the subsurface construction, the 4 experiment sites and the technical sites. The current conceptual design was considered for this step. The products chosen are used throughout the infrastructure's lifecycle.
\item EPDs acquisition: Sourcing of EPDs for each material identified, ensuring compliance with EN 15804+A2, the foundational basis for EN 17472. The materials are selected based on expert knowledge of the local environment and product availability and hence, represent state-of-the-art solutions.
\item Software tool: Selection of a certified tool compatible with French and Swiss Environmental Product Declaration - One Click LCA\textsuperscript{\textregistered}, ensuring robust and accurate calculations.
\item Data entry: Imported data into the tool and entered project-specific data, including material quantities and lifecycle phases, transport for excavated and construction material.
\item Calculations: The tool was used to estimate the carbon budget, drawing on EPDs data to evaluate GHG emissions for each lifecycle phase.
\item Result analysis: Analysed the results to pinpoint major emission sources, comparing them against benchmarks and reduction targets.
\item Formulation of recommendations: Based on the results and the identification of the key impact drivers, a set of recommendations was formulated that is to be taken into consideration in the subsequent design phase to further reduce the environmental footprint of the construction.
\end{enumerate}

\noindent For the estimation of the Scope 2 emissions, the following steps were taken:
\begin{enumerate}
\item Consumption identification: estimation of the electricity consumption for the baseload and for each operational phase and establishment of an operational schedule.
\item Procurement hypothesis: Establish a credible scenario for procuring the energy based on an external expert consultancy with a focus on sourcing energy from renewable energy sources.
\item Quantification: Quantification of the associated climate effect by using the accredited French ADEME \footnote{ADEME stands for Agence de la Transition Écologique, which translates to Agency for Ecological Transition in English. It is a French public agency under the supervision of the Ministry for the Ecological Transition and the Ministry for Higher Education and Research.} database.
\item Formulation of recommendations: Based on the results and the identification of the key impact drivers, a set of recommendations was formulated that are to be taken into consideration in the subsequent design phase to further reduce the environmental footprint of the operation.
\end{enumerate}

In both cases, the results were converted into monetary terms using the EIB-established shadow price for carbon. The results were integrated over time and discounted using the 2.8\% Social Discount Rate established for the assessment project.

\subsection{Results}

\subsubsection{Construction phase}

The LCA\cite{mauree_2024_13899160} provided a breakdown of GHG emissions across various lifecycle phases and materials.
It must be stressed that the results were obtained with the most advanced, state-of-the-art products and materials on the market.
The results are, therefore, credible, and the carbon-related climate footprint can be further reduced as more advanced technology can be included in the project.
The key emission sources are reinforced steel (14\%), precast concrete (49\%) and concrete (23\%).
The climate effects due to the use of electricity have been included in this analysis.
It is assumed that the electricity required for the entire construction process can be obtained from renewable energy sources via local suppliers in France and in Switzerland.
The official carbon footprints of today's renewable energy mix in the two host countries have been used in the study (see Table~\ref{tab:ElectricityCarbonFootprint}).
The results obtained highlight opportunities for emission reduction by establishing technical requirements for the infrastructure with carbon reduction in mind, careful selection of materials which meet the requirements, construction process optimisation, and energy efficiency improvements.
The GHG impacts of the initial and benchmark scenarios are given in Table~\ref{tab:LCA-construction-result}.

\begin{table}[ht!]
  \centering
\caption{\label{tab:LCA-construction-result} Summary of the LCA-based carbon budget of the construction process.}
  \begin{tabular}{lr}
    \toprule
    \textbf{Item} & \textbf{\makecell[r]{Footprint}} \\ \midrule
    Subsurface & 477\,390\,tCO$_2$(eq) \\ 
    4 technical sites & 17\,546\,tCO$_2$(eq) \\ 
    4 experiment sites & 31\,735\,tCO$_2$(eq) \\ \midrule
    \textbf{Total} & \textbf{526\,671\,tCO$_2$(eq)} \\ \bottomrule
  \end{tabular}
 \end{table}

The value obtained corresponds roughly to 3 years of CERN's annual carbon budget~\cite{CERN-Enviro-2021} or to one-third of the carbon budget of the Olympic Games in Paris, 2024~\cite{Paris2024}.

Further environmental performance indicators obtained with the LCA are shown in Table~\ref{tab:LCA-further}. Compared to classical, non low-carbon construction processes, the most relevant indicator is marine eutrophication, which can be linked to the use of recycled steel.

\begin{table}[ht!]
  \centering
 \caption{\label{tab:LCA-further} Summary of the LCA-based carbon budget of the construction process.}
  \begin{tabular}{lr}
    \toprule
    \textbf{Indicator} & \textbf{Value} \\ \midrule
    Potential contribution to ozone layer depletion & 19.3 kg CFC 11 eq \\ 
    Potential acidification & 929 t SO$_2$ eq \\ 
    Potential eutrophication - fresh water & 96 t (PO$_4$)$^{3-}$ \\ 
    Potential eutrophication - marine & 12 t N \\ 
    Potential eutrophication - land & 2.1 x 10$^6$ mol N \\ \bottomrule
  \end{tabular}
\end{table}

\begin{figure}[h]
    \centering
    \includegraphics[width=.8\textwidth]{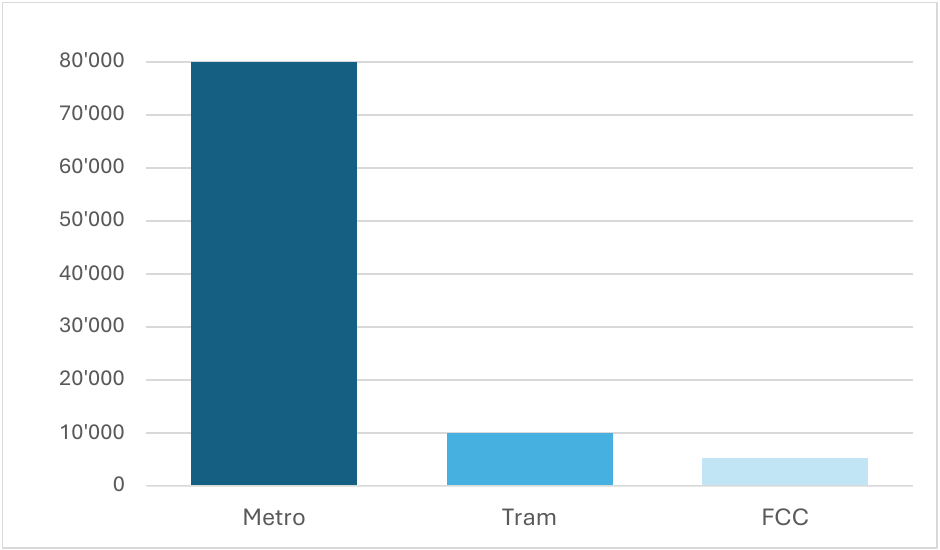}
    \caption{\label{fig:CO2-per-km} Comparison of carbon footprint between a small-scale underground metro line, a tramway line and the FCC.~\cite{mauree_2024_13899160}}
\end{figure}

The results also allow a comparison to be made of the construction of subsurface structures with conventional structures, such as road and metro tunnels and tramway lines. The latter is characterised by significantly higher carbon footprints as a result of the much stricter requirements that public use imposes. For comparison, the construction of the U5 underground line in Berlin (Germany), a typical small-scale metro line, has a carbon footprint of 80\,000\,tCO$_2$(eq)\,per\,km. The footprint of a typical tramway line has between 7\,600 and 10\,850\,tCO$_2$(eq)\,per\,km. The linear part of a future circular collider subsurface structures has a carbon footprint of about 5\,300\,tCO$_2$(eq)\,per\,km (see Fig.~\ref{fig:CO2-per-km}).

Based on the findings, the following levers were identified to reduce GHG emissions:
\begin{itemize}
\item Document technical requirements for the structural surface and subsurface elements that capture the strict minimum needs to fulfil the scientific research programme.
\item Develop an eco-design that meets the established requirements with carbon reduction in mind.
\item Make structural modifications by reducing the inner line thickness of subsurface structures by 5\,cm, leading to a reduction in the quantity of precast concrete and rebar steel of 16\%. 
\item Substitute materials, using low-impact materials wherever possible, leading to a further reduction of GHG emissions. 
\item Optimise the construction process to minimise emissions, including local sourcing of raw and recycled materials and production of construction materials.
\item Fully electrified construction processes and transport.
\item Reuse excavated materials, for instance, in concrete production. 
\end{itemize}

\subsubsection{Climate effects due to energy use during operation}

To estimate the climate effects related to the use of energy (Scope 2) during the operation phase, the electricity requirements have been estimated for the baseload and for each individual operation phase (Z, WW, ZH, $\rm t \bar{t}$).
During the operation phase the energy needed to power the infrastructures, particle accelerators and experiments will be sourced entirely from the French electricity grid, managed by the national grid operator RTE.
A portfolio of multiple electricity contracts and power purchasing agreements (PPA) facilitates the diversification of the electricity mix and sourcing it from different operators who can provide certificates of origin, leading to market-based carbon footprint reporting.
A preparatory study with two independent external consultants identified that on a short timescale, before 2030 60\% of the required energy could already be obtained from renewable energy sources and in a time frame of 15 years (around 2040) a portfolio to cover 80\% of the needs with renewable energies can be established.
For 2050, when the FCC is envisaged to operate, a coverage of 90\% is assumed with a residual supply of 10\% nuclear energy.
For the purpose of the socio-economic performance assessment, a conservative maximum use of 80\% energy from renewable energy sources has been used, thus over-reporting the carbon equivalent footprint.
To estimate the carbon intensity of the project scenario, the official values of the French National Environment and Energy Management Organisation (ADEME) were used, leading to a mix with a carbon intensity of about 15\,tCO$_2$(eq)\,per\,GWh.
For the calculations in the socio-economic performance assessment, a more conservative degressive carbon intensity starting with 26\,tCO$_2$(eq)\,per\,GWh in 2024, 19.10 in 2046 and 18.41 in 2050 assuming a constant decrease in the emission factor of 4\% have been used. Again, this approach leads to an over-reporting of the carbon equivalent footprint.
Table~\ref{tab:CO2-operation} below provides the Scope 2 indication of the carbon budget for each particle collider operation phase with beam for scientific research and the total footprint for all operation and shutdown phases for two different carbon footprint assumptions leading to a range for the expected carbon footprint: 15\,tCO$_2$(eq) per GWh and 25\,tCO$_2$(eq) per GWh.

\begin{table}[h]
  \centering
 \caption{\label{tab:CO2-operation} Summary of the operation-related Scope 2 carbon emission footprint using two different market-based carbon intensity assumptions for the electricity purchased. Note that the integral carbon footprint also covers Scope 2 emissions during shutdown periods.}
  \begin{tabular}{lrrrr}
    \toprule
    \textbf{Phase} & \textbf{Duration} & \textbf{GWh/year} & \textbf{\makecell[r]{Footprint all years\\at 15 tCO$_2$(eq)/GWh}} & \textbf{\makecell[r]{Footprint all years\\at 25 tCO$_2$(eq)/GWh}} \\ \midrule
    Z &  4 years & 1\,100 & 66\,000 tCO$_2$(eq)  & 110\,000 tCO$_2$(eq) \\ 
    WW & 2 years & 1\,300 & 39\,000 tCO$_2$(eq) & 65\,000 tCO$_2$(eq) \\ 
    HZ & 3 years & 1\,500 & 67\,500 tCO$_2$(eq) & 112\,500 tCO$_2$(eq) \\ 
    $\rm t \bar{t}$ & 5 years & 1770 & 132\,750 tCO$_2$(eq) & 221\,250 tCO$_2$(eq) \\ 
    \textbf{Total} & & 20\,350 & \textbf{305\,250 tCO$_2$(eq)} & \textbf{508\,750 tCO$_2$(eq) } \\ \bottomrule
  \end{tabular}
\end{table}

Integrating the energy consumption requirements over the years of operation leads to an integral carbon budget that corresponds to about 2.5\,years of CERN's annual carbon footprint today. For comparison, the Meta company known for its products Facebook, Instagram and WhatsApp operates currently at least three data centres in the United States with electricity consumptions between 1.2 and 1.4\,TWh per year \cite{meta-report}. Their individual, location-based Scope 2 emissions are significantly above, 350\,000\,tCO$_2$(eq) per year.

\subsubsection{Shadow cost of carbon}

The conversion of the carbon budget respects the overall goals for climate protection and temperature limits that were approved by a community of nations in the Paris Agreement. The European Investment Bank (EIB) has calculated a so-called `shadow price for carbon' that represents the costs that humankind needs to associate with climate protection measures to achieve the agreed goal.
The shadow price of carbon increases over the years, making one tonne of CO$_2$(eq) more expensive for society each year.
The discounted value of the monetary conversion of the construction and operation-related carbon footprint is about 342 million euros.
This amount is added to the integrated socio-economic assessment on the cost side of the balance and is used to estimate the net present value of the future research infrastructure at the end of its scientific operation programme.
Table~\ref{tab:shadow-cost-carbon} shows the shadow cost of carbon currently recommended for project appraisal by the European Investment Bank (EIB) and its time-adjusted value to be used in 2025.

\setlength{\tabcolsep}{4pt}

\begin{table}
\centering

\caption{Shadow cost of carbon under in currency unit per tonne CO$_2$(eq) for various jurisdictions. Rates in the first row are also published as European Union law 2021/C 280/01. Note that the 1.5\% near-term Ramsey Discount Rate is applied to the US indicated rates.}
\label{tab:shadow-cost-carbon}
\begin{tabular}{llccccccccl}
\toprule
\textbf{Organisation} & \textbf{Unit} & \textbf{2020} & \textbf{2025} & \textbf{2030} & \textbf{2035} & \textbf{2040} & \textbf{2045} & \textbf{2050} & \textbf{2060} & \textbf{Source}\\
\midrule
EIB and EU & Euro 2016 & \euro80 & \euro165 & \euro250 & \euro390 & \euro525 & \euro660 & \euro800 & n/a & \cite{EIB_economic_appraisal}\\
EIB and EU & Euro 2024$^{\dagger}$ & n/a & \euro208 & \euro316 & \euro492 & \euro663 & \euro833 & \euro1010 & n/a & \\
SGPI France & Euro 2018 & \euro87 & n/a & \euro250 & n/a & \euro500 & n/a & \euro775 & \euro1203 & \cite{SGPI:2023}\\
UBA Germany & Euro 2023 & \euro240 & n/a & \euro254 & \euro253 & n/a & n/a & \euro301 & n/a & \cite{UBA-2024}\\
UK (high) & GBP 2020 & £361 & £390 & £420 & £453 & £489 & £527 & £568 & n/a & \cite{UK_GOV_carbon:2021}\\
EPA USA & USD 2020 & \$340 & n/a & \$380 & n/a & \$430 & n/a & \$480 & \$530 & \cite{EPA2023}\\
\bottomrule
\multicolumn{11}{l}{\small{\makecell[l]{$^{\dagger}$ Adjusted based on EU-27 GDP deflator 100.0 in 2016 and 126.3 in 2023. It measures the amount to which\\ the real value of an economy's total output is reduced by inflation.}}}\\
\end{tabular}
\end{table}

\setlength{\tabcolsep}{6pt}

\section{Socio-economic performance}
\label{sustainability:results}

\subsection{Context}

The goal of the comprehensive socio-economic impact analysis including a wider set of components such as environmental externalities is to get a better understanding of the cost drivers and potential benefits.
This approach has been adopted to plan for a sustainable scenario.
The analysis can also reveal under which conditions the overall socio-economic performance can be positive, i.e., represent a long-term sustainable investment for the society.
Carrying out such analysis from the onset enables the project to incorporate the findings in the subsequent design phase, where it is easier to plan for sustainability enablers than during construction or when the infrastructure is already in operation.

The socio-economic analysis relies on a working hypothesis of the investments and resources engaged for the project by an international collaboration.
Therefore, the work is based on a total project cost estimate including all investment costs, operation costs and relevant negative and positive externalities for the entire observation period carried out in 2024.
This estimate may differ from project cost estimates carried out later or presented elsewhere, since it represents a snapshot of the estimates taken in June 2024 and it includes global personnel engagements such as those of the international collaborations and monetised negative environmental externalities.
The total costs required for the socio-economic analysis include the investment costs (capital and operation expenditures) as well as monetised externalities such as environmental costs and the shadow cost of carbon. Therefore, those costs figures must not be used to determine the financial needs to implement the project.
They represent the total cost of the science missions for society for the entire observation period.
The same holds for the reported benefits: they represent a collection of identified, quantified and monetised positive effects for society and are not limited to direct economic effects.
Neither costs nor benefits can be exhaustively covered due to uncertainties, available resources, and time limitations.
However, the herein reported results are sufficiently comprehensive and detailed to establish a stable Benefit-Cost Ratio that is unlikely to decrease further with the introduction of additional, minor cost and negative externalities.
The socio-economic performance requires periodic updates as the project is defined better and in more detail and as models and resources to estimate additional benefits become available.

A comprehensive report on the socio-economic performance analysis \cite{csil_2024_10653396} details the data, assumptions, methodologies and outcomes of the first socio-economic impact analysis for the first phase of the integrated FCC research programme: the FCC-ee lepton collider based on only two interaction points.
The analysis spans the entire lifecycle of FCC-ee, encompassing its design, the construction of the surface and subsurface structures, the creation of the technical infrastructures, the particle accelerator and experiment detector construction, the operation and the gradual beam energy upgrade phases.
The entire analysis spans a duration of 41 years (2024 to 2064), starting with pre-investment study expenses, the financial investment decision and lasting until the end of the operation phase.
The methodological approach is based on the regulatory guidelines issued by the European Commission (EC)~\cite{EC_Better_Regulation_Guidelines}, the European Investment Bank (EIB)~\cite{EIB_economic_appraisal}, the European Strategy for Research Infrastructures (ESFRI) guidelines~\cite{StrESFRI} and the French Secrétariat général pour l'Investissement (SGPI)~\cite{SGPI:2023}.
The guidelines adopted are compatible with those in other countries of the European Research Area, for instance, in the UK~\cite{UK_greenbook:2022}.
The study also considers the latest literature, works, and results of empirical research on the economic quantification of impacts associated with research infrastructures.

In evaluating the FCC-ee project, costs and benefits are expressed in comparison to a scenario where the project is not implemented.
In this counterfactual scenario, the LHC would continue its operation until its anticipated end of life around 2040, and no new particle-collider would be constructed.
CERN would continue its operation of the existing particle accelerators (e.g., AD, ELENA, LINAC4, PS) and the experimental infrastructures linked to them.
It is based on the methods of standard cost-benefit analysis (CBA).
All costs and benefits are incremental, i.e., expressed as a difference between the costs and benefits of the operation of the existing infrastructure after the shutdown of the LHC and the costs and benefits that the FCC-ee would induce
Consequently, the approach captures the net change, focusing on causally related effects that can be reliably attributed to the FCC-ee project.

\begin{figure}[h]
  \centering
  \includegraphics[width=\linewidth]{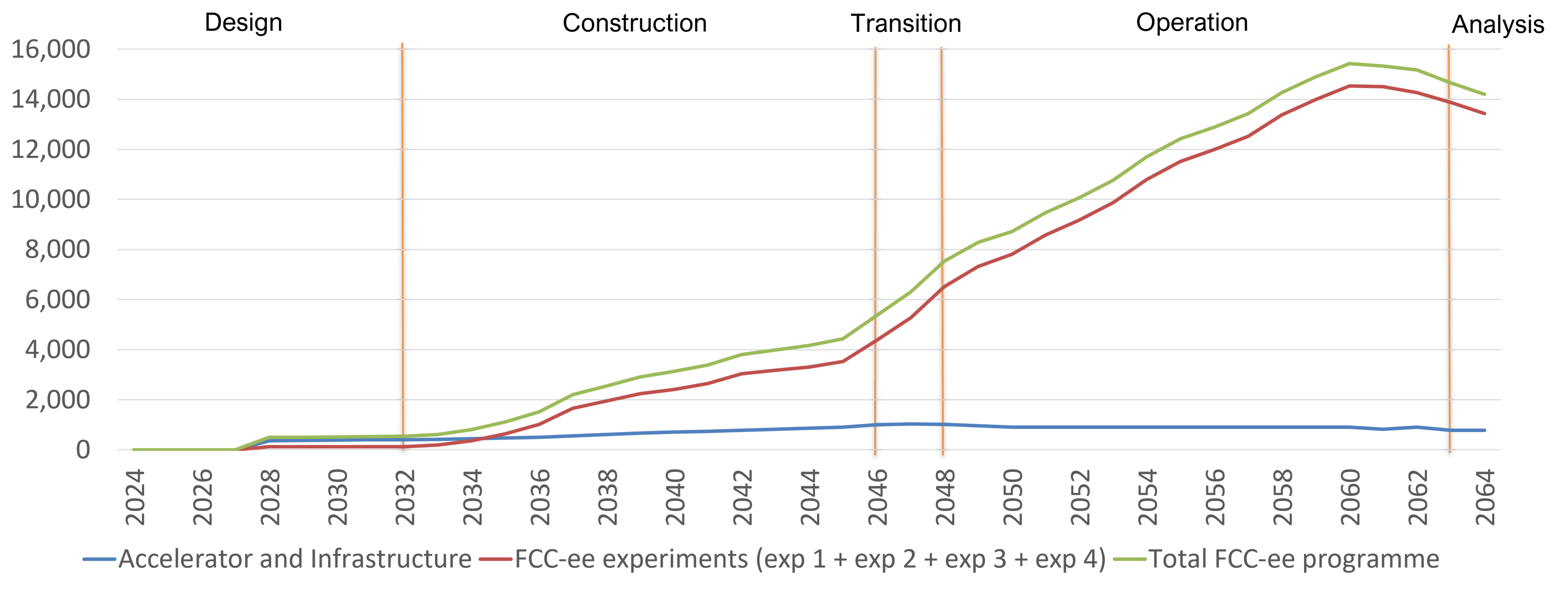}
  \caption{Number of persons engaged in the project over the years throughout all project phases.}
  \label{fig:personnel}
\end{figure}

This social cost-benefit assessment carried out assumes an FCC-ee project with four experiments, collectively involving approximately 260\,000 person-years over the project's lifespan.
The engagement of people is expected to reach its peak of about 15\,500 people during the operation phase (Fig.~\ref{fig:personnel}).
This diverse group comprises scientists, engineers, technicians, administrative staff, doctoral and post-doctoral researchers, undergraduate students, master degree level students and apprentices.
Of the total participants, approximately 11\% is assumed to be active on the particle accelerators and technical infrastructures, while the remaining 89\% will be engaged in the detectors and experimental physics.

\subsection{Results}

The initial socio-economic impact assessment based on 2 experiment collaborations, limited to the investment costs, operation costs, and core benefit pathways, yielded a positive net present value at the end of the FCC-ee operation phase \cite{csil_2024_10653396}. The subsequent complementary assessment carried out in 2024 \cite{catalano_2025_14905017} used an even more conservative calculation based on an update of the project configuration, considered only a reduced set of well-justifiable benefit pathways, and extended the analysis to wider effects and negative environmental externalities:

\begin{itemize}
\item Revised investment and operation costs, reflecting the price evolution of goods and services between 2018 and 2024.
\item Four experiment collaborations (instead of two as assumed in the initial assessment).
\item Revised social discount rate (2.8\%).
\item Introduction of noteworthy environmental negative externalities.
\item Revision of major benefit pathway monetisation including additionally gathered data.
\item Strict limitation of benefits to core benefit pathways.
\end{itemize}

\setlength{\tabcolsep}{3pt}
\begin{table}[ht]
	\centering
	\caption{Social cost-benefit assessment of the FCC-ee project expressed as an incremental benefit with respect to a counterfactual scenario in which no collider project is implemented after the end of the HL-LHC operation. The results of the 'wider' socio-economic analysis include noteworthy negative externalities and environmental benefits.}
	\label{tab:socio-economic-performance}
	\small{
	\begin{tabular}{llrr}
		\toprule
		\textbf{Cost/Benefit} &  & \textbf{Undiscounted} & \textbf{Discounted} \\ \midrule		
		\multicolumn{3}{l}{\textbf{(A) Costs}} & \textbf{19\,666 MCHF}  \\ \midrule
		\hspace{1em} Investment costs (for 4 experiments, injector and $\rm t \bar{t}$ stage) & & 16\,215 MCHF  & 10\,171 MCHF     \\ 
		\hspace{1em} Personnel costs  & & 16\,802 MCHF  & 7544 MCHF    \\ 
		\hspace{1em} Operation costs (materials, consumables, services)  & & 4410 MCHF  & 1879 MCHF      \\ 
		\hspace{1em} Dismantling costs  & & 228 MCHF  & 72 MCHF      \\ \midrule

		\multicolumn{3}{l}{\textbf{(B) Negative externalities}}& \textbf{354 MCHF}  \\ \midrule
		\hspace{1em} Shadow cost of carbon & & 634 MCHF  & 342 MCHF      \\ 
		\hspace{1em} Loss of agricultural income, biodiversity \& habitat & & 7.6 MCHF  & 4.1 MCHF      \\ 
		\hspace{1em} Social cost of project-related, induced noise & & 0.02 MCHF  & 0.02 MCHF     \\ 
		\hspace{1em} Social cost of project-related, traffic-induced air pollution & & 0.9 MCHF  & 0.6 MCHF     \\ 
		\hspace{1em} Social cost of project-related, traffic-induced GHG externalities & & 9.8 MCHF  & 7 MCHF     \\ 
		\hspace{1em} Social cost of ionising radiation & & 1.3 MCHF  & 0.6 MCHF     \\ \midrule
		
		\multicolumn{3}{l}{\textbf{(C) Core benefits}} & \textbf{23 974 MCHF}  \\ \midrule
		
		\hspace{1em} Scientific production   & & 6507 MCHF  & 2813 MCHF      \\ 
		\hspace{1em} Early career researcher training   & & 20\,687 MCHF  & 4986 MCHF      \\ 
		\hspace{1em} Industrial benefits for suppliers   & & 17\,577 MCHF  & 9569 MCHF      \\ 
		\hspace{1em} Onsite visitors   & & 4538 MCHF  & 2129 MCHF      \\ 
		\hspace{1em} Online and social media   & & 229 MCHF  & 102 MCHF      \\ 
		\hspace{1em} Open software (experiments and detectors)   & & 7428 MCHF  & 4375 MCHF      \\ \midrule
		
		\multicolumn{2}{l}{\textbf{Total costs including negative externalities}} &  \textbf{(A + B)} & \textbf{20\,020 MCHF}   \\ \midrule
		\multicolumn{2}{l}{\textbf{Total core benefits}} & \textbf{(C)} & \textbf{23\,974 MCHF}   \\ \midrule
		
		\multicolumn{2}{l}{\textbf{Reference net present value (NPV)}}  &  \textbf{(C) - (A + B)} & \textcolor{black}{\textbf{3954 MCHF} }   \\ \midrule
		
		\multicolumn{2}{l}{\textbf{Reference Benefit Cost Ratio (BCR)}}  &  & \textcolor{black}{\textbf{1.20}}   \\ \midrule 
		
	\end{tabular}
}
\end{table}
\setlength{\tabcolsep}{6pt}

\setlength{\tabcolsep}{3pt}
\begin{table}[h]
	\centering
	\caption{Additional benefit potentials not considered in the calculation of the reference net present value and benefit cost ratio.}
	\label{tab:socio-economic-performance-extended}
	\small{
	\begin{tabular}{llrr}
		\toprule
		\textbf{Cost/Benefit} &  & \textbf{Undiscounted} & \textbf{Discounted} \\ \midrule		
		\multicolumn{2}{l}{\textbf{(D) Residual asset value for a subsequent collider project}} & - 7911 MCHF  & - 2480 MCHF      \\ \midrule    
		
		\multicolumn{3}{l}{\textbf{(E) Wider benefit potentials}} & \textbf{6916 MCHF}  \\ \midrule
		
		\hspace{1em} Open information platform  & & 5053 MCHF  & 2681 MCHF      \\ 
		\hspace{1em} Open collaborative software   & & 7516 MCHF  & 3487 MCHF    \\ 
		\hspace{1em} ICT spin-offs   & & 832 MCHF  & 409 MCHF     \\ 
		\hspace{1em} Creation of renewable energy sources through contracts  & & 227 MCHF  & 117 MCHF     \\ 
		\hspace{1em} Supply of waste heat   & & 313 MCHF  & 132 MCHF     \\ 
		\hspace{1em} Avoided greenhouse gas emissions (GHG) by supply of waste heat    & & 170 MCHF  & 74 MCHF     \\ 
		\hspace{1em} Rewilding (habitats and biodiversity) & & 0.4 MCHF  & 0.2 MCHF     \\ 
		\hspace{1em} Contributions to regional emergency services   & & 31 MCHF  & 16 MCHF    \\  \midrule

		\multicolumn{2}{l}{\textbf{Net present value including residual asset value }} &  \textbf{(C) - (A + B + D)} & \textbf{6\,446 MCHF}     \\ \midrule

		\multicolumn{2}{l}{\textbf{Total core and wider benefit potentials}}  &  \textbf{(C+E)} & \textbf{30\,890 MCHF}     \\ \midrule
	\end{tabular}
}
\end{table}
\setlength{\tabcolsep}{6pt}

Table~\ref{tab:socio-economic-performance} presents the most certain and stable results of this additional social cost-benefit assessment, showing the core cost and benefit pathways. 
A set of wider benefit pathways may occur, provided that dedicated planning and implementation are incorporated into the project designs and appropriate contextual conditions are in place (see Table~\ref{tab:socio-economic-performance-extended}).
To aggregate the value of measured benefits and compare them with costs and negative externalities, a social discount rate (SDR) was established specifically for the project.
Instead of relying on existing SDRs suggested by international organisations, this project-specific rate accounts for the level of development and preferences for consumption and investments in countries contributing to the CERN budget and the very long duration of the project. The assumed SDR value is 2.8\%.
The result, strictly limited to costs, noteworthy negative externalities, and core benefits, shows a positive net present value of about 4 billion Swiss francs for the project, leading to a Benefit Cost Ratio (BCR) of about 1.20. This value must not be interpreted as a fixed and certain number.
A positive BCR under highly conservative assumptions provides confidence that the project scenario can achieve an overall beneficial contribution to the society.
Depending on the project design and implementation measures to support benefit generation and control costs and externalities, the BCR may be lower or higher.
Continuous tracking of the impact generation throughout subsequent project phases and monitoring during the project implementation and operation are required to assure an overall positive performance.

The findings of the studies carried out over a time frame of about seven years suggest a positive socio-economic net performance of the FCC-ee project.
This result serves to demonstrate the long-term social sustainability of the proposed new scientific research infrastructure. 
The analysis so far has helped identify the main impact pathways with the goal of supporting the design of the research infrastructure for sustained socio-economic impact generation.
The findings reported here are, however, not exhaustive.
Concerning costs, the estimates are likely to evolve further during the design phase until an investment decision is made provisional and subject to updates.
The approach that was taken of integrating the analysis of impacts for the environment, economy and society at large is in line with the EU practice in policy and infrastructure impact assessment.
Further complementary socio-economic analysis, given additional time and resources, would permit uncovering and quantifying additional positive impacts and strengthen the robustness of the benefits already estimated.
Studies by Flyvbjerg et al.~\cite{Flyvbjerg_2003, flyvbjerg2011case, flyvbjerg_2014} highlight a common trend of underestimating costs, overestimating revenues, and misjudging environmental and economic impacts and the unintentional introduction of bias.
Literature, therefore, supports carrying out an uncertainty analysis, especially in large-scale infrastructure projects.
To achieve this, the existing deterministic framework of the CBA can be expanded with a probabilistic model, covering costs and benefits.
This approach will provide a more comprehensive view of the project's expected outcomes.

\subsection{Cost and benefit coverage}

Good coverage of the main cost and benefit pathways has been achieved with the presented socio-economic assessment, although exhaustive coverage of a large research infrastructure is limited by the knowledge about impact pathways, models to quantify them, time, personnel and budget.
The assessment carried out includes the full investment costs for the civil structures, the injectors, the booster, and the collider up to $\rm t \bar{t}$ stage and 4 experiment detectors.
Full personnel costs comprising CERN paid and personnel supplied by the international collaborations to the programme were accounted for.
Full materials, maintenance, consumables, and service costs for the operation phase were included in the costs.
Finally, the dismantling costs of the main ring booster and collider were also included.
This part does not include the disposal of equipment that would be classified as radioactive waste.
Such an estimate can be done only once the technical designs have been completed.
It does also not consider the residual market value of the metals that would lead to a gain after dismantling.

The establishment of a proper residual value for a proposed investment is a key element for input to the financial sustainability analysis of such a project.
According to the guide to the European Commission Cost-Benefit Analysis of investment projects, the discounted value of any net future revenue after the time horizon of the project has to be included in the residual value.
The guide indicates that it is ``the present value at year $n$ of the revenues, net of operation, costs, the project will be able to generate because of the remaining service potential of fixed assets whose economic life is not yet completely exhausted''.
In line with this EC guideline (p. 45, \cite{sartori2014guide}) that economist practitioners use, the residual value is highlighted as a value with a negative sign in the cost account.
If there is a `use value' for the assets, then the residual value is greater zero.
If not, the residual value is considered zero.
Both scenarios were considered in the analysis.
Since the proposed infrastructure can serve two particle colliders in sequence, the residual asset value is considered in this assessment.
Due to the very long timespan associated with the integrated programme, this social benefit-cost assessment, however, only concerns the first phase, the lepton collider FCC-ee.
A residual asset value at the end of the FCC-ee operation phase is calculated that can be made available as a `gift' to a potential subsequent hadron collider project (FCC-hh).
The residual asset value becomes zero if no subsequent particle collider is installed that can profit from the built-up assets.
This approach has also been taken for the social cost-benefit assessment of the Large Hadron Collider \cite{FLORIO201638}.
The calculation of the reference net present value (NPV) for the lepton collider FCC-ee project is carried out in a very conservative way: it does not include the residual asset value at the end of the observation period.
For the purpose of showing that further societal impacts are possible and that an integrated programme consisting of a lepton collider (FCC-ee) followed by a hadron collider (FCC-hh) is preferable, this benefit can be included in a wider NPV.

The materialisation of negative externalities and benefits depends largely on the design of an impact-creation framework and external conditions. 

Negative externalities such as the carbon footprint and the associated shadow cost of carbon depend on the procurement actions taken for the construction, the power purchasing agreements concluded for construction and operation, and the national carbon pricing at the time of generating the carbon footprint.
The indicated shadow cost of carbon considered the construction and operation of the research infrastructure.
The costs for the accelerators and detectors could not be included at this time since they rely on the availability of a technical design and a procurement scenario.

The effects of potential disturbances such as noise and induced traffic linked to the construction depend significantly on the actual number of people that would be affected.
Additionally, certain impacts are challenging to predict at this early stage because of a lack of forecast models for certain benefit pathways, because they could not be identified, because of a lack of time and resources, or simply because of large uncertainties.
Examples include the wider positive externalities that can emerge from developing novel products and processes for re-using excavated materials, advancing technologies for low-carbon construction materials, generating local synergies with the municipalities hosting surface sites, the unanticipated creation of high-value technology spin-offs, and the invention of entirely novel technologies.

This analysis focused on the direct benefits currently known and foreseeable based on past factual evidence, for instance, from the LHC and the European XFEL projects.
It was concluded as soon as a high level of confidence was achieved in demonstrating a break-even point between benefits and total costs, including noteworthy negative externalities.
Conservative assumptions were made for all estimates.
Where a causal relationship between positive or negative effects and the project could not be reliably established, the item was entirely left out.
For cost and benefit items for which commitments remain to be formulated (e.g., the development of dedicated open software platforms to support knowledge dissemination and international collaboration, the conclusion of renewable energy contracts or PPAs, the supply of waste-heat, projects to re-create and strengthen the lost habitats, contributions to the operation of regional emergency services), the elements were labelled to be considered as a `wider' societal impact.
They were included in the calculation of a wider net present value only.

The economic benefits of visitors rely on the existence of a visit programme, dedicated visit points, infrastructures and guides. The reported benefits of onsite visitors are incremental with respect to the counterfactual scenario in which no future particle collider project is implemented. This means, it captures the likely benefits on top of the effects generated by visitors that come to CERN after the end of the HL-LHC programme. This counterfactual case is envisaged to be less impactful, since CERN without a new flagship project at a global scale is assumed to be less attractive for visitors, as recent surveys of CERN visitors carried out in the first quarter of 2025 reveal.
Some benefits are intangible in nature, such as advances in scientific knowledge, impacts related to science diplomacy, ethical considerations, and trust in science. 
Although benefits created by open software and collaborative platforms have significant value potential, it is challenging to estimate which, when and in which ways such benefits reach society without a dedicated innovation creation and transfer programme in place for such technologies.

Several further wider benefits have been envisaged, but were not further pursued at this stage due to resource and time constraints associated with unambiguously clarifying the demand.
They include, for instance, the strengthening of the local and regional electricity and communication networks, the creation of soft mobility, the contribution to the improvement of public transport, the creation of housing and schooling facilities, and the development of local businesses and services.
However, the study included the development of a framework to start elucidating the demands for such wider benefit potentials with the municipalities in the perimeter of the project.
Ongoing and future socio-economic analysis will shed more light on those opportunities once the demands are better understood.

\subsection{Cost and negative externalities}

For the cost-benefit calculations, the FCC has been considered as a design-to-cost project, with a total investment including 4 experiments and the $\rm t \bar{t}$ stage, projected at approximately 16.2 billion Swiss francs (undiscounted) with an uncertainty range of -5\% to +20\%.
This figure encompasses studies, preparatory activities, design work, civil construction works, creation of technical infrastructures, all particle accelerators (injectors, booster and collider) and all investment costs related to the four experiment detectors.

Personnel costs include all human resources involved in the project throughout all lifecycle phases, irrespective of the organisation employing them.
The total personnel costs are composed of wages, indirect costs and employer-related costs without overheads. They are averages for different personnel categories in the different countries participating in the project.
In total, personnel costs make up about 16.8 billion Swiss francs (undiscounted) and 7.5 billion Swiss francs (discounted).

Operation costs refer to all expenses (both in-kind and outflows) that are needed for running, maintaining, and repairing the research infrastructure throughout the entire scientific exploitation (e.g., electricity, water, spares, and human resources for maintenance and repair).
The likely scenario for the operational costs amounts to a total of 4.4 billion Swiss francs undiscounted with a range due to uncertain costs of resources (e.g., electricity and water), supplies and service contracts.
The discounted costs are around 1.9 billion Swiss francs.
This corresponds to an average annual undiscounted cost of about 250 million Swiss francs and about 105 million Swiss francs discounted over the operational time that spans from commissioning to the end of operation.

The cost figures used for this analysis, along with the projected number of project users must be regarded as a working hypothesis that serves for estimating the socio-economic performance, that helps to identify key impact pathways.
It also helps in understanding which benefit potentials may still have potential for further development.
They are formulated based on the understanding of the project at the time of the analysis and are expected to undergo further revisions and refinements as the design progresses.

Estimates of the following cost items covering the period from the investment decision to the end of the FCC-ee operation phase have been included in the presented analysis:

\begin{itemize}
\item Capital expenditures for civil construction, technical infrastructures, all particle accelerators (injector, booster, collider) including the upgrade to $\rm t \bar{t}$ operation and four experiments.
\item Cost of acquisition of all land for the surface sites, accesses and nature enhancement around the sites.
\item Cost of access road creation or refurbishing.
\item Creation of off-site infrastructures required for construction and operation (e.g., accesses, water supply, water treatment, local electricity connections).
\item Full personnel cost estimates (CERN and the international collaboration) required for the design, construction and operation phases covering the particle accelerators, technical infrastructures and experiments.
\item Typical operation costs (consumables, water and energy, maintenance, repair).
\item Dismantling cost of the lepton collider and its specific technical infrastructures that cannot be reused for a subsequent hadron collider.
\item Negative residual value of the infrastructures for a subsequent particle collider project.
\end{itemize}

In addition to the items mentioned so far, a portion of the investments involving reusable and durable assets such as superconducting radiofrequency systems, electricity infrastructures, civil structures, and basic technical infrastructures will last for use in a subsequent particle collider project. According to the EC guidelines on Cost-Benefit Analysis (p. 45, \cite{sartori2014guide}), this residual value is reported as a discounted value with a negative sign in the cost part. The residual value of the initial phase assets is significant and contributes to the sustainability of the integrated FCC programme. The estimated discounted value stands at approximately 2.5 billion Swiss francs, equivalent to 24\% of the total investment.

Estimates of the following negative externalities have been included in the presented analysis:

\begin{itemize}
\item Cost of managing excavated materials as part of the investment cost.
\item Shadow cost of carbon for civil construction and the energy used for operation.
\item Cost of rewilding measures in the vicinity of the surface sites (e.g., preservation and improvement of wetlands, creation of trees to compensate for forest clearance and the creation of meadows and grasslands) as part of the investment cost.
\item Economic loss (direct, indirect upstream and induced downstream) caused by the consumption of agricultural land for a period of 30 years.
\item Economic value of the cleared forest.
\item Economic value of loss of habitat and biodiversity.
\item Societal cost of construction and operation-related noise.
\item Societal cost of added traffic and air pollution induced by the transport of excavated materials.
\item Societal cost of additional ionising radiation.
\item Societal cost of radioactive waste as part of the operation cost.
\end{itemize}

\subsection{Impact pathways}

The following impact pathways have been included in the analysis either as core elements or as potential wider benefits:

\begin{itemize}
\item Value of scientific content production.
\item Increased market value of early-stage scientists and engineers.
\item Value of on-site tourism.
\item Value of online presence and social media activities.
\item Value of industrial spillovers involved in the construction of infrastructures, accelerators and experiments.
\item Market value of ICT spin-offs based on the frequency of past and current spin-off company creation at CERN.
\item Societal value of open software products.
\item Market value of waste heat supplied and the societal value of carbon emissions avoided.
\item Market value of additional renewable energy sources created as a result of long-term power purchasing agreements.
\item Market value of treated waste water.
\item Societal value of improved and recreated wetlands, meadows, grassland and forests.
\item Economic value of compensated agricultural spaces.
\item Societal value of regional emergency and fire-fighting services required for the new research infrastructure.
\end{itemize}

The following impact pathways are considered highly reliable and form, therefore, a set of core benefits on which the conservative net present value is based:

\begin{enumerate}
\item Value of produced and cited scientific publications.
\item Lifetime salary premium of early-career researchers and engineers aged under thirty.
\item Incremental value for suppliers due to follow-up contracts that result from their activity in the project.
\item Value of on-site visitors based on the travel cost method and local spending.
\item Leisure time value of people consuming project-related online and social media content.
\item Estimated value of open software developed in the area of particle detectors and experiments, used by other scientific institutes and industries.
\end{enumerate}

\begin{figure}[h]
  \centering
  \includegraphics[width=\linewidth]{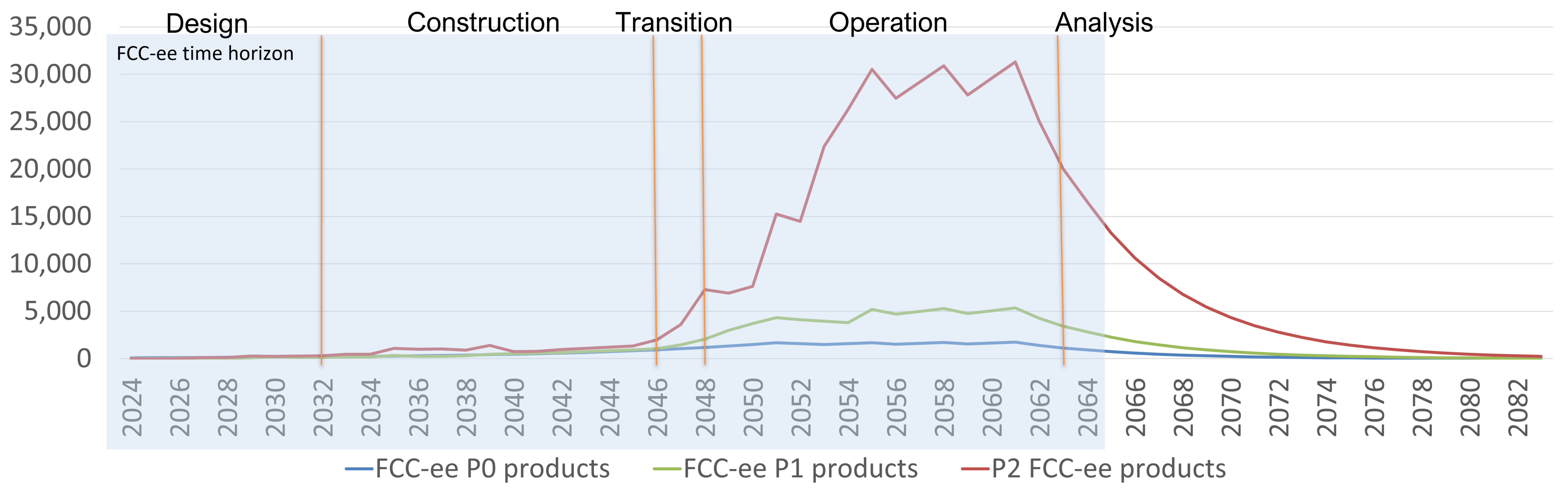}
  \caption{Expected distribution of FCC-ee scientific products over time.}
  \label{fig:scientific-products}
\end{figure}

The first impact pathway assessed is the value derived from scientific content production. It stems from the knowledge flow generated by scientists and engineers engaged in the collider and its experiments, resulting in diverse scientific products, ranging from journal articles and working papers to conference proceedings and presentations.
These products can have a lasting impact, extending beyond the high-energy and particle physics community, potentially influencing other knowledge domains and addressing broader societal challenges. Scientometric techniques were employed to estimate scientific production and its propagation from the FCC-ee project.
The methodology involved analysing historical patterns observed in comparable physics research programmes like Tevatron, LEP, and LHC.
The economic concepts of opportunity cost and value of time were employed to determine the social value of scientific products.
This approach considered the time spent by individuals, depending on their involvement and responsibilities, in an experiment collaboration for producing these outputs and valued it based on their average hourly salaries.
This method was applied to estimate the social value of scientific products produced by researchers directly involved in the research programme (so-called `tier 0' products), those citing the initial tier 0 knowledge (so-called `tier 1' products), and products citing tier 1 outputs (so-called `tier 2' products). Between 34\,000 and 38\,000 scientific products are expected to be directly produced in the course of the FCC-ee project (tier 0 publications), with an additional 538\,000 to 618\,000 tier 1 and 2 products likely to be generated based on that corpus until the year 2083 (see Fig.~\ref{fig:scientific-products}).

Considering that about 55\% of the time spent by researchers is dedicated to scientific research activities and the production of the production of scientific outputs, but only about 22\% can be attributed to FCC-ee specific scientific products, the estimated benefit from scientific production is around 6.51 billion Swiss francs undiscounted and about 2.8 billion Swiss francs discounted.
This valuation accounts for the diminishing value of tier 1 and tier 2 products compared to tier 0 products, as knowledge propagates through subsequent waves of production and the initial input from FCC-ee progressively diminishes.

\begin{figure}[h]
  \centering
  \includegraphics[width=\linewidth]{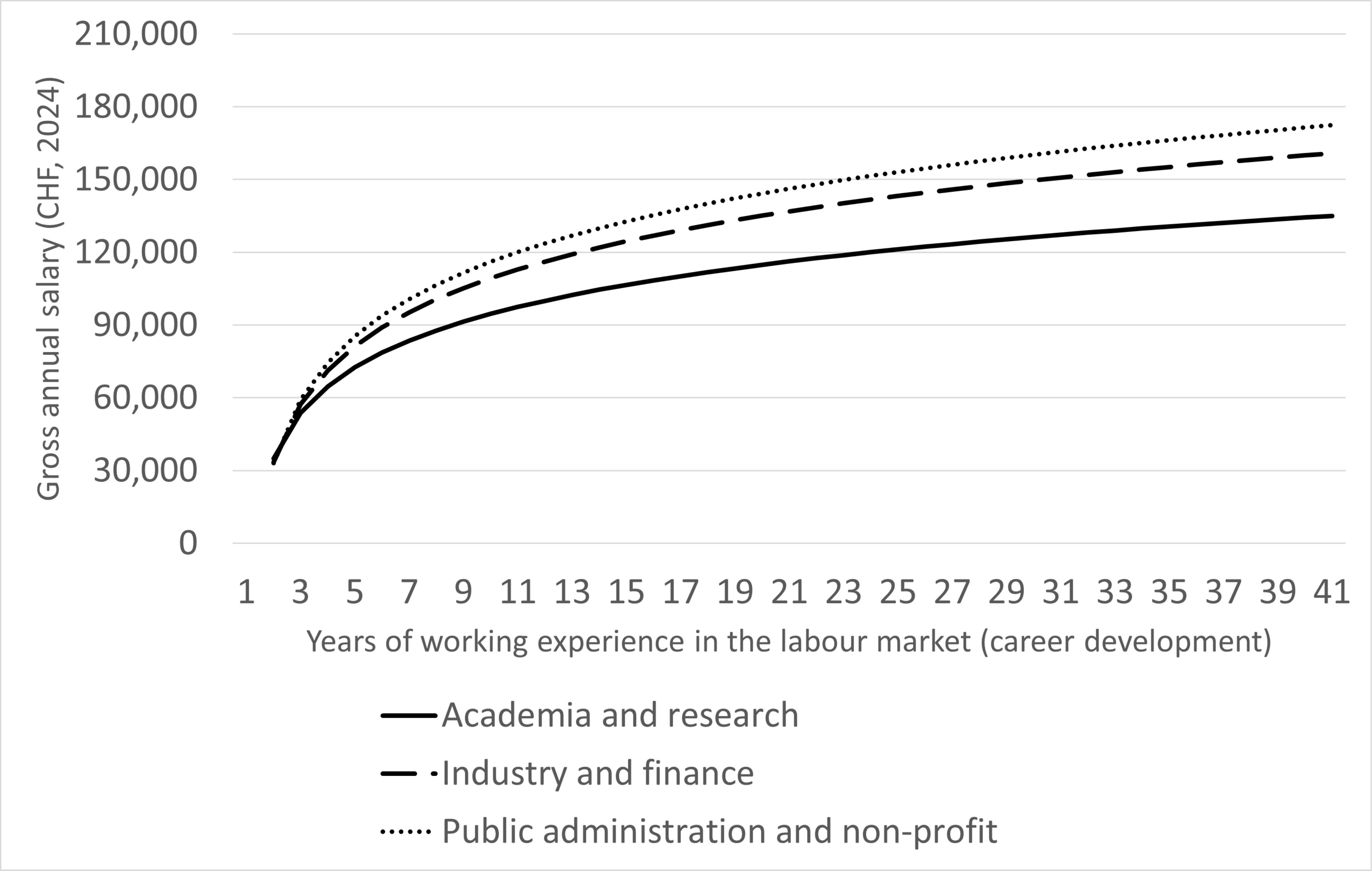}
  \caption{Lifetime salary of early-career researchers by sector of employment.}
  \label{fig:salary-early}
\end{figure}

The second impact pathway studied is the value of early career research and engineer training, which reflects the project's role in imparting knowledge, fostering skills development and building capacities for individuals actively engaged in the research programme throughout its lifecycle. The value stems from the fact that persons who were engaged in large-scale, multi-sectoral high-tech and science projects at CERN represent a higher value to their eventual employers than persons who do not have such training experience. This is reflected by higher starting wages and faster-growing salaries. The effect remains until the person retires, and it leads to an overall incremental added value with respect to peers without such an experience. The analysis specifically evaluated the benefits accrued by technical students, doctoral students, post-doctoral researchers, and associated scientific and engineering personnel up to a cut-off age of 30. The analysis did not include the training value for apprentices and other highly qualified personnel with limited-term contracts, temporary labour and contracted workers. The benefit has been quantified by estimating the lifelong career development improvements for participants upon entering the labour market after gaining work experience within the research programme. Previous studies, further validated through a survey involving approximately 2600 individuals, indicate that the lifetime salary benefit for an early-stage researcher working at FCC-ee ranges from a minimum of 2\% to 10\% for the average period of stay in the research infrastructure (3.78 years) (see Fig.~\ref{fig:salary-early}). The total undiscounted socio-economic benefit is assessed at around 20.7 billion Swiss francs, and the discounted value is around 5.0 billion Swiss francs.

\begin{figure}[h]
  \centering
  \includegraphics[width=\linewidth]{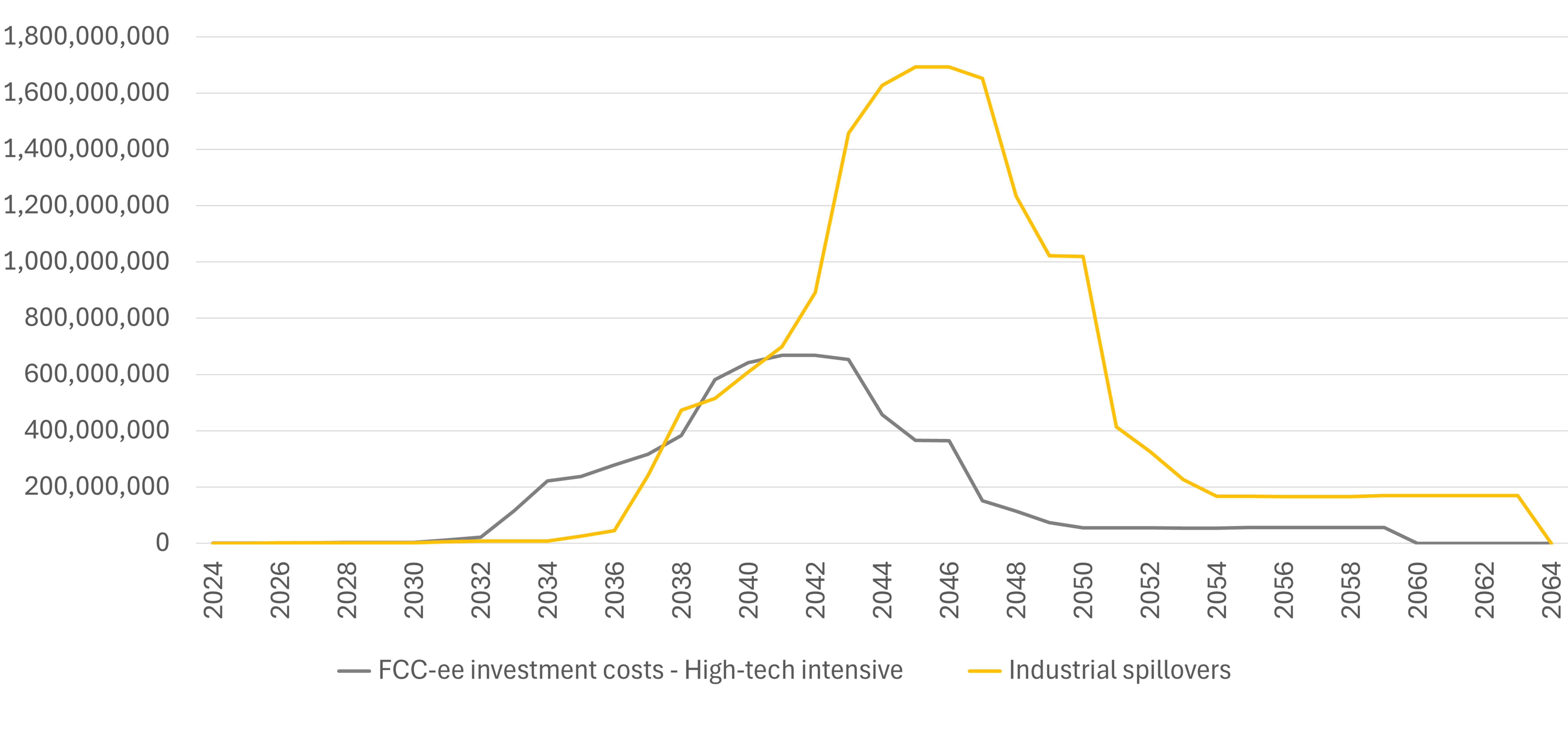}
  \caption{Time profile of FCC-ee industry benefits for suppliers, compared with the high-tech investments costs (Swiss francs, undiscounted - baseline scenario).}
  \label{fig:benefits-suppliers}
\end{figure}

The third impact pathway explores the benefits generated by the project for industrial suppliers engaged in the construction of the infrastructure, the particle accelerators, and detectors. The benefit stems from increased financial performance of the suppliers a few years after they have received contracts in the frame of the project. The effect of being a supplier in a large-scale project, in particular in domains that are characterised by project-specific designs, developments, and adaptations of off-the-shelf products and services, is linked to the gain of experience, increase in efficiency, and a broadened market access. In essence, the impact stems from a knowledge gain acquired through close collaboration with the research infrastructure. This knowledge, in turn, contributes to the creation and improvement of new processes, products, and services that suppliers can leverage in other markets and domains. It leads to additional contracts that the supplier is able to conclude because of that knowledge gained. This cause-effect-impact chain has been exhaustively analysed over some decades, and it has been found to be robust and stable. A profit multiplier has been determined, standing at 1.96 for procurement of items with low or moderate level of technology intensity and 3.06 for procurement with high  level technology intensity. Applying this multiplier to the investments that are likely to generate further industrial spillovers, the undiscounted benefits are estimated at 17.6 billion Swiss francs. The discounted benefits are estimated at 9.6 billion Swiss francs (see Fig.~\ref{fig:benefits-suppliers}). 

\begin{figure}[h]
  \centering
  \includegraphics[width=\linewidth]{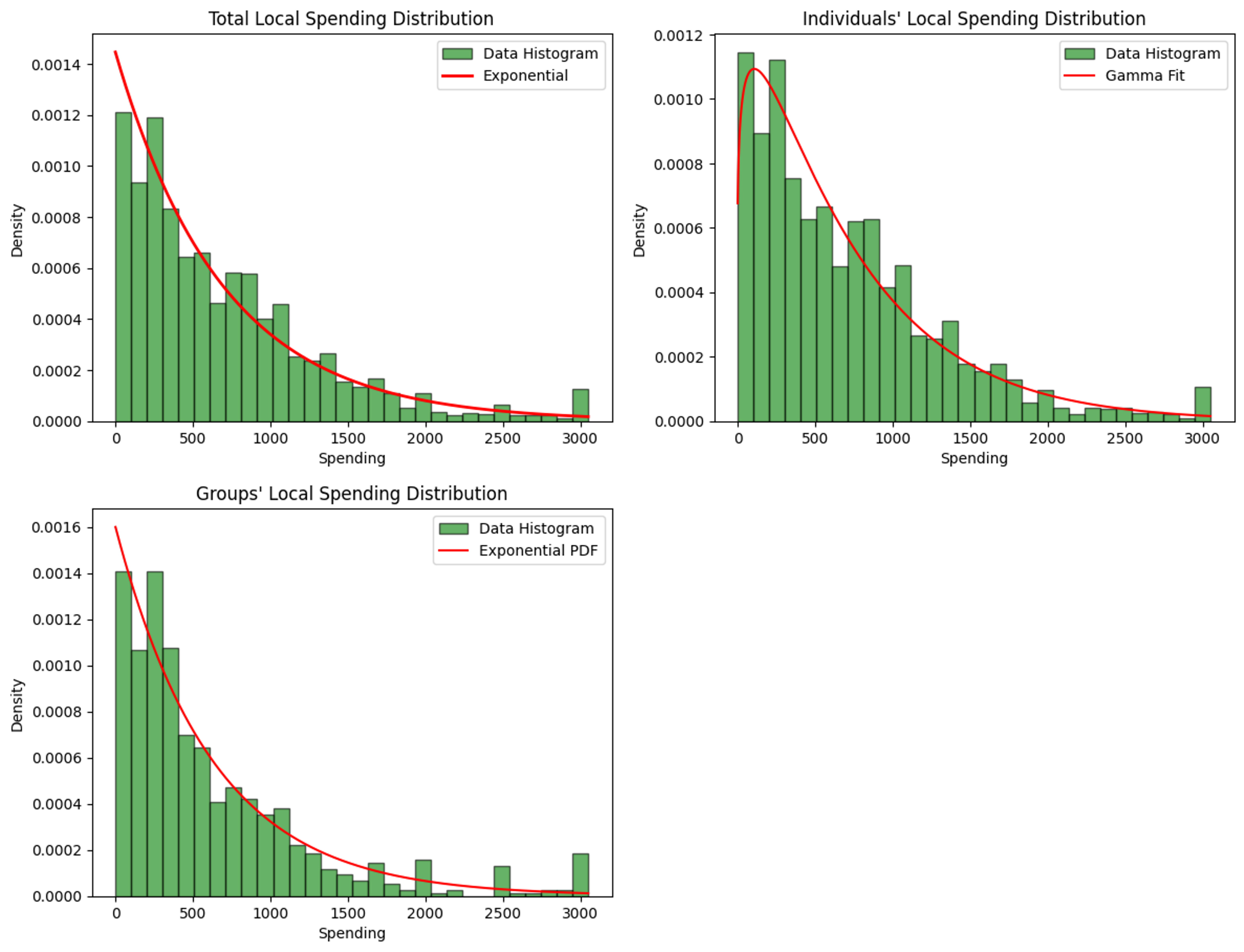}
  \caption{Distributions of the spending of all on-site visitors, individual visitors and visitors that come to CERN as part of groups.}
  \label{fig:spending_distributions}
\end{figure}

The fourth impact pathway concerns the benefits generated by on-site visitors. 
Before 2020, 150\,000 visitors per year came to CERN. In 2024, after the opening of the new Science Gateway visitor centre, the number increased to about 350\,000 external visitors (not counting visits of persons participating in CERN projects, company visits and visits of employee families, accounting to about 40\,000 additional persons).
While in the period before the Science Gateway the participation was equally distributed among group visits and individuals, the significantly increased capacity to welcome unguided visits led to a change in the distribution.
Today about 76\% of people visit CERN individually and 24\% come as part of groups that enjoy the possibility to also visit experiment facilities, in particular the ones linked to the LHC.
55\% are classified as `CERN motivated' visitors, whose primary purpose of the trip was to visit CERN, while 45\% are "region motivated" visitors, who travelled to the area for other reasons and took the chance to also visit CERN.
The absolute number of persons that are part of visit groups remained relatively stable.
Two campaigns of guided systematic interviews with on-site visitors before and after the opening of the Science Gateway permitted a robust multi-year spending behaviour of the visitors to be established (see Fig.~\ref{fig:spending_distributions}).
In total, more than 4\,000 individuals were interviewed, and only about 3\,550 responses with self-consistent and credible answers were retained for the analysis.
The average stay of persons in the region is 3.75 days, indicating that once they have visited CERN they extend their visit to further destinations in the vicinity. 

The local spending associated with the visit ranged between 625 Swiss francs for group visitors and 713 Swiss francs for individual visitors, with a mean of 691 Swiss francs for all visitors.
Based on the survey, sampling from the obtained visitor spending distribution, the sum of the local spending due to 350\,000 annual on-site visitors translates to a tangible economic benefit for the region of about 250 to 350 million Swiss francs per year.
This figure does not, however, capture the socio-economic benefit related to the FCC-ee.
Based on this current survey, only a certain fraction of future visitors can be attributed to an FCC.
As outlined initially, the analysis limits benefit potentials to the incremental effects, i.e., the difference between the effects of on-site visitors that can be attributed to the existence of the FCC and the evolution of CERN without the FCC.
For this reason, the following assumptions were applied to estimate the incremental socio-economic benefit for the FCC-ee:

\begin{itemize}
\item Conventional CERN visits: an increasing share of these visitors is attributed to FCC-ee, starting at 0\% in 2024 and gradually rising to 50\% by 2064.
\item Visitors to the FCC-ee construction sites and the four experimental sites: these visitors are fully attributed to FCC-ee, as they would not have been accounted for in a scenario without FCC-ee. 
\item Visits to the decommissioned LHC tunnel: adhering to a conservative approach, these are not attributed to FCC-ee since they would be justified even in the absence of the FCC-ee.
\item Visits to FCC experiment visitors centres.
\item About 10\% of all visitors are coming to the CERN main site.
\item Open-day visitors: these are attributed to FCC-ee only for the additional capacity created by the new accelerator; all other visitors are excluded from the count.
\item 
\end{itemize}

\begin{figure}[h]
  \centering
  \includegraphics[width=\linewidth]{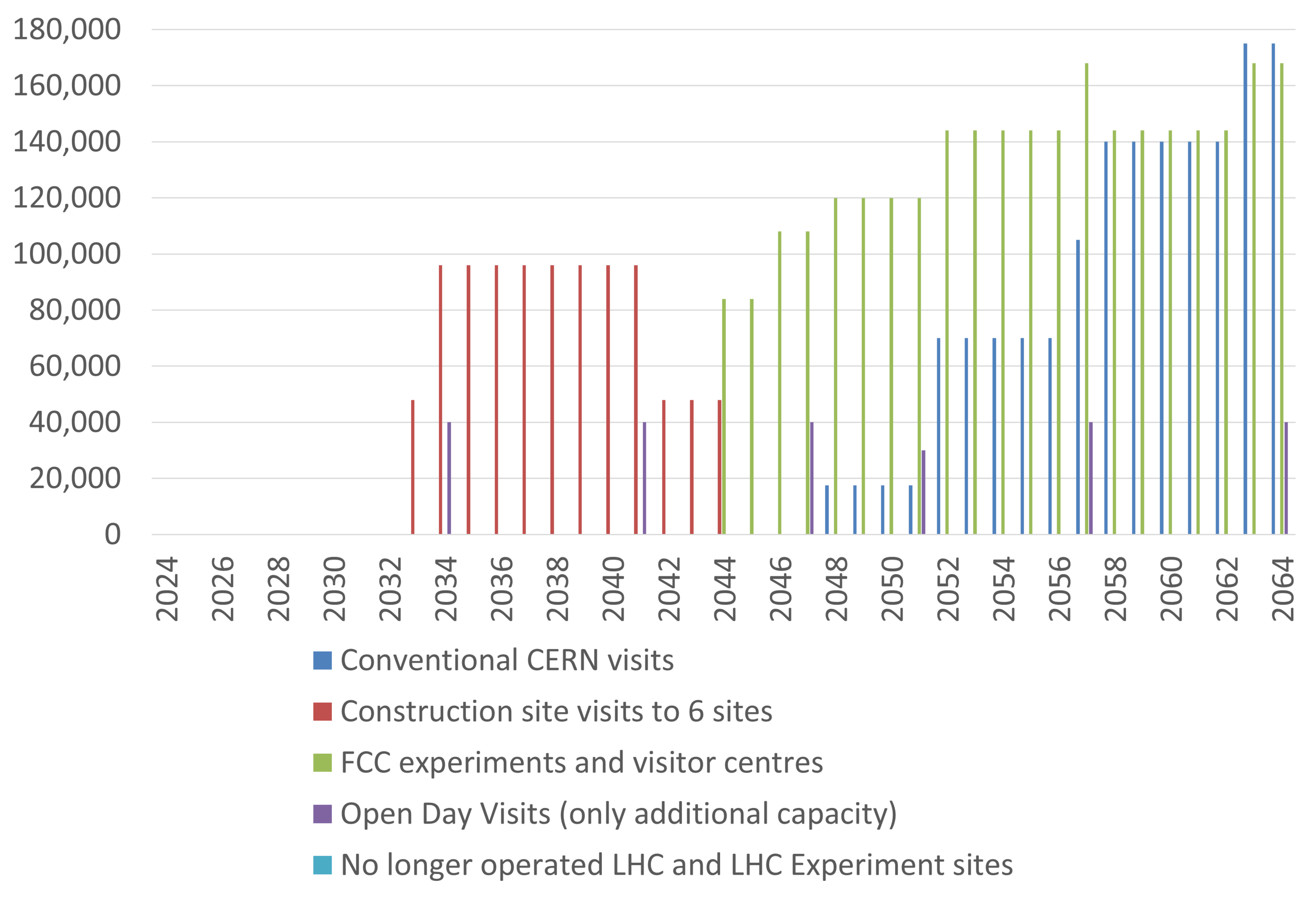}
  \caption{Number of on-site visitors attracted by FCC-ee over the period 2024-2064 used for estimating the impact estimation.}
  \label{fig:onsite_visitors}
\end{figure}

Based on these assumptions, it is estimated that out of almost 19 million visitors throughout the entire observation period, only about 5.5 million are assumed to be attracted by FCC-ee (see Fig.~\ref{fig:onsite_visitors}), while the remaining will be motivated more by the overall CERN activity and longstanding worldwide reputation of scientific excellence.

Considering that only the incremental benefit is reported here, the difference between the number of visitors that CERN would welcome without and with an FCC-ee is about 13.5 million over the observation period.

The benefit for on-site visitors was monetised using the travel cost method, which incorporates the costs borne by visitors to travel to CERN and FCC, the economic value of time spent travelling, along with local expenditures connected to their visit based on two actual surveys carried out at CERN over two years (pre- and post COVID).
The discounted benefit generated by about 5.5 million onsite visitors attributable to FCC-ee is estimated to be approximately 2.1 billion Swiss francs.

The fifth impact pathway analysed concerns virtual visitors who consume webpages and social media. Due to the limited availability of data, the analyses concerned only websites and social media channels that are directly managed by CERN. The estimate, therefore, significantly underestimates the actual benefits generated by numerous online content that would be made available on a global scale.

The estimated number of online visitors is based on historical data collected by CERN's communication group regarding visits to the main CERN website and social media accounts. Assumptions were made about the number of visits specifically related to FCC-ee. This estimate has been revised to reflect changes in the project’s timeline and to ensure consistency with the revisions made in the estimation of onsite visitors, particularly regarding the share of FCC-ee attribution. It is assumed that the highest attributable share of visits is approximately 50\%, aligning with the assumption made for CERN visitors. 

In the scenario without the FCC, it is assumed that the number of online visits to websites and social media will follow the same patterns as the on-site visitors in the absence of the FCC-ee. In the scenario with the FCC, the number of online visits corresponds to the trend of growth of scientific publications due to the projected FCC-ee research programme. The share of online visits attributable to the FCC-ee starts as a marginal share of the total CERN online visits. The main jump in online visits is expected to occur with the start of operation. This share is expected to grow further during the operative phase. The number of online visitors is expected to reach its maximum several years after 2050 and the end of operation. In these years, the share of the total CERN online visits associated with the FCC reaches conservatively about 50\% and remains constant. This assumption is in line with survey results to onsite visitors about the reason to visit CERN today. Based on these assumptions, for the FCC-ee over the 2024–2064 period, impressions are estimated at 1.7 billion, engagements at 79 million, and CERN website visits at 175 million. These estimates are highly conservative and do not take into consideration future social medial developments and online platforms are likely to appear, but whose existence can today not be anticipated.

The monetisation of these visits is based on the actual observation of an online presence of a little over 3 minutes. Distributions are applied to different media types, such as social-media interactions and consuming online video snippets related to FCC. The value of time spent is determined using the `opportunity cost' method, which suggests that time spent on social media and websites represents a missed opportunity to engage in other potentially profitable activities. Since not all virtual visitors are necessarily part of the economically active workforce, the opportunity cost of time was estimated by using per capita GDP (instead of wages), adjusted based on the geographical distribution of virtual visitors. The total undiscounted cultural benefit for online visitors is estimated to be approximately 229 million Swiss francs, which corresponds to 102 million Swiss francs discounted.

The sixth core impact pathway encompasses free and open-source software, systems, and platforms causally linked to the research programme. It is assumed that in particular software related to the experiment and detector projects generates incremental societal value due to the adoption by other science projects, research institutes, and companies with particular needs that cannot be easily satisfied by commercially available software. While the first user base is typically found in the physics, astronomy, and medical research domains, companies using such software can be very diverse. Recent examples of using software from the LHC experiments range from space-borne earth observation systems through medical imaging and mining to shipping container traffic and stock-market transaction analysis. In essence, in any application where massive amounts of data need to be processed, patterns need to be identified in background dominated environments, time-critical applications demand custom solutions, and data processing efficiency is key to success, the developments of particle physics scientists and engineers are sought after.

The estimation of this benefit pathway is based on the use of actual particle detector modelling and simulation software, for which the use outside the core community could be tracked over the recent years. It serves as a `proxy' for comparable developments for which the need during the development of FCC detectors and experiments has been confirmed. Like software package analysis, new software can find use beyond the core community, thereby generating a spillover benefit for society. A number of developments will concern the use of artificial intelligence in this area, configurable hardware, edge computing, and ever-evolving data communication technologies. Several of these developments are admittedly associated with uncertainties, but as the past has shown, it is likely that the collaborations around the FCC-ee experiments will be able to make significant contributions in the ICT domain.

For detector modelling and simulation software, approximately 50 research centres, space agencies, and companies were identified using software developed for the LHC experiments. This included 38 institutions beyond CERN that contributed in some way to the developments. Examples include but are not limited to the Fermi National Accelerator Laboratory, SLAC National Accelerator Laboratory, the Centre for Medical Radiation Physics, and the European Space Agency. Additionally, 12 other laboratories, institutes, and companies use such software without contributing to its development. They include, for example, NASA, General Electric, Philips, Siemens, Varian, Boeing, and General Motors. CERN does not systematically track the number of users and installations, but data available in 2023 suggest a current user community of about 95 contributing institutes (excluding CERN). Assuming the ratio of 3:1 between contributing and non-contributing institutions remains stable, the total number of non-contributing users is estimated to be around 30. In the absence of more reliable data, this estimate is considered reasonable and conservative.

The benefit is estimated as the avoided cost for the external users, and it is based on the cost of production of comparable new software. It was estimated that the production cost of existing detector modelling and simulation software was about 44.2 million Swiss francs up to 2013, covering the first 20 years of its development starting in 1994. CERN contributed about 50\% of this cost (22 million Swiss francs), with the remaining amount funded by other external contributors. The estimates are based on the hypothesis of comparable cost figures for FCC-ee related software and assuming CERN's contribution remains at 50\%. The avoided cost for each contributing organisation is the total estimated production cost minus their specific contribution, while the avoided cost for non-contributing users is the full production cost. 
It is assumed that new software would be first released after 5 years of development with developments continuing over the years, even during the operation phase, due to the continued interest by academia and industry.

In this scenario, the total cumulated avoided cost representing the total undiscounted benefit is estimated to be about 7.4 billion Swiss francs. The discounted benefit is about 4.4 billion Swiss francs. This benefit, in fact, would stand for a range of potential software developments that cannot be explored in detail at this early stage. It is advised that any future development foresees an open access mechanism for such software, actively promotes such software in areas outside the particle physics community in science and industry and, most importantly, includes a systematic tracking of active users to help make socio-economic impact estimates more accurate and reliable.
 
\subsection{Wider benefits}

A number of wider benefits have been analysed in detail in addition to the core benefits presented in the previous section. Due to uncertainties that are linked with the possibilities to actually turn those wider benefit potentials into tangible impacts, they are not considered in the calculation of the reference baseline net present value. The analysis and quantification of those benefit potentials is, however, based on factual observations of past effects and outcomes and is therefore solid with respect to monetisation. As with the core benefits, highly conservative working assumptions have been established, and only effects that can be credibly justified have been included. Therefore, a second net present value is presented that includes those wider benefits. To turn  such potentials into tangible impacts, the preparatory implementation phase of the future project needs to include dedicated work that plans for such benefit creation. Voluntary objectives and commitments are needed by the project owners to put instruments in place to actually turn the potentials into impacts.
Last but not least, monitoring and tracking will need to be put in place to follow up on the impact generation.

The following wider benefits have been studied and are briefly presented in this section:

\begin{enumerate}
\item Open information platform
\item Open collaborative software
\item Company spin-off generation in the information and communication technologies (ICT) sector
\item Build-up of renewable energy sources
\item Supply of waste heat
\item Avoidance of greenhouse gases by substituting traditional heat energy sources with recovered and supplied waste heat
\item Improvement and creation of habitats and increase of biodiversity
\item Strengthening of emergency services
\end{enumerate}

A future, global collaborative research project relies on a long-term, open scientific and technical information platform that permits the community to make their knowledge widely available so that it becomes findable, accessible, interoperable and reusable (FAIR principle). CERN has a long-term track record in putting such infrastructures in place, the World Wide Web being the most prominent one. This technology subsequently permitted the development and making available of additional services, such as the CERN Document Server (CDS), which is based on the in-house developed Invenio platform. This development has led to the creation of the Zenodo service, which the European Commission has endorsed as the catch-all repository for all EU-funded research. Today CERN operates this platform that is entirely part of the European Open Science Cloud (EOSC) for the European Commission. By continuing to be at the forefront of big science, CERN can have a lead role in scientific and technical information provision in Europe and at a global scale. The launch of a new large-scale science mission, the FCC, justifies not only the continuation of ongoing developments, but will likely also lead to the development and implementation of a new generation of information systems, in line with past occurrences. Such a platform may offer functionalities that go beyond pure document and data management and long-term data preservation. It may offer services that are today impossible to conceive. In order to estimate the potential value of such a future development, the value of the Zenodo platform was estimated as a conservative reference with minimum functionality offered to society today. This was possible, because historic data of sufficient quality for relevant periods could be identified. An econometric model was constructed to estimate the socio-economic impact of that virtual information repository, designed to fulfil collaborative information storage and usage requirements. The estimated monetised value is derived from measurable benefits associated with comparable repositories, encompassing data storage, online usage, and downloads, net of the present value of its development, operation and maintenance costs. 

The undiscounted value over the FCC-ee observation period of such a platform is 5 billion Swiss francs, and its discounted value is about 2.7 billion Swiss francs. To turn such an impact potential into reality, several pre-conditions apply: the global community participating in the project has to have a demand for such a platform, the community must be committed to use the platform, communities beyond the high-energy and particle physics domain must adopt the platform (as was the case with the Zenodo platform being endorsed by the European Commission). The fundamental requirements rely on an intent to play the role of developing and operating an information platform for several user communities. Long-term sustained resource engagement and a plan to disseminate the developments beyond the core community are also needed.

Another typical case for open platforms in the history of CERN's large international collaborations is the creation of tools that support collaborative work. The `Integrated Digital Conference' (Indico) event management system is one example of such development. It revolutionised the management of physical and virtual meetings, lectures, conferences and led to a comprehensive documentation of collaborative work and can be considered a key enabler of global collaborative projects. Since its start of development in 2002, the platform has spilled over to numerous academic and international organisations worldwide. A future large-scale project not only has very similar needs, but will create demands beyond those that are satisfied with this platform today. The integration of collaborative writing, sketching, AI support for meetings and minute taking, translation among different languages, collaboration management, video conferencing, recording and media publication, real-time messaging, shared workspaces, integration with document management, approval and selected information distribution processes are just a few selected examples for which the need is already starting to appear in the frame of the studies related to the FCC.

For these reasons, the current Indico platform was chosen as a conservative, representative example of a collaborative platform that can enable and improve future collaborative work in the frame of a large-scale project. The value generated from developing a new service to facilitate the worldwide collaboration involved in the FCC project in terms of meetings, calls, and event management was assessed by considering the willingness to pay (WTP) by private users for a comparable toolset, totalling 7.5 billion Swiss francs undiscounted and 3.5 billion Swiss francs discounted. Since the potential of such a new ecosystem relies on a commitment for development, the adoption by communities beyond high-energy and particle physics, the value was only considered as a wider benefit.

The knowledge-transfer office records of the last two decades have shown that a stable number of companies have been created every year by persons who were engaged in CERN's flagship particle collider project and the international experiment detector collaborations. A prominent example includes Proton AG with over 400 employees and more than 100 million customers, offering secure e-mail and VPN as an unparalleled service to other  providers. Other examples are Advacam which specialises in imaging devices for various industrial applications based on particle detector technology, LightEye working on LiDAR technology for long-range wind speed measurements to improve aviation safety and PlanetWatch which specialises in data acquisition of environmental data. Based on these factual historical data, it is assumed that FCC-ee would generate approximately two new spin-off companies in the information and computing technologies (ICT) sector alone each year from the design and preparation phase until the end of the observation period. Considering the probability of company survival each year and the annual market value of companies in the ICT sector, the socio-economic benefit is estimated to be at least 832 million Swiss francs undiscounted and 409 million Swiss francs discounted. This figure does not include the economic benefits that are generated by numerous micro enterprises and independent consultants that carry out their activities thanks to the experience and skills they acquired in the frame of large-scale particle accelerator and experiment detector projects.

Some selected environmental benefits that the project could generate were also assessed. They include the voluntary goal of supplying the infrastructure with electricity from renewable energy sources. Entering specific energy supply contracts and long-term power purchasing agreements (PPAs) can serve as a lever to build up new renewable energy sources since the power supplier can secure long-term funding of new renewable energy investment projects before making a financial investment decision. A scenario of a portfolio of complementary energy supply contracts for the FCC construction and operation has been analysed in terms of this value-generation pathway. Assuming that after 2050, society is largely de-carbonised and conservatively, no growing demand for such a funding instrument is assumed on a time horizon of more than 30 years, the reported benefit remains limited: 227 million Swiss francs discounted and 117 million Swiss francs discounted. The development of a renewable energy portfolio can, in principle, begin with an investment decision for a new particle collider project and with a commitment of the project to engage renewable energy sources. In this case, the potential benefit can rapidly turn into a tangible economic benefit. It is suggested that this benefit be revised by that time and reconsidered under the evolving project implementation boundary conditions.

A large part of the energy used to operate the particle collider and its experiments is converted into heat. Traditionally, the heat is dissipated via water-based cooling systems that connect to evaporation towers. A study has been conducted to prove the feasibility of recovering and re-using waste heat.
Recovery and supply of waste heat is a concept that is built into the design of the particle collider from the onset. 
Recovered waste heat will be supplied via district heating networks to consumers in the vicinity of the FCC-ee sites.
The environmental benefit stems from the fact that the energy used by the FCC research infrastructure is reused and supplied for heating and cooling purposes, avoiding the use of alternative, mainly non-renewable, sources with higher carbon intensity such as gas, wood, oil and conventional electricity mix.
The environmental benefit of waste heat reuse is derived from the avoided GHG emissions in the project scenario, where CERN supplies waste heat—produced using a cleaner electricity mix—to district heating networks serving consumers near the FCC-ee sites.
This is compared to the emissions from alternative heating and cooling methods typically used in the CERN region.
The total benefit is estimated at 170 million Swiss francs undiscounted, corresponding to 74 million Swiss francs after discounting.

However, turning this potential benefit into tangible socio-economic impacts requires an adaptation of the operation schedule to the demand curve, agreements with district heating operators to put networks in place and to operate them, and customers who commit to take off the heat. Since such developments are typically time-consuming, entail territorial developments that last one to two decades, and depend also on local developments around the future surface sites, the benefits are not included in the core net present value calculation.
In addition, the potential benefits emerging from the avoidance of greenhouse gas emissions by replacing conventional heat sources are uncertain for the timescale after 2050, when in principle the economy aims at being largely de-carbonised. 
Together the benefit potentials, i.e., the costs saved by using waste heat instead of conventional heat sources and avoiding greenhouse gas emissions, add up to about 313 million Swiss francs undiscounted and 132 million Swiss francs discounted. Once a decision to move forward with a project has been taken, a revision of those impacts can take place, since a stronger commitment from the project owner also enables regional stakeholders to plan for waste-heat district heating networks and this eventually will lead to significantly increased benefit potentials.

The project does only consumes land due to the development of eight surface sites, but it also opens opportunities to create new natural spaces in the immediate vicinity of the sites. Creating or improving those habitats to integrate the surface sites well in their environments can strengthen the natural spaces around the sites, protecting them from further artificialisation and helping to improve the biodiversity, partially compensating for the effects of the loss of space.
While today rewilding projects are foreseen in connection with the surface sites, they have not yet been discussed with the local stakeholders and they have not yet been designed.
Therefore, this benefit remains potential and is not included in the calculation of the core net present value.
The rewilding of habitats and biodiversity could represent an undiscounted value of 0.4 million Swiss francs corresponding to 0.2 million Swiss francs discounted.

The future particle collider will be embedded in a territory that spans an area of roughly 30 $\times$ 30 km.
Care has been taken to select locations for surface sites that are in the vicinity of major transport routes.
Nevertheless, the current concept implemented at CERN to provide emergency, rescue and fire-fighting services to surface sites of existing particle accelerators from a central fire brigade at CERN becomes unfeasible for a future infrastructure of the FCC scale.
Therefore, the safety concept foresees a strong and lasting collaboration with local emergency services.
This concept can be based on support with equipment, training, and personnel made available by the research infrastructure.
It could help to increase the expertise and skills of local emergency services, contribute with state-of-the-art equipment, strengthen the number of specialists in the region and increase the cooperation and coordination among different emergency services.
This benefit potential has an undiscounted value of about 30 million Swiss francs and represents a discounted value of about 16 million Swiss francs.
Since it relies on the development and implementation of a safety concept and plan at the territorial level which can require significant amounts of time and adjustments that make the concept implementable, the benefit is not included in the core net present value calculation today.

Furthermore, two other classes of benefits described in the subsequent sections have not been directly included in the overall formula to determine the project's net present value: the economic creation of value added and the public good value.

\subsubsection{Complementary analysis of economic value added}

The economic value added has been estimated in addition to the socio-economic impact by analysing the economic linkages (indirect, direct and induced) during the entire 30-year period covering design, construction and the operation phase of the FCC-ee using the established Input-Output Table methodology \cite{WIFO-benefits-method}.
The study estimated the economic and employment effects that are connected to the construction and operation of the new research infrastructure.
They are not included in the incremental cost-benefit analysis, since they do not represent the creation of new economic goods and services beyond the research infrastructure, but they lead to economic value added due to the activation of numerous economic sectors in the frame of the research infrastructure-related activities.

These effects arise mainly from the construction of civil structures, particle accelerators and colliders, technical infrastructures, experiments, operating expenses, and consumption related to the personnel involved. 
This economic analysis was only carried out for an infrastructure with two experiments and was not updated to the current baseline.

Using an economic input-output model, the cumulated expenditure of about 21 billion Swiss francs over a 30-year construction and research operation period could be connected to some 800\,000 person-years of employment opportunities, corresponding to almost 30\,000 jobs per year via global value-adding chains. In addition to about 6\,000 directly project-related science, engineering, administration, and management jobs globally, more than 20\,000 jobs are needed to provide the goods and services for construction and operation. The host countries, Switzerland and France, and especially the canton of Geneva and the Departments of Ain and Haute-Savoie could benefit most from the operation phase-related expenditures. In total, around 13\,000 jobs would be filled or created annually on average in France and Switzerland.

An initial, construction-related investment of 12.1 billion Swiss francs directly generates globally 5.4 billion Swiss francs of value added, generating almost 80\,000 person-years of employment, i.e., more than 8\,000 jobs per year over a ten-year investment period. Including the indirect effects in the production process, value added linked to the investment rises to 11.6 billion Swiss francs, leading to about 180\,000 person-years of employment or 18\,000 jobs per year during the investment period.
Widening the system boundaries to include depreciation (i.e., the capital stock firms need to build up or replenish in order to cope with the FCC-related production), the FCC-related value added grows to more than 14 billion Swiss francs and leads to more than 230\,000 person-years of employment opportunities or 23\,000 jobs per year during the investment phase. All types of companies, large, medium and small, can enjoy the benefits from the value added. The countries that can profit from these benefits depend on the chosen procurement strategy. The construction sector benefits most since almost half of the investment volume can be attributed to civil engineering. Along the cycle, however, its share steadily falls, and most other sectors' shares rise.

The operation of the FCC generates value added by paying wages and social security contributions and through the depreciation of the investment. No (net) operating surplus is considered since CERN, a purely scientific research organisation, is not profit-oriented. The direct value added during the operation phase is estimated at around 455 million Swiss francs annually. For operation, a mix of inputs (intermediate goods and services) is needed, whose procurement will provide suitable firms with the opportunity for sales and employment and thus generate about 165 million Swiss francs of indirect value added annually. The total direct, indirect and induced value added during the operation phase of more than 620 million Swiss francs per year supports 8\,400 jobs, the majority in France and Switzerland. The buildup and supply of renewable energy sources for the operation of the research infrastructure would raise Europe's value added by another 500 million Swiss francs, securing an additional 7\,400 person-years of employment in Europe. By economic sector, the structure of the consumption effects is markedly different from the effects of the investment and operating expenditures: real estate activities, (retail) trade, personal services and the hospitality sector are the main beneficiaries of the consumption of project-related employees.

The lower limit for annual tourism spending due to the project in the wider Geneva region is at least 130 million Swiss francs per year. Switzerland and France share the bigger part of the total effects, with around 1\,700 jobs linked to visitors only. Another 500 jobs are European; the rest – around 600 – are filled outside Europe. Globally, tourism effects would sustain about 2\,700 jobs.

The consumption of electrical energy in the frame of long-term power purchasing agreements would generate a direct value added due to the capacity build-up activities of about 200 million Swiss francs, linked to 3\,500 person-years of employment. Including intermediate inputs and investments needed to produce and install the equipment, the contribution to Europe's value added rises to 510\,million Swiss francs, securing 7\,400 person-years of employment in Europe. Worldwide added value effects related to the electricity supply sector amount to 620 million Swiss francs or 11\,600 person-years of employment.

These figures must be interpreted with some caution. Most importantly, the indirect jobs are derived under a steady-state assumption. The estimates do not project major economic variables into the future (exchange rates, price levels and productivity being the most important ones). Therefore, the effects are estimated as if the FCC was constructed and operated today. This should not be considered a shortcoming, as it helps decision makers to grasp the effects of the estimates more easily when referring to a familiar frame of reference – the economy as it is today. The resulting figures on value added are less compromised by this simplification, due to the fact that the evolution of economic key performance parameters cannot be forecast on the timescale of an FCC project with construction starting in the mid-2030s and coming into operation in the late 2040s. The employment figures linked to the expenditures represent reliable upper bounds since labour productivity is expected to rise. Even under changing economic conditions, the FCC would remain what it is today – a major undertaking for the scientific community and society at large, with likely significant scientific, technological, engineering and economic impacts.

The analysis shows that the costs that a project like the FCC entails are also connected with tangible economic impacts in terms of sales opportunities for firms and employment opportunities for scientists and non-scientists alike. By concentrating on a core set of transmission mechanisms only, these results constitute a lower bound for the expected economic effects. Therefore, even though the narrow economic linkages of the construction and operation are not larger than would be expected for a project of this size, the potential for spillovers into quite unrelated areas of technology and business are certainly much more pronounced – for example, only a few projects would have the touristic attractiveness, not to mention its technological and scientific potentials.

Also, by estimating the regional structure of the effects linked to the construction and operation of the FCC, the analysis has shown that the connection between contribution to CERN, direct contracts and indirect benefits is not always clear-cut. For example, China and the United States, which are not member states of CERN, are estimated to have sizeable economic benefits due to their prominent roles in global value chains. This information could form the basis for negotiations between CERN and countries such as China on intensifying and formalising closer collaborations in the future, which would be beneficial for both parties.

\subsubsection{Public good value}

To gauge the volume of the societal benefits, a comprehensive survey was conducted among nearly 10\,500 individuals across nine countries \cite{secci_2023_7766949}, including both CERN member and non-member states potentially contributing to the future particle collider project at CERN.
The survey aimed to assess public awareness of CERN and its research activities, to evaluate the perceived value of a new research infrastructure, like FCC, to the public, and to compare this monetised value with the per-capita annual contributions made by CERN member states. 

The public good value should not be integrated in the estimate of the project's net present value, since the perceived value of the project, i.e., the value that taxpayers associate with the project, is orthogonal to the actually estimated incremental benefits. Measured through the willingness to pay, the public good value also depends on the knowledge of the project, its costs, negative externalities and likely incremental benefits. Hence, the value of it is affected by some cognitive biases, such as, for instance, an information bias. The presence of such bias is not negative and cannot be avoided. They are part of the mechanism that permits people to associate a value to an asset to which they have no direct access.

Results indicate that 41\% of the respondents are aware of CERN and its mission, which, although lower than some other international organisations like NASA, remains generally positive. Over 80\% of respondents believe that scientific research at CERN advances our understanding of the universe and contributes to improving quality of life. The hypothetical willingness to participate financially in the development of the new research infrastructure project was assessed, revealing varying distributions by country.
Median values range from 2 Swiss francs per person per year in France to 20 Swiss francs in Switzerland, both CERN member states. For non-member states, the median willingness to pay (WTP) varies from zero in Japan to 24 in the USA (although the mean value for Japan is 10 Swiss francs, meaning that a significant fraction of the Japanese adult population values that type of scientific research).
The total value was estimated by multiplying the estimated per capita yearly WTP by a total adult population of about 380 million persons over 30 years in the CERN Member States, starting with the first relevant investments.

In all observed cases, the perceived public value in CERN’s member states is higher than CERN’s annual operational budget of 1.4 billion Swiss francs.
The average per capita contribution in these states is approximately 2.5 Euro per year or about 5 Euro per income taxpayer per year.
The total estimated WTP for a future collider project at CERN surpasses the estimated total costs of the FCC by a factor of 20 and exceeds its quantified benefits by over 11 times.

These findings robustly support the conclusion that the decision to invest in a future particle collider programme at CERN can be considered justified from a societal perspective since the people who potentially fund the endeavour assign more value to it than it costs in total.

\subsection{Conclusions}

The socio-economic impact analysis is based on a social cost-benefit assessment conducted on the first phase of the FCC programme, spanning a time frame of 40 years from the financial investment decision to the end of operation. The FCC-ee has quantified costs, negative externalities and conservative benefit potentials and wider benefit potentials across various domains. Costs of about 20 billion Swiss francs discounted and negative externalities of about 354 million Swiss francs can be compared to the benefits that this research infrastructure can generate. The total present value of the monetised core socio-economic benefits associated with the FCC-ee research infrastructures has been conservatively estimated at a discounted value of 24 billion Swiss francs. Additional wider benefits amount to about 7 billion Swiss francs discounted. The infrastructures would represent a residual value of about 2.5 billion Swiss francs for a subsequent hadron collider, made available as a `gift' to this second project.

The conservative estimate for the net present value (NPV) of the project over its entire observation period is about 4 billion Swiss francs, yielding a positive benefit-cost ratio of about 1.20.
Including the residual asset values can bring the NPV to about 6.5 billion Swiss francs.
Extending the estimates with wider benefit potentials, the project provides an opportunity to reach an even higher NPV.
However, achieving such a performance requires the design, planning, and implementation of benefits, as well as generating measures with commitments and continuous monitoring and tracking at the level of CERN and international collaboration. 
Proper risk management and cost control must be in place to manage costs and negative externalities.

Costs were based on the currently available investment cost estimates and on the experience of operating CERN's particle accelerator and collider complex over the last two decades. The most noteworthy negative externalities were identified, quantified and monetised using lifecycle analysis methodologies and guidelines at the European level for wider socio-economic impact assessment and project appraisal. Cost values were currency and time value adjusted for 2024 as the base year. 

Benefits have always been appraised through a conservative methodology, drawing on the current understanding of the project, insights garnered from analogous past research infrastructures, and socio-economic impact analyses on the LHC \cite{FLORIO201638} and the HL-LHC \cite{Bastianin:2319300} that were carried out as pre-cursors, anticipating the need to eventually produce an FCC socio-economic impact study.

Additionally, recent data collected between 2020 and 2024 has played a crucial role in shaping these estimations. To ensure a comprehensive perspective, data gathering efforts included over 16\,000 individuals through online surveys. This diverse group included members of the public, visitors to CERN, users of platforms like Zenodo and Indico, as well as former researchers. 

In a likely scenario between optimistic and pessimistic assumptions, the quantified benefits exceed the total costs associated with the design, construction, and operation of FCC-ee, resulting in a net positive socio-economic impact for the project.

Should a financial investment decision be taken to proceed with a construction project, it will be necessary to update the assessment during the coming years and to compile all the materials necessary to obtain funding from participating countries and seek authorisations from the host states. 

The approach used in this analysis intentionally entirely excludes the uncertain and unforeseeable impacts of knowledge increase generated by the science mission on society because of their inherent unpredictability.

The assessment of the public good value of the research infrastructure for the public offers an insight into the overall benefits of a future particle collider project at CERN for society, gauged through the public's perspective and willingness to financially contribute to a project's implementation.
The insight that the perceived value of a future research infrastructure project exceeds the costs of the FCC is important evidence that the public values such scientific research more than it actually costs and thus helps obtain the Social Licence to Operate (SLO). 

While recognising the need for refinement, this analysis serves as a tool to inform decision-making, optimise user engagement, identify and mitigate risks, and enhance social acceptability for the FCC-ee project. The findings emphasise that the FCC-ee project holds the promise of positive impacts not only for the scientific community but for all of society, thereby contributing to the project's long-term sustainability.

\section{Returns to participating countries}

\subsection{Overview}

Several economic analysis organisations (e.g., LSE \cite{crescenzi_2024_13166167}, WIFO \cite{wifo_project268677}) have been consulted to analyse how potentially participating nations can increase the likelihood of benefiting from financially contributing to the project. An example set of those proposals is compiled in this section.

There exists a consensus among experts that international organisations, including CERN, generate significant economic and societal impacts beyond their core activities in their host countries and their participating nations.

One existing concept to ensure a continuous return to the participating countries is the International Liaison Officer (ILO) approach for integrating national companies in the procurement processes of the international organisation. This approach was studied by economists, and it was found that for a future large-scale project, a more structured and comprehensive approach could lead to better returns for the participating nations.

The recommended measures \cite{crescenzi_2024_13166167} to ensure good returns to the financially participating nations revolve mainly around three topics:

\begin{enumerate}
\item Industry benefits
\item Training benefits
\item Cultural benefits
\end{enumerate}

One lever to ensure good returns is to increase the sustainability of spreading the benefits of procurement, by enabling more firms across the various regions to successfully participate in future collider-related procurement and to tap into the value chain created by the contracts. This concerns:

\begin{itemize}
\item Levelling the playing field so that more firms can participate in the procurement process.
\item Supporting SMEs as these are the firms that are most likely to face barriers in the procurement process.
\item Embedding more firms in the procurement value chain.
\item Decarbonising the procurement supply chain through aggregated energy supply contracts or power purchase agreements (PPAs) and energy communities that permit pooling renewable energy contracts for groups of suppliers and larger procurements.
\end{itemize}

Another lever is to foster brain circulation. CERN is an attractive place for people from across the world to study and work. However, many of these talented individuals end up remaining abroad once their project involvement ends, leading to what is often referred to as a brain drain for their country of origin. In order to ensure that more places benefit from human capital formation in a virtuous process of brain circulation, the following policies can help:

\begin{itemize}
\item Understanding the local needs for talent so that Member States and their regions can target the repatriation of people better.
\item Attracting CERN alumni through tailored initiatives.
\item Boosting connectivity so that sending and receiving regions benefit from human capital development. 
\end{itemize}

A further lever is to extend the benefits of tourism and to increase it to more regions. With the new Science Gateway, each year 300\,000 to 400\,000 people visit CERN. However, most of the benefits accrue in a small geographic perimeter in Switzerland and France, where CERN is located. One way for more regions to tap into this is to ensure that the larger FCC perimeter benefits from the on-site visits. Then, further developments can take place to turn CERN into a gateway for science tourism.

Economists have developed proposals that can be considered in the project development phase to help participating nations to reap the most from the opportunities that a new large-scale project offers. One possibility for a contributing country would be to establish a national organisation with dedicated staff that aims to develop the most suitable matches between the country's competencies and interests and the project. This activity goes beyond classical industrial liaison. It aims to integrate schools, universities, research centres, companies of all sizes and national funding agencies through development policies and concrete actions with the project. This requires in-depth knowledge about the project, ensuring that the organisation's personnel is involved long-term to be able to build up durable links and gain comprehensive knowledge of the landscape of national competency and capacity. The direct revenue from procurement contracts is only considered the tip of an iceberg of potential impacts. Long-term returns can exceed those from direct contracts by far. They include:
\begin{itemize}
    \item Repatriation of people and acquisition of highly skilled and trained personnel from the project.
    \item Training nationals at all levels for limited periods abroad.
    \item Using the project as a pilot factory, demonstrating technologies, bringing technologies to market and entering new markets.
    \item Build durable transnational value chains, using the project as a motor for national innovation leveraging national funds.
    \item Connect companies of different sizes.
    \item Break up research silos by bringing universities and institutes from different disciplines together around a single science project.
\end{itemize}

The entire concept aims to overcome the information asymmetries that privilege only a few companies and universities that benefit from historical links, which enable them to benefit from CERN's large-scale science projects.

All efforts to plan for a good return for financially participating countries rely on a systematically established analysis, tracking and regular evaluation of socio-economic performance that can adapt the project. Sufficient evidence exists about the socio-economic benefits of CERN's activities through different pathways, but there remains a gap in the understanding of what is needed to increase and sustain these effects. Making data gathering and evaluation an integral part of the FCC project can help setting goals  for data collection and the methodological approaches to measure these effects. 

The recommendations documented by LSE and WIFO for ensuring good returns for participating nations provide specific and tangible examples of successful cases for each policy recommendation based on past initiatives. A significant advantage of several proposals is their standalone nature: they can be implemented individually. Many of them can also be implemented by countries and regions  independently, allowing them to reap benefits from a project without having to involve other participating nations.

\subsection{Local employment opportunities}

A future research infrastructure with additional territorial development can lead to territorial benefits through direct employment, goods and services supply, indirect and induced job opportunities in a wider perimeter than covered by CERN's activities today. Indirect and induced impact do not only concern the value chain of regionally supplied goods and services, i.e., further materials and service suppliers to satisfy the needs of FCC suppliers and jobs that satisfy the needs that emerge from the presence of personnel associated with the FCC construction and operation activities in the vicinities of the surface sites. The federation of certain activities that relate to the construction and operation, i.e., services that are research infrastructure enablers at certain locations can also create new and durable territorial employment opportunities. Such services include, for example, during the construction project planning, project management, construction site preparation and operation, and business services (e.g., accounting, human resources, IT, land plot management). Services for the operation phase include, for instance, surface site surveillance and safety, technical operation services, technical maintenance, operation of visit facilities and general business services, as is the case of the construction phase.

The economic impact potentials, in particular the employment-related effects, are still under study and develop on the considered supplies and services as well as the locations where they would be supplied.
Preliminary analysis points to opportunities in the range of 1\,000 to 2\,000 jobs over a sustained period of time until the end of the century. The effects are mainly linked to household expenses generated by those persons, representing per year about 219 million euro for the region Auvergne-Rhône-Alpes. Direct tax-related benefits are estimated to be in the range of 2.6 (regional) to 3.3 (national) million euros per year. Indirect and induced tax-related impacts range between 3.6 (regional) and 4.1 (national) million euro. The main sectors that would profit from those benefits are trade, housing, specialised construction works, civil engineering and public works, commerce, manufacturing and specialised equipment, materials processing, financial and insurance services, chemical processing, energy production and distribution, cultural goods and services, restoration and telecommunications.

These preliminary results are to be taken with much care, since the project scenario, the construction and operation organisation are at a very early stage and subject to development.

\subsection{Opportunities for impact generation}

The following policy recommendations are examples taken from the studies carried out by independent economic research organisations to work towards generating regional returns in a global project.

\begin{itemize}
\item Participating nations build up national support organisations to connect various actors (companies, universities, research centres, innovation hubs, national funding agencies) to the new research project to leverage beyond pure contract acquisition as much as possible: \begin{itemize}
    \item Piloting technologies. 
    \item Bringing technologies to market. 
    \item Training people.
    \item Exploit the scientific research opportunities.
    \item Leverage national funding for developments motivated by the new project.
    \item Attracting highly qualified and trained people via job fairs. 
    \item Transferring knowledge from the project to universities and companies.
    \end{itemize}
\item Move to a standard tendering approach, which includes adopting standard keywords and codes and publishing tendering opportunities to a wider audience, such as in the Supplement to the Official Journal of the European Union (TED). This can reduce search costs for firms and increase the opportunities for small and medium enterprises (SMEs). 
\item Use the standard international NACE\footnote{The Statistical Classification of Economic Activities in the European Community, commonly referred to as NACE (for the French term `nomenclature statistique des activit\'{e}s \'{e}conomiques dans la Communaut\'{e} Europ\'{e}enne'), is the industry standard classification system used in the European Union.} code system to ensure that companies receive requests for participation that match their competencies well and to ensure that countries can make an efficient match of their specialisations to the project's needs.
\item Establish dedicated support centres within local business association offices in regions with concentrated clusters of firms active in the project relevant sectors. These centres will offer specialised support to regional SMEs aspiring to engage in the procurement process. 
\item Unbundle large contracts to enhance the possibility of SMEs to participate in the procurement process while making the supply chain more resilient.
\item Set aside lower value contracts for SMEs, if they meet competitiveness criteria for quality and price.
\item Conduct value chain mapping for the different technologies required to understand the spatial distribution of procurement activities and for regions to understand the local competitiveness opportunities better. 
\item Set up Local Content Units (LCUs) as separate bodies or functions to be developed within existing regional development agencies in areas where important CERN suppliers are located. They may also be established and coordinated directly by CERN as a ‘local economic impact acceleration unit’.
\item Create alliances between regions with project-relevant sectors to facilitate interregional collaboration and the sharing of knowledge. 
\item CERN could facilitate energy supply contracts or aggregate power purchase agreements (PPAs) by acting as an anchor tenant or encouraging the creation of consortia. Alternatively, energy supply contracts or PPAs can be created for larger procurements and groups of suppliers can be created in the frame of dedicated supplies. Also, energy communities that exist in other countries are an effective example of obtaining electricity for production and manufacturing purposes from renewable energy sources. Conditions can be included in procurement requirements.
\item Regions identified as talent contributors should conduct a comprehensive analysis of the type of skills that the region needs to attract with a special focus on profiling homegrown talent that has left the region to join CERN.
\item Create and promote opportunities through grants and fellowships targeting high-skilled individuals to move to their regions.
\item Provide repatriation incentives to return to their home region for individuals who stayed at CERN.
\item Engage with CERN staff and alumni abroad through knowledge exchange and entrepreneurship programmes, mentorship schemes and awards.
\item Together with tourism boards and other research facilities, develop science tour packages that feature additional scientific attractions across Europe and technology hubs in combination with other activities.
\item Establish a funding stream for sustained socio-economic impact evaluation and data collection from the outset of the FCC project.
\item Operate an Open Data portal that attracts social scientists to use CERN to continue to evaluate the socio-economic impact of investment in science.
\end{itemize}

\section{Requirements and constraints for a preparatory phase}
\label{sustainability:preparatory_phase}

The subsequent design and preparatory phase calls for a more detailed design of the technical infrastructures based on systematically identified and properly structured and documented requirements for the particle collider and the experiments. This documentation has to be used to carry out an LCA according to the standards and best practices that were already used for the analysis of the construction footprint.

Due to the absence of a detailed design, the long time scales, and the need to develop a particular procurement scenario for a study, a more high-level approach can be taken for the estimation of the potential climate effects of the particle accelerators and experiments. However, much care must be taken to avoid an approach that is too simplistic and only focuses on the generic analysis of the carbon footprint of the assumed materials. As far as possible, full systems should be considered, and where not possible, reasonable assumptions about the manufacturing, assembly, and installation process must be made and included in the analysis process.

Concerning the use of renewable energy, it is important to establish an energy procurement team that can anticipate the development of supply contracts or purchase agreements. The team needs to develop a strategy and prepare for the establishment of an energy supply portfolio in view of the test, commissioning, and operation phases. The preparation for procuring energy for the construction phase is to be started as soon as possible. A time frame of ten years is considered adequate for the preparatory and procurement process of renewable energy for the operation phase.

The potentials of waste heat supply and water consumption reduction depend on national stakeholders putting district heating networks in place and working with CERN on water treatment and supply of treated water at a regional level. Therefore, dedicated local designs, operation plans and pilot schemes will need to be put in place before the research infrastructure is assumed to enter the operation phase. A time window of 20 years is adequate, considering authorisations, financing, the establishment of local operators and the connection of consumers to the district heating network.

With respect to landscape integration, the development of rewilding projects, the use of excavated materials and the design of compensatory actions need ample time for working with local and regional stakeholders. This also requires the allocation of dedicated project-internal resources to establish contracts with external partners and to work together with them, the communities, and public administration. Several years should be allocated for building up these capacities. Hence, a schedule is needed that provides flexibility in working with the local and regional stakeholders, avoiding the build-up of too much pressure. Such pressure can be counterproductive to the achievement of mutual agreements and consensus about territorial development projects and the generation of socio-economic benefits at all levels.

\section{Recommendations for a preparatory phase project}

This section presents a set of recommendations that have been derived from the work presented in this analysis. Their purpose is to ensure that the benefits can be monitored and regularly reported so that the updated findings can be integrated in the infrastructure's design, construction, and operation phases.

The socio-economic study showed that a break-even of costs and benefits can be achieved with the FCC-ee research infrastructure with the impact pathways documented so far. However, the potential can only be fully exploited if a continuous tracking of socio-economic impacts is integrated in the infrastructure’s design, construction, and operation phases. To achieve this objective, nine recommendations have been made to the FCC project owners by the expert group of economists (CSIL, LSE, WIFO, University of Milano, Economic University of Vienna) who have been involved in the socio-economic studies so far.

\begin{enumerate}
\item \textbf{Allocate adequate personnel and material resources for the impact identification, design, planning, implementation, monitoring, and assessment} to help fully leverage the impact potentials and make FCC a socio-economically sustainable endeavour.
The personnel need to be empowered so that the results of the impact assessment and the impact generation recommendations are accommodated in the design where appropriate, that they are implemented, and that they are continuously and systematically monitored. Also, the analysis should be carried out with the help of all participating project members. This requires that the socio-economic impact assessment is implemented across the entire organisation, with the direct involvement of the highest managerial level and reporting directly to CERN's key stakeholder, the Council. This approach is expected to secure the sustainability and effectiveness of the process, as well as help enhance project acceptance among financially contributing countries.

\item \textbf{Encourage participants that scientific and engineering works are published via gold open access channels} (as opposed to self-publishing) \textbf{and reputable outlets} (as opposed to depositing information in preprint servers only). This ensures proper identification of work for tracking purposes and increases the likelihood of their uptake.
This socio-economic impact assessment study has revealed challenges in identifying all the scientific output associated with the research infrastructure activities. It is essential to track the scientific production of researchers involved in experiments and monitor citations across papers and other scientific products. However, scientific outputs lacking proper unique digital identifiers and citation references, as is often the case with workshop proceedings and presentations available on platforms like the CERN-developed Indico \cite{indico_conference_tool}, cannot be adequately identified and considered for socio-economic valuation. These types of outputs should, whenever possible, be replaced or supplemented by citable reports or papers. Establishing strategic partnerships with leading publishers, as exemplified by the SCOAP3 \cite{gentil_beccot_2014_scoap3} project, is a suitable approach to ensure the presence of citable publications. Open-access platforms used for dissemination could integrate download, citation and reference tracking if they do not already offer those functions (e.g., Zenodo \cite{nielsen_zenodo_2014} and ArXiv \cite{ginsparg_arxiv_2011} currently lack these functions). Furthermore, additional research is necessary to expand the monitoring and tracking of scientific content beyond scientific and engineering articles and presentations to include books to understand better the societal uptake of knowledge acquisition within the FCC programme.

\item \textbf{Foresee a framework to monitor the movement of people after they leave the project}, to facilitate the analysis of future career developments. 
Every individual contributing to the project for a minimum duration should be encouraged, on a voluntary basis, to engage in a comprehensive monitoring programme. Thanks to mechanisms to reconnect with them periodically, this initiative would entail long-term follow-up and periodic collection of a set of basic information concerning their current work position. 

\item To ensure the effective analysis of industrial spillovers generated during project implementation, \textbf{incorporate systematic monitoring of the incremental economic impacts on suppliers} resulting from procurement actions over multi-year periods. 
Given that these effects may not manifest immediately and can extend beyond the duration of the procurement contract, it is crucial to obtain feedback from the companies involved. This necessitates the establishment of a comprehensive procurement monitoring framework that facilitates ongoing engagement. Such a framework should include provisions for storing information on the individuals initially involved in the contract, enabling periodic follow-up after the contract's conclusion. Additionally, it should facilitate the collection of essential information regarding the spillover effects on the company generated by the procurement experience with CERN.
This can be achieved through mechanisms such as short online surveys designed to gather pertinent data. By implementing this approach, the FCC programme can gain valuable insights into the long-term economic impacts of its procurement activities, fostering continuous improvement and enhancing collaboration with industrial partners.

\item \textbf{Information and Communication Technologies (ICT) have been identified as a key impact pathway that should be better leveraged} through targeted transmission actions to society, accompanied by systematic monitoring of uptake.
CERN and the FCC represent a globally unparalleled environment for the development of software and platforms to serve the needs of global collaborations, which is in high demand across societal and industrial sectors, too. By identifying the specific requirements of such collaborations and initiating dedicated development projects, the potential for widespread adoption beyond the FCC project can be created. Examples include collaborative event management, document creation, communication (both one-to-one and one-to-many), file sharing, information management, social networks, remote operation, cybersecurity, distributed computing, and data processing. CERN and the member states should ensure that the technologies developed are accessible free of charge to users outside the high-energy and particle physics community while also guaranteeing long-term maintenance and improvement. This requires making software and tools available on platforms that support download and installation tracking. The lack of such data has been identified as one of the limiting factors hindering the accurate estimation of ICT impacts. 

\item With respect to the impact of spin-off companies, \textbf{there is a need to establish a systematic method for tracking companies founded by former participants in the FCC programme, monitoring their evolution over time, and assessing their economic value}.
The establishment of new companies leveraging knowledge acquired through the FCC programme represents an important societal benefit.  Notably, spin-offs often emerge from participants’ amalgamation of different skills and experiences rather than the direct exploitation of a single licensed technology. In particular, the technologies developed in collaborative R\&D projects cannot be attributed to CERN alone, and CERN does not track the technologies of collaboration partners. However, to accurately gauge the economic and societal benefits generated, a voluntary, systematic tracking mechanism is essential. This tracking framework should facilitate a comprehensive assessment of the impact of spin-off companies. It could include the establishment of a centralised database to record information about spin-off companies (such as their founders, country, industry sector, products/services offered), and have the possibility to periodically contact them to provide updates on their companies’ progress.  

\item The study revealed the relevance of economic benefits due to on-site visitors. To ensure sustainable monitoring of the on-site tourism impact pathway, CERN and the FCC collaboration should \textbf{establish a unified framework to continuously collect essential data on visitors to CERN’s exhibition centres, the experiments and any other relevant visit site}.
A systematic tracking mechanism for visitors is indeed necessary to accurately report on the effects of on-site visitors. Visitors should be requested to provide a small set of basic information, including their country of origin, mode of transport to Geneva, the main purpose of the visit, duration of stay, visitor spending, and feedback on the exhibitions and the visit sites (e.g., their level of satisfaction). By implementing this monitoring framework, it will be possible to effectively track and evaluate the cultural impact of on-site visitors, as well as enable continual enhancement of visitor experiences. 

\item The \textbf{identification and quantification of environmental benefit potentials and the development of dedicated projects should be intensified.}
The identification and analysis of environmental benefits are important for social acceptability and for achieving a net-zero balance of the scientific research activity for society. This requires the establishment and empowerment of economics and environmental experts in the coming project preparatory phase and intensified cooperation with industrial partners and Host State services, for a reliable identification of potential environmental benefits and costs associated with the project. In addition to the creation of renewable energy sources, the reuse of waste heat and the reuse of excavated materials, further opportunities may be uncovered through additional research and extended discussions with project stakeholders. For instance, impact pathways related to water usage, land management, biodiversity conservation, and transport infrastructure need thorough investigation. Assessing potential negative environmental impacts is equally important. Understanding and mitigating adverse effects such as pollution are critical for achieving sustainable outcomes and ensuring the responsible use of natural resources.

\item \textbf{Continuous monitoring of the so-called `common good value' has been identified as a suitable approach to validate the investments against the expectations of funders, ultimately the taxpayers}. This approach should be further intensified through surveys conducted in all potential funding countries and regularly repeated to enhance the understanding of how social acceptance can not only be maintained but also be improved.
A framework for streamlining this process has been developed within this project, enabling the activity to continue with only marginal additional resources. Expanding data collection to additional countries beyond those already surveyed can provide insights into the factors influencing the people's perception of the FCC science mission. The findings should be regularly shared with CERN management and the Council and summarised for broader dissemination to all stakeholders, including the public, through channels such as CERN's social media platforms and its main websites. 

\end{enumerate}

\clearpage

\bibliographystyle{sn-aps}
\bibliography{Assembled-FCC-MidTerm}

\addcontentsline{toc}{chapter}{References}

\end{document}